\documentclass{SciPost} 
\usepackage{multirow}
\usepackage[intoc]{nomencl} 
\usepackage{comment}
\usepackage{geometry}
\usepackage{etoolbox}
\usepackage{bm}
\usepackage{mathtools,dsfont}
\usepackage{physics}
\usepackage[most]{tcolorbox}
\usepackage[version=4]{mhchem}
\usepackage{algorithm, algpseudocode}
\usepackage{titlesec}
\usepackage[bitstream-charter]{mathdesign}
\usepackage{acronym}
\usepackage{subcaption} 
\usepackage{makecell} 
\usepackage{multicol} 
\usepackage{blindtext} 
\usepackage{imakeidx} 
\usepackage{hyperref} 
\usepackage[capitalise]{cleveref} 
\makeindex[program=makeindex, columns=2, intoc, options= -s ./addons/indexstyle.ist] 

\ifdefined\CambridgeUP
\fi

\titlespacing*{\section}{0pt}{2em}{0pt} 
\titlespacing*{\subsection}{0pt}{1em}{0pt}
\titlespacing*{\subsubsection}{0pt}{0pt}{0pt}
\titlespacing*{\paragraph}{0pt}{0pt}{8pt}

\usepackage{chngcntr}
\counterwithin{figure}{section}
\counterwithin*{footnote}{section}
\counterwithin{equation}{section}

\ifdefined\CambridgeUP
    \hypersetup{
        colorlinks,
        linkcolor={black},
        citecolor={black},
        urlcolor={black},
    }
\else
    \hypersetup{
        colorlinks,
        linkcolor={red!50!black},
        citecolor={blue!50!black},
        urlcolor={blue!80!black},
    }
\fi

\DeclareSymbolFont{usualmathcal}{OMS}{cmsy}{m}{n}
\DeclareSymbolFontAlphabet{\mathcal}{usualmathcal}

\newcommand{\estimate}[2]{\left\langle #1 \right\rangle_{#2}} 
\newcommand{\estimateE}{\mathbb{E}} 
\DeclareMathOperator{\estimationoperator}{\mathds{E}}
\newcommand*{\lossfun}{\mathcal{L}} 
\newcommand*{\class}{K} 
\newcommand*{\featnum}{m} 
\renewcommand{\i}{\ensuremath\mathrm{i}} 
\renewcommand{\Re}{\operatorname{Re}} 
\newcommand{\e}{\ensuremath\mathrm{e}} 
\renewcommand{\H}{\mathcal{H}} 
\newcommand{\spin}{\sigma} 
\newcommand{\lspace}{L^2} 
\newcommand{\Lagrange}{\mathcal{L}}
\newcommand{\id}{\mathds{1}} 
\DeclareMathOperator{\diag}{diag} 
\DeclareMathOperator*{\argmin}{arg\ min}
\DeclareMathOperator*{\argmax}{arg\ max}
\newcommand{\weight}{w} 
\newcommand{\param}{\theta} 
\newcommand{\bias}{b} 
\newcommand{\weightsvect}{\vect{\weight}} 
\newcommand{\weightsmat}{\mat{W}} 
\newcommand{\params}{\boldsymbol{\param}} 
\newcommand{\biases}{\boldsymbol{\bias}} 
\newcommand{\featmap}{\phi} 
\newcommand{\nparams}{d} 
\newcommand{\datasize}{n} 
\newcommand{\dataset}{\mathcal{D}} 
\newcommand{\normdist}{\mathcal{N}} 
\newcommand{\realset}{\mathds{R}} 
\newcommand{\domain}{\mathds{D}} 
\newcommand{\bigO}{\mathcal{O}} 
\newcommand{\loglik}{\ell} 
\newcommand{\learningrate}{\eta}

\newcommand{\transpose}{\intercal}
\newcommand{\regularization}[1]{\ell_{#1}}

\newcommand{\vect}[1]{\boldsymbol{#1}} 
\newcommand{\mat}[1]{\boldsymbol{#1}} 
\newcommand{\tensor}[1]{\mathsf{#1}} 
\newcommand{\stress}[1]{\textit{#1}}

\newcommand{\actionspace}{\mathcal{A}}
\newcommand{\statespace}{\mathcal{S}}

\newcommand{\highlight}[1]{\begin{tcolorbox}[colback=gray!20!white,colframe=gray!75!black] #1 \end{tcolorbox}} 

\newcommand\T{\rule{0pt}{2.6ex}}       
\newcommand\B{\rule[-1.2ex]{0pt}{0pt}} 

\crefname{section}{chapter}{chapters}
\crefname{subsection}{section}{sections}
\crefname{subsubsection}{section}{sections}

\newcommand{\ToggleForCUP}[2]{%
\ifdefined\CambridgeUP
    #1 
\else
    #2 
\fi
}

\begin{document}

\begin{center}{\ToggleForCUP{\LARGE \textbf{Machine learning in quantum sciences}}{\Large \textbf{Modern applications of machine learning in quantum sciences}}
}\end{center}

\begin{center}

Anna Dawid\textsuperscript{1,2,3$\star$},
Julian Arnold\textsuperscript{4$\dag$},
Borja Requena\textsuperscript{2$\dag$},
Alexander Gresch\textsuperscript{5,6$\dag$},
Marcin P\l{}odzie\'n\textsuperscript{2},
Kaelan Donatella\textsuperscript{7},
Kim A. Nicoli\textsuperscript{8,9},
Paolo Stornati\textsuperscript{2},
Rouven Koch\textsuperscript{10},
Miriam Büttner\textsuperscript{11},
Robert~Okuła\textsuperscript{12,13},
Gorka Muñoz--Gil\textsuperscript{14},
Rodrigo A. Vargas--Hernández\textsuperscript{15,16,17},
Alba Cervera-Lierta\textsuperscript{18},
Juan~Carrasquilla\textsuperscript{16},
Vedran Dunjko\textsuperscript{19},
Marylou Gabrié\textsuperscript{20},
Patrick~Huembeli\textsuperscript{21,22},
Evert van Nieuwenburg\textsuperscript{19,23},
Filippo~Vicentini\textsuperscript{21,24},
Lei~Wang\textsuperscript{25,26},
Sebastian J. Wetzel\textsuperscript{27},
Giuseppe Carleo\textsuperscript{21},
Eliška~Greplová\textsuperscript{28},
Roman Krems\textsuperscript{29},
Florian Marquardt\textsuperscript{30,31},
Michał Tomza\textsuperscript{1},
Maciej~Lewenstein\textsuperscript{2,32} and
Alexandre Dauphin\textsuperscript{2,33$\star$}
\end{center}
\ToggleForCUP{}{\vspace{-0.6cm}}
\begin{center}
{\footnotesize
{\bf 1} Faculty of Physics, University of Warsaw, Poland \\
{\bf 2} ICFO - Institut de Ciències Fotòniques, The Barcelona Institute of Science and Technology, \\08860 Castelldefels (Barcelona), Spain
\\
{\bf 3} Center for Computational Quantum Physics, Flatiron Institute, New York, USA \\
{\bf 4} Department of Physics, University of Basel, Switzerland 
\\
{\bf 5} Institute for Theoretical Physics, Heinrich Heine University Düsseldorf, Germany
\\
{\bf 6} Institute for Quantum Inspired and Quantum Optimization, Hamburg University of Technology, Germany
\\
{\bf 7} Université de Paris, CNRS, Laboratoire Matériaux et Phénomènes Quantiques, France
\\
{\bf 8} Machine Learning Group, Technische Universität Berlin, Germany
\\
{\bf 9} BIFOLD, Berlin Institute for the Foundations of Learning and Data, 10587 Berlin, Germany
\\
{\bf 10} Department of Applied Physics, Aalto University, Espoo, Finland
\\
{\bf 11} Institute of Physics, Albert-Ludwig University of Freiburg, Germany
\\
{\bf 12} International Centre for Theory of Quantum Technologies, University of Gdańsk, Poland
\\
{\bf 13} Department of Algorithms and System Modeling, Faculty of Electronics, Faculty of Electronics, Telecommunications and Informatics, Gdańsk University of Technology, Poland
\\
{\bf 14} Institute for Theoretical Physics, University of Innsbruck, Austria
\\
{\bf 15} Department of Chemistry, University of Toronto, Canada
\\
{\bf 16} Vector Institute for Artificial Intelligence, MaRS Centre, Toronto, Canada
\\
{\bf 17} Department of Chemistry and Chemical Biology, McMaster University, Hamilton, Canada
\\
{\bf 18} Barcelona Supercomputing Center, Spain
\\
{\bf 19} LIACS, Leiden University, The Netherlands
\\
{\bf 20} CMAP, École Polytechnique, France
\\
{\bf 21} Institute of Physics, École Polytechnique Fédérale de Lausanne (EPFL), Switzerland
\\
{\bf 22} Menten AI, Inc., Palo Alto, California, United States of America 
\\
{\bf 23} Niels Bohr Institute, Copenhagen, Denmark
\\
{\bf 24} CPHT, CNRS, École Polytechnique, Institut Polytechnique de Paris, F-91128 Palaiseau, France
\\
{\bf 25} Beijing National Lab for Condensed Matter Physics and Institute of Physics, \\Chinese Academy of Sciences, Beijing, China
\\
{\bf 26} Songshan Lake Materials Laboratory, Dongguan, China
\\
{\bf 27} Perimeter Institute for Theoretical Physics, Waterloo, Canada
\\
{\bf 28} Kavli Institute of Nanoscience, Delft University of Technology, NL-2600 GA Delft, The Netherlands
\\
{\bf 29} Department of Chemistry, University of British Columbia, Vancouver, Canada
\\
{\bf 30} Max Planck Institute for the Science of Light, Erlangen, Germany 
\\
{\bf 31} Department of Physics, Friedrich-Alexander Universität Erlangen-Nürnberg, Germany 
\\
{\bf 32} ICREA, Pg. Llu\'{\i}s Companys 23, 08010 Barcelona, Spain
\\
{\bf 33} PASQAL SAS, 7 Rue Leonard de Vinci, 91300 Massy, France
\\
}
\vspace{0.25cm}
${}^\dag$ {\small These authors contributed equally.}\\
\vspace{0.25cm}
${}^\star$ {\small \sf \href{mailto:a.m.dawid@leidenuniv.nl}{a.m.dawid@leidenuniv.nl}, \href{mailto:Alexandre.Dauphin@pasqal.com}{Alexandre.Dauphin@pasqal.com}
}
\end{center}

\vspace{-0.65cm}
\begin{center}
\today
\vspace{-1cm}
\end{center}

\newpage
\vspace*{8cm}
\begin{center}
\large \emph{In memory of Peter Wittek}
\end{center}
\renewcommand{\nompreamble}{\begin{multicols}{2}}
\renewcommand{\nompostamble}{\end{multicols}}
\makenomenclature

\renewcommand\nomgroup[1]{%
  \item[\bfseries
  \ifstrequal{#1}{A}{Numbers and arrays}{%
  \ifstrequal{#1}{B}{Physical constants and quantities}{%
  \ifstrequal{#1}{D}{Machine learning quantities}{%
  \ifstrequal{#1}{E}{Acronyms}{}}}}%
]}

\mbox{}

\nomenclature[A]{$A$}{random variable}
\nomenclature[A]{$a$}{scalar}
\nomenclature[A]{$\vect{a}$}{vector}
\nomenclature[A]{$\mat{A}$}{matrix}
\nomenclature[A]{$\tensor{A}$}{tensor}

\nomenclature[B]{$k_{\rm B}$}{Boltzmann constant}
\nomenclature[B]{$\estimate{x}{p}$ or $\estimationoperator[x \mid p]$}{Estimator of quantity $x$ with respect to distribution $p$}
\nomenclature[B]{$\hat{H}$}{quantum Hamiltonian}
\nomenclature[B]{$\beta$}{$1/{k_{\rm B} T}$}
\nomenclature[B]{$Z$}{partition function}
\nomenclature[B]{$\spin$}{spin variable}
\nomenclature[B]{$\hat{\sigma}$}{Pauli matrix}
\nomenclature[B]{$m$}{magnetization}
\nomenclature[B]{$\delta$}{Kronecker delta}
\nomenclature[B]{$\mathcal{H}$}{Hilbert space}
\nomenclature[B]{$H$}{classical Hamiltonian}

\nomenclature[D]{$\varsigma$}{activation function}
\nomenclature[D]{$\lossfun$}{loss (or cost/error) function}
\nomenclature[D]{$\class_i$ or $\class$}{$i$-th class or number of classes in a classification problem}
\nomenclature[D]{$\weightsvect$ or $\weightsmat$}{vector or matrix of model weights}
\nomenclature[D]{$\params$}{model parameters}
\nomenclature[D]{$\biases$}{model biases}
\nomenclature[D]{$\mat{H}$}{Hessian matrix}
\nomenclature[D]{$\params^{*}$}{converged $\params$}
\nomenclature[D]{$\hat{f}$}{model with converged $\params$}
\nomenclature[D]{$\featmap$}{feature map}
\nomenclature[D]{$\dataset$}{data set}
\nomenclature[D]{$D_{\mathrm{KL}}$}{Kullback-Leibler divergence}
\nomenclature[D]{$\eta$}{learning rate} 
\nomenclature[D]{$\pi$}{policy}
\nomenclature[D]{$\pi^*$}{optimal policy}
\nomenclature[D]{$G$}{return}
\nomenclature[D]{$r$}{reward}
\nomenclature[D]{$s$}{state}
\nomenclature[D]{$a$}{action}
\nomenclature[D]{$\regularization{n}$}{L($n$) regularization}
\nomenclature[D]{$\datasize$}{size of $\dataset$, i.e., number of training examples}
\nomenclature[D]{$\nparams$}{size of $\params$, i.e., number of model parameters}
\nomenclature[D]{$\featnum$}{dimensionality of data point $\vect{x}$, i.e., number of data features}
\nomenclature[D]{$a$}{action}

\newpage
\vspace{10pt}
\tableofcontents\thispagestyle{fancy}
\vspace{10pt}

\newpage
\section*{Preface}\label{sec:preface}
\addcontentsline{toc}{section}{Preface}
\sectionmark{PREFACE}
\begin{figure}[b!]
    \begin{center}
    \includegraphics[width=0.7\columnwidth]{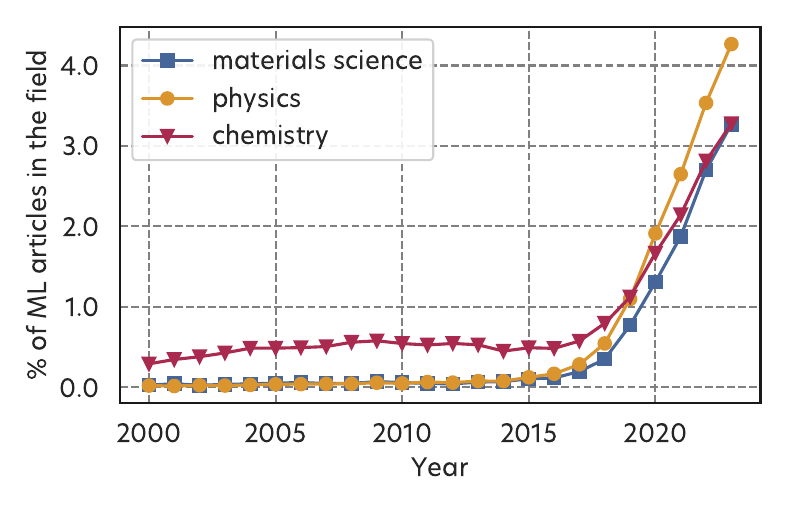}
    \end{center}
    \vspace{-0.5cm}
    \caption[Machine learning in science is booming]{The number of \ac{ML}-based publications in physics, materials science, and chemistry is growing exponentially. Adapted from \ToggleForCUP{Blaiszik, B. (2023). \textit{2022 AI/ML publication statistics and charts}. 10.5281/zenodo.7713954~\cite{blaiszikGitHub} under the MIT License.}{Ref.~\cite{blaiszikGitHub}.}}
    \label{fig:booming_ML}
\end{figure}

We live in fascinating times where scientists are starting to incorporate \ac{AI} algorithms for knowledge discovery. Advances in this booming field have led to a~rapid increase in the interest and confidence of the scientific community in these methods. This trend can be observed by tracking the percentage of \ac{ML}-based publications in physics, chemistry, and material science, shown in~\cref{fig:booming_ML}. As the number of \ac{ML} applications grows, keeping track of all advances becomes challenging. Moreover, it is difficult to find reliable intermediate-level learning material that allows one to efficiently bridge the gap between the rapidly developing field of \ac{ML} and scientists interested in incorporating \ac{ML} tools into their own research.

The idea of creating this book was born out of {\it Summer School: Machine Learning in Quantum Physics and Chemistry}, which took place between Aug 23 - Sept 03, 2021, in Warsaw, Poland. As such, its aim is to give an~educational and self-contained overview of modern applications of \ac{ML} in quantum sciences. The scientific content of this work is inspired by the topics covered by the lecturers and invited speakers of the school. We invite the reader to take a~look at the school tutorials in Ref.~\cite{OurSchoolRepo} and to reuse the figures prepared for this book, which are available in Ref.~\cite{OurFigsRepo}.

The target readership of this book is quantum scientists who want to familiarize themselves with \ac{ML} methods. Therefore, we assume a~basic knowledge of linear algebra, probability theory, and quantum information theory. We also expect familiarity with concepts such as Lagrange multipliers, Hilbert space, and Monte Carlo methods. We also assume that the reader is familiar with quantum mechanics and has a~basic grasp of the current challenges in quantum sciences.

\begin{figure}[t]
    \begin{center}
    \includegraphics[width=0.9\columnwidth]{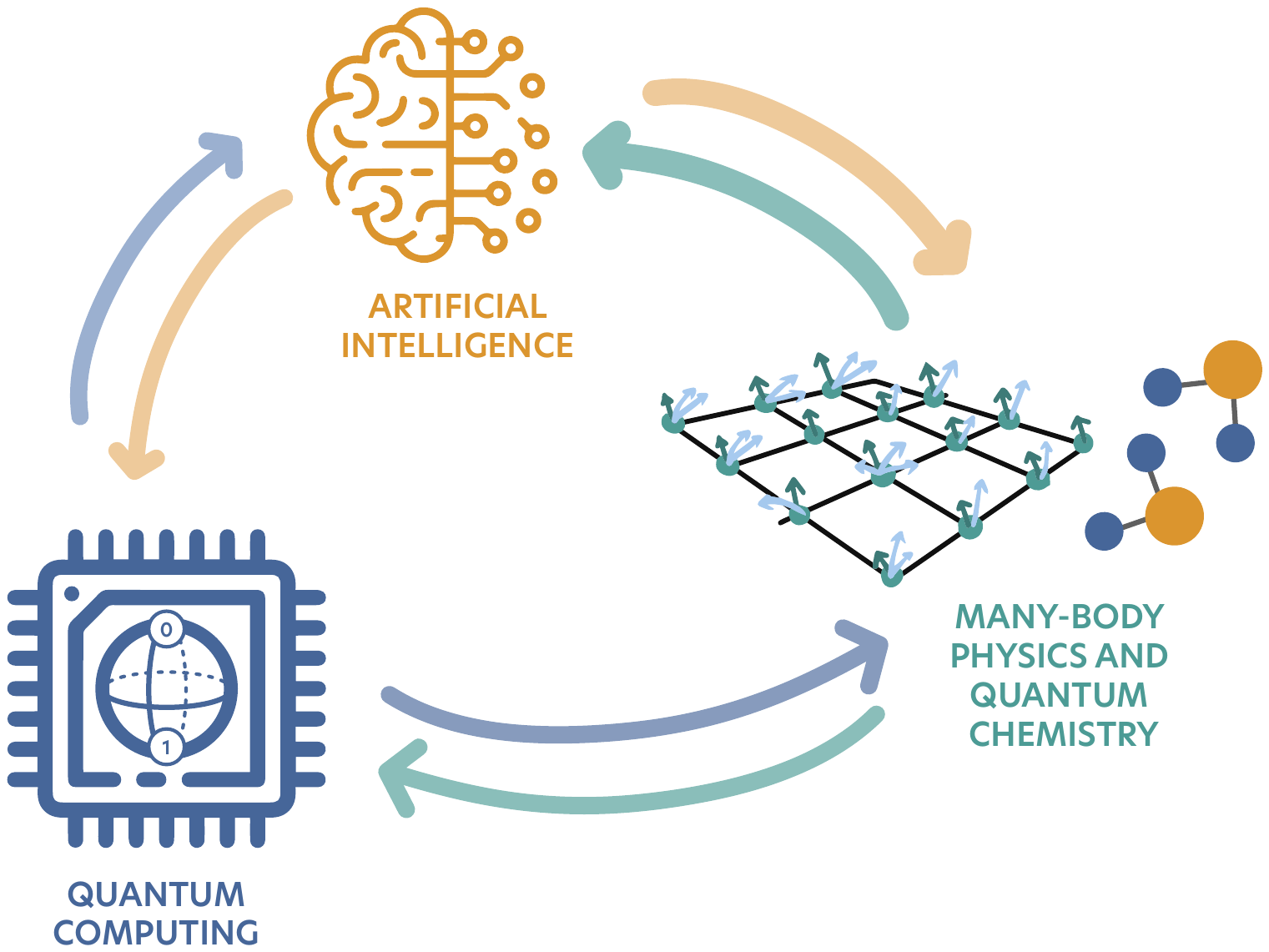}
    \end{center}
    \caption[Interplay of artificial intelligence, quantum computing, and physics]{Interplay of \ac{AI} and quantum sciences, in particular quantum computing, many-body physics, and quantum chemistry. Within this book, we focus not only on the influence of \ac{AI} on quantum sciences but also cover the reverse impact of statistical physics and quantum computing on \ac{ML}.}
    \label{fig:interplay_AI_QC_CMD}
\end{figure}

Our book is roughly divided into three parts. The first part is devoted to establishing a solid foundation of basic \acf{ML} concepts needed for understanding its applications in natural sciences. In the second part, we dive into four core application areas of \ac{ML} in quantum sciences. This covers the use of deep learning and kernel methods in supervised, unsupervised, and reinforcement learning algorithms for phase classification, representation of many-body quantum states, quantum feedback control, and quantum circuit optimization. In the third part, we introduce and discuss more specialized topics such as differentiable programming, generative models, statistical physics approaches to machine learning, and quantum machine learning. All in all, this book discusses the fruitful interplay of \ac{AI} and quantum sciences, presented schematically in~\cref{fig:interplay_AI_QC_CMD}.

We do not aim at providing an~exhaustive list of \ac{ML} applications in quantum sciences and becoming a~complete review of the field. Such reviews already exist and nicely summarize the latest achievements \cite{Dunjko2018,Carleo2019RevModPhys, Carrasquilla2020AdvPhys}. Instead, our objective is to provide the reader with enough knowledge, intuition, and tricks of the trade to start implementing \ac{ML} methods of choice in their own research. As such, we selected the \ac{ML} applications presented in this work that, we believe, are pedagogically appealing while keeping a~broad overview of the field. To this end, we focus on what a~reader could do and not only on what has been done. To fulfill this ambition, we conclude each chapter with an~outlook and open problems that we recognize as important and promising.

\ifdefined\CambridgeUP

        \section*{Acknowledgments}\label{sec:acknos}
        \sectionmark{ACKNOWLEDGMENTS}
        \addcontentsline{toc}{section}{Acknowledgments}
        We thank Hans J. Briegel, Lorenzo Cardarelli, Kacper Cybiński, and Mario Krenn for useful discussions and Fesido Studio Graficzne for the graphical design of the book.

\paragraph{Author contributions}
This manuscript is a~result of a~unique collaboration born between the participants and lecturers of \stress{Summer School: Machine Learning in Quantum Physics and Chemistry} which took place in Warsaw, Poland in August-September 2021 and which was organized by M. Tomza, A. Dauphin, A. Dawid, and M. Lewenstein. All authors of the manuscript participated in the reading and improvement of its content. In particular:
\begin{itemize}
    \item \stress{``Introduction''} was written by A.~Dawid with the help of M.~Płodzień, M.~Lewenstein, A.~Gresch, R.~Koch, B. Requena, and G. Muñoz--Gil.
    \item \stress{``Basics of machine learning''} was written by A.~Dawid, A.~Gresch, and J.~Arnold with the help of K.~Nicoli, K.~Donatella, and M.~Płodzień.
    \item \stress{``Phase classification''} was written by J.~Arnold, A.~Dawid, A.~Gresch, R.~Koch, M.~Płodzień, and S.~Wetzel based on the scientific content provided by E.~Greplová, P.~Huembeli, and S.~Wetzel.
    \item \stress{``Gaussian processes and other kernel methods''} was written by A.~Gresch, A.~Dawid, K.~Nicoli, J.~Arnold, R.~Krems, and R.~A.~Vargas-Hernández based on the scientific content provided by R.~Krems.
    \item \stress{``Neural-network quantum states''} was written by K.~Donatella, B.~Requena, and P.~Stornati with the help of R.~Okuła and M.~Płodzień based on the scientific content provided by G.~Carleo, J.~Carrasquilla, and F.~Vicentini. 
    \item \stress{``Reinforcement learning''} was written by B.~Requena, M.~Płodzień, and A.~Gresch with the help of R.~Okuła and G. Muñoz--Gil based on the scientific content provided by V.~Dunjko, F.~Marquardt, and E.~van Nieuwenburg.
    \item \stress{``Differentiable programming''} was written by J.~Arnold, \stress{``Generative models in many-body physics''} - by K.~A.~Nicoli and M.~Gabrié with the help of M.~Płodzień, K.~Donatella, and A.~Dawid, \stress{``Machine learning for experiments''} - by M.~Büttner, R.~Koch, and A.~Dawid, based on the scientific content provided by J.~Carrasquilla, E.~Greplová, and L.~Wang.
    \item \stress{``Statistical physics for machine learning''} was written by A.~Dawid with the help of M.~Płodzień based on the scientific content of M.~Gabrié, and \stress{``Quantum machine learning''} was written by P.~Stornati, A.~Dauphin, M.~Płodzień, and R.~Koch with the help of R.~Okuła , based on the lectures of A.~Cervera-Lierta and V.~Dunjko.
    \item Finally, \stress{``Conclusions''} were written by A.~Dauphin, B.~Requena, M.~Płodzień, and A.~Dawid with the help of all the co-authors.
\end{itemize}
The project was led by A.~Dawid and supervised by A.~Dauphin with the help of M.~Lewenstein and M.~Tomza.

\small{\paragraph{Funding information}
An.D. acknowledges the financial support from the National Science Centre, Poland, within the Preludium grant No. 2019/33/N/\-ST2/03123 and the Etiuda grant No. 2020/36/T/ST2/00588 as well as from the Foundation for Polish Science. The Flatiron Institute is a division of the Simons Foundation.
J.A. acknowledges financial support from the Swiss National Science Foundation individual grant (grant no. 200020 200481). 
A.G. acknowledges financial support from the Deutsche Forschungsgemeinschaft (DFG, German Research Foundation) - project number 441423094.
M.P. acknowledges the support of the Polish National Agency for Academic Exchange, the Bekker programme no: PPN/BEK/2020/1/00317.
K.A.N. acknowledges support by the Federal Ministry of Education and Research (BMBF) for the Berlin Institute for the Foundations of Learning and Data (BIFOLD) (01IS180\-37A).
R.K acknowledges financial support from the Academy of Finland Projects No. 331342 and No. 336243.
G.M-G. acknowledges support from the Austrian Science Fund (FWF) through SFB BeyondC F7102.
A.C-L. acknowledges the support by the Ministry of Economic Affairs and Digital Transformation of the Spanish Government through the QUANTUM ENIA project call - QUANTUM SPAIN project, and by the European Union through the Recovery, Transformation and Resilience Plan - NextGenerationEU within the framework of the Digital Spain 2025 Agenda.
M.G. acknowledges funding as an~Hi!Paris Chair Holder.
L.W. is supported by the Strategic Priority Research Program of the Chinese Academy of Sciences under Grant No. XDB30000000 and National Natural Science Foundation of China under Grant No. T2121001.
M.T. acknowledges the financial support from the Foundation for Polish Science within the First Team programme co-financed by the EU Regional Development Fund.
Al.D. acknowledges the financial support from a~fellowship granted by la Caixa Foundation (ID 100010434, fellowship code LCF/BQ/PR20/11770012).
This project has received funding from the European Union’s Horizon 2020 research and innovation program under the Marie Sklodowksa-Curie grant agreement No. 895439 `ConQuER'.

ICFO group acknowledges support from: ERC AdG NOQIA; MICIN/AEI (PGC2018-0910.13039/501100011033, CEX2019-000910-S/10.13039/501100011033, Plan National FIDEUA PID2019-106901GB-I00, FPI; MICIIN with funding from European Union NextGenerationEU (PRTR-C17.I1): QUANTERA MAQS PCI2019-111828-2); MCIN/AEI/10.13039/501100011033 and by the “European Union NextGeneration EU/PRTR"  QUANTERA DYNAMITE PCI2022-132919 (QuantERA II Programme co-funded by European Union’s Horizon 2020 programme under Grant Agreement No 101017733), Ministry of Economic Affairs and Digital Transformation of the Spanish Government through the QUANTUM ENIA project call – Quantum Spain project, and by the European Union through the Recovery, Transformation and Resilience Plan – NextGenerationEU within the framework of the Digital Spain 2026 Agenda.Fundació Cellex; Fundació Mir-Puig; Generalitat de Catalunya (European Social Fund FEDER and CERCA program, AGAUR Grant No. 2021 SGR 01452, QuantumCAT \ U16-011424, co-funded by ERDF Operational Program of Catalonia 2014-2020); Barcelona Supercomputing Center MareNostrum (FI-2023-1-0013); EU Quantum Flagship (PASQuan\-S2.1, 101113690); EU Horizon 2020 FET-OPEN OPTOlogic (Grant No 899794); EU Horizon Europe Program (Grant Agreement 101080086 — NeQST), National Science Centre, Poland (Symfonia Grant No. 2016/20/W/ST4/00314); ICFO Internal “QuantumGaudi” project; “La Caixa” Junior Leaders fellowships ID100010434: LCF/BQ/PI19/11690013, LCF/BQ/PI20/11760031,  LCF/BQ/PR20/11770012, LCF/\-BQ/PR21/11840013. Views and opinions expressed are, however, those of the author(s) only and do not necessarily reflect those of the European Union, European Commission, European Climate, Infrastructure and Environment Executive Agency (CINEA), nor any other granting authority.  Neither the European Union nor any granting authority can be held responsible for them.

Research at Perimeter Institute is supported in part by the Government of Canada through the Department of Innovation, Science and Economic Development Canada and by the Province of Ontario through the Ministry of Economic Development, Job Creation and Trade. We thank the National Research Council of Canada for their partnership with Perimeter on the PIQuIL.}
 
	\section*{List of acronyms}\label{sec:acronyms}
	\addcontentsline{toc}{section}{List of acronyms}
	\sectionmark{LIST OF ACRONYMS}
	\begin{multicols}{2}

\begin{acronym}

\acro{AD}{automatic differentiation}
\acro{AE}{autoencoder}
\acro{AI}{artificial intelligence}
\acro{ANN}{artificial neural network}
\acro{AR}{autoregressive}
\acro{ARNN}{autoregressive neural network}
\acro{BIC}{Bayesian information criterion}
\acro{BO}{Bayesian optimization}
\acro{CPU}{central processing unit}
\acroplural{CPU}[CPUs]{central processing units}
\acro{CE}{cross-entropy}
\acroplural{CE}[CEs]{cross-entropies}
\acro{CNN}{convolutional neural network}
\acro{DiffP}[$\partial$P]{differentiable programming}
\acro{DL}{deep learning}
\acro{DNN}{deep neural network}
\acro{DQN}{deep Q-network}
\acro{ECM}{episodic and compositional memory}
\acroplural{ECM}[ECMs]{episodic and compositional memories}
\acro{EI}{Expected Improvement}
\acro{GAMP}{generalized approximate message passing}
\acro{GAN}{generative adversarial network}
\acro{GNS}{generative neural sampler}
\acro{GP}{Gaussian process} 
\acroplural{GP}[GPs]{Gaussian processes}
\acro{GPR}{Gaussian process regression}
\acro{GPU}{graphics processing unit}
\acro{IGT}{Ising gauge theory}
\acro{KRR}{kernel ridge regression}
\acro{KL}{Kullback-Leibler}
\acro{L-BFGS}{limited-memory Broyden–Fletch\-er–Goldfarb–Shanno algorithm}
\acro{LASSO}{least absolute shrinkage and selection operator}
\acro{LE}{local ensemble}
\acro{MAE}{mean absolute error}
\acro{MAP}{maximum a posteriori estimator}
\acro{MCMC}{Markov chain Monte Carlo}
\acro{MDP}{Markov decision process} 
\acroplural{MDP}[MDPs]{Markov decision processes}
\acro{ML}{machine learning}
\acro{MLE}{maximum likelihood estimation}
\acro{MPS}{matrix product state}
\acro{MSE}{mean-squared error}
\acro{NEI}{Noisy Expected Improvement}
\acro{NF}{normalizing flow}
\acro{NIS}{neural importance sampling}
\acro{NISQ}{noisy intermediate-scale quantum}
\acro{NMCMC}{neural Markov chain Monte Carlo}
\acro{NN}{neural network}
\acro{NQS}{neural quantum state}

\acro{ODE}{ordinary differential equation}
\acroplural{ODE}[ODEs]{ordinary differential equations}
\acro{PC}{principal component}
\acro{PCA}{principal component analysis}
\acro{PES}{potential energy surface}
\acro{PI}{Probability of Improvement}
\acro{POVM}{positive operator-valued measure}
\acro{PPT}{positive under partial transposition}
\acro{PQC}{parametrized quantum circuit}
\acro{PS}{projective simulation}
\acro{QAOA}{quantum approximate optimization algorithm}
\acro{QD}{quantum dot}
\acroplural{QD}[QDs]{quantum dots}
\acro{QML}{quantum machine learning}
\acro{RBM}{restricted Boltzmann machine}
\acro{RKHS}{reproducing kernel Hilbert space}
\acroplural{RKHS}[RKHS']{reproducing kernel Hilbert spaces}
\acro{RL}{reinforcement learning}
\acro{RNN}{recurrent neural network}
\acro{RUE}{resampling uncertainty estimation}
\acro{SGD}{stochastic gradient descent}
\acro{SE}{state evolution}
\acro{SVM}{support vector machine}
\acro{t-SNE}{t-distributed stochastic neighbour embedding}
\acro{t-VMC}{time-dependent variational Monte-Carlo}
\acro{TD}{temporal-difference}
\acro{TN}{tensor network}
\acro{TNS}{tensor network state}
\acro{VAE}{variational autoencoder}
\acro{VQE}{variational quantum eigensolver}
\acroplural{VQE}[VQEs]{Variational Quantum Eigensolvers}
\end{acronym}

\end{multicols}

	\newpage
	\printnomenclature
\fi

\clearpage
\newpage

\section{Introduction}\label{sec:intro}

Making \stress{intelligent} machines, i.e., machines capable of learning and utilizing the gathered knowledge in thinking and reasoning, is a~long-lived dream of human civilization. The more we know about the human brain, intelligence, and psychology, the more challenging it seems. However, despite the many obstacles and challenges in creating \acf{AI}, the joint effort of researchers working in the natural, cognitive, mathematical, and computer sciences has produced impressive machinery that is already revolutionizing our daily life, industry, and science.

\subsection{How do computers learn?}

The ultimate goal of \ac{AI} is to endow machines with the ability to conceptualize and create abstractions. Both of these features are mechanisms that underlie learning representations of knowledge and reasoning based on experience in humans. We have multiple ways of representing ideas. For example, we can encode a~piece of music in digital format on a~computer, in analog format on a~vinyl disc, or we can write it down in a~music score. Although the representations are entirely different, the piece of music is the same. Therefore, the properties of abstract ideas do not depend on the data source.

Furthermore, conceptualization and abstraction bring the possibility of considering various levels of details within a~particular representation or the ability to switch from one level to another while preserving the relevant information~\cite{Williams1993,Dearden1997,Zucker2003, Zucker2013, Mitchell2021}. Our brain excels at extracting abstract ideas from different representations of knowledge. In our daily lives, we constantly process information from multiple sources that represent the same concept in completely different ways. For example, we can identify the concept of a~dog by seeing one, hearing or smelling it, reading the word ``dog'', painting a~snout on someone's face, or even casting shadows with our hands that resemble the shade of a~dog. This level of abstraction and conceptualization enables us to reason, connecting high-level ideas. All the properties of our brain mentioned above form what we call intelligence. Conferring these properties to a~computer would result in a~general problem-solving machine.

Today, we are at a~point in our technological advances at which the human brain and computers have a~disjoint set of tasks in which they naturally excel.\footnote{This observation was first made in the 1980s, and it is called Moravec's paradox. As Moravec wrote in 1988~\cite{Moravec1988}, ``it is comparatively easy to make computers exhibit adult level performance on intelligence tests or playing checkers, and difficult or impossible to give them the skills of a~one-year-old when it comes to perception and mobility''.}
Some tasks are easy for computers but difficult for humans. These are problems that can be described by a~list of formal, mathematical rules. Therefore, computers excel at solving logic, algebra, geometry, and optimization problems, which we can tackle with hard-coded solutions or knowledge-based \ac{AI}.
However, we would like to tackle problems that are not easy to present in a~formal mathematical way, such as face recognition, or whose exact mathematical formulation is not yet known, such as detecting new quantum phases.

\begin{figure}
    \centering
    \includegraphics[width=0.99\columnwidth]{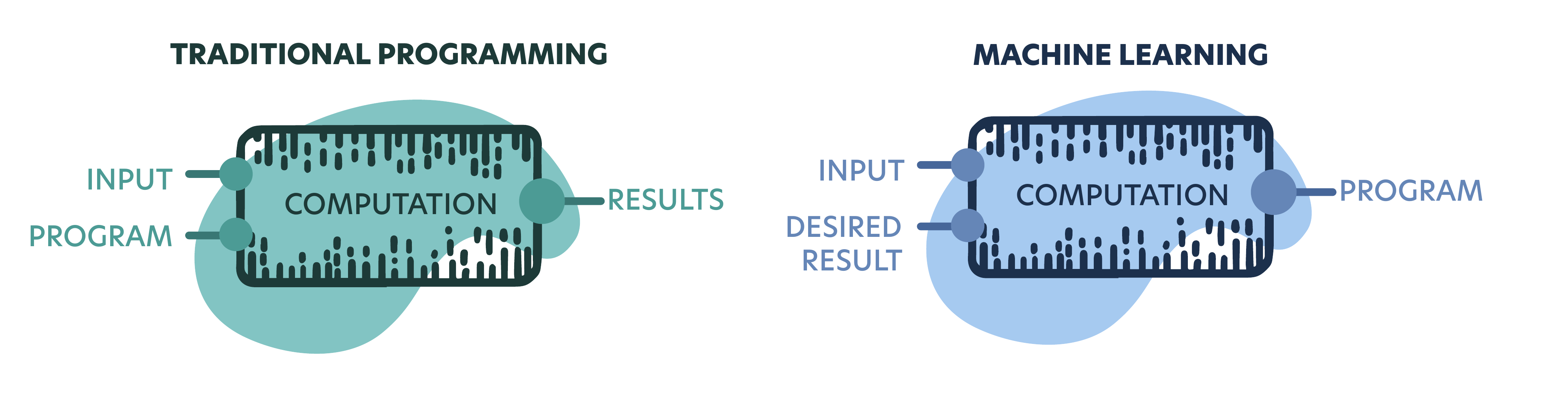}
    \caption[Traditional programming vs. machine learning based on data-driven programming]{Schematic representation of the difference between the traditional programming, based on the algorithmic approach, and the experience-based/data-driven approach, which is the backbone of the \ac{ML} paradigm. The \ac{ML} paradigm is the first step toward learning abstractions by computers through the extraction of common features from data.}
    \label{fig:intro_ml-vs-algorithm}
\end{figure}

\begin{figure}
    \begin{center}
    \includegraphics[width=\columnwidth]{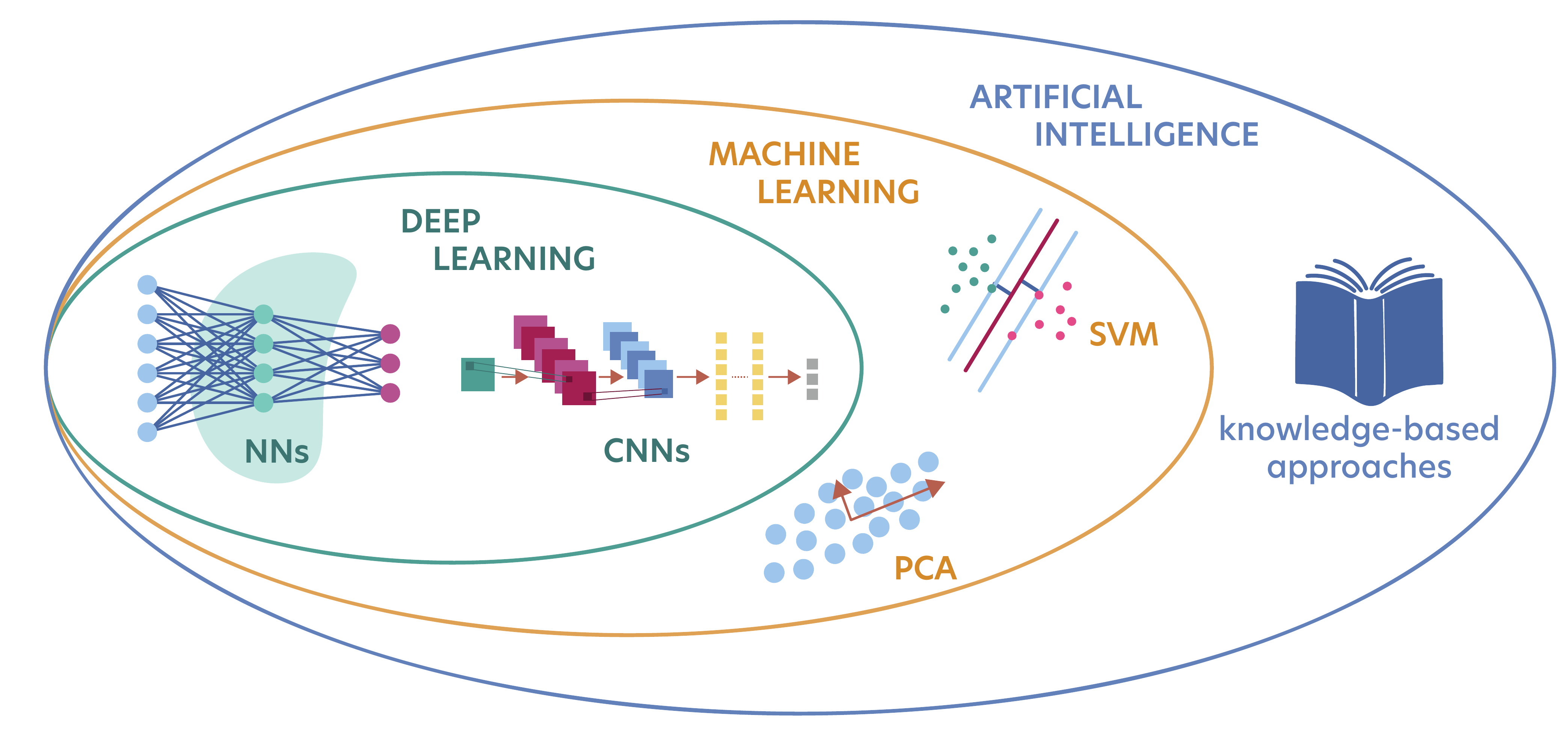}
    \end{center}
    \caption[Artificial intelligence vs. machine learning vs. deep learning]{Sketch of the relation between \ac{AI}, \ac{ML}, and \ac{DL} with examples from each field including \acfp{SVM}, \acfp{PCA}, \acfp{NN}, and \acfp{CNN}.}
    \label{fig:intro_DL_vs_ML}
\end{figure}

A~particularly exciting direction is the development of algorithms that are not explicitly programmed.
The main principle is to enable computers to learn from experience (or data). The shift toward this data-driven paradigm led to the birth of \stress{\acf{ML}}, schematically depicted in~\cref{fig:intro_ml-vs-algorithm}. This field leverages fundamental concepts of applied statistics, emphasizing the use of computers to estimate complicated functions and with a~decreased emphasis on proving confidence intervals around them~\cite{goodfellow:2016}. This trend has accelerated with the rise of \stress{\ac{DL}}, where enormous and heavily parametrized hierarchical models are used to deal with complex patterns from real-world data and do this with unprecedented accuracy.
Interestingly, many \ac{DL} architectures are designed to mimic some of the properties of the human thinking process, such as understanding correlations in visual patterns or recurrence in sound signals. 
We present a~schematic representation of the relationship between these three fields (\ac{AI}, \ac{ML}, and \ac{DL}) in~\cref{fig:intro_DL_vs_ML}.

\highlight{To make a~computer learn, we need three main ingredients:
\begin{enumerate}
    \item a~\stress{task} to solve (\cref{s:tasks}), 
    \item \stress{data} that can be considered as an~equivalent of \stress{experience}. The latter can be provided in the form of, e.g., an~interacting environment, and allows for solving the task (\cref{sec:typeoflearning}),
    \item a~\stress{model} that learns how to solve the task (\cref{sec:models}).
\end{enumerate}
To check whether a~computer successfully learns how to solve a~task, we need to define a~performance measure, which can be as simple as the comparison between the prediction of the model and the expected answer. In these terms, the learning process can be described as the iterative minimization of the model error or maximization of the model performance on the given task and data.}

\subsection{Historical view on learning machines}\label{s:history_ML}

The foundations of the theory of learning were established already in the 1940s. Its development has followed two parallel paths: a~knowledge-based approach, which dominated the \ac{AI} research field for decades, and a~data-based one, which is currently on the rise. Throughout the years, \ac{ML} has gone under various names (like cybernetics or connectionism) and experienced a~few cycles of intense popularity,\footnote{Some argue that the ``\ac{AI} winter'' is upon us unless we rethink \ac{AI} or combine it with knowledge-based approaches~\cite{Marcus2022DLHitsWall}. It is also important to remember that such hype cycles are frequent with emerging new technologies.} followed by criticism and disappointment, followed by funding cuts, followed by renewed interest years or decades later \cite{goodfellow:2016}. To give the reader some insight into the giants on whose arms we stand, we briefly present milestones in the development of \ac{ML}, following Refs.~\cite{goodfellow:2016, Sejnowski,Lecun2015,Schmidhuber2015}:

\ifdefined\CambridgeUP
\else
    \newpage
\fi
\begin{itemize}
    
\item 1943 -- Walter Pitts and Warren McCulloch create a~computer model inspired by the neural networks of the human brain called the \stress{threshold logic}. Their field of expertise is called \stress{cybernetics}.

\item 1949 -- Donald Hebb hypothesizes how learning in biological systems works and formulates \stress{Hebbian learning}. For example, if certain neurons ``fire together, they wire together''.

\item 1957 -- Frank Rosenblatt 
introduces a~\stress{Rosenblatt perceptron} modeling a~single neuron. A~perceptron is also called ``an artificial neuron'' and, after modifications in 1969 by Marvin Minsky and Seymour Papert, to this day, remains widely used as a~building block of \acp{ANN}.

\item 1962 -- David Hubel and Torsten Wiesel present, for the first time, the response properties of single biological neurons recorded with a~microelectrode.

\item 1969 -- Marvin Minsky and Seymour Papert point out the computational limitations and disadvantages of linear models, including a~single artificial neuron, contributing to the first ``\ac{AI} winter''.

\item 1986 -- David Rumelhart, Geoffrey Hinton, and Ronald Williams use \stress{backpropagation}\index{backpropagation} to train an~\ac{NN} with one or two hidden layers which, next to the revival of Hebb's ideas, causes renewed interest in the field that at this time is called \stress{connectionism}. In the same year, David Rummelhart, James McClelland, \textit{et al.} publish a~widely discussed two-volume book ``Parallel Distributed Processing'' discussing known and collecting original contributions from the field, including backpropagation and Boltzmann machines.

\item the mid-1990s -- second \ac{AI} winter whose appearance is ascribed \cite{goodfellow:2016} to exceedingly ambitious claims of the community, which led to the disappointment of investors, and the simultaneous progress of kernel methods, which require less computational resources.

\end{itemize}

Interestingly, we can see how closely the development of \ac{AI} was intertwined with neuroscience. This makes sense, as the human brain provides proof by example that intelligent behavior is possible. A~natural approach to \ac{AI} would be to try to reverse engineer the brain to reproduce its functionality. However, while the perceptron was inspired by biological neurons and some \ac{ML} models are loosely inspired by neurological discoveries, there is nowadays a~consensus that models should not be designed to be realistic simulators of biological functions~\cite{goodfellow:2016}.\footnote{Interestingly, we know that actual biological neurons compute very different functions than the perceptrons constituting our modern \acp{NN}, but greater realism has not yet led to any improvement in model performance~\cite{goodfellow:2016}.} Instead, scientists attempt to solve the mysteries of the human brain using \ac{ML}.

Since 2006, \ac{DL} has been thriving again thanks to a~breakthrough in the efficient training of deep \acp{NN}~\cite{Hinton2006deepbelief} via backpropagation\index{backpropagation}, followed by multiple analyses confirming the importance of its depth. At the same time, there has been a~rapid improvement in computational power in recent decades, which has allowed the exploration of larger \ac{ML} models. Here, the development of \acp{GPU}\index{graphics processing unit} \cite{Volkov2008GPU, Raina2009GPU} has played a~particularly important role: highly parallelizable algorithms, such as \acp{NN}, which are based on matrix and vector operations, can profit immensely from the parallel architecture of \acp{GPU} enabling them to process large amounts of data more efficiently than \acp{CPU}. Furthermore, we have started to produce and store large amounts of easily accessible electronic data throughout the world \cite{Marr2018, SeedScientific, Statista2022}, enabling data-driven programming approaches. 
Since then, progress in the field has enabled realizations of concepts known, so far, only in science-fiction literature, such as self-driving cars or robots mimicking human emotions on their artificial faces (even if we are still far from human-like intelligence~\cite{dawid2023introduction}). \ac{DL} has dominated the field of computer vision for years and has found great success in time series analysis, with applications such as stock market and weather forecasting~\cite{Dixon}. Another fruitful direction is natural language processing, where sequence-to-sequence models have achieved great feats, even combining text with images~\cite{Eisenstein,OpenAI2022}. The \ac{DL}-based algorithms obtained superhuman performance in video games~\cite{Mnih2015,Vinyals2019alphastar} and complex board games, such as Go~\cite{AlphaGo}.

Overall, the continuous progress in the field of \ac{ML} is supported by the steady increase of computational power and its easy applicability to real-world problems. The increasing amount of data produced by our society and the monetary benefit of its processing have made the largest technological companies focus enormous economic efforts on the development of \ac{ML} models. It is, hence, not a~coincidence that the most important research groups in the field are associated with such companies. Importantly, one should understand the extent to which the trends of the field are dictated by the thirst for scientific discovery or by the particular needs of one or another technological giant. In summary, \ac{ML} has become a~day-by-day tool, acting in the shades of multiple technological tools we use today~\cite{dawid2023introduction}, with the potential to solve some of the most important problems of the modern world and thus contribute to improving the quality of life of people around the world.

\subsection{Learning machines viewed by a~statistical physics}\label{s:mlstatphys}

It is also worth noting that the above-sketched developments of \ac{AI}, data science, cognitive science, and neuroscience, related to \ac{ML} and \ac{NN}, were also intertwined with the development of the statistical physics of spin glasses and \ac{NN}. A~wonderful retrospective of these developments can be found in the lecture of the late Naftali Tishby, \href{https://www.youtube.com/watch?v=BUfnIT92ukM}{``Statistical physics and \ac{ML}: A~30-year perspective''}. Therefore, here we present a~similar list of historical milestones as in \cref{s:history_ML}, but focused on statistical physics achievements:
\begin{itemize}
\item 1975 -- Philipp W. Anderson and Samuel F. Edwards formulate the Ed\-wards-Anderson spin glass model with short-range random interactions between Ising spins.
\item 1975 -- a~little later, David Sherrington and Scott Kirkpatrick formulate the Sherrington-Kirkpatrick spin-glass model with infinite-range interactions, for which the mean-field solution should be exact. They propose to solve it using the replica trick, but this approximate solution turns out to be clearly incorrect at low temperatures. 
\item 1979 -- Giorgio Parisi proposes an~ingenious replica symmetry-breaking solution of the Sherrington-Kirkpatrick model.
\item 1982 -- John J. Hopfield publishes his seminal paper on attractor \acp{NN}, where, by assuming the symmetry of interneuron coupling, he relates the model to a~disordered Ising model of $N$ spins, very much analogous to spin glasses. The maximal storage capacity is found to be $0.14\,N$. 
\item 1985 -- Daniel Amit, Hannoch Gutfreund, and Haim Sompolinski formulate the statistical physics of the Hopfield model and relate limited storage capacity to the spin-glass transition.
\item 1987 --  Marc Mezard, Giorgio Parisi, and Miguel Angel Virasoro publish the book ``Spin glass theory and beyond: An~introduction to the replica method and its applications''. Interestingly, it is one of the first works bringing together statistical physics and \acp{NN} but also putting them in a~more general context of complex systems like optimization and protein folding.

\item 1988 -- Elisabeth Gardner formulates the so-called Gardner’s program to \ac{ML}, where learning abilities are related to the relative volume in the space of those \acp{NN} that realize learning tasks and teacher-student scenarios (see \cref{sss:Marylou-perceptron}).

\item 1989 -- Daniel Amit publishes the book ``Modeling brain function: The world of attractor \acp{NN}'' where he brings closer neurophysiology and artificial \acp{NN} by introducing dynamical patterns whose temporal sequence encodes the information.

\item 1990 -- Géza Györgyi shows that sharp phase transitions from bad to good generalization can occur in learning using Gardner's program on the perceptron.

\item 1995 -- David Saad, Sara Solla, Michael Biehl, and Holm Schwarze adapt Gardner's idea to study the dynamics of gradient descent in perceptrons and simple two-layer \acp{NN} called committee machines\index{committee machine}.

\item late 2010s --  The statistical mechanics predictions for the perceptron and the committee machine start being made mathematically rigorous by Nicolas Macris, Jean Barbier, Lenka Zdeborová, and Florent Krzakala. 

\item 2010s-today -- With the explosion of \acp{DL}, interest in the statistical mechanics approach to learning is rekindled. Analyses are developed for increasingly complex models beginning to bridge the gap from perceptrons to deep \acp{NN}.

\item 2021 -- Giorgio Parisi receives the Nobel Prize in Physics ``for the discovery of the interplay of disorder and fluctuations in physical systems from atomic to planetary scales.''

\item 2024 –- John J. Hopfield and Geoffrey Hinton receive the Nobel Prize in Physics ``for foundational discoveries and inventions that enable machine learning
with artificial neural networks.''
\end{itemize} 

We discuss the intersection of statistical physics and \ac{ML} in more detail in \cref{sec:stat_phys_for_ML}.

\subsection{Examples of tasks}\label{s:tasks}

As stated above, the first ingredient needed for a~computer to learn is the notion of a~\stress{learning task}.
The archetypical \ac{ML} task is the study of a~response variable, $y(x)$, influenced by an~explanatory variable $x$.
In principle, there is no restriction on whether $y$ or $x$ or both are continuous, discrete, or even categorical.\footnote{When the inputs are, for example, words in a~sentence as they are in the field of natural language processing, we can still process them by representing words by a~suitable encoding, which can be either continuous or discrete.}
Throughout the book, we restrict both variables, possibly encoded accordingly, to be of quantitative nature.
That is, we can treat variables straightforwardly from a~numerical perspective and easily adjust them to fit our needs.

\paragraph{Regression.}
We start by considering \stress{regression}\index{regression} tasks.
In this setting, we typically assume an~immediate relationship between the two variables $x$ and $y$, which is often deterministic.
More precisely, we seek to express the variable $\vect{y}$, a.k.a. the output or target, in terms of the variable $\vect{x}$, a.k.a. the input.
In general, both variables can be multidimensional, as indicated by our notation.
The objective of regression is to find the function $f$ that yields the mapping $\vect{y} = f(\vect{x})$ for all possible tuples of $(\vect{x},\vect{y})$.
Of course, from a~practical point of view, we can neither optimize over the set of all possible functions nor over the entire domain of $\vect{x}$.
Instead, we resort to a~finite data set for which we opt to find a~model that maps every input $\vect{x}$ to its corresponding target $\vect{y}$.
Usually, the model is predefined up to some parameters,\footnote{There are also non-parametric approaches, e.g., see \cref{sec:deep_generative_models} and \cref{sss:kernel-search}.} which are tuned to fit the data set.
The most simple model assumes a~linear relationship between the input and the output.
We give more details of this model archetype in \cref{sss:intro_linear_model}.
From here, there is a~multitude of ways to extend the model by incorporating nonlinear dependencies on both the model parameters and the input $\vect{x}$.
We find interesting regression problems in a~large range of study fields, such as sociology (e.g., annual salary as a~function of years of work experience), psychology (e.g., perceived happiness relative to wealth), finance (e.g., housing market prices depending on socioeconomic factors) or, of course, (quantum) physics and chemistry.
We cover some examples in this book, for instance, the prediction of \acp{PES} in quantum chemistry in \cref{sec:BO_GPR_science}, or the estimation of the Hamiltonian's parameters given measurement data in \cref{sec:ML_for_exp}.

\paragraph{Classification.}
Another large class of tasks is \stress{classification}\index{classification}.
In this case, our goal is to use an~algorithm to assign \stress{discrete} class labels to examples. In contrast to regression, we are optimizing a~model to find a~mapping from an~input vector $\vect{x}$ to a~target $\vect{y}$, which encodes a~representation of the different possible classes.
The simplest example of this kind of task is binary classification, in which an~algorithm has to distinguish between two classes, e.g., true or false.
When the task involves more than two classes, we speak of multi-class classification. A~canonical example for such a~task is the classification of the images of handwritten digits contained in the famous MNIST \cite{lecun:1998} data set (named after the Modified National Institute of Standards and Technology) over ten classes, one for each number from zero to nine. Other famous \ac{ML} classification data sets are Iris \cite{Fisher36}, CIFAR-10 and 100 \cite{krizhevsky:2009}, and ImageNet \cite{Russakovsky2015imagenet}.\footnote{The Iris database contains 150 data points with four features\index{feature} of three species of iris. The CIFAR-10 data set consists of 60 000 32x32 color images in 10 classes and was named after the Canadian Institute for Advanced Research. Finally, the ImageNet is a~gigantic project with over 10 million labeled images whose most popular subset spans 1000 object classes.}
A~popular example from physics is the classification of different classical and quantum phases of matter, described in~\cref{sec:phase_class}. Another set of examples is provided by the classification subroutines in the automation of (quantum) experiments highlighted in \cref{sec:ML_for_exp}.

Both regression and classification tasks require a~training data set consisting of examples of inputs $\vect{x}$ together with their corresponding labels $\vect{y}$.
Nonetheless, there are also tasks that do not require explicit labels. An~example of such is \stress{density estimation}, where the aim is to infer the probability density function of the data set.
This is directly related to the field of \stress{generative problems}\index{generative models}, where the goal is to \stress{generate} new data instances that resemble some given input data.
The distinction between the two fields is that the latter does not require explicit knowledge or reconstruction of the underlying data distribution to sample new instances.
We present more details on density estimation in \cref{sec:hot-topics:DE}.

In all the previous cases, we try to infer properties of a~given pre-defined data set. However, there are other tasks that involve starting from scratch and building a~data set on the fly, from which we can then learn. 
A~paradigmatic example of such a~task is learning how to play a~game. In this case, we start tabula rasa and progressively build a~data set with the experience gathered as we play the game. From this data (or during its retrieval), our goal is to learn a~function that chooses the best possible action or move according to the current state of the game. 
In this example, we can periodically alternate between collecting experience and learning, or we can do both at the same time.

This list of tasks is, of course, not exhaustive. Other examples that do not directly fall into the previous categories include text translation, imputation of missing values, anomaly detection, and data denoising, to name a~few.

\subsection{Types of learning}\label{sec:typeoflearning}

The second learning ingredient is \stress{data}, whose accessibility also often determines the type of learning we have to consider.
It is clear, of course, that the notions of task, as presented in the previous section, and data are intertwined: certain tasks can only be solved if sufficient data is available and, in turn, a~richer data set allows to transfer from one task to another with seemingly low effort.
Although the term data is often used for a~variety of concepts across many fields, there is a~precise definition of it in the \ac{ML} community.
We usually refer to \stress{data} in terms of a~data set $\dataset$, containing a~finite amount of data instances often called \stress{data points} $\vect{x}_i$, which may be presented as is, i.e., $\dataset = \{\vect{x}_i\}$ or may be accompanied by predefined labels or targets $\vect{y}_i$, i.e. $\dataset = \{(\vect{x}_i,\vect{y}_i)\}$.
To shorten the notation, we also represent the input data points $\{\vect{x}_i\}$ by a~matrix $\mat{X}$, that can either be stacked row- or column-wise.

Although the notation is clear, there is much less convention and an~even lesser understanding of how the data should be \stress{represented}.
This is because, on one hand, the data can be arbitrarily preprocessed (for example, the data mean is often subtracted prior to any further analysis), which already provides some degree of freedom.
On the other hand, even choosing the right descriptors to characterize our object of interest is challenging: too few might not capture all relevant aspects of the object, whereas too many can lead to spurious correlations that can interfere with the conclusions that we want to draw from the data.
We refer to each element at each data point $\vect{x}_i$ as a~\stress{feature}\index{feature}. As stated before, a~central problem in \ac{ML} relates to the correct representation of the data and its features. This is the core of the field of representation learning on which we only touch, e.g., by means of \acfp{AE} and \acf{PCA} in \cref{sec:hot-topics:DE,sec:phase_class}, respectively.

Lastly, we emphasize that data can, loosely speaking, be identified with experience:
data can be produced as the result of a~repeated interaction with an~entity (such as an~experiment or a~simulation) that then leaves us with a~certain amount of experience about its underlying mechanism. In some cases, this experience may be used to further interact with such an entity and learn from it.
To this end, we set up a~model. In summary, the type of data to which we have access effectively defines the types of learning with which our model can be faced. These are usually divided into three: supervised, unsupervised, and reinforcement learning.

\paragraph{Supervised learning.}
\stress{Supervised learning}\index{supervised learning} can be seen as a~generalized notion of regression and classification, introduced in \cref{s:tasks}, and describes \ac{ML} algorithms that learn from \stress{labeled} data, i.e., $\dataset = \{(\vect{x}_i,\vect{y}_i)\}$. There exist various approaches to supervised learning, ranging from statistical methods to classical \ac{ML} and \ac{DL}, both introduced in \cref{sec:models}. 
The concept of supervised learning appears repeatedly in this book and forms the basis of many chapters, including phase classification (\cref{sec:phase_class}), Gaussian processes (\cref{sec:gp}), as well as the selected topics of \ac{DL} for quantum sciences (\cref{sec:hot-topics-physics-ml}). Importantly, some of the latter are specially suited to deal with experimental data, as, for instance, in the efficient read-out of quantum dots or the identification of Hamiltonian parameters describing quantum experimental setups. In most of these examples (but there are notable exceptions), large amounts of data are required for the training process. On top of the data, as stated above, supervised learning requires correctly labeled data. This is usually considered one of its most prominent downsides, as perfectly matching labels are not always accessible or have to be added manually by humans.

\paragraph{Unsupervised learning.} Supervised learning is not always the best option: the scarcity of labeled data is an example in which a~classical input-output design might fail. Instead, we often have access to data where no prior information, e.g., in terms of labels, is given (i.e. $\dataset = \{\vect{x}_i\}$). In this case, we can employ \stress{unsupervised learning}\index{unsupervised learning}. Unsupervised learning can either be used for preliminary preprocessing steps, such as dimensionality reduction, or for representation learning, such as in clustering\index{clustering}. In contrast, dimensionality can also be increased by adding features\index{feature} via generative models\index{generative models}. In this book, we discuss the application of unsupervised learning for phase classification in~\cref{sec:phase_class} and density estimation in~\cref{sec:hot-topics:DE}. This example is particularly interesting because it demonstrates how the choice of unsupervised learning over supervised learning can aid in the automated discovery of new physics when the interpretation of a~process, e.g., the nature of two different phases in a~transition is unknown.

\paragraph{Reinforcement learning.}
In contrast to the two previous types of learning, in \acf{RL}\index{reinforcement learning}, we usually do not have a~data set available \stress{at all}.
Instead, we have an~\stress{environment} with which we have to interact to achieve a~certain task. This interaction is augmented with feedback, i.e., some extra information on whether the action has been beneficial or harmful in achieving the task at hand. The collection of visited environment states, actions taken, and rewards or penalizations received take the role of a~data set.
Feedback is very important in \ac{RL} because we do not have a~clear-cut route in achieving our task. In fact, initially, we typically do not even know the necessary ingredients for achieving the task. Often, we only know that we \stress{achieved} a~specific goal but not \stress{why} we did it.
The field of \ac{RL} is precisely concerned with tackling the issue of how.
To introduce it properly, we devote to it \cref{sec:RL}.

\paragraph{Other types of learning.} While supervised, unsupervised, and reinforcement learning are the most common learning schemes in the \ac{ML} applications in quantum sciences, there are \ac{ML} approaches that go beyond this classification. An~interesting example is active learning\index{active learning}. This field includes selection strategies that allow for an~iterative construction of a~model's training set in interaction with a~human expert or environment. The aim of active learning is to select the most informative examples and minimize the cost of labeling~\cite{Gissin2020AL,Ren2021A}. We only touch on this topic by means of \acf{BO}\footnote{\Acf{BO} and active learning, while similar, are not the same. Active learning aims to determine optimal sampling, while \ac{BO} aims to find an extremum of a black-box function with as few function evaluations as possible.} in~\cref{ss:GPR-BO} and \acfp{LE} in~\cref{sss:hessian-based-interpret}. Another example of learning is semi-supervised learning\index{semi-supervised learning} in which a large amount of unlabeled data is explored to get better feature\index{feature} representations and improve the models trained on the small number of labeled data~\cite{vanEngelen2019semisupervised}.

\subsection{How to read this book}
This book aims at providing an~educational and self-contained overview of modern applications of \ac{ML} in quantum sciences. As such, \cref{sec:technical_intro} is devoted to the \ac{ML} prerequisites that are necessary to fully enjoy all the more advanced and further contents of this book. We discuss in detail four main \ac{ML} paradigms that have been successfully explored in quantum physics and chemistry: In \cref{sec:phase_class}, we describe how supervised and unsupervised learning can be utilized to classify phases of matter. In \cref{sec:gp} we introduce kernel methods with a~special focus on \acfp{GP} and \acf{BO}. \Cref{sec:NN_q_states} presents an~overview of various representations of quantum states based on \acfp{NN}. Finally, in \cref{sec:RL}, we dive into the foundations of \acf{RL} and how it can be applied to quantum experiments.

\begin{figure}[p]
    \begin{center}
    \includegraphics[width=0.99\textwidth]{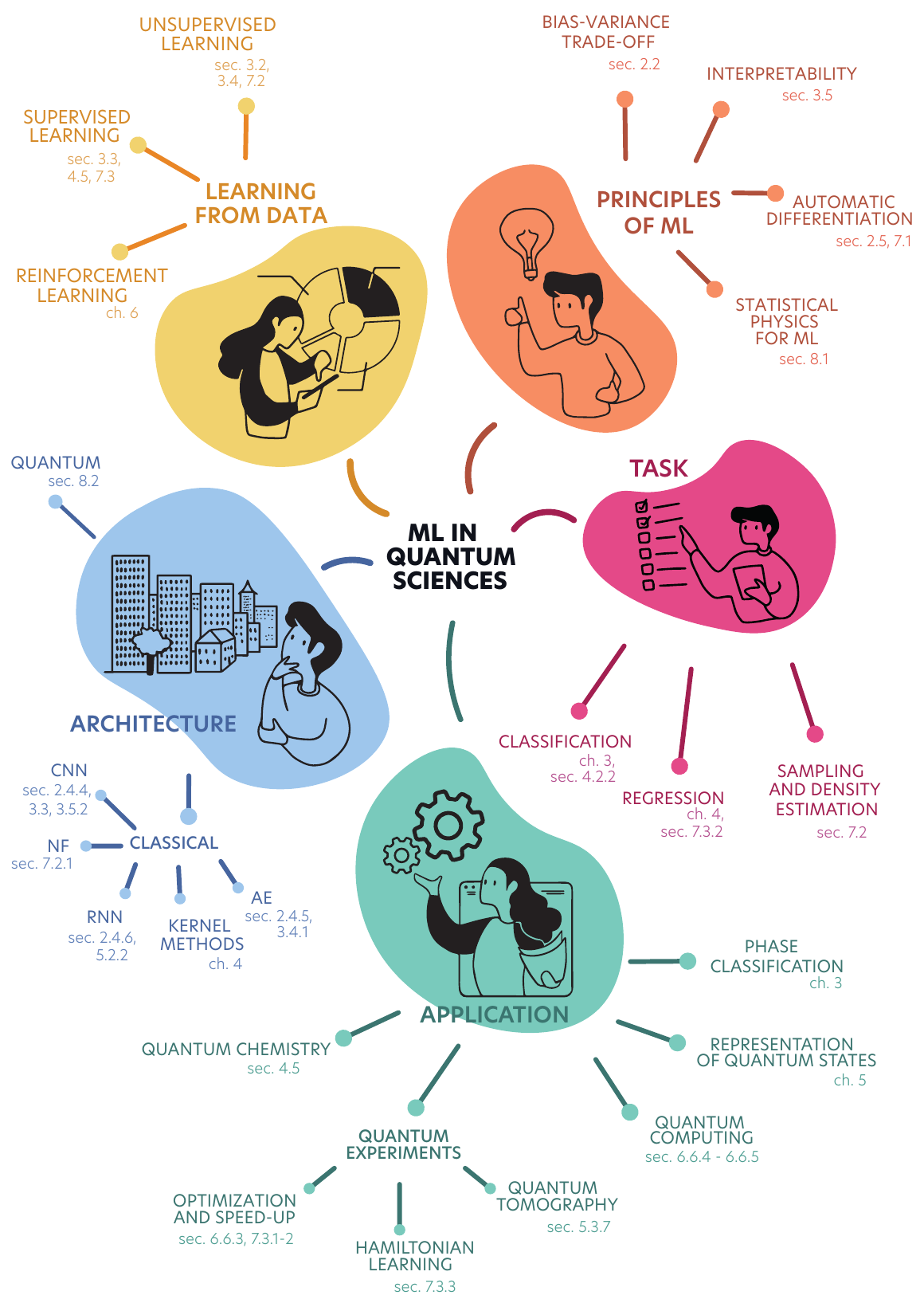}
    \end{center}
    \vspace{-0.5cm}
    \caption[Content of this book]{Content of this book. We cover three main learning schemes: supervised, unsupervised, and \acf{RL}, examples of \ac{ML} tasks like classification, regression, and density estimation, various applications in quantum sciences, quantum and classical \ac{ML} architectures. We also dive into the principles of \ac{ML}.}
    \label{fig:contents}
\end{figure}

In addition to these four pillars of \ac{ML} in quantum sciences, there exists an~exciting two-way interplay between the natural sciences and \ac{AI}.~\Cref{sec:hot-topics-ml-physics} focuses on more specialized examples of how \ac{ML}-related methods revolutionize quantum science. In particular, we introduce the paradigm of \acf{DiffP} and describe how it is becoming an~important numerical research tool. Moreover, we discuss how \ac{ML} methods assist researchers in tasks related to density estimation, as well as optimizations and speed-up of scientific experiments. There exists a~vibrant reverse influence on \ac{ML} coming from statistical physics (which we discuss in \cref{sec:stat_phys_for_ML}) and finally quantum computing. We describe the promises of \ac{QML} in \cref{s:QML}. All in all, this book discusses the fruitful interplay of \ac{AI} and physical sciences. Its content with references to relevant sections is illustrated in~\cref{fig:contents}.

\begin{figure}[p]
    \begin{center}
    \includegraphics[width=0.99\textwidth]{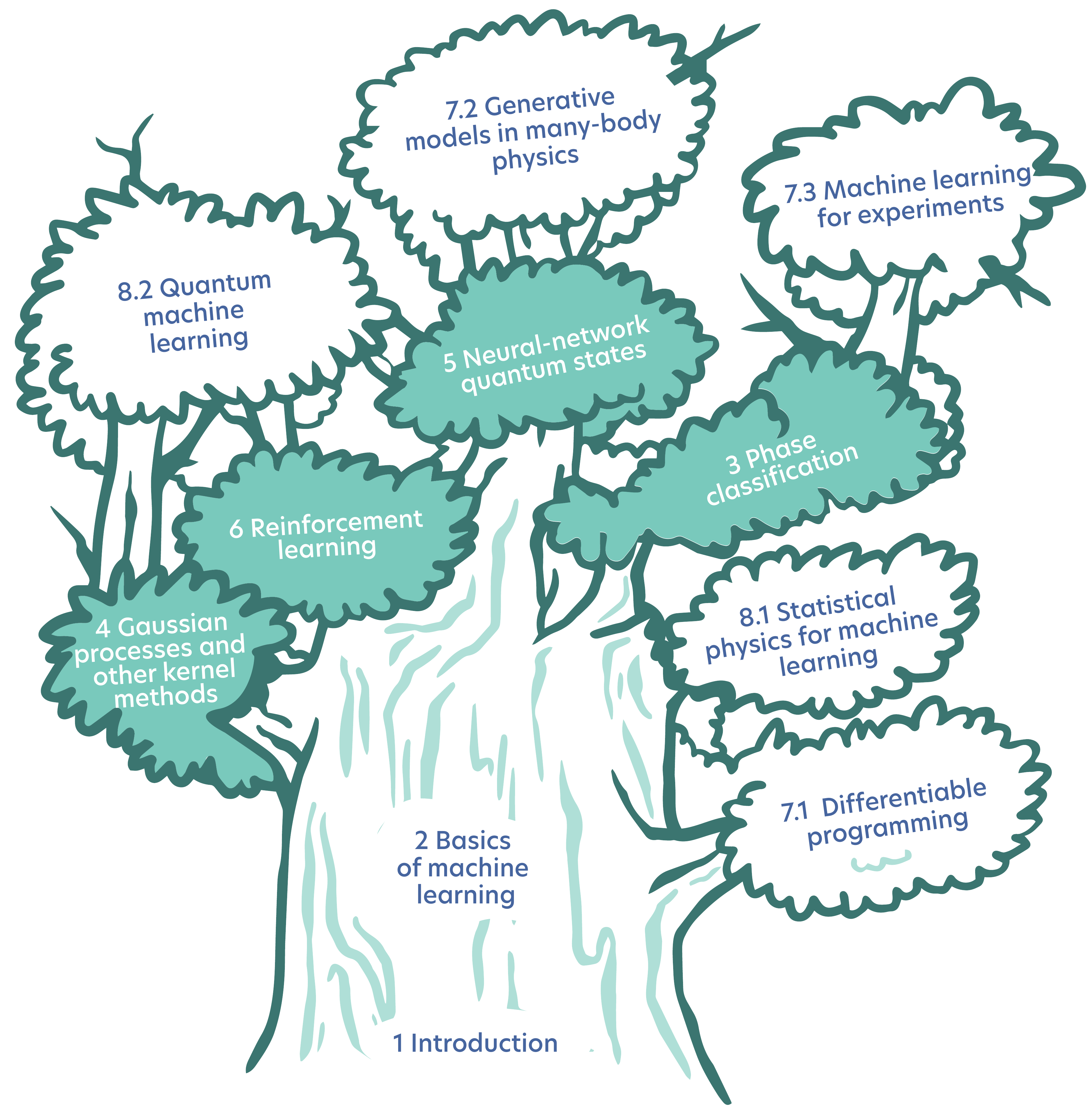}
    \end{center}
    \caption[Tree of dependencies between chapters]{Dependencies between chapters.}
    \label{fig:dependency-tree}
\end{figure}

We encourage the reader to start with the first two chapters, i.e., \textit{``Introduction''} and \textit{``Basics of machine learning''}. Then, the reader is free to wander into any of the independent \crefrange{sec:phase_class}{sec:RL} covering the four main paradigms, \cref{sec:hot-topics:dp} on differentiable programming, or \cref{sec:stat_phys_for_ML} discussing how statistical physics tackles the puzzles of \ac{ML}. \Cref{sec:hot-topics:DE} on generative models builds upon \cref{sec:NN_q_states} on \acfp{NQS}, while \cref{sec:ML_for_exp} on \ac{ML} for experiments requires knowledge of the methods used for phase classification presented in \cref{sec:phase_class}. Finally, \cref{s:QML} discussing \ac{QML} utilizes concepts introduced in \crefrange{sec:gp}{sec:RL}. The dependencies between chapters are visualized as a tree in \cref{fig:dependency-tree}.

\subsection*{Further reading}
\begin{itemize}

    \item Carleo, G. \textit{et al.} (2019). \href{https://arxiv.org/abs/1903.10563}{\textit{Machine learning and the physical sciences}}. Rev. Mod. Phys. 91, 045002. This detailed review summarizes the development of \ac{ML} in physics and its achievements till 2019~\cite{Carleo2019RevModPhys}.
    \item Carrasquilla, J. (2020). \href{https://doi.org/10.1080/23746149.2020.1797528}{\textit{Machine learning for quantum matter}}. Adv. Phys. X 5, 1. The concise review focused on phase classification and quantum state representation~\cite{Carrasquilla2020AdvPhys}.
    \item Krenn, M. \textit{et al.} (2023). \href{https://journals.aps.org/pra/abstract/10.1103/PhysRevA.107.010101}{\textit{Artificial intelligence and machine learning for quantum technologies}}. Phys. Rev. A 107(1), 010101. A perspective focusing on how quantum computing, quantum communication, and quantum simulation benefit from the \ac{ML} revolution~\cite{Krenn2023PRA}.
    \item Chollet, F. (2019). \href{https://arxiv.org/abs/1911.01547}{\textit{On the measure of intelligence}}. The review of different measures used to quantify \stress{intelligence}, which provides a perspective on \ac{AI} development \cite{Chollet2019}.
    \item Krenn, M. \textit{et al.} (2022) \href{https://www.nature.com/articles/s42254-022-00518-3}{\textit{On scientific understanding with artificial intelligence}}. Nat. Rev. Phys. 4, 761–769 \cite{Krenn2022}. A~beautiful paper discussing ways in which \ac{AI} could contribute to scientific discovery. It touches upon the philosophy of understanding and draws conclusions from dozens of anecdotes from scientists on their computer-guided discoveries.
    \item \href{https://www.youtube.com/playlist?list=PL7geCQP73UQuTRPA1NM_LP_biCrGiN2UF}{Recordings of lectures} of the \href{https://ml2021.ckc.uw.edu.pl/}{Summer School: Machine Learning in Quantum Physics and Chemistry} which took place between Aug, 23 - Sept, 03, 2021, in Warsaw, Poland.
    \item \href{https://github.com/Shmoo137/SummerSchool2021_MLinQuantum}{Jupyter notebooks} prepared as tutorials for the \href{https://ml2021.ckc.uw.edu.pl/}{Summer School: Machine Learning in Quantum Physics and Chemistry}~\cite{OurSchoolRepo}.
\end{itemize}

\clearpage
\newpage
\section{Basics of machine learning}\label{sec:technical_intro}
In this section, we describe basic \acf{ML} concepts connected to optimization and generalization. Moreover, we present a~probabilistic view on \ac{ML} that enables us to deal with uncertainty in the predictions we make. Finally, we discuss various \ac{ML} models. Together, these topics form the \ac{ML} preliminaries needed for understanding the contents of the next chapters.

\subsection{Learning as an~optimization problem}\label{sss:optimization}

We have already discussed that \ac{ML} can solve various tasks (e.g., classification or regression) and that there are different ways for the machine to access the data. The final ingredient is a~model that learns how to solve the given task with the data at hand. In general, it is a~function of the input data, $f(\vect{x})$, whose output is interpreted as a~prediction made for the input data. The form of the output depends on the task. It can be, e.g., a~class from a~discrete set of possible classes in the classification task or a~tensor from a~continuous target distribution in the regression task. Finding the function that provides the best mapping between the data and the desired outcome for a~specific task is at the heart of \ac{ML}. We start with declaring a~certain parametrization of a~model (function), e.g., $f(\vect{x}) = \weightsvect^{\transpose} \vect{x} + \biases$ with $\params \supset \{\weightsvect, \biases\}$. Then, all possible parametrizations of this function form the set of functions, i.e., the hypothesis class. \Cref{sec:models} presents specific examples of the hypothesis classes (or spaces), but for now, we focus on the learning process itself.

The mentioned learning schemes, i.e., supervised, unsupervised, or reinforcement learning, have the same underlying process of learning: finding an~optimal model $\hat{f} \equiv f_{\params^*}$ with optimal parameters $\params^*$ in the hypothesis space, which minimizes the target loss function or maximizes a~model performance. For the remainder of this section, for clarity, we focus on minimizing the loss function, $\lossfun$, which intuitively plays a~role of a~penalty for errors of a~model. 

\begin{figure}[t]
    \begin{center}
     \includegraphics[width=\columnwidth]{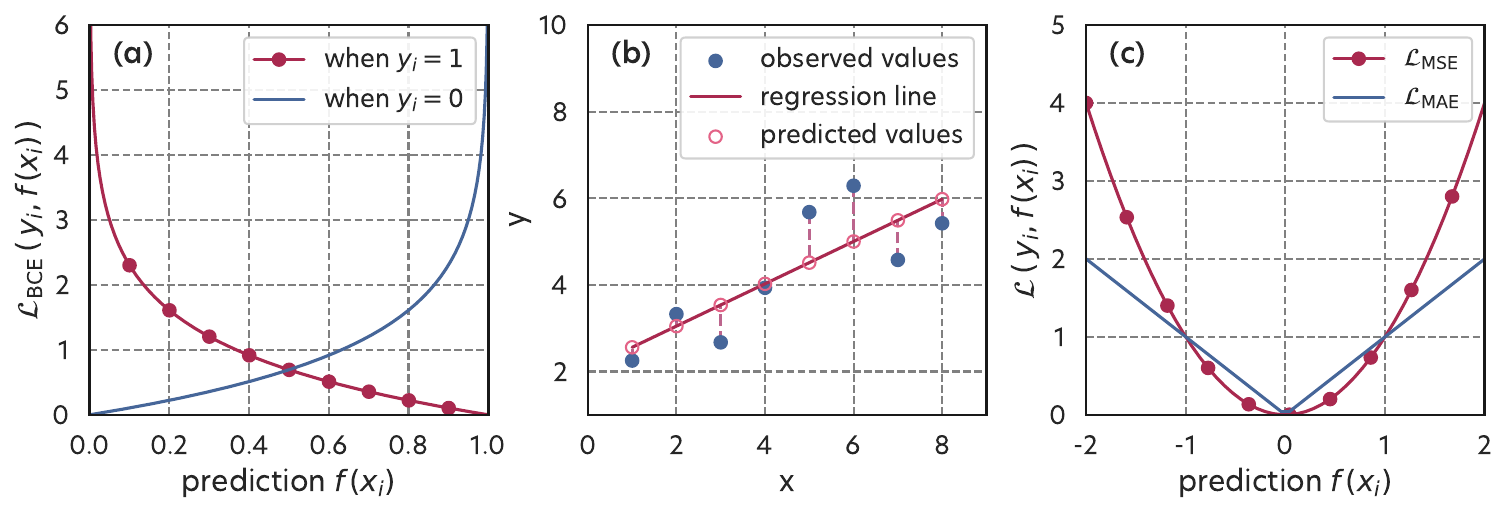}
    \end{center}
    \caption[Examples of loss functions]{Examples of loss functions. (a) Plot of the binary \acf{CE} for a~single data point, $x_i$, when the ground-truth label $y_i = 0$ (blue) or $1$ (purple). (b) The intuition behind loss functions used in regression problems. Dashed lines are differences between the labels, $y_i$, and values predicted by a~model, $f(x_i)$. (c) Plots of the \ac{MSE} (purple) and \ac{MAE} (blue) for a~single data point, $x_i$, when the ground-truth label $y_i = 0$.}
    \label{fig:lossfuns}
\end{figure}

\highlight{Machines ``learn'' by minimizing the loss function of the training data, i.e., all the data accessible to the \ac{ML} model during the learning process. The minimization is done by tuning the parameters of the model. The loss function formula varies between tasks, and there is a certain freedom in how it can be chosen. In general, the loss function compares model predictions or a~developed solution against the reality or expectations. Therefore, learning becomes an~optimization problem.}
In this book, we use the terms of loss, error, and cost functions\footnote{The literature also uses the terms of criterion or cost, error, or objective functions. Their definitions are not very strict. Following Ref.~\cite{goodfellow:2016}: ``The function we want to minimize or maximize is called the objective function, or criterion. When we are minimizing it, we may also call it the cost function, loss function, or error function. In this book, we use these terms interchangeably, though some \ac{ML} publications assign special meaning to some of these terms''. For example, the loss function may be defined for a~single data point, the cost or error function may be a~sum of loss functions, so check the definitions used in each paper.} interchangeably following Ref.~\cite{goodfellow:2016}. Popular examples of loss functions\index{loss function} include the \ac{MSE} and the \ac{CE}, used for supervised regression and classification\footnote{For classification, a~more intuitive measure of the performance could be, e.g., accuracy, which is the ratio between the number of correctly classified examples and the data set size. Note, however, that gradient-based optimization requires smooth and differentiable performance measures. These conditions distinguish loss functions from evaluation metrics such as accuracy, recall, precision, etc.} problems.
The output of the loss function depends on the model (which enters into formulas via predictions) and the data set. They are also normalized by the number of data points $\datasize$ to compare their values between problems with different data set sizes. The \ac{MSE} is a~popular loss function inherited from linear regression problems and is defined as
\begin{equation}\label{eq:mse_loss}
\lossfun_{\mathrm{MSE}} = \frac{1}{\datasize} \sum_{i=1}^{\datasize} (y_i - f\left(\vect{x}_{i}\right))^2\,.
\end{equation}
It has an~information-theoretic justification discussed in more detail in~\cref{sss:intro_linear_model,sss:generalization_regularization}. In the former, we also introduce the \ac{MAE} as another viable loss function, which is more sensitive to small errors than the \ac{MSE} as shown in \cref{fig:lossfuns}(b)-(c). \Acf{CE} is also a~concept drawn from information theory and has connections to probability theory (see \cref{sss:probability}). In the binary classification task, we can use the binary \ac{CE} (BCE), also known as the log loss (\cref{eq:bce_loss} and panel (a) in \cref{fig:lossfuns}), while for the multi-class classification, we use the categorical \ac{CE} (CCE).
They are defined as
\begin{align}
    \lossfun_{\mathrm{BCE}} &= -\frac{1}{\datasize} \sum_{i=1}^{\datasize} y_{i} \cdot \log \left(f\left(\vect{x}_i\right)\right)+\left(1-y_{i}\right) \cdot \log \left(1-f\left(\vect{x}_{i}\right)\right)\,,
    \label{eq:bce_loss} \\
    \lossfun_{\mathrm{CCE}} &= -\frac{1}{\datasize} \sum_{i=1}^{\datasize} \sum_{c=1}^{\class} y_{i,c} \cdot \log \left(f\left(\vect{x}_i\right)\right)\,,
    \label{eq:cce_loss}
\end{align}
where $\class$ is the number of classes. This formula requires representing labels in a~way called \stress{one-hot encoding}. For example, in a~$\class$-class problem, instead of having a~label with $\class$ possible values such as $y_i = 1,2,\ldots,\class$, each label is encoded as a~$\class$-element vector with all-zero elements except for one at the index corresponding to the class.
For example, $\vect{y}_i = [0,0,1,\ldots,0]$, means a~sample $i$ belongs to the third class, as only $y_{i,3}$ is non-zero.

\begin{figure}[t]
    \begin{center}
    \includegraphics[width=\columnwidth]{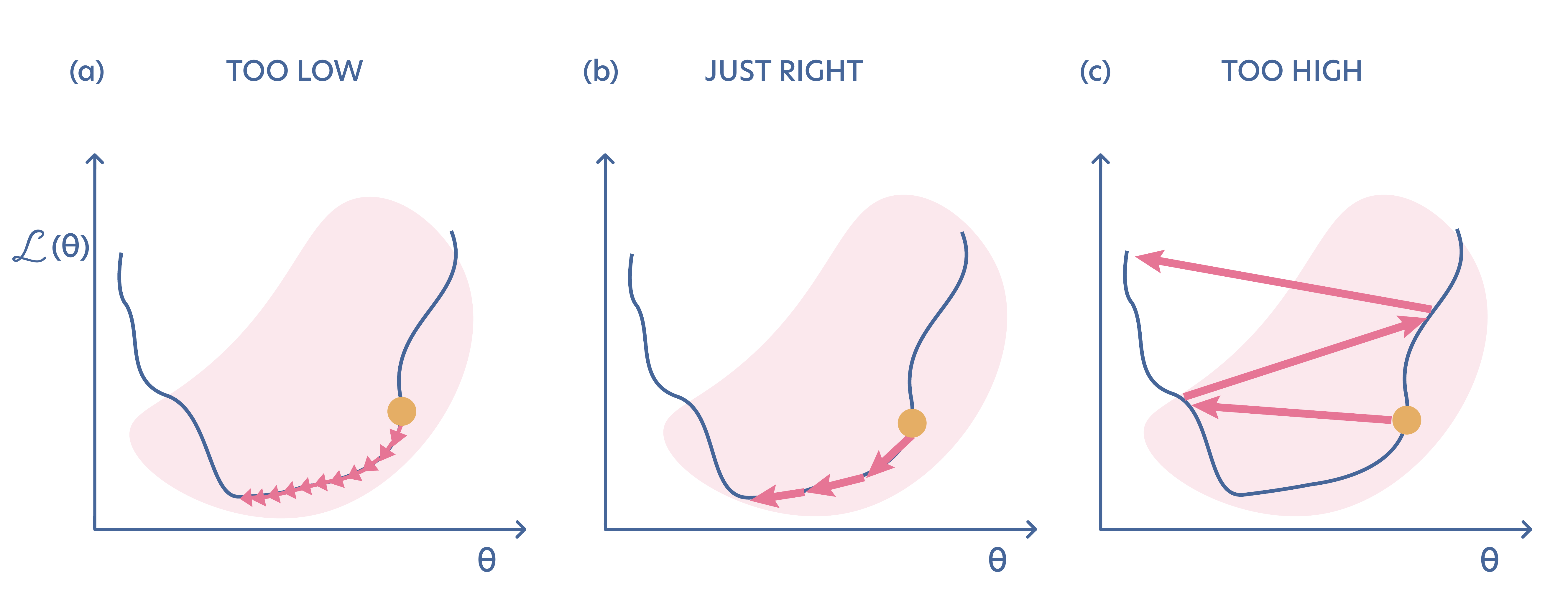}
    \end{center}
    \caption[Learning rate as a~hyperparameter]{Choosing a~learning rate has an~impact on convergence to the minimum. (a) If $\learningrate$ is too small, the training needs many epochs. (b) The right $\learningrate$ allows for a~fast convergence to a~minimum and needs to be found. (c) If $\learningrate$ is too large, optimization can take you away from the minimum (you ``overshoot''). This figure suggests that the loss function is convex which is rarely true.}
    \label{fig:learning_rate}
\end{figure}

Once we choose a~loss function, we can minimize it by varying the parameters of the \ac{ML} model, using any optimization method of our choice. In general, we can find the minimum of the loss function either via analytical construction or optimization methods that can be either gradient-based or gradient-free. A~popular example of a~gradient-based method is \stress{gradient descent}\index{gradient descent}. Optimization usually starts in a~random place within the loss landscape (meaning with a~model with randomly initialized parameters, $\params = \params_0$).\footnote{In practice, parameters are usually initialized randomly, but with the constraint to have a~mean at zero and constant variance across layers, otherwise, we may encounter problems with vanishing or exploding gradients~\cite{Bagheri2020weightinitialization}.} Using the model with $\params_0$, one makes predictions over the training data and from them computes the loss function. The next step consists of computing the gradients of the loss function with respect to each model parameter, $\param_j$. The final step is to update the parameters by subtracting the respective gradients multiplied by a~learning rate\index{learning rate}, $\learningrate$, i.e., 
\begin{equation}\label{eq_intro_GD}
    \param_j := \param_j - \learningrate \frac{\partial \lossfun}{\partial \theta_{j}}.
\end{equation}
These steps need to be repeated until the minimum is reached, and each repetition is called an~\stress{epoch}\index{epoch}. The intuition is that gradient descent updates model parameters by taking steps toward the minimum of the function (so in the opposite direction than the gradient, which indicates where the function value grows). The learning rate controls the size of these steps. \Cref{fig:learning_rate} presents in a~simplified way the importance of the $\learningrate$ choice. Both too large and too small $\learningrate$ make optimization more challenging, and only an~optimal $\learningrate$ promises efficient convergence to a~minimum. There is rarely an~obvious way of choosing $\learningrate$, which, therefore, has to be found, e.g., by trial and error. As such, the learning rate is one of the so-called \stress{hyperparameters}\index{hyperparameter} of the learning process. Hyperparameters are parameters whose values control the learning process (especially the speed of convergence and the quality of the minimum) and are chosen by a~user (in contrast to model parameters, which are derived through training). The total number of epochs or the choice of the loss function are hyperparameters, too. We encounter more examples of hyperparameters in this introductory chapter. 

To find optimal hyperparameters, we should form (in addition to the training data set) a~separate \stress{validation data set}. These data are only used to \stress{validate} the model and not for training. Then, we can set various hyperparameters and choose them in such a~way that the error on the validation set is minimized.\footnote{One can even use optimization methods to find optimal hyperparameters which minimize the validation error (a popular library is Optuna \cite{optuna19}) but a~choice of hyperparameters guided by intuition may prove to be a~faster and cheaper approach.} Dividing the data set into smaller subsets can be problematic in case of a limited number of data. Alternative approaches for model validation exist, like $k$-fold \stress{cross-validation}\index{cross-validation} \cite{goodfellow:2016}, which consists of splitting the data set into $k$ non-overlapping subsets. The validation error can then be estimated by taking the average error over $k$ trials where the $i$-th trial uses $i$-th subset as a~validation set and the rest as training data. Note that cross-validation comes at the price of increased computational cost.

Returning to the gradient descent, note that to perform it, we must first compute the gradient of the loss function with respect to the parameters to be tuned, $\nabla_{\params} \lossfun$, before each step\index{differentiation}, see \cref{eq_intro_GD}. \stress{A~priori}, there exist several different approaches to compute these derivatives. For example, one could work out the analytical derivatives by hand or approximate them numerically based on finite differences. When we are concerned with the accurate numerical evaluation of derivatives and not their symbolic form, \acf{AD} is a~good choice. \ac{AD} makes use of the fact that computer programs that compute the corresponding loss function can be decomposed into a~sequence of a~handful of elementary arithmetic operations (e.g., additions or multiplications) and functions (e.g., exp or sin). Therefore, the numerical value of the derivative of the program, i.e., the loss function, can be computed in an~automated fashion by repeated applications of basic pre-defined differentiation rules, such as the chain rule,
\begin{equation}\label{eq_intro_chain_rule}
 \frac{d f(g(x))}{dx} = f'(g(x))g'(x). 
\end{equation}
For more details on how to compute derivatives of computer programs, in particular \ac{AD}, see~\cref{sec:hot-topics:dp}.\footnote{The special case where \ac{AD} is applied in \stress{reverse-mode} to \acp{NN} is known as backpropagation\index{backpropagation} and constitutes the workhorse that enables efficient \ac{NN} training.}

The optimization procedure that we have described in the previous paragraphs and \ToggleForCUP{in~\cref{fig:learning_rate}}{in \cref{fig:learning_rate}} is very efficient when the \stress{loss landscape}\index{loss landscape}, i.e., the representation of the loss values around the parameter space of the model, is convex. 
However, especially for \ac{DL}, loss landscapes are highly non-convex and usually exhibit multiple local minima~\cite{Blum1992,Li2018}. Two immediate questions arise from this non-convexity: firstly, how can one avoid getting stuck in local minima corresponding to large loss function values or in saddle points of such landscapes? Second, are some minima better than others? Currently, such questions concerning learning dynamics are still being explored in various ongoing research directions, but some intuitions are already provided by statistical physics (see \cref{sec:stat_phys_for_ML}). A~popular approach to deal with the aforementioned problems considers a~slight modification of the gradient descent algorithm, so-called \stress{\acf{SGD}}\index{gradient descent!stochastic gradient descent}~\cite{Bottou2010SGD}. This optimization method (whose pseudocode is provided in \cref{alg:SGD}) consists of computing the loss function at each epoch on randomly selected mini-batches (subsets) of the training data. This means that during each epoch, the gradients may point in various directions. Effectively, the resulting stochasticity has been shown to help escape saddle points and narrow local minima \cite{Feng2021SGD}.\footnote{In practice, stochasticity is helpful in avoiding saddle points, but theoretical works show it is not a~necessary condition for a~proper convergence \cite{Lee2016}.} Furthermore, computing the loss function and gradients only for a~mini-batch of data instead of the whole data set provides a~nice computational speed-up for large data sets.

\begin{algorithm}[t]
\caption{Minibatch \acf{SGD}}\label{alg:SGD}
\begin{algorithmic}
\Require Learning rate $\learningrate$
\State Initialize $\params$ to random values
\For{$\mathrm{epoch}=1$ to $\mathrm{no\_epochs}$}
\State Shuffle $\dataset_{\mathrm{train}}$
\For{$\mathrm{i}=1$ to $m$ (where $m$ is a~minibatch size)}
    \State $\vect{x}_i, y_i \sim \dataset_{\mathrm{train}}$ \Comment{Draw random data point from data set without replacement}
    \State $\lossfun \gets \frac{1}{m} \sum_{i=1}^{m} \lossfun \left(y_i, f(\vect{x}_i) \right)$ \Comment{Compute loss function on the minibatch}
    \State $(\nabla \lossfun)_{j} \gets \frac{\partial \lossfun}{\partial \theta_{j}}$ \Comment{Compute gradients}
    \State $\theta_{j} \gets \theta_{j} - \learningrate \frac{\partial}{\partial \theta_{j}} \lossfun$ \Comment{Update parameters}
\EndFor
\EndFor\\
\Return $\params$
\end{algorithmic}
\end{algorithm}

Let us examine the minimum reached during the optimization of \ac{DL} models in more detail. To do that and to describe the curvature around such a~minimum, we use the Hessian\index{Hessian} of the training loss function, $\mat{H}_{\params^*} = \frac{\partial^2}{\partial \param_i \param_j} \lossfun_{\mathrm{train}} |_{\params=\params^{*}}$, i.e., the square matrix of second-order partial derivatives of $\lossfun$ with respect to the model parameters, calculated at the minimum, $\params = \params^*$. The eigenvectors of $\mat{H}_{\params^*}$ corresponding to the largest positive eigenvalues indicate the directions with the steepest ascent around the minimum. A~high curvature implies that the training data strongly determine the model parameters along that direction. What may be surprising is that the training of an~\ac{ML} model leads to a~local minimum or a~saddle point\footnote{One can wonder why we should trust a~model that does not land in the global minimum. A~series of empirical results as well as applying spin-glass theory to deep learning \cite{Choromanska2015} indicate, among others, that for large networks, most local minima are equivalent and yield similar performance on a~test set. Also, the probability of finding a~``bad'' (high value) local minimum is non-zero for small networks and decreases quickly with network size. Finally, attempting to find the global minimum on the training set (as opposed to one of the many good local ones) is not useful in practice and may lead to overfitting, i.e., much better performance on the training set than on the test set, which is equivalent to bad generalization.} \cite{Dauphin14, Sagun2016, Alain19}: the vast majority of the eigenvalues are close to zero, indicating various flat directions, and some small negative eigenvalues are also present, indicating directions with negative curvature. We present more examples of what information one can gain from $\mat{H}_{\params^*}$ in \cref{sss:hessian-based-interpret}.

Up to this point, the only gradient-based optimization method we have described is \ac{SGD}. Popular alterations to this scheme consist of, for example, including a~momentum term that takes previous update directions into account~\cite{pmlr-v28-sutskever13,Liu2020improvedSGD} or adaptive learning rates between epochs~\cite{AdaGrad} or both, culminating in the celebrated Adam optimizer~\cite{kingma2015adam, Zhang2018}.
Another different idea is to incorporate the second derivative in the update rule, as is accomplished by the \ac{L-BFGS} algorithm \cite{Zhu1997}. There are also gradient-free optimization approaches that are used, especially when the gradients or loss function itself are expensive or impossible to compute, e.g., when optimizing experiments. Examples include genetic algorithms, particle swarm optimization, random search, and simulated annealing \cite{Rios2012}. Another example we discuss in more detail in \cref{ss:GPR-BO} is \acf{BO}.

\subsection{Generalization and regularization}
\label{sss:generalization_regularization}
So far, \ac{ML} may seem like a function fitting in disguise. This changes when we go beyond simply trying to maximize the performance of a model on the available data.
\highlight{The heart of \ac{ML} lies in \stress{generalization}\index{generalization}, which is the ability to make accurate predictions on new data, never seen during the training.}

The ability to generalize can be quantified with the generalization error. The generalization error\index{generalization!generalization error} is the expected error of a model on new data drawn from the distribution of input/output pairs we expect the model to encounter in practice~\cite{goodfellow:2016}. However, such a distribution is generally inaccessible. Therefore, we approximate the generalization error of an \ac{ML} model by measuring its performance on an~additional held-out data set, commonly referred to as the \stress{test set}, composed of data points that are not used either to optimize the model parameters or to search for the best hyperparameters characterizing the learning process. The error made on the test set, called the test error, serves as a tractable measure of the generalization ability of the model and is only used to report the final performance of the model.\footnote{We need the test set because the performance as evaluated on the validation set may be overestimated because we use it to find the best hyperparameters of the learning process.} Therefore, the original data set needs to be separated into a~training, a~validation, and a~test set.\footnote{The ratio between the sizes of these sets depends on how much data is available in total, but we suggest starting with, e.g., 8:1:1.} One needs to be particularly careful in the preparation of these data sets to prevent \stress{information leakage}, i.e., the use of information in the training process that is not expected to be available at prediction time.\footnote{A~common mistake is to normalize the whole data set first and then separate it into a~training, a~validation, and a~test set. Normalization contains information about the most extreme data points, which may not even be part of the training set. This information can be exploited by the model to achieve better performance on the available data. Thus, the reported test error may not be a faithful indicator of the performance of the model on unseen data.} Also note that we always assume all data points to be drawn independently from the same distribution.\footnote{In practice, a trained model may be confronted with a sample from a distribution different than the one that generated training samples. In such a scenario, models are known to be overconfident \cite{Liu2020OOD}, which makes them unreliable. As such, the detection of out-of-distribution samples is an essential challenge in the deployment of \ac{ML} in safety-critical applications.} This ensures two things: first, that samples in our data set are uncorrelated, and second, that we can split the data set into smaller subsets.

\begin{figure}[t]
\begin{center}
\includegraphics[width=\columnwidth]{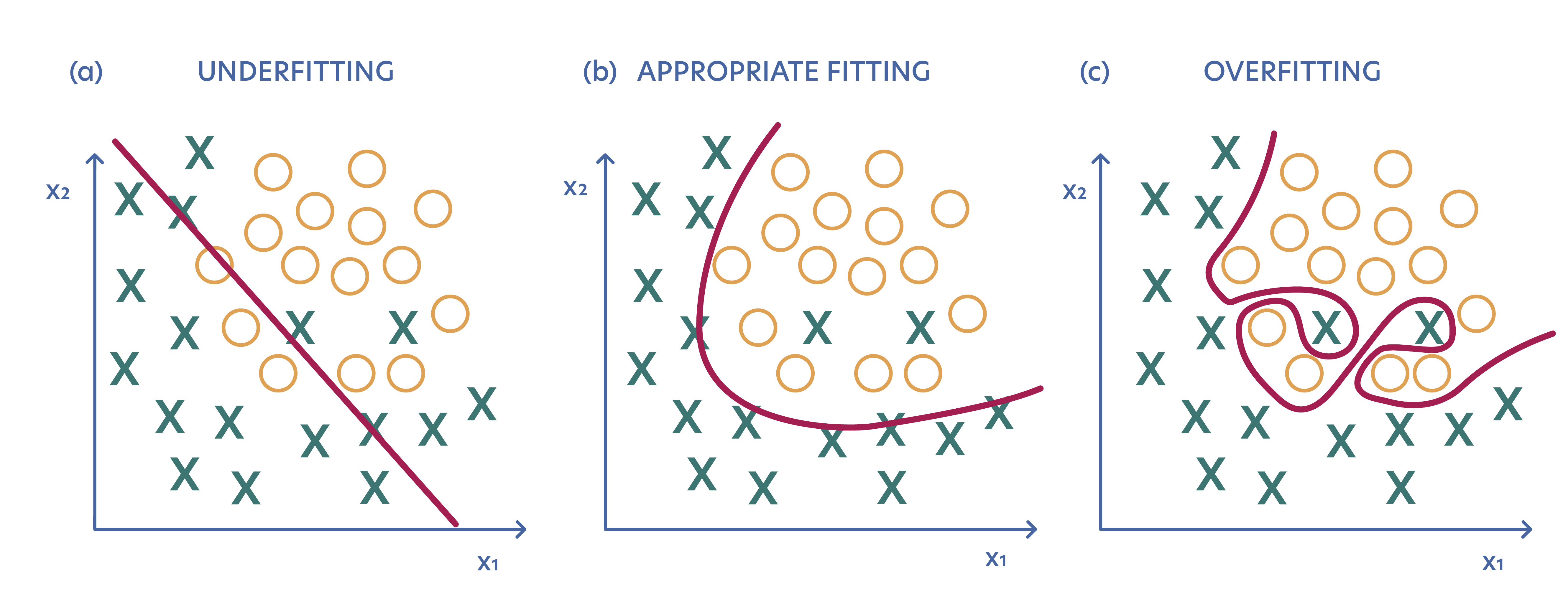}
\end{center}
\caption[Under- and overfitting]{Scheme of under- and overfitting. (a) When the model capacity is too low, the model cannot fit the training data properly. (b) With the model capacity corresponding to the task complexity, the fitting is optimal. (c) When the model capacity exceeds the task complexity, the model tends to overfit, and the generalization error increases.}
\label{fig:overfitting_underfitting}
\end{figure}

A~common feature of the training of an~\ac{ML} model is a~higher test error than the training error. Their difference is a proxy for the generalization gap, which is the difference between the training error and the generalization error.
This lower model performance on the test set compared to the training set persists even when all data points are generated by an~identical probability distribution, and it only disappears in the infinite data limit.
The main reason is the large capacity of \ac{DL} models.\footnote{\ac{DL} models are even able to fit large data sets with random labels \cite{zhang2017rethinking}!} 
\highlight{The \stress{capacity}\index{capacity} can be loosely understood as the measure of a~model's ability to fit a~variety of functions. When the capacity of the model is much higher than the required to solve the task, the model tends to \stress{overfit}\index{overfitting}, i.e., memorize all possible properties of the training set, which may not be true for the general distribution (and particularly, the test set).} In particular, the model can even fit the noise in the training data. As a~result, overfitting increases the test error while keeping the training error low (or even decreasing it). An~optimal capacity provides the lowest generalization error, minimizing the gap between the test and the training error. However, a~capacity that is too low results in an~overly constrained model that can \stress{underfit}\index{underfitting}, i.e., have a~high training error. The intuition behind the under- and overfitting is schematically shown in \cref{fig:overfitting_underfitting}. Therefore, we can improve the generalization of the model by controlling its capacity.

\highlight{Every modification of the model aiming to improve its generalization, even at the cost of increasing the training error, is called a~\stress{regularization}\index{regularization} technique.}

One can think of regularization in terms of the Occam razor.\footnote{This principle states that among competing hypotheses that explain known observations equally well, one should choose the simplest one. It is sometimes summarized as \textit{``entities should not be multiplied beyond necessity''}.} The additional motivation to use regularization is the \stress{no free lunch theorem}\index{no free lunch theorem}, which states that, when averaged over all possible data generating distributions, every classification algorithm has the same error rate when classifying previously unobserved points \cite{goodfellow:2016, Wolpert2020nofreelunch}. Therefore, no \ac{ML} model is universally better than another; and no regularization technique is universally better than another. This implies that we need to design our \ac{ML} algorithms to perform well on a~specific tasks, e.g., by regularizing it in a~way that is tailored to this task.

A~straightforward way of restricting the model's capacity is to limit the magnitude of its trainable parameters, which effectively limits the hypothesis space of a~parametrized model. This can be done by adding a~penalizing term to the training loss function, which increases with the parameters' magnitude. Such an~approach is used within the two popular regularization techniques, i.e., $\regularization{1}$ and $\regularization{2}$ regularization. In particular, $\regularization{2}$ regularization is described in more detail in \cref{sss:intro_linear_model,sec:GP_ridge_regression}.

Until now, we have discussed the relationship between a~model's complexity and its performance on the training and test set in intuitive terms. In the following, we formalize this intuition through \stress{bias-variance trade-off}\index{bias-variance trade-off}. Consider the standard situation encountered in regression\index{regression} problems: We are given an~ensemble of data points ${\dataset} = \{\vect{x}, f(\vect{x}) + \epsilon \}$ that derives from the function $f(\vect{x})$ and some noise $\epsilon$ inherent in the data. The function $f(\vect{x})$ is generally unknown, and our goal is to infer it. We do this by constructing a~regression fit of the data $\hat{f}(\vect{x})$. What we are ultimately interested in is for the test error (or generalization error\index{generalization!generalization error}) to be as small as possible. The test error is given as an~average of the loss function $\lossfun$ evaluated over test points, 
\begin{equation}
{\rm Err}_{\mathcal{T}} = \estimateE \left[ \lossfun ( \vect{y},\hat{f}(\vect{x}) ) \mid \mathcal{T} \right]\,,
\end{equation}
where $\mathcal{T}$ is a~fixed training set. This quantity is difficult to calculate, and we can instead resort to the expected prediction error obtained by averaging the generalization error over many training sets,
\begin{equation}
{\rm Err} = \estimateE \left[ \lossfun(\vect{y},\hat{f}(\vect{x})) \right] = \estimateE \left[ {\rm Err}_{\mathcal{T}} \right]\,.
\end{equation}
Let us look at the expected prediction error at a~given point $\vect{x}_{0}$
\begin{equation}\label{eq_model_selection_3}
{\rm Err}(\vect{x}_{0}) = \estimateE \left[ \left(f(\vect{x}_{0})+\epsilon - \hat{f}(\vect{x}_{0}) \right)^2 \right]\,,
\end{equation}
where, for now, we consider an~\ac{MSE}\index{mean squared error} as the loss function (\cref{eq:mse_loss}). Averaging in \cref{eq_model_selection_3} is performed over all random variables inside the expression $\estimateE \left[ \cdot \right]$, namely the noise $\epsilon$ as well as the model through the choice of different training sets. We can expand this expression as
\begin{equation}
{\rm Err}(\vect{x}_{0}) = \estimateE \left[ \left( f(\vect{x}_{0})- \hat{f}(\vect{x}_{0})   \right)^2 \right] + \estimateE \left[ 2\epsilon(f\right(\vect{x}_{0}) - \hat{f}(\vect{x}_{0})\left)\right] + \estimateE \left[ \epsilon^2 \right]\,,
\end{equation}
where $\estimateE \left[ \epsilon^2 \right] $ is the (fixed) variance of the underlying noise in the data. Next, we use the property of independent random variables $\estimateE \left[ AB \right] = \estimateE \left[ A~\right] \estimateE \left[ B \right]$ to obtain
\begin{equation}
\estimateE \left[ 2\epsilon \right(f(\vect{x}_{0}) - \hat{f}(\vect{x}_{0})\left) \right] = 2 \estimateE \left[ \epsilon \right] \estimateE \left[ f(\vect{x}_{0}) - \hat{f}(\vect{x}_{0})  \right] = 0\,,
\end{equation} 
where we assumed unbiased noise $\estimateE \left[ \epsilon \right] =0$. Thus, we are left with
\begin{equation}\label{eq_model_selection_4}
\estimateE \left[ \left(f(\vect{x}_{0})- \hat{f}(\vect{x}_{0})   \right)^2 \right] +\estimateE \left[ \epsilon^2 \right] = \estimateE \left[ \left( \hat{f}(\vect{x}_{0})\right)^2 \right] -2f(\vect{x}_{0}) \estimateE \left[ \hat{f}(\vect{x}_{0})\right] + \left( f(\vect{x}_{0})\right)^2 +\estimateE \left[ \epsilon^2 \right]\,.
\end{equation}
We modify \cref{eq_model_selection_4} by adding and subtracting $\estimateE \left[ \hat{f}(\vect{x}_{0}) \right] \estimateE \left[ \hat{f}(\vect{x}_{0}) \right]$ to get
\begin{equation}
{\rm Err}(\vect{x}_{0}) = \left(\estimateE \left[ \hat{f}(\vect{x}_{0}) \right] - f(\vect{x}_{0})\right)^2 + \left(\estimateE \left[ \left( \hat{f}(\vect{x}_{0})\right)^2 \right] - \estimateE \left[ \hat{f}(\vect{x}_{0}) \right] \estimateE \left[ \hat{f}(\vect{x}_{0}) \right] \right) +\estimateE \left[ \epsilon^2 \right]\,.
\end{equation}
We can identify the first term as the squared bias of our model
\begin{equation}
{\rm Bias}^2[\hat{f}(\vect{x}_{0})] \coloneqq \left(\estimateE \left[ \hat{f}(\vect{x}_{0})\right] - f(\vect{x}_{0})\right)^2\,,
\end{equation}
and the second as its variance
\begin{equation}
{\rm Var}[\hat{f}(\vect{x}_{0})] \coloneqq \estimateE \left[ \left( \hat{f}(\vect{x}_{0})\right)^2 \right] - \estimateE \left[ \hat{f}(\vect{x}_{0}) \right] \estimateE \left[ \hat{f}(\vect{x}_{0}) \right]\,.
\end{equation}
This results in
\begin{equation}
{\rm Err}(\vect{x}_{0}) = {\rm Bias}^2[\hat{f}(\vect{x}_{0})] + {\rm Var}[\hat{f}(\vect{x}_{0})] +\estimateE \left[ \epsilon^2 \right]\,.
\end{equation}
The average prediction error at a~given unseen test point $\vect{x}_{0}$ can therefore be decomposed into the bias of our model, its variance as well as the variance of the noise underlying our data (which is irreducible from a~model perspective).\footnote{This does not only hold for an~\ac{MSE} loss as many variations of the bias-variance decomposition are known~\cite{Domingos2000}.} \highlight{The more complex a~model $\hat{f}(\vect{x})$ is, the lower its bias is after training. However, the increased model complexity generally also results in larger fluctuations in capturing the data points, resulting in a~larger variance -- a~situation we refer to as overfitting. This is referred to as the bias-variance trade-off.}

\begin{figure}[t]
    \begin{center}
    \includegraphics[width=0.6\columnwidth]{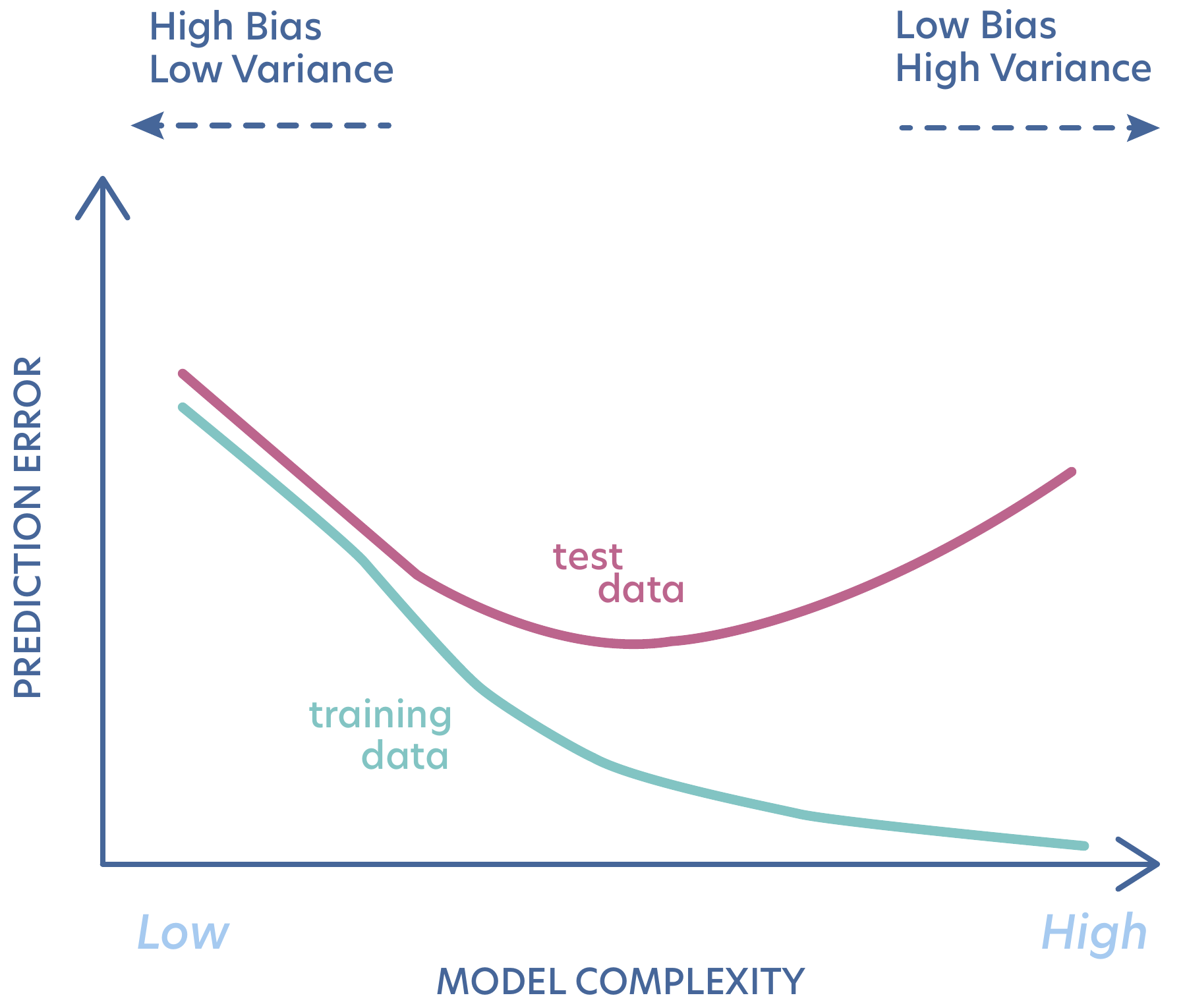}
    \end{center}
    \vspace{-0.3cm}
    \caption[The bias-variance trade-off]{Illustration of the bias-variance trade-off and its relation to the prediction error observed on training (green curve) and test sets (red curve). The ideal model, which results in the lowest test error, has both intermediate model complexity (e.g., capacity) and training error. Adapted from \ToggleForCUP{Belkin, M. \textit{et al}. (2019). \textit{Reconciling modern machine-learning practice and the classical bias-variance trade-off}. PNAS 116, 15849-15854~\cite{Belkin2019} with publisher permission.}{Ref.~\cite{Belkin2019}.}}
    \label{fig:bias_variance}
\end{figure}

\Cref{fig:bias_variance} shows an~illustration of the bias-variance trade-off, which makes clear that the ideal model realizes an~optimal trade-off between the training error and the model complexity. Interestingly, empirical studies indicate that modern large \ac{DL} models with enormous capacities can generalize very well~\cite{Kawaguchi2022generalization}. How overparametrized models can generalize so well remains a~challenging puzzle of the field\footnote{Promising observations are provided by the lottery ticket hypothesis \cite{Frankle2019LTH}.}, but some insight is provided with tools of statistical physics (see \cref{sec:stat_phys_for_ML}).

\subsection{Probabilistic view on machine learning}\label{sss:probability}

The need for a~probabilistic approach to \ac{ML} becomes apparent when we consider that this field has to tackle three sources of uncertainty (following Ref.~\cite{goodfellow:2016}). First, there may be an~inherent stochasticity of the system that generates the data we have access to (especially when dealing with quantum data). Second, we need to account for a~possible incomplete observability, i.e., an~unavoidable lack of information regarding all possible variables that influence the system.\footnote{This is, in fact, a feature and not a bug: for example, we easily understand the rotation of the earth around the sun due to its gravitational interaction. However, predicting the \stress{exact} orbit of the earth would require us to take into account all other gravitational masses in the solar system as well. Unless we intend to send a satellite into space, we are very happy to neglect these other interactions in favor of only a small error in our predictions.} In other words, we have only partial access (by means of the available data) to all relevant parts of the mechanism or distribution underlying the system. Finally, the models we use are rarely complete and need to discard some available information. An~example of incomplete modeling may be a~robot whose movement space we discretize. Such a~discretization immediately makes the robot uncertain about ``omitted'' parts of the space. To mathematically account for the uncertainty of a~model, we can follow the so-called Bayesian approach to probability, which interprets the probability as an~expectation or quantification of a~\stress{belief}.

In this section, we provide a~concise reminder of basic concepts from the probability theory which appear in the rest of this book:

\begin{itemize}
    \item Discarding any mathematical rigor, random variables are variables taking random values. If they are independent and identically distributed (i.e., drawn independently from the same probability distribution), they are called \stress{i.i.d.} random variables.
    \item A~\stress{probability distribution} is a~measure of how likely a~random variable $X$ is to take on each of its possible states $x$,\footnote{This notation is easily generalized to vector-valued random variables.} e.g., $p(X=x)$. A~probability distribution over discrete (continuous) variables is called a~probability mass function (probability density function). A~\stress{joint} probability distribution is a~probability distribution over many variables at the same time and is denoted, e.g., as $p(X=x, Y=y)$.
    When the notation is clear, we typically also drop the random variable and just write $p(X=x) \equiv p(x)$ instead.
    \item Two random variables $X$ and $Y$ are \stress{independent} if their joint probability distribution can be expressed as a~product of two factors, one involving only $X$ and one involving only $Y$:

\begin{equation}
\label{eq:independent-variables}
\forall x, y,\quad p(X=x, Y=y)=p(X=x) p(Y=y)\,.
\end{equation}

You can denote this independence by $X \perp Y$.
    \item A~vector whose elements consists of random variables is called a~\stress{random vector} and we denote it simply with $\vect{x}$.

    \item A~\stress{conditional} probability is a~probability of one event given that some other event has happened. We denote the conditional probability with $p(Y=y \mid X=x)$, meaning the probability of $Y=y$ given the observation that $X=x$. It can be calculated as:

\begin{equation}
\label{eq:conditional_probability}
p(Y=y \mid X=x)=\frac{p(Y=y, X=x)}{p(X=x)}\,.
\end{equation}

    \item Any joint probability distribution over many random variables may be decomposed into conditional distributions over only one variable each, which is called the \stress{chain rule} or \stress{product rule} of probability:

\begin{equation}
\label{eq:chain-rule-probabilities}
p\left(x^{(1)}, \ldots, x^{(\datasize)}\right)=p\left(x^{(1)}\right) \prod_{i=2}^{\datasize} p\left(x^{(i)} \mid x^{(1)}, \ldots, x^{(i-1)}\right)\,.
\end{equation}

    \item Finally, let us discuss a~situation where we know the conditional probability $p(y\mid x)$ and need to know the opposite one, $p(x\mid y)$. Fortunately, if we also know $p(x)$, we can compute the desired quantity using \stress{Bayes’ rule}:

\begin{equation}\label{eq:Bayes_rule}
p(x \mid y)=\frac{p(y \mid x) p(x)}{p(y)}\,.
\end{equation}
Bayes' rule is a~direct consequence of the definition of conditional probability in \cref{eq:conditional_probability}.
If we do not know $p(y)$, we can compute it via $p(y)=\sum_{x} p(y \mid x) p(x)$, the \stress{sum rule} of probabilities.
Coming back to the notion of \stress{belief} in Bayesian statistics, we can give the other terms of \cref{eq:Bayes_rule} a clear interpretation.
In this theory, $p(x)$ encodes our prior knowledge about a proposition $x$, i.e., modeled \stress{without} any evidence $y$ collected.
Hence, $p(x)$ is called the \stress{prior}.
Consequently, Bayes' rule gives us the recipe for how to update our beliefs using the likelihood $p(y \mid x)$ to arrive at the \stress{posterior} $p(x \mid y)$.
The posterior now models our updated knowledge about the proposition $x$ that takes the evidence $y$ collected into account.

\end{itemize}

We are now armed with enough tools to look at \ac{ML} models in a~probabilistic way. In particular, we can reformulate the definition of supervised and unsupervised learning. Unsupervised learning consists of observing some outcomes of a~random variable $X$, e.g., $x_1, x_2, \ldots, x_\datasize$, and then learning the probability distribution $p(X)$ or some of its properties.\footnote{Again, we can easily generalize this notion to random vectors.} Supervised learning is about observing instances of a~random variable $X$ and an~associated variable $Y$, e.g., $\{x_1,y_1\}, \{x_2,y_2\}, \ldots, \{x_\datasize,y_\datasize\}$, and learning to predict $y$ from $x$, usually by estimating $p(Y=y \mid X=x)$ from data. The so-called \stress{Bayes classifier} bases its predictions on the true conditional probability $p(Y=y \mid X=x)$, i.e., predicts the label $y_\mathrm{Bayes} = \argmax_{y} p(y \mid X=x)$ given the sample $x$. It is \stress{optimal} as there exists no other classifier that outperforms it in the classification task at hand (i.e., that achieves a lower misclassification probability)~\cite{devroye:1996}. However, even a Bayes classifier can be wrong and may only achieve a non-zero misclassification probability. This irreducible error (which is achieved by a Bayes classifier) is called \stress{Bayes error}\index{Bayes error} and is inherent to the classification task under consideration, i.e., is a fundamental limit independent of the choice of the predictive model. A classification problem has a non-zero Bayes error if there exist identical samples $x$ that are given distinct labels $y$, resulting in the class-conditional probabilities $p(y \mid X=x)$ being different from 0 and 1.

In the following, we now seek to combine this probabilistic view with our notion of learning as an optimization task in \cref{sss:optimization}. From our considerations above, we now understand that \ac{ML} models are used to estimate probability distributions given data. Because \ac{ML} models are typically parametrized, the concept of likelihood must enter the picture. The \stress{likelihood function} is the joint probability of the observed data as a~function of the parameters of the chosen model, $p(\dataset \mid \params)$, estimating the data-generating probability distribution.\footnote{Do not confuse likelihood and probability! 
Intuitively, probability is a~property of a~sample coming from some distribution. Likelihood, on the other hand, is a~property of a~parametrized model. In particular, if you plot $p(\dataset \mid \params)$ as a~function of possible $\params$, it does not have to integrate to one.} The likelihood provides us with the missing link between the given data for which we would like to infer, e..g, $p(\dataset)$, and the parameters of our model parameters that we want to learn. To this end, it is useful to consider how one can compare two probability distributions over the same random variable $X$, e.g., $p(x)$ and $q(x)$ with each other. An example of a~measure that one can use for such a comparison is a~\stress{relative entropy}, called the \acf{KL} divergence, $D_{\mathrm{KL}}(p||q)$\index{Kullback-Leibler divergence}. To be precise, the \acf{KL} divergence is a~measure of how the probability distribution $q$ differs from a~reference probability distribution $p$. As we typically employ it in classification tasks where $p$ and $q$ are both distributions of a discrete variable, it is defined as:
\begin{equation}
    D_{\mathrm{KL}}(p||q) = \estimate{\log\frac{p(\mathrm{x})}{q(\mathrm{x})}}{p} = \frac{1}{n} \sum_{i}^{\datasize} p(x_i) \log\frac{p (x_i)}{q (x_i)}\,.
\end{equation}
For continuous distributions, the sum has to be replaced by an~integral.
\sloppy $D_{\mathrm{KL}}(p||q)$ has some properties of distance, i.e., is non-zero and is zero if and only if $p$ and $q$ are equal.\footnote{Equal in case of discrete variables, and equal ``almost everywhere'', i.e., throughout all of the relevant space except for on a~set of measure zero, in case of continuous variables.} But it is not a~proper distance measure as it is not symmetric, $D_{\mathrm{KL}}(p||q) \ne D_{\mathrm{KL}}(q||p)$.\footnote{We recommend an~illustrative discussion of the asymmetry of $D_{\mathrm{KL}}(p||q)$ in Fig. 3.6 of~Ref.~\cite{goodfellow:2016}.} Using the properties of the logarithm, $D_{\mathrm{KL}}(p||q)$ can be expressed as
\begin{equation}
    D_{\mathrm{KL}}(p||q) = \frac{1}{n} \sum_{i}^{\datasize} p(x_i) \log p (x_i) - \frac{1}{n} \sum_{i}^{\datasize} p(x_i) \log q (x_i) \eqqcolon -{\cal S}(p) + {\lossfun_\mathrm{CE}}(p,q),
\end{equation}
where ${\cal S}(p) $ is the Shannon entropy of the reference probability distribution $p$, and as the second term we obtain the \ac{CE}, which we have already introduced in \cref{eq:bce_loss,eq:cce_loss}! We rediscover it by noting that minimizing $D_{\mathrm{KL}}(p||q)$, i.e., the difference of $p$ with respect to $q$, is equivalent to minimizing the cross-entropy because $q$ does not appear in ${\cal S}(p) $.

While the utility of comparing probability distributions is clear in the case of estimating an unknown probability distribution by a~parametrized one, it may not be immediately obvious for arbitrary \ac{ML} models. Let us discuss the case of supervised learning with a~model~$f$. Consider the labeled training data set consisting of $\datasize$ tuples $\{x_i,y_i\}$, where $x_i$ is a~given sample with label $y_i$. Each label belongs to one out of $\class$ classes. Next, we can think of each one-hot-encoded label $y_i = k$ as a~very specific probability distribution $y_i = q(x_i) = \delta_{k,j}$, where $k,j \in \{1,\dots,\class\}$ (one-hot encoding). Next, the training data $x_i$ are fed to the model~$f$, and as an~output we obtain the probability distribution $p(x_i) = f(x_i)$, which gives us the probabilities of a~given sample $x_i$ belonging to each class. In the last step, we have to compare two probability distributions, $p$ and $q$. Therefore, we rediscover the categorical cross-entropy from~\cref{eq:cce_loss}. Similarly, one can show that the \ac{MSE} loss, \cref{eq:mse_loss}, emerges naturally from a probabilistic viewpoint. We refer to the example of linear regression in~\cref{sss:intro_linear_model} for this analysis. In conclusion, we have seen that there exists a deep connection between probability theory and our optimization perspective of \ac{ML}.

\subsection{Machine learning models}\label{sec:models}

We have already described two out of three ingredients of the \ac{ML}: tasks (\cref{s:tasks}) and data (\cref{sec:typeoflearning}). 
The final element is a~model that learns how to solve a~task given some data. \ac{ML} models can be broadly divided into two classes which are \stress{standard} \ac{ML} and \ac{DL}. In~\crefrange{sss:intro_linear_model}{sec:intro-SVM}, we give an overview of the former, while the latter is explained in more depth in~\crefrange{sec:NNs}{sec:autoregressive_NNs}. Let us start by stressing the following point: \highlight{\ac{DL} is a~sub-field of \ac{ML} itself as depicted in \cref{fig:intro_DL_vs_ML}. However, it is customary to distinguish between \ac{ML} methods based on whether they use \acfp{NN}. Henceforth, in the remainder of the chapter, we refer to \stress{standard} \ac{ML} as any algorithm that does not make use of \acp{NN}.}

The distinction here becomes more subtle: in a~nutshell, what distinguishes traditional learning from \ac{DL} is the level of abstraction and the flexibility the algorithm has in extracting the features\index{feature}. In other words, traditional \ac{ML} requires very specific algorithms designed and tailored to the problem at hand. The choice of the model then often comes down to experience and further intuition of the task of interest. On the other hand, \acp{NN} are a~very flexible yet general tool whose main objective is to reproduce a~target function without any (or little) constraints on the functional class from which to search.
As a~down-side, they usually do not support an~easy interpretation of their mapping (compared to traditional \ac{ML} methods) and are often referred to as \stress{black-box} functions.
We explain to what extent this is actually the case in \cref{sec:interpretability}.
The distinction we can infer is that \ac{DL} does not require an~explicit set of instructions on how to connect the input to the output. Traditional \ac{ML} methods, on the other hand, are often constructed by geometric or information-theoretic arguments, which already provide intuition into the method by their very construction and, hence, their immediate interpretability.

\highlight{%
A~natural question that arises now is: which approach to choose, traditional \ac{ML} or \ac{DL}? As always: it depends. For instance, \ac{DL} performs at the state-of-the-art level in the big-data regime where an~\ac{NN} has enough available information to infer and perform the feature extraction. The low-data limit is where traditional \ac{ML} is still prevailing. Here, we need to incorporate as much information as possible into the algorithm of choice to make learning efficient. Thus, algorithms become much less general and more problem-dependent. %
}

Let us now focus on the standard \ac{ML} algorithms and leave the discussion about \acp{NN} for \cref{sec:NNs} and forward. Some prominent examples of traditional \ac{ML} we encounter during the rest of this book are the following: \Acf{PCA} is a~very elegant approach for the task of dimensionality reduction, i.e., for data compression. It takes multi-dimensional samples and compresses their feature space while maintaining only a~few relevant features\index{feature}. The compressed data can undergo further \ac{ML} routines (see \cref{ssec:pca}). \Acfp{GP} are another example of a~traditional \ac{ML} algorithm that deals well with learning tasks when only limited data are available. Together with \ac{BO}, they represent one of the most powerful examples of variational inference (see \cref{sec:gp}). They are furthermore an~instance of so-called \stress{kernel methods} which are as powerful as widely used. The elegance comes from the efficient application of a~feature transformation of the input data. In this way, data in a~representation that is difficult to analyze get mapped into a~domain where they are easier to analyze.

The mentioned methods are discussed in more detail throughout the book. The following sections constitute a~primer of the standard \ac{ML} models, describing basic approaches such as linear and logistic regression, linear \acfp{SVM}, and continue into the \ac{DL} regime with description of \acp{NN} with focus on \acfp{CNN} and \acfp{ARNN}.

\subsubsection{Linear (ridge) regression}
\label{sss:intro_linear_model}\index{linear regression}

Before diving into the details of the topic, let us restate the problem of regression sketched in \cref{s:tasks}.
We encounter a~labeled data set $\dataset = \{ (\vect{x}_i,y_i)\}_{i=1}^\datasize \equiv \{( \mat{X},\vect{y})\}$ of observations that are derived from an~underlying function $f$, possibly subject to some (stochastic) noise $\epsilon$.
The latter is often assumed to be sampled from an~unknown noise distribution $\mathcal{E}$, i.e., $\epsilon \sim \mathcal{E}$:
\begin{equation}
y_i = f(\vect{x}_i) + \epsilon_i \quad \forall\ (\vect{x}_i,y_i) \in \dataset\,.
\end{equation}
The function $f$ is generally unknown, and our goal is to infer it.
To this end, we build a~regression fit of the data $\hat{f}$ such that $\hat{f}(\vect{x}) \approx y $.\footnote{For the sake of simplicity, we consider one-dimensional output. The following derivations, however, can easily be extended to multi-dimensional output as well.}

Arguably, the simplest parametrized fitting method one can produce is a~linear model, where we seek to find parameters $\params\in\realset^\nparams$ that linearly connect the input variable $\vect{x}$ with the prediction $\hat{y}$, i.e.,
\begin{equation}\label{eq:intro_linear_model}
    \hat{y} = \sum_{i=1}^{\nparams - 1} \param_i x_i + b \equiv \sum_{i=0}^{\nparams - 1} \param_i x_i = \vect{x}^\transpose\params\,.
\end{equation}
To shorten the notation, we have absorbed the constant $\param_0 = b$, the so-called \stress{bias}, in the definition of the input $\vect{x}$ via setting $x_0 = 1$.
Up to now, the linear model aims to find a~hyperplane\footnote{For one-dimensional input and output, the hyperplane simply is a~line.} through the data points.
We can extend the model by a~nonlinear transformation $\featmap$ of the input, i.e., $\vect{x}\mapsto\featmap(\vect{x})$.
This is still linear regression as we maintain linearity in the parameters $\params$ that we seek to optimize.
As an~example of a~nonlinear transformation, the map $\featmap_p: x \mapsto (1,x,x^2,x^3,\dots,x^p)$ promotes our model to polynomial regression up to the $p$-th degree.
To simplify the notation in the rest of the section, we consider the case where no feature maps are applied.
The inclusion of a~feature map is a central element of \cref{sec:gp} and is discussed there to a far greater extent.

Once a~certain hyperplane is defined, by means of its parameters $\params$, we need to define a~quality measure that compares our predictions to their corresponding ground-truth values.
That is, we have to choose a~suitable loss function $\lossfun$.
The most conventional choice for the loss is the \acf{MSE}  over the data set $\dataset$ as
\begin{equation}\label{eq:intro_mse_definition}
    \lossfun_{\mathrm{MSE}}(\params\mid\dataset) \coloneqq \frac{1}{n} \sum_{i=1}^\datasize \left( y_i - \vect{x}_i^\transpose\params \right)^2 = \norm{\vect{y}-\mat{X}\params}^2\,.
\end{equation}
To attain the right-most equation, we stack all inputs $\vect{x}_i$ vertically next to each other, to form the matrix $\mat{X}\in\realset^{(\nparams-1)\times\datasize}$. The same procedure is applied to $y_i$, now to be promoted to $\vect{y}$.
The last step allows us to find the set of parameters $\params$ that minimize the \ac{MSE}.
This yields the least-squares estimator (LSE) (for the derivation, see the first half of \cref{appendix_kernel_trick})
\begin{equation}\label{eq:intro_least_squares_estimator}
    \params_\mathrm{LSE} = \left(\mat{X}^\transpose\mat{X}\right)^{-1}\mat{X}^\transpose\vect{y} = \mat{X}^+ \vect{y}\,,
\end{equation}
where the notation $\mat{X}^+$ denotes the Moore-Penrose inverse~\cite{Rao_1972_pseudoinverse}.

The \ac{MSE} as the choice of our loss function appears to be self-evident.
In fact, we can derive it by maximizing the likelihood of the labeled data given the model parameters $p(\vect{y} \mid \mat{X},\params)$.
To this end, we assume that our targets $y$ are actually sampled from a~Gaussian with a~mean given by our linear model, i.e., $\vect{x}^\transpose\params$ with some variance $\sigma$ that models the noise in the data.
We can then write the likelihood of observing the targets $\vect{y}$ given the locations $\mat{X}$ and model parameters $\params$ as
\begin{align}\label{eq:intro_data_likelihood}
    p(\vect{y} \mid \mat{X},\params) &= \normdist(\vect{y} \mid \mat{X}^\transpose\params,\sigma^2\id) \\
    &= \prod_{i=1}^\datasize \normdist(y_i \mid \vect{x}_i^\transpose\params,\sigma^2)\,.
\end{align}
In the last step, we furthermore assumed a~data set $\dataset$ of i.i.d. random variables to factorize the multivariate Gaussian.
A~common assumption is to regard the observed data set $\dataset$ as the most probable one of the underlying linear model.
Therefore, we seek to maximize the likelihood of finding the set of parameters $\params$ that have led to the most probable data.
This is the idea of \acf{MLE}.
Its estimator is defined as the argument of the maximum likelihood of \cref{eq:intro_data_likelihood}.
We can modify this estimator by including a~logarithm and obtain:
\begin{align}
    \params_\mathrm{MLE} &\coloneqq \argmax_{\params}\ p(\vect{y} \mid \mat{X},\params) \\
    \label{eq:log_trick_lin_model}
    &= \argmax_{\params}\ \log p(\vect{y} \mid \mat{X},\params) \\
    &= \argmax_{\params}\ \left( -\frac{1}{2\sigma^2} \sum_{i=1}^\datasize \left(y_i - \vect{x}_i^\transpose\params\right)^2 + \text{const.}\right)
    \label{eq:line_theta_MLE}\\
    &= \argmin_{\params}\ \left( \sum_{i=1}^\datasize \left(y_i - \vect{x}_i^\transpose\params\right)^2 \right) \equiv \argmin_{\params} \left( \lossfun_\mathrm{MSE} \right)\,.
\end{align}
The constants that appear in \cref{eq:line_theta_MLE} can be ignored since they are independent of $\params$.
From the previous results, we hence see that the assumption of i.i.d., together with the concept of \ac{MLE}, leads to the \ac{MSE} as the preferred loss function and we conclude that the \ac{MLE} $\params_\mathrm{MLE}$ coincides with the least squares estimator $\params_\mathrm{LSE}$ of \cref{eq:intro_least_squares_estimator}.

However, the estimator fully ignores the data noise modeled by $\sigma^2$, as it was also dropped out in the maximization procedure of the $p(\vect{y} \mid \mat{X},\params)$.
Thus, even if we correctly choose the model, the minimization procedure of the \ac{MSE} in \cref{eq:intro_mse_definition} generally performs well in the provided data set $\dataset$ but not on previously unencountered data points.
The reason is \stress{overfitting}\index{overfitting}, which we already introduced as a~concept in \cref{sss:generalization_regularization}. This phenomenon occurs when we incorporate the noise on the targets in our model parameters $\params_\mathrm{MLE}$.
As a~way out of this issue, we have introduced the notion of regularization.
In our linear model \eqref{eq:intro_linear_model}, we can introduce regularization by means of Bayesian inference.
This means that, instead of maximizing only the likelihood of the data in \cref{eq:intro_data_likelihood}, we encode any prior knowledge of the model into the \stress{prior} distribution $p(\params)$.
By virtue of the Bayes theorem from~\cref{eq:Bayes_rule}, we can calculate the \stress{posterior} distribution\footnote{Remember that we call it posterior because it is computed \stress{after} the observation of the data set $\dataset$.} $p(\params \mid \mat{X},\vect{y})$ over the parameters given the data set and maximize this quantity instead.
This yields the \acf{MAP} defined as
\begin{align}
    \params_\mathrm{MAP} &\coloneqq \argmax_{\params}\ p(\params \mid \mat{X},\vect{y}) \\
    &= \argmax_{\params}\ \frac{ p(\params)p(\vect{y} \mid \mat{X},\params)}{p(\mat{X},\vect{y})}\,,
\end{align}
where we have used the Bayes theorem from \cref{eq:Bayes_rule} in the second step. The denominator does not depend on $\params$ and can therefore be ignored.
For the likelihood, we keep the assumptions introduced for \cref{eq:intro_data_likelihood}.
As the prior, we now draw the parameter values from a~Gaussian distribution centered around $\vect{0}$ with some variance $\tau^2$, i.e.,
\begin{equation}\label{eq:intro_prior_dist}
    p(\params) = \normdist(\params \mid \vect{0},\tau^2\id)\,.
\end{equation}
The product of two Gaussian distributions is Gaussian itself, hence allowing us to apply the same trick with the logarithm as before in \cref{eq:log_trick_lin_model}.
We arrive at
\begin{equation}\label{eq:intro_MAP_estimator}
    \params_\mathrm{MAP} = \argmin_{\params}\ \left( \lossfun_\mathrm{MSE}(\params \mid \dataset) + \frac{\sigma^2}{\tau^2} \norm{\params}^2 \right) = \left( \mat{X}^\transpose\mat{X} + \frac{\sigma^2}{\tau^2}\id\right)^{-1} \mat{X}^\transpose \vect{y}\,.
\end{equation}
We can picture the parameter $\lambda = \sigma^2/\tau^2$ as a~signal-to-noise ratio, which effectively penalizes large magnitudes of parameter values by the additional term in the loss function.
Hence, $\lambda$ is referred to as \stress{regularization strength}.
This particular choice of the loss term is called \stress{Tikhonov} regularization.
Its corresponding \ac{MAP} is also called the linear \stress{ridge regression} estimator.

Let us compare the two estimators of \cref{eq:intro_least_squares_estimator,eq:intro_MAP_estimator}.
The additional term $\sigma^2/\tau^2\id$ in the estimator stems from the fact that we take into account both the data noise as well as a~parameter constraint.
Both are discarded in the limit of $\tau\to\infty$\footnote{This corresponds to a~uniform prior of the parameters $\params$.}, where we have $\params_\mathrm{MAP}\to\params_\mathrm{MLE}$.

\highlight{
In order to consider the underlying noise in the training data, we have to constrain the linear model. Dealing with overfitting in such a~way is usually referred to as regularization.}

Finally, the choice of the prior in \cref{eq:intro_prior_dist} is by no means unique.
In fact, there is a~plethora of regularization ideas and corresponding penalty terms~\cite{goodfellow:2016}.
An~easy variation could, for example, be to replace the $\regularization{2}$-norm with an~$\regularization{1}$-norm.
This is achieved by choosing a~Laplace distribution for the parameters as the prior.
The corresponding estimator is the result of \acf{LASSO} regression~\cite{LASSO_regression}.
Because the $\regularization{1}$-norm punishes already small parameter values severely, it favors sparse solutions for the parameters $\params$ instead.
This can, for example, be desired to detect the significant features out of a~pool of possible candidates in certain tasks~\cite{Zhou2017}.

\subsubsection{Logistic regression}\label{sec:intro-logreg}
\index{logistic regression}
In the previous section, we have discussed the linear regression problem.
The discussion can be extended to the classification task in a~very straightforward way, as we show in the following.
\highlight{The basic idea of \stress{logistic regression} is to adapt the linear model, such as to estimate the probability that a~given input falls in either one of the possible classes.}
Let us consider two classes  $\class_1$ and $\class_2$ and an~input $\vect{x}$ to classify.
We introduce the class-conditional densities $p(\vect{x}|\class_i)$ and the corresponding baseline class prior probabilities $p(\class_i)$.
Bayes' theorem of \cref{eq:Bayes_rule} immediately gives us an~expression for the posterior probability that the input belongs to class $\class_1$.
It reads as
\begin{equation}
\begin{aligned}
p(\class_1\mid\vect{x}) &= \frac{p(\vect{x}\mid\class_1) p(\class_1)}{p(\vect{x}\mid\class_1) p(\class_1) + p(\vect{x}\mid\class_2) p(\class_2)} \\
&= \frac{1}{1+\exp(-\param)} \eqqcolon \varsigma(\param) \\
\text{where } \param &\coloneqq -\log \left( \frac{p(\vect{x}\mid\class_2) p(\class_2)}{p(\vect{x}\mid\class_1) p(\class_1)} \right)
\end{aligned}
\label{eq:sigmoid_function}
\end{equation}
and equips us with the \stress{logistic sigmoid} function $\varsigma$ that maps any real-valued input $\param$ to the interval $[0,1]$.
We can now use the linear model (or any other \ac{ML} model) to yield a~value for $\param$ and map it to the corresponding posterior probability.
This additional layer turns the regression model into a~classifier.

In order to extend the situation to more than two classes, we perform a~similar reformulation as done in \cref{eq:sigmoid_function}.
In this case, one obtains the \stress{softmax} function\index{softmax}
\begin{equation}
    p(\class_k\mid\vect{x}) = \frac{\exp(\param_k)}{\sum_i \exp(\param_i)} \eqqcolon \mathrm{softmax}(\params)
    \label{eq:softmax_function}
\end{equation}
that maps the output score vector $\params$ to a~proper probability density over all classes at once.
Its name is derived from the fact that in the limiting case of $\param_i \gg \param_k\ \forall k \neq i$, the softmax converges to the maximum function, i.e., softmax $\to \max$.

In both cases, the model's parameters are trained by parsing the output scores through either \cref{eq:sigmoid_function} or \cref{eq:softmax_function} to obtain and subsequently minimize the loss in \cref{eq:bce_loss} or \cref{eq:cce_loss}, respectively.
An~interesting aspect of any classifier is how it draws a~line between data from two different phases, known as the \stress{decision boundary}.
In the case of the linear model the decision boundary is linear, which is a~simple consequence of the model choice.
Because this boundary is derived from the likelihood of the data due to the particular choice for the loss function, the model is highly prone to outliers.
One way to circumvent this issue is to take a~geometric approach in finding the decision boundary.
This is done in the next section.

\subsubsection{Support vector machines}\label{sec:intro-SVM}\index{support vector machine}

\begin{figure}[t]
\begin{center}
\includegraphics[width=\columnwidth]{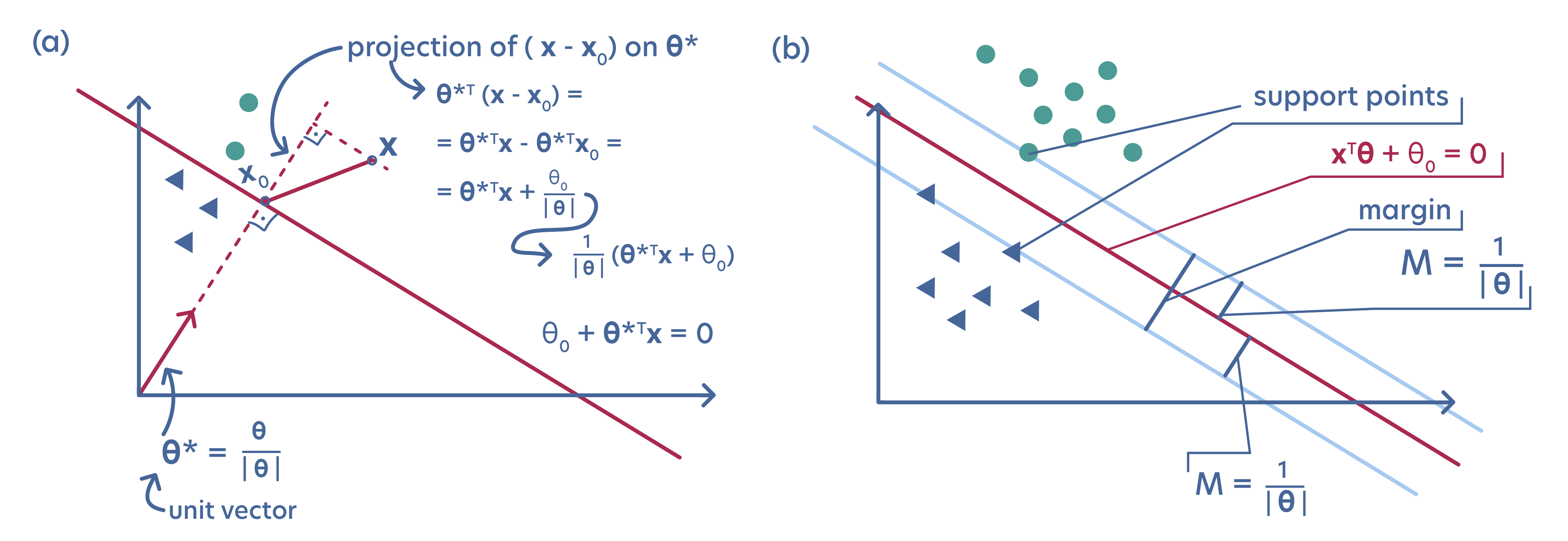}
\end{center}
\caption[Geometric construction of a~support vector machine in a~two-dimensional problem]{Geometric construction of an~\ac{SVM} in a~2D problem. (a)~Purple line, described fully by $\params$, is an~exemplary hyperplane separating two classes of data (pink and green points). (b)~The optimal hyperplane maximizes the margin, $M$, between itself and the support points, which are training data points closest to the hyperplane.}
\label{fig:SVM}
\end{figure}

An alternative approach to classification, instead of maximizing a model likelihood, is to analyze the data's geometrical properties.
This idea is encapsulated in the framework of \acfp{SVM} whose origins can be traced back to the 1960s, see Ref.~\cite{svm_chervonenkis} and the references therein.
For a visualization of the geometry of the data, take a~look at the linearly separable problem presented in \cref{fig:SVM}. In panel (a), you see that to classify two types of data, we can draw a~line (or, more generally, a~hyperplane) that separates the training data. Then, instead of making probabilistic predictions on test data, we can just check on which side of the hyperplane the test points are. Panel (a) also contains a simple geometric analysis, which shows that the equation for the hyperplane separating the data is $\params^{\mathrm{T}} \vect{x}+\param_{0}=0$. The unit vector in the direction perpendicular to the hyperplane is $\params^* = \params / |\params|$, and the shortest distance between a~point $\vect{x}$ and the hyperplane is
\begin{equation}\label{eq:SVM-distance}
d(\vect{x}, \params) = \left(\params^{\mathrm{T}} \vect{x}+\param_{0}\right) /|\params|\,.
\end{equation}

If we do not impose any additional constraints, there are many possible hyperplanes that separate the data into two classes. How do we choose the best? One way is to maximize the distance between the hyperplane and the data points~\cite{kSVM}. Therefore, let us formulate the constraint that all data points must be at least the distance $M$ away from the hyperplane. The data points separated from the hyperplane exactly by $M$, hence the closest to the hyperplane, become the support points presented in \cref{fig:SVM}(b). The classification problem boils down to finding the $\params$ that maximize the margin. From \cref{eq:SVM-distance}, we can write:
\begin{equation}
    y_{i}\left(\params^{\mathrm{T}} \vect{x}_{i}+\param_{0}\right) \geq M|\params|\,,
\end{equation}
where elements of the vector of observations $y_i$ are $\pm 1$ in order to ensure that this formulation is always positive, regardless of the class to which the data point belongs. Note that if we scale each of the $\params$ coefficients by the same factor, the above (in)equality still holds. Therefore, we can arbitrarily rescale $\params$ and $\param_0$ to have $|\params| = \frac{1}{M}$, which leads to the following canonical condition for every data point in the data set:
\begin{equation}\label{eq:SVM-constraint}
    y_{i}\left(\params^{\mathrm{T}} \vect{x}_{i}+\param_{0}\right) \geq 1\,.
\end{equation}
Therefore, to find the optimal hyperplane, we need to minimize $|\params|$, while ensuring $y_{i}\left(\params^{\mathrm{T}} \vect{x}_{i}+\param_{0}\right) \geq 1$ for every data point. This is the optimization with constraints, and we can use Lagrange multipliers for that! Minimizing $|\params|$ with constraints boils down to minimizing the following Lagrange function:
\begin{equation}\label{eq:Lagrange-function}
L=\frac{1}{2}|\params|^{2}-\sum_{i}^{n} \alpha_{i}\left[y_{i}\left(\params^{\mathrm{T}} \vect{x}_{i}+\param_{0}\right)-1\right]\,,
\end{equation}
where the Lagrange multipliers $\alpha_{i}$ are chosen such that
\begin{equation}\label{eq:Lagrange-multipiers-condition}
\alpha_{i}\left[y_{i}\left(\params^{\mathrm{T}} \vect{x}_{i}+\param_{0}\right)-1\right]=0 \text { for each } i\,.
\end{equation}
Interestingly, the loss function in \cref{eq:Lagrange-function} with the above constraints is a~so-called quadratic program as the function itself is quadratic and the constraints are linear with respect to $|\params|$. It has, therefore, a~global minimum found usually via so-called sequential minimal optimization~\cite{platt1998sequential} instead of any iterative gradient-based methods.

Also note that the condition put on the Lagrange multipliers in \cref{eq:Lagrange-multipiers-condition} implies the following:
\begin{itemize}
    \item If $\alpha_{i}>0$, then $\left[y_{i}\left(\params^{\mathrm{T}} \vect{x}_{i}+\param_{0}\right)-1\right]=0$, which means the point $\vect{x}_{i}$ lies on the boundary of the margin slab.
    \item If $\left[y_{i}\left(\params^{\mathrm{T}} \vect{x}_{i}+\param_{0}\right)-1\right]>0$, the points is outside the margin and $\alpha_{i}=0$.
\end{itemize}
Therefore, the final model coefficients are given only in terms of such points $\vect{x}_{i} := \vect{x}_{s,i}$ that lie on the boundary of the slab. These points are the support points and give the \ac{SVM} its name. The \ac{SVM} problem relies then on minimizing $L$\footnote{In practice, rather than minimizing $L$, one maximizes a~Lagrange dual, $L_D$, which provides the lower bound for $L$. We explain it in more detail in \cref{sec:kernel_SVM}.} numerically to find the coefficients $\alpha_{i}$ which are non-zero only for support points. 
\highlight{Therefore, classification with \acfp{SVM} consists of finding the optimal hyperplane separating the data by maximizing the margin between the hyperplane and the support points, which are data points closest to the decision boundary. This optimization problem with constraints is solved with Lagrange multipliers and is convex.}

With the found optimal hyperplane $\hat{f}$ we can then make predictions at an~arbitrary test point $\boldsymbol{x}^*$:
\begin{equation}
    \hat{f}(\vect{x}^*)=\params^{T} \vect{x}^*+\param_{0}= \sum_{i}^{\datasize} \alpha_{i} y_{i} \vect{x}_i^T \vect{x}^* +\param_{0} = \sum_{i} \alpha_{i} y_{i} \vect{x}_{s,i}^{T} \vect{x}^* +\param_{0}\,,
\end{equation}
where the last summation is only over support points. Finally, in order to turn this value into a~class prediction, we take the sign of $\hat{f}$ as the corresponding class label.

Until now, we only considered binary classification problems of linearly separable data sets.
There are two obvious ways of how to extend the \ac{SVM} to classification problems that have more than two classes, say $\class$ many.
The first, known as the one-to-one approach, breaks the multi-class situation down to a~binary classification between every combination of two classes, individually.
This way, we are required to train $\bigO (\class^2)$ \acp{SVM} to make predictions afterward.
This numerical overhead is eased in the second approach: one-to-rest classification.
Here, we only require a~single \ac{SVM} for each of the $\class$ classes that simply predicts whether a~test point belongs to the class or not.
As a~second extension possibility, we can ask about the classification problem that is not linearly separable. We explain this case later in \cref{sec:kernel_SVM}.

\subsubsection{Neural networks}\label{sec:NNs} 
\begin{figure}[t]
    \begin{center}
    \includegraphics[width=\columnwidth]{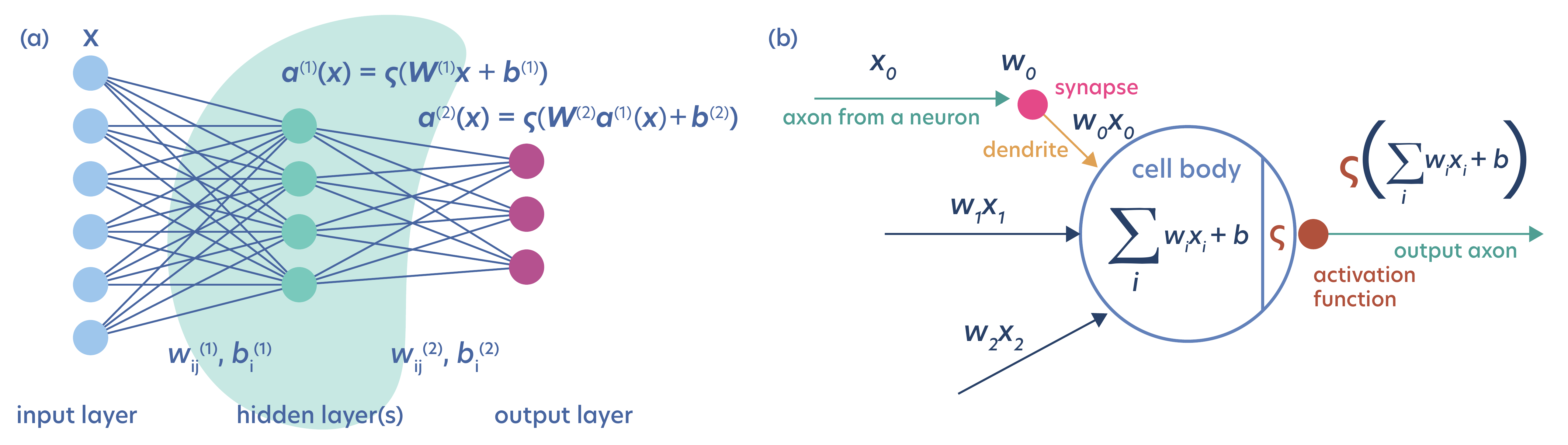}
    \end{center}
    \caption[Neural network]{Illustration of (a) a~typical fully-connected (here: two-layer) \ac{NN} and (b) one of its neurons (simple perceptron) and the computations associated with it.}
    \label{fig:NN}
    \end{figure}

    \Acfp{ANN}\index{neural network}, typically referred to as \acfp{NN}, are a~large class of models used to process data in \ac{ML} tasks. They are parametrized functions that are themselves composed of many simple functions. As the name suggests, \acp{ANN} were originally proposed by taking loose inspiration from networks of neurons that constitute our brains. They are typically composed of interconnected layers that sequentially process information, see \cref{fig:NN}. Each layer contains multiple nodes or units, also called \stress{artificial neurons} or \stress{perceptrons}.\index{perceptron}\footnote{Here and in the following, we refer to the modern perceptron introduced by Minsky and Papert~\cite{minsky:1969} which can contain smooth activation functions in contrary to the Heaviside step function utilized in Rosenblatt's original perceptron~\cite{rosenblatt:1958}.} Each node $i$ takes as input a~vector $\vect{x} = (x_{1},x_{2},\dots, x_{\featnum})\in \realset^{\featnum}$, corresponding to the activations of all nodes in the previous layer. It outputs a~scalar value $a_{i} \in \realset$ (its \stress{activation}) that is computed as $a_{i} = \varsigma(\sum_{j} \weight_{i,j} x_{j} + b_{i})$, where the parameters $\{ \weight_{i,j} \}_{j=1}^{\featnum}$ and $b_{i} \in \realset$ are the weights and bias of node $i$, respectively. The weights of a~node control the strength of its connection to the neurons of the previous layer. The function $\varsigma$ is a~nonlinear function called \stress{activation function}\index{activation function}. Common choices are the rectified linear unit (ReLU)
    \begin{equation}\label{eq:relu}
    \varsigma(z) = {\rm max}(0,z),
    \end{equation}
    the sigmoid function (\cref{eq:sigmoid_function}), or the tanh function 
    \begin{equation}\label{eq:tanh}
    \varsigma(z) = \tanh(z) = \frac{e^{z}- e^{-z}}{e^{z}+ e^{-z}}.
    \end{equation}
    
    The first layer is called the input layer, where the activations of its nodes are set according to the vector $\bm{x}$ encoding the input data. The last layer is called the output layer, and the activations of its nodes constitute the output of the \ac{NN}. All intermediate layers are called hidden layers. \acp{NN} where each node is by default connected to all nodes in the subsequent layer are referred to as \stress{fully connected}. The number of layers, nodes, and their connections is known as the \stress{architecture} of an~\ac{NN}. \acp{NN} are considered deep if they are composed of many hidden layers.\footnote{There is no clear consensus on the threshold of depth that divides shallow and deep \acp{NN}.} \Ac{ML} methods based on \acfp{DNN} as models fall under the name of \ac{DL}~\cite{goodfellow:2016}. 
    
    A~central question regarding \acp{NN} is what types of functions they can represent (recall our previous discussion on traditional \ac{ML} vs. \ac{DL}). First, consider an~\ac{NN} without its nonlinear activation functions. The function realized by such an~\ac{NN} is a~simple affine map, i.e., consists of multiplying the input by a~weight matrix and adding to it an~additional bias vector. Thus, the addition of nonlinear activation functions is crucial for \acp{NN} to be able to represent a~larger class of functions. For example, Kolmogorov and Arnold~\cite{kolmogorov1957} have shown that any arbitrary continuous high-dimensional function can be expressed as a~linear combination of the composition of a~set of nonlinear functions
    \begin{equation}
        \label{eq:universal_approx}
        f(\vect{x}) = \sum_{i=0}^{2 \featnum}\zeta_i\left(\sum_{j=1}^{\featnum}\varsigma_{i, j}(x_j)\right),
    \end{equation}
    where $\zeta_i, \varsigma_{i, j}$ are nonlinear functions that act on the individual components of the input $\vect{x} \in \realset^{\featnum}$. This means that we could represent any function $f(\vect{x})$ with a~polynomial number $O(\featnum^2)$ of one-dimensional nonlinear functions. This strongly resembles the structure of an~\ac{NN} with two hidden layers. Note, however, that the nonlinear functions must be carefully chosen depending on the target function. In \acp{NN}, the nonlinearities are typically fixed $\zeta_i = \varsigma_{i, j} \;\forall i, j$. It turns out that fully-connected \acp{NN} composed of a~single hidden layer and nonlinear activation functions are also \stress{universal function approximators}. That is, given that the target function is reasonably well-behaved, it can be approximated to any desired accuracy given that its hidden layer contains enough nodes~\cite{cybenko1989,hornik:1991,goodfellow:2016}. Note that this may still require a~hidden layer that is exponentially large in the number of nodes. This raises the question of what one can achieve with \acp{NN} that have multiple hidden layers.
    
    The universal approximation theorem guarantees that there exists an~\ac{NN}, i.e., choice of \ac{NN} architecture, as well as weights and biases, which approximates the given target function arbitrarily well. However, it does not guarantee that we are able to find this choice. It turns out that, in practice, \acp{DNN} are capable of solving many problems with much fewer nodes, i.e., trainable parameters, compared to shallow \acp{NN}. In that sense, choosing a~\ac{DNN} over a~shallow \ac{NN} yields a~useful prior over the space of functions that the \ac{NN} can approximate. 
    
    The parameters of an~\ac{NN} are typically optimized by gradient-based methods, such as \acf{SGD} or Adam, to minimize a~given loss function $\lossfun$ (see~\cref{sss:optimization}). Computing the gradient of the loss function with respect to the \ac{NN} parameter numerically is typically done by means of \stress{backpropagation}\index{backpropagation}~\cite{rumelhart1986learning} which we discuss in more detail in~\cref{sec:backprop}. In contrast, when evaluating an~\ac{NN} with a~given input, information flows forward through the networks. As such, this is called \stress{forward propagation}.

\paragraph{Convolutional neural networks}\label{sec:CNNs}
    \Acfp{CNN}\index{convolutional neural network} are a~special class of \acp{NN} where, in contrast to fully-connected \acp{NN}, not every node is connected to all nodes of the subsequent layer. Instead, convolutions replace matrix multiplications in the computation of the activations of subsequent layers. This reduction of the number of parameters per layer allows us to build and train deeper architectures. Moreover, this model architecture makes use of the spatial hierarchy typically present in input data. In image-like data, pixels that are spatially close to each other generally show more correlation than pixels that are far apart. By replacing the \stress{full} connectivity of standard \ac{NN} with multiple convolutional layers with \stress{local} connectivity, \acp{CNN} make use of this vanishing correlation at large distances.
    
    \begin{figure}[t]
        \begin{center}
        \includegraphics[width=0.8\columnwidth]{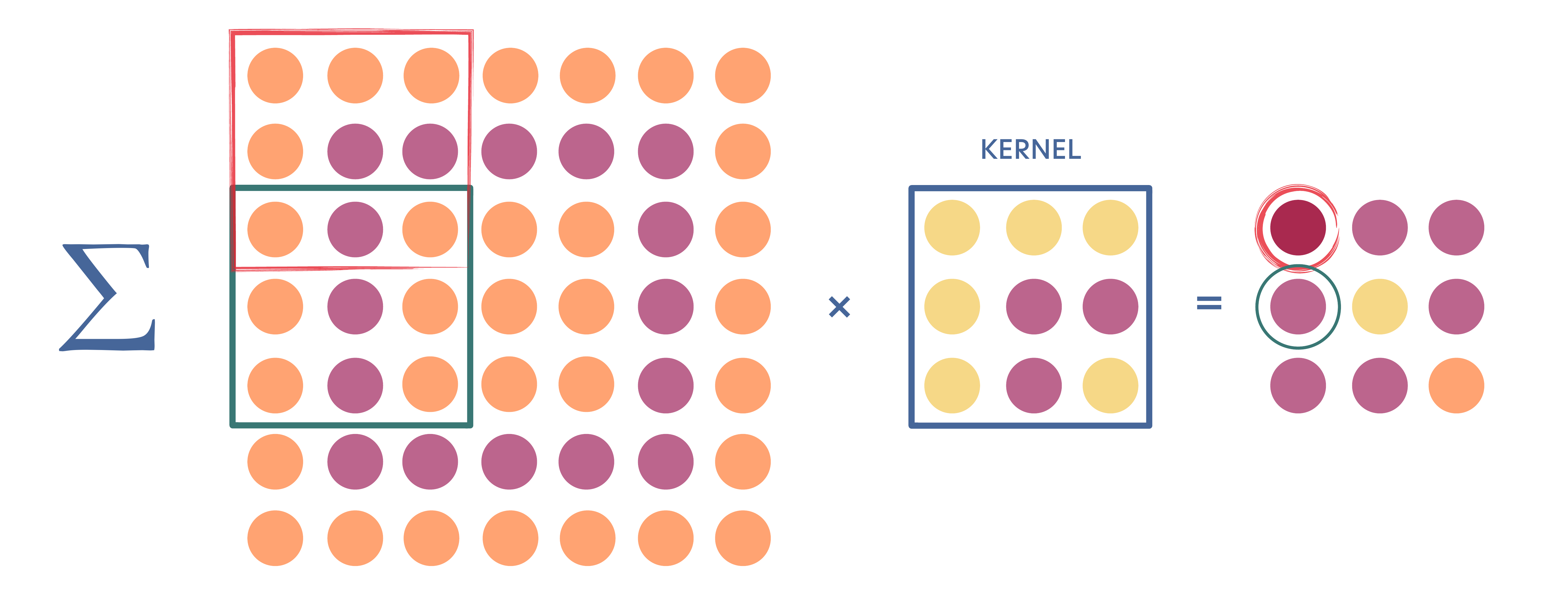}
        \end{center}
        \caption[Convolutional filter]{Schematic representation of a~convolutional layer in two dimensions: A~kernel/filter of fixed size (here 3x3) is convolved with a~two-dimensional input image. The color intensity corresponds to the magnitude of the neuron activations and kernel (filter) weights.}
        \label{fig:CNN}
        \end{figure}

    \Cref{fig:CNN} illustrates the working principle behind a~\ac{CNN} -- the \stress{convolutions}: A~filter (also called kernel) with trainable weights is slid across a~given layer. The resulting activations are then obtained by element-wise multiplication of the neuron activations and the filter's weights, followed by an~overall sum and the application of a~nonlinear activation function. This filtering causes the \ac{NN} to be only locally connected (as opposed to fully connected). Note that the number of weights, therefore, does not depend on the size of the input but rather on the size of the filter. The filter size controls the range over which spatial correlations in the input data are registered. One can build one- or two-dimensional \acp{CNN} (with filters of corresponding dimension) depending on whether the input data is naturally represented as a~vector or a~matrix. In a~typical \ac{CNN}, after the application of several such convolutional layers, the activations are flattened to a~single feature vector. This corresponds to a~lower-dimensional representation of the input data that is further processed using a~fully-connected architecture. To reduce the dimension of the data representation resulting from the application of convolutional layers, one typically also uses \stress{pooling operations}. These combine the activations resulting from applications of close-by filters, e.g., by taking the maximum or mean.
    
\subsubsection{Autoencoders}\label{sec:autoencoders}
\begin{figure}[t]
    \begin{center}
    \includegraphics[width=0.8\columnwidth]{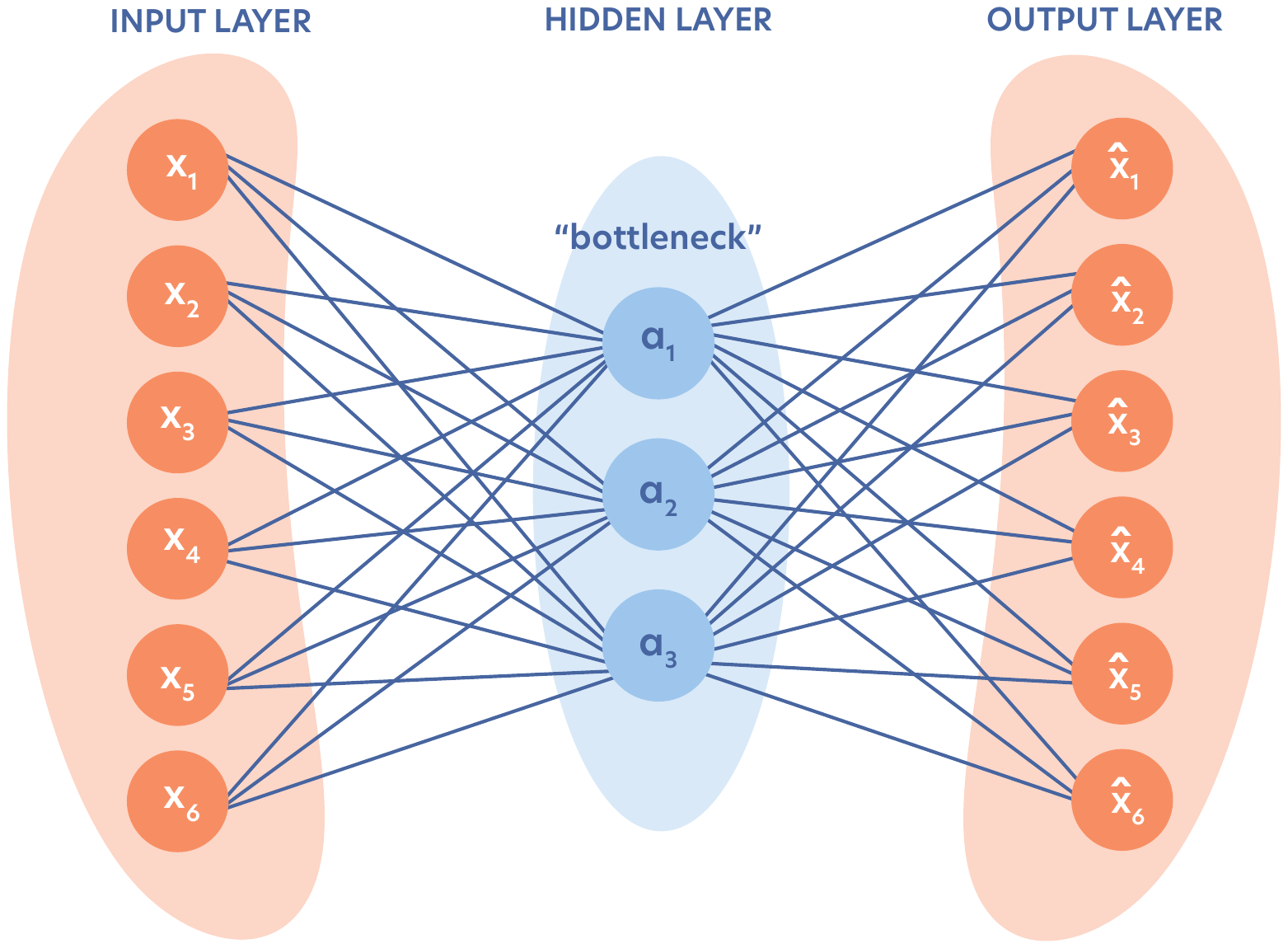}
    \end{center}
    \caption[Autoencoder]{Example of the bottleneck architecture of an~\ac{AE}\index{autoencoder}. The input is connected to the bottleneck\index{bottleneck} by an~encoder-\ac{NN} on the left while the decoder-\ac{NN} connects it with the output on the right.}
    \label{fig:VAE}
\end{figure}

\index{autoencoder}
\Acfp{AE} \cite{kingma2013auto, rezende2014stochastic} are widely used \ac{ML} tools for unsupervised learning. Unlabeled data (e.g., images, audio signals, texts) may often be high-dimensional. Hence, it is very difficult to analyze and extract any patterns when working in the data domain. However, dimensionality reduction techniques (see, e.g., \cref{ssec:pca}) represent an~advantageous approach to extract useful knowledge from such unlabeled data. In a~nutshell, the goal of \acp{AE} is to precisely \textit{encode} some knowledge, patterns, attributes of the given input data into some latent variable\footnote{A~latent variable is a~random variable that we cannot observe directly. In this case, we call variables latent because we do not observe them in the data.} on a~lower dimensional manifold. By means of a~so-called bottleneck\index{bottleneck} structure (as shown in \cref{fig:VAE}), the latent representation of the input data is then mapped back into the input space (decoding) by leveraging on the information extracted by the architecture at the time of feature extraction (encoding). This bottleneck architecture is based on two \acp{NN} performing the encoding and decoding parts. Such \acp{NN} are trained by minimizing the so-called error reconstruction loss, meaning that the optimal setup for such encoder-decoder pair is the one for which the output $\vect{x}_{\mathrm{rec}}$ is reconstructed as similar as possible to the original input data $\vect{x}$. These \acp{NN} are jointly optimized with an~iterative process. In other words, for a~given set of possible encoders and decoders, we are looking for the pair that keeps the maximum of information when encoding and, so, has the minimal reconstruction error when decoding.
This joint optimization forces the model to maintain only the variations in the data required to reconstruct the input without holding on to redundancies within the input. Henceforth, likewise in \ac{PCA}, only the most relevant features describing the data are distilled during the learning process. One important remark is that the bottleneck\index{bottleneck} is a~key attribute of such a~network design; without the presence of an~information bottleneck, our network could easily learn to simply memorize the input values by passing these values along through the network. On top of this, by relying on such a~pair of \acp{NN}, \acp{AE} are inherently more flexible yet expressive compared to standard dimensionality reduction algorithms (e.g., \ac{PCA}), which rely on a sub-manifold projection of input data through constrained linear or nonlinear transformations.

There are several types and variations of \acp{AE}, all of which share this fundamental bottleneck property as their base structure. A~concrete example of a~further development of \acp{AE} in the context of generative models\index{generative models} are \acfp{VAE}. As the name suggests, \acp{VAE}~\cite{kingma2013auto} have to do with variational inference. What they do in practice is to train the encoding-decoding pair in a~slightly more complicated way. The knowledge extracted from the data in the encoding part is nested into a~base probability density (e.g., initialized as a~Gaussian), which is trained and tuned in such a~way that it becomes a~good approximation (sampler) of the underlying data distribution. Once the training is done, the latent representation of the input data becomes thus a~probability density from which one can sample new, unseen data that resembles the one used for training, as being characterized by the same learned features. As such, the goal here is not only to reconstruct the input data from the extracted knowledge anymore but also to produce new samples as similar as possible to the training set. Further example of \acp{AE} are: sparse \acp{AE} \cite{ng2011sparse, makhzani2013k}, denoising \acp{AE} \cite{vincent2008extracting}, importance weighted \acp{AE} \cite{burda2015importance}, etc. 

\subsubsection{Autoregressive neural networks}\label{sec:autoregressive_NNs}

To complete this section, let us briefly present \acfp{ARNN}\index{autoregressive neural network}. These networks were originally inspired by autoregressive models\index{autoregressive models} in statistics and economics, which one can employ to predict future values of a~time-series (for instance, a~financial asset). \acp{ARNN} are formalized for the general task of density estimation \cite{uria2016neural}, in which the goal is to estimate a~complex, high-dimensional probability density function, see als \cref{sec:hot-topics:DE}. They are constructed to satisfy the following property on the outputs of the network, satisfying a~conditional structure 
\begin{align}
    \label{eq:arnn_function}
    f(\vect{x}) = \prod_{i=1}^{\featnum} f_i(x_i\mid x_{i-1}, \ldots,x_1)
\end{align}
with $\vect{x} = (x_1,x_2,\ldots,x_{\featnum})$ the inputs for the model. In the case of time series, the inputs $x_i$ would be values of a~variable at times $t_i$, and the model $f$ tries to predict future values based on past ones.
\begin{figure}[t]
    \centering
    \includegraphics[width=\columnwidth]{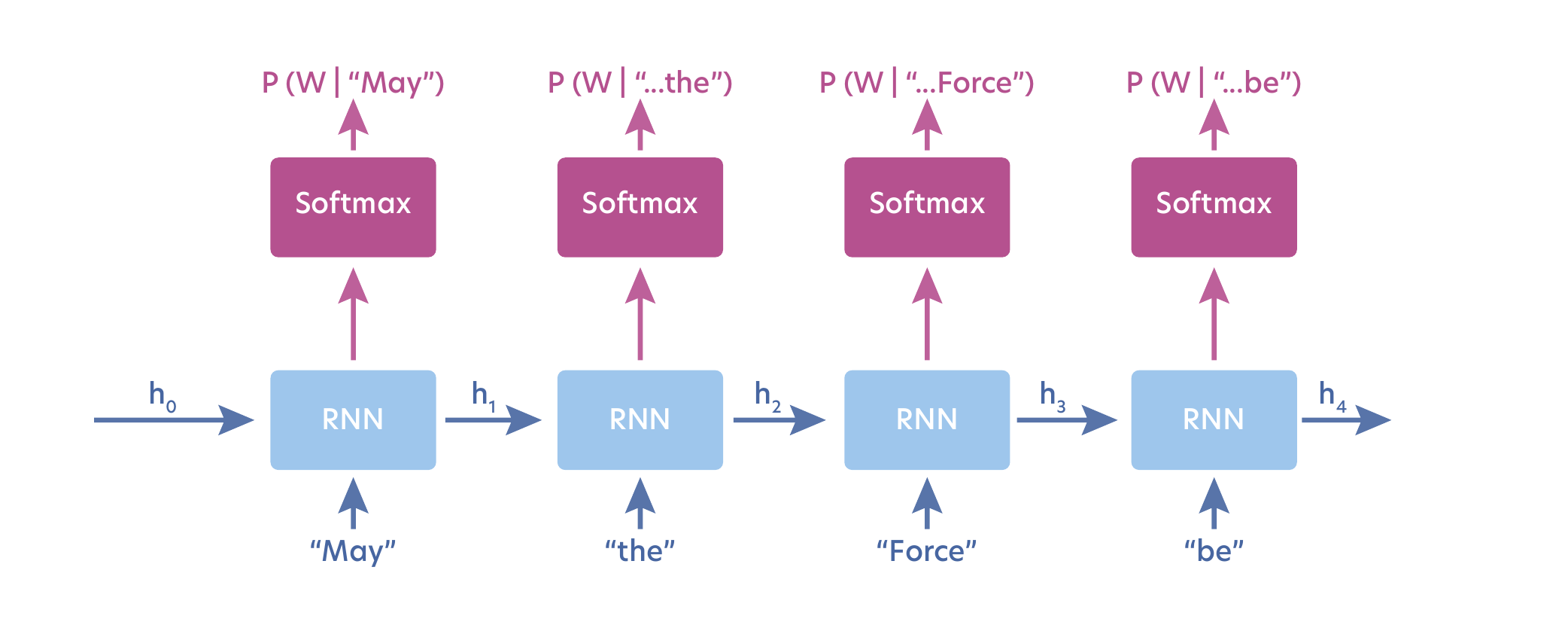}
    \caption[Recurrent neural networks]{Pictorial representation of a~\ac{RNN}. One can directly see that the model is autoregressive, as conditional probabilities only depend on the previous input data. Here, the blue box represents a~nonlinear transformation as described in the main text. The task here is to be able to generate meaningful sentences. The input data is a~sentence, and the output is the probability for the word ``W'' to be the next word in the sentence, conditioned on previous words. The $\vect{h}_i$ are the hidden vectors that take into account memory effects inherited from previous \ac{RNN} transformations.}
    \label{fig:RNN_lang}
\end{figure}
A~generic example of such networks is the \acf{RNN}\index{recurrent neural network} that was popularized in the context of natural language processing tasks. The main idea behind this class of models is that information ``loops back'' into the model, introducing correlations between different parts of the network, as opposed to feed-forward networks. Broadly speaking, a~sentence has a~causal order, but the correlations between words are not necessarily highest between words that are close together. Hence, the idea of introducing a~back loop, with a~memory, can be understood somewhat intuitively. The long-short-term memory (LSTM) is an~extension of this idea with two memory length scales (long- and short-term), and was also found to be successful for such tasks \cite{LSTMpaper}.
 A~sketch of an~\ac{RNN} is presented in \cref{fig:RNN_lang}, with an~example use-case from a~language processing task. The goal here is to predict the next word in the sentence based on previous words. The parameters of such a~network are hidden in the \ac{RNN} cell and take part in a~nonlinear transformation given by:
 \begin{equation*}
     \begin{array}{ll}
        \vect{h}^{(i)} = \varsigma(\weightsmat_h \vect{h}^{(i-1)}  + \weightsmat_x \vect{x}^{(i)})
        \end{array}
 \end{equation*} with $\weightsmat_h$ and $\weightsmat_x$ two weight matrices, $\vect{h}^{(i)}$ the $i$-th hidden vector, that represents information coming out of the previous cells, $\vect{x}^{(i)}$ the $i$-th element of the input data, and $\varsigma$ some nonlinear activation function that is applied element-wise.\footnote{One also has to choose an~initialization $\vect{x}^{(0)}, \vect{h}^{(0)}$, which are generally null vectors.} For words, $\vect{x}^{(i)}$ represents the $i$-th word of the sentence that is encoded in some form (for example, using a~one-hot encoding). Note that there exist several variants of this transformation, the most popular being the gated recurrent unit \cite{cho-etal-2014-properties}. 
The autoregressive models have been applied to different problems in physics, such as statistical mechanics \cite{wu2019solving,nicoli2020asymptotically,liu2021solving}, quantum tomography \cite{carrasquilla2019}, and ground state search \cite{sharir2020}. In~\cref{sec:NN_q_states,sec:hot-topics:DE}, we stress the advantages of using such models and present impressive results for quantum physics and chemistry that have been obtained using them.
 
\subsection{Backpropagation}\label{sec:backprop}
As already mentioned in~\cref{sss:optimization}, \Acp{NN} are typically trained via gradient-based methods. These approaches require the calculation of the loss function's derivative with respect to each trainable parameter. In principle, given a particular \ac{NN} architecture, we could derive a closed-form solution for the gradient. However, this computation would need to be performed again given different \ac{NN} architectures. Such calculations also involve some form of human input, which makes them tedious and prone to errors. Clearly, we would like to automate this gradient calculation and make it as efficient as possible. The algorithm of choice to train large \acp{NN} is \stress{backpropagation}\index{backpropagation}~\cite{rumelhart1986learning}. Backpropagation belongs to a larger class of algorithms known as \acf{AD}, which allow us to evaluate the derivative of a function represented as a computer program efficiently and in an~automated fashion.  We describe and compare these methods in detail in~\cref{sec:hot-topics:dp}.

\highlight{The basic idea behind backpropagation is to take advantage of the fact that an \ac{NN} is composed of sequences of many elementary building blocks, such as artificial neurons. Thus, we can compute derivatives through the repeated (reverse) application of the chain rule.} In order to provide some intuition, we exemplify the use of backpropagation on a simple feedforward network, where nodes in each layer are connected only to nodes in the immediate next layer. However, the main principle carries over to any other architecture, such as the ones introduced in the sections above, like \acp{CNN}, \acp{AE}, or \acp{RNN}, among others. Recall that the activations of the nodes in the $l$-th layer of a feedforward \ac{NN} are given by
\begin{equation}\label{eq:layer_backprop}
\vect{a}^{(l)} = \varsigma^{(l)}\left(\vect{z}^{(l)}\right) = \varsigma^{(l)}\left( \weightsmat^{(l)} \vect{a}^{(l-1)} + \vect{b}^{(l)}\right)\,,
\end{equation}
where $\vect{a}^{(l-1)}$ is a vector that contains the activations of the previous layer, i.e., layer $l-1$. The corresponding weight matrix is given by $\weightsmat^{(l)}$, where $\weight_{i,j}^{(l)}$ is the weight of the connection from node $j$ in layer $l-1$ to node $i$ in layer $l$, $\vect{b}^{(l)}$ is the bias vector of layer $l$, and $\varsigma^{(l)}$ is the activation function of the $l$-th layer. The function implemented by a feedforward \ac{NN} with $L$ layers ($L-1$ hidden layers and an output layer) can be obtained by stacking up multiple such layers 
\begin{equation}
    \vect{{\rm NN}}(\vect{x})=\vect{a}^{(L)}(\vect{x}) = \varsigma^{(L)}\left( \weightsmat^{(L)} \vect{a}^{(L-1)}(\vect{x}) + \vect{b}^{(L)}\right)\,,
\end{equation}
where $\vect{a}^{(0)}(\vect{x}) = \vect{x}$ is the input vector. The crucial observation is that the \ac{NN} output depends on the input $\vect{x}$ solely through the activations of the previous layer $\vect{a}^{(L-1)}$, which in turn only depends on the input through $\vect{a}^{(L-2)}$, and so on (see~\cref{eq:layer_backprop}). This simply arises from the layer-wise processing of information in a feedforward \ac{NN}.\footnote{In fact, the concept of a feedforward network can be generalized to any directed acyclic graph. In any case, the information processing occurs in a ``forwards-directed'' manner (from input nodes to output nodes).}

Eventually, we are interested in computing the derivatives of our loss function $\lossfun$ with respect to all weights $\partial \lossfun/ \partial \weight^{(l)}_{i,j}$ and biases $\partial \lossfun/ \partial b^{(l)}_{i}$. In the following, we focus only on weights. However, the procedure straightforwardly generalizes to biases. For a given training data set $\dataset = \{(\vect{x}_{i},\vect{y}_{i}) \}_{i=1}^{\datasize}$, the loss function is typically given as an average 
\begin{equation}
    \lossfun = \frac{1}{\datasize} \sum_{i=1}^{\datasize} \ell(\vect{{\rm NN}}(\vect{x}_{i}),\vect{y}_{i})\,.
\end{equation}
Here, $\ell$ measures the deviation of the prediction $\vect{{\rm NN}}(\vect{x}_{i})$ from the corresponding desired output $\vect{y}_{i}$ possibly including an additional regularization term. Thus, we have
\begin{equation}\label{eq:backprop_2}
    \frac{\partial \lossfun}{\partial w} = \frac{1}{\datasize} \sum_{i=1}^{\datasize} \frac{\partial \ell(\vect{{\rm NN}}(\vect{x}_{i}),\vect{y}_{i})}{ \partial w}\,,
\end{equation}
where $w$ is a single weight of the \ac{NN}. From~\cref{eq:backprop_2}, we see that the main task boils down to computing derivatives for a fixed input-output pair $(\vect{x}_{i},\vect{y}_{i})$ of the form
\begin{equation}\label{eq:backprop_main}
\begin{split}
    \frac{\partial \ell(\vect{{\rm NN}}(\vect{x}_{i}),\vect{y}_{i})}{ \partial w} &=  \frac{\partial \ell(\vect{{\rm NN}}(\vect{x}_{i}),\vect{y}_{i})}{\partial \vect{{\rm NN}}} \cdot \frac{\partial \vect{{\rm NN}}(\vect{x}_{i})}{ \partial w},\\
    &=  \frac{\partial \ell(\vect{a}^{(L)}(\vect{x}_{i}),\vect{y}_{i})}{\partial \vect{a}^{(L)}} \cdot \frac{\partial \vect{a}^{(L)}(\vect{x}_{i}) }{ \partial w}\,,
\end{split}
\end{equation}
The first term can be computed manually for a given choice of the loss function (see~\cref{sss:optimization}). For example, for the \ac{MSE} loss function, we have
\begin{equation}
    \ell_{\rm MSE}(\vect{{\rm NN}}(\vect{x}_{i}),\vect{y}_{i}) = \| \vect{{\rm NN}}(\vect{x}_{i}) - \vect{y}_{i}\|^2\,,
\end{equation}
resulting in 
\begin{equation}\label{eq:backprop_mse}
    \frac{\partial \ell_{\rm MSE}(\vect{{\rm NN}}(\vect{x}_{i}),\vect{y}_{i})}{\partial \vect{{\rm NN}}} = 2 (\vect{{\rm NN}}(\vect{x}_{i}) - \vect{y}_{i})\,.
\end{equation}
Therefore, the central quantity of interest is 
\begin{equation}\label{eq:backprop_central}
    \frac{\partial \vect{{\rm NN}}(\vect{x})}{ \partial w} = \frac{\partial \vect{a}^{(L)}(\vect{x})}{ \partial w}\,,
\end{equation}
which we are going to compute via repeated application of the chain rule. In the following, we drop the explicit dependence on $\vect{x}$.

The key observation for the backpropagation algorithm is the fact that, due to the layer-wise processing of information in a feedforward \ac{NN}, the only way a weight in layer $l$ influences the loss is through the next layer $l+1$. Thus, let us start by looking at the last layer $L$. Recall that $\vect{a}^{(l)} = \varsigma^{(l)}(\vect{z}^{(l)})$ from \cref{eq:layer_backprop}. Using the chain rule, we have
\begin{equation}\label{eq:backprop_main_2}
    \frac{\partial \vect{a}^{(L)}}{ \partial w} = \frac{\partial \vect{a}^{(L)}}{ \partial \vect{z}^{(L)}} \frac{\partial \vect{z}^{(L)}}{ \partial w} = \frac{\partial \vect{a}^{(L)}}{ \partial \vect{z}^{(L)}} \left( \frac{\partial \weightsmat^{(L)}}{ \partial w}\vect{a}^{(L-1)} + \weightsmat^{(L)} \frac{\partial \vect{a}^{(L-1)}}{\partial w} \right)\,,
\end{equation}
where $\partial \vect{a}^{(L)} / \partial \vect{z}^{(L)} = \mat{J}_{\varsigma}^{(L)}$ is the Jacobian matrix of $\varsigma^{(L)}$ containing the derivative of the activation functions $\left( \mat{J}_{\varsigma}^{(L)}\right)_{i,j} = \partial \varsigma_{i}^{(L)}(\vect{z})/ \partial z_{j}$. Note that $\mat{J}_{\varsigma}^{(l)}$ is diagonal, e.g., for ReLUs (\cref{eq:relu}), but not for the softmax function (\cref{eq:softmax_function}). From~\cref{eq:backprop_main_2}, if $w$ is a weight of layer $L$, i.e., $\weight=\weight_{i,j}^{(L)}$, we have
\begin{equation}\label{eq:backprop_occuring_L}
    \frac{\partial \vect{a}^{(L)}}{ \partial w} = \mat{J}_{\varsigma}^{(L)} \vect{e}^{(L)}_{i} a_{j}^{(L-1)}\,.
\end{equation}
Here, $\vect{e}^{(L)}_{i}$ is an activation vector of layer $L$ where the activation of all nodes is zero except for the $i$-th node whose activation is one. Otherwise, we have
\begin{equation}\label{eq:backprop_not_occuring_L}
    \frac{\partial \vect{a}^{(L)}}{ \partial w} = \mat{J}_{\varsigma}^{(L)} \weightsmat^{(L)} \frac{\partial \vect{a}^{(L-1)}}{\partial w}\,.
\end{equation}
To evaluate this expression, one needs to go further back in the layers and compute the derivatives $\partial \vect{a}^{(L-1)}/\partial w$ given by
\begin{equation}
\frac{\partial \vect{a}^{(L-1)}}{\partial w} = \frac{\partial \vect{a}^{(L-1)}}{ \partial \vect{z}^{(L-1)}} \frac{\partial \vect{z}^{(L-1)}}{ \partial w} = \mat{J}_{\varsigma}^{(L-1)} \left( \frac{\partial \weightsmat^{(L-1)}}{ \partial w}\vect{a}^{(L-2)} + \weightsmat^{(L-1)} \frac{\partial \vect{a}^{(L-2)}}{\partial w} \right)\,.
\end{equation}
Again, if $w$ is part of layer $L-1$, i.e., $\weight=\weight_{i,j}^{(L-1)}$, we have
\begin{equation}
    \frac{\partial \vect{a}^{(L-1)}}{ \partial w} = \mat{J}_{\varsigma}^{(L-1)} \vect{e}^{(L-1)}_{i} a_{j}^{(L-2)}\,.
\end{equation}
Otherwise, we have
\begin{equation}\label{eq:backprop_3}
    \frac{\partial \vect{a}^{(L-1)}}{ \partial w} = \mat{J}_{\varsigma}^{(L-1)} \weightsmat^{(L-1)} \frac{\partial \vect{a}^{(L-2)}}{\partial w}\,.
\end{equation}
Recognizing the recursive nature of the computation, we have the following relation
\begin{equation}\label{eq:backprop_linear}
    \frac{\partial \vect{a}^{(l)}}{ \partial w} = \mat{J}_{\varsigma}^{(l)} \weightsmat^{(l)} \mat{J}_{\varsigma}^{(l-1)} \weightsmat^{(l-1)}\dots \mat{J}_{\varsigma}^{(l'+1)} \weightsmat^{(l'+1)} \frac{\partial \vect{a}^{(l')}}{\partial w}\,,
\end{equation}
given that $w$ is not a weight of the layers $l'$ through $l$. If $w$ is part of layer $l'$, i.e., $\weight=\weight_{i,j}^{(l')}$, we instead have
\begin{equation}\label{eq:backprop_linear_2}
    \frac{\partial \vect{a}^{(l)}}{ \partial w} = \mat{J}_{\varsigma}^{(l)} \weightsmat^{(l)} \mat{J}_{\varsigma}^{(l-1)} \weightsmat^{(l-1)}\dots \mat{J}_{\varsigma}^{(l'+1)} \weightsmat^{(l'+1)} \mat{J}_{\varsigma}^{(l')} \vect{e}^{(l')}_{i} a_{j}^{(l'-1)}\,.
\end{equation}

Finally, we have all the ingredients to formulate the backpropagation algorithm. Recall that our goal is to compute the derivative in~\cref{eq:backprop_main} with respect to all tunable weights. To do that efficiently, we first initialize the following ``deviation'' at the output layer
\begin{equation}
    \vect{\Delta}^{(L)} = \frac{\partial \ell(\vect{a}^{(L)}(\vect{x}_{i}),\vect{y}_{i})}{\partial \vect{a}^{(L)}} \odot \mat{J}_{\varsigma}^{(L)}\,,
\end{equation}
where $\odot$ denotes an element-wise (Hadamard) product.\footnote{It simplifies to a regular scalar product given a single output node.} This intermediate quantity turns out to be useful throughout the computation. Taking the \ac{MSE} loss as an example, this would correspond to (see~\cref{eq:backprop_mse})
\begin{equation}\label{eq:backprop_MSE_last_Delta}
    \vect{\Delta}^{(L)} = 2(\vect{a}^{(L)} - \vect{y}_{i}) \odot \mat{J}_{\varsigma}^{(L)}\,.
\end{equation}
From~\cref{eq:backprop_occuring_L}, it follows that the contributions to the derivative of the loss function with respect to a weight $\weight_{i,j}^{(L)}$ in layer $L$ are given by
\begin{equation}\label{eq:backprop_MSE_last_der}
    2(\vect{a}^{(L)} - \vect{y}_{i}) \odot \mat{J}_{\varsigma}^{(L)} \vect{e}^{(L)}_{i} a_{j}^{(L-1)} =  \vect{\Delta}^{(L)} \vect{e}^{(L)}_{i} a_{j}^{(L-1)}\,.
\end{equation}
The final derivative $\partial \ell(\vect{a}^{(L)}(\vect{x}_{i}),\vect{y}_{i}) / \partial \weight_{i,j}^{(L)}$ is then obtained by summing up all components of this vector
\begin{equation}
    \partial \ell(\vect{a}^{(L)}(\vect{x}_{i}),\vect{y}_{i}) / \partial \weight_{i,j}^{(L)} = \sum_{k} \left( \vect{\Delta}^{(L)} \vect{e}^{(L)}_{i} a_{j}^{(L-1)}\right)_{k}\,,
\end{equation}
i.e., all individual contributions to the inner product given in~\cref{eq:backprop_main}. Having computed the derivative with respect to all weights in layer $L$, we move one layer backward, hence the name \stress{back}propagation. From~\cref{eq:backprop_not_occuring_L}, it follows that the contributions to the derivative of a weight $\weight_{i,j}^{(L-1)}$ in layer $L-1$ are given by
\begin{equation}\label{eq:backprop_MSE_2nd_last_der}
    \vect{\Delta}^{(L-1)} \vect{e}^{(L-1)}_{i} a_{j}^{(L-2)}\,,
\end{equation}
where
\begin{equation}\label{eq:backprop_MSE_2nd_last_Delta}
\vect{\Delta}^{(L-1)} = \vect{\Delta}^{(L)} \weightsmat^{(L)} \mat{J}_{\varsigma}^{(L-1)}\,.
\end{equation}
Notice the intimate connection between the above procedure and the expressions in~\cref{eq:backprop_linear} and \cref{eq:backprop_linear_2}. Thus, through recursion we have
\begin{equation}\label{eq:backprop_recursive_main}
    \vect{\Delta}^{(l-1)} = \vect{\Delta}^{(l)} \weightsmat^{(l)} \mat{J}_{\varsigma}^{(l-1)}\,.
\end{equation}
This process is repeated until one arrives at the first layer. At the end of this reverse pass through the \ac{NN}, one has computed the desired derivative (\cref{eq:backprop_main}) with respect to all tunable weights. During the backpropagation algorithm, the value of all activations $\{ \vect{a}^{(l)}(\vect{x}_{i}) \}_{l=0}^{L}$ and the derivatives of the corresponding activation function evaluated at that activation $\{ \mat{J}_{\varsigma}^{(l)}(\vect{x}_{i}) \}_{l=1}^{L}$ must be known. In order to avoid any recomputation, one performs an evaluation of the network for the given input $\vect{x}_{i}$, i.e., a forward pass, and caches all the required intermediate computation results before executing the backpropagation algorithm, i.e., the reverse pass.

\begin{figure}
    \begin{center}
    \includegraphics[width=\textwidth]{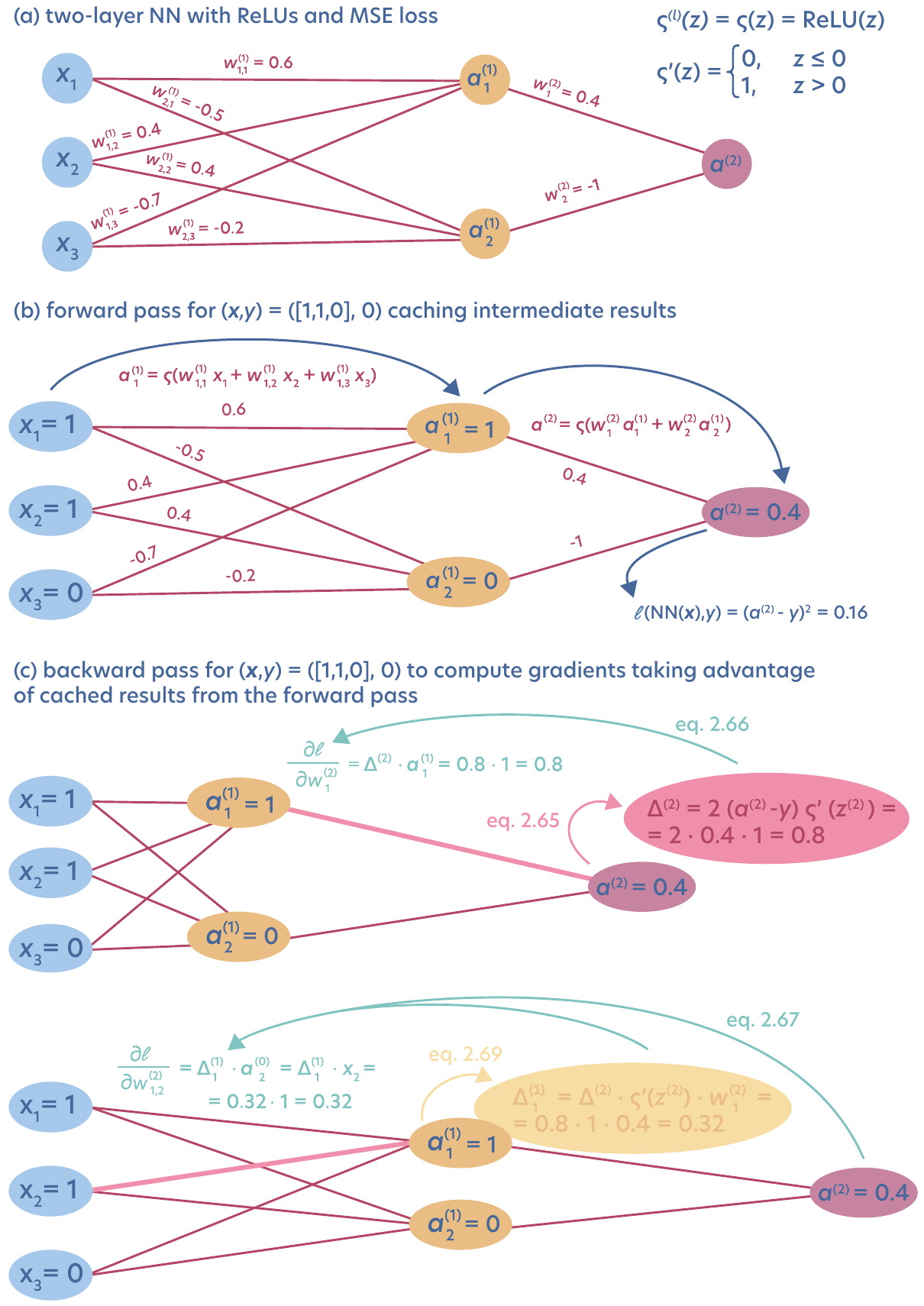}
    \end{center}
    \caption[Backpropagation on a simple feedforward \ac{NN}]{Backpropagation on a simple example of a two-layer feedforward \ac{NN} presented in (a). (b) Forward pass through the network. (c) Reverse pass calculating derivatives.}
    \label{fig:backpropagation}
\end{figure}

To further illustrate how backpropagation works, let us calculate both the forward and reverse passes \stress{explicitly} on the example of a simple two-layer \ac{NN} with the \ac{MSE} as the loss function and ReLUs as activation functions, $\varsigma^{(l)}(z) = \varsigma(z) = \mathrm{ReLU}(z) = \mathrm{max}(0,z)$ which act element-wise. The derivative of ReLU is 1 for $z > 0$ and 0 otherwise.\footnote{Formally, ReLU is non-differentiable at $z=0$. In numerical practice, the derivative at $z=0$ is usually set to 0.} We randomly initialize the weights of this \ac{NN} and ignore biases, see panel (a) of \cref{fig:backpropagation}. The forward pass for an exemplary input-output pair is presented in \cref{fig:backpropagation}(b). Importantly, the intermediate computation results are cached. This includes the activations of all nodes, $\vect{a}^{(1)}$ and $a^{(2)}$, and the derivatives of the corresponding activation functions. Then, the backward pass starts in panel (c) of \cref{fig:backpropagation}. Here, we only focus on the calculation of the derivative of $\ell$ with respect to two weights coming from different layers, $\weight^{(2)}_1$ and $\weight^{(1)}_{1,2}$. The first step is to compute the deviation $\Delta^{(2)}$ on the last layer from \cref{eq:backprop_MSE_last_Delta}, after which $a^{(2)}$ can be erased from memory. Then, using \cref{eq:backprop_MSE_last_der}, we can calculate the derivatives of $\ell$ with respect to any weight $\vect{w}^{(2)}$ in the last layer, and activations $\vect{a}^{(1)}$ can be discarded. The next step is to compute the deviations $\Delta^{(1)}_1$ and $\Delta^{(1)}_2$ on the second-last layer following \cref{eq:backprop_recursive_main}. After this computation, $\Delta^{(2)}$ can be discarded. Using $\Delta^{(1)}_1$ and $a^{(0)}_2 = x_2$ and following \cref{eq:backprop_MSE_2nd_last_der}, we can compute $\partial \ell / \partial \weight^{(1)}_{1,2}$. In the case of a two-layer \ac{NN}, this concludes the calculation of the gradient. Note that at every step memory can be freed by erasing cached results from the forward pass.

At this point, an interesting question might arise. Why do we compute the derivative in a reverse pass instead of a forward pass? To answer this question, we first have to formulate the corresponding ``forward-propagation algorithm''. For each weight $\weight_{i,j}^{(l)}$ with respect to which one wants to compute a derivative (\cref{eq:backprop_main}), one first computes 
\begin{equation}
    \frac{\partial \vect{a}^{(l)}}{\partial \weight_{j,i}^{(l)}} = \mat{J}_{\varsigma}^{(l) }\vect{e}^{(l)}_{i}  a_{j}^{(l-1)}.
\end{equation}
Now, one can directly make use of the relation in~\cref{eq:backprop_linear} to obtain $\partial \vect{a}^{(L)}/ \partial \weight_{i,j}^{(l)}$, from which the final derivative can be computed via~\cref{eq:backprop_main}. Notice that the difference between the backpropagation and forward-propagation algorithm amounts to evaluating the expression in~\eqref{eq:backprop_linear_2} from left-to-right (backward) or right-to-left (forward), respectively.

The forward-propagation algorithm also requires knowledge of the activations and derivatives of the activation functions. In this case, however, one does not need to cache any of the results. Instead, the computation of the derivatives can be carried out in parallel with the forward pass (i.e., evaluation of the \ac{NN}). This is because information flows forwards from layer-to-layer and no information from the earlier layers is explicitly required at later stages. Besides this difference in memory cost, one can identify two key distinctions between an algorithm based on forward propagation and the backpropagation algorithm. First, in forward propagation algorithms, lots of redundant computations are performed, given that the derivatives in later layers have to be computed each time. This redundancy is not present in backpropagation. Secondly, forward propagation involves unnecessary intermediate calculations where the derivatives of individual nodes are computed. Ultimately, this culminates in the fact that the number of passes through the \ac{NN} in the forward-propagation algorithm scales with the number of tunable weights, whereas this is not the case in the backpropagation algorithm. This is why backpropagation is generally preferred over forward-propagation-based algorithms for computing gradients in \acp{NN}, in particular, \acp{DNN} featuring a large number of tunable parameters. As such, backpropagation has played a key role in the success of \ac{DL} and enabled the widespread application of \acp{NN}. For a more general in-depth discussion of these concepts, including implementation details, see~\cref{sec:hot-topics:dp}.
\subsection*{Further reading}

\begin{enumerate}

    \item Bishop, C. M. (2006). \href{https://www.microsoft.com/en-us/research/uploads/prod/2006/01/Bishop-Pattern-Recognition-and-Machine-Learning-2006.pdf}{\textit{Pattern Recognition and Machine Learning}}. Springer ``Information Science and Statistics'' series. The standard book about standard \ac{ML}~\cite{bishop:2006}.
    
    \item Goodfellow, I., Bengio, Y. \& Courville, A. (2016). \href{https://www.deeplearningbook.org/}{\textit{Deep Learning}}. An~MIT Press book. One of the best textbooks on \ac{DL} with an explanation of all preliminaries~\cite{goodfellow:2016}.
    
    \item Mehta, P. \textit{et al.} (2019). \href{https://doi.org/10.48550/arXiv.1803.08823}{\textit{A~high-bias, low-variance introduction to machine learning for physicists}}. Phys. Rep. 810, 1-124. For the physicist-friendly introduction to \ac{ML}~\cite{mehta:2019}.
    
    \item Zhang, A. \textit{et al.} (2021). \href{https://d2l.ai/}{\textit{Dive into Deep Learning}}. Interactive \ac{DL} book with code, math, and discussions. Implemented with NumPy/MXNet, PyTorch, and TensorFlow~\cite{zhang2021dive}.
    
    \item Recordings of lectures on \href{https://pad.gwdg.de/s/Machine_Learning_For_Physicists_2021}{``\ac{ML} for physicists''} from 2020/21 and \href{https://pad.gwdg.de/s/2021_AdvancedMachineLearningForScience}{``Advanced \ac{ML} for physics, science, and artificial scientific discovery''} from 2021/22 by Florian Marquardt.
    
    \item Introductory \ac{ML} course developed specifically with STEM students in mind: \href{https://ml-lectures.org/docs/index.html}{ML-lectures.org} and accompanying content: Neupert, T. \textit{et al.} (2021). \href{https://doi.org/10.48550/arXiv.2102.04883}{\textit{Introduction to machine learning for the sciences}}. arXiv:2102.04883~\cite{Neupert21}.
    
    \item Carrasquilla, J. \& Torlai, G. (2021). \href{https://doi.org/10.48550/arXiv.1803.08823}{\textit{How to use neural networks to investigate quantum many-body physics}}. PRX Quantum 2, 040201. Tutorial on \ac{ML} for selected physical problems with code~\cite{Carrasquilla2021PRXQuantum}.
\end{enumerate}

\clearpage
\section{Phase classification}
\index{phase classification}
\label{sec:phase_class}
    One of the fields in physics where \acf{ML} and, in particular, \acfp{NN}, could be especially useful is condensed matter physics~\cite{Carleo2019RevModPhys}, which revolves around the study of the collective behavior of interacting particles. The difficulties associated with describing such systems arise due to the rapid growth of the number of degrees of freedom as the particle number grows, leading to a~larger configuration space. The ``standard'' approach to circumvent these challenges is to find suitable order parameters -- quantities that represent the important ``macroscopic'' degrees of freedom in a~system without keeping track of all the microscopic details. The order as quantified by these order parameters naturally separates matter into different states, i.e., phases~\cite{sachdev:2011,goldenfeld:2018}. For some systems, the order parameter is quite simple: in ferromagnets, for example, the order parameter simply corresponds to the magnetization, which is given by a~sum of \stress{local} magnetic moments. In general, however, the identification of order parameters and the classification of matter into distinct phases are difficult tasks. Topological phases of matter, for example, are characterized by topological properties that are intrinsically \stress{non-local}. The identification of order parameters represents a~crucial first step toward understanding the physics that underlies a~many-body system, and identifying an~appropriate order parameter for novel phases of matter typically requires lots of physical intuition and educated guessing.
    
    On the other hand, in fields such as computer vision, it has been demonstrated that \acp{NN} can be trained to correctly classify intricate sets of labeled data naturally living in high dimensions (see MNIST~\cite{lecun:1998} or CIFAR~\cite{krizhevsky:2009}). This motivates us to explore \ac{ML} techniques as a~novel tool to probe the enormous state space of relevant many-body systems that are currently intractable with other algorithms~\cite{Carrasquilla2020AdvPhys}. Among all potential applications of \ac{ML} to condensed-matter physics, learning phases from (simulated or experimental) data is a~particularly intriguing one: It could allow us to discover new phases and new physics without prior human knowledge or supervision. In what follows, we aim to give the reader a~first introduction to the field of phase classification using \ac{ML}.

\subsection{Prototypical physical systems for the study of phases of matter} 
    In the following, we briefly describe the two prototypical physical systems for which we demonstrate the task of phase classification\index{phase classification} in the next sections: the Ising model ~\cite{onsager:1944}, which exhibits a~\stress{symmetry-breaking} phase transition and can be characterized by a~simple local order parameter, as well as the \acf{IGT} ~\cite{wegner:1971}, which shows a~\stress{topological} phase without a~local order parameter.\footnote{In the Landau paradigm of phase transitions~\cite{landau1:1937,landau2:1937}, changes between phases of matter are fundamentally connected to changes in the underlying symmetries. Interestingly, Landau's symmetry-breaking theory of phase transitions breaks down for topological phases of matter~\cite{wen:1990}.} 
    \subsubsection{Ising model}\label{sec:ising_model}
\begin{figure}[t]
        \begin{center}
        \includegraphics[width=\columnwidth]{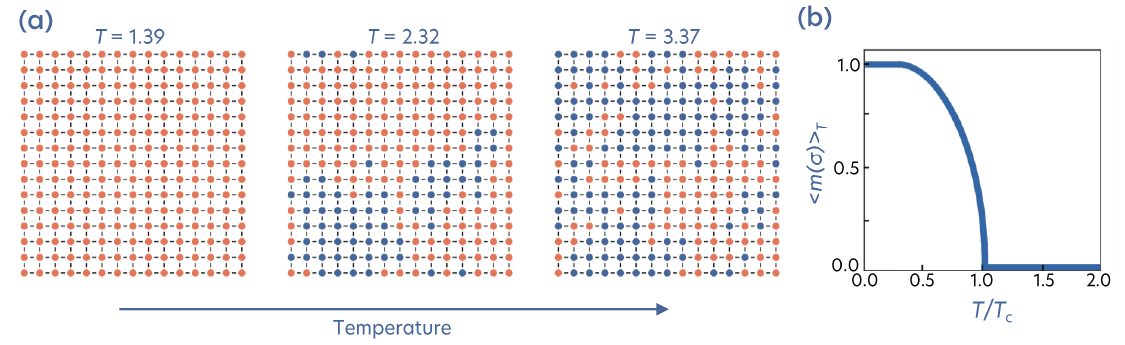}
        \end{center}
        \caption[Ising model]{(a) Example spin configuration samples of the Ising model with $J=1$ and $k_{\rm B}=1$ at various temperatures, where $T_{\rm c} \approx 2.27$ (see \cref{eq:ising_Tc}). Here, the blue (orange) colored dots on each lattice site denote the value of the spin variable at that site $\sigma_{k}=1$ ($\sigma_{k}=-1$). Panel reproduced from Ref.~\cite[Notebook A1]{OurSchoolRepo}. (b) Mean magnetization per site $\estimate{m(\vect{\sigma})}{T}$ of the Ising model as a~function of the temperature $T$.}
        \label{fig:Ising_model_samples}
        \end{figure}

        We consider the two-dimensional square-lattice ferromagnetic Ising model, which is one of the simplest classical statistical models to show a~phase transition and serves as a~simple description of ferromagnetism. Ferromagnetism arises when a~collection of spins aligns, yielding a~net magnetic moment that is macroscopic in size. In the Ising model , for each lattice site $k$ there is a~discrete (classical) spin variable $\sigma_{k} \in \{+1,-1 \}$ leading to a~state space of size $2^{N}$ given $N$~lattice sites. The energy of a~spin configuration is specified by the following Hamiltonian
        \begin{equation}\label{eq:ising_H}
        H(\vect{\sigma}) = - J \sum_{\langle i,j\rangle} \sigma_{i}\sigma_{j},
        \end{equation}
        where the sum runs over nearest-neighboring sites (with periodic boundary conditions), and $J$~is the interaction strength $J>0$ (ferromagnetic interaction).\footnote{The case $J<0$ corresponds to the two-dimensional square-lattice \stress{antiferromagnetic} Ising model which exhibits a~phase transition at the same critical temperature.} Let us assume that the system is at equilibrium at an~inverse temperature $\beta = 1/k_{\rm B}T$, where $k_{\rm B}$ is the Boltzmann constant and $T$ the temperature. Then, the probability of finding the system in a~state with a~spin configuration $\vect{\sigma}$ is described by the Boltzmann distribution
        \begin{equation}\label{eq:boltzmann_distr}
        P_{T}(\vect{\sigma}) = \frac{e^{-\beta H(\vect{\sigma})}}{Z_{T}}.
        \end{equation}
        Here $Z_{T} = \sum_{\vect{\sigma}} e^{-\beta H(\vect{\sigma})}$ is the partition function, where the sum runs over all possible spin configurations. Example spin configurations of the Ising model at various temperatures are shown in \cref{fig:Ising_model_samples}(a). Using~\cref{eq:boltzmann_distr} the expectation value of a~given observable $O(\vect{\sigma})$ can be expressed as
        \begin{equation}
        	\estimate{O(\vect{\sigma})}{T} = \sum_{\vect{\sigma}} P_{T}(\vect{\sigma}) O(\vect{\sigma}).
        \end{equation}
        For example, the observable corresponding to the magnetization per site is given by
        \begin{equation}
        m(\vect{\sigma}) = \frac{1}{N} \sum_{i} \sigma_{i}.
        \end{equation}

        In 1944, Onsager~\cite{onsager:1944} obtained the following analytical expression for the critical temperature
        \begin{equation}\label{eq:ising_Tc}
        T_{\rm c} = \frac{2J}{k_{\rm B} \ln(1+\sqrt{2})},
        \end{equation}
        at which a~phase transition between a~\stress{high-temperature paramagnetic} (disordered) phase and a~\stress{low-temperature ferromagnetic} (ordered) phase occurs, see \cref{fig:Ising_model_samples}. For temperatures below the critical temperature $T_{\rm c}$, spontaneous magnetization occurs, i.e., the interaction is sufficiently strong to cause neighboring spins to spontaneously align, leading to a~non-zero mean magnetization. At temperatures above $T_{\rm c}$, thermal fluctuations completely dominate over any alignment of spins, and a~zero magnetization is observed. As such, the magnetization serves as an~order parameter, which is zero within the disordered (paramagnetic phase) and approaches one in the ordered (ferromagnetic phase), see \cref{fig:Ising_model_samples}(b).
        
    \subsubsection{Ising gauge theory}\label{sec:ising_gauge_theory}
    
        \begin{figure}[t]
        \begin{center}
        \includegraphics[width=0.8\columnwidth]{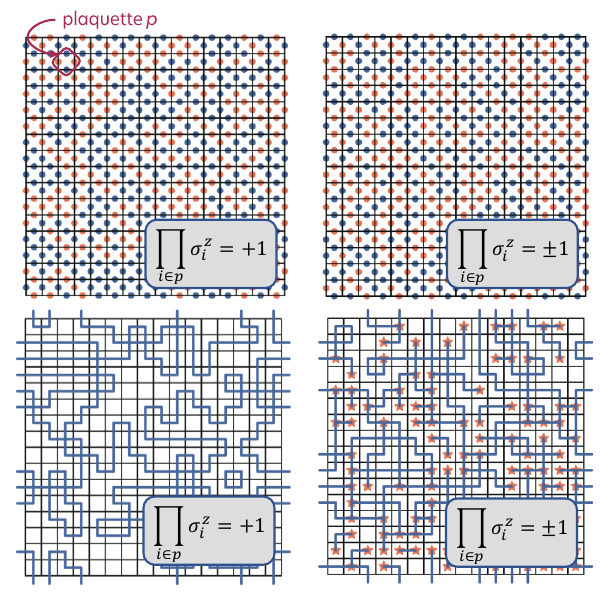}
        \end{center}
        \caption[Ising gauge theory]{The upper panels show example spin configuration samples of \ac{IGT} at $T=0$ (left panel) and $T\rightarrow \infty$ (right panel) where local constraints are satisfied for all (some) plaquettes, respectively. An~exemplary plaquette is marked in red. The lower panels show the corresponding dual representation, where the stars highlight loop breakage. Reproduced with Ref.~\cite[Notebook A1]{OurSchoolRepo}.}
        \label{fig:IGT_samples}
        \end{figure}
    
         One of the most exciting research areas is the classification of phases that do not have a~local order parameter but rather a~global one. Examples of systems that exhibit such phases are band topological insulators and topological superconductors~\cite{bernevig2013topological}. Detecting topological phases is a~challenging task from the experimental point of view because, in general, experimentalists have access only to local observables. In this context, \ac{ML} techniques can be of great help~\cite{Kaming2021,Zhai2018PRB,Zhai2018PRL,1901.03346, Holanda2020MLtopo,10.21468/SciPostPhysCore.6.4.087,PhysRevB.97.134109}.
     
        The \acf{IGT} ~\cite{wegner:1971} is the prototypical example of a~system which exhibits a~topological phase of matter. Like the Ising model, the \ac{IGT} is also a~classical spin model ($\sigma_{k} \in \{+1,-1 \}$) defined on a~square lattice (with periodic boundary conditions). Here, however, the spins are placed on the lattice bonds. It is described by the following Hamiltonian
        \begin{equation}
        	H(\vect{\sigma}) = - J \sum_{p} \prod_{i \in p} \sigma_{i},
        \end{equation}
        where $p$ refers to plaquettes on the lattice, see \cref{fig:IGT_samples}. The ground state of this Hamiltonian is a~highly degenerate manifold spanned by all states that meet the local constraint that the product of spins along each plaquette is $\prod_{i \in p} \sigma_{i}=1$. As such, this ground state corresponds to a~topological phase of matter. In systems of finite size, the violations of the local constraints are strongly suppressed, and the system exhibits a~slow transition from the low-temperature topological phase to the high-temperature phase with violated constraints. This allows for the definition of a~crossover temperature $T_{\rm c}$ defined by the first appearance of a~violated local constraint.\footnote{Note that as we increase the system size $T_{\rm c}\rightarrow 0$, i.e., the crossover temperature vanishes in the thermodynamic limit. As such, the \ac{IGT} does not exhibit a~phase transition at a~non-zero temperature.} 
        
        There exists an~interesting representation that highlights the topological character of the ground state of the \ac{IGT}: connect the edges of the lattice that contain spins with the same orientation and form loops. The ground-state phase is then characterized by the property that all these loops are closed; the violation of a~constraint results in the appearance of an~open loop, see \cref{fig:IGT_samples}. Looking at typical spin configuration samples of the \ac{IGT} makes clear that its phases are hard to distinguish visually without prior knowledge of the local constraints or the corresponding dual representation. As such, \ac{IGT} and other systems characterized by non-local and long-range correlations pose a~hard problem for any phase classification algorithm.

\subsection{Unsupervised phase classification without neural networks}\label{sec:unsup_phase_no_NN}\index{unsupervised learning}
    Having introduced the Ising model and the \ac{IGT}, let us discuss how we can classify their respective phases of matter. In particular, we are concerned with unsupervised\index{phase classification!unsupervised phase classification} \ac{ML} algorithms. They work with training data that do not need to be labeled (see \cref{sec:typeoflearning}). Unsupervised learning algorithms must \stress{by itself} discover the relevant patterns in a~training data set. As such, these algorithms represent a~primary candidate for the autonomous discovery of new phases as they do not require prior labeling of the samples by the phase they belong to.
    
    In particular, we discuss algorithms that perform a~dimensionality reduction\index{dimensionality reduction}. In dimensionality reduction, we are concerned with projecting the input data into a~lower-dimensional space. While any dimensionality reduction necessarily leads to an~information loss, one aims to discard only information in the input data that is less relevant to the problem at hand. In particular, it is believed that real-world data often resides on a~low-dimensional manifold within the original space~\cite{fefferman:2016}. For example, one expects that the set of images one would like to classify constitutes a~small subspace of all possible images. In this case, the data can be effectively described by fewer degrees of freedom. Clearly, such an~approach lends itself naturally to distinguishing between different phases of matter and detecting phase transitions in condensed matter systems: we want to discard the information-rich but complicated microscopic description of the system for the sake of a~simpler macroscopic description, e.g., in the form of an~order parameter.
    
    Once we have performed the dimensionality reduction, we may already learn a~lot about the given problem by visualizing the data within the low-dimensional representation space\index{representation space}. We tend to think that samples from the same phase of matter should be more similar to each other than to samples from another phase. If the dimension reduction technique preserves some of this similarity, we expect this to reveal itself in the data visualization. However, this is not guaranteed to work in general. We see an~example of such a~failure in the following. 
    
    Going beyond visualization, we can process the data further, e.g., using \stress{clustering}\index{clustering} methods. Clustering is one of the most fundamental unsupervised learning methods used to group unlabeled data into clusters of similar data points, where the similarity is assessed by a~distance measure. In our context, the clusters would ideally correspond to the different phases of matter present in the data. There exist many different clustering algorithms suited for different types of data, with $k$-means clustering being one of the simplest (see Ref.~\cite{mehta:2019} for further details).
    
    Clustering\index{clustering} can, in principle, be performed without dimensionality reduction as a~pre-processing step. However, dimensionality reduction may help in several aspects~\cite{wang:2016,mehta:2019}. Firstly, clustering typically relies on the Euclidean distance being a~good measure of similarity.\footnote{In general, whether clustering succeeds or not depends on whether the choice of distance measure (be it Euclidean or not) is a~good measure of similarity.} The distance between two data points in the original high-dimensional representation may, however, not be particularly relevant as it is believed to often reside on a~non-Euclidean manifold. Dimensionality reduction techniques can allow for the identification of a~low-dimensional Euclidean representation of the data. The Euclidean distance between data points within this representation is often physically more meaningful, resulting in a~better clustering. Secondly, performing dimensionality reduction as a~pre-processing step helps to alleviate the problems of the curse of dimensionality experienced when clustering data in high-dimensional spaces. Finally, identifying a~low-dimensional representation also helps to better visualize and understand the clustering that is eventually obtained.
    
    \subsubsection{Principal component analysis} \label{ssec:pca}\index{unsupervised learning}
        As an~example, we consider \acf{PCA}\index{principal component analysis}, which is a~common method to perform dimensionality reduction. \Ac{PCA} identifies mutually orthogonal directions, called \acfp{PC}, in the data space along which the linear correlation in the data vanishes. We rank each \ac{PC} based on the variance of the data along the corresponding direction. To reduce the dimensionality of our space, we discard the \acp{PC} along which the data shows the least variance. As such, in \ac{PCA} directions along which the data exhibits a~\stress{large variance} are considered to contain the most \stress{important information}. In our case, ideally, the data (raw spin configuration samples) naturally splits into different clusters corresponding to the individual phases of the system when displayed in their low-dimensional representation.
    
        \begin{figure}[t]
        \begin{center}
        \includegraphics[width=\ToggleForCUP{0.8\textwidth}{\columnwidth}]{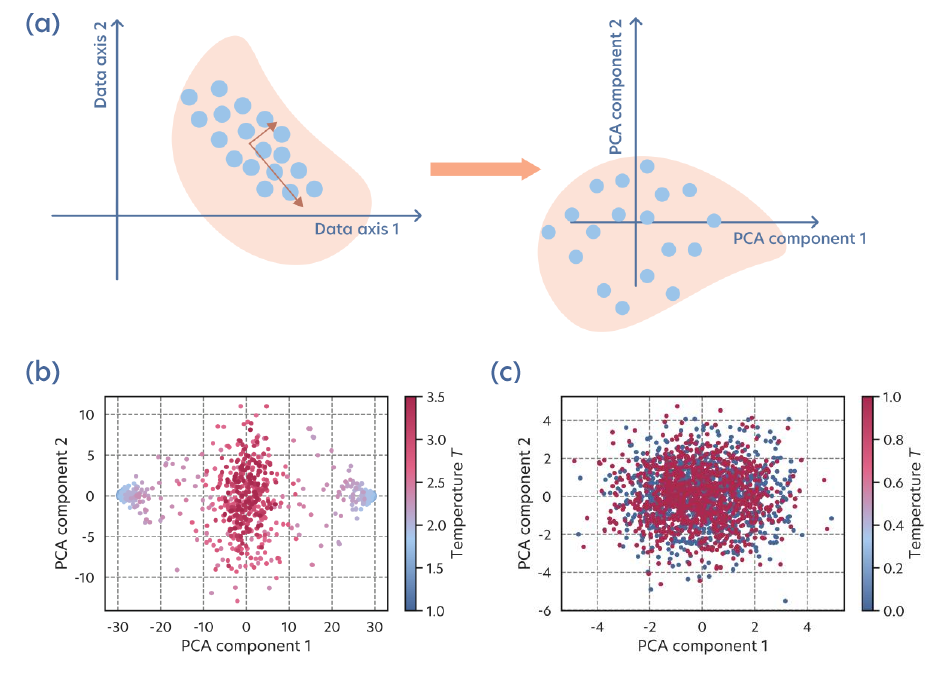}
        \end{center}
        \caption[Phase classification using principal component analysis]{(a) Illustration of the principle behind \ac{PCA} applied to data points (blue) living in a~two-dimensional feature space. Orange vectors denote the first two principal components on which we project the data (right panel). \ac{PCA} applied to the spin configuration samples of (b) the Ising model and (c) \ac{IGT} with $k=2$. For the Ising model, the data consists of 50 spin configurations (linear lattice size $L=30$) sampled using Monte Carlo methods at temperatures $T$ ranging from $T_{1}=1$ to $T_{20}=3.5$ in equidistant steps. For the IGT, the data consists of 1000 spin configurations (linear lattice size $L=16$) drawn within the topological phase and the disordered phase at high temperatures. Panels (b) and (c) are reproduced from Ref.~\cite[Notebook A1]{OurSchoolRepo}.}
        \label{fig:PCA}
        \end{figure}
        
        To be more precise, we consider the case where we are given $\datasize$ data points $\{ \vect{x}_{1},\vect{x}_{2},...,\vect{x}_{\datasize} \}$ each living in a~$\featnum$-dimensional feature space $\realset^{\featnum}$ with zero mean $\bar{\vect{x}}=\frac{1}{\datasize} \sum_{i=1}^{\datasize} \vect{x}_{i}=0$. Note that real-life data typically does not have zero mean. In this case, the data first needs to be transformed by subtracting the mean element-wise. We define the $\datasize \times \featnum$ design matrix $\mat{X} = [\vect{x}_{1},\vect{x}_{2},...,\vect{x}_{\datasize}]^{\rm T}$. The symmetric $\featnum \times \featnum$ empirical covariance matrix is then given as $\mat{\Sigma} = \frac{1}{\datasize} \mat{X}^\transpose \mat{X}$. Here, the $i$-th diagonal entry of the covariance matrix $\mat{\Sigma}_{ii}$ corresponds to the variance of the $i$-th feature over the entire data and the off-diagonal entries $\mat{\Sigma}_{ij}$ correspond to the covariance between feature $i$ and feature $j$. The basis in which the correlations between features vanish corresponds to the eigenbasis of $\mat{\Sigma}$ in which $\mat{\Sigma}$ appears diagonal. Consequently, the problem of finding directions along which the linear correlation in the data vanishes reduces to diagonalizing $\mat{\Sigma}$, i.e., finding its eigenvectors (or \acp{PC}) $\{ \vect{v}_{1},\vect{v}_{2},...,\vect{v}_{\featnum} \}$ and eigenvalues $\{ \lambda_{1},\lambda_{2},...,\lambda_{\featnum} \}$. Here, the eigenvalue $\lambda_{i}$ corresponds to the variance of the data along the direction given by $\vect{v}_{i}$. We denote $\tilde{\lambda}_{j} = \lambda_{j}/\sum_{i=1}^{\featnum} \lambda_{i}$ as the ratio of explained variance contained in the $j$-th \ac{PC}. We refer to the appendix for a~mathematical derivation of the procedure. Dimensionality reduction is then performed by selecting the first $k$ \acp{PC} with the largest ratios of explained variance $\tilde{\lambda}$ and projecting the data into this space of reduced dimensionality. The projection is performed by the linear transformation $\tilde{\mat{X}}=\mat{X}\tilde{\mat{V}}$, where $\tilde{\mat{V}}=[\vect{v}_{1},\vect{v}_{2},...,\vect{v}_{k}]$ and $\tilde{\mat{X}}$ is the projected design matrix. Note that one has to choose $k$, the number of \acp{PC} to keep. This can be done in an~\stress{ad-hoc} fashion that may be problem-specific or, e.g., by choosing the minimal number of \acp{PC} such that $\sum_{i=1}^{k} \tilde{\lambda}_{i} \geq \tilde{\lambda}_{\rm thresh}$, where $\tilde{\lambda}_{\rm thresh}$ is the desired threshold explained variance ratio.
        The procedure is summarized in \cref{alg:pca}. For an~intuitive understanding of the procedure, we refer to \cref{fig:PCA}(a) \& (b): in this example, the data resides in a~two-dimensional feature space. After subtracting the data mean, \ac{PCA} identifies the first \ac{PC} that contains the largest proportion of the data variance.
        \Ac{PCA} can not only be understood as variance maximization but also as a~minimization of a~reconstruction error of a~linear transformation.
        The proof of this equivalence can be found in \cref{appendix_PCA}. For further details, see, e.g., Ref.~\cite{mehta:2019}.
        
        \begin{algorithm}
        \caption{\Acf{PCA}}\label{alg:pca}
        \begin{algorithmic}
        \Require Hyperparameter $k$ (dimensionality of the projected data)
        \Require Design matrix $\mat{X} \in \realset^{\datasize \times \featnum}$
        \State $\mat{X} \gets \mat{X} - \mathrm{mean}(\mat{X})$ \Comment{Remove mean element-wise}
        \State $\mat{\Sigma} \gets \mat{X}^\transpose \mat{X} / \datasize$ \Comment{Construct empirical covariance matrix}
        \State $\mat{V} \gets \mathrm{Eigenvectors}(\mat{\Sigma})$ \Comment{Find eigenvectors and order them by descending eigenvalue}
        \State $\tilde{\mat{V}} \gets \mat{V}[:,:k]$ \Comment{Keep only first $k$ eigenvectors} \\
        \Return $\tilde{\mat{X}} \gets \mat{X}\tilde{\mat{V}} \in \realset^{\datasize \times k}$
        \end{algorithmic}
        \end{algorithm}

        Now, we can readily apply \ac{PCA} to our spin configuration samples.
        \Cref{fig:PCA}(b,c) shows the results of \ac{PCA} applied to spin configuration samples of the Ising model and \ac{IGT}, respectively. For the Ising model, \ac{PCA} separates the data into three clusters -- a~high-temperature cluster corresponding to the disordered phase, as well as two low-temperature clusters corresponding to the ordered phase with either positive or negative magnetization. Further analysis shows that the first \ac{PC} corresponds to the magnetization~\cite{wang:2016}. By drawing a~vertical decision boundary (perpendicular to \ac{PC}1), which separates the high-temperature cluster and a~low temperature cluster a~rough estimate of the critical transition temperature can be obtained as $T_{\rm c,{\rm PCA}}\approx 2.3$ which is in agreement with the Onsager solution. In the case of the \ac{IGT}, \ac{PCA} fails to cluster the data into the two prevalent phases [see \cref{fig:PCA}(c)]. This is because \ac{PCA} is restricted to linear transformations of the input data. While this is sufficient to encode simple local order parameters~\cite{wang:2016,wetzel:2017,hu:2017}, such as the magnetization in the case of the Ising model, linear transformations are not sufficient to compute topological features, i.e., non-local correlations in the data~\cite{hu:2017}.
    
        As illustrated by the failure of \ac{PCA} in the case of the \ac{IGT}, the restriction of \ac{PCA} to linear transformations of the input space severely limits its performance. That is, one may not be able to find the optimal set of directions to perform dimensionality reduction using \ac{PCA}. In particular, the low-dimensional manifold on which the data resides within the original space may not necessarily be parametrized by linear transformations of the original coordinates. In such cases, a~dimensionality reduction using \ac{PCA} does not preserve the relative pairwise distance, or similarity, between data points with respect to the manifold. However, this is a~desired property for any algorithm that aims at performing dimensionality reduction. This problem is tackled by nonlinear dimensionality reduction techniques, such as the kernel \ac{PCA} (k\ac{PCA}) ~\cite{6790375} (see \cref{sec:gp} on the kernel trick), the \ac{t-SNE}~\cite{vandermaaten:2008}, or uniform manifold approximation and projection (UMAP)~\cite{mcinnes:2018}. In the following section, we briefly describe \ac{t-SNE}. 
        
        \subsubsection{t-Distributed stochastic neighbor embedding}\label{ss:tSNE}\index{unsupervised learning}
        
Stochastic neighbor embedding~\cite{Hinton2002}\index{stochastic neighbor embedding (SNE)} and its variant called \acf{t-SNE}~\cite{vandermaaten:2008}\index{stochastic neighbor embedding (SNE)!t-distributed SNE} are techniques for nonlinear dimensionality reduction, which aim to preserve the local structure of the original data. That is, points that are close in the high-dimensional data set tend to be close to one another in the low-dimensional representation.

    \begin{algorithm}[bth!]
        \caption{\acf{t-SNE}}\label{alg:tsne}
        \begin{algorithmic}
        \Require Hyperparameters: $d$ (dimensionality of the projected data), perplexity $P$, learning rate $\eta$
        \Require Original data set of $\datasize$ points in $\featnum$ dimensional space $\mat{X} \in \realset^{\datasize \times \featnum}$
        \Require Random set of $\datasize$ points in lower dimensional $d_{\mathrm{red}}<\featnum$ representation $\mat{Y}^{(0)} \in \realset^{\datasize \times d_{\mathrm{red}}}$
            \For{each $\vect{x}_i$}
            \For{each $\vect{x}_j$}
            \State Calculate pairwise conditional probability distribution $p_{i|j}$ with fixed perplexity $P$
        \EndFor
        \EndFor
        \State Calculate probability distribution $p_{ij}$ 
        \For{$t = 1$ to $T$ }
            \For{each $\vect{y}^{(t-1)}_i \in \mat{Y}^{(t-1)} $}
            \For{each $\vect{y}^{(t-1)}_j \in \mat{Y}^{(t-1)} $}
                \State Calculate probability distribution $q_{ij}$
                \EndFor
                \State Calculate gradients of the \ac{KL} divergence with respect to coordinates of each $\vect{y}^{(t-1)}_i(z_1^{(i)},z_2^{(i)},\dots,z_{d_{\mathrm{red}}}^{(i)}) \in \mat{Y}^{(t-1)}$, i.e.,  $\frac{\partial D_{KL}(p||q)}{\partial \vect{z}^{(i)}}$
            \EndFor
            \State $\mat{Y}^{(t)} \gets \mat{Y}^{(t-1)}  - \eta  \frac{\partial D_{KL}(p||q)}{\partial \vect{z}}$ \Comment{Update the coordinates of each point}
            
        \EndFor
        \end{algorithmic}
        \end{algorithm}
        
Let us consider an~initial $\featnum$-dimensional space with $\datasize$ points, i.e., $\vect{x}_i \in \realset^{\featnum}$. We define the conditional probability $p_{i|j}$ that two points $\vect{x}_i$ and  $\vect{x}_j$ are similar (i.e., close to one another) as
\begin{equation}
    p_{i|j}  = \frac{e^{-||\vect{x}_i-\vect{x}_j||^2/2\sigma_i^2}}{\sum_{k\ne l} e^{-||\vect{x}_k-\vect{x}_l||^2/2\sigma_i^2}},
\end{equation}
where $||\vect{x}_i-\vect{x}_j||$ is the Euclidean distance between the two points. The fact that Gaussian likelihoods are used in $p_{i|j}$
implies that only points near $\vect{x}_i$ contribute significantly to its probability.
The variance $\sigma^2_i$ depends on the \stress{perplexity}\index{perplexity} defined as
\begin{equation}
    P_i = 2^{-\sum_{j=1}^{\datasize} p_{j|i}\log_2 p_{j|i} },
\end{equation}
which is a~measure based on Shannon entropy.
In the first step of the \ac{t-SNE} algorithm, the variances $\sigma^2_i$ are optimized for each point $\vect{x}_i$  to have a~fixed perplexity value $P_i=\text{const}$. Points in regions of high density have a smaller variance, while regions of low density have a larger variance.
In practice, the perplexity is usually set between 5 and 50.
Note that $p_{i|j} \ne p_{j|i}$ due to the dependence on $\sigma^2_i$.  To recover a~symmetric relation $p_{i|j} = p_{j|i}$, we define the joint probability distribution as
\begin{equation}
    p_{ij} = \frac{p_{i|j} + p_{j|i}}{2 \datasize}.
\end{equation}

The objective of the \ac{t-SNE} algorithm is to find another set of points in lower dimensional representation $\vect{y}_i \in \realset^{\datasize \times d_{\mathrm{red}}}, d_{\mathrm{red}} < \featnum$ and corresponding  probability distribution $q_{ij}$ in a~new representation for which the \ac{KL} divergence
\begin{equation}\label{KL_divergence}
    D_{KL}(p||q) = \sum_{i,j} p_{ij}\log\frac{p_{ij}}{q_{ij}}
\end{equation}
is minimal.

The procedure starts with randomly sampling $\datasize$ points $\vect{y}_i(z_1,z_2,\dots,z_{d_{\mathrm{red}}})$ in a~$d_{\mathrm{red}}$-dimensional space. For each point, we define the probability distribution $q_{ij}$ in a~similar way as in the high-dimensional space but using the t-Student probability distribution instead of Gaussian distributions:
\begin{equation}
    \vect{q}_{ij} = \frac{(1+||\vect{y}_i - \vect{y}_j||^2)^{-1}}{\sum_{k\ne l} (1+||\vect{y}_k - \vect{y}_l||^2)^{-1} }.
\end{equation}
In the last step, we minimize the \acf{KL} divergence from \cref{KL_divergence} (see \cref{sss:probability}) by optimizing the position $(z_1, z_2, ..., z_{d_{\mathrm{red}}})$ of each point $\vect{y}_i = \vect{y}_i(z_1, z_2, ..., z_{d_{\mathrm{red}}})$ in the $d_{\mathrm{red}}$-dimensional space 
which eventually yields a~low-dimensional data representation. The \ac{t-SNE} algorithm is summarized in \cref{alg:tsne}.

The low-dimensional data representation preserves the local structure of the original data set, i.e., similar points in the original data set are now clustered in the $d_{\mathrm{red}}$-dimensional representation space. However, the distance between the resulting clusters loses its meaning in representation space. 
        
        \highlight{In general, clustering in combination with dimensionality reduction works elegantly for simple problems, such as the Ising model. However, such approaches typically do not perform well when applied to more difficult phase classification tasks, e.g., in the presence of topological phases such as in the \ac{IGT}~\cite{greplova:2020}, or when a~large number of phases is present~\cite{arnold:2021}.}

\subsection{Supervised phase classification with neural networks}\label{sec:supervised_phase_class}
\begin{figure}[t]
        \begin{center}
        \includegraphics[width=\ToggleForCUP{0.9\textwidth}{\columnwidth}]{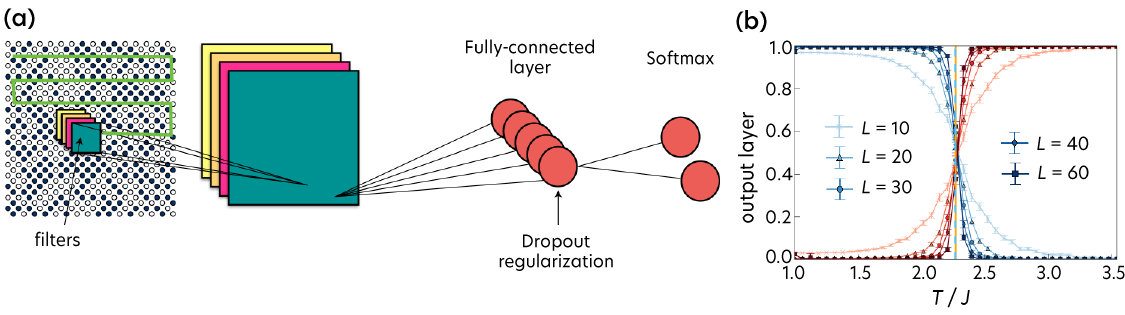}
        \end{center}
        \caption[Phase classification with supervised learning]{Supervised phase classification with an~\ac{NN} performed on the Ising model of varying linear lattice size $L$ ($N=L^2$) (see~\cite{carrasquilla:2017} for further details). (a) (Convolutional) \ac{NN} applied to a~configuration sample of the Ising model. (b) The average activations of the two nodes in the output layer are given by blue and red curves, respectively. The predicted critical temperature is marked by their crossover and is in good agreement with the Onsager solution [\cref{eq:ising_Tc}] depicted as an~orange vertical line. Adapted from \ToggleForCUP{Carrasquilla, J. \& Melko, R. (2017). \textit{Machine learning phases of matter}. Nat. Phys. 13, 431–434~\cite{carrasquilla:2017} with permission from Springer Nature.}{Ref.~\cite{carrasquilla:2017}.}}
        \label{fig:supervised_phase_classification}
    \end{figure}

\index{phase classification!supervised phase classification}
    One may wonder whether the issues encountered by clustering methods introduced in the previous section can be tackled by making use of the powerful machinery of \acp{NN} introduced in \cref{sec:NNs}. The idea is the following~\cite{carrasquilla:2017}. We train an~\ac{NN} to take spin configuration samples as input and correctly label them by the phase they belong to, see \cref{fig:supervised_phase_classification}(a). Typically, the label is encoded as a~binary bit string in a~\stress{one-hot encoding}\index{one hot encoding}. In case of the Ising model this would correspond to the label 1 for all samples drawn within the ordered phase ($T<T_{\rm c}$) or the label 0 for all samples drawn within the disordered phase ($T>T_{\rm c}$).\footnote{Of course, the opposite choice for labeling the two phases with label 0 for the ordered phase and 1 for the disordered phase is equally good.} To ensure that the output of the \ac{NN} can be used to predict a~binary label, we choose the output layer to be composed of two nodes to which we apply the softmax activation function introduced in~\cref{eq:softmax_function} over the activations $\vect{x}_{j}$ of all nodes within the output layer. This ensures that the output layer encodes a~valid probability distribution over the classes. The predicted label is then typically chosen based on the node which yields the maximum probability. For training, one typically employs the binary cross-entropy (see~\cref{eq:bce_loss}) which, for a~fixed input $\vect{x}$ is given as
    \begin{equation}\label{eq:celoss}
        \lossfun = -\sum_{j} p_{j}(\vect{x})\log({\rm NN}(\vect{x})_{j}).
    \end{equation}
    Here, ${\rm NN}(\vect{x})$ denotes the output of an~\ac{NN}, which contains a~softmax activation function in its last layer, applied to the input $\vect{x}$. The sum runs over all output nodes, i.e., the number of distinct classes. $p_{j}(\vect{x})$ is the true label of the input $\vect{x}$ as specified by the one-hot encoding. For example, given two classes and an~input whose true label is 0, we have $p_{0}(\vect{x})=1$ and $p_{1}(\vect{x})=0$ such that $\sum_{j} p_{j}(\vect{x})=1$. In \cref{eq:celoss}, this is compared to ${\rm NN}(\vect{x})_{j}$ which is the activation of the $j$-th output node and corresponds to the predicted probability of the input $\vect{x}$ to belong to class $j$.
    
    In our example, the training set consists of labeled spin configuration samples for a~wide range of temperatures far above and below $T_{c}$, whereas the test set is chosen over the entire temperature range. After training the \ac{NN} (see \cref{sec:NNs}) on the training set, it is evaluated on the test set. In particular, we average the activation of the two nodes in the output layer, which encode the probability of the input sample belonging to phase 0 or 1, respectively, over the test set. Remarkably, \cref{fig:supervised_phase_classification}(b) shows that these activations cross over precisely at $T_{\rm c}$ enabling us to extract the correct critical temperature. Similarly, this method is capable of correctly identifying the crossover temperature in the \ac{IGT}~\cite{carrasquilla:2017}. The fact that \acp{NN} can generalize to unseen input data can, for example, be exploited as follows. An~\ac{NN} trained on configurations for the square-lattice ferromagnetic Ising model can also highlight the critical temperature of the Ising model with a~different lattice geometry, such as a~triangular lattice~\cite{carrasquilla:2017}. Note that the ferromagnetic Ising model on a~triangular lattice is an~typical example of a~frustrated system.

\subsection{Unsupervised phase classification with neural networks}\label{sec:unsupervised_phase_class}
In \cref{sec:supervised_phase_class}, we showed that \acp{NN} can perform supervised phase classification. Due to its supervised nature this approach requires partial knowledge of the phase diagram of the system. One can determine the critical temperature (through ``interpolation'') if one knows the labels of samples deep within two neighboring phases. Ideally, in order to discover new phases of matter a~phase classification algorithm should not rely on such \stress{a priori} knowledge about the phases, i.e., it should be unsupervised in that regard. While clustering is unsupervised, we have seen that its power can be limited. In the following, we discuss three methods that use \acp{NN} to perform unsupervised phase classification.

    \subsubsection{Learning with autoencoders}\label{sec:learning_with_AEs}

\begin{figure}[t]
        \begin{center}
        \includegraphics[width=\ToggleForCUP{0.7\textwidth}{0.98\columnwidth}]{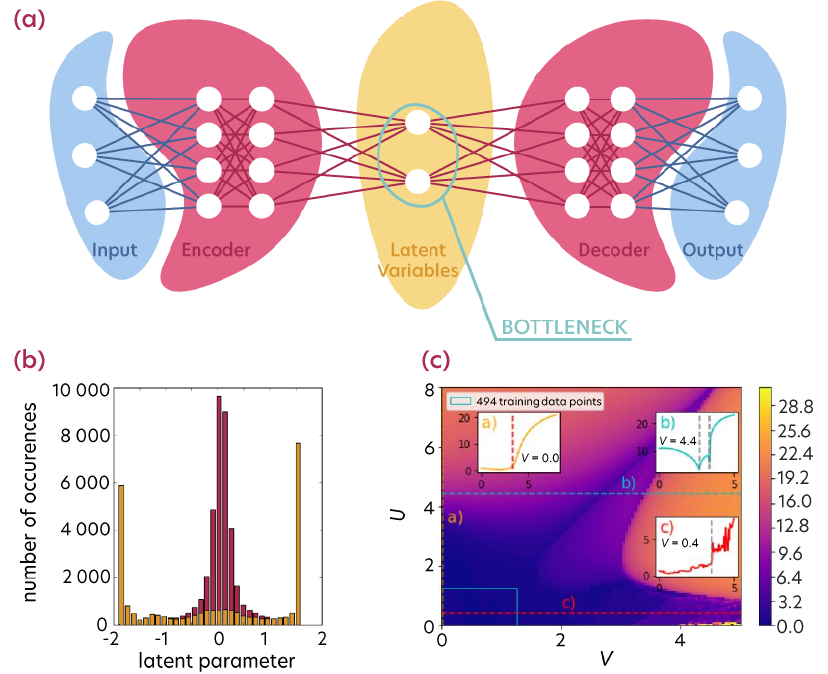}
        \end{center}
        \caption[Unsupervised learning with autoencoders]{(a) Illustration of a~natural bottleneck (here two neurons) in an~\ac{AE} architecture. (b) Analysis of bottleneck neurons of an~\ac{AE} trained to reconstruct spin configurations of a~two-dimensional Ising model. Latent representation of Ising configurations clusters into two phases visible as a~histogram. (c) Anomaly detection\index{anomaly detection} scheme allows for the recovery of the phase diagram from the reconstruction loss of an~\ac{AE} trained on one phase (blue box in the bottom left). Panel (b) is taken from \ToggleForCUP{Wetzel, S.~J. (2017). \textit{Unsupervised learning of phase transitions: From principal component analysis to variational autoencoders}. Phys. Rev. E 96, 022140~\cite{wetzel:2017},}{Ref.~\cite{wetzel:2017},} panel (c) is from \ToggleForCUP{Kottmann, K. \textit{et al}. (2020). \textit{Unsupervised phase discovery with deep anomaly detection}. Phys. Rev. Lett. 125, 170603~\cite{Kottmann2020PRL}.}{Ref.~\cite{Kottmann2020PRL}.}}
        \label{fig:AE_unsupervised}
    \end{figure}

    A~natural \ac{NN}-based unsupervised method is based on the analysis of the latent data representation given by an~\acf{AE}\index{autoencoder}. As we have briefly explained in \cref{sec:autoencoders}, \acp{AE} are \acp{NN} with a~bottleneck\index{bottleneck} in their center, which are trained to reconstruct the input at the output. The architecture of a~typical \ac{AE} is depicted in~\cref{fig:AE_unsupervised}. Due to the bottleneck, the information passing through the network needs to get compressed at the bottleneck, and then decompressed to recover the input. As a~consequence of the compression, some information may be lost.\footnote{In general, it is possible that \acp{NN} could compress more dimensions into a~single neuron. However, in practice \acp{NN} tend to learn smooth functions, which penalizes this behavior.} However, the retained information in the bottleneck should ideally contain everything relevant for the reconstruction of the input. Therefore, the bottleneck forms a~\stress{latent space} that contains a~compressed representation of the input data. This is akin to the dimensionality reduction schemes we discussed previously (see~\cref{sec:unsup_phase_no_NN}), which preserve the most important features for the reconstruction. As such, we can analyze the latent representation of the input data in a~similar way as the lower-dimensional representation obtained by \ac{PCA} in~\cref{ssec:pca}.\footnote{The quantum versions of \acp{AE} are also being developed and applied to phase classification \cite{Kottmann2021PRR} and clustering of subspaces of the Hilbert space \cite{Szoldra2022}. For more details, see \cref{par:quantumAEs}.}
    
    Let us apply an~\ac{AE} to reconstruct Monte Carlo samples of the two-dimensional Ising model~\cite{wetzel:2017}. Clearly, this represents an~unsupervised phase classification scheme because we do not provide any labels. The relevant loss function to be minimized is given by the reconstruction error between the input and output spin configurations (e.g., \acf{MSE}). If we look at how the latent representation of the spin configurations in the trained \ac{AE} change with the temperature (see~\cref{fig:AE_unsupervised}[b]), we can immediately observe a~clustering of the latent parameters. The clusters correspond to the two phases of the Ising model.\footnote{Beware, clustering of data in the latent space according to the phases present in the system is not a~general property of \acp{AE}. The clustering occurs when input data causes distinctive activations in the bottleneck, which \stress{often} corresponds to different phases.} Red points correspond to the high-temperature paramagnetic phase, while yellow points correspond to the low-temperature ferromagnetic phase. Note the two large yellow bins at the edges of the histogram in \cref{fig:AE_unsupervised}(b). These are formed due to the degeneracy of the ground state, which has either all spins pointing up, or all spins pointing down.

    Analysis of the \ac{AE} latent representation of the input data is not the only way of an~\ac{AE}-based unsupervised phase classification. Another successful and robust scheme based on anomaly detection\index{anomaly detection}\footnote{The \ac{AE}-based anomaly detection scheme was also successfully applied to quantum dynamics problems \cite{Szoldra2021PRB}.} was presented in Ref.~\cite{Kottmann2020PRL}. The basic idea is as follows. Imagine training an~\ac{AE} to reconstruct states coming from one phase. Then, the \ac{AE} is used to reconstruct states coming from the rest of the phase diagram. Such a~task is difficult because the training data is limited only to one phase, and the \ac{AE} is bound to make reconstruction errors in other phases. Moreover, we expect that the error is lower for phases that are similar to the ``training'' phase and higher for phases that contain states that look very different. Finally, the quantum states from the transition regimes are usually distinctive and the most unique from the rest of the phase diagram. Altogether, the reconstruction error across the phase diagram, made by an~\ac{AE} trained to reproduce states from one phase, is expected to vary according to the phases and the phase boundaries in the system. This scheme enables the discovery of phases in a~fully unsupervised way. The authors of Ref.~\cite{Kottmann2020PRL} used this scheme based on anomaly detection to recover a~full phase diagram of the extended Bose-Hubbard model in one dimension at exact integer filling. This result is presented in panel (c) of~\cref{fig:AE_unsupervised}. Interestingly, their work also revealed within the phase diagram a~phase-separated region\footnote{This phase-separated region is located between supersolid and superfluid phases, for more details see Ref.~\cite{Kottmann2020PRL}.} with unexpected properties which may be one of the first fully unsupervised discoveries in the \ac{ML}-guided phase classification.
    
    \subsubsection{Learning by confusion}\label{sec:lbc}
\begin{figure}[t]
    \begin{center}
    \includegraphics[width=0.6\columnwidth]{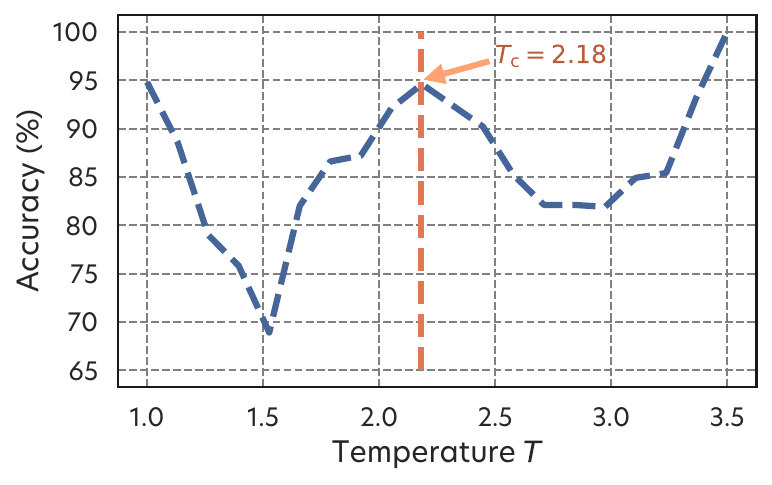}
    \end{center}
    \caption[Phase classification with learning by confusion]{Result of the learning by confusion scheme applied to the Ising model. The data consists of 100 spin configurations (linear lattice size $L=30$) sampled using Monte Carlo methods at temperatures $T$ ranging from $T_{1}=1$ to $T_{20}=3.5$ in equidistant steps. The data set is split into equally sized training and test sets (such that 50 spin configurations are present at each sampled temperature). The blue curve shows the classification accuracy on the test set for various choices of bi-partitions. It has a~characteristic W-shape whose middle peak is at $T\approx 2.3$, which is in good agreement with the Onsager solution. Reproduced with Ref.~\cite[Notebook A3]{OurSchoolRepo}.}
    \label{fig:LBC}
\end{figure}

\index{learning by confusion}
    \stress{Learning by confusion}~\cite{van:2017} is another \ac{NN}-based unsupervised method that works as follows. We start by partitioning the temperature range into two regions with distinct labels. Based on these labels, we perform supervised learning over the entire temperature range as described in \cref{sec:supervised_phase_class} and keep track of the final overall classification accuracy of the model. This classification accuracy is associated with the guess for the critical temperature located at the boundary of the two regions. We repeat this procedure systematically for multiple bi-partitions of the temperature range, i.e., guesses for the critical temperature. Finally, we plot the classification accuracy against the guessed critical temperature.  This procedure is summarized in \cref{alg:lbc}. Note that each partitioning requires the training of a~separate \ac{NN}.\footnote{Retraining a~model for each choice of a~bi-partition can become computationally expensive, in particular when increasing the resolution of the method. There has been an~extension of the learning-by-confusion scheme that uses two \acp{NN}~\cite{liu:2018} to try to circumvent this issue by choosing bi-partition points one at a~time in a~guided manner.} The results of this algorithm applied to the Ising model are depicted in \cref{fig:LBC}. We observe that the classification accuracy is \stress{W-shaped}. The high classification accuracy at the extremes of the temperature range arises due to the fact that, in these cases, almost all samples are assigned the same label. In particular, in the extreme case where all samples are assigned the same label a~classification accuracy of 1 can be achieved trivially because the \ac{NN} simply needs to learn to output the same label independent of the input. The middle peak, however, is non-trivial and corresponds to the predicted critical temperature of the method. Here, the predicted critical temperature is in good agreement with the Onsager solution. The presence of this middle peak can be explained as follows. Let us assume that the data can naturally be classified into two distinct groups realized by a~particular choice for the bi-partition of the temperature range. Then, the closer our choice of bi-partition matches the ``correct'' bi-partition underlying the data, the larger the classification accuracy of our algorithm.

    \begin{algorithm}[bth!]
\caption{Learning by confusion}\label{alg:lbc}
\begin{algorithmic}
\Require Data set of (spin configuration) samples $\mathcal{D}_0=\{\mat{x} \}$, guesses for critical temperature $ \mathcal{T} =  \{ T_1, ..., T_{\rm max} \}$ 
\For{$T_{\rm c}^* \in \mathcal{T}$}
    \State Partition data set $\mathcal{D}_0$ into two regions with $T\leq T_{\rm c}^*$ and $T> T_{\rm c}^*$
    \State Set label $\vect{y}$ of all samples in region with $T\leq T_{\rm c}^*$ as 0 and $T\leq T_{\rm c}^*$  as 1
    \State Split resulting data set into training and test set
    \State Perform supervised learning on the training set, i.e., train an~\ac{NN} to minimize loss in \cref{eq:celoss}
    \State Evaluate classification accuracy on the test set
\EndFor
\State Plot accuracy vs. $T_{\rm c}^*\; \forall T_{\rm c}^* \in \mathcal{T}$ (see \cref{fig:LBC}) \Comment{Critical temperature $T_{\rm c}^*$ at which the accuracy peaks corresponds to the best guess for the location of the phase transition}
\end{algorithmic}
\end{algorithm}

    Here, we have discussed the case where there are precisely two distinct phases present in the parameter range under consideration. In this case, the accuracy ideally displays a~characteristic W-shape, see~\cref{fig:LBC}. If multiple phases are present, this characteristic W-shape is modified. The shape of the signal (in particular, the number of obtained peaks) could then be used to identify the number of different phases present in the data~\cite{van:2017,lee:2019,maska:2022}.
    
    \subsubsection{Prediction-based method}\label{sec:pbm}\index{prediction-based method}
    
        \begin{figure}[t]
        \begin{center}
        \includegraphics[width=\ToggleForCUP{0.9\textwidth}{0.99\columnwidth}]{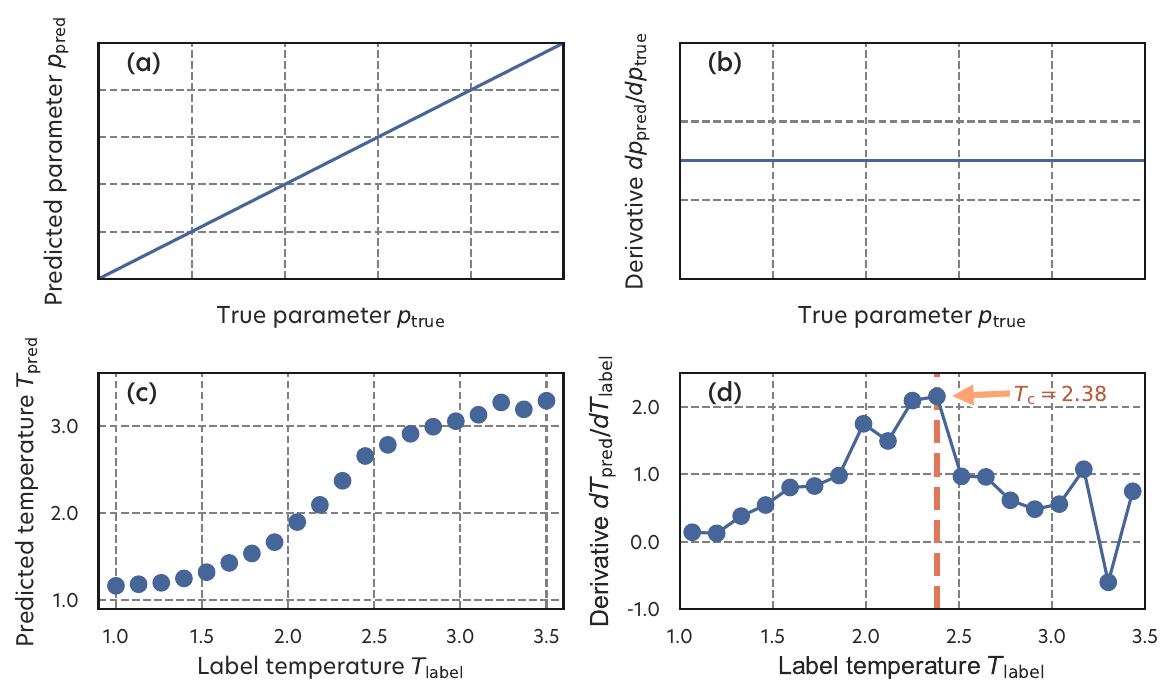}
        \end{center}
        \caption[Phase classification with the prediction-based method]{(a,b) Illustration of the output of the prediction-based method if the system does not undergo any phase transition. (c,d) Result of the prediction-based method applied to the Ising model. The data consists of 100 spin configurations (linear lattice size $L=30$) sampled using Monte Carlo methods at temperatures $T$ ranging from $T_{1}=1$ to $T_{20}=3.5$ in equidistant steps. The data set is split into equally sized training and test sets (such that 50 spin configurations are present at each sampled temperature). (c) Average predicted temperature for the test data as a~function of the true underlying temperature. (d) Derivative of the average predicted temperature for the test data as a~function of the true underlying temperature that peaks at the critical temperature of the Ising model. Panels (b) and (c) reproduced from Ref.~\cite[Notebook A3]{OurSchoolRepo}.}
        \label{fig:PBM}
    \end{figure}
    
    The learning by confusion scheme is difficult to efficiently extend to high dimensional parameter spaces, which may feature several distinct phases.\footnote{In Ref.~\cite{liu:2018} an~approach to extend the scheme to two-dimensional parameter spaces featuring two distinct phases is presented.} Moreover, it has been shown that the learning by confusion scheme has difficulties in correctly identifying the crossover in the \ac{IGT}~\cite{greplova:2020}. These limitations can be circumvented through the so-called \stress{prediction-based method}~\cite{schaefer2019,greplova:2020,arnold:2021}, which works as follows.
    
    We train an~\ac{NN} to predict the tuning parameter (here, the temperature) for each configuration sample. The value of the tuning parameter at which a~given configuration sample has been generated is readily available both in experiment and simulation. If the system does not undergo any phase transition, the predicted tuning parameter is \stress{linearly} dependent on the true tuning parameter as shown in \cref{fig:PBM}(a). Consequently, the derivative of the predicted tuning parameter with respect to the true tuning parameter is constant, see \cref{fig:PBM}(b). For systems that exhibit a~phase transition, the situation is different. In this case, the tuning parameter cannot be predicted with perfect accuracy resulting in a~\stress{nonlinear} relationship between the predicted and the true value of the tuning parameter. \Cref{fig:PBM}(c) illustrates this in the case of the Ising model. Consequently, the derivative is not constant, and at the critical tuning parameter, the largest rate of change occurs, see \cref{fig:PBM}(c) and (d). In other words, the parameter value for which the \ac{NN} predictions are most susceptible identifies the position of the phase transition.
    
    Let us elaborate on this point: While the tuning parameter only changes marginally in the vicinity of the phase transition, the system's state and its corresponding order parameter change dramatically as the tuning parameter crosses its critical value. As a~result, the \ac{NN} is the best at distinguishing samples originating from two different phases, whereas it has difficulties in distinguishing samples from within the same phase. That is, its predictions change the most as the tuning parameter is swept across its critical value. \Cref{fig:PBM}(c) shows that the predictions start to saturate deep within each phase, whereas they vary most strongly with the tuning parameter around the transition point.

So far, we have seen that phase classification methods based on \acp{NN} are capable of locating the phase transition of the two-dimensional Ising model. To date, these methods have successfully revealed a~plethora of other phase transitions in various physical systems.\footnote{There exist various other \ac{ML} methods for detecting phase transitions and classifying phases of matter~\cite{ronhovde:2011,ronhovde:2012,Vargas-Hernandez2018,shirinyan:2019,Mazaheri2019PRM,balabanov:2020,arnold:2021}. For example, in Ref.~\cite{Vargas-Hernandez2018} phase transitions can be inferred by training an~\ac{ML} model to fit the properties within one phase and extrapolating toward other regions in parameter space. Here, the model is based on a~\acf{GP} utilizing kernels. We discuss kernel methods, including \acp{GP}, in detail in~\cref{sec:gp}.} This fact highlights that these methods are \stress{generic} and have been formulated in a~system-agnostic fashion. There may exist various physical observables (such as order parameters) that can be used to identify a~given phase transition. However, finding these quantities is typically a~hard task and requires a~deep understanding of the physical system at hand. Remarkably, the \ac{NN}-based methods we showcased here can successfully classify different phases of matter in an~automated fashion without \stress{a priori} knowledge of the underlying physics. Note that there exist similar system-agnostic tools which do no rely on \ac{ML}, such as the specific heat for thermal phase transitions or the fidelity susceptibility~\cite{gu:2010} for quantum phase transitions.\footnote{A~quantum phase transition~\cite{sachdev:2011} corresponds to non-analytic behavior of the ground-state properties at the critical value of the tuning parameter $p_{\rm c}$, where the system Hamiltonian is $H(p)$. It emerges due to the competition of individual terms in the Hamiltonian, which depends on the tuning parameter.} However, these tools can still fail for a~given system and can be expensive to compute or difficult to measure in an~experiment. For example, the specific heat fails to locate the crossover temperature in the \ac{IGT}. In case of the fidelity susceptibility, one investigates the change in the overlap $\langle \Psi_{0}(p) | \Psi_{0}(p + \epsilon)\rangle $, where $| \Psi_{0}(p)\rangle$ is the ground state of the Hamiltonian $H(p)$, $p$ is the tuning parameter, and $\epsilon$ is an~infinitesimal perturbation. Because one typically does not have access to the full wave function, the fidelity susceptibility typically remains difficult to evaluate. The \ac{NN}-based methods we discussed constitute alternative tools. In particular, they can \stress{in principle} be applied using various properties of the system's state at different values of the tuning parameter as input. This allows them to identify phase transitions based on experimentally accessible measurement data~\cite{Rem19,bohrdt:2021, Kaming2021}.

While the phase classification methods we discussed up to now are capable of locating phase transitions, we have not yet gained any insights into the specific type of phase transition that the system undergoes. The crucial question is whether one can extract physical insights from the \acp{NN} concerning the underlying phase classification tasks. In particular, one can ask whether it is possible to extract novel order parameters from such \acp{NN}, which is an~ultimate goal of the interpretable \ac{ML} applied to phase classification problems.

\subsection{Interpretability of machine learning models}\label{sec:interpretability}\index{interpretability} 
As seen in the previous sections, \acp{NN} are powerful tools to identify phases in physical data. Now imagine applying these methods to a~novel physical system whose phases and corresponding order parameters are not yet known. The natural questions that arise in this scenario are: Can we trust the \ac{NN} predictions? In particular, how can we know that the model correctly located a~phase transition in the parameter space? Moreover, assuming that the methods correctly classified the data into different phases of matter, how can we gain physical insights into the problem at hand? For instance, can we analyze the trained \acp{NN} to determine what types of phase transitions the system undergoes? Or would it even be possible to extract novel order parameters from them? When using \ac{ML} (and especially \ac{DL}) models, answers to these questions are not easy to find. Such challenges are being addressed by the research on the \ac{ML} reliability and interpretability.\footnote{Note that the formal definitions of these terms are not agreed upon in either the physical or computer science community~\cite{Lipton2018}. To circumvent the problem, here we provide intuitions about the meanings of these terms.} 
\highlight{Reliability\index{reliability} is about trusting our \ac{ML} model predictions. Our trust in the model is increased, e.g., when we have access to the uncertainty of model predictions. Interpretability\index{interpretability} is about understanding what an~\ac{ML} model learns and how it makes its predictions. As such, these two ideas are closely intertwined.}
\noindent Both concepts are particularly important on our way toward scientific discovery using \ac{ML}. If we are not able to understand what an~\ac{NN} learns when given a~problem, \stress{our} understanding of the problem remains limited!\footnote{We can imagine a~non-interpretable black-box \ac{NN} that after training can give insights to the problem, e.g., Ref~\cite{Bohrdt2019}. However, the model still needs to be reliable so we can trust the new insights, and we need to have previous deep insights into the problem.}

We have already mentioned that \stress{a priori}, \ac{DL} models are usually neither reliable nor interpretable. As such, they largely serve as black-box models that provide us with suitable predictions (from which we, e.g., can locate phase transitions in the underlying input data). There are several reasons for that: firstly, their learning dynamics are largely opaque and not well understood.\footnote{We show you how some of these questions can be answered with tools from statistical physics in \cref{sec:stat_phys_for_ML}.} Secondly, the direct analysis of trained \acp{NN} is challenging, as we explain in the next section. In particular, the ``reasoning'' of \acp{NN} does not necessarily have to be based on the same observations on which a~human would base its decisions. Tackling these challenges is important for all \ac{ML} applications, but especially crucial, e.g., for medical diagnosis or insurance and hiring decisions.

\subsubsection{Difficulty of interpreting parameters of a~model}
When looking at a~\ac{DNN} with possibly billions of trainable parameters, it is hard for us humans to decipher what the \ac{NN} is really doing under the hood. It may be that an~\ac{NN} actually computes a~simple, physically relevant function, such as an~order parameter, to make its predictions. Recognizing whether that is the case is hard because the computation and relevant information are spread over the multiple layers containing a~large number of neurons each. However, if an~\ac{NN} is sufficiently small, a direct interpretation by looking at its trainable parameters may be possible. Consider the limiting case of a~single-layer \ac{NN} without any nonlinear activation function. This corresponds to a~simple linear regression model, described in \cref{sss:intro_linear_model}:
\begin{equation}\label{eq:linear_model}
    \vect{\hat{y}} = \weightsvect \vect{x} + \biases,
\end{equation}
where $\weightsvect$ is a~vector of weights and $\biases$ is a~vector of biases. Evidently, such a~linear model allows for a~direct interpretation in terms of its weights: the larger the magnitude of a~given weight (connection), the more important the corresponding normalized feature for solving the problem at hand. For an~example of weight interpretation in the context of phase classification, see Ref.~\cite{zhang:2020}.

However, a reduction in depth and loss of nonlinearity comes at the cost of expressivity. For such a~model to be accurate, it generally requires highly pre-processed inputs $\vect{x}$ whose processing takes care of the necessary nonlinearities. Moreover, the importance of a~given feature has more meaning if it is already present in a~compact, physically relevant form. This largely limits the domain of applicability of small predictive models to problems of which we (at least) have partial knowledge.

\paragraph{Reducing the number of effective parameters via regularization.}\index{interpretability}
One way to obtain \acp{NN} with a reduced number of effective parameters is regularization\index{regularization} -- in particular, the addition of a~$\regularization{1}$ regularization term in the loss function given by $\regularization{1} = \lambda \sum_{i} |w_{i}|$, where $\lambda$ parametrizes the regularization strength, and the sum runs over all (trainable) weights within the \ac{NN}. This term forces the weights to vanish, i.e., for connections to be cut. Ideally, this results in a~sparser, and thus effectively smaller, \ac{NN} which enables interpretability. In Ref.~\cite{cranmer:2020}, for example, the authors could extract analytical expressions for force laws and dark matter distributions from graph \acp{NN} trained to predict planetary and dark matter dynamics. This was achieved by performing symbolic regression on the corresponding sparse networks.\footnote{A graph \ac{NN} is similar to a \ac{CNN} in the sense that the spatial location of the input is crucial for the meaning of the input. While in \acp{CNN}, neighbors are determined by their position of the input data grid, in graph \acp{NN}, neighbors can be defined much more broadly through custom connections between graph nodes. \Ac{NN} layers can act on each node or through message-passing between nodes.} Regularization is also important when interpreting linear models as in \cref{eq:linear_model}. Often, learning problems do not have a~unique solution. This means that the weights can vary given the same data and optimization procedure, which would result in different ``interpretations'' of the \ac{NN}'s inner workings. Regularization terms help to remove the remaining degrees of freedom of the weights and enforce Occam's razor.

\paragraph{Extracting order parameters with \acfp{SVM}.}\index{interpretability}\index{support vector machines!interpretability}
A large class of \ac{ML} algorithms that allow for a direct interpretation in terms of model parameters are \acp{SVM}, see \cref{sec:intro-SVM} and \cref{sec:kernel_SVM}. \acp{SVM} were first proposed for solving phase classification tasks in Ref.~\cite{ponte2017kernel} and were later expanded and applied to higher-order spin systems in Refs.~\cite{greitemann2019probing,liu2019learning}. As discussed above, while these algorithms might not be as powerful as \acp{NN}, a major advantage is the possibility of having an interpretable decision function from which order parameters can be inferred.\footnote{The decision function determines the distance of a given sample $\vect{x}$ from the hyperplane.}

\subsubsection{Interpretability via bottlenecks}\index{interpretability}
As we have explained in the previous section, interpretability is an~inherent characteristic of small models. Fortunately, there are alternative approaches to interpretability that are not limited to simple small models. What we can do in large architectures is to identify bottlenecks in the information flow and focus our attention there. A~bottleneck\index{bottleneck} in an~\ac{NN} is just a~layer with fewer neurons than the layer before and after it. An~example of a~\ac{NN} with a~natural bottleneck has already appeared in~\cref{sec:learning_with_AEs} and in~\cref{fig:AE_unsupervised}(a), namely an~\acf{AE}. Its bottleneck forces the \ac{NN} to distill the relevant information within the inputs such that it can flow through this constriction. As such, the \ac{NN} performs a~dimensionality reduction and finds a~suitable low-dimensional feature representation. While the entire \ac{NN} architecture can be large and have many trainable parameters, the bottleneck itself can have as little as a few parameters. Because all the relevant information for the predictions of the \ac{NN} must eventually flow through the bottleneck, we can limit our analysis to the small number of trainable parameters of the bottleneck as opposed to the entire \ac{NN}. 

In particular, we can perform a~regression on the output of such bottleneck neurons and extract the mapping between the input features and the activations of the bottleneck neurons. There is a~natural bottleneck in almost every \ac{NN} -- its output neuron. However, performing a~regression on it without imposing any additional bottlenecks is challenging because you need to take into account all input features, which can grow quickly in number. Apart from the output neuron, other types of bottlenecks can appear naturally in \acp{NN} architectures, such as in \acp{AE} (see \cref{sec:autoencoders})~\cite{wetzel:2017, iten:2020} and \acp{CNN} (see \cref{sec:CNNs})~\cite{wetzel2:2017}. However, we can also introduce bottlenecks into our architecture on purpose to have more interpretable \ac{ML} models. This idea gave birth to, e.g., Siamese \acp{NN}~\cite{wetzel:2020}. We look at these approaches in more detail in the next paragraphs and see what information on physical systems we can extract using them.

\begin{figure}[t]
\begin{center}
\includegraphics[width=0.7\columnwidth]{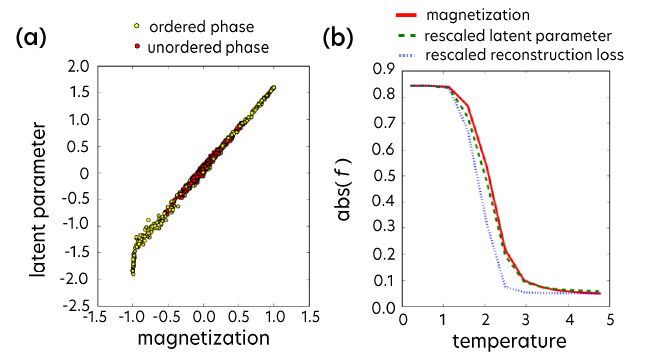}
\end{center}
\caption[Interpretation of an~autoencoder for a~two-dimensional Ising model]{Analysis of bottleneck neurons of an~\ac{AE} trained to reconstruct spin configurations of a~two-dimensional Ising model. (a) Dependence of latent space parameter on the magnetization. The red (yellow) color corresponds to samples from the low(high)-temperature regime. (b) Absolute magnetization, absolute rescaled values of the latent parameter, and reconstruction loss averaged for fixed temperature. Adapted from \ToggleForCUP{Wetzel, S.~J. (2017). \textit{Unsupervised learning of phase transitions: From principal component analysis to variational autoencoders}. Phys. Rev. E 96, 022140~\cite{wetzel:2017}.}{Ref.~\cite{wetzel:2017}.}}
\label{fig:AE_interpretation}
\end{figure}

\paragraph{Interpretability with \acfp{AE}.}\index{interpretability}\index{autoencoder!interpretability}
As we have explained in \cref{sec:autoencoders,sec:learning_with_AEs}, \acp{AE}\index{autoencoder} are \acp{NN} with a~bottleneck in the middle that are trained to reconstruct the input at the output. We have already shown in~\cref{sec:learning_with_AEs} and in~\cref{fig:AE_unsupervised}(b) that we can obtain clustering in the latent representation corresponding to phases present in the input data, achieving the unsupervised phase classification. Moreover, thanks to the bottleneck, which ideally should contain all relevant information for the reconstruction, we can also extract additional information and interpret what property of the input data is preserved by an~\ac{AE}. In particular, if we plot the latent parameter against the magnetization of the respective two-dimensional Ising spin configuration, as in \cref{fig:AE_interpretation}(a), we see a~linear dependence. This suggests that the compressed representation learned by the \ac{AE} is connected to the magnetization. To be more precise, the behavior deviates from a~strict linear dependence at values of the magnetization close to $-1$. However, we still can make the statement that the \ac{AE} learned a~property related to the magnetization, given that the mapping between the latent parameter and the magnetization is bijective. For example, such a~statement would hold even if the latent parameter as a~function of the magnetization would vary according to a~sigmoid function. 

\begin{figure}[t]
\begin{center}
\includegraphics[width=\ToggleForCUP{0.7\textwidth}{0.8\columnwidth}]{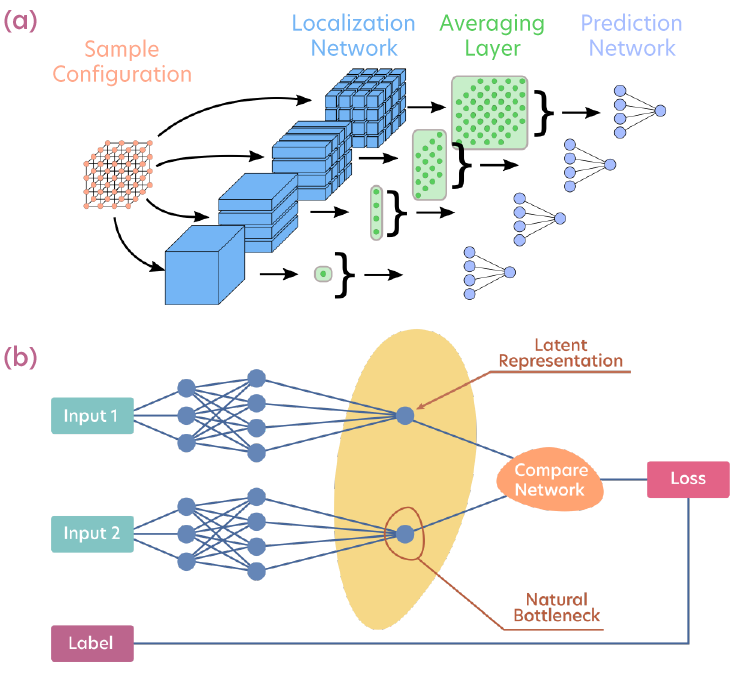}
\end{center}
\caption[Interpretation of neural networks by bottlenecks]{(a) Interpretation Net where the size of the receptive field is systematically reduced in each training step. The localization network consists of one or more subsequent convolutional layers. The averaging layer collapses convolved features to a~single number. (b) Scheme of a~Siamese \ac{NN}\index{Siamese neural network}, whose two subnets are identical and share the same tunable parameters. The input is a~pair of data points, and the resulting label is either ``same'' or ``different''. Panel (a) is adapted from \ToggleForCUP{Wetzel, S.~J. \& Scherzer, M. (2017). \textit{Machine learning of explicit order parameters: From the Ising model to SU(2) lattice gauge theory}. Phys. Rev. B 96, 184410~\cite{wetzel2:2017}.}{Ref.~\cite{wetzel2:2017}.}}
\label{fig:CNN_Siamese}
\end{figure}

Another example of analysis of latent space of \acp{AE} is work by Iten \textit{et al.} \cite{iten:2020}. They used a~special \ac{AE} architecture with a~\stress{question neuron}, i.e., an~additional neuron connected to the first decoding layer after the bottleneck. The input of this question neuron is provided by a~user. You can think of it as an~alternative way of providing data to the network. The authors showed that you can train such a~special \ac{AE} in a~way that a~user can ask a~question via the question neuron, and the answer is encoded in the latent space.

As you see, \acp{AE} are to some degree inherently interpretable by virtue of their low-dimensional latent space. However, the analysis of latent space does not give us any hint about the order parameter or important features. We can only compare it against the quantities or features we suspect to be important. If you look for a~more automated way of detecting order parameters, we can turn to very special \acp{CNN}.

\paragraph{Extracting order parameters with \acfp{CNN}.}\index{interpretability}\index{convolutional neural network!interpretability}
We have already mentioned that \acp{CNN} have natural bottlenecks in their architectures. These bottlenecks are their filters or kernels, i.e., the structures with which they ``scan'' the data. Their size can be thought of as a~receptive field size and tells us how many neighboring features (e.g., pixels) the network can analyze at the same time. Of course, if you have multiple convolutional layers with multiple kernels of different sizes intertwined with pooling layers, their analysis is still challenging. But if you consider a~simple \ac{CNN} with only one or few subsequent convolutional layers with kernels of a~fixed size and only one averaging layer at the end of the architecture, such a~regression becomes tractable.\footnote{It remains non-trivial and involves careful zeroing of weights, Fourier series, and other tricks. If you are interested, see Ref.~\cite{wetzel2:2017}.} The mentioned architecture was proposed by Wetzel \textit{et al.} (2017)~\cite{wetzel2:2017} and is called \stress{Interpretation Net} or \stress{Correlation-Probing \ac{NN}}, see~\cref{fig:CNN_Siamese}(a). Such an~architecture allows us to perform a~regression on the output neuron with features extracted by kernels. Eventually, we obtain an~analytical expression for the \ac{CNN} decision function. If applied to a~phase classification problem, such a~decision function could unravel the order parameter. It seems, however, that such a~decision function, and therefore the order parameter that may potentially be discovered through the \ac{CNN}, depends on the choice of kernel size. What is the appropriate choice of kernel size and, thus, decision function? Occam's razor tells us we should be interested in the simplest decision function. That is, one should aim to take into account only a~small number of input features. Crucially, this also makes the task of symbolic regression easier.\highlight{Therefore, the idea of the Interpretation Net is to systematically reduce the size of the kernel (and thus the input dimension for the symbolic regression task) by cutting connections until there is a~significant drop in the \ac{CNN} performance. This drop corresponds to the \ac{CNN} becoming ``blind'' to the correlations that are crucial for detecting and distinguishing different phases.}

Imagine starting from a~large kernel whose size corresponds to the size of the entire input image, e.g., $28\times28$. We train our Interpretation Net with such $28\times28$ kernels and see that it yields good results. Now, we reduce the receptive field size, e.g., to $20\times20$, retrain, and observe the performance. We repeat this process of reducing the kernel size and retraining until we see a~significant drop in the \ac{CNN} performance. Such a~drop occurs as soon as the \ac{CNN} gets blind to correlations in the system that are crucial for the phase classification, e.g., next-nearest-neighbor correlations. Finally, we can perform a~regression on the output neuron of the \ac{CNN} with the smallest kernel size that still yields good performance. The decision function we recover in the process is ideally connected to the underlying order parameter. With this approach, the Interpretation Net is capable of successfully classifying the phases of the two-dimensional Ising model\footnote{Interestingly, in their paper, the quantity which leads to the better \ac{CNN} performance in case of the Ising model is the expected energy per site ($\frac{-J}{N} \sum_{\langle i,j \rangle } \sigma_i \sigma_j$) which can still be detected with a~$2\times1$ kernel, not the magnetization ($\frac{1}{N} \sum_{i} \sigma_i$) which can still be detected with a~$1\times1$ kernel.} or $SU(2)$ lattice gauge theory and allows for extracting the corresponding decision functions. We stress that \stress{the learned decision function can strongly depend on the choice of the network architecture and training procedure as well as the available data}. Therefore, various networks can detect different order parameters, e.g., in the Ising model, they can detect the expected energy per site, magnetization, or a~scaled combination of those.

\begin{figure}[t]
\begin{center}
\includegraphics[width=0.99\columnwidth]{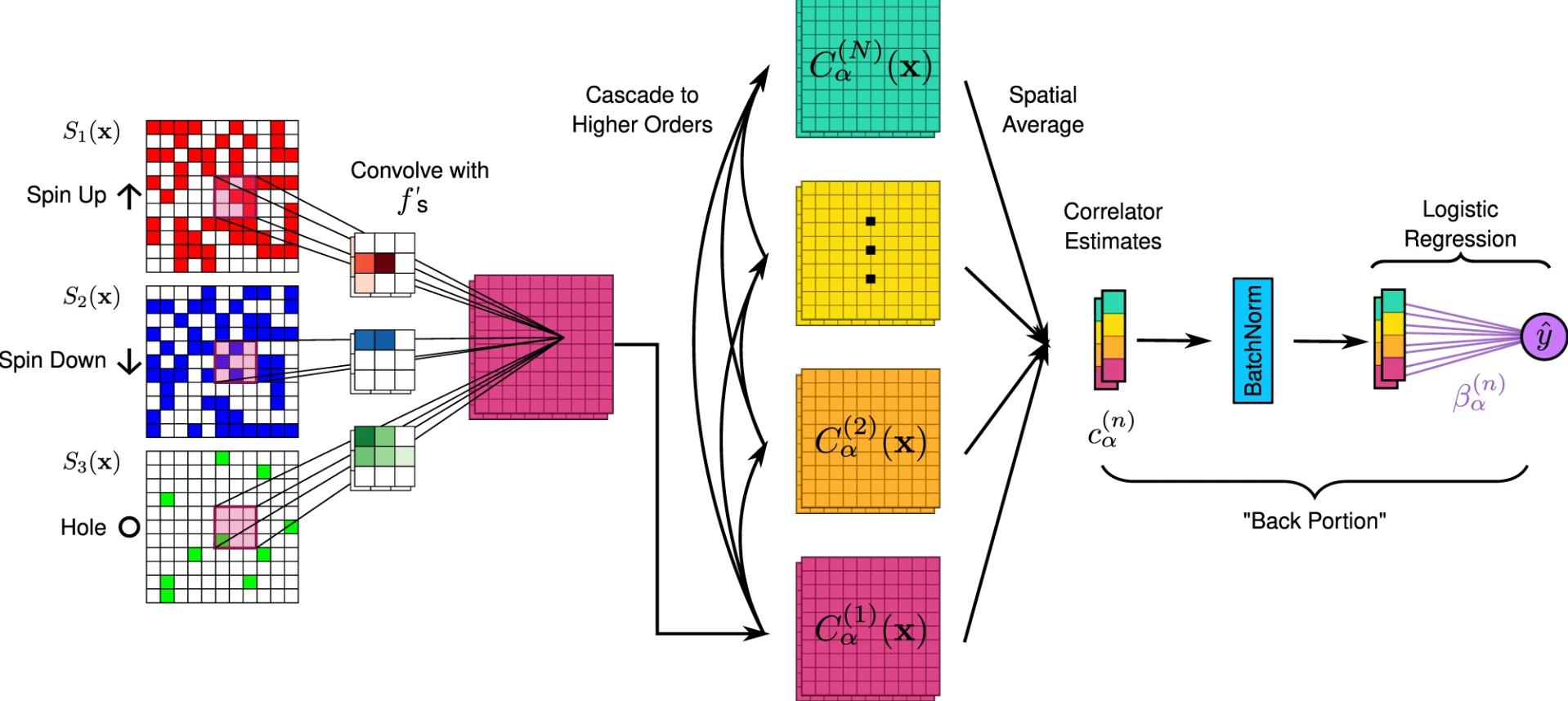}
\end{center}
\caption[Correlator Net]{The Correlator \ac{CNN} has learnable filters that can be activated and deactivated through regularization. The input is first convolved with learned filters to produce a set of convolutional maps from which information about higher-order local correlations can then be recursively constructed. Taken from \ToggleForCUP{Miles, C. \textit{et al.} (2021). \textit{Correlator convolutional neural networks as an interpretable architecture for image-like quantum matter data}. Nat. Commun. 12, 3905~\cite{miles:2021} with permission from Springer Nature.}{Ref.~\cite{miles:2021}.}}
\label{fig:CCNN}
\end{figure}

A~related approach was used by Miles \textit{et al.} (2021) \cite{miles:2021} when designing a~so-called \stress{Correlator \ac{CNN}} depicted in \cref{fig:CCNN}. This \ac{CNN} performs automatic feature engineering by probing for one-, two-, and higher-body correlations in the first few layers. The subsequent layers are designed to check which correlations are the most important for the classification. Again, the network is rendered blind to certain features by tuning what the \ac{NN} can learn. In contrast to the previous section, the authors penalize the learning of certain filters through regularization. By increasing the regularization strength, it is possible to successively disable certain features depending on their importance for the prediction accuracy, thus leading to a hierarchy of important correlations corresponding to the underlying physics.

The Correlator \ac{CNN} was used to detect the key many-body correlators differentiating between two theoretical quantum models serving as two candidate theories approximating the doped Fermi-Hubbard model  \cite{miles:2021}. As such, it explained the results of another work \cite{Bohrdt2019}, where a \ac{CNN} was trained to differentiate between numerically generated snapshots of the quantum system following two candidate theories. Then the trained \ac{CNN} was tested on experimental snapshots and indicated which which of the two theoretical quantum models described those snapshots better. This represents one of the first examples of scientific discovery with \acp{NN}.

\paragraph{Interpretability with Siamese neural networks.}\index{interpretability}\index{Siamese neural network}

One can consider even more complex architectures with artificial bottlenecks allowing for an~interpretation via symbolic regression. An~interesting example of this is the~Siamese \ac{NN}~\cite{wetzel:2020}\index{Siamese neural network} (sometimes also called twin \ac{NN}), presented in \cref{fig:CNN_Siamese}(b). It takes two input data points at the same time and is composed of two twin subnetworks with the same parameters and architecture. Their output neurons form two bottlenecks, which, in turn, are inputs for the third subnetwork, the aim of which is to connect and compare the twin outputs. The task of the network is to determine whether two input data points are similar or not.\footnote{A quantum version of Siamese \acp{NN} was developed in Ref.~\cite{Radha2022}, and, interestingly, it goes beyond distance-based notions of similarity.} This means that Siamese \acp{NN} are able to perform a~multiple-class classification without fixing the number of classes \stress{a priori} and with relatively little training data per class. Moreover, by analyzing the bottlenecks, we can extract what the \ac{NN} learns in a~problem.

These networks provide a powerful tool to discover phase transitions in an unsupervised manner as outlined in Refs.~\cite{patel2022unsupervised,han2022simple}. Learning phase transitions with Siamese \acp{NN} is very similar to using \ac{CNN}s. However, due to the unsupervised nature, one is not able to supply phase labels. Hence, the labels are initialized such that all data pairs sampled from the same point in the phase diagram get the \emph{same} label, while pairs from different points in the phase diagram obtain the \emph{different} label. After training, the Siamese \ac{NN} can be used to detect phase boundaries by sampling pairs from adjacent points in the phase diagram and predicting whether they are \emph{similar} or \emph{different}. A spike in dissimilarity then marks the phase transition.

Before we conclude, let us look at a few other applications beyond classifying phases of matter where the interpretation of \acp{NN} via bottlenecks comes in handy. The authors of Ref.~\cite{wetzel:2020} applied the Siamese \ac{NN}\index{Siamese neural network}, e.g., to the motion of a~particle in a~central potential. The task was to learn whether two observations of a particle correspond to the same particle trajectory (or two distinct trajectories). After successful training, we can perform a~polynomial regression on the bottleneck with respect to the input data features. In this case, the features are comprised of the~position of the particle in two-dimensional space and its two-dimensional velocity vector. By analyzing the dominant regression terms, they observed that the result of the regression is proportional to the angular momentum of the particle. Such an~analysis of bottlenecks of a successfully trained Siamese \ac{NN} indicates that the network learns conserved quantities and invariants.\footnote{It is much easier to detect invariants that can be represented as polynomial functions of input features.} Similar results can be obtained for problems in special relativity and electromagnetism~\cite{wetzel:2020}.\footnote{The process of finding symmetry invariants and conserved quantities with \ac{ML} has emerged as its own subfield, and many improved methods to detect these have been devised in Refs.~\cite{Tegmark2021,liu2022ai,ha2021discovering}.}

So far, we have learned that we can interpret \ac{ML} models by analyzing bottlenecks in their architecture. These bottlenecks can either appear naturally, such as in \acfp{AE}, or be imposed explicitly, such as in \acp{CNN} where the kernel size is systematically reduced or in Siamese \acp{NN}). Another approach towards interpretability is based on the analysis of the minimum of the training loss function reached by a~model during the optimization. Because such an analysis is based on the minimum, it is generic and can be applied independent of the particular choice of \ac{ML} model architecture or learning procedure.

\subsubsection{Hessian-based interpretability}\label{sss:hessian-based-interpret}\index{Hessian}

\begin{figure}[t]
\begin{center}
\includegraphics[width=0.99\columnwidth]{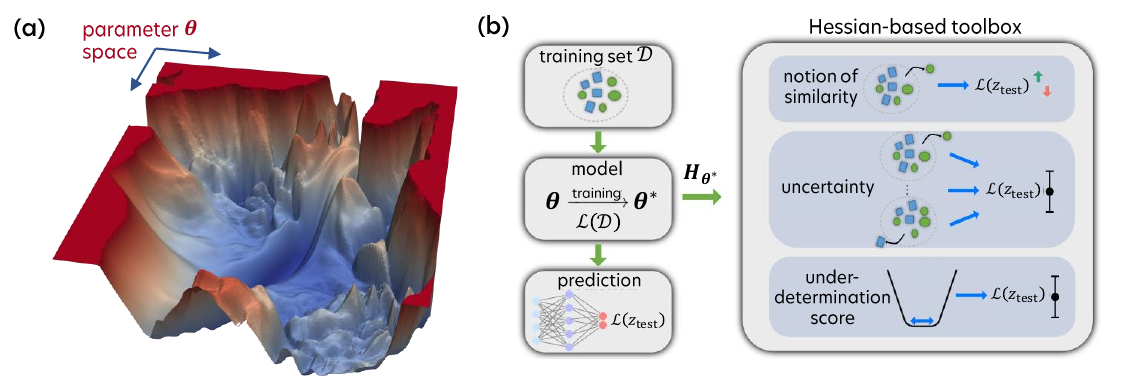}
\end{center}
\caption[Hessian-based interpretability]{(a) Low-dimensional visualization of a~non-convex loss landscape\index{loss landscape} of a~\ac{DNN} called VGG-56 trained on CIFAR-10. Taken from \ToggleForCUP{Li, H. \textit{et al.} (2018) \textit{Visualizing the loss landscape of neural nets}. Adv. Neural Inf. Process. Syst. 31 (NeurIPS 2018)~\cite{Li2018} under the MIT license.}{Ref.~\cite{Li2018}.} (b) Hessian-based toolbox to increase the interpretability and the reliability of a~trained \ac{ML} model. It is based on the Hessian of the training loss at the minimum (or an~approximation thereof). Adapted from \ToggleForCUP{Dawid, A. \textit{et al.} (2021) \textit{Hessian-based toolbox for reliable and interpretable machine learning in physics}. Mach. Learn.: Sci. Technol. 3 015002~\cite{dawid:2021} under the \href{https://creativecommons.org/licenses/by/4.0/}{CC BY 4.0 DEED} license.}{Ref.~\cite{dawid:2021}.}}
\label{fig:hessian-interpretability}
\end{figure}

As described in \cref{sss:optimization}, \ac{ML} models learn by minimizing a~training loss function $\lossfun$ describing the problem through the variation of their parameters $\params$. The training loss landscape\index{loss landscape} of deep \acp{NN} is, however, highly non-convex. This renders the optimization problem difficult, e.g., due to the presence of many local minima [see \cref{fig:hessian-interpretability}(a)]. Moreover, these minima may not have equally good generalization abilities, and it seems that these abilities are connected to the curvature around a~minimum.\footnote{There is a~general consensus that wide, flat minima generalize better than sharp minima~\cite{Keskar2017,Wu2017,Izmailov2018,He2019}. Keep in mind that flatness is not a~well-developed concept in non-convex landscapes of deep models ~\cite{Dinh2017}.} This connection is an instance where the shape of the reached minimum can tell us something useful about trained \ac{ML} models. The shape or curvature around the minimum $\params=\params^{*}$ is described by a~Hessian matrix calculated at the minimum, i.e.:

\begin{equation}\label{eq:hessian}
    \mat{H}_{\params^{*},ij} = \frac{\partial^2}{\partial \param_i \param_j} \lossfun_{\mathrm{train}} |_{\params=\params^{*}}\,.
\end{equation}

The knowledge of the curvature\index{Hessian} around the minimum also allows us to approximate how our \ac{ML} model (and, as a~result, its predictions) would change upon some action. Possible actions could be the removal of a~single training point or a~slight modification of the model parameters $\params^* \rightarrow \tilde{\params}$ toward an~adjacent minimum with identical training error. The study of how a~model reacts to such actions is at heart of the Hessian-based toolbox summarized in~\cref{fig:hessian-interpretability}(b), which contains influence functions~\cite{Koh17}, the \ac{RUE}~\cite{Schulam2020}, and \acp{LE}~\cite{Madras2020} whose conceptual ideas we introduce in the following.

\stress{Influence functions}\index{influence functions} are an~approximation of the procedure known as leave-one-out training\footnote{Leave-one-out training for \ac{DL} models with non-convex loss landscapes is tricky because if we land in a~different local minimum, we cannot make any claims on the perturbation caused by the removal of a~single training point. This is why we usually retrain carefully, starting from the minimum reached by the original model.} and estimate how the model prediction on a~test point $\vect{x}_\mathrm{test}$ change if a~certain training point $\vect{x}_\mathrm{R}$ is removed from the training set. You can imagine three outcomes of such a~removal: (1) the prediction stays the same because the removed training point does not influence on the model prediction, (2) the prediction gets better (i.e., it leads to a~lower test loss for $\vect{x}_\mathrm{test}$), so $\vect{x}_\mathrm{R}$ is a~``harmful'' training point for making a~prediction on $z_\mathrm{test}$, (3) the prediction gets worse (i.e., it leads to a~larger test loss for $\vect{x}_\mathrm{test}$), so $\vect{x}_\mathrm{R}$ is a~``helpful'' training point for making a~prediction on $z_\mathrm{test}$, and its removal made the task of constructing an~accurate model harder. With such an~analysis we can determine how influential training data points are to predictions at test points, which may give us a~hint at how the model reasons. We can even go a~step further and say that if two data points strongly influence each other, it is because they are very similar from the model's perspective.\footnote{This argument is well based on the geometric interpretation of influence functions, for details see section 2.3.3 in Ref.~\cite{dawid:2021}.} This concept of similarity learned by a~\ac{ML} model can be understood as a~distance between data points in the internal model representation and is a~powerful tool for \stress{detecting additional phases in mislabeled data}~\cite{dawid:2020, Kaming2021}, \stress{detecting influential features}~\cite{Kaming2021}, and \stress{detecting anomalies}\index{anomaly detection}~\cite{dawid:2021}.\footnote{We stress that \stress{similarity}, which is arguably the central concept behind classification tasks, has various meanings when it comes to Siamese \acp{NN}, influence functions, and kernel methods (which will be covered in the next chapter).}

Another tool in the Hessian-based toolbox is the \stress{\acf{RUE}}\index{reliability!resampling uncertainty estimation}~\cite{Schulam2020}. It aims to estimate the uncertainty\index{uncertainty} of model predictions. It is an~approximation of the classical procedure known as bootstrapping. You start with your original training set containing each training data point once. Imagine now that you create $b$ new training sets by drawing samples uniformly with replacement from the original data set. Due to the replacement, your new sets contain some training points in more than one copy, and some points have been omitted. Now you can train $b$ models on these $b$ training sets and make $b$ predictions on the same test point, $z_\mathrm{test}$. These predictions generally vary due to the distinct nature of the training sets. Computing the variance of these predictions on $z_\mathrm{test}$ gives us an~estimate for the uncertainty of the original model prediction. A~small variance signals that one can trust the prediction of the original model because small random modifications to the training set do not change its prediction too much. A~large variance signals that the prediction is based on a~small number of training points and is therefore not reliable. In Ref.~\cite{dawid:2021}, you can see how such error bars indicate the \stress{sharpness of quantum phase transitions}.

Finally, \stress{\acfp{LE}}\index{reliability!local ensembles}~\cite{Madras2020} allow us to detect the \stress{underspecification}\index{reliability!underspecification} of a~given model at the test point. A~trained model is underspecified at a~test input if
many different predictions at that input data are all equally consistent with the constraints posed by the training data and the learning problem specification (i.e., the model architecture and the loss function). As described in~\cref{sss:optimization}, the minimum reached within the optimization is usually surrounded by a~mostly flat landscape. This means that if the model had ended up in one of these flat neighboring points, the training error would have stayed exactly the same. Thus, such changes should not impact the predictions -- unless a~prediction is \stress{underdetermined}, i.e., unstable and not well-explained by the training data. Therefore, we can again create multiple models by shifting the parameters of the original model by small amounts. As such, these new models explore the flat landscape around the original minimum. Eventually, we make predictions with these new models. If a~prediction on a~test point $z_\mathrm{test}$ changed due to such modifications, this point may be an~out-of-distribution point, i.e., a~point coming from a~distribution that is significantly different from the distribution underlying the training data. \Acp{LE} allow for the detection of such out-of-distribution test points, which increases the reliability of the \ac{ML} model. Moreover, the authors of Ref.~\cite{Madras2020} successfully used \acp{LE} for active learning\index{active learning}, i.e., they built a~much smaller, yet similarly informative training data set by iteratively adding to it test points with the largest underspecification score detected by \acp{LE}.

\highlight{Therefore, we have answered our initial questions regarding interpretability\index{interpretability}: It is indeed possible to look inside the black box of \ac{ML} models. If you focus your attention on the bottlenecks present in \ac{NN} architectures, you can determine which quantities dominate in the \ac{NN} prediction using regression methods. Further, it is possible to interpret the filters in the early layers of neural networks. If these quantities are physically relevant, we can argue that the \ac{NN} indeed bases its predictions on physically relevant quantities. Additionally, you can turn your attention to the curvature around the minimum of the training loss. It contains information on the similarity learned by a~model and allows for estimating the uncertainty.}

\subsubsection{A probabilistic view on phase classification}
In Sec.~\ref{sss:probability}, we have introduced a probabilistic view on \ac{ML}. This viewpoint turns out to be particularly useful when applying \ac{ML} to the task of classifying phases of matter. In the following, we focus on the simple case of supervised learning (\cref{sec:supervised_phase_class}). However, the idea also generalizes to other \ac{NN}-based phase classification methods, such as learning by confusion (\cref{sec:lbc}) or the prediction-based method (\cref{sec:pbm}), see Refs.~\cite{arnold:2022,arnold:2023}. Let us first consider the scenario where we would like to distinguish between images of cats and dogs in a supervised setting. Recall from Sec.~\ref{sss:probability} that a Bayes classifier outputs
\begin{equation}
    y_\mathrm{Bayes}(\vect{x}) = \argmax\ \{ p({\rm{cat}} \mid \vect{x}),\; p({\rm{dog}} \mid \vect{x})\}\,,
\end{equation}
where $p(\rm{cat}\mid\vect{x})$ and $p(\rm{dog}\mid\vect{x})$ are the probabilities that the given sample $\bm{x}$ is a cat or a dog, respectively. The \stress{Bayes classifier} is optimal as it outperforms any other classifier in the classification task at hand, i.e., achieves the lowest possible misclassification probability. However, for most real-world data sets, such as images of cats and dogs, the ground-truth class-conditional probabilities are inaccessible, abstract quantities, and we do not know them (or their form) \stress{a priori}. One way to tackle the classification task is thus to parametrize these conditional probabilities by an \ac{NN} and train it to minimize the misclassification probability (i.e., the corresponding loss function). Because the \ac{NN} is a universal function approximator, its predictions are expected to approach those of a Bayes classifier as we make the \ac{NN} more expressive, train it better, and increase the size of our data set. Similarly, the misclassification probability of our \ac{NN} is expected to approach the Bayes error from above.

As discussed, in realistic scenarios, it is typically impossible to construct a Bayes classifier. Interestingly, this can change when we move to the realm of physics, particularly statistical physics and quantum physics, where the data underlying the task of classifying different phases of matter resides. To illustrate this, let us consider the case of supervised learning, where we want to distinguish between two phases: phase A and phase B. The optimal outputs of a Bayes classifier are then given as
\begin{equation}
    y_\mathrm{Bayes}(\vect{x}) = \argmax\ \{p({\rm phase \; A}\mid\vect{x}),\;p({\rm phase \; B}\mid\vect{x}) \}\,.
\end{equation}
Using Bayes' rule, \cref{eq:Bayes_rule}, we have
\begin{equation}
    p({\rm phase \; A}\mid\vect{x}) = \frac{p(\vect{x}\mid{\rm phase \; A})\,p({\rm phase \; A})}{p(\vect{x})}\,,
\end{equation}
and similarly for $p({\rm phase \; B}\mid\vect{x})$. Now assuming both phases are represented equally in terms of their labels in our data set, we have $p({\rm phase \; A})=p({\rm phase \; B})=\frac{1}{2}$. Moreover, we have
\begin{equation}
    p(\vect{x}) = p({\rm phase \; A})p(\vect{x}\mid{\rm phase \; A}) + p({\rm phase \; B})p(\vect{x}\mid{\rm phase \; B})\,,
\end{equation}
which, inserting in Bayes' rule, yields
\begin{equation}
    p({\rm phase \; A}\mid\vect{x}) = \frac{p(\vect{x}\mid{\rm phase \; A})}{p(\vect{x}\mid{\rm phase \; A}) + p(\vect{x}\mid{\rm phase \; B})}\,.
\end{equation}
Hence, we can compute the class-conditional probability (corresponding to the optimal prediction of a Bayes classifier) if we know the probability of drawing the sample $\vect{x}$ in either of the two phases, i.e., if we know $p(\vect{x}\mid{\rm phase \; A})$ and $p(\vect{x}\mid{\rm phase \; B})$.\footnote{We might apply Bayes' rule in the same fashion to the task of classifying of images of cats and dots. In this case, we would need to know $p(\vect{x}\mid{\rm cat})$ and $p(\vect{x}\mid{\rm dog})$ to construct the class-conditional probabilities; quantities that are abstract and inaccessible and thus evade this discussion.} Let us denote the physical parameter we vary to get from phase A to phase B as $\gamma$ and assume that we sample phase A/B at distinct points $\{\gamma \in {\rm phase \;A/B}\}$. Then, we have 
\begin{equation}
    p(\vect{x}\mid{\rm phase \; A}) = \frac{1}{|\{\gamma \in {\rm phase \;A}\}|}\sum_{\gamma \in {\rm phase \;A}} p_{\gamma}(\vect{x})\,,
\end{equation}
where $|\{\gamma \in {\rm phase \;A}\}|$ is the number of distinct sampled points in phase A (and similarly for phase B). This shows that we can express $p(\vect{x}\mid{\rm phase \; A/B})$ (and thus the class-conditional probability) based on the probability distribution $p_{\gamma}(\vect{x})$ underlying the physical system of interest sampled at distinct values of the tuning parameter $\gamma$. These probability distributions have a clear physical meaning, and we often know them completely or at least partially.\footnote{In experimental scenarios, we may not be able to access the distribution directly. However, physicists strive to isolate and characterize the classical statistical ensemble or quantum state they realize in their experiment as best as possible.} For example, when we analyze systems at thermal equilibrium at various temperatures, such as in the case of the Ising model (\cref{sec:ising_model}) or Ising gauge theory (\cref{sec:ising_gauge_theory}), we know that the underlying distribution is Boltzmann, \cref{eq:boltzmann_distr}. Similarly, when studying quantum phases, we may have access to the wave function of the quantum state that governs the measurement statistics. 

From this perspective, the task of (optimal) phase classification boils down to characterizing the probability distributions underlying the physical system at hand. Phase transitions then manifest themselves as rapid changes in these probability distributions. This offers an alternative avenue for classifying data into distinct phases of matter (namely by estimating $p_{\gamma}(\vect{x})$ first rather than trying to estimate $p(\vect{x}\mid{\rm phase \; A})$ and $p(\vect{x}\mid{\rm phase \; B})$ directly) and sheds a different light on ``traditional'' \ac{NN}-based phase classification. For a more complete overview of this probabilistic view on phase classification, see Refs.~\cite{arnold:2022,arnold:2023}.
\subsection{Outlook and open problems}
Over the last five years, there have been many works applying supervised and unsupervised phase classification algorithms, including supervised learning (\cref{sec:supervised_phase_class}), learning by confusion (\cref{sec:lbc}), and the prediction-based method (\cref{sec:pbm}), to models with well-known phases. However, only a few works have applied unsupervised phase-classification methods to experimental data. Moreover, the discovery of a~novel phase of matter using unsupervised phase classification methods still remains to be demonstrated. This would constitute a~major step toward the automation of scientific discovery.

While there has been significant progress regarding the interpretability of phase classification methods in recent years, we still lack a~deeper understanding of these methods. In particular, it remains difficult to tell when and why a~given method fails or succeeds~\cite{arnold:2022}. With the goal of automated scientific discovery in mind and having demonstrated that phase classification methods are capable of dealing with a~vast range of physical systems, addressing these gaps in knowledge and developing corresponding interpretability tools is of crucial importance.

\subsection*{Further reading}
\begin{itemize}
    \item Carleo, G. \textit{et al.} (2019).
    \href{https://journals.aps.org/rmp/abstract/10.1103/RevModPhys.91.045002}{\textit{Machine learning and the physical sciences}}. Rev. Mod. Phys. 91, 045002.
    An~overview of the current state of the phase classification landscape is presented in section 4C~\cite{Carleo2019RevModPhys}.
    \item Neupert, T. \textit{et al.} (2021). \href{https://arxiv.org/pdf/2102.04883.pdf}{\textit{Lecture notes: Introduction to machine learning for the sciences}}. An~introduction to fundamentals of \ac{ML} and clustering algorithms for scientists~\cite{mehta:2019}.
    \item Molnar, C. (2019).  
     \href{https://christophm.github.io/interpretable-ml-book/}{\textit{Interpretable Machine Learning: A~Guide for Making Black Box Models Explainable}}.
     An~introductory book on interpretable \ac{ML}~\cite{molnar:2019}.
    \item \href{https://zenodo.org/record/3759432}{Jupyter notebook} on phase classification~\cite{OurSchoolRepo}.
\end{itemize}

\clearpage
\section{Gaussian processes and other kernel methods}
\label{sec:gp}

This section deals with the so-called kernel methods, of which \acfp{SVM} and \acfp{GP} are prominent examples. We point the interested reader to the exemplary and definitely non-exhaustive collection of Refs.~\cite{muller2018introduction, Scholkopf2018kernels} for a deep dive into the foundations of kernel methods in \ac{ML} and to Ref.~\cite{overview_kernel_methods} for a more high-level overview.
These methods are particularly well suited in the case of low availability of labeled data. This usually happens when the creation of a~large data set is expensive in terms of money, time, effort, etc. As a~second advantage, the predictions of \acp{GP} are, \stress{by construction}, accompanied by their uncertainties which other methods do not readily provide. We see how this property arises from the design choice of the models in \cref{sec:kernel_methods}. 

But first, we have to introduce the notion of the \stress{kernel}\index{kernel} that is an integral part of all the methods discussed in this chapter. The introduction of the kernel allows us to extend the range of problems we are able to tackle substantially. Before properly defining the mathematical foundation of kernels, we start by providing some intuition on how to use them in practice based on the \stress{kernel trick}\index{kernel trick} and its implications. Afterward, we show how to extend the aforementioned methods via this kernel trick and discuss how to train each of them given data. For example, it turns out that \acp{GP} can be approached from an~information-theoretic perspective. Furthermore, we explain how we can make use of concepts from information theory for a~guided data acquisition procedure, as well as to select a~good model among various possible ones. We end the chapter by showcasing the power of these methods in tackling quantum problems in \cref{sec:BO_GPR_science}.

\subsection{The kernel trick}\label{sec:kernel-trick}

As we have seen in \cref{sec:models}, simple approaches such as the linear regression model or the linear \ac{SVM} have severe limitations with respect to the properties of data. They are applied to the input data \stress{as is}, i.e., they are bound to the given representation of the input data. In order to avoid confusion later on, we refer to this data space as the \stress{input space}. In this input space, it can, for example, happen that the given input data are not linearly separable. One possible remedy consists of extending the classification power of the model by first transforming the input data into an~alternative \stress{feature space}\index{feature space}. In contrast to what we have said earlier in \cref{sec:intro}, we explicitly distinguish between the input and the feature space in the following.\footnote{Each element of a~data point is still called a~\stress{feature}\index{feature} -- in this chapter, we are, however, only interested in finding the most convenient representation of the data whose space we hence call the \stress{feature} space.} Ideally, in this new representation, the data possesses a~more convenient structure compared to the original representation. For example, data that was initially not linearly separable may be linearly separable in this new space. In particular, it is often useful to transform into a~higher-dimensional feature space in which our data is now nested on a~manifold that (ideally) possesses beneficial additional structure. 

One may think that this makes machine learning extremely costly (or even infeasible if the dimension of the feature space approaches infinity). However, we can make use of the fact that the predictions of our \ac{ML} algorithms of interest are often formulated in terms of distances between data in the input space.
\highlight{This is where the \stress{kernel trick} comes in: instead of explicitly transforming the data into the feature space and then calculating the distance therein, we start from the other end and provide a closed-form expression for the distance in terms of the data representation in the input space. This typically is much more efficient from a numerical perspective.}
As we will see later on, we can always associate a unique feature space with any valid distance function.
Thus, we shift the focus from finding a suitable representation to choosing a suitable distance function.
This trick is called the \stress{kernel} trick because the kernel is the mathematical object we associate with such a~feature space. As such, the kernel trick allows one to retain all the benefits of high-dimensional feature spaces at a~manageable computational cost.
Moreover, the mathematical foundation of kernels allows us, as we see in the next section, to enrich our motivation with rigorous, analytical validity.
Especially important from a~practical point of view is the \stress{representer theorem}: so far, we have set out to find a~suitable transformation, i.e., function to simplify our task at hand. However, it is unclear how to optimize over functions instead of parameters. The representer theorem endows us with both: in essence, it assures that the optimization over the function space is equivalent to optimizing the coefficients of a~closed form solution, which, in turn, allows us to devise feasible numerical optimization routines.

In the following, we start with an~intuitive example to illustrate why and how the transformation into the feature space can be beneficial.
Afterward, we properly introduce the mathematical notion of kernels, which gives us the analytical tools required to understand the representer theorem.

\subsubsection{Intuition behind the kernel trick}\label{ssec:kernel-trick-sketch}

\begin{figure}[t]
\centering
\includegraphics[width=0.9\textwidth]{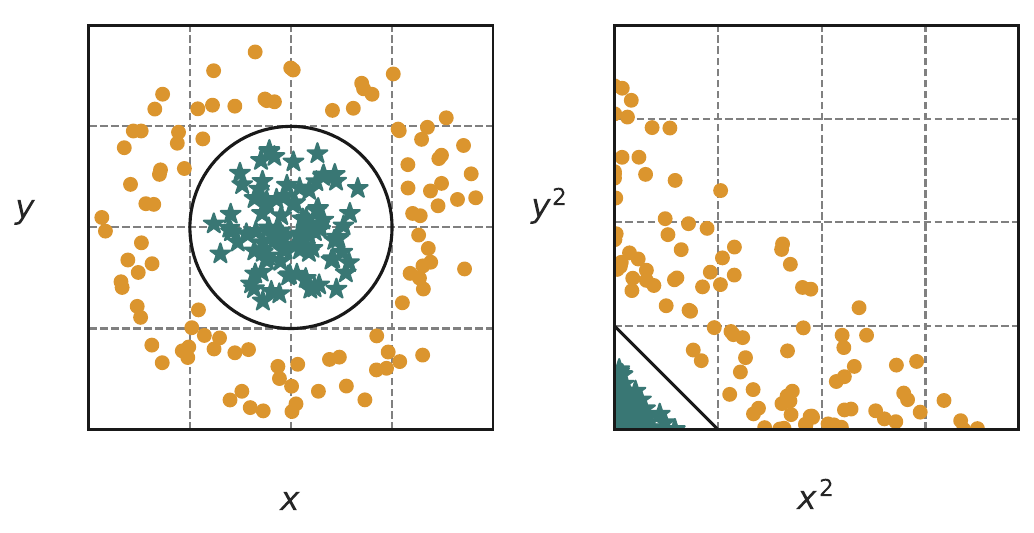}
\caption[Toy example of a~two-dimensional data set in the input and feature space]{Toy example of a~labeled two-dimensional data set. The data points are labeled according to their position with respect to the decision boundary indicated by the black circle in (a).
In this \stress{input space}, such a~data set is not linearly separable.
After a~transformation of the input variables into a~nonlinear \stress{feature space}, however, the data becomes linearly separable as indicated by the black line in (b).}
\label{fig:kernel_trick_motivation}
\end{figure}

To gain some intuition, let us consider a~labeled two-dimensional data set as depicted in \cref{fig:kernel_trick_motivation}.
In this toy example, the black line indicates the underlying decision boundary, i.e., the line that separates input data with different labels.
In higher dimensions, the decision boundary generalizes to a~hyperplane.
In our example, one label refers to the center of the data cloud and the other to its outskirts. A~label distribution is said to be separable if one can find at least one such hyperplane separating the two class sets.
If, furthermore, this decision boundary is linear, the data is called linearly separable.\footnote{Whether a~given data set actually is (linearly) separable or not is not easily detectable. In practice, we can at least run algorithms such as an~\ac{SVM} explained in \cref{sec:intro-SVM}, which are guaranteed to find the corresponding separating hyperplane if it exists.} Clearly, our toy data set is not linearly separable in the input space. As a~consequence, we cannot find a~straight line that fully separates the two data classes by means of a~simplistic linear classifier. Additionally, other linear methods such as \ac{PCA} (see \cref{sec:phase_class}) fail to cluster this data.\footnote{As described in \cref{sec:phase_class}, \ac{PCA} itself is not a clustering algorithm. However, we have also seen that it can be used to provide a low-dimensional representation of the data in which the data is split into different clusters.}

However, as shown in the right panel in \cref{fig:kernel_trick_motivation}, the data set becomes linearly separable if we transform the data in the appropriate way.
Here, we have applied the transformation $\featmap: (x_1,x_2) \mapsto (x_1^2,x_2^2)$ to map the input data nonlinearly to the so-called feature space. Hence, the map $\featmap$ is called the feature map.
There are two important caveats:
firstly, finding a~useful feature map is a~highly-non-trivial task.
In our toy example, the labeling procedure considers the data points in polar coordinates and uses the radial distance $r$ to the origin as the label criterion.
Using the connection $r^2 = x_1^2 + x_2^2$, we motivate our feature map $\featmap$, whose choice is by no means unique.
Secondly, even if we had found a~good feature map, it could be infinite-dimensional.\footnote{In a~sense, we do the opposite of a~dimensionality reduction method such as \ac{PCA} in \cref{ssec:pca} -- usually, we drastically increase the dimension of the feature space to rearrange the data most conveniently. Another difference is that we do not lose any information in the data by embedding them into a~high-dimensional space.} In that case, it would not be feasible to transform the data with the feature map. In our example, we only squared the input variables to achieve linear separability. However, there can be instances where the polynomial expansion has to be taken to infinite order. We encounter such an~example where the most suitable feature space is infinite-dimensional later on when we come back to our toy example at the end of \cref{sec:kernel_SVM}.

\highlight{Fortunately, it turns out that many \ac{ML} algorithms can be expressed only in terms of inner products between data points $\vect{x}$, $\vect{y}$.\footnote{We remind the reader that these data points are, in general, vectors, i.e., they may live in a~high-dimensional space, see also \cref{sec:typeoflearning}} As such, we do not need to consider the individual inputs $\vect{x}$ and their representation in feature space $\featmap (\vect{x})$ explicitly.
The \stress{kernel trick}\index{kernel trick} simply consists of exchanging the inner product 
$\estimate{\vect{x},\vect{y}}{} = \vect{x}^\transpose \vect{y}$ in the corresponding algorithms by the function 
$K(\vect{x},\vect{y}) = \featmap (\vect{x})^\transpose \featmap (\vect{y})$.}
The function $K$ can be efficiently evaluated, in particular, for an~infinite-dimensional feature space to which $\featmap$ maps and yields a~single real non-negative number regardless of the data under consideration or the dimension of the feature space.\footnote{In many cases, it is easier to work with kernels yielding infinite-dimensional feature spaces than with finite ones. Indeed, the finite dimensionality may often be a problem. We refer to \cref{sec:qml_kernels} for quantum kernel methods and the references therein regarding this issue.}
As we see later, the function $K$ is referred to as the kernel function.
This way, the kernel trick allows for a~feasible nonlinear extension of a~variety of \ac{ML}-algorithms, such as ridge regression, \acp{SVM}, or \ac{PCA}.
We detail the mathematical foundation of kernels in the next sections.

\subsubsection{The function space as a~Hilbert space}\label{sss:kernel-Hilber-space}

As we have sketched above, we want to find a~suitable choice for our feature map, i.e., search for a~function that lives in some high-dimensional space of functions.
A~mathematical space is a~set of elements that obey common rules specifying the relationship between them.
As we want to search in a~function space, it is reasonable to assume it to be a~vector space. In such a~space, we are allowed to add functions to each other or rescale them by a~constant without leaving the function space.
This assumption is necessary to express the unknown target function by a~weighted sum of other known functions. Fortunately, tuning the weights is something that we are familiar with in \ac{ML} problems.
As a~second ingredient, we require some distance measure or, equivalently, a~measure of similarity between functions.
This is achieved by equipping our vector space with an~inner product $\estimate{\cdot , \cdot}{}$, turning the mathematical space into a~Hilbert space.
Different choices of inner products lead to different notions of distance, i.e., different spaces altogether.
We will see why we need this similarity measure shortly.

Consider, for example, a space of real-valued \stress{square-integrable} functions called the $\lspace$-space. When equipped with an inner product, it is 
akin to the more intuitive Euclidean space.~\footnote{Of course, not all functions we are interested in finding are necessarily members of this function space. Because we are always presented with a~finite data set, however, we usually do not care what happens very far outside this regime. Even though our target function might not be a~member of this space, we should be able to find a~member that resembles the target function in the region of interest, nevertheless. This is why we can restrict ourselves with the $\lspace$-space in the first place.}
Square-integrability refers to the fact that the integral over the full domain $D$ remains finite, i.e., $\int_{\domain} f^2(\vect{x})\ d\vect{x} < \infty\ \forall f \in \lspace$. The domain $\domain$ is often given by $\realset^\featnum$ with some dimension $\featnum$ in typical \ac{ML}-scenarios as it corresponds to our input space. The inner product of an 
$\lspace$-space can be defined as
\begin{equation}
    \estimate{f,g}{} = \int_{\domain} f(\vect{x})g(\vect{x})\ d\vect{x}
    \label{eq:kernel_inner_prod_l2}
\end{equation}
which provides the notion of orthogonality as $\estimate{f,g}{} = 0$, and the~norm $\norm{f}_{\lspace}^2 = \estimate{f,f}{}$.
Since a~Hilbert space is a~vector space, one can find an~orthogonal basis set $\{ \featmap_n(\vect{x}) \}$ that spans the space. In case of the $\lspace$-space it is of infinite (but countably infinite) dimension.
This basis allows for any function $f \in \lspace$ to be decomposed as
\begin{equation}
    \label{eq:kernel_basis_decomposition}
    f(\vect{x}) = \sum_n a_n \featmap_n (\vect{x})\,,
\end{equation}
with expansion coefficients $\{a_n\}$.

Any two Hilbert spaces have the same geometric structure, regardless of their respective elements -- this is the reason we can draw the analogies between the $\lspace$ and the intuitive Euclidean space in the first place.
This can be made more formal by the representation theorem of Riesz~\cite{Bachman2000_functional_analysis}, which essentially boils down to this:  
one can express certain linear functionals by means of the Hilbert space's inner product. The functional of interest is the \stress{evaluation} of an arbitrary function $f$ living in a Hilbert space $\H$ at any point $\vect{x} \in \domain$, i.e., $f \in\H \mapsto f(\vect{x})\in\realset$.
This evaluation property will become essential once we are trying to learn a target function.
Riesz' theorem further tells us that there exists a unique function $K_{\vect{x}}$ such that any function $f\in\H$ can be evaluated at $\vect{x}$ by its inner product with $K_{\vect{x}}$.

\subsubsection{Reproducing kernel Hilbert spaces}\label{sss:kernel-RKHS}

As anticipated, we want to use the Riesz theorem to connect the function space to the evaluation of its members.
That is, we require $K_{\vect{x}}$ to assert that point evaluation is of the form
\begin{equation}
    f(\vect{x}) = \langle f, K_{\vect{x}} \rangle\,.
\label{eq:reproducing_kernel}
\end{equation}
Because of \cref{eq:kernel_inner_prod_l2} and with the definition $K_{\vect{x}} \eqqcolon K(\vect{x},\cdot)$, this leads to the integral transform
\begin{equation}
    f(\vect{x}) = \int_{\domain} K(\vect{x},\vect{x'}) f(\vect{x'})\ d\vect{x'} \quad \forall f\in\H\,.
\label{eq:kernel_definition}
\end{equation}
Here, $K$ acts as the \stress{kernel} of the integral transform.
It can be seen that the kernel $K$ of this transform must be symmetric, i.e. 
$K(\cdot,\vect{x}) = K(\vect{x},\cdot)$.
Moreover, due to its role in \cref{eq:kernel_definition}, we call the kernel $K$ a \stress{reproducing} kernel as it faithfully reproduces any function $f$ in our Hilbert space.
If point evaluations of every function in a Hilbert space can be represented as a reproducing kernel, the Hilbert space is called a \acf{RKHS}.

We will refer to $K(\vect{x}, \vect{x'})$ as the \stress{kernel function}. Furthermore, we will restrict our discussion to functions $K(\vect{x}, \vect{x'})$ that are positive-semidefinite. Due to Mercer's theorem~\cite{mercer:1909}, any symmetric, positive-semidefinite function $K(\vect{x}, \vect{x'})$ can be represented as
\begin{equation}\label{eq:kernel_Mercer_theorem}
    K(\vect{x},\vect{x'}) = \sum_n \lambda_n \featmap_n (\vect{x}) \featmap_n(\vect{x'})
\end{equation}
with non-negative coefficients $\{\lambda_n\}$.
Moreover, for an~orthonormal basis set, the coefficients $\lambda_n$ must be all equal to one in order to yield $K(\vect{x},\vect{x'})$ satisfying Eq.(\ref{eq:kernel_definition}).
We can prove this by choosing $f = \featmap_k$ for some integer $k$.
Then from \cref{eq:kernel_definition} and the decomposition of the kernel, we require that
\begin{equation}
    \featmap_k(\vect{x}) = \sum_n \lambda_n \featmap_n(\vect{x}) \int_{\domain} \featmap_n(\vect{x'}) \featmap_k(\vect{x'})\ d\vect{x'} = \lambda_k \featmap_k(\vect{x})\,.
\end{equation}
Since $k$ was chosen arbitrarily, we find that $\lambda_n = 1\ \forall n$ if and only if we select an~orthonormal basis set.
We can also convince ourselves that this kernel representation actually performs the integral transformation in \cref{eq:kernel_definition} via
\begin{equation}
\ToggleForCUP{\begin{aligned}
    \int_{\domain} K(\vect{x},\vect{x'})f(\vect{x'})\ d\vect{x'} &= \sum_{n,k} a_n \featmap_k(\vect{x}) \int_{\domain} \featmap_k(\vect{x'}) \featmap_n (\vect{x'})\ d\vect{x'} \\
    &= \sum_n a_n \featmap_n(\vect{x}) = f(\vect{x})
\end{aligned}}{\int_{\domain} K(\vect{x},\vect{x'})f(\vect{x'})\ d\vect{x'} = \sum_{n,k} a_n \featmap_k(\vect{x}) \int_{\domain} \featmap_k(\vect{x'}) \featmap_n (\vect{x'})\ d\vect{x'} = \sum_n a_n \featmap_n(\vect{x}) = f(\vect{x})}
\end{equation}
as intended.

One may recognize \cref{eq:kernel_Mercer_theorem} with $\lambda_n = 1~\forall n$ as a basis-set representation of a delta-function $\delta(\vect{x}-\vect{x'})$.
Obviously, this is not the most useful choice of the kernel function because, in \ac{ML}, the task is generally to estimate the target function $f(\vect{x})$ at values of $\vect{x}$ different from the positions of the training points $\vect{x}_i$ by means of equations written in terms of the kernel functions $K(\vect{x},\vect{x}_i)$.  The question thus becomes: can the above arguments be extended to any arbitrary function $K(\vect{x},\vect{x'})$ that is symmetric and positive-semidefinite? In particular, can the reproducing kernel yielding \cref{eq:reproducing_kernel} be defined for any symmetric, positive-semidefinite kernel function?

To answer this question, consider the eigenvalue decomposition of a~positive-semidefinite function $K(\vect{x},\vect{x'})$ 
\begin{equation}\label{eq:ev_eq}
    \int_{\domain} K(\vect{x},\vect{x'}) \featmap_n (\vect{x'})\ d\vect{x'} = \lambda_n \featmap_n(\vect{x}) \quad \forall n\,,
\end{equation}
where $\lambda_n \geq 0$.
Using \cref{eq:kernel_basis_decomposition}, we can rewrite the inner product \cref{eq:kernel_inner_prod_l2} as
\begin{equation}
\ToggleForCUP{\begin{aligned}
    \estimate{f,g}{} &= \int_{\domain} f(\vect{x}) g(\vect{x})\ d\vect{x} = \sum_{m,n} a_m b_n \underbrace{\int_{\domain} \phi_m(\vect{x}) \phi_n(\vect{x}) \ d\vect{x}}_{= \estimate{\phi_m,\phi_n}{} = \delta_{m,n}} = \sum_n a_n b_n  \\
    &= \sum_n \int a_m \phi_m(\vect{x}) \phi_n(\vect{x})\ d\vect{x} \int b_k \phi_k(\vect{x'}) \phi_n(\vect{x'})\ d\vect{x'} \\
    &= \sum_n \estimate{f,\phi_n}{} \estimate{g,\phi_n}{}\,.
\end{aligned}}{\begin{aligned}
    \estimate{f,g}{} &= \int_{\domain} f(\vect{x}) g(\vect{x})\ d\vect{x} = \sum_{m,n} a_m b_n \underbrace{\int_{\domain} \phi_m(\vect{x}) \phi_n(\vect{x}) \ d\vect{x}}_{= \estimate{\phi_m,\phi_n}{} = \delta_{m,n}} = \sum_n a_n b_n  \\
    &= \sum_n \int a_m \phi_m(\vect{x}) \phi_n(\vect{x})\ d\vect{x} \int b_k \phi_k(\vect{x'}) \phi_n(\vect{x'})\ d\vect{x'} = \sum_n \estimate{f,\phi_n}{} \estimate{g,\phi_n}{}\,.
\end{aligned}}
\label{eq:kernel_l2_inner_prod_rewritten}
\end{equation}
However, in order for $K$ to be a~valid kernel function of an \ac{RKHS}, it has to give rise to \cref{eq:reproducing_kernel} as well.
Plugging this into the first step of \cref{eq:kernel_l2_inner_prod_rewritten} together with Mercer's decomposition, we see that
\begin{equation}
\begin{aligned}
    \estimate{f,g}{} &= \int f(\vect{x}) g(\vect{x})\ d\vect{x} = \int \estimate{f,K_{\vect{x}}}{} g(\vect{x})\ d\vect{x} \\
    &= \sum_n \lambda_n \int \phi_n(\vect{x}) \estimate{f,\phi_n}{} g(\vect{x}) = \sum_n \lambda_n \estimate{f,\phi_n}{} \estimate{g,\phi_n}{}.
\end{aligned}
\label{eq:kernel_l2_inner_prod_mercer}
\end{equation}
Comparing the two previous results, \cref{eq:kernel_l2_inner_prod_rewritten} and \cref{eq:kernel_l2_inner_prod_mercer}, we see a~discrepancy in terms of the prefactors $\lambda_n$. 
In order to compensate for this, we can redefine the inner product as
\begin{equation}
    \estimate{f,g}{\H} = \sum_{n=1}^\infty \frac{\estimate{f,\featmap_n}{} \estimate{g,\featmap_n}{}}{\lambda_n}\,,
    \label{eq:kernel_inner_prod_rkhs}
\end{equation}
which is equivalent to the previous inner product in \cref{eq:kernel_inner_prod_l2} if and only if $\lambda_n = 1\ \forall n$. With this definition of the inner product, we have
\begin{equation}
    \estimate{f,K_{\vect{x}}}{\H} = \sum_{n=1}^\infty \frac{\estimate{f,\featmap_n}{} \estimate{K_{\vect{x}},\featmap_n}{}}{\lambda_n} = \sum_{n=1}^\infty \estimate{f,\featmap_n}{} \featmap_n(\vect{x}) = f(\vect{x})\,.
\end{equation}
Here, we have used~\cref{eq:ev_eq} in the second step, and the last equality follows from the fact that the set $\{ \featmap_n \}$ forms a~complete basis, i.e., \cref{eq:kernel_basis_decomposition}.
This now fulfills \cref{eq:reproducing_kernel} as intended and, furthermore, renders the Hilbert space $\H$ an~\ac{RKHS}, and shows that $K$, any arbitrary symmetric positive-semidefinite function of two arguments, is indeed a~kernel function.
We note that this definition of the inner product -- which is unique for every kernel function -- is crucial. It exemplifies what is known as 
the Moore-Aronszajn theorem~\cite{aronszajn:1950}, which states that every \ac{RKHS} is associated with a~unique positive-semidefinite kernel, and vice versa.

\subsubsection{The representer theorem}
\label{sss:kernel_representer_theorem}
If we were to represent our target function as a basis set expansion (\cref{eq:kernel_basis_decomposition}), determining the function would require finding an -- in principle -- infinite number of the expansion coefficients. The representer theorem~\cite{scholkopf:2001}\index{representer theorem}, which plays a central role in kernel methods of \ac{ML}, allows one, however, to express the target function by a finite sum:  

Given a~loss function $\lossfun$ (including a~regularization term) and $\datasize$ training samples $\{ (\vect{x}_i,y_i) \}_{i=1}^{\datasize}$, the theorem implies that
\begin{equation}\label{eq:representer_theorem}
    f^\ast(\vect{x}) \coloneqq \argmin_f \lossfun\Big(\ \big\{ (f(\vect{x}_i),y_i) \big\}_{i=1}^\datasize\ \Big) = \sum_{i=1}^\datasize a_i K(\vect{x},\vect{x}_i),
\end{equation}
i.e., the loss is minimized by a function that can be written as a \stress{finite} sum over the kernel function evaluated at one of the arguments set to the position of training data points.
Moreover, one only requires the knowledge of the kernel function $K$ -- no explicit feature mapping is required.
\highlight{
The representer theorem guarantees that we can formulate the search for a~function $f^\ast$ that minimizes a~specific loss function over an~infinite-dimensional function space as a~search over $n$ kernel coefficients $\{a_{i} \}_{i=1}^{n}$. Thus, it significantly reduces the complexity of the minimization problem at hand and renders it computationally tractable.
}
Finally, our mathematical efforts have come to fruition: the reproducing property of the kernel in \cref{eq:kernel_definition} allows us to \stress{implicitly} embed the input data in a~(possibly) high-dimensional feature space in which we calculate the similarities to a~test point $\vect{x}$.
Because we are only required to calculate the similarity measure between data points, we do not lose efficiency here.
We see how to practically do this in the following when we extend some of the models of \cref{sec:models} via the kernel trick in \cref{sec:kernel_methods}.

\subsubsection{Consequences of the kernel trick}

We have gone into some detail on the mathematics behind kernels, in particular, concerning the \ac{RKHS}.
Despite the perceived detour through Hilbert spaces and a~redefinition of the inner product in the \ac{RKHS}, this leg work equips us with the necessary foundation for the theory of kernels: the reproducing feature of \cref{eq:kernel_definition} is not a~mere mathematical curiosity but has straightforward implications in terms of the representer theorem.
Secondly, the kernel formulation allows us to solve our initial problem (efficiently finding the unknown target function) by identifying the proper kernel.
The kernel can be understood as a~similarity measure between feature vectors representing data.
This similarity between inputs can be directly calculated even if the underlying feature space is high- or even infinite-dimensional.

One important consequence of \stress{implicitly} switching from the input to the feature space by means of the kernel trick is that we have to rethink our intuition of regularization:
now, we have to perform the regularization of the learned function in the function space given by the \ac{RKHS}\index{reproducing kernel Hilbert space} as discussed in the previous section. Thus, we have to start from the inner product in the \ac{RKHS}, i.e., \cref{eq:kernel_inner_prod_rkhs}.
The coefficients $\lambda_n$ correspond to the weights of the basis functions of the kernel and are non-negative by construction.
In particular, they depend on the actual choice of the kernel function $K$.
For instance, our kernel decomposition could include zero entries for some of the basis functions.
This is not an~issue per se: the function $f$ we are interested in might still lie entirely in the \ac{RKHS}.
Using \cref{eq:kernel_basis_decomposition}, we can decompose it as $f(\vect{x}) = \sum_{n=1}^\infty a_n \featmap_n(\vect{x})$.
Its corresponding $\lspace$-norm is $\norm{f}_2^2 = \sum_n |a_n|^2$.
Due to \cref{eq:kernel_inner_prod_rkhs}, this translates to a~norm in the \ac{RKHS} as
\begin{equation}\label{eq:regularization_term}
    \norm{f}_{\H}^2 = \sum_{n=1}^\mathrm{dim(RKHS)} \frac{|a_n|^2}{\lambda_n}.
\end{equation}
Thus, we see that we potentially run into trouble in case of vanishing $\lambda_n$ as our norm may diverge.
It remains finite if and only if the corresponding function coefficient $a_n$ is equal to zero at the same time.
This is the case if the function is entirely in the \ac{RKHS} corresponding to the kernel function.
If not, i.e., when choosing the wrong kernel function, we cannot regularize our model and cannot expect to learn the unknown target function $f$ entirely.

\highlight{%
The theory of \acp{RKHS} can give us an~intuition on why certain choices of kernel function seem to work while others fail:
the target function $f$ has to fully lie in the \ac{RKHS} uniquely defined by $K$.
If not, our approach to learning the function is doomed to fail from the start.
It is possible to develop strategies to build optimal kernels (which provide an optimal \ac{RKHS} for a particular problem), as, for example, discussed in \cref{sss:kernel-search}.
}

\subsection{Kernel methods}\label{sec:kernel_methods}

In \cref{sec:kernel-trick}, we have presented the mathematical foundation of kernels. In short, we want to map our data to a~feature space that possesses a~more suitable structure for the task at hand. Instead of explicitly defining a~feature map $\featmap$, we introduce a~kernel function $K$, which provides a~similarity measure between data points in the underlying feature space. As such, we exchange the problem of searching for a~(potentially) high-dimensional feature map for finding an~optimal kernel function, which is rigorously easier. This constitutes the kernel trick\index{kernel trick}.

Kernel methods correspond to all classification and regression methods that take advantage of the kernel trick. The validity of these approaches is ensured by the representer theorem, see \cref{sss:kernel_representer_theorem}.
The first step of every kernel method is to choose a~kernel function.
As discussed previously, any symmetric, positive-semidefinite function of two arguments can be used as a~kernel function.
Typically, one starts by assuming some functional form (see examples in \cref{tab:kernel-functions}). These functions are parametrized by a~few parameters, such as $\param$, or $\param_1$ and $\param_2$ in these examples. Having chosen a~particular functional form of the kernel function, one varies these parameters to find the best kernel in the corresponding functional ansatz class. This makes the optimization already easier as we now have to optimize over a~set of parameters and not over a~set of functions. Also note that given a~set of kernels, there exist many transformations which yield new valid kernels. For example, any linear combination of kernel functions $\sum_{i} c_{i} K_{i}\left(\vect{x}, \vect{x'}\right)$ with coefficients $c_{i}$ constitutes a~valid kernel. For a~more exhaustive list of techniques for constructing new kernels, see~\cite{bishop:2006}. We explicitly make use of these rules in~\cref{sss:kernel-search}, where we discuss how to construct good kernels systematically through compositional kernel search.

\begin{table}[t]
\centering
\caption[Examples of kernel functions]{Examples of kernel functions, where $\param$, $\param_{1}$ and $\param_{2}$ are free parameters~\cite{Vargas-Hernandez2018}.}
\label{tab:kernel-functions}
\centering
\ToggleForCUP{\begin{tabular}{|p{0.25\textwidth}|p{0.675\textwidth}|}
\hline
Kernel function & Mathematical form \T \B \\ \hline
Linear                  & $\quad K_{\rm LIN}\left(\vect{x}, \vect{x'}\right)= \vect{x}^\transpose \vect{x'} + \param$ \T\B  \\ 
Radial basis              & $\quad K_{\rm RBF}\left(\vect{x}, \vect{x'}\right)=\exp \left(-\frac{1}{2 \param^2} \|\vect{x}-\vect{x'} \|^2\right)$ \T\B \\ 
Matérn 5/2            & $\begin{aligned} \quad K_{\rm MAT}\left(\vect{x}, \vect{x'}\right)&= \left(1 + \frac{\sqrt{5}}{\param} \|\vect{x}-\vect{x'} \| + \frac{5}{3} \|\vect{x}-\vect{x'} \|^2 \right) \\ &\phantom{=} \times \exp \left(-\frac{\sqrt{5}}{\param^2} \|\vect{x}-\vect{x'} \| \right)\end{aligned}$ \T\B \\ 
Rational quadratic              & $\quad K_{\rm RQ}\left(\vect{x}, \vect{x'}\right)= \left(1 + \frac{\|\vect{x}-\vect{x'} \|}{2\param_{1}\param_{2}^2} \right)^{-\param_{1}}$ \B \\ \hline
\end{tabular}}{\begin{tabular}{|l|l|}
\hline
Kernel function & Mathematical form \T \B \\ \hline
Linear                  & $\quad K_{\rm LIN}\left(\vect{x}, \vect{x'}\right)= \vect{x}^\transpose \vect{x'} + \param$ \T\B  \\ 
Radial basis              & $\quad K_{\rm RBF}\left(\vect{x}, \vect{x'}\right)=\exp \left(-\frac{1}{2 \param^2} \|\vect{x}-\vect{x'} \|^2\right)$ \T\B \\ 
Matérn 5/2            & $\quad K_{\rm MAT}\left(\vect{x}, \vect{x'}\right)= \left(1 + \frac{\sqrt{5}}{\param} \|\vect{x}-\vect{x'} \| + \frac{5}{3} \|\vect{x}-\vect{x'} \|^2 \right) \times \exp \left(-\frac{\sqrt{5}}{\param^2} \|\vect{x}-\vect{x'} \| \right)$ \T\B \\ 
Rational quadratic              & $\quad K_{\rm RQ}\left(\vect{x}, \vect{x'}\right)= \left(1 + \frac{\|\vect{x}-\vect{x'} \|}{2\param_{1}\param_{2}^2} \right)^{-\param_{1}}$ \B \\ \hline
\end{tabular}}
\end{table}

We turn to three prominent kernel methods in the remainder of this subsection: \acf{KRR}, \acfp{SVM}, and \acfp{GP}.
As we focus on supervised kernel methods, we do not elaborate on a kernel extension for unsupervised methods such as \ac{PCA} from \cref{sec:unsup_phase_no_NN}~\cite{6790375,scholkopf1997kernel}.

\subsubsection{Kernel ridge regression}
\label{sec:GP_ridge_regression}
\Acf{KRR}\index{kernel ridge regression} is an~extension of ridge regression (presented in \cref{sss:intro_linear_model}) to nonlinear regression problems~\cite{Saunders1998_KRR}. The functional we want to minimize is very similar to the one of ridge regression, but this time, the model $f$ lives in the \ac{RKHS} $\H$ corresponding to the particular choice of the kernel function (however, note the use of \ac{MSE}!):
\begin{equation}\label{eq:KRR}
\lossfun_\mathrm{KRR} = \sum_{i}^{\datasize}\left(y_{i}-f\left(\vect{x}_{i}\right)\right)^{2}+\lambda\|f\|_{\H}^{2} = \lossfun_\mathrm{MSE} + \lossfun_\mathrm{reg}
\end{equation}
with a~regularizing term introduced in \cref{eq:regularization_term}. Here, $f$ is not restricted to a~linear function (as in linear ridge regression). Instead, $f$ can, in principle, be arbitrary. As such, \ac{KRR} is capable of building highly expressive models given an~appropriate choice of kernel.
Increasing the data-efficiency of a~\ac{ML}~problem, and consequently the accuracy of the resulting model given a fixed, finite data set, translates to finding the optimal kernel. This is not immediately apparent from \cref{eq:KRR}. To illustrate the role of kernels, recall that the model $f(\vect{x})$ can be written as the following sum over the training data:
\begin{equation}\label{eq:kernel_model}
f(\vect{x}) = \sum_{j=1}^{\datasize} \alpha_j K \left(\vect{x}, \vect{x}_{j} \right) \,.
\end{equation}
This formulation is an~instance of the already discussed representer theorem, see \cref{eq:representer_theorem}. Apart from $K$, whose mathematical form we know (or assume), we also have here coefficients $\alpha_{j}$ of our kernel model $f(\vect{x})$ which we need to find. We can express \cref{eq:KRR} with \cref{eq:kernel_model} in matrix form:
\begin{gather}
\lossfun_\mathrm{MSE} = \sum_{i}^{\datasize}\left(y_{i}-f\left(\vect{x}_{i}\right)\right)^{2}=(\vect{y}-\mat{K} \vect{\alpha})^\transpose(\vect{y}-\mat{K} \vect{\alpha}), \\
\lossfun_\mathrm{reg} = \lambda\norm{f}_{\H}^{2}=\lambda \vect{\alpha}^\transpose \mat{K} \vect{\alpha},
\end{gather}
where the matrix $\mat{K}$ is called the \stress{kernel matrix}.
It is a~positive-semidefinite, square $\datasize \times \datasize$ matrix with elements $K\left(\vect{x}, \vect{x'}\right)$ with training points $\vect{x}$ and $\vect{x'}$ belonging to the training set: $\vect{x}_1, \vect{x}_2, \vect{x}_3, \dots, \vect{x}_\datasize$. The vector $\vect{y}$ represents the targets for the corresponding training input $\vect{x}$.  Finally, if we set the derivative of the sum of these two components equal to zero, we can find a~solution for $\vect{\alpha}$ which is:
\begin{equation}
    \hat{\vect{\alpha}}=[\mat{K}+\lambda \id]^{-1} \vect{y}\,.
\end{equation}
Given $\hat{\vect{\alpha}}$, we can write the estimator of the model $\hat{f}$ at a~test point $\vect{x}^{\ast}$ as
\begin{align}\label{eq:krrfhat}
\hat{f}\left(\vect{x}^{\ast}\right)=\vect{k}^\transpose\left(\vect{x}^{\ast}\right) \vect{\hat{\alpha}}=\vect{k}^\transpose\left(\vect{x}^{\ast}\right)[\mat{K}+\lambda \id]^{-1} \vect{y}\,,
\end{align}
where $\vect{k}\left(\vect{x}^{\ast}\right)=[k \left(\vect{x}^{\ast}\right)_i]=\left[ K\left(\vect{x}^{\ast}, \vect{x}_{i}\right)\right]$.
The analogous and thorough derivation of the kernel trick on the example of \ac{KRR} is provided in \cref{appendix_kernel_trick}. Finally, we can see that the prediction of the output for an~unseen input $\vect{x}^\ast$ can be written in terms of:
\begin{itemize}
    \item The target vector $\vect{y}$,
    \item The kernel matrix $\mat{K}$, whose elements are the kernel function values $K\left(\vect{x}_{i}, \vect{x}_{j}\right)$, which takes advantage of the kernel trick,
    \item The column vector $\vect{k}\left(\vect{x}^{\ast}\right)$, whose elements are the kernel function values $K\left(\vect{x}^{\ast}, \vect{x}_{i}\right)$, which also takes advantage of the kernel trick,
    \item The regularization term of magnitude $\lambda$.
\end{itemize}
As anticipated earlier in the chapter, successfully applying \ac{KRR} boils down to finding the appropriate kernel function $K$ and tuning its corresponding hyperparameters, which -- for \ac{KRR} -- is most often done by cross-validation. This should be contrasted with how the kernel function parameters are estimated for \ac{GPR}, discussed below. 

\subsubsection{Support vector machines}\label{sec:kernel_SVM}
In the previous section, we have discussed how to use kernel methods for regression problems (in particular ridge regression). In this section, we show how one can use them for classification. In this context, the intuition behind the kernel approach is to embed the input space into the feature space in such a~way that the data becomes linearly separable with a~hyperplane (as described already in \cref{ssec:kernel-trick-sketch}). The most common \ac{ML}-classification method utilizing the kernel trick is \acfp{SVM}\index{support vector machine!kernel support vector machine}\cite{Smola1998_kernel_methods}.\footnote{There is a~variant of this approach designed for regression called support vector regression that is almost identical with \ac{KRR} but minimizes a~different form of a~loss function.}

\acp{SVM} have been introduced already in \cref{sec:intro-SVM} as geometric linear classifiers. Before we see how kernels enter \acp{SVM}, let us recall how linear \acp{SVM} work and rephrase the optimization problem that we have described in~\cref{sec:intro-SVM}. The problem there is to find an~optimal hyperplane separating data from different classes. The optimal hyperplane is defined as the one with the maximal distance between the hyperplane and the data points. In other words, we can say that all data points need to be at least the distance $M$ away from the hyperplane. The data points that are separated from the hyperplane exactly by $M$, so are the closest to the hyperplane, become support points, $\vect{x}_{s,i}$. The classification problem boils down to finding such a~hyperplane described by $\params$ that maximizes the margin between the hyperplane and support points $\vect{x}_{s,i}$ (see \cref{fig:SVM}). As we can rescale the hyperplane in an~arbitrary way, we can have $|\params| = 1/M$. Then maximizing a~margin, becomes minimizing $\params$ that in turn comes down to minimizing the Lagrange function $L$ in \cref{eq:Lagrange-function}, which we restate here for readability:
\begin{equation}\label{eq:Lagrange-function2}
L=\frac{1}{2}|\params|^{2}-\sum_{i=1}^{\datasize} \alpha_{i}\left[y_{i}\left(\params^{\transpose} \vect{x}_{i}+\param_{0}\right)-1\right]\,,
\end{equation}
where the Lagrange multipliers $\alpha_{i}$ are chosen such that
\begin{equation}
\alpha_{i}\left[y_{i}\left(\params^{\transpose} \vect{x}_{i}+\param_{0}\right)-1\right]=0\quad\forall i=1,\dots,\datasize\,.
\end{equation}
As we have already discussed, $\alpha_i$ is non-zero (and positive) only for $\vect{x}_{s,i}$. In practice, rather than minimizing $L$, we go for the dual formulation of the problem, and we maximize a~Lagrange dual, $L_D$, which provides the lower bound for $L$. $L_D$ remains a~quadratic program similarly as $L$ as we have discussed in \cref{sec:intro-SVM}. To express the problem via $L_D$, we first take the derivative of $L$ with respect to $\params$ and $\param_0$ and set it to zero. We arrive at:
\begin{equation}
\begin{aligned}
    \params &=\sum_{i=1}^{\datasize} \alpha_{i} y_{i} \vect{x}_{i} \\
    0 &=\sum_{i=1}^{\datasize} \alpha_{i} y_{i}\,.
\end{aligned}
\end{equation}
We can see that the coefficients $\params$ are given by the Lagrange multipliers $\alpha_{i}$, which can be found numerically. When we plug these equations back into the Lagrange function in \cref{eq:Lagrange-function2}, we arrive at the Lagrange dual:
\begin{equation}\label{eq:Lagrange-dual}
    L_{D}=\sum_{i=1}^{\datasize} \alpha_{i}-\frac{1}{2} \sum_{i=1}^{\datasize} \sum_{j=1}^{\datasize} \alpha_{i} \alpha_{j} y_{i} y_{j} \vect{x}_{i}^{\transpose} \vect{x}_{j} \quad \text { subject to } \alpha_{i} \geq 0\,.
\end{equation}

Finally, to put kernels in the picture, we change the notation from $\vect{x}_{i}^{\transpose} \vect{x}_{j}$ to $\estimate{\vect{x}_i,\vect{x}_j}{}$. For now, the \ac{SVM} remains a~linear model. To deal with nonlinearities in the input space, we can introduce a~feature map, $\vect{x}_{i} \to \Phi(\vect{x}_{i})$,
which gives us
\begin{equation}
L_{D}=\sum_{i=1}^{\datasize} \alpha_{i}-\frac{1}{2} \sum_{i=1}^{\datasize} \sum_{j=1}^{\datasize} \alpha_{i} \alpha_{j} y_{i} y_{j}\estimate{\Phi\left(\vect{x}_{i}\right), \Phi\left(\vect{x}_{j}\right)}{}\,.
\end{equation}
So we finally see our kernel function, $K(\vect{x}_{i}, \vect{x}_{j}) = \estimate{\Phi\left(\vect{x}_{i}\right), \Phi\left(\vect{x}_{j}\right)}{}$, appearing. In this kernel formulation, the margin we maximize is between the hyperplane and the support points \stress{in the feature space}~\cite{kSVM}.
Therefore, the \ac{SVM} problem boils down to maximizing $L_{D}$ numerically to find the coefficients $\alpha_{i}$, e.g., using sequential minimal optimization \cite{platt1998sequential}.
\begin{figure}[t]
\begin{center}
\includegraphics[width=\ToggleForCUP{0.7\textwidth}{0.8\columnwidth}]{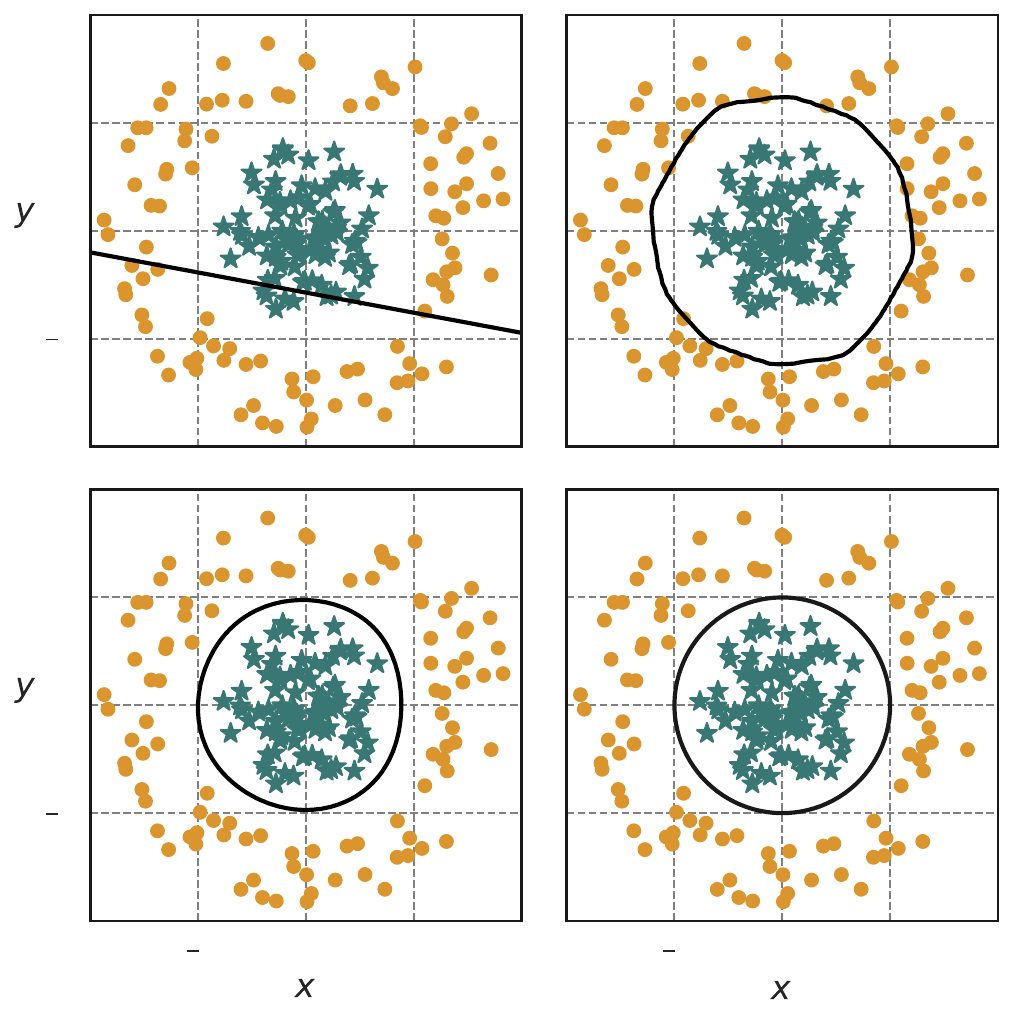}
\end{center}
\caption[Classification using support vector machines with different kernels]{The kernel form makes a~difference! The same data as in \cref{fig:kernel_trick_motivation} is classified using an~\ac{SVM} with different kernel choices. The black line corresponds to each underlying decision boundary. (a) As the data is not linearly separable, a~linear kernel does not work. (b) A~polynomial kernel up to degree $50$ does the trick at the expense of signs of overfitting. (c) Due to the rotation symmetry, an~RBF kernel is best suited for classifying this data set whose decision boundary closely resembles the actual underlying decision boundary of the data shown in (d).}
\label{fig:SVM-nonlinear_2}
\end{figure}
Afterwards, the parameters of the kernel function $K$ need to be validated using, for example, the held-out validation set or cross-validation.
As with \acp{NN}, this requires a retraining of the \ac{SVM} for each trial choice of kernel parameters.
Once both parameter sets are known, the hyperplane separating the classes in the typically high-dimensional space is also known. With the optimal hyperplane $\hat{f}$ we can then make predictions at an~arbitrary test point $\vect{x}^\ast$:
\begin{equation}
\ToggleForCUP{\begin{aligned}
    \hat{f}(\vect{x}^\ast)=\params^{\transpose} \vect{\featmap}(\vect{x})+\param_{0}&= \sum_{i} \alpha_{i} y_{i}\estimate{\featmap\left(\vect{x}_{i}\right), \featmap\left(\vect{x}_{i}\right)}{}+\param_{0} \\
    &= \sum_{i} \alpha_{i} y_{i} K\left(\vect{x}_{i}, \vect{x}\right)+\param_{0}\,.
\end{aligned}}{\hat{f}(\vect{x}^\ast)=\params^{\transpose} \vect{\featmap}(\vect{x})+\param_{0}= \sum_{i} \alpha_{i} y_{i}\estimate{\featmap\left(\vect{x}_{i}\right), \featmap\left(\vect{x}_{i}\right)}{}+\param_{0}= \sum_{i} \alpha_{i} y_{i} K\left(\vect{x}_{i}, \vect{x}\right)+\param_{0}\,.}
\end{equation}
Finally, in order to turn this value into a~class prediction, we take the sign of $\hat{f}$ as the corresponding class label. Note that the choice of the kernel function here matters, as discussed in the opening of \cref{sec:kernel_methods}. We visualize this problem in \cref{fig:SVM-nonlinear_2}.

While we can put lots of effort into finding a~kernel function that renders our problem linearly separable, we can also relax the problem by allowing some misclassification. Let us thus move to the \stress{problems that are not linearly separable}. The derivations above still hold with one modification: we now allow a~number of data points to be on the wrong side of the margin, as shown in \cref{fig:SVM-nonlinear}. This modifies our constraint from \cref{eq:SVM-constraint} to $y_{i}(\params^{\transpose} \vect{x}_{i}+\param_{0}) \geq 1 - \xi_i$, where $\xi_{i}=0$ if the data point is on the correct side of the margin. The variables $\xi_i$ are often referred to as \stress{slack variables}. We can incorporate the control over how ``wrong'' the hyperplane can be by adding another constraint, i.e., $\sum_{i} \xi_{i}<C=$ const, which also adds terms to the Lagrange function:
\begin{figure}[t]
\begin{center}
\includegraphics[width=0.75\columnwidth]{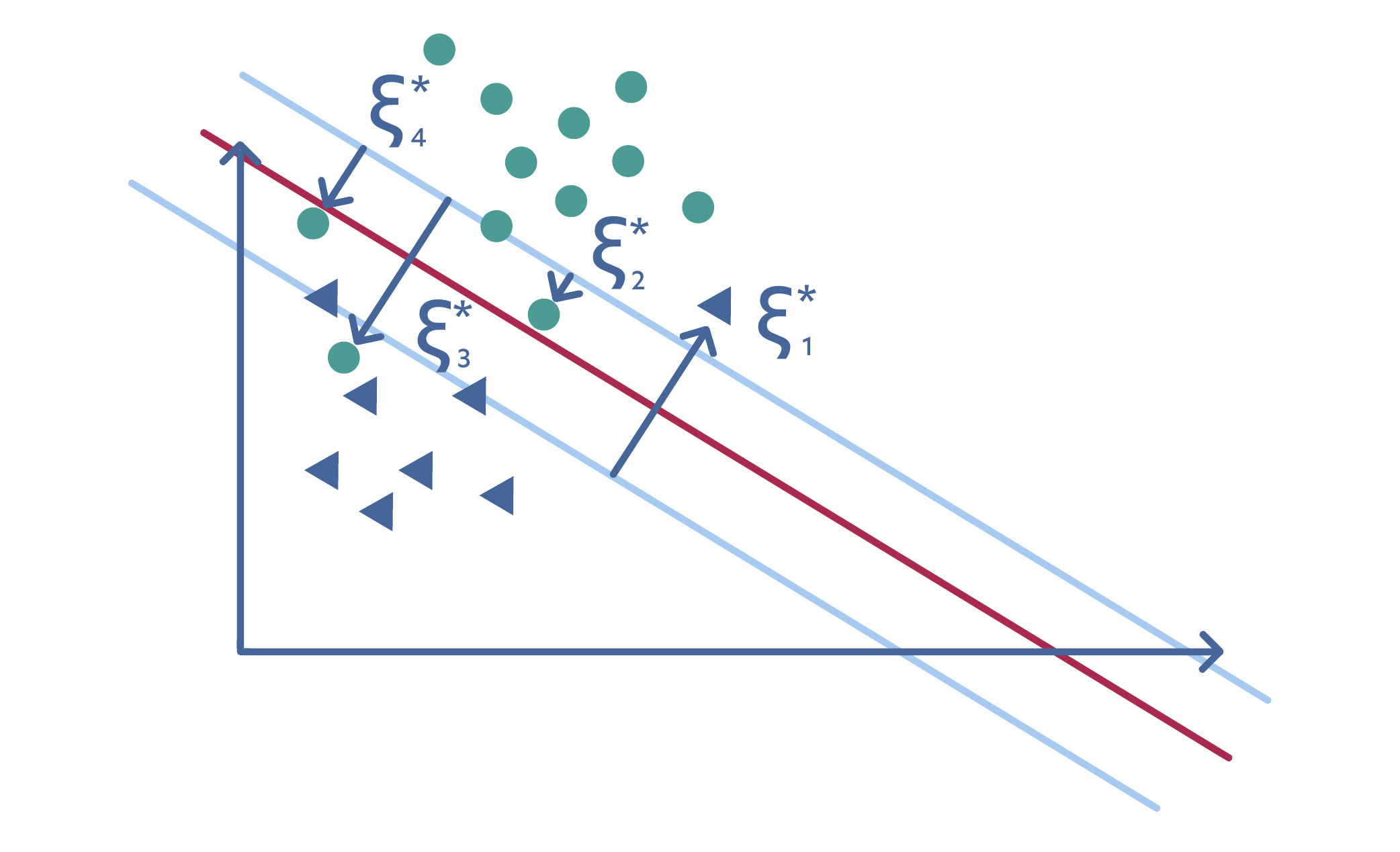}
\end{center}
\vspace{-0.4cm}
\caption[A~linear support vector machine applied to a~data set that is not linearly separable] {A~linear \ac{SVM} applied to a~data set that is not linearly separable. The \ac{SVM} tries to minimize the total distance of all misclassified data points $\xi_i$ called slack variables. $\xi_i \geq 1$ for misclassified $\vect{x}_i$ (here, $i=1,3,4$) and $0 < \xi_i < 1$ for points on the correct side of the decision boundary but within the margin (here, $i=2$).}
\label{fig:SVM-nonlinear}
\end{figure}
\begin{equation}
    L=\frac{1}{2}|\params|^{2}-\sum_{i} \alpha_{i}\left[y_{i}\left(\params^{\transpose} \vect{x}_{i}+\param_{0}\right)-\left(1-\xi_{i}\right)\right]+C \sum_{i} \xi_{i}.
\end{equation}
Now, we maximize the margin while minimizing the violation of the margin constraints. This loss function is still a~quadratic program but now has also a~largely increased number of optimization variables (one slack variable per data point). As previously mentioned, instead of minimizing $L$ you can maximize $L_D$. Note that in this case, you get an~additional regularization term with magnitude $C$. Large $C$ allows for more misclassified data points but promotes simpler decision boundaries.
This is because, in this case, the \ac{SVM} focuses on a minimal number of relevant data points to draw a decision boundary.
Since this number of relevant data points is usually much smaller than the total number of given training data points, this is referred to as a \stress{sparse} solution.
In contrast, a small $C$ forces the model to better fit training data, sometimes at the expense of the validation data.\footnote{Beware of various definitions and notations regarding regularization strength, particularly in \acp{SVM}. For example, in Scikit-learn, decreasing a~hyperparameter $C$ corresponds to more regularization.}
In this case, the solution is less sparse but can tend to be overfitted to the training data.

Finally, we can make a~connection between \acp{SVM} and \ac{KRR}. We started by saying that we need to minimize $|\params|^{2}$ subject to the following conditions:
\begin{equation}
\begin{aligned}
    y_{i}\left(\params^{\transpose} \vect{x}_{i}+\param_{0}\right) &\geq 1-\xi_{i}\text{ with } \xi_{i} \geq 0 \\
    \sum_{i} \xi_{i} &< C=\mathrm{const}\,.
\end{aligned}
\end{equation}
If we simply write
\begin{equation}
    \xi_{i} \geq 1-y_{i}\left(\params^{\transpose} \vect{x}_{i}+\param_{0}\right)
\end{equation}
and vary $\params$ and $\param_{0}$ to find the minimum of the following function:
\begin{equation}
L\left(\params, \param_{0}\right) = \frac{1}{2}|\params|^{2}+C \sum_{i}\left[1-y_{i}\left(\params^{\transpose} \vect{x}_{i}+\param_{0}\right)\right]_{+}\,,
\end{equation}
the problem is equivalent to minimizing the following function:
\begin{equation}
L\left(\params, \param_{0}\right)=\frac{1}{2 C}|\params|^{2}+\sum_{i=1}^{\datasize}\left[1-y_{i}\left(\params^{\transpose} \vect{x}_{i}+\param_{0}\right)\right]_{+}\,,
\end{equation}
which is the same as
\begin{equation}
L\left(\params, \param_{0}\right)=\frac{1}{2} \lambda|\params|^{2}+\sum_{i=1}^{\datasize} \max \left[0,1-y_{i} f\left(\vect{x}_{i}\right)\right]\,.
\end{equation}
Finally, let us compare it to the functional of \ac{KRR} from \cref{eq:KRR}: the regularization term is the same, the main difference is that \ac{KRR} uses the squared error loss, while \acp{SVM} uses a~function called the Hinge loss.
The difference, of course, stems from the fact that we try to solve two different tasks, regression for \ac{KRR} vs. classification for \acp{SVM}.
It is possible, however, to modify \acp{SVM} such that they can be applied to regression tasks.
In this case, \ac{KRR} has the advantage of being computationally more efficient, especially for small data sets.
The advantage of the modified \acp{SVM}, again, is their sparsity to yield potentially less overfitted solutions.

\subsubsection{Gaussian processes}
At this point, we have already covered powerful and general regression tools. We learned about the kernel trick and what it means to learn in feature spaces rather than in the input space. Moreover, we have seen how this is useful, particularly for high-dimensional feature spaces, and examined in more depth some tools that manifestly use the kernel trick to perform good learning tasks. In this section, we cover an~additional tool, namely \acfp{GP}\index{Gaussian process}. Since we have already introduced powerful regression models such as \ac{KRR}, a~natural question to ask is: \stress{why do we need another regression model?}. The short answer is that \acf{GPR} does all that \ac{KRR} offers but also allows one to calculate the Bayesian uncertainty of the predictions. As we shall see, this can be used for Bayesian optimization \cite{bayesoptbook_2022}. \Ac{GPR} is also well suited for algorithms aiming to build kernels well aligned with data in a setting where data is very limited. In the remainder of this section, we describe what \acp{GP} are and how \ac{GPR} works.
\begin{figure}[t]
        \begin{center}
        \includegraphics[width=0.75\columnwidth]{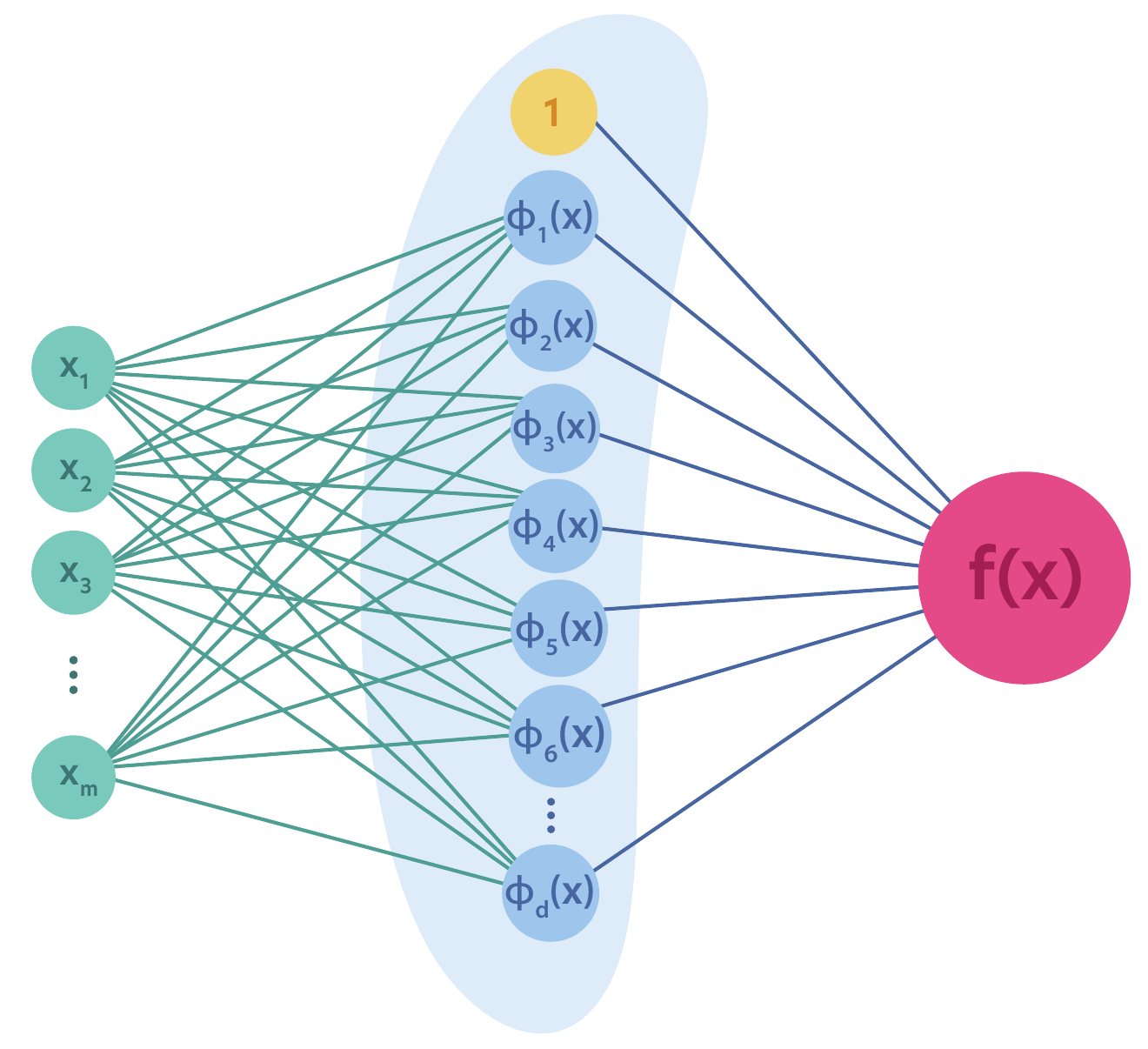}
        \end{center}
        \caption[Bayesian neural network]{Sketch of a~Bayesian \ac{NN}. The features in the hidden layer (blue) are multiplied by random variables as shown in \cref{eq:bayesianNN}. This way, the \ac{NN} is no longer deterministic, and its output $f(\vect{x})$ given the input $\vect{x}$ is itself a~random variable.}
        \label{fig:bayesianNN}
        \end{figure}

Consider the regression problem of finding a function $f: \realset^\featnum \to \realset$.
In the spirit of this chapter, we want to map the input to a suitable $\nparams$-dimensional space using nonlinear functions $\featmap_i$.
Then, as in \cref{sss:intro_linear_model}, we apply a linear model, i.e.,
\begin{equation}\label{eq:bayesianNN}
f(\vect{x}) = \left( \param_0, \param_1 \dots \param_{\nparams} \right)^{\transpose} \left( 1, \featmap_1(\vect{x}), \dots , \featmap_{\nparams}(\vect{x}) \right)\,.
\end{equation}
Here, each of the functions $\featmap_i(\vect{x})$ is some parametrized nonlinear function of $\vect{x} \in\realset^\featnum$ such as
\begin{equation}
\featmap_i(\vect{x}) = \tanh{\left[w_i \sum_{j=1}^{\featnum} x_j + \bias_i \right]}\,.
\end{equation}
\Cref{eq:bayesianNN} can be visualized as the \ac{NN} from \cref{fig:bayesianNN}. If all the weights of the network are fixed, the network maps any given data point $\vect{x}$ to a~single value $f(\vect{x})$. However, if we now assume the parameters $\param_0, \dots \param_{\nparams}$ as well as the hyperparameters of the activation functions $(w_i,b_i)_i$ to be samples of random variables, distributed according to some distribution, the output of the \ac{NN} for a~fixed $\vect{x}$ becomes a~random variable itself. Such an~\ac{NN} becomes a~Bayesian \ac{NN}.
Here, all tunable parameters of the \ac{NN} have been promoted to random variables. As a consequence, we have to average over the parameter values to obtain, e.g., the expectation value of $f(\vect x)$ and similar quantities of interest.
In \cref{eq:bayesianNN}, there are $\nparams +1 + 2\nparams = 3\nparams +1$ many continuous random variables, $\param_i,w_i,b_i$. Hence, taking averages requires us to solve a $(3\nparams +1)$-dimensional integral.
This is just not feasible and potentially impossible analytically, depending on the distributions involved for each of the random variables.
There is a way out of this conundrum, luckily: assuming that all $\param_i$ are i.i.d. random variables, the limit of $\nparams \to \infty$, by the central limit theorem\footnote{The central limit theorem states that the sum of many independent random variables is approximately normally distributed. For example, if you roll two six-sided dice multiple times, the sum of the obtained results converges to the Gaussian distribution centered at seven in the limit of an~infinite number of rolls.
}, yields a normal distribution for the output \cite{neal2012bayesian}. Because the distribution always remains Gaussian for any given input $\vect{x}$, $f(\vect{x})$ is our~first example of a~\ac{GP}. In general, a~\ac{GP} can be viewed as a~normal distribution over functions. Its output is an~instantiation of a~random variable, distributed according to a~multivariate density. The upshot of this preliminary example is that by incorporating randomness into the model, under the assumption of infinitely many independently sampled \ac{NN}-parameters, the output of the model remains Gaussian. The advantage of all this, compared to what we have previously analyzed, is that it is possible to obtain closed-form expressions for the log marginal likelihood and the predictive distribution later on.

Let us come back to our problem setting: we want to infer the~function $f$ that describes a~given data set. The data are generally noisy. We model this data noise by a Gaussian distribution with zero mean and variance $\sigma^2$:
\begin{equation}\label{eq:gaussianassump}
y = f(\vect{x}) + \epsilon, \quad \epsilon \sim \normdist(0, \sigma^2)\,.
\end{equation}
For the sake of simplicity and clarity, we first consider only the~linear model without any nonlinear feature map, i.e., $\featmap$ is just the identity map in this case, which reduces \cref{eq:bayesianNN} to
\begin{equation}
f(\vect{x}) = \params^\transpose\vect{x}\,,
\end{equation}
similar to \cref{sss:intro_linear_model}.
Again, this includes the bias term such that we have $\nparams = \featnum + 1$ many random variables in this case.
We assume that each data point $\vect{x}_i$ is independent of the others. For any given point, the model likelihood is expressed as $p(\vect{y}_i \mid \params, \vect{x}_i)$. Due to the independence of the data points, for a~labeled training data set $\dataset=(\mat{X}, \vect{y})$ the joint likelihood reads (see also \cref{sss:probability}):
\begin{equation}\label{eq:gaussianld}
p(\vect{y} \mid \params, \mat{X}) = \prod_{i=1}^\datasize p(\vect{y}_i \mid \params, \vect{x}_i)\,.
\end{equation}
Because of our assumption of Gaussian distributed noise (see \cref{eq:gaussianassump}), we can explicitly rewrite each term of the product from the right-hand side of the equation above as:
\begin{equation}
p(\vect{y}_i \mid \params, \vect{x}_i) = \frac{1}{\sqrt{2\pi}\sigma} \exp\left[-\frac{(y_i - \params^T\vect{x}_i)^2}{2\sigma^2} \right]\,.
\end{equation}
Hence, the model likelihood given the data $\dataset$ is a~product of Gaussians. Thus, \cref{eq:gaussianld} can be explicitly rewritten as:
\begin{equation}
p(\vect{y} \mid \params, \mat{X}) = \prod_{i=1}^\datasize p(\vect{y}_i \mid \params, \vect{x}_i) =
\frac{1}{(2\pi\sigma^2)^{\datasize/2}} \exp\left[-\frac{|\vect{y} - \mat{X}\params|^2}{2\sigma^2} \right]\,.
\end{equation}
Our goal now is to use the Bayes' theorem to calculate the posterior over the weights $\params$ (see \cref{sss:probability}). Using \cref{eq:Bayes_rule}), we obtain
\begin{equation}\label{eq:bayestheorem}
    p(\params \mid \vect{y}, \mat{X}) = \frac{p(\vect{y} \mid \params,\mat{X}) p(\params)}{p(\vect{y} \mid \mat{X})}
\end{equation}
where $p(\vect{y} \mid \mat{X})$ is a~normalization constant not depending on $\params$ and $p(\params)$ represents our prior (again see \cref{sss:probability}). In general, one has freedom over choosing the prior. Nevertheless, it would always be better in practice to choose this prior wisely. Hence, one should choose it according to the prior knowledge of the problem at hand. For instance, when thinking about physical problems, one can use some context or prior knowledge of the system to set the prior appropriately. Of course, the better the prior is chosen, the more effective the model is.
While these are useful guidelines in general, we completely discard them at this point and
set the prior to be
\begin{equation}
    \params \sim \normdist(0, \mat{\Sigma}_{\nparams})
    \label{eq:gp_Gaussian_prior}
\end{equation} which is a~joint normal distribution with zero mean and covariance matrix $\mat{\Sigma}_{\nparams}$. In particular, under the assumption of independent parameters, $\mat{\Sigma}_{\nparams}$ can be chosen to be a~diagonal matrix. As we shall argue below, we are even able, without loss of generality, to choose $\mat{\Sigma}_{\nparams}$ to be simply the identity matrix, as the specific choice of $\mat{\Sigma}_{\nparams}$ (and hence also any prior knowledge of the problem at hand) can be rolled into the definition of the kernels.
The choice of zero mean for the prior ensures that the covariance function of a \ac{GP} is indeed the kernel function.
We refer to \cref{appendix_CovMat} for further details on this.

Getting back to the posterior, we can collect all the $\params$-independent terms under a~normalization constant $A$ and rewrite the likelihood using \cref{eq:gaussianld} and \eqref{eq:bayestheorem} as
\begin{equation}\label{eqn:gp_likelihood}
\ToggleForCUP{\begin{aligned}
    p(\params \mid \vect{y}, \mat{X}) &= \frac{p(\vect{y} \mid \params,\mat{X}) p(\params)}{p(\vect{y} \mid \mat{X})} \\
    &= \frac{1}{A}\exp\left[-\frac{|\vect{y} - \mat{X}\params|^2}{2\mat{\sigma}^2} \right] \exp\left[ -\params^\transpose\mat{\Sigma}_{\nparams}^{-1}\params\right]\,.
\end{aligned}}{%
    p(\params \mid \vect{y}, \mat{X}) = \frac{p(\vect{y} \mid \params,\mat{X}) p(\params)}{p(\vect{y} \mid \mat{X})} =
    \frac{1}{A}\exp\left[-\frac{|\vect{y} - \mat{X}\params|^2}{2\mat{\sigma}^2} \right] \exp\left[ -\params^\transpose\mat{\Sigma}_{\nparams}^{-1}\params\right]\,.}
\end{equation}
Rearranging the expressions in the exponents, the posterior becomes:
\begin{equation}\label{eq:posterior}
    p(\params \mid \vect{y}, \mat{X}) = \frac{1}{A}\exp\left[-\frac{1}{2} (\params - \vect{\mu})^\transpose\mat{C}(\params - \vect{\mu}) \right]
\end{equation}
where we used the following definitions
\begin{align}
     \vect{\mu} &= (\mat{X}^\transpose\mat{X}+\sigma^{2}\mat{\Sigma}_{\nparams}^{-1})^{-1} \mat{X}^\transpose\vect{y} \\
    \mat{C}^{-1} &= (\mat{X}^\transpose\mat{X}+\sigma^{2}\mat{\Sigma}_{\nparams}^{-1})^{-1}
\end{align}
with $\vect{\mu}$ being the posterior mean and $\mat{C}^{-1}$ the posterior covariance matrix.\footnote{Even though we arrived at these results from a Bayesian perspective, similar results have already been developed in the geostatistics community in the 1960s. There, \ac{GPR} is more often known as \stress{kriging}~\cite{Cressie1990_kriging}.}
Finally, \cref{eq:posterior} gives us access to the analytical form for the distribution of the model parameters $p(\params \mid \vect{y}, \mat{X})$ given the data $\dataset$. This, again, is a~normal distribution with updated mean and covariance. These new values intrinsically possess the information about the training data as inferred from their analytical forms. Hence, $p(\params \mid \vect{y}, \mat{X})\sim \normdist(\vect{\mu}, \mat{C}^{-1})$. Once we have our updated distribution over our model's parameters, the next step is to predict the output $y^\ast$ for a~previously unseen data point $\vect{x}^\ast$. To do this, we need to multiply the posterior by the probability for $y^\ast$ and integrate over all possible parameters. This yields:
\begin{equation}
\ToggleForCUP{\begin{aligned}
    &\hphantom{=} p(y^\ast \mid \vect{x}^\ast) = \int_{\realset^{\nparams}} p(y^\ast \mid \vect{x}^\ast, \params) p(\params \mid \vect{y}, \mat{X})\ \text{d}\params \\
    &\propto \int_{\realset^{\nparams}} \exp\left[-\frac{(y^\ast - \params^\transpose\vect{x}^\ast)^2}{2\sigma^2} \right]\exp\left[-\frac{1}{2} (\params - \vect{\mu})^\transpose\mat{C}(\params - \vect{\mu}) \right]\ \text{d}\params\,.
\end{aligned}}{%
\begin{aligned}
    p(y^\ast \mid \vect{x}^\ast) &= \int_{\realset^{\nparams}} p(y^\ast \mid \vect{x}^\ast, \params) p(\params \mid \vect{y}, \mat{X})\ \text{d}\params \\
    &\propto \int_{\realset^{\nparams}} \exp\left[-\frac{(y^\ast - \params^\transpose\vect{x}^\ast)^2}{2\sigma^2} \right]\exp\left[-\frac{1}{2} (\params - \vect{\mu})^\transpose\mat{C}(\params - \vect{\mu}) \right]\ \text{d}\params\,.
\end{aligned}}
\end{equation}
It is easy to notice that this distribution is again going to be Gaussian. We can also analytically compute the conditional mean and variance:\footnote{with $\mat{C} = [\sigma^{-2}\mat{X}^\transpose\mat{X}+\mat{\Sigma}_{\nparams}^{-1}]$ already defined above}
\begin{align}
     \hat{\mu} &= \vect{x}^{\ast\transpose} \vect{\mu} = \vect{x}^{\ast\transpose} \sigma^{-2}\mat{C}^{-1}\mat{X}^\transpose\vect{y} \\
    \hat{\sigma}^2 &= \vect{x}^{\ast\transpose} \mat{C}^{-1} \vect{x}^{\ast}\,. 
\end{align}
The mean can be used to make predictions, while the variance gives the uncertainty over such estimation. It is now time to compare these results with previously introduced methods:
\begin{align}
    \text{Linear Regression:}\ \vect{x}^{\ast\transpose}\hat{\params} &= \vect{x}^{\ast\transpose} (\mat{X}^\transpose\mat{X})^{-1} \mat{X}^\transpose\vect{y}\\
    \text{Linear Ridge Regression:}\ \vect{x}^{\ast\transpose}\hat{\params} &= \vect{x}^{\ast\transpose} (\mat{X}^\transpose\mat{X} + \lambda \id)^{-1} \mat{X}^\transpose\vect{y}\\
    \text{Linear \ac{GPR}:}\ \hphantom{\vect{x}^{\ast\transpose}}\hat{\mu} &= \vect{x}^{\ast\transpose} (\mat{X}^\transpose\mat{X}+\sigma^{2}\mat{\Sigma}_{\nparams}^{-1})^{-1}\mat{X}^\transpose\vect{y}.
\end{align}\index{linear regression}
This entire derivation, which we have been doing for the linear case, can easily be generalized to nonlinear regression.
To this end, we revisit our Bayesian \ac{NN} from \cref{eq:bayesianNN} above.
Here, we map $\vect{x} \mapsto \featmap = \featmap(\vect{x})$, embedding our input data with a~feature map into a~potentially high-dimensional space. It follows that $\mat{X} \mapsto \mat{\Phi} = \featmap(\mat{X})$
and the conditional mean and variance can be derived accordingly. 
Substituting the terms appropriately and doing a~little bit of math adjustments, under the simplification of unit variance $\mat{\Sigma}_{\nparams} = \id$ from above, we obtain:
\begin{align}
    \hat{\mu} &= \featmap^{\ast\transpose}\mat{\Phi}^\transpose \left[\mat{\Phi}^\transpose\mat{\Phi} + \sigma^2\id \right]^{-1} \vect{y}  \\
    \hat{\sigma}^2 &= \featmap^{\ast\transpose}\featmap^{\ast}-\featmap^{\ast\transpose}\mat{\Phi}^\transpose \left[\mat{\Phi}^\transpose\mat{\Phi} + \sigma^2\id \right]^{-1} \mat{\Phi}\featmap^\ast\,.
\end{align}
These expressions can be re-written in terms of the kernel function $K$:
\begin{align}\label{eq:gpmuhat}
    \hat{\mu} &= \vect{k}^\transpose\left(\vect{x}^{\ast}\right)[\mat{K}+\sigma^2 \id]^{-1} \vect{y} \\
    \hat{\sigma}^2 &= K(\vect{x}^\ast, \vect{x}^\ast) - \vect{k}^\transpose\left(\vect{x}^{\ast}\right) \left[\mat{K} + \sigma^2\id \right]^{-1} \vect{k}\left(\vect{x}^{\ast}\right)
\end{align}
where we used the following definitions:
\begin{align}
    \featmap^{\ast\transpose}\featmap^\ast &= K(\vect{x}^\ast, \vect{x}^\ast) \label{eq:featmap}\,,\\
    \left(\mat{\Phi}\featmap^\ast\right)_i &= \vect{k}\left(\vect{x}^{\ast}\right)_i = K(\vect{x}_i, \vect{x}^\ast)\,,\\
    \left(\mat{\Phi}^\transpose\mat{\Phi}\right)_{ij} &= \mat{K}_{ij} = K(\vect{x}_i,\vect{x}_j).
\end{align}
\sloppy 
While choosing $\mat{\Sigma}_{\nparams} = \id$ may appear as unnecessarily restrictive, we note that one can always rewrite the identities above such that they take the covariance matrix into account (e.g., \cref{eq:featmap} becomes $\featmap^{\ast\transpose}\mat{\Sigma_{\nparams}}\featmap^{\ast\transpose} = K(\vect{x}^\ast, \vect{x}^\ast)$). Since the kernel function is to be defined anyway, the choice of $\mat{\Sigma_{\nparams}} = \id$ does not lose generality.
We detail in \cref{appendix_CovMat} that the kernel trick does not interfere with the Gaussian prior assumption of \cref{eq:gp_Gaussian_prior} for the parameters $\params$.

Thus, looking at the conditional mean from \cref{eq:gpmuhat} one can directly see the analogy with the result of \ac{KRR} from \cref{eq:krrfhat} where our regularization strength $\lambda$ can be seen to correspond to the data noise assumption, parametrized by $\sigma^2$.\footnote{The difference in the two constants actually only arises from our choice of the covariance matrix. If we had chosen $\mat{\Sigma}_{\nparams} = \tau^2 \id$ instead, the difference would vanish completely, and the two methods would yield the same estimator.}
This comes to no surprise as the conditional mean from \cref{eq:gpmuhat} corresponds to the \ac{MAP}, \cref{eq:intro_MAP_estimator} introduced in \cref{sss:intro_linear_model}.
However, \ac{GPR} also yields the uncertainty of the prediction.
Furthermore, we see, once again, the consequence of the representer theorem\index{representer theorem} in \cref{eq:representer_theorem} on the form of the conditional mean $\hat{\mu}$, i.e., $\vect{\alpha} = (\mat{K} + \sigma^2\id)^{-1} \vect{y}$.

\subsubsection{Training a~Gaussian process}\label{sec:traingp}
In the previous section, we have seen what a~\ac{GP} is, how one constructs it, and how it allows one to obtain a closed-form expression for the conditional mean and the Bayesian uncertainty for the output over one (or more) unseen data point(s) $\vect{x}^\ast$. We understand that the performance of the \ac{GP} strongly relies on the choice of the prior and on the amount of data the model is exposed to. So, at this point, one natural question that might arise is: how do we train a~\ac{GP}? 
First and foremost, we need to choose an~appropriate kernel function $K$, which defines the kernel matrix $\mat{K}_{ij} = K(\vect{x}_i, \vect{x}_j)$ accordingly. The parameters of this function (e.g., as in \cref{tab:kernel-functions}) are tuned in order to maximize the so-called marginal likelihood $p(\vect{y} \mid \mat{X})$. This name comes from the fact that this quantity is obtained from the Bayes' theorem \cref{eq:bayestheorem} when marginalizing over the model parameters (i.e., taking the integral over $\params$). Here, our goal is to express the marginal likelihood of a~\ac{GP} in terms of the kernels. To this end, we choose the covariance matrix of the \ac{GP} to be the kernel matrix such that $\text{Cov}(\vect{x}, \vect{x}')= \mat{K}$\footnote{On the notation: when providing the inputs of the covariance matrix we use $\text{Cov}$ while when they are implicit we just use $\mat{C}$.} which turns out to be equivalent to the choice of the prior made previously $\params_{\mathrm{prior}} \sim \normdist(0, \mat{\Sigma}_{\nparams})$. A mathematical justification for this is provided in \cref{appendix_CovMat}.
Our ultimate goal is to evaluate the marginal likelihood
\begin{equation}
  p(\vect{y} \mid \mat{X}) = \int_{\realset^{\nparams}} p(\vect{y} \mid \params, \mat{X}) p(\params \mid \mat{X})\ \text{d}\params\,.
  \label{eq:gp_marginal_integral}
\end{equation}
Given that the prior’s covariance is the kernel function and the integrand is a~product of two Gaussians, it is possible to express $p(\vect{y} \mid \mat{X})$ in terms of the kernel matrix $\mat{K}$. Working with the logarithm of the marginal likelihood, it follows from \cref{appendix_CovMat} that our objective of the training process is
\begin{equation}\label{eq:logmarginal}
    \log p(\vect{y} \mid \mat{X}) = -\frac{1}{2}\vect{y}^\transpose \left(\mat{K} + \sigma^2\id \right)^{-1}\vect{y} - \frac{1}{2}\log \left|\mat{K} + \sigma^2\id \right| - \frac{n}{2}\log 2\pi\,.
\end{equation}
When training a~\ac{GP} one aims to find the parameters of the kernel function that maximize the logarithm of marginal likelihood from \cref{eq:logmarginal}.
 As can be seen, training a~\ac{GP} requires the inversion of the kernel matrix (and the calculation of its determinant), whose dimension is determined by the size of the training set. This already gives an~intuition why \ac{GP}s are the tool of choice in a~regime of~few data points where they can be very effective and relatively cheap. GP models can be difficult to train for problems with a lot of training data. 

\subsection{Bayesian optimization}\label{ss:GPR-BO}
In the previous section, we have discussed \acf{GPR} and how to train a~\acf{GP}. Furthermore, we have shown that \acp{GP} yield a~closed-form expression for the estimate of the output for a~test data point $\vect{x}^\ast$, conditioned by a~set of given data $\dataset$, in a~similar fashion to \ac{KRR}. Unlike \ac{KRR}, \ac{GPR} also comes with a~prediction uncertainty. This is of a~great relevance as it can be used for \acf{BO}\index{Bayesian optimization}~\cite{bayesoptbook_2022}. 

\begin{algorithm}[t]
\caption{\Acf{BO}}\label{alg:BO}
\begin{algorithmic}
\Require initial data set $\dataset_0=\{( \mat{X}, \vect{y})\}$
\Require initial surrogate model \ac{GP} trained on $\dataset_0$ with mean and variance $\hat{\mu}, \hat{\sigma}^2$
\Require acquisition function to be maximized
\For{iteration $t  < T_{\mathrm{max}}$}
    \State Sample a~set of candidate points $\mat{X}_\mathrm{cand}$ \Comment{batch optimization}
    \State $\vect{x}^\ast \gets \max_{\vect{x}\in\mat{X}_\mathrm{cand}} a(\vect{x}, \hat{\mu}, \hat{\sigma})$
    \State $y^\ast\gets f_{\rm BB}(\vect{x}^\ast)$ at $\vect{x}^\ast$
    \State $\dataset  \gets \dataset_t \bigcup \{(\vect{x}^\ast, y^\ast)\}$ \Comment{update data set}
    \State Update the surrogate model's parameters $\hat{\mu}$ and $\hat{\sigma}^2$
\EndFor\\
\Return $\hat{\mu}$ and $\hat{\sigma}^2$ \Comment{prediction and uncertainty of the surrogate model}
\end{algorithmic}
\end{algorithm}

\ac{BO} is a~technique used for the optimization of expensive black-box functions where gradients can not be easily computed or estimated (e.g., time-consuming experiments). Here, the term \textit{expensive} is very important. Indeed, optimizing black-box functions is the general goal of a~big set of \ac{ML}-models and techniques. Such optimization usually relies on efficient computation, arbitrarily large amounts of data, and so forth. However, this might not always be the case. Sometimes, we might face problems where the amount of data is rather limited, or the routine to extract additional experimental data is very expensive. In this context, the interplay between \ac{GP}\index{Gaussian process} and \ac{BO} becomes extremely important. It is also worth mentioning that contrary to the optimization methods used in most \ac{ML}-approaches we have encountered so far (in particular \acp{NN}), \ac{BO} is a~gradient-free method. Therefore, it is particularly well suited for functions that are very difficult or expensive to evaluate.

\begin{figure}[t]
        \begin{center}
        \includegraphics[width=0.9\columnwidth]{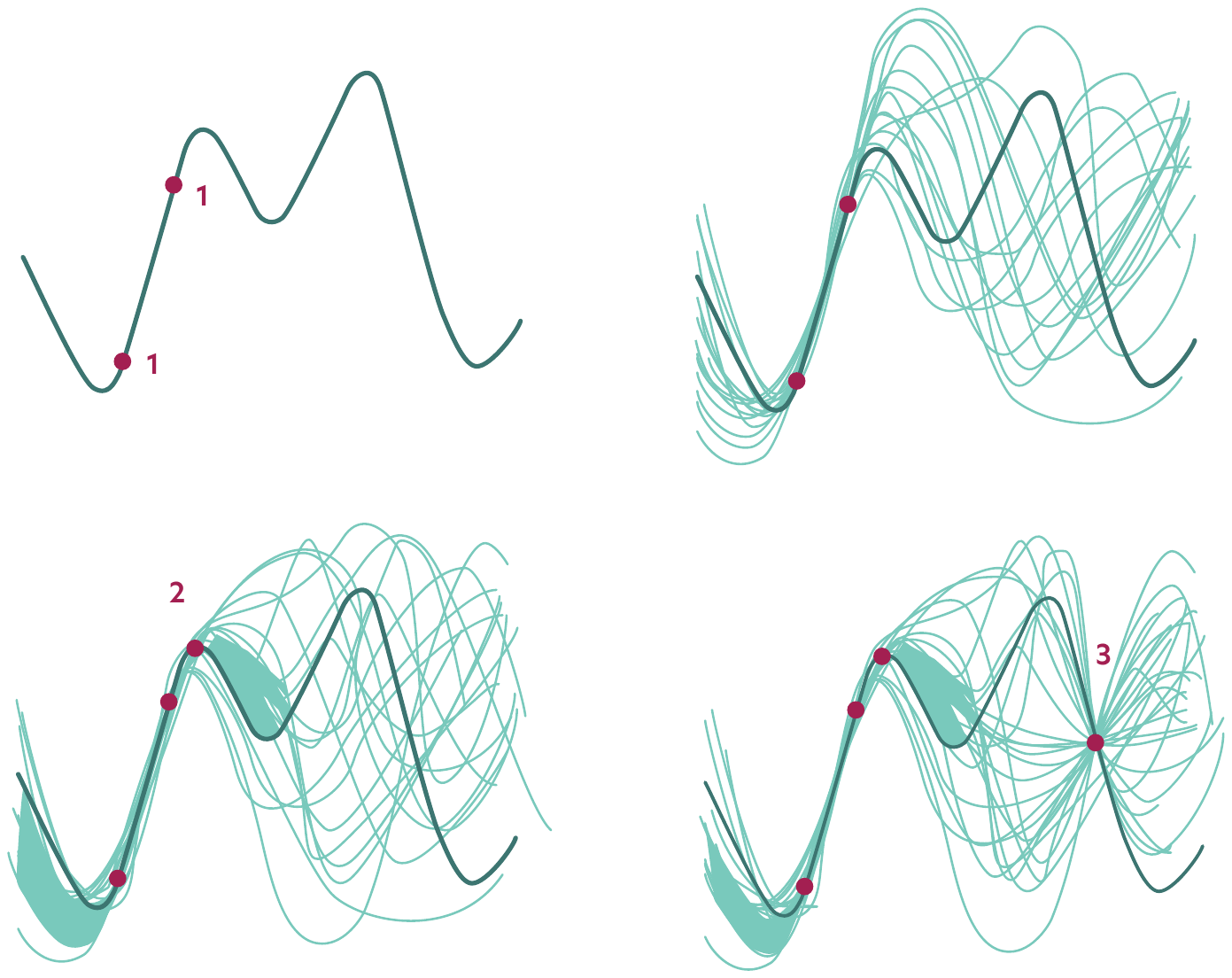}
        \end{center}
        \caption[Bayesian optimization]{Example of how \ac{BO} can be applied to \ac{GPR}. A~\ac{GP} represents a~surrogate model (light green lines) trained on an~initial set of observations (red dots in the upper left panel). At each \ac{BO}-step, a~new point is added to the previous set of observations such that the surrogate model becomes increasingly certain as more points are added. The dark green line represents the underlying black box function.}
        \label{fig:bo}
\end{figure}
The way \ac{BO} works in the context of \ac{GPR} is that \ac{BO} takes \acp{GP} \stress{as surrogate models of the black-box function to be optimized.} Recalling the results of the previous section, \ac{GPR} gives us access to a~conditional prediction along with an~estimate for its uncertainty\index{uncertainty}. This said, the next important thing to notice is that \ac{BO} is an~iterative process. This process works as follows (see \cref{alg:BO} for the pseudo-code): The \ac{BO} procedure starts with a~few evaluations of the black box function at some random locations of the input space. We refer to this initial data set as $\dataset_0=\{( \mat{X}, \vect{y})\}$. These evaluations are used to train the first version of a~\ac{GP}. Hence, we can think of those as our training data.\footnote{Since we work in a~Bayesian setting, the data noise is taken into account in \cref{eq:gaussianassump}, and all predictions are based on top of that assumption.} Once we have our first surrogate model, we introduce the so-called \stress{acquisition function}\index{Bayesian optimization!acquisition function}. The acquisition function is typically a~function of both $\hat{\mu}$ and $\hat{\sigma}$ and essentially tells us where to perform the next evaluation $x$ in order to maximize the knowledge we gain about the underlying black-box function. In the next section, we see what the acquisition function looks like. For now, we can just think of it to be an~arbitrary function $a(\vect{x}, \hat{\mu}, \hat{\sigma})$. The prediction and its uncertainty are fixed given a~surrogate \ac{GP} prior. Thus, the acquisition function is only a~function of a~new candidate point $\vect{x}$. Our goal is now to find such a candidate point that is as informative as possible. As such, the next point to evaluate is determined by maximizing the acquisition function, i.e., $\vect{x}^\ast \coloneqq \max_{\vect{x}\in D}a(x, \hat{\mu}, \hat{\sigma})$, where $D$ is the domain of $\vect{x}$. Once the new target location $\vect{x}^\ast$ is found, the next step is to evaluate the black box function such that $y^\ast=f_{\rm BB}(\vect{x}^\ast)$. The result of the evaluation is appended to the training set for \ac{GP} such that $\dataset = \dataset_0 \bigcup \{(\vect{x}^\ast, y^\ast)\}$ and a~new, less uncertain, surrogate model is trained on the updated $\dataset$. From this point on, the iteration starts over: every time we update the surrogate model, we have new predictions and uncertainties, hence a~new acquisition function. At each step of the \ac{BO}, a~new point is thus added to $\dataset$, and the entire process goes on until a~maximum number of iterations $T_\mathrm{max}$ is reached or some convergence criterion is met. The plot in \cref{fig:bo} shows how three subsequent steps of \ac{BO} result in an~increasingly more certain surrogate model of the underlying black box function (dark green line). This illustrates how \ac{BO} can be used in the context of \stress{active learning}\index{active learning}, where the training data set is built step-by-step with the aim of minimizing the number of training points while maximizing the information it contains. However, \ac{BO} should not be confused with active learning as they serve different purposes. The former aims to optimize the target function with as few evaluations as possible. The latter, instead, tries to sample the input space as efficiently as possible to target more accurate prediction models.

\highlight{\ac{BO} with \ac{GPR} has the following advantages over other optimization methods:
\begin{itemize}
    \item smaller number of function evaluations\footnote{A~suitable choice for the kernel can be used to lower the number of function calls. Moreover, it is preferable to have smaller training sets for this method. \Cref{eq:gpmuhat} shows that the kernel matrix needs to be inverted for each trial kernel, which adds a~computational constraint.},
    \item gradient-free.
\end{itemize}}

\subsubsection*{The acquisition function}
In the previous section, we have briefly described the idea of \ac{BO} and how it operates combined with \acp{GP}. In this context, we have introduced the acquisition function\index{Bayesian optimization!acquisition function}. This quantity is very important as it represents a~mathematical technique that guides the exploration of the entire parameter space during the \ac{BO}-routine. We have previously defined the acquisition function as a~general function of $\vect{x}$, the surrogate's prediction and its uncertainty. There are different kinds of acquisition functions, and most of the time, the choice is problem-dependent. However, most importantly, its mathematical form should always incorporate the trade-off between exploration and exploitation. In other words, the goal of the acquisition function is to evaluate the usefulness of the next data location to look at in order to achieve the maximization of the surrogate model of our black box function and, thus, to approximate the target function with lower uncertainty.
As such, the ultimate goal in \ac{BO} is to find the next point to evaluate by maximizing such acquisition function. 
One example, commonly used and easy to interpret, is the Upper Confidence Bound (UCB):
\begin{equation}\label{eq:acqfunc}
    a_{\rm UCB}(\vect{x}, \hat{\mu}, \hat{\sigma}) =\hat{\mu}(\vect{x}) + \beta\hat{\sigma}(\vect{x})
\end{equation}
where $\beta\geq 0$ is an~arbitrary parameter that ideally should be tuned during the optimization routine. Here, the first term drives the exploitation, while the second drives the exploration. In the remainder, we refer to those as the \stress{exploitation} and the \stress{exploration} terms, respectively.

\begin{figure}[t]
        \begin{center}
        \includegraphics[width=0.99\columnwidth]{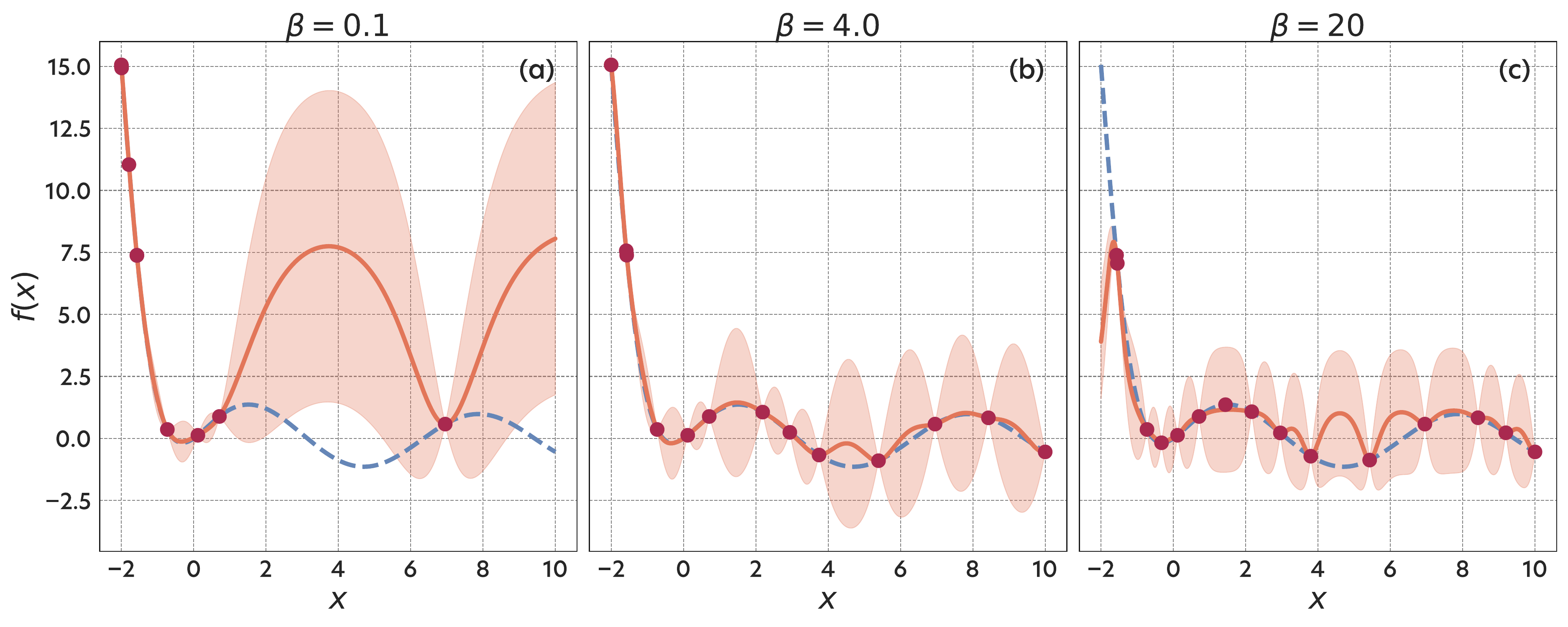}
        \end{center}
        \caption[Acquisition function example]{Selection of new candidate points via \acf{BO} using the Upper Confidence Bound acquisition function. The target function is represented by the blue dashed line. The solid orange line is the surrogate model (\ac{GP}), while the orange shading represents its uncertainty. The leftmost plot exhibits exploitative behavior, i.e., the most selected points are around the peak(s). Contrarily, in the rightmost plot, the parameter choice for $\beta$ heavily enforces exploration. As such, new sampled points (red dots) are evenly distributed through some part of the domain (e.g., $x \sim 2$ would require more exploitation). The middle plot shows a~trade-off between exploration and exploitation: the sampled candidates are well-distributed across the entire domain, thus approximating the target function efficiently even around the peak(s) and boundaries.}
        \label{fig:acqfunc}
\end{figure}

Depending on the value of $\beta$, the exploration term might dominate in the maximization. By looking at \cref{eq:acqfunc} it is immediately clear that a~new candidate point with the higher variance is preferred as the model rewards the evaluation of currently unexplored regions of the domain. That is not surprising as the model seeks to explore what it does not know yet. With respect to the mean, according to the UCB, higher values for the mean are preferred. That is because, by definition, we are seeking for an~upper bound, hence enhancing sampling in the upper quartile of the surrogate model. In other words, in the extreme case where $\beta\gg0$, the exploration dominates, hence regions of higher variance are preferred (see \cref{fig:acqfunc} leftmost plot). 

When instead $\beta \to 0$, the acquisition function becomes far more conservative, hence samples aggressively around the best solution, i.e., exploiting the region where the surrogate model feels confident as visible in the rightmost panel of \cref{fig:acqfunc}. In \cref{fig:acqfunc}, the middle plot shows a~good balance between the exploration and the exploitation. Hence, for two candidate points with comparable predicted mean, the one with higher uncertainty is preferred. As a~consequence, the acquisition function, at least at the beginning of the optimization, prefers to explore rather than exploit.

Moreover, looking at the analytical form from~\cref{eq:gpmuhat} (which appears as the first term in \cref{eq:acqfunc}), the acquisition function might not always be easy to maximize (minimize) in practice. Therefore, one needs to leverage efficient numerical optimization routines. As acquisition functions are highly non-convex, what is done in practice is to do \stress{batch optimization}. At each \ac{BO} step, starting points $\vect{x}_\mathrm{cand}$ are randomly sampled over a~specified domain $D$. Then, one takes the best one of the sampled points (that maximizes the acquisition function) as the actual candidate. Other prominent examples of widely used acquisition functions are: \ac{EI}, \ac{NEI}, \ac{PI}. For a~deeper yet more detailed overview of other types of acquisition functions we refer to Ref.~\cite{frazier2018tutorial}.

\subsection{Choosing the right model}\label{ss:right-kernel-model}
Having introduced the powerful toolbox of kernels, a~natural question arises: Suppose we are given a~set of ``noisy'' data points, forming a~data set $\dataset$, and two distinct models $\mathcal{M}_{1}$ and $\mathcal{M}_{2}$ possibly based on two different kernels, which one should we choose? This is the central question behind \stress{model selection}. To answer this question, we again take a~Bayesian approach (see~\cref{sss:probability}). Applying Bayes' theorem from \cref{eq:Bayes_rule} to each model $\mathcal{M}_{i}$ yields
\begin{equation}
p(\mathcal{M}_{i} \mid {\dataset}) = \frac{p({\dataset} \mid \mathcal{M}_{i})p(\mathcal{M}_{i})}{p({\dataset})}\,.
\end{equation}
We combine the expressions of the two models to obtain
\begin{equation}\label{eq_model_selection_1}
\frac{p(\mathcal{M}_{1} \mid {\dataset})}{p(\mathcal{M}_{2} \mid {\dataset})} = \frac{p({\dataset} \mid \mathcal{M}_{1})}{p({\dataset} \mid \mathcal{M}_{2})}\ \frac{p(\mathcal{M}_{1})}{p(\mathcal{M}_{2})}\,.
\end{equation}
If we have no prior knowledge of the model performance, we must set the priors for the two models to be the same. In this case, the ratio of the posterior probabilities $P(\mathcal{M}_{i} \mid {\dataset})$ is equal to the ratio of the prior probabilities $P(\mathcal{M}_{i})$ times the so-called \stress{Bayes factor}
\begin{equation}\label{eq_model_selection_2}
\frac{p({\dataset} \mid \mathcal{M}_{1})}{p({\dataset} \mid \mathcal{M}_{2})}\,.
\end{equation}
\Cref{eq_model_selection_1} gives us a~first answer to our question: In a~Bayesian framework, the ratio of posterior probabilities can be used to decide which model is superior given the data at hand, i.e., the model with the larger posterior probability is superior. In scenarios where we do not know anything about the data, we can set the prior probabilities equal to each other, which leaves us with the Bayes factor
\begin{equation}
\frac{p(\mathcal{M}_{1} \mid {\dataset})}{p(\mathcal{M}_{2} \mid {\dataset})} = \frac{p({\dataset} \mid \mathcal{M}_{1})}{p({\dataset} \mid \mathcal{M}_{2})}\,.
\end{equation}

To calculate the Bayes factor in~\cref{eq_model_selection_2}, we need to compute $p({\dataset} \mid \mathcal{M}_{i})$ for each model which can be viewed as ~\stress{marginal likelihood}, i.e., a~likelihood function in which all variables except the type of the model have been marginalized (integrated out). Let us define the likelihood as $p({\dataset} \mid \params,\mathcal{M}_{i})$, such that the marginal likelihood can be obtained as
\begin{equation}
p({\dataset} \mid \mathcal{M}_{i}) = \int_{\realset^{\nparams}} p({\dataset}|\params,\mathcal{M}_{i})p(\params \mid \mathcal{M}_{i})\ d\params\,,
\end{equation}
where we integrate over the distribution of model parameters $\params\in \realset^{\nparams}$ given by $p(\params \mid \mathcal{M}_{i})$. Unfortunately, marginal likelihoods are typically hard to compute as they involve high-dimensional integrals. Choosing a~kernel with $\nparams$ parameters results in a~$\nparams$-dimensional integral for its marginal likelihood.

Having encountered this problem, let us take a~step back: When we train a~model, we minimize a~loss function\index{loss function} (or equivalently, we maximize the log-likelihood)\index{maximum likelihood}. Therefore, why not simply choose the model that gives the lowest loss or largest likelihood? Intuitively, this leads to \stress{overfitting}\index{overfitting}. This intuition is formalized by the \stress{bias-variance trade-off}\index{bias-variance trade-off} (see~\cref{sss:generalization_regularization}). In particular, the bias-variance trade-off makes it clear that the ideal model realizes an~optimal balance between the training error and the model complexity. Rather than choosing the model that results in the lowest loss during training, we thus need to take its complexity into account.
        
\subsubsection{Bayesian information criterion}\label{sss:BIC}
A~computationally tractable criterion for model selection which takes model complexity into account is the \acf{BIC}~\cite{schwarz:1978} defined as\index{Bayesian information criterion}
\begin{equation}\label{eq_BIC_main}
{\rm BIC} = -2 \max(\loglik)   + \nparams \log(\datasize),
\end{equation}
where $\max(\loglik)$ is the maximum of the log likelihood, $\datasize$ is the number of training points, and $\nparams$ is the number of model parameters. The lower the \ac{BIC}, the better the model. Clearly, the \ac{BIC} reflects the trade-off between bias, here given by $\max(\loglik)$, and the model complexity as measured by $\nparams \log(\datasize)$. Moreover, it turns out that the \ac{BIC} approximates the logarithm of the marginal likelihood in the large $\datasize$-limit~\cite{stoica:2004}:
\begin{equation}\label{eq_BIC}
\log p({\dataset} \mid \mathcal{M}_{i}) \approx \log p({\dataset} \mid \params^{\ast}, \mathcal{M}_{i})-\frac{d}{2} \log(n)\,,
\end{equation}
where $\params^{\ast}$ are the model parameters that maximize the likelihood. This expression reveals that the model selection criterion given in \cref{eq_model_selection_1} based on the Bayesian approach does indeed take the model complexity into account. Moreover, we see that the criterion can be used to estimate the posterior probability of a~model $\mathcal{M}_{i}$ as
\begin{equation}
p_{i} = \frac{\exp(-\frac{1}{2} {\rm BIC}_{i})}{\normdist}\,.
\end{equation}
Here, the normalization constant $\normdist = \sum_{i} e^{-{\rm BIC}_{i}/2}$ ensures that each model $\mathcal{M}_{i}$ is assigned a~valid probability $p_{i}$ to enable comparability. As such, the \ac{BIC} gives us a~tractable way to select models according to the criterion given in~\cref{eq_model_selection_1}. In fact, \ac{BIC} is asymptotically consistent as a~model selection metric: Given a~family of models, including the model underlying the data, the probability that \ac{BIC} correctly selects the model underlying the data approaches one as $\datasize\rightarrow \infty$.\footnote{While the \ac{BIC} criterion approximates the log marginal likelihood in the large $\datasize$-limit, it can still be applied as a~heuristic model selection criterion at low values of $\datasize$ and can be confirmed empirically to often still yield good results. There exist many other model selection criteria (see~\cite{stoica:2004} for a~review), a~popular one being the Akaike information criterion~\cite{akaike:1974} which closely resembles the \ac{BIC}. The crucial difference between the \ac{BIC} and many other methods is its asymptotic consistency. One may question the importance of asymptotic consistency due to the fact that the ground-truth model typically is not present in the candidate set of models in practice.}

Inspired by these findings, we can adapt the criterion to \ac{GPR}\index{Gaussian process} based on the log marginal likelihood, which is optimized during training (at fixed kernel parameters) and the number of kernel parameters in the \ac{GPR}. This criterion is computationally tractable and thus allows one to select between different kernels in the regression task using \acp{GP}.

\subsubsection{Kernel search}\label{sss:kernel-search}
Choosing the right kernel\index{kernel search} is crucial when using a kernel-based method, as we have seen, e.g., for the performance of \acp{SVM} for different kernels in \cref{fig:SVM-nonlinear_2}. When performing \acf{GPR} in a~\stress{naive} manner, we simply select a~fixed kernel from a~set of conventional kernels such as listed in \cref{tab:kernel-functions}. We then optimize their hyperparameters by maximizing the marginal likelihood during training. There are now several possible routes toward achieving a~more accurate model. Clearly, we may improve the model accuracy by providing more training points. However, keeping the number of training points low is one of the main advantages of \ac{GPR} compared to other methods and constituted our main initial motivation. At a~fixed number of training data, the result from \ac{GPR} can only be improved through a~better kernel. Moreover, while \ac{BO}\index{Bayesian optimization} is guaranteed to converge, the exact number of iterations may vary drastically. The choice of a~good kernel can significantly speed up the convergence of \ac{BO}.

\begin{figure}[t]
        \begin{center}
        \includegraphics[width=0.9\columnwidth]{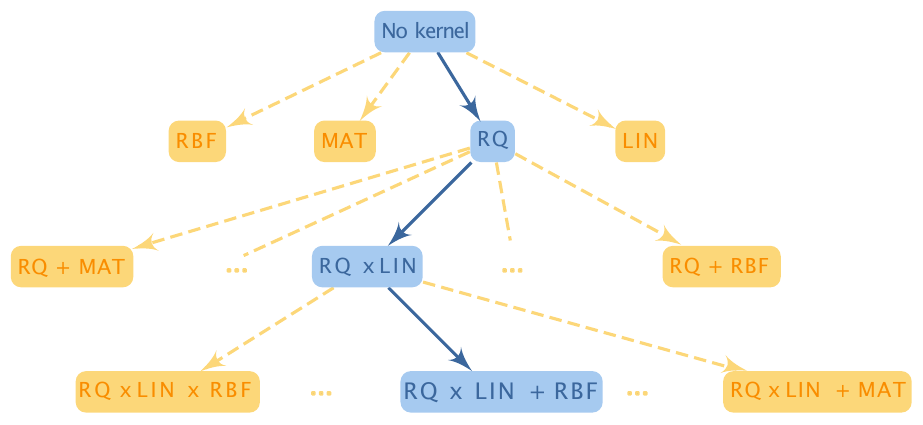}
        \end{center}
        \caption[Algorithm for optimal kernel construction]{Illustration of the search tree behind the algorithm for the optimal kernel construction in \ac{GPR}. It utilizes the \ac{BIC} for the model selection introduced in \cref{eq_BIC_main}. For an~overview of possible kernel functions and corresponding abbreviations, see~\cref{tab:kernel-functions}. Adapted from \ToggleForCUP{Vargas-Hernández, R. A. \textit{et al.} (2018). \textit{Extrapolating quantum observables with machine learning: Inferring multiple phase transitions from properties of a single phase}. Phys. Rev. Lett. 121, 255702~\cite{Vargas-Hernandez2018}.}{Ref.~\cite{Vargas-Hernandez2018}.}}
        \label{fig:search_tree}
\end{figure}

The construction of good kernels ultimately boils down to a~(possibly high-dimensional) optimization problem~\cite{duvenaud:2013}. This happens, for example, when constructing a~good kernel through optimization of the kernel hyperparameters itself. The key challenge is posed by the fact that the parametric form of the kernel must be proposed by the user itself. This is a~non-trivial task that relies on trial and error -- even for experts. In Refs.~\cite{duvenaud:2011,duvenaud:2013,Vargas-Hernandez2018} the kernel learning problem was reframed as a~search tree problem (see~\cref{fig:search_tree}): the space of parametric forms of kernels is constructed as a~tree which can be searched systematically in an~automated fashion, where the powerful \ac{BIC}\index{Bayesian information criterion} is used for the kernel selection and new kernels are proposed via composition.

We start by selecting each kernel from a~set of conventional kernels and training a~\ac{GP} for each of them on the same data set. Then, we select the one that achieves the lowest value of \ac{BIC} as given by \cref{eq_BIC_main} (highlighted in blue). This kernel serves as the base kernel for the subsequent round, where it is combined with the various kernels from the starting set to create new candidate kernels by forming products or combining them linearly. Again, the best one is selected according to the \ac{BIC}, and the process is repeated. The complexity of the model, i.e., of the composite kernel, increases as one progresses in the search tree. Eventually, increasing the kernel complexity further leads to overfitting and, hence, does not improve the \ac{BIC} value compared to the kernel of the previous round, and the algorithm is stopped. The algorithm can also be stopped prematurely if the number of kernel parameters becomes large and the associated training simply takes too long to be practical. Other than greedily searching the tree, the reformulation of the kernel construction as a~search tree problem opens up the possibility for more advanced strategies which could yield better kernels more efficiently~\cite{dai:2020,vargas:2021_GPSD}.

\subsection{Applications in quantum sciences}\label{sec:BO_GPR_science}
\begin{figure}[t]
\begin{center}
\includegraphics[width=0.7\columnwidth]{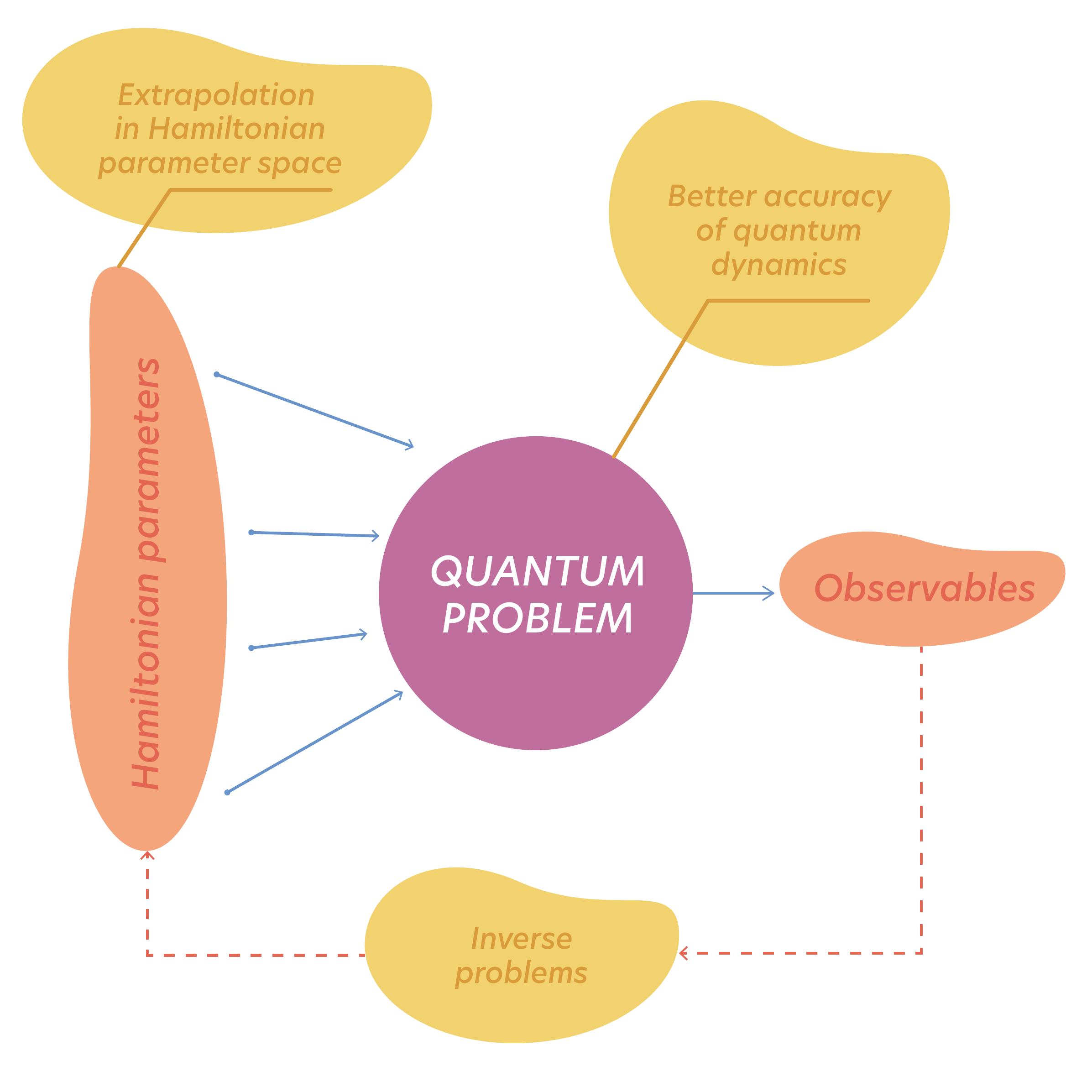}
\end{center}
\caption[Three main classes of quantum problems successfully tackled with Bayesian optimization and Gaussian processes]{Illustration of the three main classes of problems in quantum sciences (marked in yellow) that have been successfully tackled with \ac{BO} and \acp{GP}.}
\label{fig:GPRforscience}
\end{figure}

In the previous sections, we have motivated \acfp{GP}\index{Gaussian process} and \acf{BO}\index{Bayesian optimization} as powerful methods that together allow us to build expressive \ac{ML} models.
Importantly, they are equipped with an~intrinsic measure for uncertainty\index{uncertainty} and can be trained using a~small amount of training data. In this section, we discuss how these two methods can be useful in the context of quantum sciences, as sketched in \cref{fig:GPRforscience}. In particular, \ac{GP} and \ac{BO} can be used to tackle inverse problems, extrapolate in Hamiltonian parameter spaces, and increase the accuracy of quantum dynamics calculations.

\subsubsection{Inverse problems}

As explained in \cref{ss:GPR-BO}, \ac{BO} is very useful when you need to optimize black-box functions that are expensive to evaluate. This property proves extremely useful in inverse quantum problems aiming at finding a theoretical description of the system by experimentally measuring its observables. The idea is related to a~popular experimental approach known as \stress{optimal control}. The optimal control approach aims to design external field parameters that yield the desired quantum dynamics. It is usually achieved by a~feedback loop, which iteratively modifies experimental parameters such that they yield system dynamics advancing to the target one. 

We can imagine applying a~similar feedback loop for the inverse quantum problems. It would consist of iterative modifications of parameters of the theoretical description (such as Hamiltonian parameters) till the observables predicted theoretically agree with those measured experimentally. However, solving the iterative inverse quantum problem is challenging. Each iteration requires an~additional run of theoretical calculations, e.g., the numerical solution of the Schrödinger equation, which is time-consuming. The optimization itself is also difficult as we do not explicitly know the range of parameters that needs to be explored. Finally, the curse of exponential scaling of the Hilbert space dimension with the complexity of the quantum systems definitely does not help. How to make it more feasible? Both inverse quantum problems and optimal control become easier when the expensive black box (either the experimental set-up or the theoretical calculations) is replaced by a~trained surrogate \ac{ML} model such as a~\ac{GP}. Finally, instead of a~blind search for the optimal parameters, we can employ \ac{BO}.

\begin{figure}[t]
\begin{center}
\includegraphics[width=0.8\columnwidth]{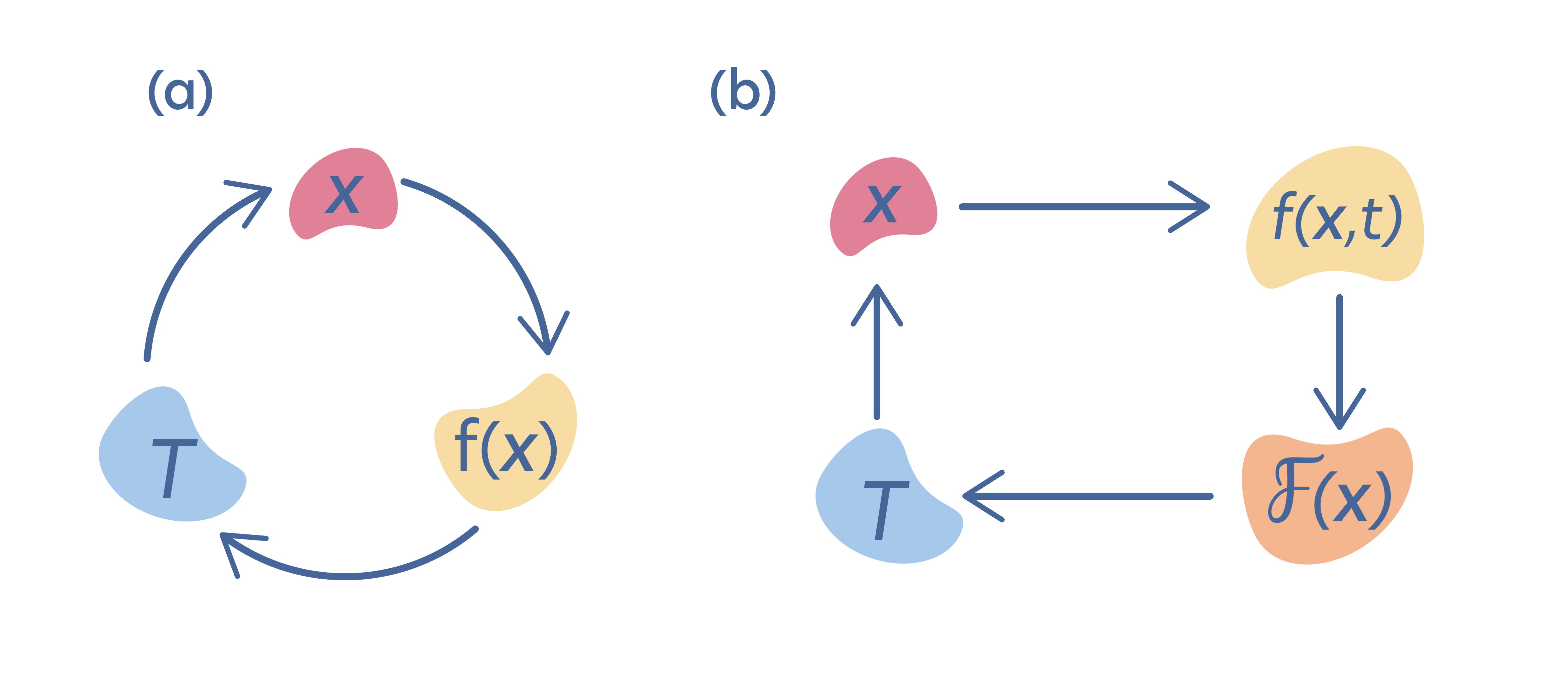}
\end{center}
\caption[Gaussian processes and Bayesian optimization for feedback loops.]{Examples of feedback loops whose optimization becomes feasible when implemented with \ac{BO} and \acp{GP}. (a) $\vect{x}$ corresponds to \ac{PES}, $f(\vect{x})$ is quantum scattering calculations taking the \ac{PES} as an~input, $T$ is the difference between reaction probabilities calculated by $f(\vect{x})$ and measured in the experiment across various collisional energies. A search for the optimal $\vect{x}$ would require minimization of $T$ via optimization of $f(\vect{x})$. It becomes feasible when we surrogate $f(\vect{x})$ with one \ac{GP} and $\vect x$ with another \ac{GP} and apply \ac{BO}. (b) $\vect{x}$ are Hamiltonian parameters, $f(\vect{x},t)$ is the time-dependent observable $f$ (e.g., molecular orientation or alignment), $\mathcal{F}(\vect{x})$ is the time-dependent Schrödinger equation, $T$ the difference between calculated and measured time-dependent observable $f$. When $\mathcal{F}(\vect{x})$ is surrogated by a~\ac{GP}, \ac{BO} is used to minimize $T$ and find $\vect{x}$ of the underlying Hamiltonian.}
\label{fig:GP_feedback_loops}
\end{figure}
\begin{figure}[t]
\begin{center}
\includegraphics[width=0.95\columnwidth]{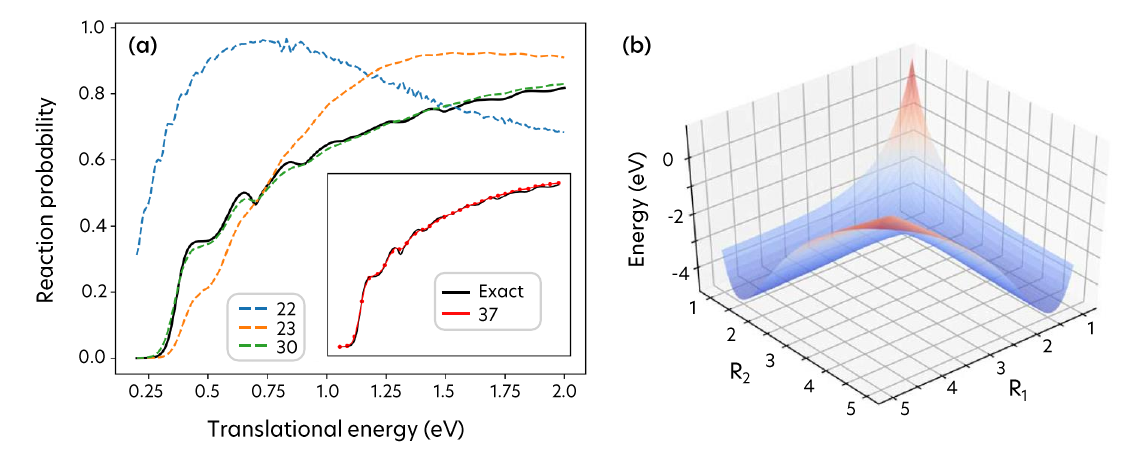}
\end{center}
\caption[Gaussian process as a~potential energy surface for quantum calculations.]{(a) The reaction probability for the \ce{H2 + H -> H + H2} reaction as a~function of the collision energy. The black solid curve represents calculations from Ref.~\cite{Su2015} based on the surface with 8701 \textit{ab initio} points. The dashed blue/orange/green/red curves are calculations based on the \ac{GP} \ac{PES} obtained with 22/23/30/37 \textit{ab initio} points. (b) \ac{GP} model of the \ac{PES} for the \ce{H3} reaction system constructed with 30 \textit{ab initio} points. $R_1$ and $R_2$ are the distances between atoms 1 and 2 and atoms 2 and 3, respectively. Adapted from \ToggleForCUP{Vargas-Hernández, R.~A. \textit{et al.} (2019). \textit{Bayesian optimization for the inverse scattering problem in quantum reaction dynamics}. New J. Phys. 21, 22001~\cite{Vargas2019} under the \href{https://creativecommons.org/licenses/by/3.0/}{CC BY 3.0 DEED} license.}{Ref.~\cite{Vargas2019}.}}
\label{fig:GP_PES}
\end{figure}
As a~practical example of the inverse problems solved with \ac{BO} and \acp{GP}, let us consider the application to \stress{scattering experiments}. The outcomes of such experiments are determined by the microscopic interactions between scattered particles. We have a~quantum theory that describes these interactions and can predict the outcome of such scattering events. Therefore, our aim, in case of an~inverse problem, may be to infer these microscopic interactions from the experiment. More concretely, the authors of Ref.~\cite{Vargas2019} aimed to recover a~global \acf{PES}\footnote{A~\acf{PES} describes interactions between some particles. As a~result, it models landscapes of chemical reactions, which can be used to predict reactive pathways and final products. Traditionally, it is constructed as an~analytic fit to many, usually costly {\it ab initio} quantum-chemical calculations of the potential energy for reactants for various relative positions.} governing the chemical reaction \ce{H + H2 -> H2 + H}, using as few experimental measurements of the reaction rates (depending on the constituents' translational energy) as possible. The feedback loop that needed to be solved is presented in \cref{fig:GP_feedback_loops}(a). Firstly, they trained a~\ac{GP} to surrogate a~quantum scattering theory on a~series of \acp{PES} and predicted reaction rates. Secondly, they modeled the \ac{PES} with another \ac{GP}. Finally, they used \ac{BO} to find the three-dimensional \acfp{PES}, recovering the measured reaction rates. Only eight iterations of \ac{BO} (where every iteration rebuilds the \ac{PES} completely) were required to reach the accuracy of conventional approaches! Moreover, in this case, a~traditional approach of building a~\ac{PES} requires around 8 700 points -- their \ac{GP} was modeled based only on 30 points!\footnote{Remember, these are not any 30 points, but points indicated by \ac{BO} as needed for the optimal description.} This impressive scaling is presented in~\cref{fig:GP_PES}. As a~result, they successfully surrogated two complex models (\ac{PES} and quantum scattering calculations using \ac{PES} as an~input) with two \acp{GP} trained on a much smaller number of data points than needed to build the original complex models. They also used this approach for a~six-dimensional \ac{PES} of \ce{OH + H2}, where \ac{BO} beat the traditional approach with 290 points compared to 17 000 points.\footnote{The number of points needed for efficient \ac{BO} scales roughly like 10 times the number of \ac{PES} dimensions. We can raise an~interesting point, which is, how are we even sure that we faithfully reproduce the \ac{PES} if we build it only from reaction probabilities? It may happen that we capture the reactive chemical channels accurately, but the remaining parts of the surface are unconstrained and as a~result may be nonphysical. One can argue that we ultimately do not need a~complete faithful reproduction of the underlying \ac{PES}. We only need a~\ac{PES} that allows us to accurately predict what we are interested in, here reaction probabilities. Note, however, that if we take a~\ac{PES} built from a~particular set of observables and we use it to calculate another observable, the result may be wrong.}

Another example of an~inverse quantum problem is the task of inferring molecular properties from time-dependent observables. Authors of Ref.~\cite{Deng2020} tackled the reconstruction of molecular polarizability tensors from the observed time evolutions of the orientation or alignment signals of \ce{SO3} and propylene oxide induced by strong laser pulses. The feedback loop that was solved is drawn in~\cref{fig:GP_feedback_loops}(b). They used a~\ac{GP} with a~vector output whose elements corresponded to a~prediction of a~chosen observable (orientation or alignment) in a~different time step. The \ac{GP} was trained to surrogate the numerical integration of the time-dependent Schrödinger equation given the Hamiltonian parameters. Interestingly, the authors showed what we discussed already in \cref{ss:right-kernel-model}: that a~proper choice of the kernel can result in a~two times faster convergence of \ac{BO}. Analogous approaches were used for the reconstruction of scattering matrices of molecules from molecular hyperfine experiments~\cite{Cantin2020} and for optimizing the reaction conditions of an organic chemistry experiment~\cite{Sugisawa2021}.

\subsubsection{Improving quantum dynamics, physical models, and experiments}
\acp{GP} and \ac{BO} can also be used for transfer learning\index{transfer learning} in the context of quantum dynamics calculations. These are typically very difficult, and one quickly has to rely on approximations. The authors of Ref.~\cite{Jasinski2020} proposed to apply \acp{GP} to correct such approximate quantum calculations for computing cross-sections for molecular collisions. The idea is to train a~model on a~small number of exact results and a~large number of approximate calculations, resulting in \ac{ML} models that can generalize exact quantum results to different dynamical processes.

Moreover, as the minimization of any function using \ac{BO} bypasses the need for computing gradients \cite{bayesoptbook_2022}, successful applications of \ac{BO} include optimization of parameters of physical models. Most models do not have a~closed-form solution and conventionally have to be approximated numerically using finite differences. For example, Refs. \cite{BO_DFT_jpca_2020, BO_DFT_jpca_jctc_2019} showed that \ac{BO} could efficiently optimize density functional models to improve their accuracy and minimize the energy of the Ising model~\cite{BO_Ising_2018}.
Furthermore, \ac{BO} was used to generate low-energy molecular conformers \cite{pmlr-v48-carr16,Chan2019}, tuning the parameters of various models used to simulate \textit{cis–trans} photoisomerization of retinal in rhodopsin \cite{multi-BO_photoisomer_ravh_2021}, and the optimization of lasers \cite{BO-lasers_PRL_2020,BO-lasers_PRL_2021, BO-lasers_NatComm_2020}. \ac{BO} has also been impactful in material science in chemical-compounds screening~\cite{COMBO_2016,Jalem2018,BO_PhysRevX_2017,BO_crystal_plasticity_2021,cBO_Lobato_2020,deshwal_simon_doppa_2021} and optimization of experimental setups~\cite{Phoenics_AAG_2018,Gryffin_AAG_2021,multiBO_ferroelectric_2021,NEXTorch_BOChem_2021,Golem_BO_AAG_2021, Vendeiro22PRR}.

\subsubsection{Extrapolation problems}
The second class of problems that seems suitable for \acp{GP} are extrapolation tasks: given some function values for data points in one regime, the goal is to accurately predict the function values of data points in different regimes. This section touches upon two possible applications that are (1) learning \ac{PES} from a~possibly smallest number of \textit{ab initio} calculations in one regime and (2) predicting the existence of quantum phases without knowledge of the full phase diagram\index{phase classification}.

An~example of a~successful extrapolation in the case of \ac{PES}-learning was shown in Ref.~\cite{dai:2020} where authors studied the six-dimensional \ac{PES} of \ce{H3O+}. They trained \ac{GP} models on 1000 \textit{ab initio} geometries from a~low-energy regime (up to $\approx 7000$ cm$^{-1}$) and checked that the model predictions in higher energy regimes match the full calculations with a~high level of accuracy. If you doubt it, this result can be reproduced using the published code and data~\cite{OurSchoolRepo}. It gets better! You can get similarly accurate extrapolations from a~\ac{GP} model trained on 5000 molecular geometries of a~51-dimensional problem of a~protonated imidazole dimer, which contains 19 atoms~\cite{Sugisawa2020}. The scaling of the extrapolation accuracy with respect to the number of training points seems to be even more favorable for large molecules: high accuracy was reached already for 1000 randomly sampled geometries of the 57-dimensional aspirin.\footnote{More precisely, the test energy mean absolute error was 0.177 kcal/mol. Is this error small? Are these \acp{PES} accurate enough for modern spectroscopic applications? Spectroscopy discerns two kinds of accuracies. The spectroscopic accuracy is 1 cm$^{-1}$, while the chemical accuracy is 1 kcal/mol $\approx 350$ cm$^{-1}$. Modern spectroscopic applications need a~spectroscopic accuracy, and this result has an~error that is 60 times larger than spectroscopic accuracy - so no, it is not yet good enough.
But it is more than enough, e.g., for simulations of molecular dynamics or reactions, especially at room temperature \cite{Puzzarini2019}.}

\begin{figure}[t]
\begin{center}
\includegraphics[width=0.95\columnwidth]{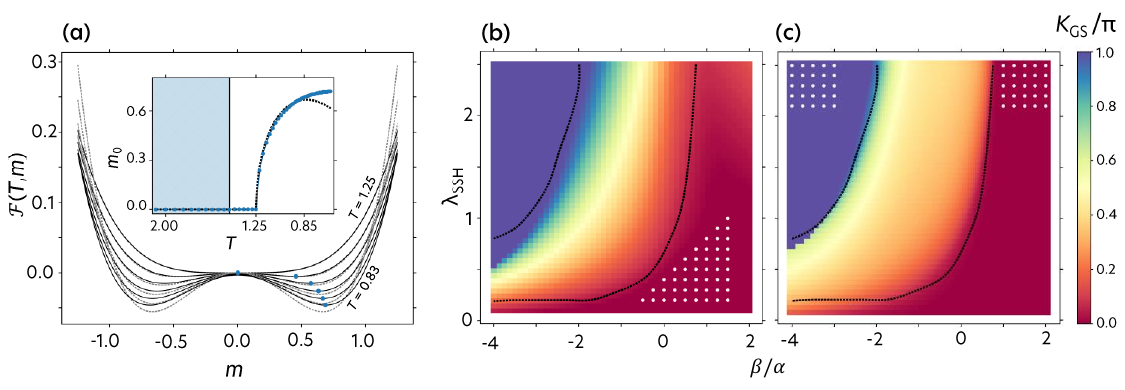}
\end{center}
\caption[Gaussian process for extrapolation to non-seen quantum phases]{\acp{GP} for extrapolation to non-seen quantum phases. (a) The mean-field Heisenberg spin model in the nearest-neighbor approximation, where black dots are mean-field results and blue dots are \ac{GP} predictions. \ac{GP} was trained on the high-temperature shadowed regime of the data. (b)-(c) generalized lattice polaron model with different \ac{GP} training regimes (marked with white dots). In both cases, \ac{GP} correctly predicted the phase transitions (color map) as compared to the quantum calculations (black lines). Adapted from \ToggleForCUP{Vargas-Hernández, R. A. \textit{et al.} (2018). \textit{Extrapolating quantum observables with machine learning: Inferring multiple phase transitions from properties of a single phase}. Phys. Rev. Lett. 121, 255702~\cite{Vargas-Hernandez2018}.}{Ref.~\cite{Vargas-Hernandez2018}.}}
\label{fig:GPRforextrapolation}
\end{figure}

Another example of extrapolation, this time in the space of Hamiltonians, is the task of inferring properties of other phases of a~system given knowledge of one particular phase. In Ref.~\cite{Vargas-Hernandez2018}, the authors proposed to train a~\ac{GP} on one phase of the system and expected the model to predict phase transitions and properties of other phases. Let us start by discussing how they achieved this for the mean-field Heisenberg spin model in the nearest-neighbor approximation. They trained the \ac{GP} on the free energy of the system in the high-temperature regime, where the average spin magnetization is zero, far from the phase transition point. Then the trained \ac{GP} was asked to extrapolate within the low-temperature regime, and it predicted correctly both the location of the phase transition as well as the free energy (and consequently non-zero magnetization) of the system as presented in \cref{fig:GPRforextrapolation}(a).

The authors also applied this approach to a~much more complex system\footnote{The studied system was a~generalized lattice polaron model \cite{Herrera13PRL} describing an~electron in a~one-dimensional lattice with $N \rightarrow \infty$ sites coupled to a~phonon field. The interaction between an~electron and a~phonon field was a~combination of two qualitatively different terms: the Su-Schrieffer-Heeger (SSH) electron-phonon coupling and the breathing-mode model with the Holstein coupling.} whose Hamiltonian can be written in the following generic form:
\begin{equation}
H=H_{0}+\alpha H_{1}+\beta H_{2}\,,
\end{equation}
where $\alpha$ and $\beta$ are tunable parameters along which phase transitions occur. They trained a~\ac{GP} in some parameter regime of the Hamiltonian and were able to successfully extrapolate to others.\footnote{How is this even possible? The intuition behind it is that the evolution of physical properties that are given to the \ac{ML} model as input should somehow reflect the fact that there is a~phase transition. The model probably picks up on the prevalent correlations within one phase, and it observes that these correlations change when crossing to other phases.} This approach proves to be useful for such a~class of Hamiltonians for another reason. Usually, we are able to easily compute or measure the eigenspectrum in certain limits of $\alpha$ and $\beta$, but not at arbitrary points within the parameter space. We can then train \acp{GP} in these limits and can expect them to extrapolate successfully to other parameter regimes where the direct calculation is more difficult. Finally, in the same system, the authors of Ref.~\cite{Vargas2019} studied the importance of choosing the kernel\index{kernel search}. They compared the results from the original work~\cite{Vargas-Hernandez2018} obtained for kernels found with the \ac{BIC} as described in \cref{sss:BIC} and \cref{sss:kernel-search} to the results obtained for kernels with the same complexity (that is, at the same search tree level, see \cref{fig:search_tree}) but chosen at random. Predictions of such \acp{GP} were much worse and were prone to overfitting, which stresses the power of the \ac{BIC} as a selection criterion for kernels. The appropriate choice of the kernel is, therefore, crucial as it determines how far the model can accurately extrapolate.
\subsubsection{Bayesian optimization of variational quantum algorithms}\label{sec:BO_for_VQE}

Another suitable application for \ac{BO} is within the context of near-term quantum computing, where a computational advantage is sought by the use of \ac{NISQ} devices, see also \cref{sec:qml_nisq_era}.
One popular strategy is \acp{VQE}:
These algorithms are hybrid quantum-classical algorithms that are suited for finding the ground state of a given Hamiltonian. Applications of \acp{VQE} can be found in several domains, see e.g., Refs.~\cite{deglmann2015application,cao2019quantum,mccaskey2019quantum,ciavarella2022preparation,banuls2020simulating}.
In these algorithms, parameter optimization happens classically.
Treating the quantum circuit as a parametrized black-box, \ac{BO}, hence, presents itself as a gradient-free optimization tool.


Recent works have demonstrated the resource efficiency of \ac{BO} in optimizing \acp{VQE}~\cite{iannelli2021noisy,Mueller2022AcceleratingNV,nicoli2023physics}, i.e., the number of calls of the quantum algorithm. In particular, Ref.~\cite{nicoli2023physics} showed that kernel methods are a natural choice for this task by incorporating physical prior knowledge directly into the kernel itself. To achieve this, the authors derive a new general type of kernel with the same functional form as the target black-box function one seeks to minimize, e.g., using \ac{BO}.
Thus, optimizing the parameters of a variational quantum circuit is aided by the prior knowledge one has about the quantum circuit at hand. In addition to that, the authors introduce a new type of acquisition function. The pivotal feature of this acquisition function is to look for the next point to measure by using the level of confidence the model has with respect to the current choice of parameters. 
Should the confidence of the \ac{GP} for a given candidate point be high, i.e., small variance, it is likely to be skipped. Therefore, leveraging a sequential minimal optimization scheme \cite{nakanishi20,Platt1998SequentialMO}, i.e., sequentially optimizing one parameter after the other, this new acquisition function achieves better convergence to the global optimum. The results shown in Ref.~\cite{nicoli2023physics} show that the novel optimization scheme introduced therein is capable of outperforming the state-of-the-art~\cite{nakanishi20} (at the time of writing). 
While the kernel in Ref.~\cite{nicoli2023physics} has been proposed within the context of finding the ground state of a given Hamiltonian, it can be combined with any other acquisition function and used for any other optimization task. Similarly, the novel acquisition function serves as a general framework suited for any other optimization tasks when combined with the kernel mentioned above.

\subsection{Outlook and open problems}
\begin{itemize}
    \item While \acp{GP} successfully surrogate \acp{PES} and need a much smaller number of \textit{ab initio} calculations, it is challenging to reach the spectroscopic accuracy with this approach. What is stopping us from achieving such accuracy levels with \acp{GP}? The major limitation is the number of training data. In practice, it is often observed that the error during learning eventually decreases by a~factor of $1/n$, where $n$ is the number of training points. As such, the number of training points required to reach a~level of accuracy on the order of 10 cm$^{-1}$ for a~57-dimensional surfaces is still manageable. However, reaching spectroscopic accuracy requires an~excessive amount of training data. In particular, the size of the training data set grows beyond the regime where \ac{GP} are useful~\cite{Asnaashari2021}. The high-accuracy limit may be obtained if one incorporates some knowledge of the system into the kernel. An~open question of how to do that remains to be answered.
    \item In \cref{sec:BO_GPR_science}, we have presented how \ac{BO} and \acp{GP} can be used to tackle optimization of expensive setups where gradients are not accessible. Such is also the case of quantum \acp{NN} or \acp{VAE}. Therefore, this approach may prove useful in the optimization of a~quantum model!
    \item An~interesting research direction is combining the power of \acf{AD}\index{differentiation!automatic differentiation}, described in \cref{sec:hot-topics:dp}, and kernel methods. Already, \ac{AD} has played a~major role in developing more robust kernel functions for \ac{GP} models. 
    For example, Ref.~\cite{pmlrv51wilson16} showed that by maximizing the log marginal likelihood, (\cref{eq:logmarginal}), one could jointly optimize the weights and biases of a~deep \ac{NN} combined with any parameter of a~standard kernel function. A~more recent work~\cite{pmlr-v80-sun18e} also showed that learning the composition of kernels is differentiable under the \ac{AD} framework, and more complex kernels could be parametrized. 
    Currently, there are two main ecosystems for \acp{GP} based on \ac{AD}, \texttt{GPytorch} \cite{gardner2021gpytorch,JMLR:v22:20-275}, and \texttt{GPflow} \cite{GPflow2017}. 
    
    \item The training procedure of \acf{KRR} could also be differentiated using \ac{AD} bypassing the need of using a~cross-validation scheme ~\cite{blondel:2021}.
    
    \item With the advent of quantum extensions of classical \ac{ML}-methods for near-term quantum devices, there are several paths on how to encode a~data point $\vect{x}$ in a~Hilbert space as $\ket{\vect{x}}$.
    As a~consequence, the kernel function has to be promoted to its quantum version. Interestingly, there is a~provable advantage of such kernels based on measurement results of the quantum state~\cite{Huang2020}. We give a~bit more detail in \cref{s:QML}.
\end{itemize}

\subsection*{Further reading}
\begin{enumerate}
    \item Rasmussen, C. E. \& Williams, C. K. I. (2006). \href{http://www.gaussianprocess.org/gpml/}{Gaussian Processes for Machine Learning.} The MIT Press. The standard go-to reference on kernel methods and \acp{GP} in particular~\cite{rasmussen:2006}.
    \item Bishop, C. M. (2006), \href{https://www.microsoft.com/en-us/research/uploads/prod/2006/01/Bishop-Pattern-Recognition-and-Machine-Learning-2006.pdf}{Machine Learning and Pattern Recognition}. Another go-to reference for \acp{GP}~\cite{bishop:2006}.
    \item Krems, R. V. (2019). \href{https://doi.org/10.1039/C9CP01883B}{\textit{Bayesian machine learning for quantum molecular dynamics}}. PCCP, 21(25), 13392–13410. Discusses various applications of \acp{GP} for quantum molecular dynamics~\cite{Krems2019}.
    \item \href{https://github.com/Shmoo137/SummerSchool2021_MLinQuantum/tree/master/B\%20-\%20Gaussian\%20Process\%20Regression\%20(by\%20Roman\%20Krems)}{Jupyter notebook} allowing to faithfully reproduce a~six-dimensional \ac{PES} with a~\ac{GP} and \ac{BO} including optimal kernel search using the \ac{BIC} criterion for the \ce{H3O+}~\cite{OurSchoolRepo}.
    \item Vargas-Hernández, R. A., \& Krems, R. V. (2020). \href{https://doi.org/10.1007/978-3-030-40245-7_9}{\textit{Physical extrapolation of quantum observables by generalization with Gaussian processes}}. Lect. Notes Phys., 968, 171–194. In-depth review of possible applications of \acp{GP} and \ac{BO} for extrapolation problems in quantum sciences~\cite{Vargas-Hernandez2020}.
    \item Huang, H. et al. (2021). \href{http://dx.doi.org/10.1126/science.abk3333}{\textit{Provably efficient machine learning for quantum many-body problems}}. Science 377(6613). It introduces quantum-measurement-inspired kernels for a provable advantage of kernel methods over classical methods that do not use measurement data~\cite{huang:2021}.
\end{enumerate}

\clearpage
\section{Neural-network quantum states}
\label{sec:NN_q_states}

In the early days of quantum mechanics, it soon became clear that approximation methods would be needed to solve most relevant real-world problems~\cite{Dirac1929}.
Indeed, in most cases, the Schrödinger equation cannot be exactly solved for systems with more than a few interacting particles.
This came to be referred to as the \stress{quantum many-body problem}\index{quantum many-body problem}.
In this chapter, we show how \acfp{NN} have been introduced to tackle this problem \cite{Carleo_2017}, in a~variety of applications, including ground-state and quantum dynamics of interacting quantum systems.
For simplicity, we mainly focus our discussion on spin systems and discuss applications to fermions \cite{Choo2020} and bosons \cite{Saito2017} only toward the end.

According to the axioms of quantum mechanics, the state of an~isolated quantum system is encoded into a~complex-valued vector of probability amplitudes commonly known as the wave function.
In the case of a~single spin-$\frac{1}{2}$, the wave function in the computational z-basis $\hat{Z}$ is $|\Psi\rangle = C_{\uparrow}|\uparrow\rangle + C_{\downarrow}|\downarrow\rangle$.
The coefficients $C_{\uparrow}$ and $\ C_{\downarrow}$ are the complex probability amplitudes of the spin being aligned along ($C_{\uparrow})$ or opposite to  ($C_{\downarrow})$ the direction of the computational basis, and they are subject to the normalization condition $|C_{\uparrow}|^2 + |C_{\downarrow}|^2=1$.
For many-body quantum systems of $N$ spins, where $N$ can be any large number from tens to the order of the Avogadro number $\sim10^{23}$, the number of coefficients in the wave function scales as $2^N$.
Following up on the spin example, the wave function can be expressed as follows:
\begin{equation}
\begin{split}
    |\Psi\rangle &= C_{\uparrow\uparrow\cdots\uparrow}|\uparrow\uparrow\cdots\uparrow\rangle + C_{\uparrow\uparrow\cdots\downarrow}|\uparrow\uparrow\cdots\downarrow\rangle + \dots + C_{\downarrow\downarrow\cdots\downarrow}|\downarrow\downarrow\cdots\downarrow\rangle \\
    &= \sum_{\spin_1,\spin_2, \cdots,\spin_N}  C_{\spin_1,\spin_2, \dots,\spin_N}  |\spin_1\rangle \otimes  |\spin_2\rangle \otimes\cdots \otimes  |\spin_N\rangle\,,
    \label{eq:spin_wf}
\end{split}
\end{equation}
where the $|s\rangle = |\spin_1\rangle \otimes  |\spin_2\rangle \otimes \cdots \otimes |\spin_N\rangle$ are the basis vectors of the Hilbert space that describes the $N$ spin system, and $C_{\spin_1,\spin_2, \dots ,\spin_N}$ are their associated amplitudes.

The \stress{quantum many-body problem}\index{quantum many-body problem} originates from the exponential scaling of the number of the basis elements, which leads to an~exponential computational complexity in the system size.
In particular, the memory required to naively store the wave function of just 60 spins is $16 \cdot 2^{60} \approx 18$ exabytes, about 500 times more than what is available on the world's largest supercomputer as of 2022.

Nevertheless, while the Hilbert space of many-body quantum systems is exponentially large, physically -relevant states are typically confined to a~corner of the Hilbert space that is of limited dimension.
For instance, many physical Hamiltonians only contain {\stress{local}} interactions, which significantly constrains the form of the associated many-body wave functions.

\highlight{The main idea behind variational methods is to find a~computationally efficient representation of the physically relevant quantum states within the Hilbert space of interest.}
Variational methods circumvent the issue of an~exponential complexity by encoding the complex amplitudes of the wave function onto a~parametrized function (often called the \textit{ansatz})\index{ansatz}, which depends on a~set of parameters $\params$.
If the number of parameters is polynomial in the system size, the state can be efficiently stored with limited computational resources.
In general, the variational state\index{variational state} $|\Psi_{\params}\rangle$ can be expanded onto the computational basis as 
\begin{equation}
    |\Psi_{\params}\rangle = \sum_{s=1}^{2^N}\Psi_{\params}(\vect{s})|s\rangle,
    \label{eq:parametrized_wave_function}
\end{equation}
where $\Psi_{\params}(\vect{s})=\langle s|\Psi_{\params}\rangle$ denotes the probability amplitude corresponding to the state $|s\rangle$.
The task is then to find the parametrization $\params$ that best describes our desired quantum state of interest, such as the ground state of a~given Hamiltonian. 

\subsection{Variational methods}
\label{sec:NQS_variational_methods}

 Even when using variational states\index{variational state}, computing expectation values can still be of exponential complexity since one must perform sums over all the basis elements of the Hilbert space for these calculations.
Among the variational states that are practically usable, there are two possible approaches which distinguish two families of variational ansätze: those that can be used to compute expectation values exactly with a~polynomial cost, and those that do so only approximately, with an~accuracy improvable at a~polynomial cost in system size.
In the former, the only source of error in the expectation value of observables comes from the truncation of (exponentially large) regions of the Hilbert space, limiting its ability to represent wave functions.
In the latter, an~additional source of error typically comes from sampling, which, however, does not necessarily add a~systematic error and can be improved upon for an~additional computational cost. 

The third category consists of parameterized quantum states whose cost for computing expectation values scales exponentially with system size. In practical applications, for example, in the case of tensor networks in two dimensions, approximate algorithms for computing expectation values are introduced. Strictly speaking, however, these are not variational methods, as we cannot compute expectation values to arbitrary accuracy in polynomial time, and they introduce a~systematic bias that goes beyond the pure variational error. 

\subsubsection{Variational states with exact expectation values}
\label{sec:local_ansatz}

In the first kind of variational states\index{variational state}, we mainly encounter locally constrained ansätze\index{ansatz}, for which mean-field and matrix-product states are notable examples. 

\paragraph{Mean-field Ansatz} Mean-field states\index{ansatz!mean-field} are one of the simplest variational quantum states. With these, we model our variational wave function by the mean-field approximation, that is, as the tensor product of single-spin wave functions 
\begin{equation}
\begin{split}
    |\Psi_{\params}\rangle &= |\phi_1(\param_{\uparrow}^{(1)},\param_{\downarrow}^{(1)})\rangle\otimes|\phi_2(\param_{\uparrow}^{(2)}, \param_{\downarrow}^{(2)})\rangle\otimes\dots\otimes|\phi_N(\param_{\uparrow}^{(N)}, \param_{\downarrow}^{(N)})\rangle \\ &= \bigotimes_{i=1}^N |\phi_i(\param_{\uparrow}^{(i)}, \param_{\downarrow}^{(i)})\rangle\,,
    \label{eq:nqs_mean_field_ansatz}
\end{split}
\end{equation}
where $|\phi_i\rangle$ are the single-spin wave functions at site $i$. They are subject to the orthogonality condition $\langle\phi_i|\phi_j\rangle=\delta_{ij}$, with $\delta_{ij}$ denoting the Kronecker delta. This way, $|\phi_i\rangle$ has only two coefficients corresponding to the probability amplitudes of the spin being up or down, which we take as variational parameters
\begin{align}
    &|\phi_i\rangle = \param_{\uparrow}^{(i)}|\uparrow\rangle + \param_{\downarrow}^{(i)}|\downarrow\rangle \\
    &\left|\param_{\uparrow}^{(i)}\right|^2 + \left|\param_{\downarrow}^{(i)}\right|^2 = 1\,,
\end{align}
resulting into $2N$ complex parameters in total, i.e., $\params=\left\{\param_{\uparrow}^{(i)}, \param_{\downarrow}^{(i)}\big| i=1,2,\dots,N\right\}$. 

With this family of wave functions, we can compute expectation values of quantum Hamiltonians exactly. This is a~consequence of the fact that we can exploit the tensor product structure of our wave function to simplify the expectation values over many-body states to the expectation over the corresponding single-body ones. For example, the expectation value of the $\sigma_i^x$ Pauli operator acting on the $i$-th site can be obtained as $\langle\Psi_{\params}|\sigma_i^x|\Psi_{\params}\rangle = \langle\phi_i|\sigma_i^x|\phi_i\rangle$. The calculation is straightforward, as $|\phi_i\rangle$ is a~two-dimensional vector and $\sigma_i^x$ is a~$2\times2$ matrix. 

\paragraph{Tensor network states} However, mean-field states are not able to capture correlations between local degrees of freedom. \Acfp{TNS}\index{tensor networks} are a~family of quantum states that improve upon such a~limitation, and a~subset of \acp{TNS} also allow to compute expectations exactly. One of the most broadly used \acp{TNS} with this property are \ac{MPS}\index{tensor networks!matrix product states}, which predominate in the study of one-dimensional systems. 

Let us consider the coefficients $C_{\spin_1,\spin_2,...,\spin_n}$, defined in \cref{eq:spin_wf}. We can consider $C_{\spin_1,\spin_2,...,\spin_N}$ as a~tensor with $N$ indexes, which we can always express as the contraction of tensors $A^{\spin_i}$, such that:
\begin{equation}\label{eq:ansatz}
C_{\spin_1,...,\spin_N}= \sum_{\alpha , \beta, \, \dots \, , \gamma } A^{\spin_1}_{\alpha , \beta} A^{\spin_2}_{ \beta , \delta} \, ... \, A^{\spin_N}_{  \gamma , \alpha}
\end{equation}
where the maximal dimension of the Greek indices $\alpha , \beta, \, \dots$ is the bond dimension $\chi$. This way, an~exact representation of $C_{\spin_1,...,\spin_N}$ requires an~exponentially large number of parameters. This means that the bond dimension, $\chi$, must increase exponentially with $N$. The idea of the \ac{MPS} ansatz\index{ansatz} resides in the truncation of the dimension of the indices of the tensors $A^{\spin_i}$. With the truncation, we reduce the number of parameters of our ansatz to be $\bigO (dN\chi^2)$, where $d$~is the local Hilbert space dimension, e.g., two for a~spin-$1/2$ particle. We usually truncate the bond dimension in an~elegant and controlled way using the singular value decomposition of the tensors $A^{\spin_i}$, which has a~strict connection with the maximal entanglement entropy the \ac{MPS} can contain,\footnote{The bond dimension is, in fact, the rank of the Schmidt decomposition of the quantum state.} as we discuss in~\cref{sec:NQS_capacity_entanglement}.

As a~final remark, let us mention an~important algorithm proposed by S. White for the energy minimization of variational quantum states, known as density matrix renormalization group~\cite{White1992}. This algorithm is particularly well-suited for \acp{MPS}, and their combination is the current state-of-the-art technique to compute the ground state wave function of one-dimensional systems. However, the description of this algorithm falls out of the scope of this book. We refer to \cite{Schollwock2011} for a~complete review of the use of \ac{MPS} and to \cite{Orus2014} for a~review of methods based on \acp{TN}.

\subsubsection{Variational states with approximate expectation values}

The second family of variational states\index{variational state} we encounter are known as \stress{computationally tractable states}\index{ansatz!computationally tractable states} \cite{nest2010simulating}.
\highlight{Variational ans\"atze must satisfy two conditions to be computationally tractable:
\begin{itemize}
    \item Amplitudes for arbitrary single basis elements $\Psi_{\params}(\vect{s})=\langle s|\Psi_{\params}\rangle$ can be computed efficiently.
    \item It is possible to efficiently generate samples $s$ from the Born distribution $P(s)=\frac{|\langle s|\Psi_{\params}\rangle|^2}{\langle\Psi_{\params}|\Psi_{\params}\rangle}$.
\end{itemize}}

If these two conditions are met, we can efficiently estimate expectation values of arbitrary $k$-local operators, and the statistical error due to stochastic sampling can be rigorously controlled by increasing the number of samples.
Therefore, the computational time to compute expectation values is polynomial in both the system's size and accuracy.  

\highlight{A~$k$-local operator\index{k-local operator} is an~operator that contains terms acting on at most $k$ local quantum numbers at the same time. For instance, a~nearest-neighbour Hamiltonian is a~$2$-local Hamiltonian because it contains terms acting on 2 qubits.}

In general, given a~variational state\index{variational state} $\ket{\Psi_{\params}}$, we can obtain the expression for the expectation value of an~operator $\hat{O}$ as follows
\begin{align}
    \langle \hat{O} \rangle &= \frac{\langle \Psi_{\params}|\hat{O}|\Psi_{\params}\rangle}{\langle\Psi_{\params}|\Psi_{\params}\rangle} \\
    &= \frac{\sum_{s,s'}\langle\Psi_{\params}|s\rangle\langle s|\hat{O}|s'\rangle\langle s'|\Psi_{\params}\rangle}{\sum_s |\langle\Psi_{\params}|s\rangle|^2} \\
    &= \frac{\sum_s \langle\Psi_{\params}|s\rangle \frac{\langle s|\Psi_{\params}\rangle}{\langle s|\Psi_{\params}\rangle}\sum_{s'}\langle s|\hat{O}|s'\rangle\langle s'|\Psi_{\params}\rangle}{\sum_s |\langle\Psi_{\params}|s\rangle|^2} \\
    &= \frac{\sum_s |\langle\Psi_{\params}|s\rangle|^2 \sum_{s'}\langle s|\hat{O}|s'\rangle\frac{\langle s'|\Psi_{\params}\rangle}{\langle s|\Psi_{\params}\rangle}}{\sum_s |\langle\Psi_{\params}|s\rangle|^2}\,,
\end{align}
where we have added two identities of the form $\sum_s \ketbra{s}{s} = \id$ in the numerator, and one in the denominator. Then, we have multiplied by $\frac{\langle s|\Psi_{\params}\rangle}{\langle s|\Psi_{\params}\rangle}$ in the numerator.\footnote{Notice that this manipulation is always valid, since amplitudes with $\langle s|\Psi_{\params}\rangle=0$ never appear in the summation over $s$.}
We identify two main terms: 
\begin{align}
    P(\vect{s}) &= \frac{|\langle\Psi_{\params}|s\rangle|^2}{\sum_s |\langle\Psi_{\params}|s\rangle|^2} \\
    O_{\text{loc}}(\vect{s}) &= \sum_{s'}\langle s|\hat{O}|s'\rangle\frac{\langle s'|\Psi_{\params}\rangle}{\langle s|\Psi_{\params}\rangle}\,,
    \label{eq:NQS_Oloc}
\end{align}
where $O_{\text{loc}}(\vect{s})$ is the so-called \textit{local estimator} of $\hat{O}$.
Therefore, we can write the quantum expectation value of an~observable $\hat{O}$ as the statistical expectation value of its local estimator $O_{\text{loc}}$ over the probability distribution $P(\vect{s})$:
\begin{equation}
    \langle\hat{O}\rangle = \sum_s P(\vect{s})O_{\text{loc}}(\vect{s}) = \estimate{O_{\text{loc}}(\vect{s})}{P}\,.
\end{equation}
Let us stress that these calculations only hold for operators with the property that the number of states $\vect{s}'$ such that $|\langle s|\hat{O}|s'\rangle|\neq 0$, for arbitrary $\vect{s}$ is at most polynomial in the number of spins. For example, it is easy to convince oneself that $k$-local operators satisfy this property. Conversely, evaluating $O_\text{loc}(\vect{s})$ would not be tractable, given that the sum over $s'$ in~\cref{eq:NQS_Oloc} would be over an~exponential number of elements.

The procedure described above allows to obtain a~controlled, stochastic estimate of the expectation values by directly sampling a~series of states, $\vect{s}^{(1)}, \vect{s}^{(2)}, \dots, \vect{s}^{(M)}$, from $P(\vect{s})$, and approximating $\langle\hat{O}\rangle$ with the following arithmetic mean
\begin{equation}
    \langle\hat{O}\rangle \approx \frac{1}{M}\sum_{i=1}^M O_{\text{loc}}(\vect{s}^{(i)})\,.
\end{equation}
The statistical error associated with such an~estimate is $\varepsilon = \sqrt{\sigma^2/M}$, and it is bounded as long as the variance $\sigma^2$ of $O_{\text{loc}}$ is finite. For example, when $\hat{O}$ is a~k-local spin operator with bounded coefficients, its variance is strictly finite since it can be shown that $\sigma^2=\langle\hat{O}^2\rangle - {\langle\hat{O}\rangle}^2$. \footnote{It is also simple to prove that, when $|\psi_\theta\rangle$ approaches an~eigenstate of $\hat{O}$, the variance vanishes. 
Consequently, considering $\hat{O} = \hat{H}$, the statistical error vanishes as we approach the ground (or any excited) state.}
Therefore, the error in the estimate of expectation values decreases as $\varepsilon\sim 1/\sqrt{M}$, which allows us to reach arbitrary accuracy in the estimation by increasing the number of samples $M$, given that $\lim_{M\to\infty}\varepsilon = 0$.
However, generating a~set of samples according to the Born distribution, $\{\vect{s}^{(i)}\}\sim P(\vect{s})$, is in general a~non-trivial computational task in the case where the variational ansatz, $\Psi_{\params}(\vect{s})$, is parameterized by an~efficiently computable, yet arbitrary function. One of the most commonly adopted strategies to sample from $P(\vect{s})$ is through \stress{\acf{MCMC}}\index{Markov chain Monte Carlo} methods, including the Metropolis-Hastings method, which generate a~sequence of correctly distributed samples $\vect{s}^{(i)}$.

Metropolis-Hastings methods construct a~markovian stochastic process which satisfies the \stress{detailed balance} relation for the target probability distribution
\begin{equation}
    P(\vect{s})\mathcal{T}(\vect{s}\rightarrow \vect{s}') = P(\vect{s}')\mathcal{T}(\vect{s}'\rightarrow \vect{s})\,,
\end{equation}
where $\mathcal{T}(\vect{s}^{(i)}\rightarrow \vect{s}^{(i+1)})$ is the probability that the state $\vect{s}^{(i)}$ at step $i$ transitions to the state $\vect{s}^{(i+1)}$ at the following step. 
As the process is Markovian, the transition probability at every step depends exclusively on the current configuration.
The detailed balance condition ensures that regardless of the initial configuration $\vect{s}^{(0)}$, the sequence eventually converges to the correct distribution $P(\vect{s})$ in the long time limit.

One possible choice of the transition probability $\mathcal{T}$ is given by the \stress{Metropolis-Hastings algorithm}~\cite{1970Bimka..57...97H}. 
The main idea is to express $\mathcal{T}$ in terms of a~local transition kernel $T$ and an~acceptance probability $A$ such that
\begin{equation}
    \mathcal{T}(\vect{s}\rightarrow \vect{s}') = T(\vect{s}\rightarrow \vect{s}')A(\vect{s}\rightarrow \vect{s}')\,.
\end{equation}
This way, we split the global stochastic process into the product of two local subprocesses that we can compute efficiently.
For instance, it is very easy to find a~normalized local transition kernel that allows us to modify only a~few degrees of freedom, like flipping a~single spin in a~given configuration.
Conversely, it is hard to find a~normalized global kernel that would act on all spins.

The acceptance probability to go from a~configuration $\vect{s}$ to $\vect{s}'$ through a~local transition is defined as
\begin{equation}
    A(\vect{s}\rightarrow \vect{s}') = \min\left(1, \frac{P(\vect{s}')T(\vect{s}'\rightarrow \vect{s})}{P(\vect{s})T(\vect{s}\rightarrow \vect{s}')}\right)\,.
\end{equation}
Notice that the normalization of the Born probabilities cancels out, giving the expression
\begin{equation}
    \frac{P(\vect{s}')}{P(\vect{s})} = \left|\frac{\langle s'\mid\Psi_{\params}\rangle}{\langle s\mid\Psi_{\params}\rangle}\right|^2\,,
\end{equation}
which allows us to consider unnormalized variational ans\"atze.
Additionally, if the variational state\index{variational state} is computationally tractable, the transition probability also has a~tractable complexity, provided it only acts on the basis elements.

Choosing a~valid transition rule $T(\vect{s}\rightarrow \vect{s}')$ is not trivial, and we must take special care in the case of systems with symmetries.
For example, if the total magnetization along the direction of the computational basis is known, we might want to fix it and use a~transition rule that does not project the Markov chain outside of a~certain region.
In general, a~computationally expensive yet effective choice for the transition kernel is to use the Hamiltonian itself:
\begin{equation}
    T(\vect{s}\rightarrow \vect{s}') = \frac{\abs{\langle s| \hat{H}|s'\rangle}(1-\delta_{\vect{s},\vect{s}'})}{\sum_{'\neq s} \abs{\langle s| \hat{H}|s'\rangle}}\,,
\end{equation}
which is known as the Hamiltonian transition rule \cite{Carleo_2017}.

\highlight{This way, with the Metropolis-Hastings algorithm, starting from a~random configuration $\vect{s}^{(0)}$, we can sample from $P(\vect{s})$ by iteratively proposing local modifications $\vect{s}'$ according to $T(\vect{s}\rightarrow \vect{s}')$, and accepting them according to $A(\vect{s}\rightarrow \vect{s}')$.} 

Nonetheless, this sampling procedure is imperfect, and it can fail to converge for a~reasonable number of iterations if the sampled distribution is too complex. In addition, the procedure suffers from the fact that the samples are correlated since we flip spins iteratively. See \cref{alg:metropolis_hastings} for further details. 

\begin{algorithm}
\caption{Metropolis-Hastings algorithm}\label{alg:metropolis_hastings}
\begin{algorithmic}
\State $\vect{s} \gets$ uniform$\in[1,2^N]$ \Comment{sample initial state uniformly at random}
\For{i = 1 to M}
    \State propose $\vect{s}'$ according to $T(\vect{s}\rightarrow \vect{s}')$
    \State $A \gets \frac{P(\vect{s}')T(\vect{s}'\rightarrow s)}{P(\vect{s})T(\vect{s}\rightarrow \vect{s}')}$ \Comment{calculate acceptance probability}
    \State $\xi\gets$ uniform$\in[0, 1]$
    \If{$\xi \leq A$}
        \State $\vect{s} \gets \vect{s}'$ \Comment{update state}
    \EndIf
\EndFor
\end{algorithmic}
\end{algorithm}

\subsection{Representing the wave function}
\label{sec:NQS_representing_wave_function}

Now that we have seen how to compute the quantities of interest using parametrized quantum states, let us dive into how to devise expressive variational states\index{variational state} in practice. The main idea is that we need to represent high-dimensional functions with a~parametrization that is flexible and general enough to describe physical systems while involving only a~polynomial amount of parameters.

Traditionally, researchers have relied on physically-inspired variational ans\"atze.
The Jastrow wave function~\cite{Jastrow1955PR,Manousakis1991RMP} stands out as one of the most successful and widely used ones.
It is based on the assumption that two-body interactions are the most physically relevant, and it assigns a~trainable potential to every interacting pair. 
Formally,
\begin{equation}
    \Psi_{\params}(\vect{s}) = e^{-\frac{1}{2}\sum_{i\neq j}\param_{ij}\spin_i\spin_j}\,,
    \label{eq:NQS_jastrow_ansatz}
\end{equation}
where the sum runs over all possible spin pairs, and $\param_{ij}$ are the parameters encoding pairwise spin correlations. Therefore, for a~system of $N$ spins, the resulting wave function has $\mathcal{O}(N^2)$ parameters. Moreover, in translationally invariant systems, the parameters $\param_{ij}$ can be made depend exclusively on the distance between $i$ and $j$, resulting in a~reduced number, $\mathcal{O}(N)$, of parameters.

The \acp{ANN} have taken over more traditional ans\"atze to approximate the wave function itself~\cite{Carleo_2017}.
This family of variational states\index{variational state} is known as \acfp{NQS}\index{neural quantum states}.
For instance, we can write a~parametrized wave function as a~feed-forward \ac{NN}.
In this case, $\Psi_{\params}(\vect{s})$ corresponds to the output of a~\ac{NN} that takes the configuration $\vect{s}$ as input in the form of a~vector.

In a~feed-forward neural network of depth $D$, every layer $l$ consists of a~nonlinear activation function $g^{(l)}$ that acts, component-wise, on a~vector resulting from applying the weight matrix $\mat{W}^{(l)}$ to the output of the previous layer.
This way, it is possible to write the variational state\index{variational state} as the composition of operations $g^{(l)}\cdot\mat{W}^{(l)}$, where ``$\cdot$'' indicates point-wise operation, such that 
\begin{equation}
    \Psi_{\params}(\vect{s}) = g^{(D)}\cdot\mat{W}^{(D)}\dots g^{(2)}\cdot\mat{W}^{(2)}g^{(1)}\cdot\mat{W}^{(1)}\vect{s}\,.
\end{equation}
Hence, the output is a~scalar, complex or real, representing the probability amplitude of configuration $\vect{s}$.  

From a~mathematical perspective, these ansätze are of great interest given that \acp{NN} are subject to universal representation theorems~\cite{kolmogorov1957}, as we explain in~\cref{sec:NNs}.
According to~\cref{eq:universal_approx}, we could represent the many-body wave function with a~polynomial number $\mathcal{O}(N^2)$ of one-dimensional nonlinear functions, with $N$ denoting the number of spins. 

However, these results hold for arbitrary nonlinear functions, $\zeta_{i},\,\varsigma_{i,j}$ in~\cref{eq:universal_approx}, that must be appropriately found in order to represent the target function. 
In practice, \acp{NN} use a~fixed nonlinear activation, and we can only adjust the number of operations.
In these cases, the number of neurons does not have a~strict polynomial scaling, and it can be, in the worst case, exponential in $N$~\cite{cybenko1989}.
Nevertheless, the state-of-the-art results in computer vision and natural language processing~\cite{GPT32020NeurIPS,Caron2021DINO,Nichol2021glide} should be sufficient motivation to employ similar techniques to represent quantum states.
Note that the \ac{NN} representation of quantum states does not preserve the Hilbert space structure, which means that for two NN representations $\ket{\psi_1}$ and $\ket{\psi_2}$ it is not possible to construct a~valid wave function $\ket{\psi} = \ket{\psi_1} + \ket{\psi_2}$ represented by a~\ac{NN} of the same size as the ones representing $\ket{\psi_1}$ and $ \ket{\psi_2}$ by simply adding up the parameters together, as the ansatz is generally nonlinear.

\subsubsection{Restricted Boltzmann machines}
\label{sss:rbm}

\begin{figure}
    \centering
    \includegraphics[width=0.7\textwidth]{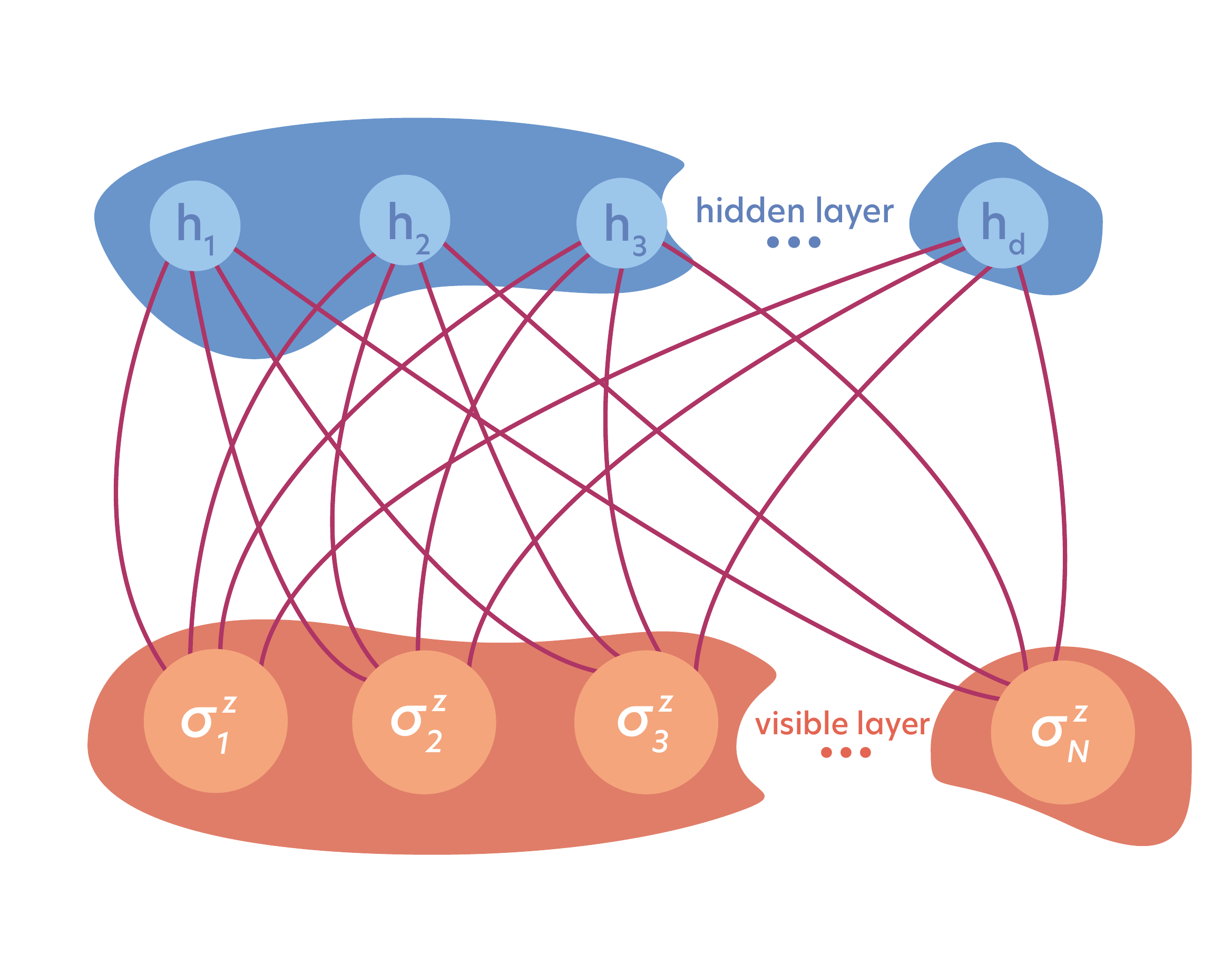}
    \caption[Scheme of a~restricted Boltzmann machine]{Pictorial representation of a~\acf{RBM} that represents the wave function of an~$N$-spin system, with $\vect{s} = (\spin_1,\spin_2,\dots,\spin_N)$ and $\vect{h} = (h_1,h_2,\dots,h_d)$ the hidden units.}
    \label{fig:RBM}
\end{figure}

\Ac{NQS} were first introduced using \stress{\acp{RBM}}~\cite{Carleo_2017}\index{restricted Boltzmann machine}. \Acp{RBM} are shallow models featuring two fully-connected layers: a~\stress{visible} layer, consisting of $N$ units, and a~\stress{hidden} layer, consisting of $M$ units. A~scheme of an~\ac{RBM} architecture is presented in \cref{fig:RBM}.
The wave function amplitudes of an~\ac{RBM} ansatz are given by:
\begin{equation}
    \Psi_{\params} (\vect{s}) = \sum_{\vect{h}}e^{\biases_v^\dagger\vect{s} + \biases_h^\dagger\vect{h} + \vect{h}^\dagger\mat{W}\vect{s}}.
    \label{eq:psi_rbm_1}
\end{equation}
where $\vect{s},\vect{h}$ represent the visible and hidden units, respectively, and the parameters $\params = \{\biases_v,\biases_h,\mat{W}\}$ represent the visible and hidden biases and the weight matrix, respectively. In the \ac{NN} picture, the \ac{RBM} is a~single-layer nonlinear feed-forward \ac{NN}, with the visible units serving as inputs and the exponential serving as the activation function. While it is common to have biases for the hidden layer (see \cref{sec:NNs}), \acp{RBM} also have somewhat unusual biases connected to the input values, which is explained in the next paragraph. 

By construction, \ac{RBM}s are designed in such a~way that computing the summation over hidden units, as in \cref{eq:psi_rbm_1}, can be done analytically. To see this, we can rewrite \cref{eq:psi_rbm_1} in a~tractable form considering binary hidden units $h_i\in\{-1, 1\}$, leading to 
\begin{equation}
    \Psi_{\params}(\vect{s}) = e^{\biases_v^\dagger \vect{s}}\prod_{i=1}^{M}2\cosh\left(\biases_{h,i} + \mat{W}_{i\cdot}\vect{s}\right),
    \label{eq:NQS_rbm_amp}
\end{equation}
where $\biases_{h,i}$ and $\mat{W}_{i\cdot}$ denote the $i$-th hidden bias and weight matrix row, respectively.
To treat spin systems, the visible units will represent the $N$ physical spins. Thus the input of the \ac{RBM} is simply the spin configuration $\vect{s}$.
In this way, we obtain an~analytical expression to evaluate the amplitude for a~given spin configuration, and thus represent the full wave function with this ansatz.
One can also interpret the hidden units as $M$ hidden spins, and in this picture, the \ac{RBM} can be thought of as an~interacting spin model with interaction strengths $\mat{W}_{ij}$.
Moreover, we can treat an~RBM as a~model with an~associated energy depending on its parameters, input, and hidden spin values. This is known as an~energy-based model and explains why input biases are present in~\cref{eq:psi_rbm_1}.
In fact, the \ac{RBM} is equivalent to a~Hopfield network, a~type of spin glass~\cite{Barra_2011}. For more details on this view, see~\cite{Montufar_2018}.

Being the first to be introduced in this context, most of the early works about \ac{NQS} employ \acp{RBM}, but other architectures have been systematically explored in more recent years. The capacity of \acp{RBM} and its relationship to quantum entanglement has been examined in various works \cite{Deng2017, Chen2018PRB}.
An~extension of this architecture, the deep \acp{RBM}, has also been introduced to solve more complex problems \cite{Gao2017}, which consists of stacking more than two fully connected layers.

\subsubsection{Autoregressive and recurrent neural networks}\label{sss:arnn_nqs}

\begin{figure}[t]
    \centering
    \includegraphics[width=\columnwidth]{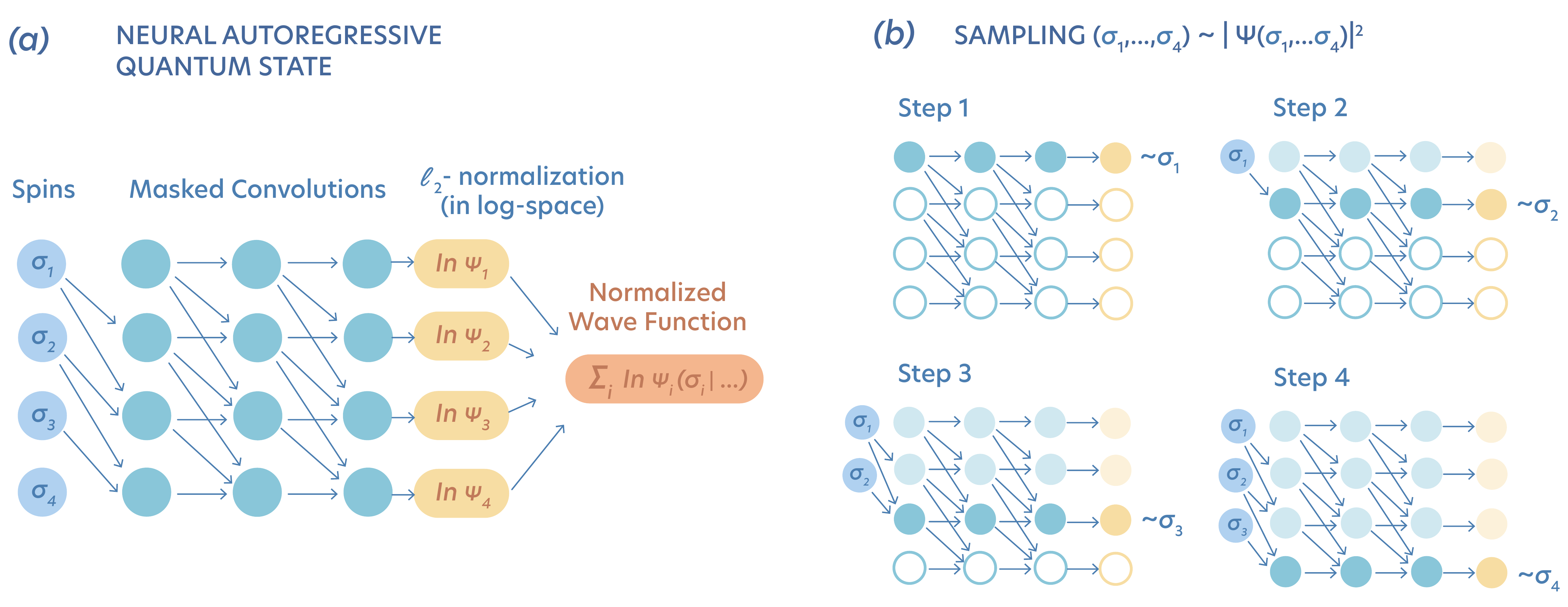}
    \caption[Autoregressive neural quantum state]{Example of an~\ac{ARNN}\index{autoregressive neural network} quantum state for four spins. (a) Pictorial representation of the network. The arrows representing the weights of the model are skewed in order not to break the conditional structure of the output probability distribution. These layers are ``masked'', due to some connections being deleted. (b) Sampling algorithm. One samples consecutive spins using direct sampling on the conditional probabilities at each step. Adapted from \ToggleForCUP{Sharir, O. \textit{et al.} (2020). \textit{Deep autoregressive models for the efficient variational simulation of many-body quantum systems}. Phys. Rev. Lett. Phys. Rev. Lett. 124, 02050~\cite{sharir2020}.}{Ref.~\cite{sharir2020}.}}
    \label{fig:AR_Carleo}
\end{figure}

\Acp{ARNN}\index{autoregressive neural network}, as presented in~\cref{sec:autoregressive_NNs}, can also be used for constructing \ac{NQS}, as introduced in Ref.~\cite{sharir2020} and later applied to both quantum~\cite{Luo_2021} and classical problems~\cite{wu2019solving}. 
Their main advantage is that their Born probability distribution is normalized, allowing for direct (autoregressive) sampling, which is easier to parallelize than \acf{MCMC}\index{Markov chain Monte Carlo}.
A~pictorial representation of both the network and the sampling algorithm is presented in \cref{fig:AR_Carleo}. 

Analogously to~\cref{eq:arnn_function}, we express the many-body wave function in terms of a~product of conditional complex amplitudes:
\begin{equation}
    \Psi_{\params}(\vect{s}) = \prod_{i=1}^N \phi_i(\spin_i\mid\spin_{i-1},\ldots,\spin_{1})\,,
\end{equation}
which is subject to the normalization condition $\sum_{\spin}\big|\phi_i(\spin\mid\spin_{i-1},\dots,\spin_1)\big|^2=1$.
With this architecture, we can compute expectation values by directly sampling state configurations instead of building a~Markov chain through the Metropolis-Hastings algorithm, for example (see~\cref{alg:metropolis_hastings}).
We sample state configurations by iteratively sampling one spin after the other: we start sampling the first spin $\spin_1$ from the reduced probability distribution $|\phi(\spin_1)|^2$.
Then, we sample the second one $\spin_2$ according to the conditional probability distribution $|\phi_2(\spin_2\mid\spin_1)|^2$, then the next one $|\phi(\spin_3\mid\spin_2,\spin_1)|^2$, and so on until $\spin_N$. This sampling procedure is embarrassingly parallel.\footnote{We can use the intermediate conditional probabilities to draw samples for a~low computational cost, e.g., use the probabilities for $N-1$ spins and sample from the last one, to obtain new samples; with \ac{MCMC} we cannot do this.}

This sampling procedure yields independent, identically distributed samples. Conversely, \ac{MCMC} methods may suffer from highly correlated consecutive samples,\footnote{\Acf{MCMC} methods such as the Metropolis-Hastings algorithm generally rely on performing modifications to the spin configurations to sample subsequent states. Therefore, this process could yield highly correlated consecutive samples that may have a~negative impact on the results. In order to compute expectation values, we need to estimate the autocorrelation time to draw uncorrelated samples from the resulting chain. Moreover, when approaching a~phase transition points, such methods suffer from critical slowing down, making the sampling of uncorrelated configurations unfeasible in many situations. } which is problematic for complex probability distributions, e.g., that are far from Gaussian.
Consider a~quantum state that spans several separated regions in the Hilbert space, where the probability is concentrated.
In this case, Markov chains generally remain stuck in one of the regions, given that it must take several penalizing steps to travel from one to another, resulting in a~highly inaccurate sampling. In contrast, the direct sampling procedure can seamlessly draw spin configurations belonging to all the regions according to the probability distribution, yielding much better samples.

\begin{figure}
    \centering
    \includegraphics[width=0.98\columnwidth]{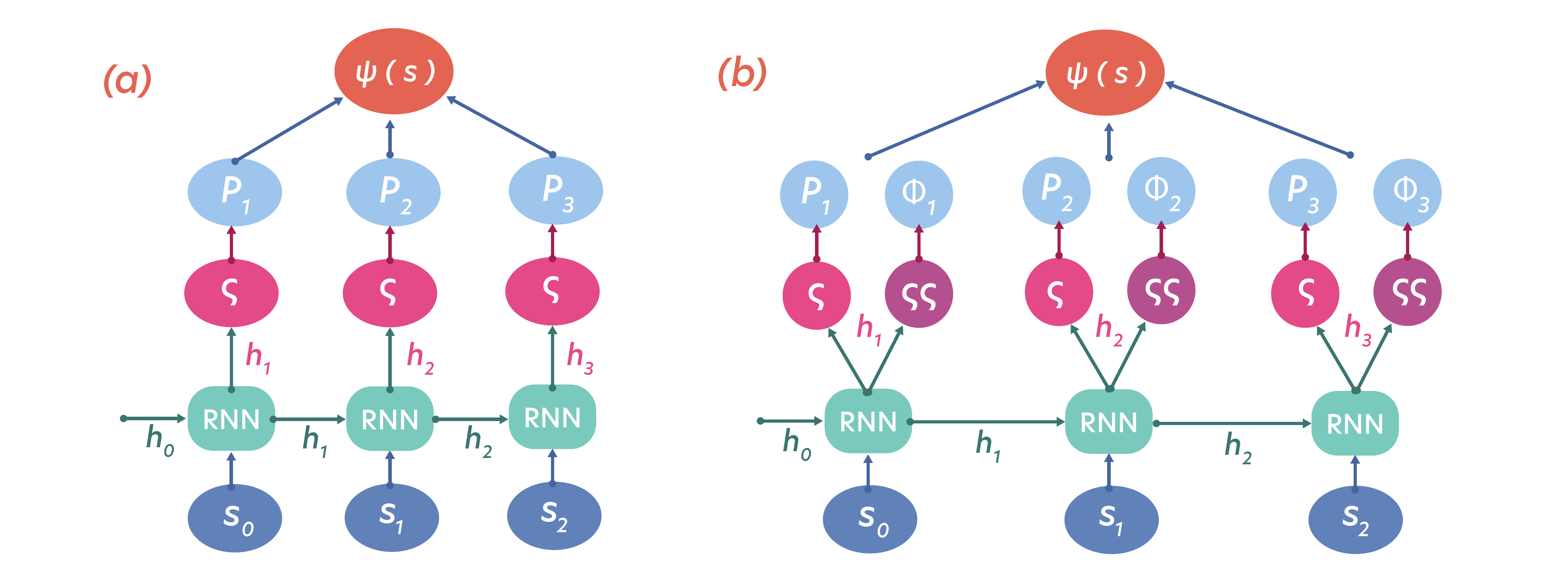}
    \caption[Recurrent neural-network architecture as a~neural quantum state]{Pictorial representation of an~\ac{RNN} architecture for \ac{NQS}. Panel (a) is for real-valued wave functions, which can be relevant for a~certain class of problems, and panel (b) is for complex-valued wave functions. In both schemes, a~local spin configuration $s_i$ and a~hidden vector $h_i$ are fed into an~\ac{RNN} cell, which performs a~nonlinear transformation. Then an~activation function ($\varsigma$, for softmax and/or $\varsigma\varsigma$, for softsign) is applied to obtain the final probability and/or phase corresponding to the configuration. In the end, the probabilities (and phases) are combined to obtain the final wave function amplitudes $\psi(\vect{s})$.}
    \label{fig:NQS_rnn_scheme}
\end{figure}

While the first autoregressive models\index{autoregressive models} used in quantum physics were built from masked dense or convolutional layers, mimicking the so-called PixelNet architecture \cite{PixelNet}, recurrent neural networks\index{recurrent neural network} were later introduced~\cite{hibat2020recurrent}
\acp{RNN}, inspired by natural language processing models, are also generative models\index{generative models}. We can draw a~simple analogy between correlations in sentences, with their elements living in a~large ``word space'', and spin configurations.
Considering spin systems and supposing some hidden structure, quantum states are correlated, and their base elements are elements of the Hilbert space. Following this analogy, Hibat-Allah \textit{et al.} introduced \ac{RNN} wave functions \cite{hibat2020recurrent}, obtaining impressive results even for frustrated systems. An~example of such an~architecture is shown in~\cref{fig:NQS_rnn_scheme}.
Clearly, many different \ac{NN} architectures can work.
A~plethora of different architectures have been implemented as \ac{NQS} in recent years, such as \acp{CNN} \cite{Schmitt_2020}, and group \acp{CNN}~\cite{roth2021group}, which can conveniently implement certain symmetries, as we describe in more detail in~\cref{sec:NQS_symmetries}.

\subsubsection{Capacity and entanglement}
\label{sec:NQS_capacity_entanglement}

As we show in~\cref{sec:NQS_variational_methods,sec:NQS_representing_wave_function}, there is a~whole plethora of methods to represent quantum many-body wave functions.
For instance, only in \acp{NQS}, we already encounter substantial differences between ans\"azte based on different \ac{NN} architectures. Hence, a~natural question arises regarding their expressive capacity and how they compare to each other.

\begin{figure}[t]
    \centering
    \includegraphics[width=0.7\columnwidth]{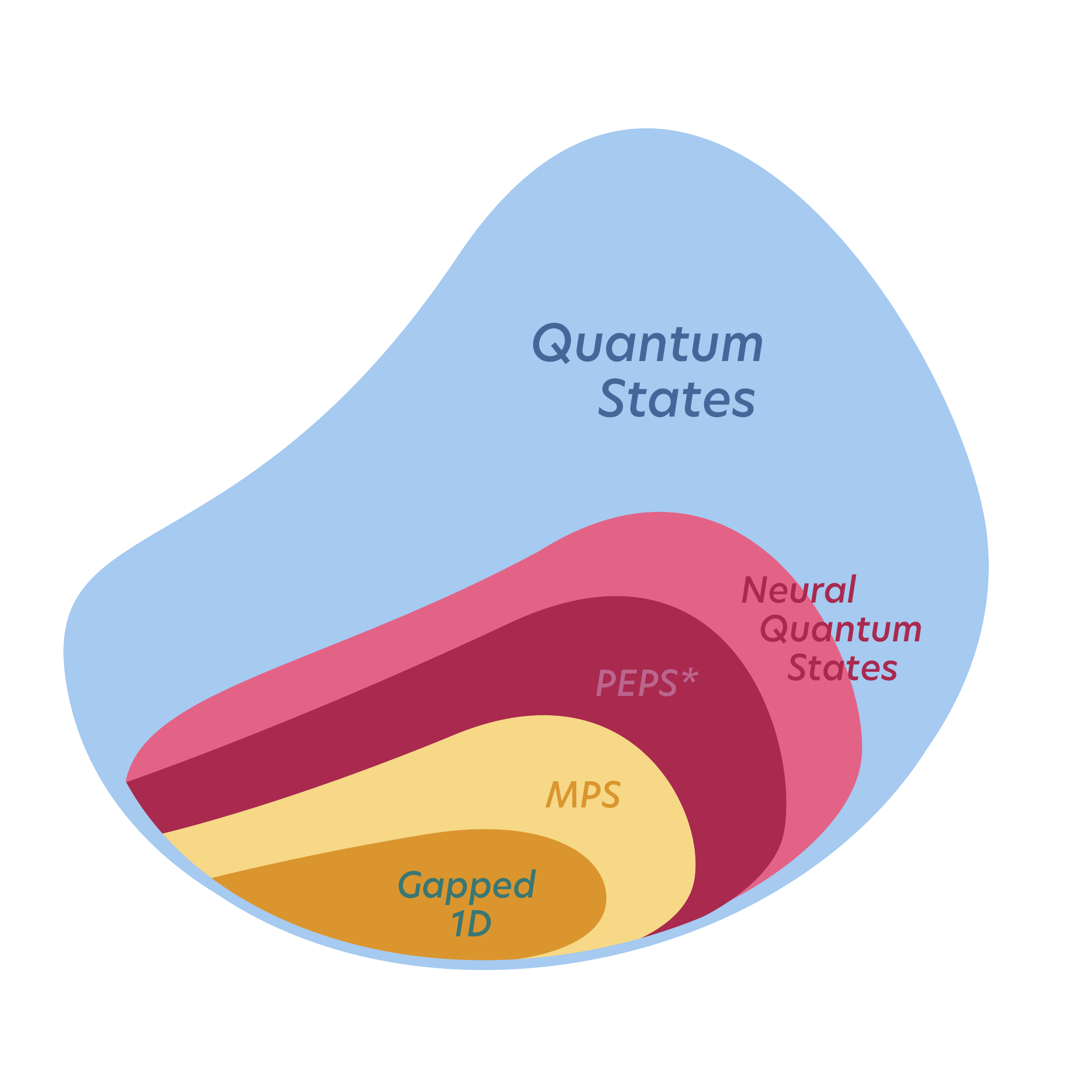}
    \caption[Expressive capacity of neural quantum states]{Expressive capacity of different classes of variational states, as explicitly proven in Ref.~\cite{sharir2021} by mapping \acp{TNS} to \ac{NQS}. PEPS* refers to a~sub-class of projected entangled pair states\index{tensor networks!projected entangled pair states}, a~generalization of \acp{MPS}. Adapted from \ToggleForCUP{Sharir, O., Shashua, A., \& Carleo, G. (2022). \textit{Neural tensor contractions and the expressive power of deep neural quantum states}. Phys. Rev. B 106, 205136~\cite{sharir2021}.}{Ref.~\cite{sharir2021}.}}
    \label{fig:NQS_exp_power}
\end{figure}

\Acp{TN} have been a~recurrent tool to perform this kind of studies, provided that they are well established and characterized, and they constitute a~theoretical language to study quantum many-body phenomena. For this reason, there has been a~significant community effort to study the relationship between \acp{TNS} and \acp{NQS}~\cite{Glasser2018PRX,Chen2018PRB,sharir2021}
, which provides insight about the expressive capacity of \acp{NQS}~\cite{levine2019}. Following the first introduction of \acp{NQS} implementing \acp{RBM}~\cite{Carleo_2017}, early works focused on finding direct relationships between various kinds of \ac{RBM}-based states and \acp{TNS}~\cite{Glasser2018PRX,Chen2018PRB}. It has been proven that \acp{NN} can efficiently approximate, in logarithmic space-complexity, all efficiently contractible \acp{TN} with arbitrary precision. Therefore, for every \ac{TNS} there exists an~equivalent \ac{NQS} of polynomial size. Conversely, there are quantum states that can be efficiently described by \acp{NQS}, whose representation in terms of \acp{TNS} requires an~exponential amount of parameters. Hence, \acp{TNS} are a~subset of \acp{NQS}~\cite{sharir2021}, as depicted in~\cref{fig:NQS_exp_power}.

As a~measure of expressive capacity, we often rely on the entanglement that the different ans\"atze can capture. For instance, the mean field ansatz\index{ansatz} is, by construction, a~product state (recall~\cref{eq:nqs_mean_field_ansatz}). Hence, it cannot capture entanglement, while \acp{TNS} and \acp{NQS} do not have such strong local limitations. This way, \acp{TNS} and \acp{NQS} have higher expressive capacity than the mean field ones.

More precisely, we study the entanglement scaling captured by the different ans\"atze. In a~generic quantum many-body system with density matrix $\rho$, the entanglement entropy is defined as
\begin{equation}
\label{eq:entanglement_entropy}
S(\rho)=-\Tr\left[\rho \log_2{\rho} \right]\,,
\end{equation} 
which is zero for any pure state. Let us consider a~partition of the system in two subsets: $I$~and its complementary $O$, as well as the reduced density matrix $\rho_I=\Tr_O[\rho]$. In general, $\rho_I$ represents a~mixed state, which can have nonzero von Neumann entanglement entropy. For a~generic quantum state, the entanglement entropy of $\rho_I$ grows with the volume of the cut. Thus, it corresponds to a~\stress{volume-law} scaling. \Acp{NQS} can efficiently capture such scaling with architectures ranging from very basic shallow ones, such as \acp{RBM}~\cite{Deng2017}, to more modern and deeper approaches, such as \acp{CNN} or \acp{RNN}~\cite{levine2019}. Some traditional ans\"atze, such as the Jastrow wave function (see~\cref{eq:NQS_jastrow_ansatz}), can also capture volume-law entanglement.\footnote{The Jastrow ansatz is, indeed, a~specific case of \ac{RBM} wave function with $N(N-1)/2$ hidden neurons~\cite{Glasser2018PRX}.}

However, there is a~subclass of states in which the entanglement entropy grows, at most, as the boundary area between two regions. This is known as \stress{area-law} scaling, and it is a~property of ground states of local and gapped Hamiltonians~\cite{Calabrese2004entanglement}. Due to their local nature, \acp{TNS} can efficiently capture area-law entanglement~\cite{RevModPhys.82.277}. For instance, in a~one-dimensional chain, the area of the cut between two subsystems is constant, meaning that the entanglement entropy is a~constant, and not an~extensive quantity in the infinite volume limit. For an~\ac{MPS} with bond dimension $\chi$, the von Neumann entanglement entropy of any possible bi-partition of the system is bounded from above as $ S \leq \mathcal{O}(\log_2{\chi})$, thus making the \ac{MPS} ansatz\index{ansatz} an~excellent candidate to study one-dimensional systems.

\begin{figure}[t]
    \centering
    \includegraphics[width=0.7\columnwidth]{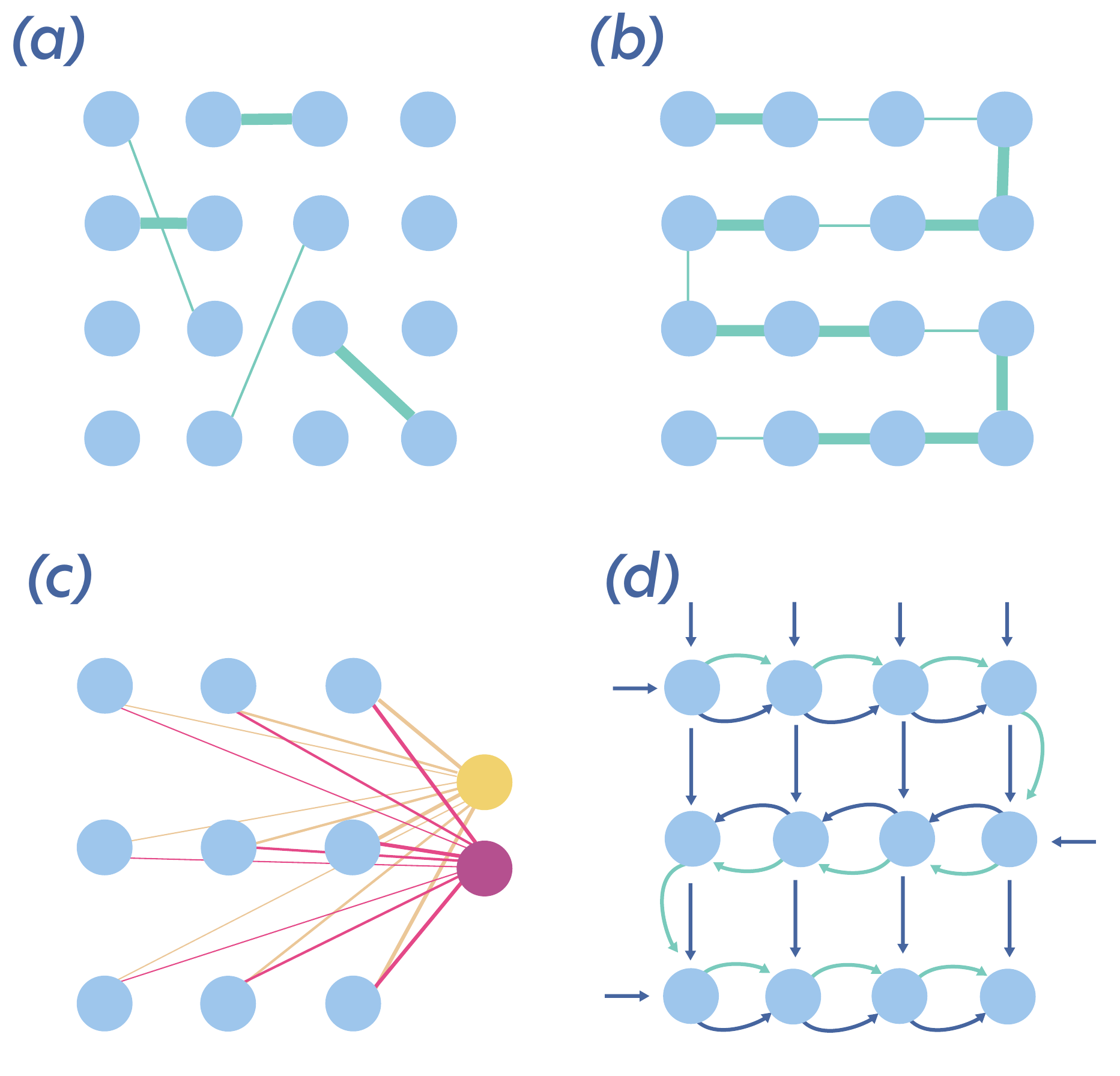}
    \caption[Schematic representation of the of various ans\"atze]{Schematic representation of various ans\"atze inspired by Refs.~\cite{Glasser2018PRX, hibatallah2021variational}. (a) The Jastrow ansatz draws connections between all possible pairs of sites.  (b) The \ac{MPS} ansatz draws connections between nearest-neighbor sites along a~line. (c) The \ac{RBM} ansatz connects all the sites to every hidden neuron, illustrated in different colors. (d) The \ac{RNN} ansatz processes the state sequentially, following the green arrows. The dark blue arrows indicate the flow of information within the model. Arrows without a~starting site correspond to free parameters.}
    \label{fig:NQS_schemes}
\end{figure}

We can understand most differences between the ans\"atze at an~intuitive level by, simply, looking at how they are built. In~\cref{fig:NQS_schemes}, we provide a~pictorial representation of the different connections that some ans\"atze can draw in a~bi-dimensional system. Clearly, the \ac{MPS} ansatz, depicted in~\cref{fig:NQS_schemes}(b), is the most locally restricted one, as it can only account for nearest neighbor connections in a~snake-like pattern. This effectively limits the entanglement that \ac{MPS} can capture. The \ac{RNN} ansatz\index{ansatz}, illustrated in~\cref{fig:NQS_schemes}(d), while it is limited to parse the state in the same pattern as the \ac{MPS}, it has the freedom to account for additional information, allowing it to capture richer correlations.

In contrast, other ans\"atze such as the Jastrow or \ac{RBM} wave functions, respectively illustrated in~\cref{fig:NQS_schemes}(a) and (c), can draw connections between arbitrary sites. The Jastrow ansatz can account for all possible pairs in the system, regardless of the distance. Then, the \ac{RBM} ansatz is a~generalization of the Jastrow by means of an~auxiliary hidden layer of variable size. Through the hidden neurons, the ansatz is no longer limited to pairs, and it can actually consider up to all-to-all connections. This non-local character allows them to capture volume-law entanglement. 

\subsubsection{Implementing symmetries}
\label{sec:NQS_symmetries}

Encoding symmetries in \acp{NQS} allows us to reduce the number of parameters in the \ac{NN}, restricting the region of the Hilbert space that our ansatz can cover to a~subspace of interest, thus improving the accuracy of the results. Let us first explain what we mean by symmetry in this context.
Consider a~group of linear transformations: if the Hamiltonian is invariant under those transformations, meaning that they all commute with the Hamiltonian, then the Hamiltonian is symmetric under that group. Some of the most common symmetries in lattice models are the translation symmetry, the rotation symmetry in two or higher dimensions, the inversion or reflection symmetries, and all the compositions of those.

It is possible to show that if the Hamiltonian commutes with a~set of operators $\mathcal{T} = \{\hat{T}_k\}_{k=1}^K$, its ground state must also be left invariant under those transformations.
Therefore, the amplitude for two configurations $\ket{s}$ and $\ket{s(k)} = \hat{T}_k\ket{s}$ must be invariant for any $\hat{T}_k$: $\Psi_{\params}(\vect{s})=\Psi_{\params}(\hat{T}_k\vect{s})\, \forall k$.\footnote{Up to a~phase on the right-hand side, but let us ignore it for convenience.} One way to introduce symmetries in our \ac{NQS} is to take, as output, the sum of the ansatz evaluated on the set of symmetry-invariant configurations $\{\vect{s}(k)\}$. This way, the output is invariant by construction. However, we have not improved the performance of our model with this approach.

A~more efficient approach is to build a~dense layer at the beginning of the \ac{NQS} model that fulfills the symmetry condition \cite{Schmitt_2020}.
We can use this technique to encode any symmetry group isomorphic to a~polynomially large permutation group.
This usually comprises the set of all \emph{lattice symmetries} (translations, rotations, reflections...), global discrete symmetries, such as a~global spin-flip, but it cannot deal with continuous symmetries, such as $\text{SU}(2)$. 
For instance, we can implement translation symmetries through a~convolution with a~kernel as wide as the system itself.
Since the convolution is translationally invariant by definition, it's easy to see that the output of the layer is symmetry-invariant.

In the case of \acp{RBM}, we can rearrange the terms of~\cref{eq:psi_rbm_1} to make it invariant under the elements of a~symmetry group. Let us denote the transformation of local spins as $\spin_j(k) = \hat{T}_k \spin_j$. We can write our symmetry-invariant amplitude as: 
\begin{align}
    \nonumber \Psi_{\params} (\mathbf{s}) = \sum_{\mathbf{h}} \exp\Bigg( \sum_{f=1}^{\alpha} b_{v}^{f} \sum_{k=1}^{K} \sum_{j=1}^{N} s_{j}(k) &+ \sum_{f=1}^{\alpha} b_{h}^{f} \sum_{k=1}^{K} h_{f,k}\\ &+ \sum_{f=1}^{\alpha} \sum_{k=1}^{K} h_{f,k} \sum_{j=1}^{N} W_{j}^{f} s_{j}(k) \Bigg),
\end{align}

where we have explicitly written the matrix products as sums. The important point here is that $\vect{b}^{f}_v, \vect{b}^{f}_h$ are now vectors in a~feature space with $f = 1, \ldots, \alpha$, and the matrix $\mat{W}^f$ is now of size $\alpha \times N$.\footnote{Note that this expression is equivalent to \cref{eq:psi_rbm_1} with $M = K\times \alpha$ hidden variables.} If we consider translational invariance, the corresponding symmetry group is made of $N$ translation operators. In this case, $\mat{W}^f$ can be seen as a~kernel acting over configurations to which we have applied the translation operators. 

There are, in fact, many ways to directly encode symmetries in \acp{NQS}. For more details, we refer to \cite{Choo2018} for general feedforward networks, or \cite{Schmitt_2020} for an~example with \acp{CNN}. 

\subsubsection{Limitations}

Similar to many \ac{ML} methods, \acp{NQS} suffer from an~interpretability problem, as we have discussed extensively in~\cref{sec:interpretability} for generic \ac{ML} approaches. However, there has been substantial progress since the seminal paper from Carleo \& Troyer~\cite{Carleo_2017}. 
For instance, a~recent work introduced an~interpratable \ac{RBM} ansatz, in which the authors add some correlation terms to the expression of the probability distribution given by \cref{eq:psi_rbm_1}. With this, one can look at the magnitude of the trained parameters to understand which correlations are more important for the given physical problem \cite{valenti2021correlationenhanced}.

Another route to gain further understanding of \acp{NQS} is through the mapping of \ac{NQS} architectures to other known ansätze, such as \acp{TNS}. By exploiting this idea, works have shown \Acp{NQS} to be capable of describing volume-law states, as opposed to \acp{TNS}, as we show in~\cref{sec:NQS_capacity_entanglement}.
In terms of expressive capacity, \acp{NQS} can efficiently represent the ground states of one-dimensional gapped Hamiltonians, all the \acp{TNS} that are efficiently contractable in classical computers, and volume-law states~\cite{sharir2021}. Furthermore, there have been found exact \ac{NQS} representations of several interesting phases of matter, such as topological states and stability codes~\cite{Deng2017,Gao2017,Glasser2018PRX,Carleo2018NatCommun,Kaubruegger2018PRB,Zheng2019PRB,Lu2019PRB,HuangPRL2021}. However, not all quantum states can be efficiently represented in terms of \acp{NQS}. For instance, we cannot represent random states since they do not have structure.

Another important aspect is choosing the right \ac{NN} architecture and training strategy for the problem. For instance, we may be interested in implementing certain symmetries, as we have discussed in~\cref{sec:NQS_symmetries}. However, on a~given problem, a~certain \ac{NQS} ansatz may be well-suited for the task, but the training procedure can fail numerically. Some works have analyzed the training procedure involving stochastic reconfiguration \cite{Park2020}. Others have found that states involved in the dynamics of non-integrable systems are not representable by various architectures, but their entanglement structure can be recovered, hinting at a~different limit from the built-in limitation on entanglement in \ac{TN}-based ansätze \cite{lin2021scaling}.

These findings, along with state-of-the-art results, point toward a~superior expressive power of \ac{NQS} over existing simulation methods, but many research routes have to be taken to fully understand their capabilities, much like many \ac{ML} methods discussed in this book.

\subsection{Applications}
In this section, we present various applications of \ac{NQS}, ranging from the ground state search to quantum state tomography, featuring real-time dynamics, quantum circuits, and fermionic systems. In addition to presenting how the methods described previously apply to such problems, we provide results for each application and compare them to other state-of-the-art methods. By doing this, we hope to show both the potential and versatility of \ac{NQS} approaches, which is still a~young field of research.

\subsubsection{Finding the ground state}\label{sec:NQS_GS}

As common in many \ac{ML} tasks, we define a~loss function $\lossfun$ that depends on the trainable parameters of the \ac{NN}. In this situation, this corresponds to the variational energy, i.e., the expectation value of the Hamiltonian in the variational state\index{variational state}:
\begin{equation}
    \lossfun(\params) = E(\params) = \langle \Psi_{\params} | \hat{H} | \Psi_{\params} \rangle.
\end{equation}This choice of the loss function is naturally introduced since it follows from the variational principle in quantum mechanics.
\highlight{The \stress{variational principle} states that given an~Hamiltonian $\hat{H}$, the energy $E(\theta)$ of a~variational wave function $|\Psi_\theta\rangle$ is greater or equal than the exact ground state energy, i.e.,
\begin{equation}
    E(\params) = \frac{\mel{\Psi_{\params}}{\hat{H}}{\Psi_{\params}}}{{\langle\Psi_{\params}|\Psi_{\params}\rangle}} \geq E_0.
\end{equation}
Therefore the energy is a~valid loss function, as the \stress{lower} the expectation value of the energy, \stress{the better the approximation is}.\footnote{We stress that the principle is only valid when computing expectation values exactly. When the energy is computed as a~stochastic average, its estimated average can be lower than the exact energy.
Nevertheless, as discussed previously, the increase in the number of samples and the use of an~efficient sampling approach systematically reduce fluctuations below the exact energy.}}
In fact, having a~loss function strongly rooted in a~principle of physics is crucial since it also allows us to compare different methods. By looking at the variational energy, we can for example understand how a~method performs at solving a~given problem: if the resulting approximate ground state energy is significantly lower than what was found by alternative techniques, we can be reasonably sure that the solution found is of better quality.
Following the general discussion on expectation values of operators, the variational energy can be stochastically approximated as
\begin{equation}\label{eq:var_energy}
E(\params) \approx \frac{1}{M} \sum_i^M E_{\text{loc}} (\vect{s}^{(i)}),
\end{equation}
where $E_{\text{loc}}$ is the local estimator and is defined as $E_{\text{loc}}(\vect{s}) = \sum_{\vect{s}'} \langle s | \hat{H} | s \rangle \frac{\langle s' | \Psi \rangle}{\langle s | \Psi \rangle}$.
We aim to minimize this loss function by means of gradient-based optimization algorithms.
The energy gradients can also be written in terms of expectation values\footnote{Computationally speaking, one does not need to store in memory the full Jacobian matrix $O_p(\vect{s})$, but can compute this gradient directly through the vector-Jacobian product (reverse-mode differentiation) of the vector $v=E_{\text{loc}} (\vect{s}) - \langle E_{\text{loc}} (\vect{s}) \rangle )$ and the Jacobian $O_p(\vect{s})$.
This approach considerably lowers the memory and computational cost. For more details, see~\cref{sec:hot-topics:dp}.}
\begin{align}
\frac{\partial E(\params)}{\partial \param_p} &= 2 \Re \left[\langle E_{\text{loc}} (\vect{s}) O_p^*(\vect{s}) \rangle - \langle E_{\text{loc}} (\vect{s}) \rangle \langle O_p^*(\vect{s})\rangle\right] \\
&= 2 \Re \left[\langle (E_{\text{loc}} (\vect{s}) - \langle E_{\text{loc}} (\vect{s}) \rangle ) O_p^*(\vect{s}) \rangle \right]
\label{eq:NQS_GS_0}
\end{align}
where we have assumed that the parameters are real\footnote{The requirement of real parameters is not actually necessary. For complex parameters, the expression is very similar, though care has to be taken in order to consider non-holomorphic ansätze. 
Note that many common ansätze, particularly most autoregressive ones, are not holomorphic.
Discussion of this can be found in the appendix of Ref.~\cite{Vicentini:2021pcv}.} and that $\param_p$ is the $p$-th parameter of the \ac{NQS}. The diagonal operator $\hat{O}_p$ is defined as
\begin{equation}\label{eq:NQS_Oks}
O_p(\vect{s}) = \frac{\partial}{\partial \param_p} \log \langle s | \Psi_{\params} \rangle = \langle s | \hat{O}_p | s \rangle.
\end{equation}
We also remark that the expression used in eq. (\ref{eq:NQS_GS_0}) has the form of a~covariance, and therefore is particularly stable with respect to sampling noise. Most notably, when the wave function is close to the exact ground state, statistical fluctuations in the local energy are suppressed, implying that also statistical fluctuations of the gradients are small because of the covariance structure.  

The learning algorithm is thus straightforward.
First, we initialize the weights $\params^{(0)}$.
Next, at each step a~sequence of $M$ configurations is sampled according to the Born distribution: $P(\vect{s}; \params^{(s)}) \sim \vect{s}^{(1)} \ldots \vect{s}^{(M)}$.
This can be done with a~Markov chain or with direct sampling techniques as explained above.

The next step is to compute the mean of the local energy $E(\params)$, which gives us the estimate of the expectation value of the Hamiltonian.
Additionally, the gradients can also be calculated as shown in \cref{eq:NQS_GS_0}.
For the last step, we can use a~gradient-based optimizer of our choice, to update the parameters for the next step, i.e., $\param_p^{(s+1)} = \param_p^{(s)} -\learningrate \frac{\partial E(\params)}{\partial \param_p}$ for vanilla gradient descent where $\learningrate$ is the learning rate.

The procedure is repeated until it converges to a~minimum of the energy landscape.
Here, there is no training data set as the approach is not based on any supervised learning method.
The presented task is in fact to determine the optimal (unknown) wave function by drawing samples from the associated Born distribution and using a~\ac{NN} to model the state itself.
These steps are summarized in \cref{alg:NQS_GS_search}.
Note that this algorithm is not the most commonly used, as it is less accurate than imaginary-time evolution, which is presented in \cref{sec:NQS_imagtim}.

\begin{algorithm}
\caption{Ground state search with \ac{NQS}}\label{alg:NQS_GS_search}
\begin{algorithmic}
\State Initialize $\params$ randomly
\For{i = 1 to $n_{\mathrm{steps}}$}
    \State Generate $M$ samples according to some algorithm (usually a~Markov chain)
    \State Calculate the gradient of the energy $\partial E(\params)/ \partial \param_p$
    \State Update parameters as $\param_j \gets \param_j - \learningrate \partial E(\params)/ \partial \param_j$ (or with a~more advanced update rule) 
\EndFor\\
\Return Optimized parameters $\params$
\end{algorithmic}
\end{algorithm}

\subsubsection{Real-time evolution}

\Acp{NQS} can also be used to variationally perform real-time evolution \cite{Yuan_2019} through a~procedure known as \ac{t-VMC} ~\cite{carleo_localization_2012,Carleo_2017, Schmitt_2020,gutierrez2021real}.
This is of particular interest for non-equilibrium quantum dynamics of closed, interacting quantum systems.
Studying these problems enables one to understand critical properties, entanglement spectra, and many other physical quantities of interest in complex many-body quantum systems.
The problem one wants to solve is to integrate the time-dependent Schrödinger equation ($\hbar = 1$ in the following) in time, using a~parametrized wave function $\ket{\Psi_{\params}(t)}$:
\begin{align}\label{eq_NQSdyn0}
    \i\frac{d\ket{\Psi_{\params}(t)}}{dt} = \hat{H}\ket{\Psi_{\params}(t)},
\end{align}
i.e., find the correct form of $\ket{\Psi_{\params}(t)}\ \forall t$. Expanding \cref{eq_NQSdyn0} at first order in $\delta$ and taking the inner product with $\langle s |$, we obtain:
\begin{align}\label{eq_NQSdyn1}
    \Psi_{\params} (t+\delta) (\vect{s}) &= 1 - \i\delta\langle s |\hat{H}|\Psi_{\params}(t)\rangle+ O(\delta^2)\\
    &= 1- \i \delta E_{\mathrm{loc}}(\vect{s}) + O(\delta^2),
\end{align}
where we used $E_{\mathrm{loc}}(\vect{s})$ as defined in \eqref{eq:var_energy} in the previous section. In order to get a~good variational approximation of the state at the next time step, $t+\delta$, it is natural to define the cost function $\lossfun(\tilde{\params})$:
\begin{align}\label{eq_NQSdyn2}
     \lossfun(\tilde{\params})=\text{ dist}\left(|\Psi_{\tilde{\params}}\rangle, | \Psi_{\params} (t+\delta) \rangle\right),
\end{align} 
with $\params$ the variational parameters at the previous time step, and $\tilde{\params}$ variational parameters to be determined. The loss function can be minimized analytically, if the time step is sufficiently small. One starts by noticing that $\tilde{\params} =  \params + \delta\dot{\params} +\bigO(\delta^2)$. One can therefore expand the variational state $|\Psi_{\tilde{\params}} \rangle$ at first order and take its inner product with $\langle s |$, much like we did for \cref{eq_NQSdyn1}:
\begin{align}\label{eq_NQSdyn3}
    \Psi_{\params+\tau\dot{\params}}(\vect{s}) = \left(1 - \delta\dot{\params}\partial_{\params} \Psi_{\params}(\vect{s})\right)\Psi_{\params}(\vect{s}) + \bigO(\delta^2).
\end{align}
We need to consider a~distance measure between the two states  $|\Psi\rangle$ and $|\phi\rangle$ which can be efficiently sampled.
There is a~certain freedom in this choice, which can lead to slightly different variational principles. 
For an~extensive discussion of these issues, see~\cite{Yuan_2019}.
By considering the infidelity, keeping in mind that for many~\ac{NQS} architectures the quantum states are unnormalized, we have~\footnote{Rigorously, one should consider the Fubini-Study metric, but taking this distance leads to the same equations.}:
\begin{align}\label{eq_NQSdyn4}
    \text{dist}\left(|\Psi\rangle , |\phi\rangle \right) = 1 - \frac{\langle\phi|\Psi\rangle \langle \Psi|\phi\rangle}{\langle\phi|\phi\rangle\langle\Psi|\Psi\rangle}.
\end{align} By plugging \cref{eq_NQSdyn3} and \cref{eq_NQSdyn1} into the distance of \cref{eq_NQSdyn4}, minimizing it, and keeping the leading terms in $\delta$ one obtains an~equation giving the time derivative of the variational parameters $\dot{\params}$, enabling high-order integration methods such as Runge-Kutta integration:
\begin{align}\label{eq_NQSdyn5}
    \mat{S} \dot{\params} = -\i \vect{f}
\end{align}
with the \stress{quantum geometric tensor} $\mat{S}$ and the vector $\vect{f}$, whose elements are given by:
\begin{align}
    S_{pp'} &= \langle O_p^* O_{p'}\rangle - \langle O_p^*\rangle \langle O_{p'} \rangle \\
    f_p &= \langle E_{\mathrm{loc}} O^*_{p}\rangle - \langle E_{\mathrm{loc}}\rangle \langle O^*_{p} \rangle
\end{align} with the $O_p$s given by \cref{eq:NQS_Oks} and $E_{\mathrm{loc}}$ is the local energy.
The vector $\vect{f}$ is the gradient of the local energy with respect to the variational parameters and, in analogy with classical mechanics, it is often called the vector of \stress{forces}. The spectrum of the geometric tensor instead encodes the (linearized) curvature of the variational space, akin to the Hessian discussed in \cref{sec:BO_GPR_science}.
For a~full derivation and an~in-depth discussion of the time-dependent variational principles, see Ref.~\cite{Yuan_2019}.
The spectrum  of $\mat{S}$ has been extensively studied in the case of ground state optimization with \acp{RBM} \cite{Park2020}, where it has been connected to the different regimes of considered physical system. In practice, solving the linear system \cref{eq_NQSdyn5} implies either using an~iterative solver (for example, conjugate gradient) or a~direct solver (for example, QR factorization). An~important pratical numerical issue is that the matrix $\mat{S}$ is often singular. Some techniques have been found to regularize $\mat{S}$ and obtain more stable dynamics \cite{Schmitt_2020, hofmann2021role}. In all cases, since only stochastic averages for both $\mat{S}$ and $\vect{f}$ are available, stable and accurate long time dynamics are still a~challenge for \ac{NQS} \cite{hofmann2021role}.
\begin{figure}
    \centering
    \includegraphics[scale=0.25]{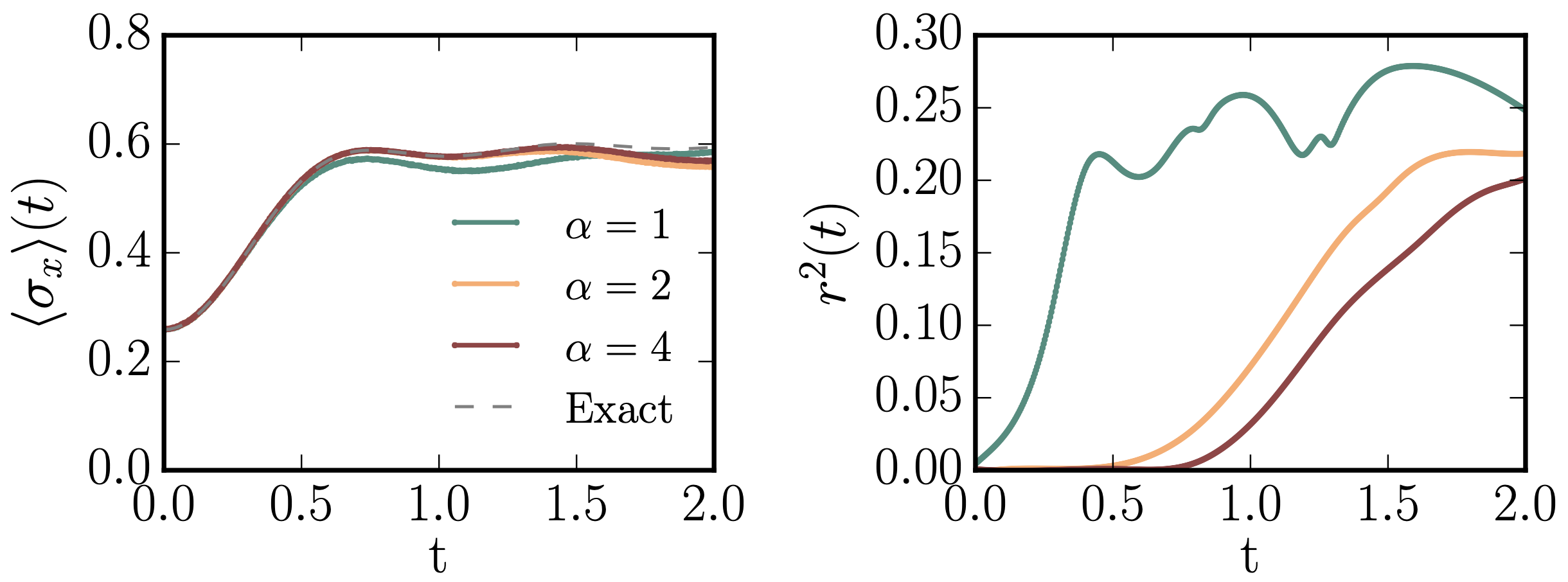}
    \caption[Dynamics of the neural quantum state]{Critical quench dynamics with an~\ac{RBM}, preparing the system in the ground state of $\hat{H}_{\mathrm{TFI}}$ for $h_i/J = 1/2$, then suddenly quenching to $h_f/J = 1$. This excites many eigenstates of the system at criticality (which exhibit long-range correlations, making the dynamics difficult to capture). Left panel: average magnetization along $x$ for different values of the density of hidden neurons $\alpha$ of the \ac{RBM}. Right panel: integrated error, systematically reduced by increasing $\alpha$. Taken from \ToggleForCUP{Carleo, G. \& Troyer, M. (2017). \textit{Solving the quantum many-body problem with artificial neural networks}. Science 355, 602–606~\cite{Carleo_2017}.}{Ref.~\cite{Carleo_2017}.}}
    \label{fig:NQS_timeevolution}
\end{figure}
As an~example, in \cref{fig:NQS_timeevolution} we show the quench dynamics of a~one-dimensional spin chain, subject to the Ising Hamiltonian with a~transverse field:
\begin{equation}\label{eq_NQSdyn6}
    \hat{H}_{\mathrm{TFI}} = -J\sum_j \hat{\sigma}_j^z\hat{\sigma}_{j+1}^z + h\sum_j\hat{\sigma}^x_j.
\end{equation} Here, $J$ is the nearest-neighbor coupling, and $h$ is the transverse field strength. This model exhibits a~second-order phase transition in one dimension at $h=J$, that separates a~ferromagnetic (for $J>0$, or antiferromagnetic for $J<0$) phase from a~paramagnetic phase, with all spins aligned along the transverse-field for $h \gg J$. The critical quench dynamics can be investigated by preparing the system in an~eigenstate of the Hamiltonian for some value of $h = h_i$, then suddenly switching the Hamiltonian parameters to $h_f/J = 1$. As seen in \cref{fig:NQS_timeevolution}, an~\ac{RBM} captures the dynamics up to about $Jt = 1.5$, and increasing the number of hidden layers $\alpha$ systematically improves the precision. As mentioned, more recent results have also been obtained using a~\ac{CNN} on a~two-dimensional system, whose dynamics are a~challenge for \ac{TN} methods \cite{Schmitt_2020}.

\subsubsection{Imaginary-time evolution}\label{sec:NQS_imagtim}
The first-order optimization scheme presented in~\cref{sec:NQS_GS} to estimate the ground state of many-body systems can be improved to yield more accurate results. For this purpose, it is useful to consider an imaginary-time evolution through Wick's rotation $t\to i\tau$:
\begin{equation}\label{eq:NQS_imagtime}
    |\psi_{\params}(\tau) \rangle = \exp (-\tau \hat{H}) |\psi_{\params}(0) \rangle,
\end{equation}
where $\hat{H}$ is the Hamiltonian, and $\tau$ is a~real number. It can be shown that $\lim_{\tau \rightarrow \infty} |\phi (\tau) \rangle = |\phi_0 \rangle$, with $|\phi_0 \rangle$ being the exact ground state of the Hamiltonian and provided $|\langle \psi_{\theta}(\tau)|\phi_0\rangle| \neq 0$. Furthermore, it can be shown that the convergence of imaginary-time evolution toward the exact ground state is exponentially fast with $\tau$, thus offering a~systematic way to find the ground state. Analogously to real-time evolution, imaginary-time evolution can also be performed variationally. This leads to the same type of equation as \cref{eq_NQSdyn5}:
\begin{equation}\label{eq_NQSdyn7}
     \mat{S} \dot{\params} = - \vect{f},
\end{equation}
with the factor $\i$ missing due to the form of the exponent in \cref{eq:NQS_imagtime}. Hence, a~very similar procedure is obtained as for the real time evolution in which we can update the weights according to the update given by the equation above. 
To summarize both real and imaginary time, the algorithm for variational time evolution is given in~\cref{alg:NQS_real_imag_evo}. In the case of imaginary-time evolution, the algorithm is typically modified in such a~way that the S matrix is regularized by adding a~constant, $\Lambda>0$, proportional to the identity: $\mat{S} \rightarrow \mat{S} + \Lambda \mat{I}$. In this case, one recovers the stochastic reconfiguration method, as originally introduced by S.~Sorella \cite{sorella_green_1998,becca_sorella_2017}. 
\begin{algorithm}
\caption{Real ($\xi = \i)$ or imaginary ($\xi = 1$) time evolution algorithm for \ac{NQS}}\label{alg:NQS_real_imag_evo}
\begin{algorithmic}
\State $\params \gets$ random initialization
\For{i = 1 to $n_{\mathrm{steps}}$}
    \State Calculate $S_{pp'}$ and $F_p$
    \State Get $\dot{\params}$ by inverting the equation  $\sum_{p'} S_{pp'}\dot{\param}_{p'} = - \xi F_p $ (and possibly regularizing)
    \State Update $\params$ using an~\acs{ODE} integrator
\EndFor
\end{algorithmic}
\end{algorithm}
\begin{figure}
    \centering
    \includegraphics[scale=0.2]{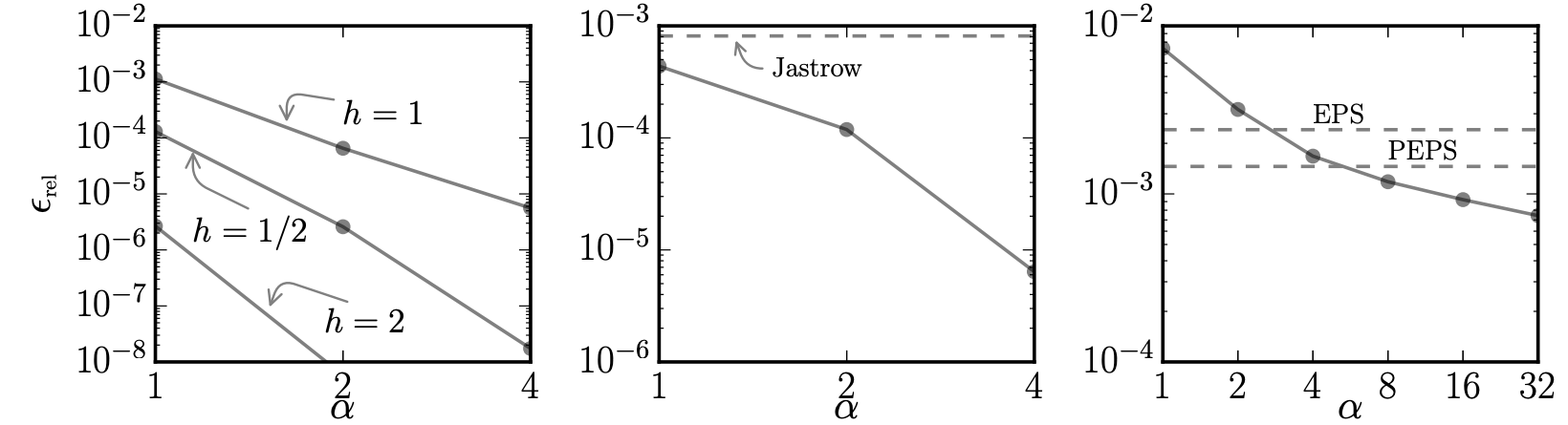}
    \caption[Ground state results with a~Restricted Boltzmann Machine ansatz]{Relative error of the ground state search using imaginary time evolution with respect to the exact solution. Left panel: relative error for the ground state of the transverse-field Ising model, for various values of $h/J$ as a~function of the number of $\alpha = M/N$, $M$ being the number of hidden units for $N=80$ spins. Center panel: Same plot for the ground state of the antiferromagnetic Heisenberg model, compared to another variational ansatz, the Jastrow wave function. Right panel: comparison with state-of-the-art \ac{TN} results, showing that \ac{NQS} perform at least as good. Taken from \ToggleForCUP{Carleo, G. \& Troyer, M. (2017). \textit{Solving the quantum many-body problem with artificial neural networks}. Science 355, 602–606~\cite{Carleo_2017}.}{Ref.~\cite{Carleo_2017}.}}
    \label{fig:NQS_imaginarytimeevolution}
\end{figure}
In \cref{fig:NQS_imaginarytimeevolution}, results are shown for imaginary-time evolution performed on the transverse-field Ising model and the Heisenberg model \cite{Carleo_2017}. These results show two important features: (i) using an~\ac{RBM} ansatz, the relative error can be systematically reduced by increasing $\alpha = M/N$, with $M$ the number of hidden units and (ii) the results achieve a~higher precision than state-of-the-art \ac{TN} methods.

\subsubsection{Fermionic systems}

Fermions constitute one of the two fundamental types of elementary subatomic particle. As such, fermionic systems are ubiquitous in many-body quantum physics, high energy physics, as well as chemistry. However, the classical simulation of fermionic systems is difficult because fermionic operators obey anticommutation relations, which constrain the wave functions of fermionic systems to be antisymmetric under particle exchange. Due to the infamous sign problem~\cite{Hangleiter_2020}, all the known quantum Monte Carlo methods become extremely expensive computationally for fermionic systems. 

Variational approaches with \ac{NQS} for fermions may be divided into two classes: (i) using the first quantization, one may impose antisymmetry on the wave function by constructing it as a Slater determinant or (ii) going to the second quantization and mapping the fermionic many-body Hamiltonian to a spin Hamiltonian. The first approach builds on a formulation of the problem in terms of a continuous state space. Many impressive results for realistic systems have been obtained by employing such an approach~\cite{Luo_2019, hermann2020deep, pfau2020ab}. For further reading, we recommend Ref.~\cite{Hermann2022review} which is a recent review on the topic. In the following, we will focus our discussion on the second approach based on the second quantization.

As explained, a convenient approach to simulate fermionic systems is based on mapping the fermionic degrees of freedom to spins. A~generic protocol is the Jordan-Wigner transformation, which enables us to map fermionic problems to interacting spin problems. Note that there are many other possible transformations that have mostly been developed in the context of quantum simulation, such as the Bravyi-Kitaev encoding~\cite{Bravyi2002}. Historically, this technique has been used to solve spin models~\cite{Jordan1928}. Here we do the opposite: we map fermionic operators to spin operators in order to use \acp{NQS} and the techniques presented throughout the chapter to solve the corresponding many-body problem. This approach does not suffer from the sign problem, because the antisymmetry is directly encoded in the terms of the Hamiltonian. However, ultimately the approach is limited by the difficulty of the resulting spin problem that may include complicated, nonlocal interactions.

Let us consider the creation and annihilation fermionic operators acting on site $j$, $\hat{c}^\dagger_j$ and $\hat{c}_j$, respectively. The Jordan-Wigner transformation prescribes:
\begin{align}
    \hat{c}_j &= \left(\prod_{k=1}^{j} \hat{\sigma}^z_k \right) \hat{\sigma}^-_j\,,\\
    \hat{c}^\dagger_j &= \left(\prod_{k=1}^{j} \hat{\sigma}^z_k \right) \hat{\sigma}^+_j\,,
\end{align} 
where $\hat{\sigma}_j^{\pm} = \frac{1}{2}\left(\hat{\sigma}_j^x\pm i\hat{\sigma}_j^y\right)$ denote the spin raising and lowering operators. The first term in the transformation provides a~phase that can be $\pm1$ depending on whether the number of occupied fermionic modes is even or odd in sites $k=1,\dots,j$. We can conveniently rewrite this term using the relations $\hat{\sigma}^z_j = 2\hat{\sigma}^+_j\hat{\sigma}^-_j - 1$, and $\hat{\sigma}^+_j\hat{\sigma}^-_j=\hat{c}^\dagger_j\hat{c}^{\phantom{\dagger}}_j=n_j$. This ensures that the resulting operators fulfill fermionic anticommutation relations.

For example, using this transformation, we can map a Hamiltonian describing free fermions in one dimension
\begin{equation}
    \hat{H} = -\frac{1}{2}\sum_j \hat{c}_j\hat{c}^\dagger_{j+1} + \hat{c}^\dagger_j\hat{c}_{j+1}
\end{equation}
to an~interacting spin Hamiltonian of the form
\begin{equation}
    \hat{H} = -\frac{1}{2} \sum_j \hat{\sigma}^+_j\hat{\sigma}^-_{j +1}+ \hat{\sigma}^+_{j+1}\hat{\sigma}^-_{j}.
\end{equation}
In this form, we can implement all the methods described throughout~\cref{sec:NN_q_states}. 

The main issue with this transformation is that it does not generalize well to arbitrary dimensions. In higher dimensions, the Jordan-Wigner transformation results in a nonlocal spin Hamiltonian which cannot be tackled with most standard techniques. Different mappings for fermionic degrees of freedom that work in higher dimensions have been proposed. These are not general mappings, but instead are tailored to specific problems. For example, we can map local Hamiltonians in more than one dimension to local bosonic Hamiltonians for certain specific gauge theories~\cite{Zohar_2018,PhysRevLett.124.120503}. Another approach to avoid nonlocal spin Hamiltonians in high dimensions based on using an ancillary system as been considered in Ref.~\cite{Nys_2022}. In this case, the degrees of freedom of the ansatz are separated into a main system and an auxiliary system. By doing so, one can build a local spin Hamiltonian from a local fermionic Hamiltonian at the expense of having a larger Hilbert space. 

Calculating the electronic structure of molecules is a timely fermionic problem that is important for applications in chemistry~\cite{Choo2020}. It is one of the oldest instances of a quantum many-body problem\index{quantum many-body problem} first mentioned by Dirac in 1929~\cite{Dirac1929}. In this context, one is generally interested in finding the ground state energy as a~function of some physical parameter, such as the distance between two nuclei for a~diatomic molecule. This way, by looking at the minimum of the energy, one can find out what the stable geometry of the molecule of interest is.
\begin{figure}
    \centering
    \includegraphics[width=\textwidth]{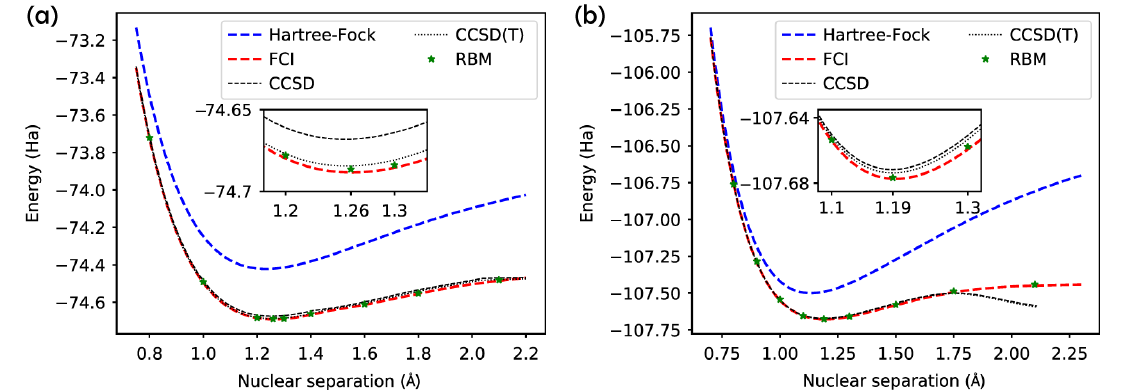}
    \caption[Electronic structure calculations with neural quantum states]{Ground state energies of a (a) C$_2$ and (b) N$_2$ molecule measured in Hartree (Ha) as a~function of nuclear separation, given by various techniques compared with results obtain by an~\ac{RBM} ansatz with $M = 40$ hidden units. CCSD(T): coupled-cluster approaches, FCI: full-configuration interaction. Taken from \ToggleForCUP{Choo, K., Mezzacapo, A., \& Carleo, G. (2020). \textit{Fermionic neural-network states for ab-initio electronic structure}. Nat. Commun. 11, 2368~\cite{Choo2020} under the \href{https://creativecommons.org/licenses/by/4.0/}{CC BY 4.0 DEED} license.}{Ref.~\cite{Choo2020}.}}
    \label{fig:NQS_molecules}
\end{figure}
Usually, the interacting fermionic Hamiltonian is defined on a lattice and takes the following form
\begin{equation}
    \hat{H} =  \sum_{i,j}t_{ij} \hat{c}_i^\dagger \hat{c}^{\phantom{\dagger}}_j +  \sum_{i,j,k,l} U_{ijkl}\hat{c}_i^\dagger \hat{c}_j^\dagger \hat{c}^{\phantom{\dagger}}_k \hat{c}^{\phantom{\dagger}}_l,
\end{equation} where $t_{ij}$ is a~single-body hopping term, $U_{ijkl}$ is a~two-body interaction strength, and $\hat{c}^{\phantom{\dagger}}_i$ ($\hat{c}_i^\dagger$) is a~fermionic annihilation (creation) operator for mode $i$.\footnote{We can formulate it in a~real-space or a~momentum-space basis, which we leave unspecified for the sake of generality.} The Jordan-Wigner transformation changes this Hamiltonian to the form
\begin{equation}
    \hat{H} = \sum_r a_r \mathbf{S}_r,
\end{equation} 
where $a_r$ are scalar coefficients and $\mathbf{S}_r$ are Pauli strings composed of elements of the set of single-qubit operators $\{\mathbb{1}, \hat{\sigma}^x, \hat{\sigma}^y, \hat{\sigma}^z\}$. In other words, we now have an~interacting spin problem and, while the resulting Hamiltonian is not necessarily $k-$local, it can be shown that the local energy can still be estimated efficiently. Therefore, these mappings are amenable to variational searches using \ac{NQS}.

Another recent improvement in the field is the construction of explicit autoregressive ansätze, as presented in section~\ref{sec:autoregressive_NNs}, for fermions~\cite{Barrett_2021}. Here, authors consider a basis set of spin-orbitals consisting of $M$ spatial orbitals, each existing for upward and downward spins, i.e., spanned by $\ket{\mathbf{x}_k} = \ket{x_k^{1\uparrow},x_k^{1\downarrow},x_k^{M\uparrow},x_k^{M\downarrow}}$. Each spatial orbital can be treated as a single unit that can take on four possible values, denoted $v_k^i$. Therefore, the logarithm of the corresponding wave function coefficients takes on the following form:
\begin{equation}
    \ln \psi_k = \sum_{j=1}^M \ln |\psi_j(v^j_k \mid v^1_k\ldots v^{j-1}_k)| + \mathrm{i}\phi_j(v^1_k\ldots v^M_k).
\end{equation}
This ensures a proper normalization of the wave function provided the conditional amplitudes $\psi_i$ are normalized, therefore a direct sampling scheme which leads to improved results.

Finally, in \cref{fig:NQS_molecules}, we show physical results obtained using \ac{NQS} for fermionic systems. The dissociation curves (ground state energies) as a~function of the nuclear separation for the molecules \ce{C2} and \ce{N2} provided by various numerical methods are displayed. By using an~\ac{RBM} ansatz and a~simple Jordan-Wigner transformation, one is able to recover results that are competitive with recent full configuration interaction calculations, which demonstrates the versatility and power of \acp{NQS}.

\subsubsection{Classical simulation of quantum circuits}

Another promising direction for \acp{NQS} is the classical simulation of quantum circuits, which we introduce in \cref{sec:gbqc}. Indeed, current classical simulation methods for large quantum circuits (of the order of at least $50$ qubits) rely on \ac{TN} methods that are explicitly restricted by entanglement. In particular, \acp{TNS} cannot capture volume-law entanglement scaling, which quickly arises in quantum circuits, whereas certain \ac{NQS} architectures such as deep \acp{CNN} can \cite{sharir2021}. In this context, \acp{NQS} can be investigated in two somewhat orthogonal directions: one could use quantum circuits to probe the limits of their capacity and trainability, and \acp{NQS} can be used to push the classical simulation limits of quantum hardware.

Let us consider a~quantum circuit defined by a~set of $D$ gates $\mathcal{G} = \{\hat{G}_i\}_{i=1}^D$, each gate being defined by a~unitary operator $\hat{G}_i$. After each gate, the variational state\index{variational state} must be updated so as to capture the application of the previous gate. The following variational distance must therefore be minimized for each gate:\begin{align}\label{NQScircuit1}
    \lossfun(\tilde{\params})=\text{ dist}\left(|\Psi_{\tilde{\params}}\rangle, \hat{G}_i| \Psi_{\params}\rangle\right),
\end{align} with $\tilde{\params}$ the parameters to be optimized, $\params$ the parameters of the previous variational state\index{variational state}, and $\hat{G}_i$ the unitary operator corresponding to gate $G$ (for instance, for a~NOT gate, $\hat{G}_i = \hat{\sigma}^x$). One can apply this procedure for each gate $G$ and obtain the output state at the end of a~circuit, after $D$ optimizations. Note that $\hat{G}_i$ must be a~$k-$local gate, or else the minimization procedure cannot be carried out (this is reminiscent of ground state search). This is rarely a~problem, since universal gate sets can be constructed with only single- and two-qubit gates. With this condition, minimizing \cref{NQScircuit1} closely resembles the ground state optimization. One can develop this expression using the infidelity, see~\cref{eq_NQSdyn4}], and obtain:
\begin{align}
    \lossfun(\tilde{\params})&=1 - \langle G_{\mathrm{loc}}(\params, \tilde{\params}) \rangle_{|\Psi_{\tilde{\params}}|^2}  \langle G_{\mathrm{loc}}(\tilde{\params}, \params) \rangle_{|\Psi_{\params}|^2} \\ \text{with}\ &G_{\mathrm{loc}}(\params, \tilde{\params})(s)= \sum_{s'} \frac{\mel{s}{\hat{G}_i}{s'}}{\Psi_{\tilde{\params}}(s)}\Psi_{\params}(s'),
\end{align}where we have defined $G_{\mathrm{loc}}$ as a~local estimator, similarly to the procedure described for a ground state search. As often in \ac{ML}, the minimization of $\lossfun(\tilde{\params})$ may be inaccurate, which reduces the overall fidelity of the simulation. Using an~\ac{RBM} architecture, however, not all gates need to be approximated through minimization of the loss above. Some gates can be applied "analytically", i.e., it is possible to find the exact update on the parameters of the network so as to match the applied gate. In general, it is impossible to realize these exact updates for all gates of a~universal gate set, or else one could simulate any quantum circuit with an~\ac{RBM} with infinite precision by obtaining the exact parameter update for each gate.
For example, let us consider a~$Z$ gate acting on spin $s_j$ defined by the operator $\hat{G} = \hat{\sigma}^z_j$. The action of such an~operator on a~basis state $\ket{s}$ is simply $\hat{G}\ket{s} = (-1)^{s_j}\ket{s}$. The \ac{RBM} parameters before the gate are defined as $\params = (\vect{b}_v, \vect{b}_h, \mat{W})$ and the parameters after the gate are defined as $\tilde{\params} = (\tilde{\vect{b}}_v, \tilde{\vect{b}}_h, \tilde{W})$. The parameter update is given by the solution of the following equation (with $C$ a~constant):
\begin{align}\label{NQScircuit2}
    \Psi_{\tilde{\params}}(s) &= C\mel{s}{\hat{G}}{\Psi_{\params}}\\
    e^{\tilde{b}_{v,j}s_j} &= C(-1)^{s_j}\e^{b_{v,j}s_j},
\end{align} which is simply $\tilde{b}_{v,j} = b_{v,j} +\i\pi$ for $C=1$. The simplification in the previous equation is due to the fact that this gate acts trivially on the other parts of the \ac{RBM} amplitude, defined in \cref{sss:rbm}. Details of how to apply other gates analytically can be found in Ref.~\cite{jonsson2018neuralnetwork} and \cite{Medvidovi__2021}. In this last reference, authors classically simulate the circuit corresponding to the \ac{QAOA}~\cite{farhi2014quantum}\footnote{The \acf{QAOA} is a variational quantum algorithm designed to tackle combinatorial optimization problems.} using an~\ac{RBM} ansatz. This quantum algorithm enables one to access the solution of a~certain class of combinatorial optimization problems. The corresponding circuit, which is quite shallow, can be implemented on current hardware \cite{google2021}. In \cref{fig:NQS_circuits}, one can see that the results obtained by simulating the quantum circuit with the \ac{RBM} ansatz closely match the result of the exact simulation, enabling one to find the solution of the optimization problem for large systems. Authors also estimate a~significant advantage over \ac{TN} methods.\footnote{Based on an~extrapolation of numerical simulation data.} Indeed, in the left panel of \cref{fig:NQS_circuits}, one can see that the required bond dimension required to reach the same results as that of the \ac{RBM} would quickly become dauntingly large when using an~\ac{MPS}.

Alternatively, authors of Ref.~\cite{PhysRevA.104.032610} have proposed to simulate quantum circuits using a~transformer architecture. A~transformer is a~deep learning model that adopts the mechanism of self-attention, differentially weighting the significance of each part of the input data \cite{AttentionMechanism2017}. Using this framework, a~practical algorithm to simulate quantum circuits using a~transformer ansatz responsible for the most recent breakthroughs in natural language processing was introduced in Ref.~\cite{PhysRevA.104.032610}. This framework allows for the simulation of circuits that build Greenberger-Horne-Zeilinger and linear graph states of up to 60 qubits.
\ToggleForCUP{
\begin{figure}[p]
    \centering
    \includegraphics[scale=0.23]{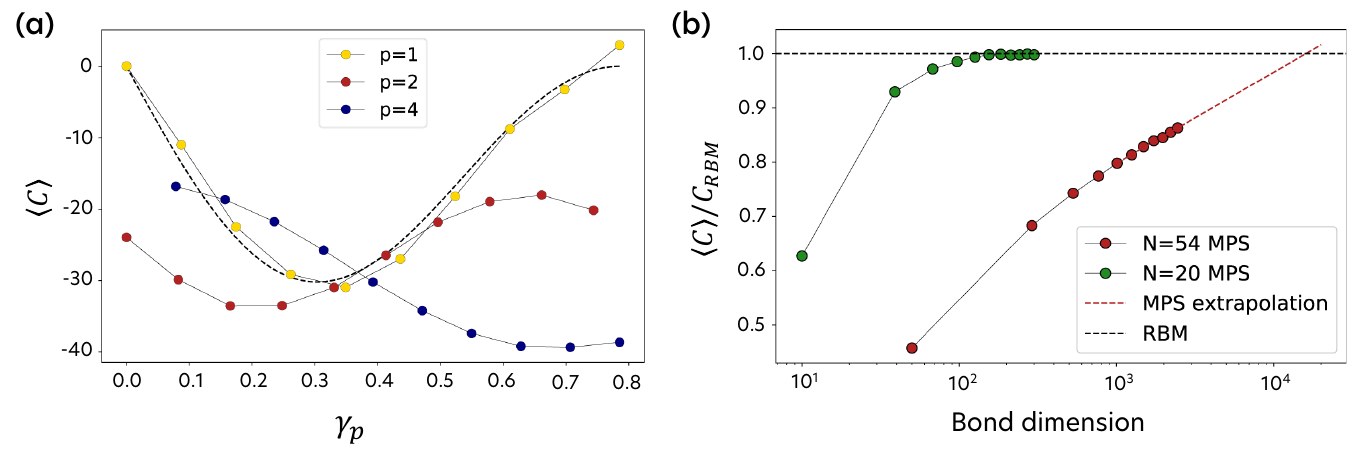}
    \caption[Quantum Approximate Optimization Algorithm with neural quantum states]{Results for the classical simulation of \ac{QAOA} for a~3-planar graph. Panel (a) shows $\langle \mathcal{C} \rangle$, the approximated cost function for various values of $p$, the depth of the quantum circuit. The dashed line corresponds to the exact simulation of the $p=1$ quantum circuit which the \ac{RBM} simulation accurately reproduces for this task. (b) Estimation of the required bond dimension in a~Matrix Product State (MPS) simulation of the \ac{QAOA} circuit to match the accuracy of the \ac{RBM}. For $N=54$ qubits, the required bond dimension is of about $10^4$, which amounts to using billions of parameters, whereas the \ac{RBM} uses a~few hundred. Taken from \ToggleForCUP{Medvidovi\'c, M. \& Carleo, G. (2021). \textit{Classical variational simulation of the quantum approximate optimization algorithm}. npj Quantum Inf. 7, 101~\cite{Medvidovi__2021} under the \href{https://creativecommons.org/licenses/by/4.0/}{CC BY 4.0 DEED} license.}{Ref.~\cite{Medvidovi__2021}.}}
    \label{fig:NQS_circuits}
\end{figure}}
{\begin{figure}[h]
    \centering
    \includegraphics[width=0.95\columnwidth]{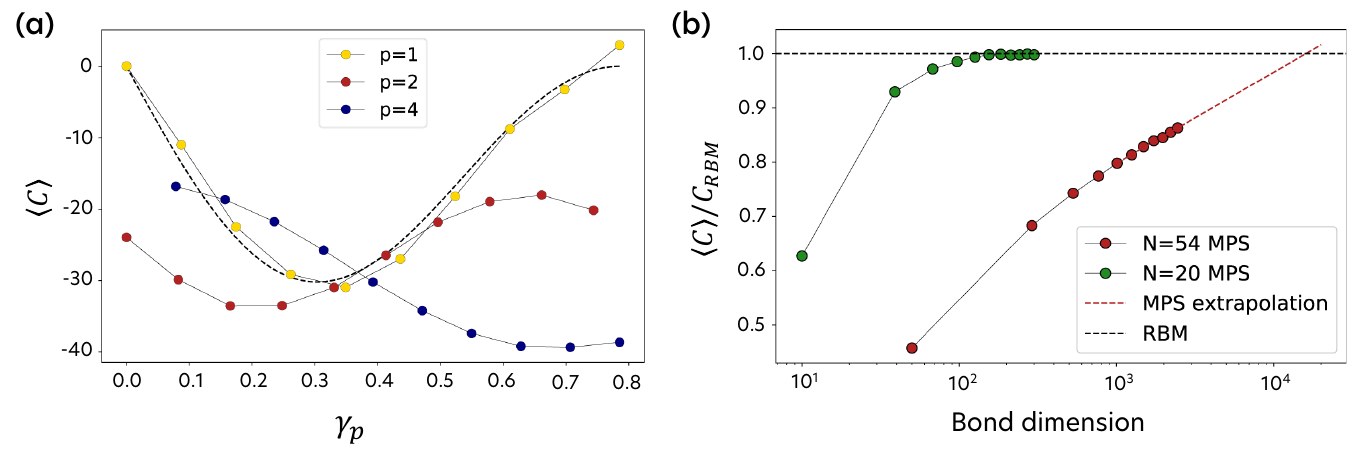}
    \caption[Quantum Approximate Optimization Algorithm with neural quantum states]{Results for the classical simulation of \ac{QAOA} for a~3-planar graph. Panel (a) shows $\langle \mathcal{C} \rangle$, the approximated cost function for various values of $p$, the depth of the quantum circuit. The dashed line corresponds to the exact simulation of the $p=1$ quantum circuit which the \ac{RBM} simulation accurately reproduces for this task. (b) Estimation of the required bond dimension in a~Matrix Product State (MPS) simulation of the \ac{QAOA} circuit to match the accuracy of the \ac{RBM}. For $N=54$ qubits, the required bond dimension is of about $10^4$, which amounts to using billions of parameters, whereas the \ac{RBM} uses a~few hundred. Taken from \ToggleForCUP{Medvidovi\'c, M. \& Carleo, G. (2021). \textit{Classical variational simulation of the quantum approximate optimization algorithm}. npj Quantum Inf. 7, 101~\cite{Medvidovi__2021} under the \href{https://creativecommons.org/licenses/by/4.0/}{CC BY 4.0 DEED} license.}{Ref.~\cite{Medvidovi__2021}.}}
    \label{fig:NQS_circuits}
\end{figure}}

\subsubsection{Open quantum systems}\label{NQS_sec:OQS}

The idea of using \acp{NN} to represent quantum states was also applied to open quantum systems. An~open quantum system is a~physical system that interacts with an~environment, for example an~array of atoms interacting with an~electromagnetic field. Rather than describing the full system+environment ensemble, one is generally only interested in the properties of the system (in the example above, the atoms) and one only keeps an~effective description of its interaction with the environment (the field). This description enables one to understand effects such as decoherence. In the Born-Markov approximation, the time evolution of an~open quantum system is given by the Lindblad master equation \cite{Breuer2007}
\begin{equation}\label{eq_NQSOpenSys1}
	\partial_t\hat{\rho} = -\text{i}[\hat{H},\hat{\rho}] + \sum_{i=1}^D\big(\hat{J}_i\hat{\rho}\hat{J}_i^\dagger - \frac{1}{2}\lbrace\hat{J}_i^\dagger\hat{J}_i,\hat{\rho}\rbrace\big) \equiv \mathbf{L}[\hat{\rho}],
	\end{equation}
where $\hat{\rho}$ is the system density operator, $\hat{H}$ is the system Hamiltonian ($\hbar = 1$), and $\hat{J}_i$ are so-called jump operators, that describe the system-environment interaction. We have also defined $\mathbf{L}$, the Liouvillian, which is to open quantum systems what the Hamiltonian is to closed quantum systems (up to an~$i$) -- their time evolution generator. Many works focus on finding the steady state that corresponds to the state $\hat{\rho}$ which satisfies $\mathbf{L}[\hat{\rho}] = 0$, or the dynamics of particular systems, which means one must in general integrate \cref{eq_NQSOpenSys1} in time, analogously to non-equilibrium dynamics of closed systems.
A~first difficulty one faces is finding a~correct representation for $\hat{\rho}$ in terms of an~\ac{NN}. Indeed, a~density matrix is harder to represent than a~wave function, because it has to be Hermitian, semi-positive, and of trace one. In fact, a~general method to encode a~density matrix into arbitrary \acp{NN} has still not been found. The key point of these works is that one can always purify a~density matrix, and write its elements as 
\begin{align}\label{eq_NQSOpenSys2}
    \mel{s}{\hat{\rho}}{s'} = \sum_{s'}\Psi(s,s')\Psi^*(s,s'),
\end{align}
with ${\Psi(s,s')}$ the purification that belongs to the joint Hilbert space composed of the system and an~imaginary ancilla (whose Hilbert space is of at least the same dimension as the system's Hilbert space).
With an~\ac{RBM} architecture, one can encode the purification $\Psi_{\params}(s,s')$ with an~\ac{NN}, and the \ac{RBM} architecture enables one to trace out the ancilla analytically spins $s'$ without explicitly performing the summation which in general requires exponentially many operations. For more details, see Refs.~\cite{Yoshioka_2019, Nagy_2019, Vicentini_2019, Hartmann_2019}.

A~second more recent approach proposes to view the density matrix as a~probability distribution over \acp{POVM}, and represent the resulting distribution using models employed in general density estimation, such as \acp{RNN} and \acp{ARNN}. In this formalism, the density matrix is simply written as:
\begin{align}\label{eq_NQSOpenSys3}
    \hat{\rho}_{\params} = \sum_a p_{\params}(a) \hat{M}_a,
\end{align}
with $\hat{M_a}$ \acp{POVM} that belong to a~chosen complete set of \acp{POVM}. This could, for example, be all the operators composed as tensor products of the Pauli operators and the identity. In this picture, one simply needs to encode the probability distribution $p_{\params}(a)$ with an~\ac{NN}. This \ac{POVM}-based representation is motivated by the fact that the density matrix can be viewed as an~ensemble of $4^N$ measurements, which is naively how experimentally one performs tomography to reconstruct the density matrix. This method alleviates the constraint on using an~\ac{RBM} for open systems, but does not guarantee positivity of the density matrix, which can lead to unphysical states. However, a~certain number of results using the \acp{POVM} encoding are promising \cite{luo2021autoregressive, reh2021timedependent}, and understanding in which regime they work best is a~key research direction. 
\ToggleForCUP{
\begin{figure}[p]
    \centering
    \includegraphics[scale=0.4]{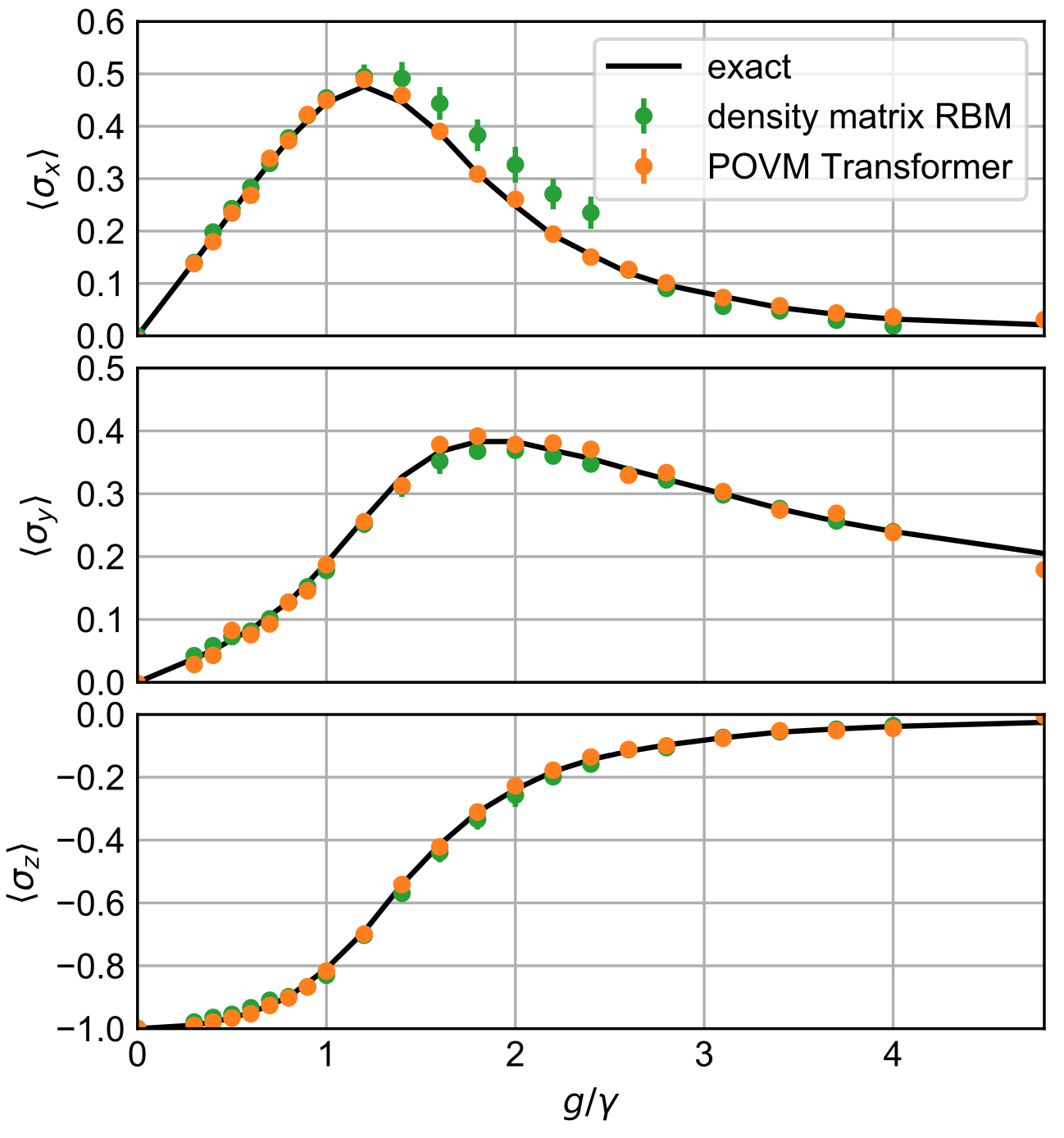}
    \caption[Open quantum systems with neural quantum states]{Expectation values of observables $\hat{\sigma}_k = 1/N \sum_i \hat{\sigma}_i^k$, $k \in \{x,y,z\}$ at the steady state of the open system described by the transverse-field Ising Hamiltonian and jump operators $\hat{J}_i = \sqrt{\gamma}\hat{\sigma}^-_i$ as a~function of $g/\gamma$, with $g$ the magnetic field strength. Results are shown for both the purified \ac{RBM} approach and the \ac{POVM} approach with a~Transformer network. Taken from \ToggleForCUP{Luo, D. \textit{et al.} (2022). \textit{Autoregressive neural network for simulating open quantum systems via a probabilistic formulation}. Phys. Rev. Lett. 128, 090501~\cite{luo2021autoregressive}.}{Ref.~\cite{luo2021autoregressive}.}}
    \label{fig:NQSopensys1}
\end{figure}}
{\begin{figure}[h]
    \centering
    \includegraphics[scale=0.4]{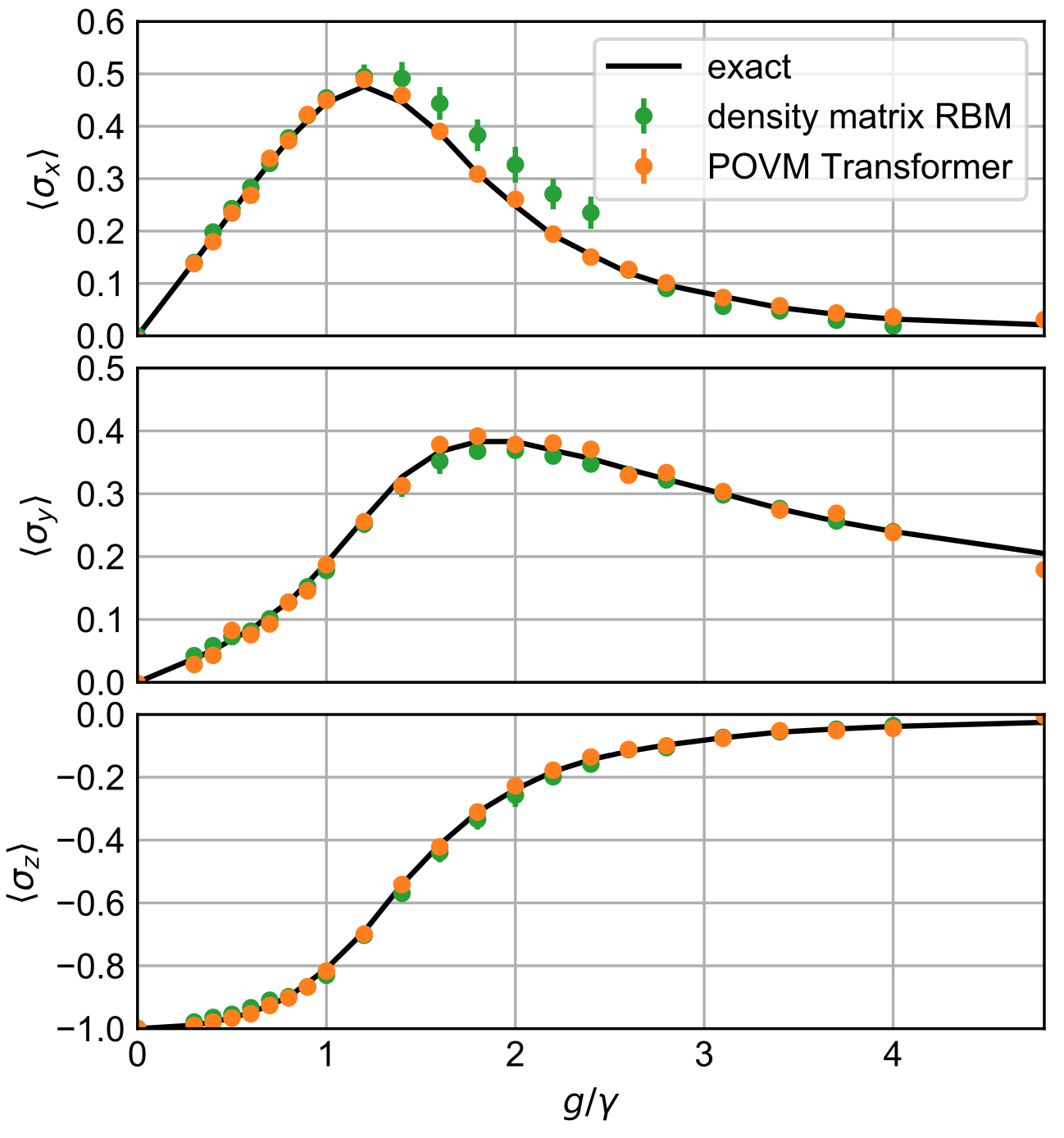}
    \caption[Open quantum systems with neural quantum states]{Expectation values of observables $\hat{\sigma}_k = 1/N \sum_i \hat{\sigma}_i^k$, $k \in \{x,y,z\}$ at the steady state of the open system described by the transverse-field Ising Hamiltonian and jump operators $\hat{J}_i = \sqrt{\gamma}\hat{\sigma}^-_i$ as a~function of $g/\gamma$, with $g$ the magnetic field strength. Results are shown for both the purified \ac{RBM} approach and the \ac{POVM} approach with a~Transformer network. Taken from \ToggleForCUP{Luo, D. \textit{et al.} (2022). \textit{Autoregressive neural network for simulating open quantum systems via a probabilistic formulation}. Phys. Rev. Lett. 128, 090501~\cite{luo2021autoregressive}.}{Ref.~\cite{luo2021autoregressive}.}}
    \label{fig:NQSopensys1}
\end{figure}}

Finding the steady-state(s) of open quantum systems is both challenging, due to the daunting size of the Liouvillian one would need to diagonalize ($4^N$ for a~spin system of $N$~spins), and interesting, for example, for the study of dissipative phase transitions \cite{Minganti2018}. Once a~parametrization $\hat{\rho}_{\params}$ of the density matrix is constructed, one can simply minimize the following cost function:
\begin{equation}\label{eq_NQSOpenSys4}
    \lossfun(\params) = \langle \hat{\rho}_{\params}  \mathbf{L}^\dagger \mathbf{L}\hat{\rho}_{\params}\rangle\,.
\end{equation}The obtained state corresponds to the zero eigenvalue is zero of the Liouvillian, which is the steady-state. For details about the procedure and how to retrieve the gradients, see \cite{Vicentini_2019, Hartmann_2019, Nagy_2019, Yoshioka_2019}. As one can see in \cref{fig:NQSopensys1}, this method has been applied to the dissipative version of the transverse-field Ising model, with good results for both a~POVM approach and an~\ac{RBM} approach. In the former case, the expressive power of the network is higher, but the positivity of the density matrix is not enforced. A~clear picture of when each approach fails or succeeds is still lacking, and is an~important research direction. 
The dynamics of open quantum systems is not described in detail here, but stochastic reconfiguration can also be used for open systems in both the \ac{RBM} \cite{Hartmann_2019} and POVM \cite{reh2021timedependent} approaches.

\subsubsection{Quantum state tomography}\index{quantum state tomography}\label{sss:quantum_tomography}

The future quantum technologies are fueled by quantum resources such as coherence, entanglement, or Bell nonlocality. One of the main challenges is the experimental certification of such properties for a given unknown quantum state \cite{GUHNE20091,PhysRevLett.107.210404,PhysRevLett.125.150503,PRXQuantum.2.010201,Friis2019,Eisert2020,Sotnikov2022,9996689,PRXQuantum.3.010317,10.21468/SciPostPhysCore.6.2.028,Hangleiter_2017,Frerot_2023}.
Information about quantum resources is encoded in the density matrix of the state, which can only be reconstructed based on finite-statistic measurements - this process is known as quantum state tomography \cite{PhysRevLett.74.4101,PhysRevLett.83.3103,PhysRevLett.92.220402, Haffner2005,PhysRevLett.105.150401,5714248,PhysRevLett.105.250403,Moroder_2012,Cramer2010,PhysRevLett.111.020401,Lanyon2017}. Density matrix reconstruction is a~challenging task - with increasing system size, the number of required measurements scales exponentially.
The field of quantum state tomography has been entered by artificial neural
networks proposing supervised deep-learning approaches  \cite{adri1,Pan2022,koutny2022PRA,Shahnawaz2021,ma2023attentionbased, palmieri2023enhancing}. The following paragraphs introduce basic concepts of neural networks-assisted quantum state tomography.
    \highlight{Quantum state tomography is the process of density matrix reconstruction based on finite-statistic measurement data. Due to the exponential growth of the Hilbert space with increasing system size, the number of required measurements also scales exponentially. However, with the help of deep neural networks, the density matrix reconstruction can be done with a~polynomial number of measurements.}

    Let us consider the task of reconstructing a~wave function $| \psi \rangle$ from a~limited number of snapshots $|\psi(s)|^2$ obtained by performing projective measurements in some basis spanned by $\ket{s}=\ket{s_1,s_2,\ldots,s_N}$, with $s_i$ some local quantum numbers and $N$ the size of the system. Then, the task, in the \ac{NQS} language, is simply to minimize:
    \begin{equation}\label{eq_NQStomo1}
        \underset{{\params}}{\min}\text{ dist}\left(|\psi_{\params}\rangle, |\psi\rangle \right)
    \end{equation} with ${\params}$ being some variational parameters, and $| \psi_{\params} \rangle$ the variational state to optimize, parametrized by a~neural network. The architecture of this network is left unspecified here, and all architectures work provided training can be performed efficiently. Many distances can be considered, but we here focus on the \acf{KL} divergence\index{Kullback-Leibler divergence} (see \cref{sss:probability}) as was first presented in the work by Torlai \textit{et al.} \cite{Torlai_2018}. It is defined as:
    \begin{equation}\label{eq_NQStomo2}
        \mathrm{D_{\mathrm{KL}}}(p||q) = \sum_{x \in \mathcal{P}} p(x)\log\frac{p(x)}{q(x)}
    \end{equation} for two probability distributions $p$ and $q$, defined on the same space $\mathcal{P}$. The application to quantum states is straightforward, as one can obtain probability distributions from the Born rule, i.e., $p_{\params}(s) = |\psi_{\params}(s)|^2$, $q(s) = |\psi(s)|^2$. 
    By taking $x$ to be configurations $s$ in some set $S$ of snapshots, one can simply minimize:
    \begin{equation}\label{eq_NQStomo3}
        \mathrm{D_{KL}}({\params}) = \sum_{s \in S} |\psi_{\params}(s)|^2\log\frac{|\psi_{\params}(s)|^2}{|\psi(s)|^2}
    \end{equation} which concludes one possible approach. 
    
    However, recall that a~quantum state is not simply a~probability distribution. A~probability distribution can \textit{always} be defined from a~quantum state, but \textit{not the reverse}. More explicitly, we want to reconstruct the full quantum state whose amplitudes are $\psi(s) = \sqrt{Q(s)}e^{i\phi(s)}$. By minimizing \cref{eq_NQStomo3}, \stress{information about the phase, $\phi(s)$, is lost.} This difference is crucial and is at the heart of many issues in learning quantum states. As mentioned in \cref{sec:NN_q_states}, learning the phase of a~frustrated quantum state is challenging \cite{Szabo2020}.
    The elegant solution to this problem is to \stress{consider measurements performed in different bases}. Indeed, the form of the quantum state in a~different basis involves the interference between amplitudes in different bases. Hence, matching the probability distribution defined by snapshots in different measurement bases leads to the correct quantum state as long as the bases contain enough information about the quantum state. Mathematically, one can simply replace \cref{eq_NQStomo3} by:
    \begin{equation}\label{eq_NQStomo4}
        \mathrm{D_{KL}}({\params}) = \sum_B\sum_{s \in S_B} |\psi^B_{{\params}}(s)|^2\log\frac{|\psi^B_{{\params}}(s)|^2}{|\psi^B(s)|^2}
    \end{equation} where $S_B$ is the set of snapshots of the quantum state in basis $B$, and $\psi^B(s) = \bra{s}\hat{U}_B\ket{\psi}$ with $\hat{U}_B$ a~unitary operator. Then, gradients are found as usual, either with automatic differentiation or analytically with simple models such as \acp{RBM}.  
    
    In \cref{fig:MLEXP_tomographyRNN}, various observables are shown for a~synthetic state and a~reconstructed state. The synthetic state approximates the ground state of the Heisenberg model in a~triangular lattice, that authors of the corresponding work generated with tensor network simulations~\cite{carrasquilla2019}. The reconstructed state was obtained employing the ideas presented in this section with an~\ac{RNN} architecture (for more details, see the Introduction or~\cref{sec:NN_q_states,sec:hot-topics:DE}), using a~\ac{POVM} representation of quantum states (for more details, see~\cref{NQS_sec:OQS}). Note that the approach has been extended to reconstruct mixed states \cite{schmale2021scalable}, although additional care must be taken to avoid issues related to positivity of the reconstructed density matrix, similar to what was presented above.
    
\ToggleForCUP{
\begin{figure}[p]
    \centering
    \includegraphics[width=0.95\columnwidth]{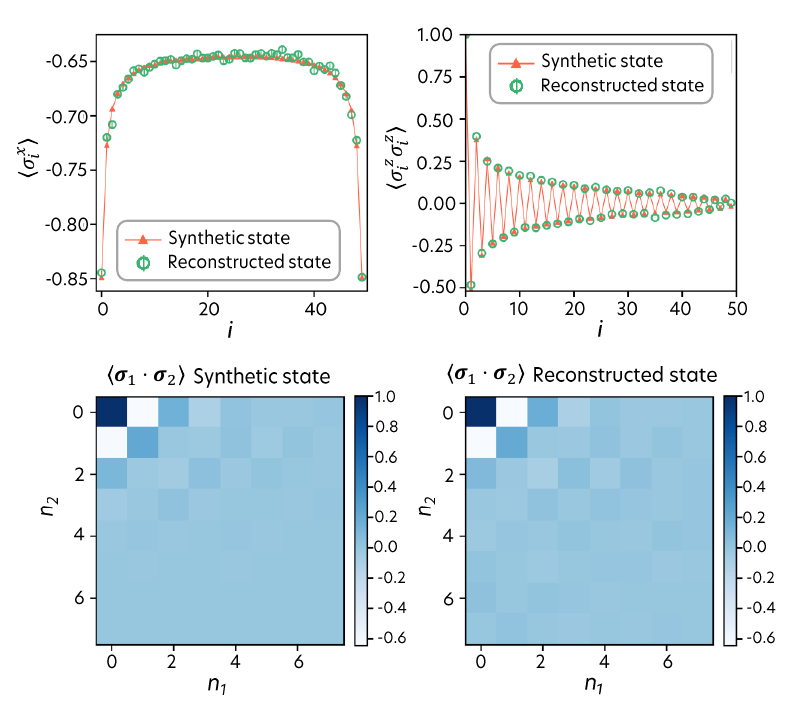}
    \caption[Quantum tomography with a~recurrent neural network]{Various observables corresponding to the ground state of the Heisenberg model on a~triangular lattice with $N=50$ spins. (a) Average magnetization along x for each spins $i$. (b) Spin-spin correlation function between the spin at site $1$ with spins at site $i$. (c) and (d) Average spin-spin correlation between the first and the $i$-th spin. One can see that all observables are reproduced with a~very high precision. Adapted from \ToggleForCUP{Carrasquilla, J. \textit{et al.} (2019). \textit{Reconstructing quantum states with generative models}. Nat. Mach. Intell. 1, 155–161~\cite{carrasquilla2019} with permission from Springer Nature.}{Ref.~\cite{carrasquilla2019}.}}
    \label{fig:MLEXP_tomographyRNN}
    \end{figure}
    }{
    \begin{figure}[h]
    \centering
    \includegraphics[width=0.95\columnwidth]{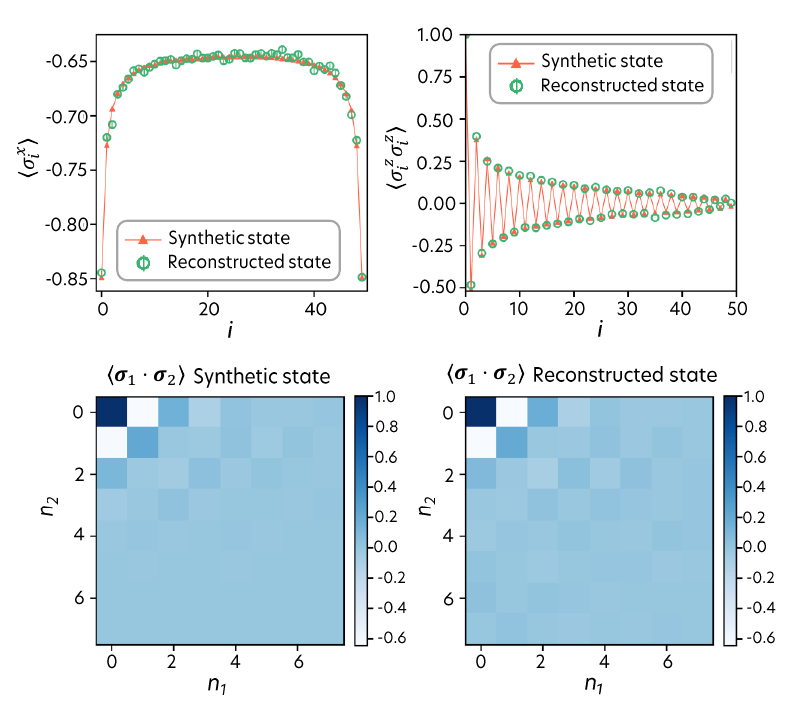}
    \caption[Quantum tomography with a~recurrent neural network]{Various observables corresponding to the ground state of the Heisenberg model on a~triangular lattice with $N=50$ spins. (a) Average magnetization along x for each spins $i$. (b) Spin-spin correlation function between the spin at site $1$ with spins at site $i$. (c) and (d) Average spin-spin correlation between the first and the $i$-th spin. One can see that all observables are reproduced with a~very high precision. Adapted from \ToggleForCUP{Carrasquilla, J. \textit{et al.} (2019). \textit{Reconstructing quantum states with generative models}. Nat. Mach. Intell. 1, 155–161~\cite{carrasquilla2019} with permission from Springer Nature.}{Ref.~\cite{carrasquilla2019}.}}
    \label{fig:MLEXP_tomographyRNN}
    \end{figure}
    }

    Experimentally, one can implement this strategy by applying rotations with, for instance, laser pulses, and then measure the system repeatedly. In Ref.~\cite{Torlai2019}, authors demonstrate the first state reconstruction from experimental data from a~programmable array of Rb atoms, using an~\ac{RBM} architecture. Here snapshots of the wave function in the $\hat{\sigma}^z$ basis are obtained through site-resolved fluorescence imaging. A~challenge that arises when using real data is that noise is introduced, which comes from measurement errors, leading to a~set of snapshots $|\psi(s)|^2$ that imperfectly match the state of the system. This is taken care of in this work by adding a~noise layer to the neural network, with which the snapshots are transformed to filter out the noise during training. At the expense of increasing the total number of parameters in the network, this is a~quick and easy strategy to deal with experimental noise, enabling high-fidelity state reconstruction. Since quantum state tomography with \ac{NQS} has been proposed, substantial efforts have been made to implement it in real-world experiments. For more details on the experimental challenges of such proposals, see for instance, Refs.~\cite{Lohani_2020, Lohani2021_2, Lohani2021, Lohani2022, Danaci_2020}.
    The underlying principle behind these approaches is that with a~polynomial number of bases $B$ and a~polynomial number of snapshots, one should be able to reconstruct states belonging to a~certain class (not fully random states, for instance, which contain almost no structure). As underlined previously, this class is not exactly known, and is the subject of current research. To draw an~analogy with images, images are not fully random; they contain a~lot of hidden structure, that can be learned by a~properly designed and trained neural network. The hope is that the same is true for quantum states, and investigating the limits of such techniques could also help us understand in more detail their hidden structure, beyond what has been found with entanglement properties through the study of \acp{TN}.
    
    Finally, we mention the randomized measurement techniques that allow predicting selected properties of spin-$1/2$ quantum systems without reconstructing the full quantum state, the so-called ``shadow tomography\index{quantum tomography!shadow tomography}'' (see Refs.~\cite{aaronson2018shadow,  Huang2020, 10.1145/3313276.3316378,altepeter20044,  10.1145/2897518.2897544,Koh2022classicalshadows,PhysRevLett.125.200501,  Elben2023}).
    Shadow tomography allows estimation of the expectation value of the given observable based on data collected during repetitive measurements prepared in a randomly chosen basis of each spin separately. However, the number of required measurements scales exponentially with the locality of the operator averages to be reconstructed, albeit with an a~runtime that is typically better than the naive direct measurement of the observables from the data.  In contrast to these approaches, in this section we have instead considered the task of training a~low-dimensional representation of the full wave function from a~limited number of measurements.

\subsection{Outlook and open problems}

We hope to have provided enough material to stimulate further research in the growing field of \acp{NQS}. Here is a~non-exhaustive list of open problems and challenges related to the above discussion:
\begin{itemize}
    \item \textbf{Capacity of \ac{NQS}}. Some works have proven the capability of \acp{NQS} to represent volume-law entanglement, which means they could outperform \acp{TNS} for strongly correlated and two- and three-dimensional systems \cite{Deng2017, sharir2021, levine2019}. Others have proven the equivalence of \acp{RBM} with matrix-product states, meaning that the former cannot represent more states than the latter \cite{Chen2018PRB}. Even though general theorems have been found, knowledge about specific architectures is still rare, and understanding which architectures perform better on which problems is a~crucial point. In addition, proving representativity does not mean that the models can be efficiently trained, thus understanding how the training of \ac{NQS} models works is key.
    \item \textbf{Long-time dynamics}. Long-time dynamics remains a relatively untouched area for \ac{NQS}, due to stability issues of stochastic reconfiguration~\cite{hofmann2021role}. However, progress has been made thanks to regularization techniques \cite{Schmitt_2020}. Some works proposed infidelity minimization~\cite{gutierrez2021real, donatella2023dynamics}, which enables going beyond stochastic reconfiguration for regimes where its performance is poor. In Ref.~\cite{sinibaldi2023unbiasing}, a systematic bias that appears when performing time evolution was explained, which should stimulate progress in long-time dynamics, where ample results on large lattices are still lacking with \ac{NQS}.
    \item \textbf{Open quantum systems}. No general method of encoding a~density matrix into an~arbitrary neural network has been found yet; one is either forced to use an~\ac{RBM}, which has known limitations, or one can use a~POVM approach, which may fail due to non-positive density matrices.
    \item \textbf{Frustrated systems}. Finding the ground state of frustrated systems with an~\ac{NQS} approach has proved to be challenging \cite{Szabo2020}, and understanding exactly how one can improve the optimization of the procedure to learn the phase (which has a~nontrivial sign structure) is of particular interest.
    \item \textbf{Simulation of quantum circuits}. Few results have been obtained with networks other than \acp{RBM}, and investigating how different circuits affect the accuracy of the chosen ansatz can lead to results in two ways: understanding the complexity of a~given circuit and the limitations of the chosen ansatz.
    \item \textbf{Quantum state tomography.} Quantum state tomography based on neural networks is still in its infancy. So far, it has only been explored numerically on toy models and small experimental settings where traditional quantum state tomography is still feasible. It is likely that its real benefits may emerge in the context of estimation of difficult quantities in quantum simulation. In this setting, the complexity of estimation arises because even simple quantities, such as energy and other correlation functions, can have high variance. This implies that some of these quantities have a~sample complexity, which can grow quickly with the size of the system.
    
    \item \textbf{Quantum resources certification.}
    The generation of quantum resources can be performed dynamically by means of the one-axis twisting protocol~\cite{PhysRevA.47.5138,wineland1994squeezed}. One-axis twisting can be implemented with ultra-cold atoms in optical lattices to generate many-body entanglement and many-body Bell correlations~\cite{Plodzien2020, Plodzien2022, Plodzie2023generation, Hernandez2022,Dziurawiec2023,yanes2023spin}. The challenge for this technique is to verify the quantum resources generated in many-qubit systems, which can be done with the help of \ac{DL}\cite{palmieri2023enhancing}.
    
    \item \textbf{Extension to continuous Hilbert spaces and bosonic systems.} For now most techniques and works have focused on systems with discrete degrees of freedom (such as spins). Extensions to continuous Hilbert spaces have been addressed, for example, in the context of quantum chemistry \cite{hermann2020deep,pfau2020ab,Hermann2022review} and nuclear matter \cite{adams_variational_2021}. Efficient encodings for bosonic Hilbert spaces would also be of particular interest for photonic systems, for example, which are usually treated with mean-field-like approaches.
    \item \textbf{Applications in quantum information.} As mentioned previously, \ac{NQS} have been used to simulate quantum circuits. They have also been applied to quantum codes \cite{Bausch_2020} for quantum error correction and quantum communication. In this paper, the authors demonstrate that efficient quantum codes can be learned by $\ac{NQS}$ according to which noise channels a physical system is subject to. $\ac{NQS}$ have not yet been widely used for quantum information, and we expect them to be useful tools for this field in the coming years.
    
\end{itemize}
\subsection*{Further reading}
\begin{itemize}
    \item Carleo, G. \& Troyer, M. (2017). \href{https://www.science.org/doi/full/10.1126/science.aag2302}{\textit{Solving the quantum many-body problem with artificial neural networks}}. The original paper by Carleo and Troyer that introduced \ac{NQS}~\cite{Carleo_2017}.
    \item Becca F. and Sorella, S. (2017). \href{https://www.cambridge.org/core/books/quantum-monte-carlo-approaches-for-correlated-systems/EB88C86BD9553A0738BDAE400D0B2900}{\textit{Quantum Monte Carlo Approaches for Correlated Systems}}. A~comprehensive book that includes details on quantum Monte-Carlo methods, and variational states\index{variational state} \cite{becca_sorella_2017}.
    \item Vicentini, F. \textit{et al.} (2021). \href{https://arxiv.org/abs/2112.10526}{\textit{NetKet 3: Machine learning toolbox for many-body quantum systems}}. The paper accompanying the open-source library NetKet 3, which contains an~extensive discussion of how to implement several algorithms introduced in this chapter, as well as a~collection of tutorials showing how to solve some benchmark problems with \ac{NQS} \cite{Vicentini:2021pcv,Vicentini2021Nat}.
    \item Carrasquilla, J. \& Torlai, G. (2021). \href{https://journals.aps.org/prxquantum/abstract/10.1103/PRXQuantum.2.040201}{\textit{How to use neural networks to investigate quantum many-body physics}}. A~recent tutorial by Carrasquilla and Torlai that includes interesting applications and code snippets can help anyone who wants to start in the field \cite{Carrasquilla2021PRXQuantum}.
    \item Carleo, G. (2017). \href{https://gitlab.com/nqs/ucas_workshop}{Repository for example codes presented at the ``Machine Learning and Many-Body Physics'' workshop}. Notes, exercises, and code produced for the 2017 Beijing workshop on Machine Learning and Many-Body Physics~\cite{carleo_Beijing_nots}.
\end{itemize}

\clearpage
\section{Reinforcement learning}
\label{sec:RL}

So far, we have encountered multiple \ac{ML} scenarios featuring supervised or unsupervised learning problems where we want to infer some labels, predict certain values, or find patterns in the data. In this chapter, we describe a~different approach: \stress{learning strategies}.

In the supervised learning framework, we can think of a~student who learns from a~teacher who knows the correct answers to all possible questions within a~given domain. 
In this scheme, the student is limited by the teacher's knowledge and can never surpass it or address questions outside the teacher's expertise. 
To overcome this limitation, in \acf{RL}\index{reinforcement learning}, we remove the teacher and let the student try things out and learn from the resulting experience.
We refer to the student as the \stress{agent}\index{agent}, as it can actively take actions.
Just like us humans, the agent learns from the interaction with an \stress{environment}\index{environment}, understands the consequences of its actions, and finds strategies to achieve particular goals.

For instance, let us consider the case in which we teach an~agent to play chess. 
A~supervised learning approach would consist of training an~\ac{ML} model to reproduce the moves of recorded chess games by the best players in the world. 
In this setting, given a~state of the game, i.e., the position of the remaining pieces on the chessboard, the model predicts the move such reference players would make. 
However, this approach suffers from some major shortcomings.
For example, there is no single optimal move for every situation, and the moves strongly depend on the game strategy adopted by the players. 
As a~result, the agent may be unable to consistently execute a~strategy through various actions. 
Additionally, the agent's performance is ultimately limited by the quality of the training data, meaning that it may be impossible to outperform the reference players. We refer to~\cref{sec:rl_go_atari} for a~related example.

Instead, we can let the agent play chess games, either against various opponents or even against itself, without providing any additional knowledge besides the rules. 
In that case, it develops its own understanding of the game and devises its own strategies. 
The resulting agent's potential is far superior to the previous one, as it is not limited by its teacher. 
Nevertheless, learning from experience may be challenging, provided that the quality of the actions is only assessed at the very end of the game when the outcome is decided: victory or loss.\footnote{In some cases, we may be tempted to add intermediate rewards, such as a~bonus for taking out a~piece from the opponent. However, in doing so, we effectively change the game and its goal, and, as a~consequence, we might fail to find the optimal strategy of the original problem.} 
Hence, the agent must develop a~deep understanding of the long-term consequences of the actions based on the sparse feedback from the environment.

Framing problems as games to discover strategies has countless applications. In particular, control problems naturally fit this framework. However, we can design games to obtain any protocols or algorithms of interest, from new quantum experiments~\cite{melnikov2018active} to faster matrix multiplication or sorting algorithms~\cite{Fawzi2022Nature,Mankowitz2023Nature}. Here, we show how to tackle some paradigmatic problems in the field of quantum technologies with \ac{RL}.

In this chapter, we introduce the field of \ac{RL}. We start with an~intuitive view on the concept of learning from experience and its mathematical foundations in~\cref{sec:rl_foundations}. 
Then, we present two main approaches: value-based \ac{RL} in~\cref{sec:rl_value_based_methods}, and policy gradient in~\cref{sec:rl_policy_gradient}. In~\cref{ssec:rl_actor_critic}, we combine the two paradigms, introducing actor-critic algorithms. Then, we provide an~alternative approach to \ac{RL}, projective simulation, in~\cref{sec:rl_projective_simulation}. Finally, we present a~series of application examples of \ac{RL} in~\cref{sec:rl_examples}, featuring superhuman performance in games as well as various problems in quantum technologies.

\subsection{Foundations of reinforcement learning}
\label{sec:rl_foundations}

\begin{figure}[t]
    \centering
    \includegraphics[width=0.7\columnwidth]{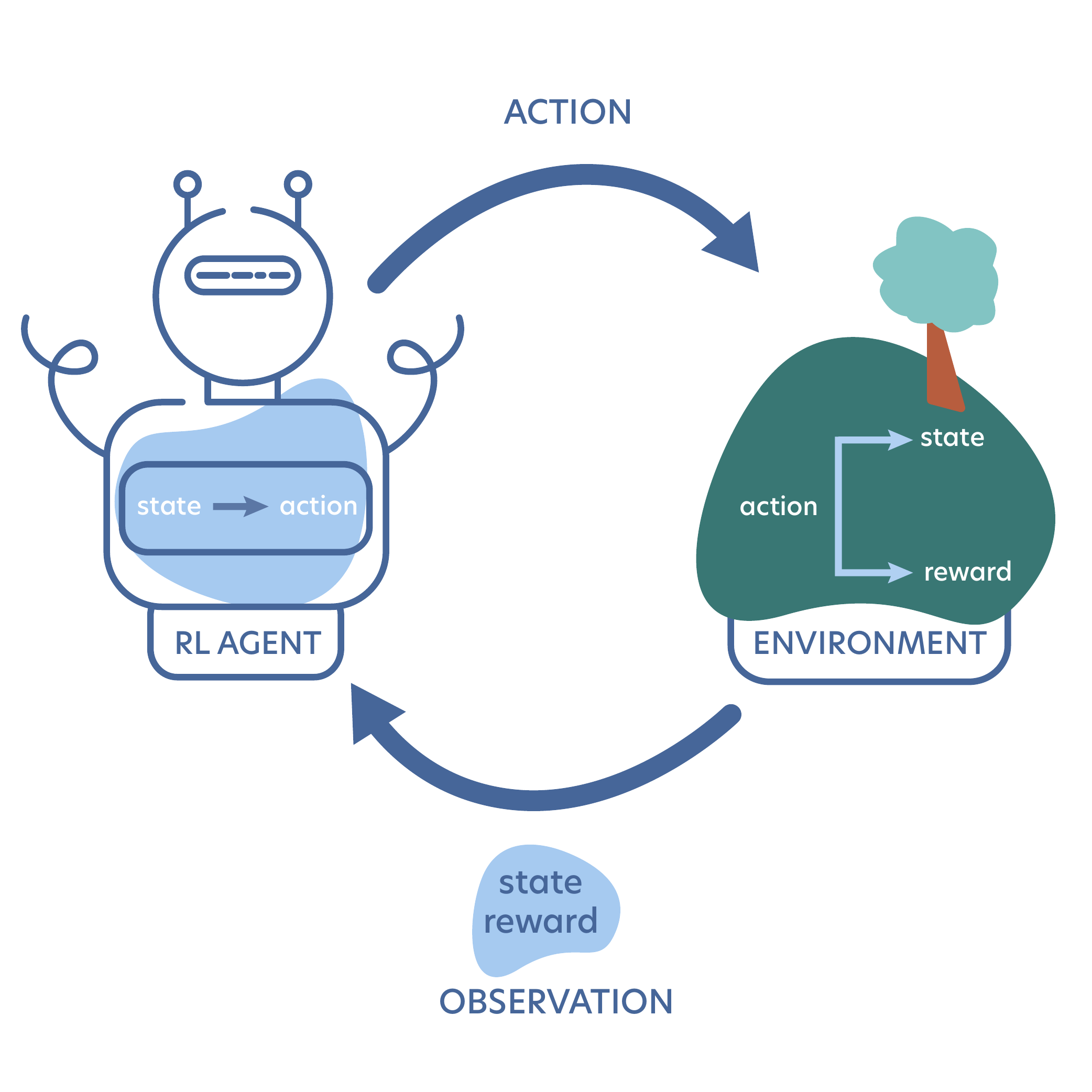}
    \caption[Overview of the basic reinforcement learning setting]{Overview of the basic \ac{RL} setting. The agent receives an~observation from the environment. Given the observation, it chooses the next action according to its policy. The environment determines the outcome of the action, and it returns an~observation to the agent consisting of the new state and a~potential reward.}
    \label{fig:rl_overview}
\end{figure}

The general setting of any \ac{RL} problem consists of two main elements: an~\stress{agent}\index{agent}, and an~\stress{environment}\index{environment} that it interacts with, as illustrated in \cref{fig:rl_overview}. 
The environment contains all the information defining the problem at hand, e.g., the rules of a~game, and it provides the agent with observations and feedback according to its \stress{actions}\index{action}.
The environment defines the set of all possible \stress{states}\index{state (\ac{RL})}, $s\in\statespace$, which can range from an~empty set, in the case of a~stateless environment (see the first example in \cref{sec:policy_gradient_toy_examples}), to a~multi-dimensional continuous space. 
For example, these could be all the possible configurations of a~board game or all the possible combinations of joint angles in a~robot. 

The agent can observe (sometimes only partially) the state $s$ of the environment, and it can choose an~action $a$ to perform, which may include the possibility of remaining idle. 
The action is chosen from the set of possible actions, $a\in\actionspace$, which is defined by the environment and can be state-dependent. 
For instance, the action of pushing forward a~pawn in chess is only possible if there is a~free position in front of it. 
The actions may alter the state in which the environment is found, and they can have deterministic or stochastic outcomes.
In the chess example, all the actions are deterministic. 
In contrast, in the case of a~walking robot, the action to move forward may have different results: it can succeed in doing so, the robot may trip, or it may even remain idle with a~certain probability due to a~hurdle or malfunctioning. 
This information is encoded in the environment, and the agent may not have access to it.

Nevertheless, every time the agent performs an~action, the environment provides it with an~observation of the new state together with a~feedback signal called \stress{reward}\index{reward}, $r$. 
The reward can take any numerical value. 
It may depend on the previous state, the new state, and the action that was taken. 
The main purpose of the agent is to maximize the obtained rewards by the end of the task, and it is, therefore, the quantity that defines the objective task. 
Hence, the agent obtains higher rewards when accomplishing the objective task or progressing toward the goal, e.g., winning a~game, while it might receive penalties when performing harmful or bad actions, e.g., losing a~game.

\highlight{%
The central objective of any \ac{RL} problem is to learn the \stress{optimal policy}\index{policy!optimal policy}, $\pi^*$, that maximizes the obtained rewards. 
A~policy\index{policy}, $\pi$, dictates which actions to take given the observations and thereby defines the strategy followed by the agent.}

In general, the policy can take any form, as we show in forthcoming sections. 
For example, it can be a~table assigning the best possible action to every possible state or an \ac{ML} model that, given a~state, provides a~probability distribution over all the possible actions. 
However, the learned policy is specific to the problem. 
We summarize the introduced key elements of the \ac{RL} setting in~\cref{fig:rl_overview}.

Let us provide some insight on the main elements of the \ac{RL} setting with a~couple of examples. 
In the case of the chess game from \cref{sec:RL}, the agent is one of the players. 
The environment models the game's rules, the opponent,\footnote{The opponent could be the same agent, which would play against itself, but each agent would perceive the other as part of their respective environment. This is known as \stress{self-play}, and it helps explore new strategies faster.} and its states that correspond to the piece positions on the board.\footnote{The state for chess can also contain extra information, such as whether castling is still possible. For the purpose of this example, and to keep it simpler, we restrict ourselves only to the piece positions here} 
The state space\index{Markov decision process}\index{Markov decision process!state space} contains all the possible board configurations that can be reached within a~game, e.g., excluding those where one of the kings is missing. 
The action space\index{Markov decision process!action space} corresponds to all the possible legal moves that can be made at every turn. 
In this case, the agent does not obtain rewards until the game is resolved.
At this point, the agent receives a~positive or negative reward upon victory or defeat, respectively. In case of a~draw, the final reward could be zero or even negative. The goal is to learn the policy that yields the highest possible number of victories.

As a~second example, we consider a~robot trapped in a~maze. The robot can only see its immediate surroundings and has to maneuver to reach a~target location. In this case, the agent is the robot, and the environment models the maze, its walls, and the target location. The state is the current position of the agent plus its immediate surroundings, and the state space comprises all the reachable locations. The action space contains the moves in all possible directions, and the environment ensures that the agent does not cross the walls. Hence, moving into a~wall would leave the agent in the same position and, therefore, would not modify the state. As a~reward, we can provide the agent with a~constant negative reward after every move in order to encourage it to take the least amount of steps toward the goal. 

\subsubsection{Delayed rewards}
\label{sec:rl_delayed_rewards}

As we have previously introduced, the reward\index{reward} $r$ is a~key concept in \ac{RL}. The agent learns to maximize the reward, and therefore, the quantity defines the problem. At a~given discrete time $t$, the agent observes a~state $s_t$ and performs an~action $a_t$ according to its policy. Then, the environment presents the agent with a~new state $s_{t+1}$ and a~reward $r_{t+1}$. Hence, $r$~is time-dependent, and it may depend on any of the other three quantities $r_t = r\left(s_{t-1}, a_{t-1}, s_t\right)$ (see~\cref{sec:MDP} for further details).

So far, we have briefly talked about maximizing the rewards. In order to formalize the \ac{RL} objective, we need to introduce the notion of \stress{delayed rewards}. They introduce the idea of ``looking ahead'' to the agent, allowing it to account for the future rewards obtained along a~trajectory through the state space. However, we can penalize the rewards that are far into the future with a~discount factor\index{discount factor} $\gamma \in [0,1]$.

\highlight{%
The discount factor weights the rewards according to their temporal separation. This way, immediate rewards have larger weights than those far into the future.
The \ac{RL} objective is to maximize the \stress{discounted return}\index{return!discounted return}, defined as the weighted sum of future rewards
\begin{equation}
    G_t = \sum_{k=0}^{T-t-1} \gamma^{k}r_{t+k+1},
    \label{eq:rl_definition_return}
    \end{equation}
which accounts for the rewards obtained starting at time $t$ until the final time $T$.\footnote{In \ac{RL}, we typically consider finite trajectories. However, a~discount factor $0\leq\gamma < 1$ allows us to consider infinite trajectories $T=\infty$ with finite returns.}}

Notice that the return presents a~recursive form that is essential for many \ac{RL} algorithms
\begin{equation}
    \label{eq:rl_return_recursive}
    G_t = r_{t+1} + \gamma G_{t+1}\,.
\end{equation}

\begin{figure}[t]
    \centering
    \includegraphics[width=0.9\columnwidth]{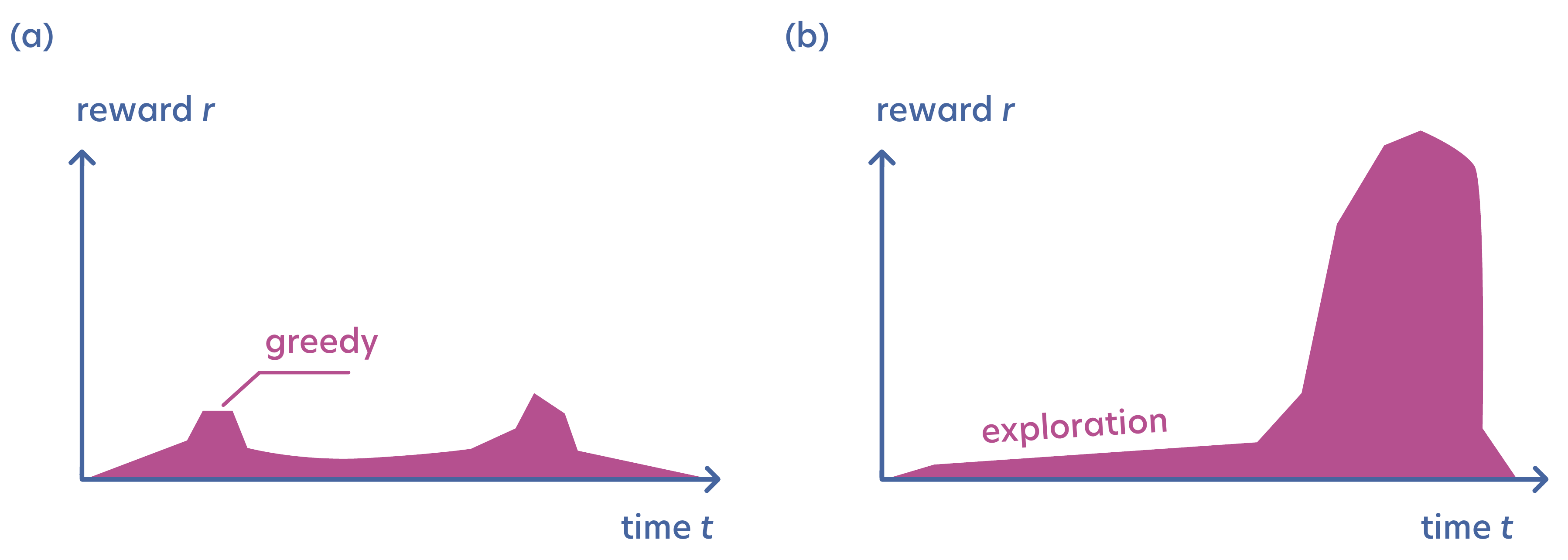}
    \caption[Short-term and long-terms rewards in reinforcement learning algorithms]{Impact of the discount factor, $\gamma$, in \ac{RL} algorithms. (a) A~myopic algorithm ($\gamma\to0$) may settle for a~greedy policy that leads to early immediate rewards, even if they are smaller than possible latter ones. (b) However, a~long-term oriented algorithm ($\gamma\to1$) might sacrifice early rewards in favor of larger late ones.}
    \label{fig:rl_delayed_rewards}
\end{figure}

This concept draws inspiration from human psychology, and it mimics our daily observation that far-term rewards, even if high, are less desired than near-term ones, e.g., we favor procrastinating instead of reading this book. We can distinguish two limits: for a small discount factor, $\gamma\to0$, the return\index{return} becomes myopic, i.e., immediate rewards predominate over any other possible future ones. On the other hand, large discount factors, $\gamma\to1$, result in equal weights for early and late rewards, which encourage long-term-oriented strategies. This includes, in particular, the deliberate choice to perform a few seemingly sub-optimal choices in the beginning that, however, result in a far greater final return. We depict the two cases in \cref{fig:rl_delayed_rewards}.

Hence, the discount factor strongly affects the resulting policy. In fact, it defines the $\ac{RL}$ task, as the agent aims to maximize the return, introduced in~\cref{eq:rl_definition_return}. Nevertheless, we often rely on trial-and-error methods to find the discount factor that best suits our needs.

\subsubsection{Exploration and exploitation}
\label{sec:rl_exploration_exploitation}

In \ac{RL} we encounter a~trade-off between exploration\index{exploration} and exploitation\index{exploitation}. In order to maximize the return, the agent must \stress{exploit} its knowledge about good strategies. However, the agent must \stress{explore} other different actions in order to improve them or even discover better strategies in the future. 

However, a~learning algorithm cannot rely on exploration alone, as it would be reduced to a~brute-force search algorithm. Conversely, in a~case of pure exploitation, the agent would blindly commit to the first working strategy that it found, even if it was highly sub-optimal. Hence, we need to find a~balance between both regimes in which the agent can try several actions and progressively favor the best ones. This way, the exploration is conducted around the most promising areas of the state and action spaces, heavily reducing the amount of experience that the agent must gather in order to find the optimal policy. 

A~common strategy to balance exploration and exploitation is the so-called \stress{$\varepsilon$-greedy policy}\index{policy!$\varepsilon$-greedy policy}. In this case, the agent follows its policy to perform actions (exploits), and it may take a~random action (explores) with probability $\varepsilon\in[0, 1]$ at any point. This approach encompasses both paradigms: for $\varepsilon = 1$, we have full exploration, whereas we have full exploitation for $\varepsilon = 0$. By tuning $\varepsilon$, we interpolate between both regimes. A~common practice is to start with high $\varepsilon$, to enforce early exploration, and decrease it during the training process.

\subsubsection{Markov decision processes}
\label{sec:MDP}

All \ac{RL} problems are modeled by the same underlying mathematical structure: \Acp{MDP}\index{Markov decision process}. They constitute a~general framework to model environments with a~notion of sequentiality between states. In such environments, the future is independent of the past, given the present. This is known as the \stress{Markov property}\index{Markov property}.

\highlight{%
In essence, the \stress{Markov property} means that the current state is a~sufficient statistic containing all the required information relevant to the possible evolution of the environment. In particular, we do not have any memory effects from previously visited states. Formally, at any time step $t$, 
\begin{equation}
    p(s_{t+1} | s_0, \dots, s_t) = p(s_{t+1} | s_t ).
    \label{eq:rl_Markov_property}
\end{equation}}

Mathematically, an~\ac{MDP} is a~tuple $(\statespace, \actionspace, p, G, \gamma)$, respectively denoting the state space $\statespace$\index{Markov decision process!state space}, the action space $\actionspace$\index{Markov decision process!action space}, the \stress{dynamics} $p$, the set of total returns $G$, and the discount factor $\gamma$. In this formalism, the return $G$, together with the discount factor $\gamma$, determines the objective, and $p$ describes the dynamics of the environment\index{environment!environment dynamics},
\begin{equation}
p (s', r|s, a) =  p(s_{t+1}=s',r_{t+1}=r| s_t = s, a_t = a)\,,
\label{eq:rl_MDP_dynamics}
\end{equation}
which corresponds to the joint probability of observing a~new state $s'$ and obtaining a~reward $r$ by performing action $a$ in state $s$. For fully deterministic environments, $p(s',r|s,a)$ is either zero or one. 

From \cref{eq:rl_MDP_dynamics} we can derive all the relevant information about the environment. For instance, \stress{state-transition probabilities} are a~central quantity in many \ac{RL} algorithms:
\begin{equation}
    p(s'|s,a) = \sum_{r}p(s',r|s,a)\,.
\end{equation}
Furthermore, it allows us to determine the reward functions. In \cref{sec:rl_delayed_rewards}, we briefly introduce the reward function $r(s,a,s')$. In the most general form, the reward is jointly determined with the state $s'$, as shown in~\cref{eq:rl_MDP_dynamics}.\footnote{In stochastic environments, the reward can be inherently sampled from a~probability distribution. Consider the game of blackjack: with the same hand (state), the action of settling may have different rewards depending on the opponent's hand (environment). Hence, the reward is stochastic.} However, in many cases, we may need to consider the expected rewards for state$-$action pairs and state$-$action$-$next-state triplets:
\begin{align}
    r(s,a) &= \sum_{r}\sum_{s' \in \statespace} r p(s',r|s,a)\,, \\
    r(s,a,s') &= \sum_{r}r\frac{p(s',r|s,a)}{p(s'|s,a)}\,.
    \end{align}

In the iterative interaction between the agent and environment, the agent chooses the actions according to a~\stress{policy}\index{policy}. The policy is a~mapping from states to the probability of performing each possible action
\begin{equation}
    \pi(a|s) = p(a_t = a~| s_t = s)\,.
    \label{eq:rl_definition_policy}
\end{equation}
In the limit of deterministic policies, $\pi(a|s)$ is one for a~single action and zero for the rest. 

The policy is improved with the experience gathered from interacting with the environment to achieve the goal. This interaction generates \stress{trajectories}\index{Markov decision process!trajectory} of the form $$s_0, a_0, r_1, s_1, a_1, r_2, s_2, a_2, \dots, s_T\,,$$ where all states, actions and rewards are random variables. This way, the agent performs a~trajectory through the state-action space $\tau=a_0,s_1,a_1,\dots,s_T$ with probability 
\begin{equation}
    p(\tau) = \prod_{t=0}^{T-1} p(s_{t+1}|s_t,a_t)\pi(a_t|s_t)\,,
    \label{eq:rl_trajectory}
\end{equation}
starting from an~initial state $s_0$. We denote the discounted return \index{return!discounted return} associated to the trajectory as $G(\tau)=\sum_{t=0}^{T-1} \gamma^{t}r_{t+1}$.

This entire formalism holds assuming the Markov property from~\cref{eq:rl_Markov_property}, which implies that the environment is memory-less.
However, we may encounter situations in which the environment has certain memory effects, such as games in which the execution of a~sequence of actions yields an~additional effect at the end. In these cases, we may recover the Markov property by considering an~extended state space that already includes the memory.
In return, this implies that even deterministic Markovian dynamics on the full state space can give rise to non-deterministic and non-Markovian dynamics on the smaller state space.\footnote{An~analogous situation is encountered in the discussion of open quantum systems: non-unitary dynamics in the subsystems arise despite a~global unitary evolution of the system and its bath.}

\subsubsection{Model-free vs. model-based reinforcement learning}

We can distinguish between two main paradigms in \ac{RL}: model-free\index{reinforcement learning!model-free reinforcement learning} and model-based \ac{RL}\index{reinforcement learning!model-based reinforcement learning}. In the first setting, the agent does not have any kind of information about the underlying mechanisms of the environment, and it must purely learn by trial and error. In the second one, the agent either has access to a~\stress{model of the environment} or builds one from the gathered experience. Then, the agent can use this model to plan ahead, inferring the result of a~sequence of actions before executing any of them, in order to choose the best possible ones.  

Although we focus on model-free \ac{RL} in the remainder of the chapter, we briefly elaborate on how to exploit the knowledge of a~model. Building a~model of the environment provides the agent with an~enhanced understanding of the problem and can potentially help it face new situations. For example, in a~case where an~agent juggles a~set of balls, if it has a~good model of the laws of physics, it is much easier for it to learn to juggle a~new set of balls with different shapes and weights.

These models can take various forms, but a~general formulation are fully characterizable \acp{MDP}. This way, the model approximates the dynamics of the underlying \ac{MDP} of the problem. In some situations, the true model is too complex to be grasped, and we may simply try to approximate the parts of the dynamics that are the most relevant to the problem. An~example of a~simple model would be a~\ac{ML} algorithm that predicts both the expected next state and the reward $(s_{t+1}, r_{t+1})$ given the current state and an~action $(s_t, a_t)$ at any time step $t$. Such a~model allows us to predict the outcome of a~series of future actions given the current state, and we can train it in a~supervised way directly from the experience gathered by the agent.

In continuous-action spaces, the model provides a~direct connection between the input action and the received reward, allowing us to employ backpropagation methods to maximize the return instead of mere sampling from the environment. See~\cref{sec:hot-topics:dp} for examples illustrating the process. In the case of discrete-state spaces, the model typically takes the form of a~search tree that we can explore to our advantage. Models are especially convenient when the interaction cost with the environment is very high, such as realizing a~physical or chemical experiment. In these cases, we try to augment our dataset of actual samples from the environment with artificial samples drawn from the model in order to minimize the total sampling costs.

However, we do not always have access to a model, or building one may not be in our interest. Building models is costly, especially in cases where we have limited knowledge about the environment, and they are only helpful when accurate. Furthermore, models are often tailored to specific problems. On the contrary, model-free \ac{RL} algorithms come with the advantage that they are agnostic to the problem at hand and, thus, they are more versatile. Therefore, we focus on model-free \ac{RL} for the rest of the chapter for pedagogical purposes, as they prove useful on the full range of \ac{RL} tasks. In particular, we provide an~introduction to policy-based and value-based \ac{RL} in \cref{sec:rl_policy_gradient} and \cref{sec:rl_Q_learning}, respectively.

\subsubsection{Value functions and Bellman equations}
\label{sec:rl_value_functions}

As we have mentioned in the previous sections, the goal in \ac{RL} is to find the optimal policy $\pi^*$ that maximizes the return, introduced in~\cref{eq:rl_definition_return}. Such a~clear objective allows us to define \stress{value functions}\index{value function} that estimate how convenient it is for the agent to be in a~given state or to perform a~certain action to accomplish the task. For instance, consider the case in which we are looking for a~treasure on a~map. Being one step away from the treasure is, overall, much better than being ten steps away. However, not all actions in the close position are equally good, provided that one leads to the treasure, but the others move away from it. This is quantified by the expected future return that the agent may obtain, given the current conditions. However, given that the future rewards strongly depend on the actions that the agent will take, value functions are defined with respect to the policy. 
\ToggleForCUP{\highlight{The \stress{state-value function}\index{value function!state-value function}, $V_\pi(s)$, of a~state $s$ under the policy $\pi$ is the expected return when starting at state $s$ and following the policy $\pi$ thereafter. We formally define it as  
 \begin{equation}
     V_\pi(s) =  \estimationoperator[G_t|s_t = s, \pi] = \estimationoperator\left[\sum_{k=0}^{T-t-1} \gamma^k r_{t+k+1} \Bigg|s_t = s, \pi\right]
     \label{eq:rl_state_value_function}
 \end{equation}
In a~similar way, the \stress{action-value function}, $Q_\pi(s,a)$\index{value function!action-value function}, is the expected return when starting at state $s$, performing action $a$, and then following the policy $\pi$:
 \begin{equation}
 \begin{split}
     Q_\pi(s,a) =  \estimationoperator[G_t|s_t = s,\ a_t = a, \pi] = \\
     = \estimationoperator\left[\sum_{k=0}^{T-t-1} \gamma^k r_{t+k+1} \Bigg|s_t = s, a_t = a,\pi\right]
     \label{eq:rl_action_value_function}
 \end{split}
 \end{equation}}
 \highlight{The \stress{advantage}\index{value function!advantage}, $A_\pi(s,a)$, is the additional expected return obtained by following an~action $a$ at state $s$, over the expected policy behavior:  
 \begin{equation}
     A_\pi(s,a) = Q_\pi(s,a) - V_\pi(s)\,.
     \label{eq:rl_advantage}
 \end{equation}}}{
 \highlight{The \stress{state-value function}\index{value function!state-value function}, $V_\pi(s)$, of a~state $s$ under the policy $\pi$ is the expected return when starting at state $s$ and following the policy $\pi$ thereafter. We formally define it as  
 \begin{equation}
     V_\pi(s) =  \estimationoperator[G_t|s_t = s, \pi] = \estimationoperator\left[\sum_{k=0}^{T-t-1} \gamma^k r_{t+k+1} \Bigg|s_t = s, \pi\right]
     \label{eq:rl_state_value_function}
 \end{equation}
In a~similar way, the \stress{action-value function}, $Q_\pi(s,a)$\index{value function!action-value function}, is the expected return when starting at state $s$, performing action $a$, and then following the policy $\pi$:
 \begin{equation}
 \begin{split}
     Q_\pi(s,a) =  \estimationoperator[G_t|s_t = s,\ a_t = a, \pi] = \\
     = \estimationoperator\left[\sum_{k=0}^{T-t-1} \gamma^k r_{t+k+1} \Bigg|s_t = s, a_t = a,\pi\right]
     \label{eq:rl_action_value_function}
 \end{split}
 \end{equation}The \stress{advantage}\index{value function!advantage}, $A_\pi(s,a)$, is the additional expected return obtained by following an~action $a$ at state $s$, over the expected policy behavior:  
 \begin{equation}
     A_\pi(s,a) = Q_\pi(s,a) - V_\pi(s)\,.
     \label{eq:rl_advantage}
 \end{equation}}
 }
The value functions fulfill a~recursive relationship that is exploited by many \ac{RL} algorithms, which stems from the recursive nature of the return~\cref{eq:rl_return_recursive}. This allows us to write the state-value function $V_\pi(s)$ as a~function of the next states
\begin{equation}\label{eq:rl_bellman_V_pi}
\begin{split}
    V_\pi(s) & = \estimationoperator[G_t|s_t = s,\pi] = \estimationoperator[r_{t+1} + \gamma G_{t+1}|s_t = s,\pi] \\
    & = \sum_{a}\pi(a,s)\sum_{s',r}p(s',r|s,a)\left( r + \gamma \estimationoperator[G_{t+1}|s_{t+1} = s',\pi]\right) \\
    & = \sum_{a}\pi(a,s)\sum_{s',r}p(s',r|s,a)\left( r + \gamma V_\pi(s') \right) \\
    & = \estimationoperator[r_{t+1} + \gamma V_{\pi}(s_{t+1}) | s_t = s, \pi]\,.
\end{split}
\end{equation}
We can do the analogous derivation for the action-value function $Q_\pi(s,a)$
\begin{equation}\label{eq:rl_bellman_Q_pi}
\begin{split}
    Q_\pi(s, a) & = \estimationoperator[G_t|s_t = s, a_t = a, \pi] = \estimationoperator[r_{t+1} + \gamma G_{t+1}|s_t = s, a_t = a, \pi] \\
    & = \sum_{s',r}p(s',r|s,a)\left( r + \gamma \estimationoperator[G_{t+1}|s_{t+1} = s',\pi]\right) \\
    & = \sum_{s',r}p(s',r|s,a)\left( r + \gamma V_\pi(s') \right) \\
    & = \estimationoperator[r_{t+1} + \gamma V_{\pi}(s_{t+1}) | s_t = s, a_t = a,\pi]\,,
\end{split}
\end{equation}
from which the relationship $V_\pi(s)=\sum_a \pi(a|s)Q_\pi(s,a)$ becomes evident. These are the \stress{Bellman equations}\index{Bellman equations} for the value functions, and they lie at the core of \ac{RL} as they define the relation between the value of a~state $s$ and its successors $s'$, recursively capturing future information. 

These concepts introduce the notion of partial ordering between policies. A~policy $\pi$ is better than another policy $\pi'$ if it yields a~higher return. Hence, $\pi>\pi'$ if and only if $V_\pi(s) > V_{\pi'}(s)\ \forall s\in \statespace$. Therefore, the optimal policy $\pi^*$ is such that it is better than or equal to all the other possible policies.\footnote{The ordering operator is not always defined between policies. Two policies $\pi$, $\pi'$ cannot be ordered iff $\exists\ s,s'\in\statespace :\ V_\pi(s) > V_{\pi'}(s),\ V_\pi(s') < V_{\pi'}(s')$. However, for \acp{MDP} there always exist an~optimal policy $\pi^*$ s.t. $\pi^* \geq \pi\ \forall \pi$~\cite{SuttonBarto2018}.} Hence, the optimal policy maximizes the value function. Taking the Bellman equations,~\cref{eq:rl_bellman_V_pi,eq:rl_bellman_Q_pi}, $\pi^*$ is such that 
\begin{equation}\label{eq:rl_bellman_V_pi_opt}
        \begin{split}
            V_{\pi^*}(s) & = \max_{a}\estimationoperator[G_t | s_t = s, a_t = a,\pi^*] \\
                         & = \max_{a}\estimationoperator[r_{t+1} + \gamma V_{\pi^*}(s_{t+1}) | s_t = s, a_t = a,\pi^*] \\
                         & = \max_{a}Q_{\pi^*}(s,a)\,.
        \end{split}
    \end{equation}
Notice that in this new Bellman equation there is a~maximization over the first action, as opposed to the expectation over actions from~\cref{eq:rl_bellman_V_pi}. This is because the value of a~state under the optimal policy must be equal to the expected return for the best action. 
In a~similar way, we can find the Bellman equation for the action-value function $Q_\pi(s,a)$ for an~optimal policy $\pi^*$. Together, they define the set of the \stress{Bellman optimality equations}\index{Bellman equations!optimal Bellman equations}:
\begin{equation}\label{eq:rl_bellman_opt_eqns}
    \begin{split}
        V_{\pi^*}(s) = & \max_{a}\sum_{s',r}p(s',r|s,a)\left[r + \gamma V_{\pi^*}(s')\right]\\
        Q_{\pi^*}(s,a) = & \sum_{s',r}p(s',r|s,a)\left[r + \gamma \max_{a'}Q_{\pi^*}(s',a')\right]
    \end{split}
\end{equation}
These equations fulfill
\begin{equation}
    \begin{split}
      Q_{\pi^*}(s,a) &= \max_{\pi}Q_\pi(s,a) \\
        V_{\pi^*}(s) & = \max_{\pi} V_\pi(s) =  \max_{a} Q_{\pi^*}(s,a)\,.
    \end{split}
\end{equation}
\highlight{%
We can define the optimal policy $\pi^*(a|s)$ and action $a^*$ at a~given state $s$ as:
\begin{equation}
    \label{eq:rl_bellman_optimal_policy}
    \begin{split}
        \pi^*  & = \argmax_{\pi} V_{\pi^*}(s)\\
        a^* &= \argmax_{a} Q_{\pi^*}(s,a)\,.
    \end{split}
\end{equation}
The optimal policy $\pi^*$ corresponds to the deterministic choice of the best action $a^*$ for a~given state $s$ according to the optimal action-value function $Q_{\pi^*}(s,a)$. Due to the recursive nature of the value functions, a~greedy action according to $V_{\pi^*}$ or $Q_{\pi^*}$ is optimal in the long term.}
The Bellman optimality equations \cref{eq:rl_bellman_opt_eqns} are, indeed, a~system of equations with one for every state. In order to solve them directly, we need to explicitly use $p(s',r|s,a)$.\footnote{Due to the maximization step in \cref{eq:rl_bellman_opt_eqns}, this is a~nonlinear optimization problem.} If $p(s',r|s,a)$ is known, we know the underlying model of the system, and thus we deal with model-based \ac{RL}, as discussed in the previous section. In a~general model-free \ac{RL} scenario, it is unknown and, as such, we need additional methods to solve them, such as the ones we introduce in the following sections.

\subsection{Value-based methods}
\label{sec:rl_value_based_methods}

In value-based \ac{RL}, the goal is to obtain the optimal policy $\pi^*(a|s)$ by learning the optimal value functions, as in~\cref{eq:rl_bellman_optimal_policy}. This way, we start with an~initial estimation of the value function for every state, $V_\pi(s)$, or state$-$action pairs, $Q_\pi(s,a)$. Then, we progressively update them with the experience gathered by the agent following its policy.

Given that the value functions are defined with respect to a~policy (recall~\cref{sec:rl_value_functions}), we need to define a~fixed policy for this family of algorithms. A~common choice is an~$\varepsilon$-greedy policy, as introduced in~\cref{sec:rl_exploration_exploitation}, provided that the optimal policy is greedy with respect to the optimal value function. Hence, learning the value function for such policy provides us with the optimal one in the greedy limit.

One of the most straightforward and naive approaches to learn the value function would be to sample trajectories $\tau\sim p(\tau)$ (\cref{eq:rl_trajectory}), and then use the return $G_t$ to update our value function estimation\footnote{The return is an~unbiased estimator for the expectation $V_\pi(s_t)=\estimationoperator[G_t|s_t,\pi]$ from~\cref{eq:rl_state_value_function}. This is known as a~sample update, as we only use a~single sample to determine the expectation.} for every visited state $s_t$ along the way:
\begin{equation}
    V_\pi(s_t) = V_\pi(s_t) + \learningrate(G_t - V_\pi(s_t))\,,
\end{equation}
where $\learningrate$ is a~learning rate. We can do an~analogous process for every visited state and action along the trajectory to learn $Q_\pi(s,a)$ instead.

However, with this approach we can only learn at the end of each trajectory, also known as \stress{episodes}, which can be very inefficient in problems involving long episodes, or even infinite ones. On the contrary, \ac{TD} algorithms\index{temporal-difference learning} exploit the recursive nature of the value functions,~\cref{eq:rl_bellman_V_pi,eq:rl_bellman_Q_pi}, to learn at every time step:
\begin{equation}
    \label{eq:rl_TD0}
    V_\pi(s_t) = V_\pi(s_t) + \learningrate\left(r_{t+1} + \gamma V_\pi(s_{t+1}) - V_\pi(s_t)\right) \,.
\end{equation}
Notice that, while $V_\pi(s_t)$ is an~estimate, $V_\pi(s_{t+1})$ is also an~estimate. This is known as a~\stress{bootstrapping} method, as the update is partially based on another estimate. Nevertheless, it is proven to converge to a~unique solution. The term in brackets is known as \stress{\ac{TD} error}\index{temporal-difference learning!temporal-difference error}.

The algorithm implementing~\cref{eq:rl_TD0} is known as \ac{TD}$(0)$, which is a~special case of the \ac{TD}$(\lambda)$ algorithms~\cite{Sutton1988TD}. The analogous algorithm for the action-value function is known as SARSA~\cite{Rummery1994sarsa,Seijen2014TDlambda}:
\begin{equation}
    Q_\pi(s,a) = Q_\pi(s, a) + \learningrate\left(r + \gamma Q_\pi(s',a') - Q_\pi(s,a)\right)\,,
\end{equation}
\sloppy where we have recovered the notation $s',\,a',\,r$ to denote the next state, action and reward. Replacing the term $Q_\pi(s',a')$ by an~expectation over the next possible actions, such as $\sum_{a'}\pi(a'|s')Q_\pi(s',a')$, we obtain the expected SARSA algorithm~\cite{John1994ExpSarsa}. If, instead, we take a~maximization, as in~\cref{eq:rl_q_learning_update} below, we obtain Q-learning~\cite{Watkins1992Qlearning}, for which we provide a~detailed introduction in the following~\cref{sec:rl_Q_learning}. 

\subsubsection{Q-learning}
\label{sec:rl_Q_learning}

Q-learning\index{Q-learning} is one of the most widely used \ac{TD} algorithms due to its desirable properties~\cite{Watkins1992Qlearning}. Most of the \ac{TD} algorithms that we introduce in the previous section learn the value functions for their given policies, mainly $\varepsilon$-greedy policies. These include exploratory random actions (recall~\cref{sec:rl_exploration_exploitation}) that have an~impact on the learned value functions. Therefore, the policy determines the result, and we must adjust $\varepsilon$ during the training process to ensure their proper convergence toward the optimal value functions. However, Q-learning always learns the optimal action-value function regardless of the policy followed during the training.\footnote{Q-learning is an~off-policy algorithm, which means that the policy it learns (optimal $\pi^*(a|s)$) is different from the one it follows in the training episodes. Algorithms like SARSA are on-policy, and learn the value function that corresponds to the policy with which they generate the training data.} 

\highlight{The goal is to directly learn the optimal Q-values, $Q_{\pi^*}(s,a)$, hence the name Q-learning, in order to obtain $\pi^*(s|a)$ by performing greedy actions over them, as in~\cref{eq:rl_bellman_optimal_policy}.} We start by arbitrarily initializing our estimates $Q_\pi(s,a)\ \forall s\in\statespace,a\in\actionspace$, which are typically stored in a~table (see~\cref{sec:rl_deep_Q_learning} for an~implementation with \acp{NN}). Then, we sample trajectories $\tau\sim p(\tau)$ according to the policy to progressively update our estimates with the relation
\begin{equation}
    \label{eq:rl_q_learning_update}
    Q_\pi(s,a) = Q_\pi(s, a) + \learningrate\left(r + \gamma \max_{a'}Q_\pi(s',a') - Q_\pi(s,a)\right)\,.
\end{equation}
We illustrate the process in~\cref{alg:Q-learning}.

\begin{algorithm}
\caption{Q-learning}\label{alg:Q-learning}
\begin{algorithmic}
\Require learning rate $\learningrate$, maximum time $T$, policy parameter $\varepsilon$
\State Initialize $Q(s,a)\ \forall s\in\statespace, a\in\actionspace$
\While{not converged}
\State Initialize $s_0$
\For{$t=0$ to $T-1$}
    \State $\xi\gets$ uniform$\in[0, 1]$
    \State $a\gets$ uniform $a$ \textbf{if} $\xi\leq\varepsilon$ \textbf{else} $\argmax_{a}Q_\pi(s,a)$ \Comment{$\varepsilon$-greedy policy}
    \State Move to next state $s'$ and obtain reward $r$ 
    \State $Q(s,a) \gets Q(s,a) + \learningrate \left(r + \gamma \max_{a'} Q(s',a') - Q(s,a)\right).$
\EndFor
\EndWhile\\
\Return $Q(s,a)$ \Comment{Optimal action-value function for all states and actions}
\end{algorithmic}
\end{algorithm}

This method is guaranteed to converge to the optimal action-value function as long as all possible state$-$action pairs continue to be updated. This is a~necessary condition for all the algorithms that converge to the optimal behavior and it can become an~issue for fully deterministic policies. However, with Q-learning, we can have an~$\varepsilon$-greedy policy with $\varepsilon\neq0$ that ensures that this condition is fulfilled.     

The key element is that, while the policy determines which states and actions are visited by the agent, the Q-value update is performed over a~greedy next action, as shown in~\cref{eq:rl_q_learning_update}. This way, the learned Q-values are those corresponding to the greedy policy over them, which is the one fulfilling the Bellman optimality equations~\cref{eq:rl_bellman_opt_eqns}.  

\subsubsection{Double Q-learning}
\label{sec:rl_double_q_learning}

Most of the \ac{TD} algorithms suffer from a~maximization bias that results in an~overestimation of the Q-values, which can harm the performance. Especially, in Q-learning, we encounter two maximizations: one in the $\varepsilon$-greedy policy and one in the greedy target policy (\cref{eq:rl_q_learning_update}). This way, we use a~maximum overestimated value (see below) to update the maximum Q-value, which corresponds to the greedy action taken by the policy, potentially incurring into a~significant positive bias for $Q_\pi(s,a)$. 

\sloppy The maximization over next possible actions in~\cref{eq:rl_q_learning_update} is a~sample estimate for the maximum expected value $\max_{a'}\estimationoperator[Q_\pi(s',a')]$. However, it is a~positively biased estimator, provided that the sample estimate actually corresponds to the expected maximum value $\estimationoperator[\max_{a'}Q_\pi(s',a')]$~\cite{Smith2006overestimates}. In Ref.~\cite{SuttonBarto2018} they provide a~simple example to develop intuition on the matter: suppose that the true Q-values for all actions in a~state are zero and that our estimates $Q_\pi(s,a)$ are distributed around them taking positive and negative values. The maximum value is positive and, hence, it is an~overestimation. The overestimation of the Q-values can prevent the algorithm from learning the optimal policy~\cite{Trhun1993overestimate}.

We overcome this issue with double Q-learning~\cite{vanHasselt2010doubleQlearning}\index{Q-learning!double Q-learning}. This way, instead of learning a~single set of Q-values, we learn two: $Q_\pi^A(s,a)$, and $Q_\pi^B(s,a)$. However, in order to update one, we use the other to estimate the value of its corresponding next greedy action:
\begin{equation}
    \label{eq:rl_double_q_learning}
    Q_\pi^A(s,a) = Q_\pi^A(s,a) - \learningrate\left(r + \gamma Q_\pi^B\left(s',\argmax_{a'}Q_\pi^A(s',a')\right) - Q_\pi^A(s,a)\right) \,, 
\end{equation}
where $A,\,B$ are interchangeable. This approach avoids using the same estimate to determine both the maximizing action and its value, yielding an~unbiased estimate.

We learn both sets of values by randomly updating one at a~time at every time step. The only additional difference with respect to standard Q-learning is that we take actions following an~$\varepsilon$-greedy policy that combines the information of both $Q_\pi^A(s,a)$ and $Q_\pi^B(s,a)$, e.g., using their sum or mean. With double Q-learning, we overcome a~major limitation of Q-learning at the price of doubling the memory requirements.  

\subsubsection{Implementing Q-learning with a~neural network}\label{sec:rl_deep_Q_learning}
In Q-learning, as we have introduced it in~\cref{sec:rl_Q_learning}, we store the Q-values, $Q_\pi(s,a)$, for every possible state$-$action pair. This approach allows us to find the exact optimal action-value function. However, it is only viable for small problems, as the memory requirement quickly becomes unfeasible for moderately large ones.

In these cases, we must rely on an~efficient way to represent $Q_\pi(s,a)\ \forall s\in\statespace,a\in\actionspace$. \Acp{NN} are a~prominent candidate to approximate the action-value function, as introduced in Ref.~\cite{Mnih2015}, with significantly less parameters than state$-$action pairs (recall ~\cref{sec:NNs}). Using \acp{NN} to learn the Q-values is known as \stress{deep Q-learning}\index{Q-learning!deep Q-learning} and the network is commonly referred to as \ac{DQN}\index{deep Q-network}. \Acp{DQN} take a~representation of state in the input layer $\featmap(s)$, and have as many neurons as possible actions in the output layer, which encode $Q_\pi(s,a;\params)\ \forall a\in\actionspace$. Here, $\params$ denotes the set of learnable parameters of the neural network. This way, the \ac{DQN} provides the Q-value of all possible actions given a~state. 

Nevertheless, \acp{DQN} may become highly unstable when directly applying~\cref{alg:Q-learning} with an~update rule for the network parameters:
\begin{equation}
    \params = \params + \learningrate\left(r + \gamma\max_{a'}Q_\pi(s',a';\params) - Q_\pi(s,a;\params)\right)\nabla_{\params}Q_\pi(s,a;\params)\,,
\end{equation}
which is analogous to a~regression problem in which we minimize the \ac{MSE} loss (\cref{eq:mse_loss}) between the target, $r + \gamma\max_{a'}Q_\pi(s',a';\params)$, and the prediction, $Q_\pi(s,a;\params)$, through gradient descent. The instabilities are mainly due to correlations in consecutive observations along the trajectories, correlations between target and prediction, and significant changes in the data distribution due to small variations in the parameters. The latter happen because the agent follows an~$\varepsilon$-greedy policy, and small changes in the parameters may change the actions that have the maximum Q-value for the states, abruptly altering the course of the trajectories.\footnote{Consider the case of two separate paths that lead to different treasures. We initialize the Q-values arbitrarily, and the $\varepsilon$-greedy policy mainly takes the path with the highest one, while casually following the other with small probability $\varepsilon$. However, if the second one leads to a~bigger treasure, its Q-value will eventually become the highest, and the data distribution will suddenly change to mainly sample this path and casually take the other.} We overcome these limitations with \stress{experience replay}~\cite{Lin1993RL4robots}\index{experience replay}, and introducing a~\stress{target network}\index{target network}.
    
With experience replay\index{experience replay}, instead of learning at every time step, we store the experience gathered along the episodes in a~memory, which keeps the information of every transition $(s,a,r,s')$. Then, once the agent has gathered enough experience, it replays a~randomly sampled batch of transitions in its memory to compute the loss and update the \ac{DQN} parameters. This way, the agent alternates between episodes to gather experience and replaying it to perform the learning process. This technique removes the correlation between training samples and mitigates the sudden changes in data distribution. Furthermore, it allows the agent to reuse the experience to prevent forgetting and re-learning.\footnote{This is specially valuable when the experience is costly to obtain. For instance, if a~robot receives severe damage, having a~memory allows it to keep learning from the situation without receiving further injuries.}

In order to remove the correlation between target and prediction, we consider a~target network\index{target network}, which is a~clone of the \ac{DQN} that we update at a~different rate. While we update the \ac{DQN} parameters, $\params$, at every iteration, we only update the parameters of the target network, $\params^{-}$, copying $\params$ every few iterations. Then, we use it to predict the target term $\max_{a'}Q_\pi(s',a';\params^{-})$, hence the name of the network. This ensures that the prediction, $Q_\pi(s,a;\params)$, and the target are uncorrelated.  

Additionally, we can go a~step further and use the target network for double Q-learning (see~\cref{sec:rl_double_q_learning}) in order to prevent the \ac{DQN} from overestimating the action-value function, as introduced in Ref.~\cite{vanHasselt2016doubleDQN}. Thus, the overall implementation consists of gathering experience by following an~$\varepsilon$-greedy policy on the Q-values, $Q_\pi(s,a;\params)$. Then, the agent replays randomly selected transitions from the experience to compute the \ac{MSE} loss function between the target and the prediction, but using a~target network to perform double Q-learning:
\begin{equation}
    \label{eq:rl_double_DQN_loss}
    \lossfun = \frac{1}{n}\sum_{i=1}^n\left(r_i + \gamma Q_\pi\left(s'_i,\argmax_{a'}Q_\pi(s'_i,a';\params);\params^{-}\right) - Q_\pi(s_i,a_i;\params)\right)^2\,,
\end{equation}
where $i$ denotes the index in a~batch of $n$ randomly sampled transitions from the memory. Then, we perform a~gradient descent step over the loss in~\cref{eq:rl_double_DQN_loss} to update $\params$. Finally, every few iterations, we update the target network $\params^{-}\gets\params$. 

\subsection{Policy gradient methods}
\label{sec:rl_policy_gradient}
 
The main goal of \Ac{RL} is to find the optimal policy $\pi^*(a|s)$ that maximizes the expected return for a~given task. In policy gradient algorithms\index{policy gradient} we try to directly find the optimal policy by proposing a~parametrized ansatz $\pi_{\params}(a|s)$ and optimizing its parameters $\params$, similar to the variational wave functions from~\cref{sec:NN_q_states}. Hence, finding the optimal policy $\pi^*(a|s)$ is equivalent to finding the optimal set of parameters $\params^*$ that best approximates it $\pi_{\params^*}(a|s)\approx\pi^*(a|s)$. This parametrization can take several forms, such as a~\ac{NN}, and controlling the shape of the policy may allow us to leverage prior knowledge about the task to obtain better results. Furthermore, the policies are stochastic, which have a~natural exploratory character and the flexibility to also approximate deterministic policies. 

In order to optimize the parameters, we use an~objective function $O_\pi$ that we aim to maximize. This can be any figure of performance, such as the state-value function $V_{\pi}$, the action-value function $Q_{\pi}$, or the return $G$. Having continuous parametrized policies, the objective function changes smoothly with changes in the parameters, which allows us to compute their derivatives. We approach the optimization by a~gradient ascent method: we compute the gradient of the expectation value $\nabla_{\params} \estimationoperator[O_{\pi}|\pi_{\params}]$, and perform a~small update of the parameters $\params$. The expectation value is taken over the trajectories $\tau$ sampled according to the policy (recall~\cref{eq:rl_trajectory}).

Directly evaluating the gradient is not straightforward because it depends on the stationary distribution of the states, to which we do not have access in model-free \ac{RL}. Hence, it is difficult to estimate the effect of the policy update on the state distribution. However, the \stress{policy gradient theorem}\index{policy gradient!policy gradient theorem} \cite{Marbach2001policygrad,Sutton2000policygrad} provides us with an~analytical form for the gradient of the objective function that does not involve the derivative over the state distribution. 
\highlight{
\textbf{Policy gradient theorem:}\index{policy gradient!policy gradient theorem}
For any differentiable policy $\pi_{\params}(a|s)$ and objective function $O_\pi$, the gradient of its expectation value $\nabla_{\params} \estimationoperator [O_{\pi}|\pi_{\params}]$ can be expressed in terms of derivatives acting exclusively on the logarithmic policy $\nabla_{\params}\log\pi_{\params}(a|s)$. The term $\nabla_{\params}\log\pi_{\params}(a|s)$ is often referred to as the \stress{score function}\index{score function}.
}

To get some additional intuition on the above theorem, let us consider an~example with the total return $G(\tau)$ as objective function (see~\cite{SuttonBarto2018} for an~extended proof with $V_\pi(s)$). Thus, we are interested in maximizing the expectation value $\estimationoperator[G|\pi_{\params}]$, which is performed over the trajectories $\tau\sim p_{\params}(\tau)$. We restate~\cref{eq:rl_trajectory} to explicitly show the parameter dependence
\begin{equation}
    \label{eq:rl_trajectory_param}
    p_{\params}(\tau) = \prod_{t=0}^{T-1} p(s_{t+1}|s_t,a_t)\pi_{\params}(a_t|s_t).
\end{equation} 
Therefore, we can write the expectation as 
\begin{equation}
    \estimationoperator[G|\pi_{\params}] = \sum_\tau p_{\params}(\tau)G(\tau).
\end{equation}

In order to take the gradient, let us first recall the property of logarithmic derivatives $\nabla_{\params} p_{\params} = p_{\params}\nabla_{\params}\log p_{\params}$, which we apply in the following derivation:
\begin{equation}
    \label{eq:rl_nabla_log}
    \begin{split}
        \nabla_{\params}\estimationoperator[G|\pi_{\params}] &= \sum_\tau G(\tau)\nabla_{\params} p_{\params}(\tau) \\ 
        &= \sum_\tau G(\tau)p_{\params}(\tau)\nabla_{\params}\log p_{\params}(\tau).
    \end{split}
\end{equation}
Then, from~\cref{eq:rl_trajectory_param}, we see that the only dependence on $\params$ from $p_{\params}(\tau)$ is in the policy. Therefore, 
\begin{equation}
    \nabla_{\params} \log p_{\params}(\tau) = \sum_{t=0}^{T-1} \nabla_{\params}\log\pi_{\params}(a_t|s_t),
\end{equation}
which, combined with~\cref{eq:rl_nabla_log}, we obtain the expression
\begin{equation}
    \label{eq:rl_nabla_expect}
    \begin{split}
        \nabla_{\params}\estimationoperator[G|\pi_{\params}] &= \sum_\tau p_{\params}(\tau)G(\tau)\sum_{t=0}^{T-1}\nabla_{\params}\log\pi_{\params}(a_t|s_t) \\
        &= \estimationoperator\left[G(\tau)\sum_{t=0}^{T-1}\nabla_{\params}\log\pi_{\params}(a_t|s_t)\Bigg|\pi_{\params}\right]
    \end{split}
\end{equation}
The importance of the policy gradient theorem lies in the fact that it yields a~closed form for the gradient as an~expectation value. As a~consequence, we can estimate it via Monte-Carlo sampling over different trajectories $\tau$. Furthermore, the gradient of the objective function is independent of the initial state $s_0$, as it does not depend on the policy. 

\subsubsection{REINFORCE}
\label{ssec:rl_reinforce}

The REINFORCE\index{policy gradient!REINFORCE} algorithm~\cite{Williams1992} is one of the most commonly used policy gradient algorithms and it uses the return as objective $O_\pi = G(\tau)$.\footnote{In~\cref{sec:rl_value_functions} we mention that the optimal policy maximizes $V_\pi(s)\ \forall s\in\statespace$. Taking $V_\pi(s)$ as objective, the gradient is $\nabla_{\params}\estimationoperator[V_\pi(s)|\pi_{\params}]=\estimationoperator[Q_\pi(s,a)\nabla_{\params}\log\pi_{\params}(a|s)|\pi_{\params}]$ (see~\cite{SuttonBarto2018}). In REINFORCE, $G_t$ acts as an~unbiased estimator of $Q_\pi(a_t,s_t)$ to find the optimal policy, since $Q_\pi(a_t,s_t)=\estimationoperator[G_t|s_t,a_t,\pi_\theta]$ from~\cref{eq:rl_action_value_function}.} \highlight{The main principle of REINFORCE is to directly modify the policy to favor series of actions within the agent's experience that lead to a~high return.
This way, previously beneficial actions are more likely to happen the next time the agent interacts with the environment.}

Formally, we solve the optimization problem $\params^* = \argmax_{\params}\estimationoperator[G|\pi_\theta]$. We find $\params^*$ via an~iterative update rule in which we compute the gradient $\nabla_{\params}\estimationoperator[G|\pi_{\params}]$ and perform a~gradient ascent step in its direction. In practice, we estimate it by sampling a batch of $n$ trajectories $\tau\sim p_{\params}(\tau)$, also known as \stress{episodes}, to approximate the expectation value from~\cref{eq:rl_nabla_expect}. This way, at learning iteration $k$,
\begin{align}
    \Delta\params_k &\approx \frac{1}{n} \sum_{i=1}^n G(\tau_i)\sum_{t=0}^{T_i-1}\nabla_{\params}\log\pi_{\params}(a_t|s_t)
    \label{eq:rl_reinforce_update}\\ 
    \params_{k+1} &= \params_k + \learningrate\nabla\params_k,
\end{align}
where $\learningrate$ is the learning rate.\footnote{In some cases, it is beneficial to compute the expectation of the gradient as a weighted sum over the trajectory returns. In this case, rather than dividing by $n$, we divide by $\sum_\tau G(\tau)$, which makes the update rule independent of the scale of the returns. This approach disregards trajectories with zero return, which do not contribute to the gradient and would dilute the information, yielding very small updates.} We illustrate the procedure in \cref{alg:REINFORCE}.

\begin{algorithm}
\caption{REINFORCE}\label{alg:REINFORCE}
\begin{algorithmic}
\Require learning rate $\learningrate$, number of trajectories $n$, maximum time $T$
\Require randomly initialized differentiable policy $\pi_{\params}(a|s)$ 
\While{not converged}
\For{$i = 1$ to $n$} 
 \State Initialize $s_0$
 \For{$t = 0$ to $T-1$}
  \State Take action $a_t\sim\pi_{\params}(a_t|s_t)$ and store $\nabla_{\params}\log\pi_{\params}(a_t|s_t)$
  \State Move to next state $s_{t+1}$ and store reward $r_{t+1}$
 \EndFor
   \State $G^{(i)} \gets \sum_{t} \gamma^t r_{t+1}$
   \State $z^{(i)} \gets \sum_{t}\nabla_{\params}\log\pi_{\params}(a_t|s_t)$
\EndFor
\State $\Delta\params \gets (1/n)\sum_{i}G^{(i)}z^{(i)}$
\State $\params \gets \params + \learningrate\Delta\params$
\EndWhile\\
\Return $\params$ \Comment{Optimal policy parameters}
\end{algorithmic}
\end{algorithm}

However, the trajectory sampling introduces significant fluctuations to the expected quantities that result in large training variances, which is a~general problem with any Monte-Carlo-based approach. Some episodes may be quite successful whereas some others could be a~complete failure with very low returns. Such high variance results into unstable policy updates, which increase the convergence time toward the optimal policy. A~common technique to tackle this issue is to introduce a~\stress{baseline}\index{baseline} into the returns, which reduces the variance of the method without incurring any bias, and therefore \stress{should always be used}. 

In order to provide a~better description of the baseline, let us first rewrite~\cref{eq:rl_nabla_expect} in a~more convenient way, and omitting the condition $\estimationoperator[\cdot|\pi_{\params}]$ for the rest of the chapter:
\begin{equation}
    \begin{split}
    \label{eq:rl_nabla_return}
        \nabla_{\params}\estimationoperator[G|\pi_{\params}] &= \estimationoperator\left[\left(\sum_{t'=0}^{T-1}\gamma^{t'}r_{t'+1}\right)\sum_{t=0}^{T-1}\nabla_{\params}\log\pi_{\params}(a_t|s_t)\right] \\ 
        &= \estimationoperator\left[\sum_{t'=0}^{T-1}\gamma^{t'}r_{t'+1}\sum_{t=0}^{t'}\nabla_{\params}\log\pi_{\params}(a_t|s_t)\right] \\ 
        &= \estimationoperator\left[\sum_{t=0}^{T-1}\nabla_{\params}\log\pi_{\params}(a_t|s_t)\sum_{t'=t}^{T-1}\gamma^{t'}r_{t'+1}\right] \\
        &= \estimationoperator\left[\sum_{t=0}^{T-1}\gamma^{t}G_t\nabla_{\params}\log\pi_{\params}(a_t|s_t)\right],
    \end{split}
\end{equation}
where in the first equation we write the explicit form of $G(\tau)$. In the second equation we use the relation
\begin{equation}
    \begin{split}
    \nabla_{\params}\estimationoperator[G|\pi_{\params}] &= \nabla_{\params}\estimationoperator\left[\sum_{t'=0}^{T-1}\gamma^{t'}r_{t'+1}\right] = \sum_{t'=0}^{T-1}\nabla_{\params}\estimationoperator_{\tau_{t'}}\left[\gamma^{t'}r_{t'+1}\right] \\
    &= \sum_{t'=0}^{T-1} \estimationoperator_{\tau_{t'}}\left[\gamma^{t'}r_{t'+1}\sum_{t=0}^{t'}\nabla_{\params}\log\pi_{\params}(a_t|s_t)\right] \\ 
    &= \estimationoperator\left[\sum_{t'=0}^{T-1}\gamma^{t'}r_{t'+1}\sum_{t=0}^{t'}\nabla_{\params}\log\pi_{\params}(a_t|s_t)\right],
    \end{split}
\end{equation}
where $\estimationoperator_{\tau_{t'}}$ denotes expectation over trajectories up to time $t'$. Then, in the third line of~\cref{eq:rl_nabla_return}, we rearrange the terms in the summations and we find the explicit form of $G_t$ offset by a~$\gamma^t$ factor. In the final expression, it becomes clearer how past rewards in the trajectories do not contribute to the gradient of the policy from a~given time onwards, which recovers the Markov property.

We can reduce the variance in the gradient by introducing a~state-dependent baseline $b(s_t)$ in~\cref{eq:rl_nabla_return} such that 
\begin{equation}
    \label{eq:rl_nabla_baseline}
    \nabla_{\params}\estimationoperator[G|\pi_{\params}] = \estimationoperator\left[\sum_{t=0}^{T-1}\gamma^{t}\left(G_t-b(s_t)\right)\nabla_{\params}\log\pi_{\params}(a_t|s_t)\right].
\end{equation}
Any baseline is appropriate as long as it does not depend on the actions. This way, we do not introduce any bias, given that
\begin{equation}
    \label{eq:rl_baseline_expect}
    \begin{split}
        &\estimationoperator\left[b(s_t)\nabla_{\params}\log\pi_{\params}(a_t|s_t)\right] = \estimationoperator_{\tau_t}\left[b(s_t)\estimationoperator_{\tau_{t:T}}\left[\nabla_{\params}\log\pi_{\params}(a_t|s_t)\right]\right] \\
        &= \estimationoperator_{\tau_t}\Bigg[b(s_t)\sum_{a_t}\pi_{\params}(a_t|s_t)\nabla_{\params}\log\pi_{\params}(a_t|s_t)\underbrace{\sum_{s_{t+1}}p(s_{t+1}|s_t,a_t)}_{1}\underbrace{\sum_{\tau_{t+1:T}}p_{\params}(\tau_{t+1:T})}_{1}\Bigg] \\
       &= \estimationoperator_{\tau_t}\Bigg[b(s_t)\nabla_{\params}\underbrace{\sum_{a_t}\pi_{\params}(a_t|s_t)}_{1}\Bigg] = \estimationoperator_{\tau_t}\left[b(s_t)\cdot 0\right] = 0 \,,
    \end{split}
\end{equation}
where $\tau_{t:T}$ indicates a~trajectory from time $t$ until the end $T$. We move from the second to the third line using the property of logarithmic derivatives, as in~\cref{eq:rl_nabla_log}. Notice that the expectation remains unbiased even if the baseline depends on $\params$. 

While the expectation is unaffected, the baseline can have a~major impact in the variance.\footnote{Recall that $\text{Var}[x]=\estimationoperator[x^2] - \estimationoperator[x]^2$. Hence, adding a~term with null expectation does not affect the second term but it does have an~impact on the first one $\text{Var}[x-b]=\estimationoperator[(x-b)^2] - \estimationoperator[x-b]^2=\estimationoperator[(x-b)^2] - \estimationoperator[x]^2$.} Let us consider the case of a~state-independent baseline. We can find the optimal baseline that minimizes the variance in the gradient for each parameter. In order to simplify the notation, let $z_k$ and $b_k$ be the $k$-th components of the score function $z_k=\partial_{\params_k}\log\pi_{\params}(a|s)$ and a~state-independent baseline vector, respectively. Hence, the goal is to minimize the variance of the term $(G_t-b_k)z_k$,\footnote{In this case, we take the approximation $\text{Var}\left[\sum_t X_t\right]\approx\sum_t \text{Var}\left[X_t\right]$} which is the argument of~\cref{eq:rl_nabla_baseline}. Formally, we aim to find $b_k^*=\argmin_{b_k}\text{Var}\left[(G_t-b_k)z_k\right]$, that is such that $\partial_{b_k^*}\text{Var}\left[(G_t-b_k)z_k\right]=0$. Therefore,
\begin{align}
    \text{Var}\left[(G_t-b_k)z_k\right] &= \estimationoperator[((G_t-b_k)z_k)^2] - \estimationoperator[G_t z_k]^2 \\
    \partial_{b_k}\text{Var}\left[(G_t-b_k)z_k\right] &= -2\estimationoperator[(G_t-b_k)z_k^2] \\
    b_k^* &= \frac{\estimationoperator[G_t z_k^2]}{\estimationoperator[z_k^2]}\,,
\end{align}
where in the first equation we have used~\cref{eq:rl_baseline_expect} to remove $b_k$ in the second term. 

There are several other valid baselines that we can consider, besides the state-independent example above, with which we may obtain better results. For instance, an~estimation of the value function $\hat{V}_{\pi}(s_t)\approx\estimationoperator[G_t|s_t]$ is a~common state-dependent baseline. This can either be learned, either directly from $G_t$ or as we show in~\cref{ssec:rl_actor_critic}, or it can be estimated through sampling in self-critic schemes (see~\cite{Rennie2017IEEE}). With such baseline, actions that lead to returns higher than expected with the current policy are reinforced, while those that lead to lower rewards are penalized. This is equivalent to weighting the score function by the advantage. Given that $\estimationoperator[G_t |s_t,a_t] = \estimationoperator[Q_\pi(s_t,a_t)|s_t,a_t]$, from~\cref{eq:rl_action_value_function}, subtracting a~baseline $b(s_t)=V_\pi(s_t)$, we obtain the expectation of the advantage (recall~\cref{eq:rl_advantage}). Hence, $\nabla_{\params}\estimationoperator[G|\pi_\theta]= \estimationoperator\left[\sum_t\gamma^{t}A(s_t,a_t)\nabla_{\params}\log\pi_{\params}(a_t|s_t)\right]$. Directly estimating the advantage provides the least possible variance, see~\cite{Schulman2016ICLR} for further reference on this matter.

Another common practice is to \stress{whiten} the return. This consists of subtracting the mean of the return along all the time steps of a~trajectory and dividing by its standard deviation $\bar{G_t} = (G_t - \estimate{G}{})/\sigma_G$. Since this is not exactly a~baseline, this method \stress{does} introduce a~bias.

\subsubsection{Implementing REINFORCE with a~neural network}
\label{sec:rl_reinforce_nn}

The parametrized policy $\pi_{\params}$ is a~central quantity in policy gradient methods and it can take any form as long as it is differentiable with respect to its parameters. One of the most common approaches in discrete action spaces is to define action probabilities according to a~\stress{softmax} distribution:
\begin{equation}
    \pi_{\params}(a|s) = \frac{e^{x(s,a)}}{\sum_{a'\in\actionspace}e^{x(s,a')}}\,,
    \label{eq:rl_softmax_policy}
\end{equation}
where $x(s,a)$ is the \stress{action preference}\index{action!action preference} for action $a$ in state $s$. 

The simplest way to define action preferences is through a~set of linear parameters $\params$ applied to a~feature representation of the state and action $\featmap(s, a)$, such that $x(s,a) = \params^T\featmap(s,a)$. However, this approach may lack the expressive power to approximate the optimal policy $\pi^*$ in complex problems. 

In these cases, we may need to use a~deep \ac{NN} to parametrize the action preferences. \Acp{NN} are a~natural generalization of the linear parameter approach that we can tune to increase the expressive power by, e.g., increasing the number of hidden layers or their size. This way, the \ac{NN} parametrizing the policy takes a~state representation in the input layer $\featmap(s)$, and has as many neurons as possible actions in the output layer, which encode $x(s,a)\ \forall a\in\actionspace$. Applying a~softmax activation function in the output layer (see~\cref{eq:softmax_function}), we obtain $\pi_\theta(a|s)\ \forall a\in \actionspace$, as in~\cref{eq:rl_softmax_policy}.

The training process is analogous to training a~supervised classifier on the experience gathered by the agent. Implementing REINFORCE with gradients from~\cref{eq:rl_nabla_baseline} is equivalent to performing gradient descent with a~modified categorical cross-entropy loss (recall~\cref{eq:cce_loss}):
\begin{equation}
    \label{eq:rl_reinforce_loss}
    \lossfun = - \frac{1}{n}\sum_{i=1}^{n}\sum_{t=0}^{T-1}\gamma^{t}(G_{ti}-b(s_{ti}))\log\pi_{\params}(a_{ti}|s_{ti})\,,
\end{equation}
where $i$ denotes the index in a~batch of $n$ trajectories. This way, the procedure is analogous to training an~\ac{NN} classifier in which the actions act as state labels. The main difference with supervised classification problems is that, given a~state, we do not know the true probability distribution of the actions (true labels), as that would be given by optimal policy. Instead, we assign the obtained return $G_t$ as true label for the taken action $a_t$.\footnote{The standard categorical cross entropy would be $\lossfun = -\frac{1}{n}\sum_n\sum_{k}p(a_{k})\log\pi_{\params}(a_{k}|s)$, where $p(a_{k})$ is the true probability distribution that we want to learn. In standard classification problems, this is typically $1$ for the true label and $0$ for the rest. Here, it corresponds to the optimal policy  $p(a_k)=\pi^*(a_k|s)$. Since we do not have access to $\pi^*$ (it is our goal!), we use the return $G_t$ for the chosen action in its place, as $\pi^*$ would favor actions with high returns. This effectively removes the expectation over actions, and we make the sum over time explicit in~\cref{eq:rl_reinforce_loss}.} Intuitively, in classification problems we aim to enhance the probability that the \Ac{NN} provides the right label, whereas here we reinforce the actions with high returns.

In many situations, actions can take a~range of continuous values rather than a~discrete set of categories. For instance, a~robotic arm may rotate by a~certain angle or we can tune various continuous parameters in an~experimental setup. Sometimes, we can discretize the action space into small intervals at the cost of a~loss in precision and an~increasing amount of actions. Nevertheless, this may not always be possible depending on the problem requirements and the resulting number of actions. 

In these cases, we model the stochastic continuous actions with a~mean $\mu$ and a~standard deviation $\sigma$, such that 
\begin{equation}
    a~= \mu + \sigma\xi\,,
\end{equation}
where $\xi$ is a~random normal variable with unit variance. Analogously to the action preferences above, we can parametrize $\mu_{\params}(s),\sigma_{\params}(s)$ in various ways, ranging from a~set of linear parameters, e.g., $\mu_{\params}(s) = \params^T\featmap(s)$, to an~\ac{NN} with two output neurons that determine both $\mu_{\params}(s)$ and $\sigma_{\params}(s)$ for the given observation. Formally, 
\begin{equation}
    \pi_{\params}(a|s) = \frac{1}{\sigma_{\params}(s)\sqrt{2\pi}}\exp\left(-\frac{1}{2}\left(\frac{a - \mu_{\params}(s)}{\sigma_{\params}(s)}\right)^2\right)\,.
\end{equation}

In many cases, as the learning advances, and the agent becomes better at taking the right actions (choosing $\mu_{\params}(s)$), the deviations decrease and we obtain a~quasi-deterministic policy.

\subsection{Actor-critic methods}
\label{ssec:rl_actor_critic}
 
In~\cref{sec:rl_value_based_methods}, we introduce value-based \ac{RL}, featuring the Q-learning algorithm in~\cref{sec:rl_Q_learning}. These methods excel at dealing with discrete state$-$action spaces, and their \ac{TD} character makes them data efficient and allows them to tackle continuing tasks (infinite episodes). However, they experience difficulties to deal with large state$-$action spaces, and can't deal with their continuous version. Furthermore, they are bound to implement deterministic greedy policies, while many problems present stochastic optimal policies. Finally, small changes in the value functions can cause large variations in the policy, which may cause instabilities in learning.  

On the other hand, we introduce policy-gradient methods in~\cref{sec:rl_policy_gradient}, featuring the REINFORCE algorithm in~\cref{ssec:rl_reinforce}. These algorithms overcome the aforementioned limitations of value-based methods, provided that they can deal with continuous (infinite) state$-$action spaces, and they are based on continuous stochastic policies, which ensure smooth changes in the policy throughout the learning process, and can become deterministic when needed. However, the learning happens at the end of the episodes, once we know the return, which is an~issue for long trajectories or continuing tasks. 

\highlight{Actor-critic\index{actor-critic} algorithms combine value-based and policy-based methods in order to obtain the best of both approaches. We can understand actor-critic methods as the \ac{TD} version of policy gradient, with which we retain all its advantages and overcome its major limitation. It features two main elements: the \stress{actor}, a~parametrized policy that dictates the decisions, and the \stress{critic}, a~model that evaluates them.}

The presence of the critic allows the agent to immediately learn from each action without waiting for the outcome at the end of the episode. Evaluating the policy mainly consists of learning its value functions, which allows the critic to assess whether the actions are more or less favorable. In~\cref{ssec:rl_reinforce}, we introduce the state-value function, $V_\pi(s)$, as the optimal baseline to reduce the variance in policy gradient. Although, in this case, we only look at $V_\pi(s)$ of the initial state in in the transitions, which does not allow us to evaluate the actions.\footnote{In order to determine the quality of an~action, we need to compare the initial and final positions. In a~game, an~action that escapes from the brink of a~loss toward a~less disadvantageous position may be more valuable than one that moves from an~already favorable position to a~slightly better one, despite the latter providing a~higher final state-value function.}

However, we sow that, with such baseline, we can compute the gradient in terms of the advantage $A(s,a)$, introduced in~\cref{eq:rl_advantage}. The explicit form of the advantage sets the foundation for actor-critic methods~\cite{Barto1983IEEE,Konda1999NIPS,Degris2012ICML}:
\begin{equation}
    \label{eq:rl_advantage_explicit}
    A(s_t,a_t) = \estimationoperator[r_{t+1} + \gamma V_\pi(s_{t+1}) - V_\pi(s_t)]\,,
\end{equation}
which is derived from~\cref{eq:rl_advantage,eq:rl_bellman_Q_pi}. This expression lies at the core of \ac{TD} algorithms, as it corresponds to the \ac{TD} error from~\cref{eq:rl_TD0}. 

In~\cref{eq:rl_advantage_explicit}, we use $V_\pi(s)$ to evaluate both the initial and final states of a~given transition, thus constituting a~critic of the action. This allows the agent to learn from every time step in REINFORCE, processing states, actions and rewards as they occur, like the \ac{TD} algorithms from~\cref{sec:rl_value_based_methods}. Nevertheless, this advantage comes as the cost of learning two models: the policy $\pi_{\params}(a|s)$, and the state-value function $V_\pi(s;\vect{w})$, which are usually parametrized with \acp{NN} with parameters $\params$ and $\vect{w}$, respectively. The \ac{NN} parametrizing the state-value function takes a~feature representation of the state, $\featmap(s)$, in the input layer, and has a~single output neuron encoding $V_\pi(s;\vect{w})$. The policy parametrization is the same as in~\cref{sec:rl_reinforce_nn}. We train both models simultaneously by following~\cref{alg:actor_critic}.

\begin{algorithm}
\caption{Actor-critic}\label{alg:actor_critic}
\begin{algorithmic}
\Require learning rates $\learningrate_{\params}, \learningrate_{\vect{w}}$, maximum time $T$
\Require randomly initialized differentiable policy $\pi_{\params}(s|a)$ 
\Require randomly initialized differentiable state-value function $V_\pi(s;\vect{w})$
\While{not converged}
\State Initialize $s_0$
\For{$t=0$ to $T-1$}
    \State Take action $a\sim\pi_{\params}(a|s)$
    \State Move to next state $s'$ and obtain reward $r$
    \State $A\gets r + \gamma V_\pi(s';\vect{w}) - V_\pi(s;\vect{w})$
    \State $\params\gets\params + \learningrate_{\params}\gamma^{t} A\nabla_{\params}\log\pi_{\params}(a|s)$ \Comment{Update actor}
    \State $\vect{w}\gets\vect{w} + \learningrate_{\vect{w}} A\nabla_{\vect{w}}V(s;\vect{w})$ \Comment{Update critic}
\EndFor
\EndWhile\\
\Return $\params,\, \vect{w}$ \Comment{Optimal actor and critic parameters}
\end{algorithmic}
\end{algorithm}

We train the actor with the methods from~\cref{sec:rl_policy_gradient}, and the critic using the principles from~\cref{sec:rl_value_based_methods}. Hence, all the methods in both sections apply to this algorithm. The parameter updates in~\cref{alg:actor_critic} come from performing gradient ascent with~\cref{eq:rl_nabla_baseline} on the actor, and an~analogous update rule to~\cref{eq:rl_q_learning_update} for the critic, using $V_\pi(s)$ instead of $Q_\pi(s,a)$. The process is equivalent to perform gradient descent on the losses $\lossfun_{\params} = \frac{1}{n}\sum_n\sum_t \gamma^{t}A(s_t,a_t;\vect{w})\log\pi_{\params}(a_t|s_t)$, and $\lossfun_{\vect{w}} = \frac{1}{n}\sum_n A(s,a;\vect{w})^2$, respectively, in which we omit the index for the sum over $n$ samples. They are based on the same principles as the ones in~\cref{eq:rl_reinforce_loss,eq:rl_double_DQN_loss}.

This method is often referred to as advantage actor-critic (A2C). It has been further enhanced using asynchronous actors, giving raise to the asynchronous advantage actor-critic (A3C) algorithm~\cite{Mnih2016ICML}. Other improvements rely on implementing more advanced optimization techniques, such as the natural gradient~\cite{Amari1998natgrad}, as in natural policy gradient~\cite{Kakade2002NIPS}, natural actor-critic~\cite{Peters2008Neurocomputing,Bhatnagar2009Automatica}, or more involved parameter updates such as trust-region~\cite{Schulman2015ICML,Wu2017NIPS} or proximal policy optimization algorithms~\cite{Schulman2017PPO}.

\subsection{Projective simulation}
\label{sec:rl_projective_simulation}

In recent years, there have been introduced novel approaches to \ac{RL} that explore techniques beyond the prototypical value-based and policy gradient methods that we introduce in~\cref{sec:rl_value_based_methods,sec:rl_policy_gradient}. Among those, \acf{PS}~\cite{briegel2012projective} is of particular interest for the physics community, due to its numerous applications in the field.

\begin{figure}
    \centering
    \includegraphics[width=\ToggleForCUP{\columnwidth}{0.95\columnwidth}]{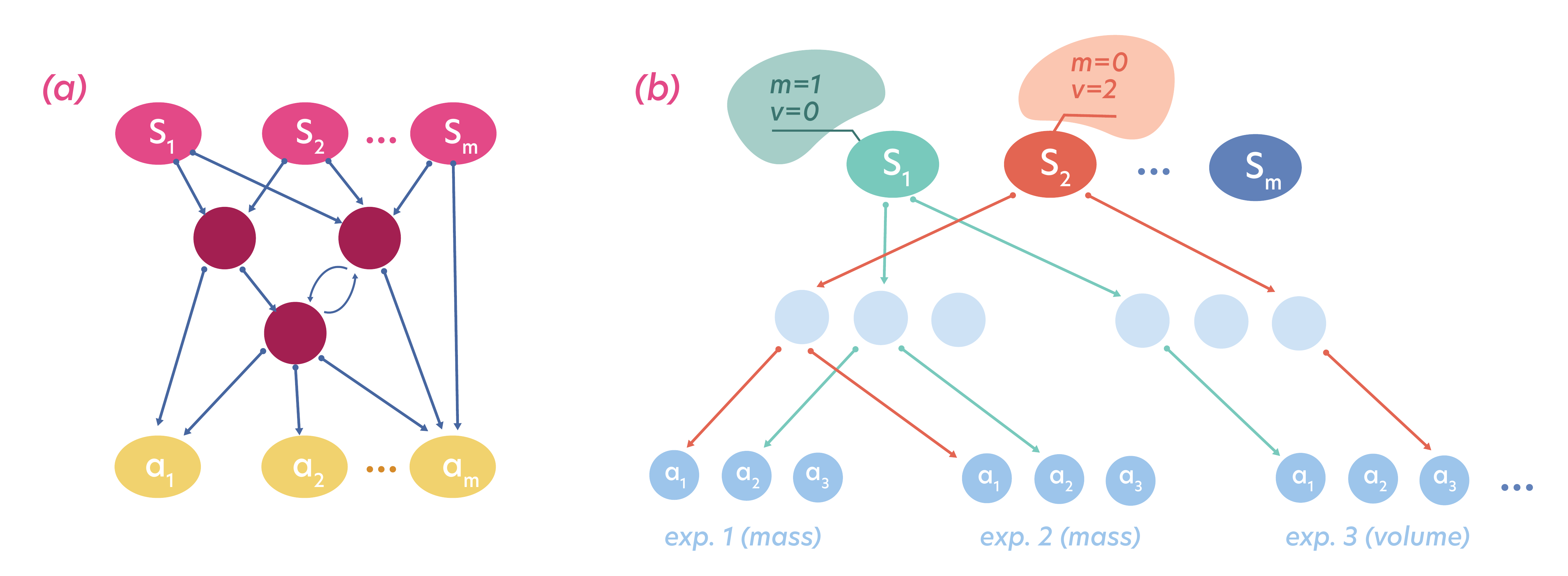}
    \caption[Schematic representation of the episodic and compositional memory of various projective simulation agents]{Schematic representation of the \acf{ECM} of various \acf{PS} agents. (a) A~multi-layer \ac{ECM} with $m$ state nodes (pink), three hidden nodes (red) and $m$ action nodes (yellow). The connectivity of the graph can also be set as a~(b) A~three-layer \ac{ECM}, used to demonstrate the possibility of feature extraction by the \ac{PS} model. See the main text for details.}
    \label{fig:rl_ps}
\end{figure}

\ac{PS} considers an~agent based on an~\acf{ECM}, a~mathematical object capable of storing the information about visited states and actions, and drawing connections between them. This way, the ECM is continuously updated as the agent gathers experience, and it ultimately determines the policy at any given state, as we show below. Usually, the \ac{ECM} is represented as a~directed weighted graph, as shown in \cref{fig:rl_ps}(a). The nodes, defined here as \emph{clips}, represent either visited states, actions, or hidden information learned by the agent. As the agent explores, clips corresponding to new visited states are added to the graph. Similarly, an~agent may create additional ones to accommodate new actions, e.g., the combination of two actions, or hidden information. The edges are weighted, and every new node is initialized with uniform edge weights. The weights determine the transition probability between clips, and they are updated as the agent gathers rewards.

As we have previously mentioned, the \ac{ECM} defines the policy of the \ac{PS} agent. In the vanilla version of \ac{PS}, given an~observed state, the agent performs a~weighted random walk through the \ac{ECM} starting on the corresponding state clip. The walk ends as soon as it lands in an~action node, and the corresponding action is chosen. The probability to jump from one clip, $c_i$, to another, $c_j$, can be any normalized function of the edge weights $h(c_i,c_j)$, such as
\begin{equation}
    P(c_i, c_j) = \frac{h(c_i, c_j)}{\sum_{j\in \mathcal{I}}h(c_i, c_j)}\,,
\end{equation}
where $\mathcal{I}$ is the set of edges of $c_i$. Other transition functions have also been introduced, such as softmax transitions, which allow us to have arbitrary $h$-values.

Following the previous scheme, training a~\ac{PS} agent consists of updating the \ac{ECM} by adding new nodes, and learning the edge weights. The goal is that, for every state clip, the path through the \ac{ECM} leads to the correct action with high probability. Thus, the training can then be reduced to the update of the $h$-values at every time-step via
\begin{equation}
\label{eq:rl_hvalues}
    h(c_i, c_j) \leftarrow h(c_i, c_j) + \gamma (h(c_i, c_j)-1) + r
\end{equation}
where $c_i$ and $c_j$ represent the clips traversed during the random walk through the \ac{ECM}, $\gamma$~is a~damping parameter, and $r$ is the reward given by the environment after performing the chosen action.

With this update rule, for every agent’s decision, i.e., every time it performs a~walk from a~state node to an~action node, all $h$-values of the visited edges are updated. In this way, the $h$-values along the walk are always damped by a~factor $\gamma$, and, in the case that they led to a~rewarded action, they also increase their value by a~factor $r$.

In many practical scenarios, rewards are obtained at the end of a~long series of actions, e.g., performing various steps in a~grid-world to reach a~target. Hence, it is important to ``\stress{backpropagate}'' such reward through the sequence of all the actions that led to it. For instance, in \ac{TD} algorithms, this is achieved by considering the expected value of future states to perform the updates, as we introduce in \cref{sec:rl_value_based_methods}. To accommodate such property, we can generalize the update rule from~\cref{eq:rl_hvalues} by introducing the concept of an~\stress{edge glow}: every time an~edge is traversed, it starts to glow decaying with time. This feature allows the agent to update all the edges in the \ac{ECM} involved in the decisions to describe a~trajectory $\tau=a_0,s_1,a_1,\dots$\footnote{Be careful to not confuse the trajectories through the \ac{ECM} with the trajectories through the state and action spaces. Given a~state $s_t$, the \ac{PS} agent chooses the action $a_t$ by performing a~trajectory through the \ac{ECM} that starts on the corresponding $s_t$ node until it reaches an~action node. Then, the corresponding action $a_t$ is performed to move toward the next state $s_{t+1}$.} which led to a~certain reward. The update rule can then be rewritten as 
\begin{equation}
\label{eq:rl_hvalues_glow}
    h(c_i, c_j) \leftarrow h(c_i, c_j) - \gamma (h(c_i, c_j)-1) + g(c_i, c_j) r\,,
\end{equation}
where $g$ is the glow value.

Each time a~certain edge is visited, its corresponding glow value is set to 1. Then, at every step, all the glow values are dampened via
\begin{equation}
\label{eq:rl_glow_update}
g(c_i, c_j) \leftarrow g(c_i, c_j)(1-\eta)\,,
\end{equation}
effectively decreases all of them with a~rate $\eta$. This means that edges that have been recently visited and led to a~reward $r\neq 0$ are strengthened, while those visited earlier on received a~lesser update, analogous to \ac{TD} algorithms. We refer to ~\cite{mautner2015projective, melnikov2018benchmarking} for an~in-depth and practical description of the usage of the \ac{PS} models.

The presented approach to \ac{PS} is a~tabular method, similarly to Q-learning from~\cref{sec:rl_Q_learning}, as the agent's deliberation is saved in the adjacency matrix of the \ac{ECM}, namely the $h$-matrix. As commented previously, tabular methods have strong limitations when dealing with large action and state spaces.
The non-tabular approaches for \ac{PS} have been proposed~\cite{jerbi2021quantum}. In that case, a~neural network (and more precisely, an~energy-based model) is trained to output the $h$-value for a~certain state-action pair, analogously to how \acp{DQN} are used to predict Q-values, as we introduce in  \cref{sec:rl_deep_Q_learning}.

An~important feature of the \ac{PS} model is its transparency and potential interpretability\index{interpretability} power, in contrast to other approaches such as Q-learning. In the latter, the Q-values encode the expected reward received from an~action-state tuple. As the policy relies on performing the action with largest Q-value, there is little to no room for interpretability, aside from such maximization. Conversely, \ac{PS} constructs a~visible graph encoding the probabilities to hop between nodes, which may represent both direct information from the \ac{RL} task, i.e., actions and states, but also hidden information extracted by the agent. For instance, as we describe in \cref{sec:rl_qexperiments}, the authors of Ref.~\cite{melnikov2018active} were able to interpret the hidden structure of the \ac{ECM}, related in that example to different optical devices. Interestingly, the \ac{PS} agent was able to create useful optical gadgets composed of multiple devices by composing actions together into new joint nodes (see~\cite{briegel2012projective}). Nonetheless, when working in the so-called two-layer \ac{PS} (one layer of nodes for the states and one for the actions), \ac{PS} reduces to a~very similar model to Q-learning. Indeed, recent works have extensively compared both approaches~\cite{boyajian2020convergence}. However, we can introduce further \textit{hidden} nodes to build deeper \ac{PS} models, as shown in \cref{fig:rl_ps}(a). 

There have been multiple efforts to build such deep \ac{PS} architectures and to show that they are indeed able to extract relevant hidden features from the environment or the task at hand~\cite{melnikov2017projective,ried2019minimal}. An~enlightening example is shown in Ref.~\cite{ried2019minimal}, which we schematically reproduce in \cref{fig:rl_ps}(b). In this work, an~agent is given a~set of objects with different physical properties, such as mass, charge and volume. For simplicity, these quantities can take only one of three values: 0, 1 or 2. The agent has access to different experiments, which measure each of this quantities separately. The states are then different objects with certain properties, e.g., in \cref{fig:rl_ps}(b), $S_2$ is an~object of mass 0 and volume 2, obviously in arbitrary units or categories. On the other hand, the actions are the predictions over the various experiments. For instance, $a_1$ corresponds to the prediction that the object has the lowest value measured by experiment one (related in this case to mass), $a_2$ to an~intermediate value of that same experiment, etc. The authors show that the \ac{PS} agent would assign the hidden nodes to meaningful features of the problem. In particular, each hidden node would represent a~particular value of a~physical quantity, as shown in \cref{fig:rl_ps}(b). Such interesting feature is not only a~valuable sign of the interpretability of the \ac{PS} model, but also was shown to increase its generalization performance.
 
\subsection{Examples and applications }\label{sec:rl_examples}

In this section, we showcase a~series of prominent applications of \ac{RL}. Between all the examples, we find instances of each \ac{RL} paradigm that we discuss in the previous sections. We start with two toy examples to settle the theoretical foundations of policy gradient, as they have analytical solutions.
Then, we briefly comment on some of the most famous examples of \ac{RL}: Atari video games, and Go.
Finally, we highlight a~few applications of \ac{RL} to quantum physics, more precisely, in the context of future quantum technologies such as quantum circuits, error correction, and certification.

\subsubsection{Toy examples}
\label{sec:policy_gradient_toy_examples}

Let us illustrate the REINFORCE algorithm, from~\cref{ssec:rl_reinforce}, by solving a~couple of toy examples. These simple scenarios allow us to solve all the equations analytically in order to lay down the foundations and become familiar with the basic concepts.

\paragraph{The random walker} Consider an~agent that can move along a~one-dimensional path with only two actions: move up or down. Every time the agent goes up, it receives a~positive reward $r_t=+1$ and every time it goes down it receives a~negative reward $r_t=-1$. Considering the undiscounted case, $\gamma=1$, the return of a~trajectory of $T$ steps is the final position $G(\tau)=x_T$. We can also express it in terms of the number of times the agent has taken the actions to go up or down $G(\tau)=n_{\text{up}} - n_{\text{down}} = 2n_{\text{up}} - T$. Clearly, the optimal policy is to always go uphill regardless of the current position.

In such a~simple scenario, there is no notion of a~state for the agent. Therefore, the policy only depends on the action. Furthermore, since there are only two possible actions, we can define the parametrized policy for one, e.g., $\pi_{\params}(\text{up})\in[0,1]$, and take the other as $\pi_{\params}(\text{down})=1-\pi_{\params}(\text{up})$. Let us consider the parametrized sigmoid policy
\begin{equation}
    \label{eq:rl_random_walker_policy}
    \pi_{\param}(\text{up}) = \frac{1}{1+e^{-\param}}, \ 
    \pi_{\param}(\text{down}) = \frac{1}{1+e^{\param}} \,,
\end{equation}
which determine the probability to move upwards or downwards, respectively, in terms of the single parameter $\param$. Their score functions are 
\begin{equation}
    \label{eq:rl_random_walker_scorefun}
     \nabla_{\param}\log\pi_{\param}(\text{up}) = \pi_{\param}(\text{down}), \ \nabla_{\param}\log\pi_{\param}(\text{down}) = -\pi_{\param}(\text{up})\,.
\end{equation}

With~\cref{eq:rl_random_walker_policy,eq:rl_random_walker_scorefun}, we can compute the parameter update rule from~\cref{eq:rl_reinforce_update} analytically. We can express each of its terms as a~function of $\pi_{\param}(\text{up})$:
\begin{equation}
    \begin{split}
        \estimationoperator\left[G(\tau)\sum_{t=0}^{T-1}\nabla_{\param}\log\pi_{\param}(a_t)\right] &= \estimationoperator\left[\left(n_{\text{up}} - n_{\text{down}}\right)\left(n_{\text{up}}\pi_{\param}(\text{down}) - n_{\text{down}}\pi_{\param}(\text{up})\right)\right] \\
        &= \estimationoperator\left[\left(2n_{\text{up}}-T\right)\left(n_{\text{up}}-T\pi_{\param}(\text{up})\right)\right] \\
        &= 2\estimationoperator\left[\left(n_{\text{up}}-T/2\right)\left(n_{\text{up}} - \estimate{n_{\text{up}}}{\pi_{\param}}\right)\right] \\
        &= 2\text{Var}\left[n_{\text{up}}\right] = 2T\pi_{\param}(\text{up})\left(1-\pi_{\param}(\text{up})\right)\,,
    \end{split}
\end{equation}
where we have taken $T\pi_{\param}(\text{up})$ as the expected number of upwards moves $\estimate{n_{\text{up}}}{\pi_{\param}}\,$. With this, we are able to reach a~closed analytical form for the parameter update rule in this simplistic scenario, which is not the usual case in \ac{RL}. This allows us to understand the way that actions are reinforced. For instance, the term $n_{\text{up}} - \estimate{n_{\text{up}}}{\pi_{\param}}$ reinforces actions that lead toward higher upwards moves than expected following the policy, and it penalizes those that lead to fewer. 

\begin{figure}[t]
    \centering
    \includegraphics[width=\ToggleForCUP{\columnwidth}{0.9\columnwidth}]{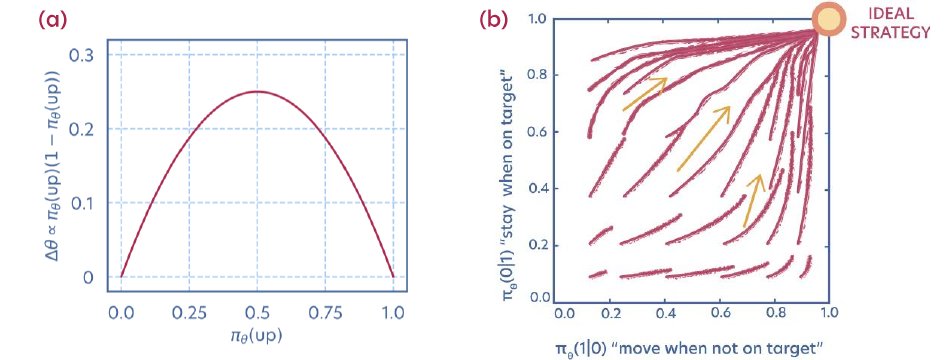}
    \caption[Walkers and reinforcement learning]{Walkers and \ac{RL}. (a) Parameter update from \cref{eq:rl_random_walker_update} for the random walker. (b) Evolution of various policies trained on the walker with target.}
    \label{fig:toy_model_RL}
\end{figure}

The parameter update is a~quadratic function on the policy, such that
\begin{equation}
    \label{eq:rl_random_walker_update}
    \Delta\param\propto\pi_{\param}(\text{up})(1-\pi_{\param}(\text{up}))\,,
\end{equation}
which is minimal either close to the optimal policy $\pi_{\param}(\text{up})\simeq 1$ or far from it $\pi_{\param}(\text{up})\simeq 0$, as shown in~\cref{fig:toy_model_RL}(a). We can understand this in a~very intuitive way: if the agent is already prioritizing the action to move upwards, it has very little to learn from there on. Conversely, if it barely takes this action, it cannot learn that it is the right choice. Hence, the agent learns the most whenever it takes both of actions at a~similar rate. This also reflects the importance of the initialization. If we initialize the policy to $\pi_{\param_0}(\text{up})\simeq0$, the agent takes much longer to converge to the optimal policy than with $\pi_{\param_0}(\text{up})\simeq0.5$.

\paragraph{The walker with target} Consider now a~slightly more complex situation in which the agent moves along a~one-dimensional path and has to stop at a~target location. In this case, the two actions are to move forward, or to stay. The agent receives a~reward every time step it stays at the target location. In this example, the optimal policy is to move forward until the agent reaches the target, and then stop.

In contrast to the previous example, the agent is no longer blind, and the policy does depend on the state. Notice that, even though the agent moves in space, the actual position is completely irrelevant to the problem, and the only important information is whether the agent is in the right position or not. Therefore, we encode this information with Boolean indicators, assigning $s=1$ when the agent is at the target location, and $s=0$ elsewhere. Hence, despite the agent moving in real space, it only navigates in a~two-state \ac{MDP}.\footnote{We emphasize that, when we frame a~problem as an~\ac{RL} instance, we only need to model and encode the information that is relevant to the problem. Hence, the resulting state and action spaces do not need to correspond directly to those in the ``real world''. The simpler the \ac{MDP}, the easier it is be for the agent.} 
Then, we denote the actions ``stay'' and ``move'' with $a=0$ and $a=1$, respectively. This way, the optimal policy always takes the action to move when not in target, and to stay when in target. We illustrate the optimal policy in~\cref{tab:rl_walker_target_policy}. Additionally, we illustrate the convergence of various policies to the optimal one with REINFORCE in~\cref{fig:toy_model_RL}(b).

\begin{table}[t]
    \centering
    \begin{tabular}{ccc}
    \hline
         & \makecell{stay\\ $a=0$} & \makecell{move\\ $a=1$} \\
        \cline{2-3}
        \makecell{out of target\\ $s=0$} & 0 & 1 \\
        \makecell{on the target\\ $s=1$} & 1 & 0 \\
    \hline
    \end{tabular}
    \caption{Optimal policy $\pi^*(a|s)$ for the walker with target example.}
    \label{tab:rl_walker_target_policy}
\end{table}

\subsubsection{Go and Atari games}
\label{sec:rl_go_atari}
 
Games are one of the most natural applications for \ac{RL}, and they serve as a~benchmark for the state of the art methods. Most games involve long-term strategies, and early actions may lead to completely different outcomes, even in short time scales. Furthermore, many games involve vast state spaces, or even infinite ones. Overall, they pose a~great challenge that has motivated some of the greatest advances in the field.  

The first applications of \ac{AI} to games were board games. The first superhuman performance was demonstrated in chess when, in 1997, a~knowledge-based system \textit{Deep Blue}~\cite{Campbell2002deepblue} beat Garry Kasparov, the highest-rated chess player in the world at the time. A~more recent breakthrough has been achieving superhuman performance in the game of Go~\cite{AlphaGo}. Go is a~Chinese board game which is over 3,000 years old. Two players take turns to place stones on the board. The goal is to conquer as much space as possible, either by strategically surrounding empty spaces, or capturing the opponent's stones by surrounding them. Once all stones are allocated, the player with the largest captured territory wins. Even with this simple set of rules, there are $10^{172}$ possible board configurations, making this game order of magnitudes more complex than chess \cite{GoVsChess}. 

The computer program developed by DeepMind, AlphaGo~\cite{AlphaGo}, combines a~technique called Monte Carlo tree search~\cite{Coulom2006MCTS} with deep \acp{NN}. With this approach, the goal is to progressively build a~search tree of the state space that grows as the agent gathers experience. In the tree, each edge contains the learned action-value function $Q(s,a)$, which partially determines the policy, similar to Q-learning from~\cref{sec:rl_Q_learning}. However, since the state space is virtually infinite, they implement two \acp{NN} that guide the search through the regions outside of the tree: a~parametrized policy that guides the exploration, and a~parametrized value function that predicts the probability to win from each state. See~\cite{AlphaGo} for a~detailed explanation. 

Initially, they train the policy network by supervised learning, taking example moves from expert games. This provides them with an~early advantage with respect to starting tabula rasa to build the search tree from already functional strategies. However, they then proceed to train the whole pipeline through \stress{self-play}, i.e., playing against itself, further refining the policy via policy gradient, as shown in~\cref{sec:rl_policy_gradient}. This model defeated the world champion of Go in 2015.
\begin{figure}
    \centering
    \includegraphics[width=0.7\columnwidth]{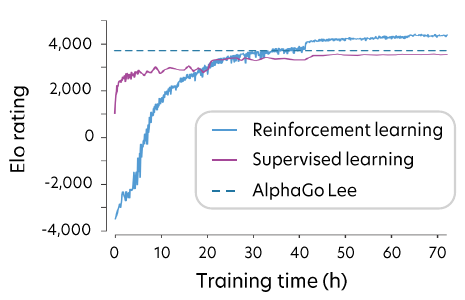}
    \caption[Performance of AlphaGo and AlphaGo Zero]{Performance comparison between AlphaGo (initial supervised learning)~\cite{AlphaGo}, and AlphaGo Zero (pure \ac{RL})~\cite{AlphaGoZero} at the game of Go. Initially, AlphaGo has an~advantage thanks to the initial supervised training. However, it limits its capabilities and it is quickly outperformed by AlphaGo Zero. The horizontal dashed line corresponds to the Elo rating of the AlphaGo version that defeated Lee Sedol, the winner of 18 international titles, in March 2016, being a~reference point for the supervised/pure \ac{RL} performance. Taken from \ToggleForCUP{Silver, D. \textit{et al.} (2017). \textit{Mastering the game of Go without human knowledge}. Nature 550, 354~\cite{AlphaGoZero} with permission from Springer Nature.}{Ref.~\cite{AlphaGoZero}.}}

    \label{fig:rl_alpha_go}
\end{figure}

This approach has been improved by removing the initial supervised training over expert human games, and purely training through self-play from scratch. This algorithm is known as AlphaGo Zero~\cite{AlphaGoZero}. This new version defeated the previous one by a~hundred games to zero. In~\cref{fig:rl_alpha_go}, we see the performance of AlphaGo and AlphaGo Zero with training time in terms of Elo rating.\footnote{The Elo rating system, named after its creator Arpad Elo, is a~method to calculate the relative skill level of players in zero-sum games. After every game, the winning player takes points from the losing one. The difference in rating between players determines the total number of points gained or lost after a~game. If the higher-rated player wins, only a~few rating points are taken from the lower-rated player. However, in the opposite case, the lower-rated player takes many points from the higher-rated one.}
Initially, AlphaGo has a~substantial advantage thanks to the previous supervised learning phase. However, this pre-training ultimately limits its capabilities, and AlphaGo Zero outperforms it in just a~few hours of training. Furthermore, while these algorithms are generally tailored to the specific game, more general and recent approaches, defeated the previous benchmarks in chess, shogi, and Go at the same time~\cite{Silver2018generalRL}. 

Another exciting avenue for \ac{RL} in games are video games. One of the first applications were Atari games, achieving superhuman performance with deep Q-learning~\cite{Mnih2015}, as we explain in~\cref{sec:rl_deep_Q_learning}. In this case, the state space is also infinite and the agent receives the screen pixels as input, together with the current score. However, the action space is limited by the game controller, which is very convenient for Q-learning. This approach achieved superhuman performance in forty nine different games with the same algorithm.

Some other recent outstanding results in video games include competitive performance in StarCraft II~\cite{Vinyals2019alphastar}, Dota 2~\cite{OpenAI2019Dota}, and Minecraft~\cite{Guss2019minerl}. Furthermore, advances in model-free \ac{RL} have motivated the research on planning with model-based algorithms, with which some of the benchmarks that we introduce above have been bested~\cite{MuZero}. This approach does not even require the explicit encoding of the game rules, as it builds a~model of them while playing.

\begin{figure}[t]
    \centering
    \includegraphics[width=0.7\columnwidth]{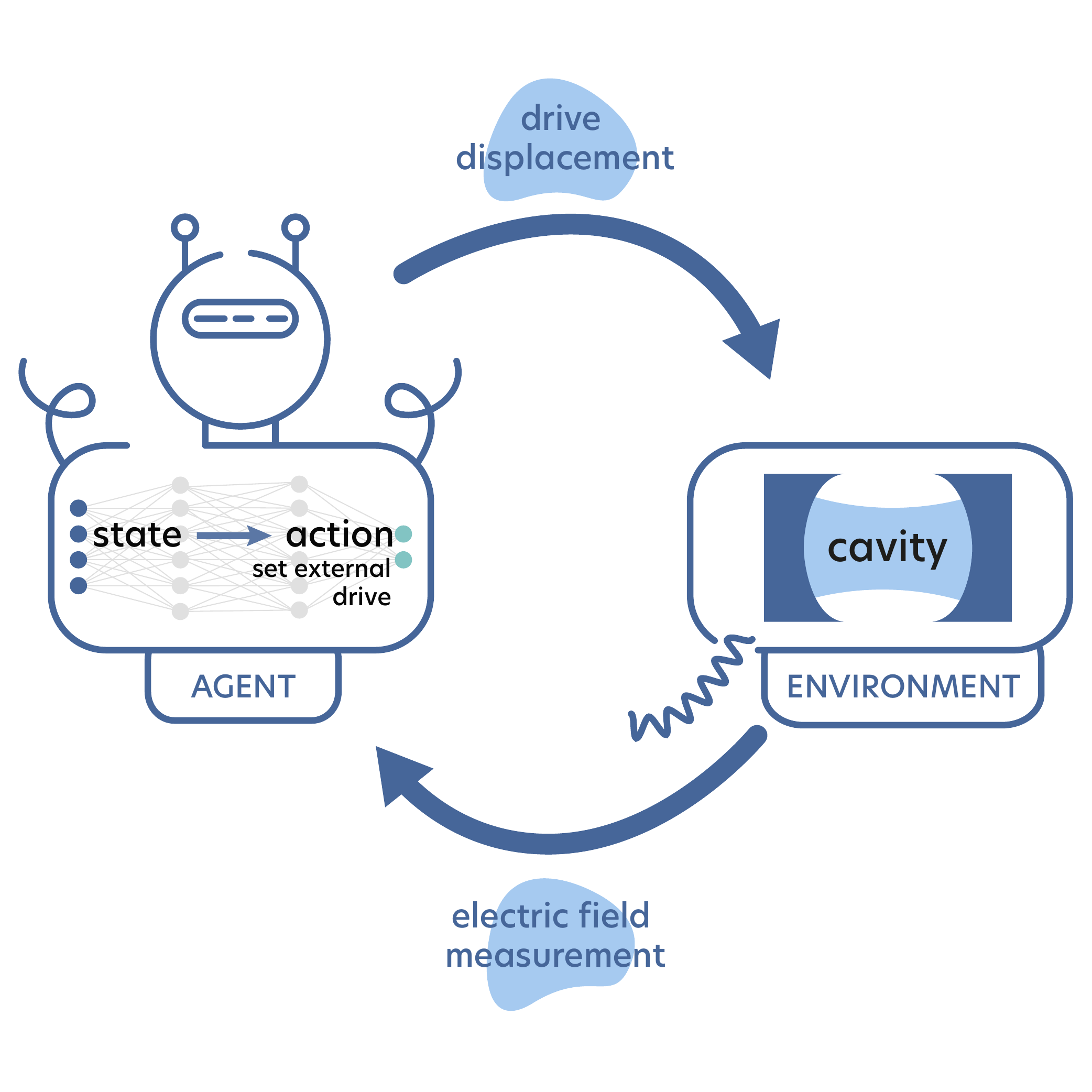}
    \caption[Reinforcement learning for quantum feedback of an~optical cavity]{Driven single mode microcavity as an~\ac{RL} environment proposed in Ref.~\cite{Marquardt2021_RL_Lecture_notes}. The mode decays from the cavity. Its measurement serves as the observation for the agent represented by an~\ac{NN}. The network converts a~measurement trace into probabilities for all the available actions, which give feedback for the displacement drive of the cavity.}
    \label{fig:rl_quantum_feedback_1}
\end{figure}

\subsubsection{Quantum feedback control}\label{sss:RL_qcontrol}

Quantum control is a~research direction in quantum technologies that aims to improve the initialization and stabilization of a~desired quantum state. Deep \ac{RL} algorithms have already been successfully employed in a~wide range of applications for quantum feedback control \cite{QuantumFeedback2021,Foesel2018errorcorrection,Twamley2021,Nguyen2021}. In general, the quantum system is controlled by an~\ac{RL} agent with a~feedback loop with some measurements periodically performed on the system. In this way, the agent drives the control scheme based on the measurement results. In Ref.~\cite{Foesel2018errorcorrection}, the authors consider a~single-mode quantum cavity. The cavity mode is leaking, and this signal can be measured. The goal is to adjust an external drive amplitude of a~beam entering the cavity to create and stabilize a~cavity quantum state with a~single photon as depicted in \cref{fig:rl_quantum_feedback_1}.

The agent observes the measured electric field that leaked from the cavity, which is the state of the environment. Given the observation, the agent can set the value of the driving laser amplitude. The system evolves under the set parameters for a short period of time. Then, the leaked electric field is measured again and the process is repeated until a time limit is reached.
The agent is trained with a policy gradient approach, as introduced in \cref{sec:rl_policy_gradient}. During training, the agent eventually finds a~strategy to compete losses with the proper drive, ending up in the stabilized target cavity state.

\subsubsection{Quantum circuit optimization}\label{sssec:rl_qcircuit_optimization}

Quantum computing based on quantum gates requires designing a~quantum circuit for a~specific quantum algorithm. However, there can be many different sequences of quantum gates implementing the same algorithm. Additionally, due to the fact that quantum gates have non-perfect fidelity, the more gate operations are performed, the more errors appear during the algorithm execution. As such, quantum circuits should be designed in the most optimal way, implementing the least possible number of quantum gates.  This is especially important for \ac{NISQ} devices, which currently allow for $> 100$ qubits \cite{Preskill2018quantumcomputingin} but, at the same time do not allow for high-level logical quantum error correction.\footnote{In fact, we show how to employ \ac{RL} methods to tackle quantum error correction in~\cref{sec:RL_qerror_correction}}
Quantum circuit optimization utilizes the fact that there exist certain sets of transformation rules that allow us to replace sequences of quantum gates by others that yield the same output. For example, these transformations could involve swapping the position of two gates, or moving one gate to a~different position relative to another. Furthermore, some sequences of gates can be shortened by merging gates without changing the output.

\begin{figure}[t]
    \centering
    \includegraphics[width=0.7\columnwidth]{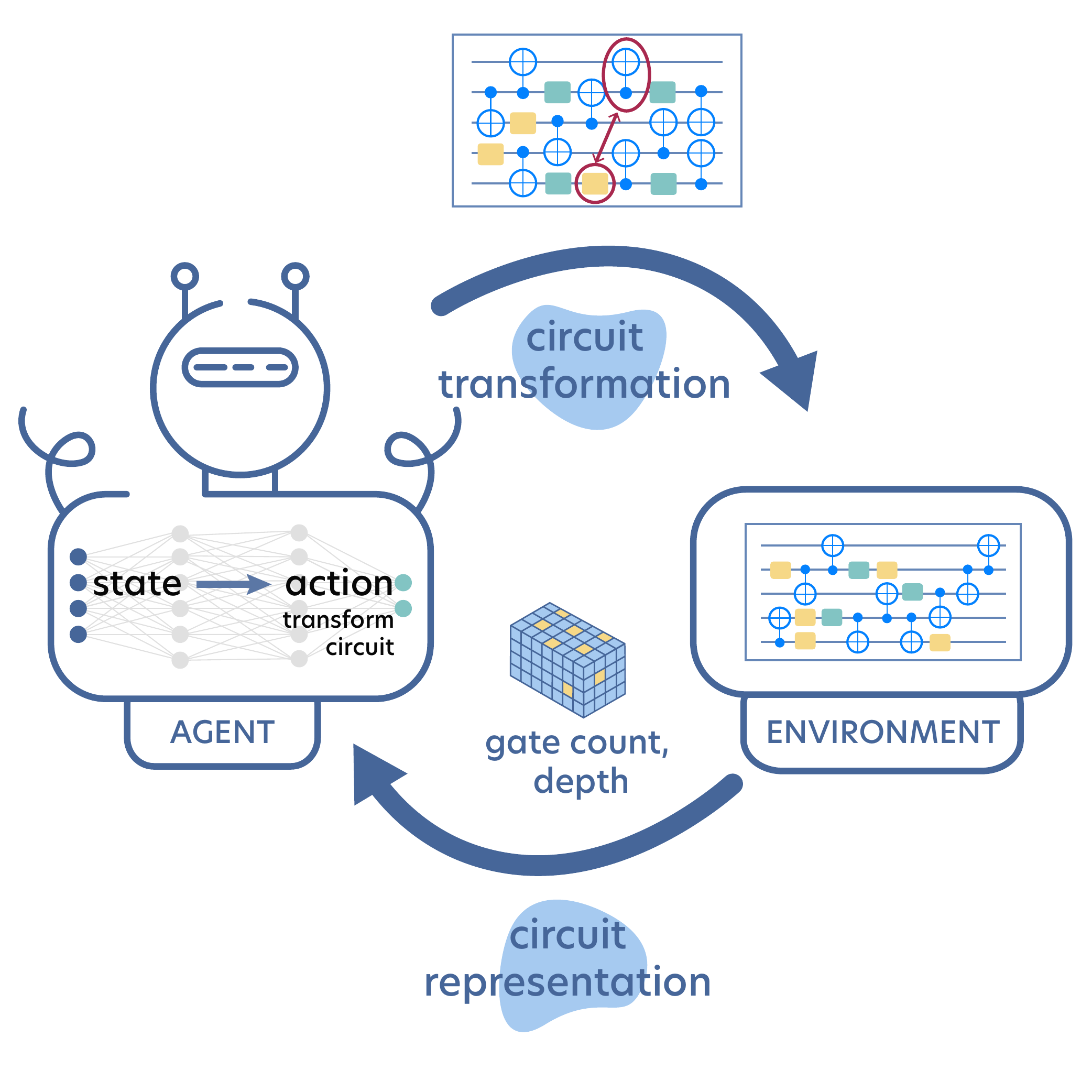}
    \caption[Reinforcement learning for circuit optimization]{Schematic representation of the \ac{RL} framework for circuit optimization proposed in Ref.~\cite{Foesel2021circuitoptim}. The agent observes a~representation of a~quantum circuit given by the environment. Then, it can choose to perform a~modification to the circuit. The environment calculates a~reward depending on the gate count (or another metric) of the resulting circuit, and it provides the agent with the new circuit and the reward.}
    \label{fig:RL-QCO-2}
\end{figure}

We can naturally formulate quantum circuit optimization as an~\ac{RL} problem~\cite{Foesel2021circuitoptim}.
In the resulting framework, depicted in~\cref{fig:RL-QCO-2}, the environment holds the quantum circuit, containing information about the different gates, such as their error rates. Te agent can observe a~representation of the quantum circuit, which corresponds to the state, and it can decide to perform a~transformation to the circuit from a~set of possible transformation rules.
The environment can evaluate the resulting circuit after the transformation, and provide the agent with a~reward. The reward can account for various aspects, such as the reduction in the total gate count, the reduction in depth (the time needed for the circuit to run), or the combination of both. Additionally, the reward function can also depend on a~decoherence estimate for the whole circuit, based on the decoherence that happens on each the gates.

This way, the resulting circuit optimization is an~autonomous process that can account for specific information about the hardware when chosing the actions, e.g., some gates involve longer execution times, or a~given qubit may be prone to further errors than others. In the future, quantum compilers will be able to optimize circuits tailored to the hardware specifications and native gate implementation.

\subsubsection{Quantum error correction} \label{sec:RL_qerror_correction}

Whenever we perform any kind of computation, we have to ensure that it is performed flawlessly. In both classical, and quantum computation, we need mechanisms to mitigate any possible effect of errors occurring during computations.
Whereas classical error corrections methods have long been established, the current quantum error correction schemes come with a~daunting overhead in the number of qubits.
Moreover, classical correction schemes cannot be transferred directly to the quantum case, since we can neither simply copy arbitrary quantum states (known as the no-cloning-theorem\cite{no_cloning_theorem}) nor measure the quantum computer's state arbitrarily to find possible errors, as we would erase the state's superposition.
Some error correction implementations tackle these challenges using \ac{RL} methods.
Here, we discuss two different approaches.

\begin{figure}[t]
    \centering
    \includegraphics[width=0.7\columnwidth]{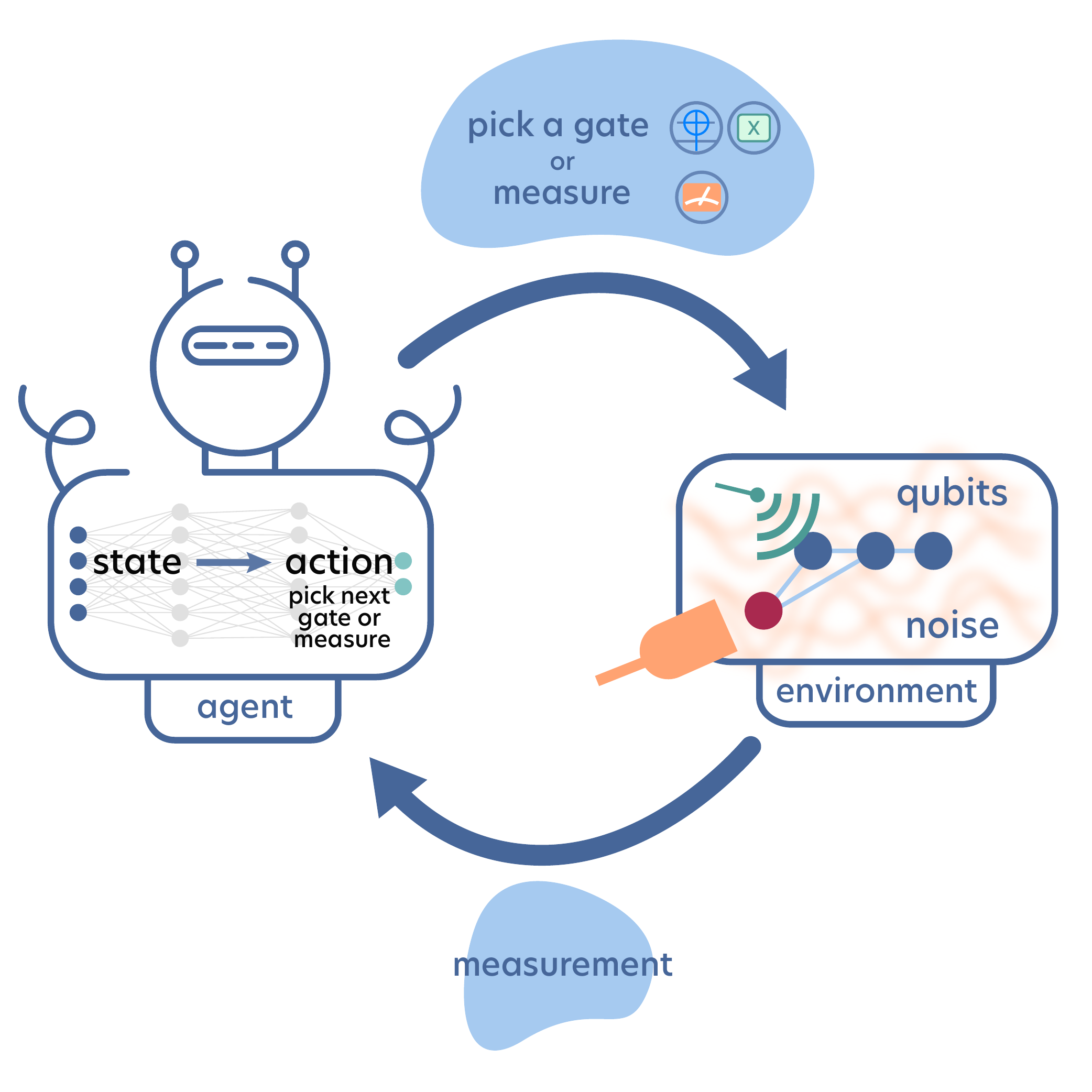}
    \caption[Reinforcement learning for quantum error correction]{Schematic representation of the \ac{RL}-based error correction framework proposed in Ref.~\cite{Foesel2018errorcorrection}. The agent can choose the next gate or measurement to be applied to an~ensemble of a~few, possibly error-affected, qubits in order to  protect a~single target qubit.}
    \label{fig:rl_error_correction}
\end{figure}

\paragraph{Error correction with qubit interaction} The first one proposes a~suitable error correction scheme from scratch, simply interacting with a~collection of qubits~\cite{Foesel2018errorcorrection}, as sketched in \cref{fig:rl_error_correction}.
This approach treats the actual hardware as a~black-box, and therefore it is versatile regarding the hardware's constraints, as it does not require any prior knowledge about the task.
In this setting, the goal is to preserve an~arbitrary single-qubit state, $\ket{\phi(0)} = \alpha \ket{0} + \beta \ket{1}$, over time.
In order to do so, the agent can choose to apply gates from a~given set, or to perform measurements on auxiliary qubits.
This way, any hardware limitation can readily be incorporated by a~suitable choice of the avilable gates, which conforms the action space. Then, we can measure the performance in terms of the fidelity $F = |\braket{\phi(T)}{\phi(0)}| \in [0,1]$ after some arbitrary, but fixed time $T$.

However, a~naive \ac{RL} approach is bound to fail when we only consider the fidelity as the reward.
Almost all possible circuit transformations reduce the fidelity, thus making random strategies worse than remaining idle. This happens even when considering the fidelity after each new gate or measurement, as the optimal scheme initially decreases the fidelity, and applies a~recovery sequence to restore it afterwards. Hence, the  chance of finding the right gate sequence to protect the state vanishes for large times $T$.
In order to overcome these challenges, the authors in Ref.~\cite{Foesel2018errorcorrection} propose a~two-stage learning scheme, and a~more convenient reward function.

\highlight{%
The two-stage learning consist of training two models. First, we train an~\ac{RL} agent which has access to enhanced information with respect to what it is available in an~actual device, such as a~full description of the multi-qubit state. This also allows us to use a~more convenient reward function: the \stress{recoverable quantum information}
\begin{equation}
\label{eq:rl_recoverable_quantum_info}
    r_t = \frac{1}{2} \min_{\Vec{n}} \norm{\hat{\rho}_{\Vec{n}}(t) - \hat{\rho}_{-\Vec{n}}(t) }_1\,,
\end{equation}
where $\Vec{n}$ denotes the vector in the Bloch sphere corresponding to the initial state. This reward uses the idea that $\Vec{n}$ and $-\Vec{n}$ are orthogonal to provide a~reward at every time-step that guides the agent toward the optimal gate sequence. See~\cite{Foesel2018errorcorrection} for details.

Then, we train the second model using the first one as a~teacher. The second model only has access to the information available in a~real device, such as the gates it applies, and the occasional measurement outcomes. Instead of using \ac{RL}, we train it in a~supervised way to mimic the behavior of the first one. The process is analogous to training a~supervised classifier in which the labels are the actions of the first model.
}

The overall process of two-stage learning is way faster, and much less computationally demanding than directly solving the original problem. The main limitation is that the teacher model requires a~full state description of the multi-qubit system, which limits the application to just a~few qubits, and requires a~well-characterized noise map of the device that might not be known, in practice.

\paragraph{Error correction with stabilizer codes} Whereas the first approach aims at discovering the best error correction scheme from direct qubit interaction, the second one implements a~quantum code to represent logical qubits~\cite{Shor1995PRA,Steane1996PRL,stabilizer_codes_review}. In the example we consider here~\cite{Sweke2021_errorcorrection}, the authors use stabilizer codes to achieve error correction via redundancy.
In order to properly understand the process, let us build some basic intuition about the stabilizer formalism.  Consider a~precursory code to correct arbitrary single bit flips of the physical qubits, $\ket{0} \leftrightarrow \ket{1}$ with the encoding
\begin{equation*}
    \ket{0_L} = \frac{1}{\sqrt{2}} \left( \ket{000} + \ket{111} \right)
\end{equation*}
for a~single logical qubit state $\ket{0_L}$ in terms of three physical qubits.
We can jointly measure subsets of qubits without changing the state with \stress{stabilizer operations}.
In this case, we can apply the operations $Z_1Z_2$, and $Z_2Z_3$ without altering the qubit state: $Z_1Z_2 \ket{0_L} = \ket{0_L} = Z_2Z_3 \ket{0_L}$.
Moreover, we can use these operators to detect bit-flip errors on one of the physical qubits, as the stabilizer operators are designed to not alter the erroneous state either.\footnote{In practice, we would first devise a~set of stabilizer operators, and then, we would define the logical 0 and 1 states as the simultaneous eigenstates of all of the stabilizers.} The stabilizer measurements have an~outcome of $\pm 1$, and applying them successively we can identify whether any qubit suffered an~error to proceed with the correction.
The series of outcomes is known as the \stress{syndrome}, and, in practice, these can also have errors.

In this example, we can deal with single bit-flip errors, but not with phase errors represented by $Z_i$ operators.
For the error correction of arbitrary single-qubit errors, we need five physical qubits with four stabilizer operations~\cite{stabilizer_codes_review}.
The amount of qubit overhead grows quickly with the number of qubit error classes to cover.
The stabilizer code in Ref.~\cite{Sweke2021_errorcorrection}, is a~surface code to protect a~single logical qubit against arbitrary errors affecting up to $d$ qubits while using, at most, $d^2$ physical ones.

To properly perform error correction, we need a~combination of accuracy, scalability, and  speed to detect and correct errors.
We can formulate this as an~\ac{RL} task~\cite{Andreasson2019quantum,Fitzek2020_errorcorrection,Sweke2021_errorcorrection,Theveniaut2021_errorcorrection_neat} implementing the full toolbox introduced in this chapter.
In the setting from~\cite{Sweke2021_errorcorrection}, the environment tracks the underlying quantum state, accounting for possible stochastic errors on the physical qubits in the form of depolarizing and bit-flip noise. The agent can choose to perform single-qubit $X$-, $Y$-, or $Z$- rotations,\footnote{Given that $\mathrm{XZ} = \i \mathrm{Y}$, we can reduce the action space in certain cases.} or to perform syndrome measurements. Then, the environment provides the agent with the (possibly faulty) measurement outcome, and a~reward, from which the agent can decide the new set of actions to perform. The environment employs a~\stress{referee decoder} that checks whether the multi-qubit state after the agent's actions leads to the same logical qubit state. If it is the case, the reward is positive, otherwise, it is negative and the episode terminates. 

\begin{figure}
    \centering
    \includegraphics[width=\ToggleForCUP{0.75\textwidth}{0.7\textwidth}]{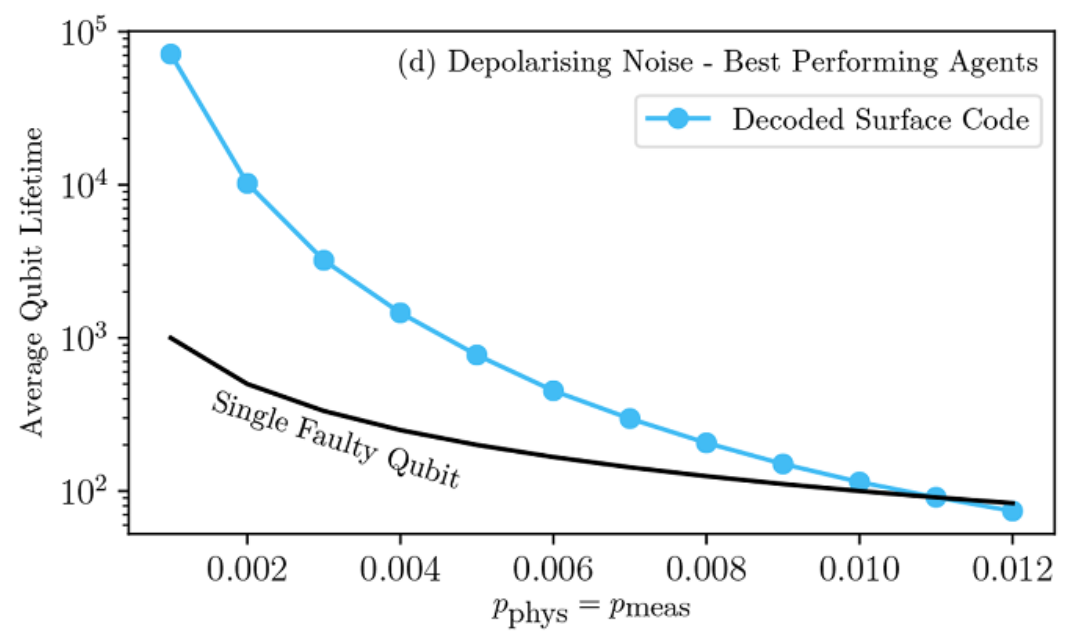}
    \caption[Increased qubit lifetime due to reinforcement learning]{Average lifetime of the logical qubit encoded with the surface code. The physical qubits are affected by depolarising noise with parameter $p_\mathrm{phys}$ that drastically decreases the unprotected qubit lifetime (black line). With agents trained on various noise levels, $p_\mathrm{phys}$, we can dramatically increase the qubit lifetime (blue dots). Taken from \ToggleForCUP{Sweke, R. \textit{et al.} (2021). \textit{Reinforcement learning decoders for fault-tolerant quantum computation.} Mach. Learn.: Sci. Technol. 2, 025005~\cite{Sweke2021_errorcorrection} under the \href{https://creativecommons.org/licenses/by/4.0/}{CC BY 4.0 DEED} license.}{Ref.~\cite{Sweke2021_errorcorrection}.}}
    \label{fig:rl_stabilizers_depol}
\end{figure}

\highlight{%
The authors in Ref.~\cite{Sweke2021_errorcorrection} consider an~agent based on a~\ac{DQN} which they train with Q-learning algorithm, as we explain \cref{sec:rl_deep_Q_learning}.
After training, the average lifetime of the encoded logical qubit can be extended drastically as shown in \cref{fig:rl_stabilizers_depol}.

Furthermore, they implement an~alternative genetic algorithm within the described framework that results into significantly smaller \ac{ML} models, which are better suited to run in actual devices.}

\subsubsection{Quantum experiment design}
\label{sec:rl_qexperiments}
The design of new experiments is key for the development of the quantum sciences. The more complex the applications become, the harder it is to find suitable setups to test our ideas. In the context of quantum physics, this can be illustrated in an~optical experiment, where we combine different components such that the final quantum state has certain desired properties. For instance, finding the appropriate set of components to create multipartite entanglement in high dimensions is a~non-trivial task, and usually relies on sophisticated previous knowledge on the states, and involved mathematical approaches \cite{Erhard2020NatRevPhys}. Nonetheless, such states are of great importance in applications of quantum information and computation, and hence they are highly coveted.

In Ref.~\cite{melnikov2018active}, the authors propose an~autonomous approach to build experiments with \ac{RL}, using the \ac{PS} algorithm that we introduce in~\cref{sec:rl_projective_simulation}. The goal is to create high-dimensional many-particle entangled states, based on the orbital angular momentum of light. To do so, the agent has access to a~set of optical elements, and the actions consist on placing one of such components in the optical table. The states are the different configurations of optical components in the table. After each placement, the environment analyzes the resulting quantum state generated by the setup. If it corresponds to the desired quantum state, it provides the agent with a~reward and the episode ends. If not, the agent continues placing more elements. It is important to note that, due to the presence of noise in optical setups, the more elements, the harder it becomes to correctly find the target quantum state. Hence, the agent is given a~maximal number of elements to reach its goal, after which the episode ends and the table resets.

From a~technical point of view, the agent has a~2-layer \ac{ECM}: one representing the table configurations (states), and one representing the optimal components (actions). An~interesting feature of \ac{PS} is \textit{action composition}: the agent can create new composite actions from simpler ones that where found useful in previous episodes. In the current context, if the agent finds a~particular profitable action sequence leading to a~reward, the actions can be added combined as a~new single one in the \ac{ECM}, hence allowing the agent to access rewarded experiments in a~single decision step. This way, the agent can distill combinations of components that lead to well known setups, such as optical interferometers, as well as completely novel ones, such as a~non-local version of the Mach-Zender interferometer.

Hence, we can divide the general task of generating quantum states in two: finding the simplest optical configuration leading to the target state, and finding as many experiments as possible that produce it. The former is crucial in terms of practical applications of quantum technologies, as shorter experiments are less noisy, and usually easier to implement. The latter allows us to explore to the full extent all possibles solutions to the problem, which may lead to the discovery of new approaches to create the desired quantum states.

The automated design of quantum optical experiments has also been tackled with non-\ac{RL} approaches \cite{Krenn2016PRL,Krenn2021PRX,Krenn2020NatRevPhys}. We describe them in more detail in \cref{sss:MLforexpdesign}.

\subsubsection{Building optimal relaxations}\label{sec:RL_relaxation}

In physics we often encounter optimization tasks that we cannot solve in a~reasonable amount of time. In these cases, we rely on approximate methods to obtain solutions that are as close as possible to the exact one. There are two paradigmatic approaches: variational and relaxation methods. In the former, we parametrize a~family of solutions with the hope that it contains the exact one, such as the variational quantum states introduced in \cref{sec:NN_q_states}. In the latter, we build a~relaxed (easier) version of the problem in order to provide the optimization process with desirable properties, such as convexity.

\begin{figure}[t]
    \centering
    \includegraphics[width=0.7\columnwidth]{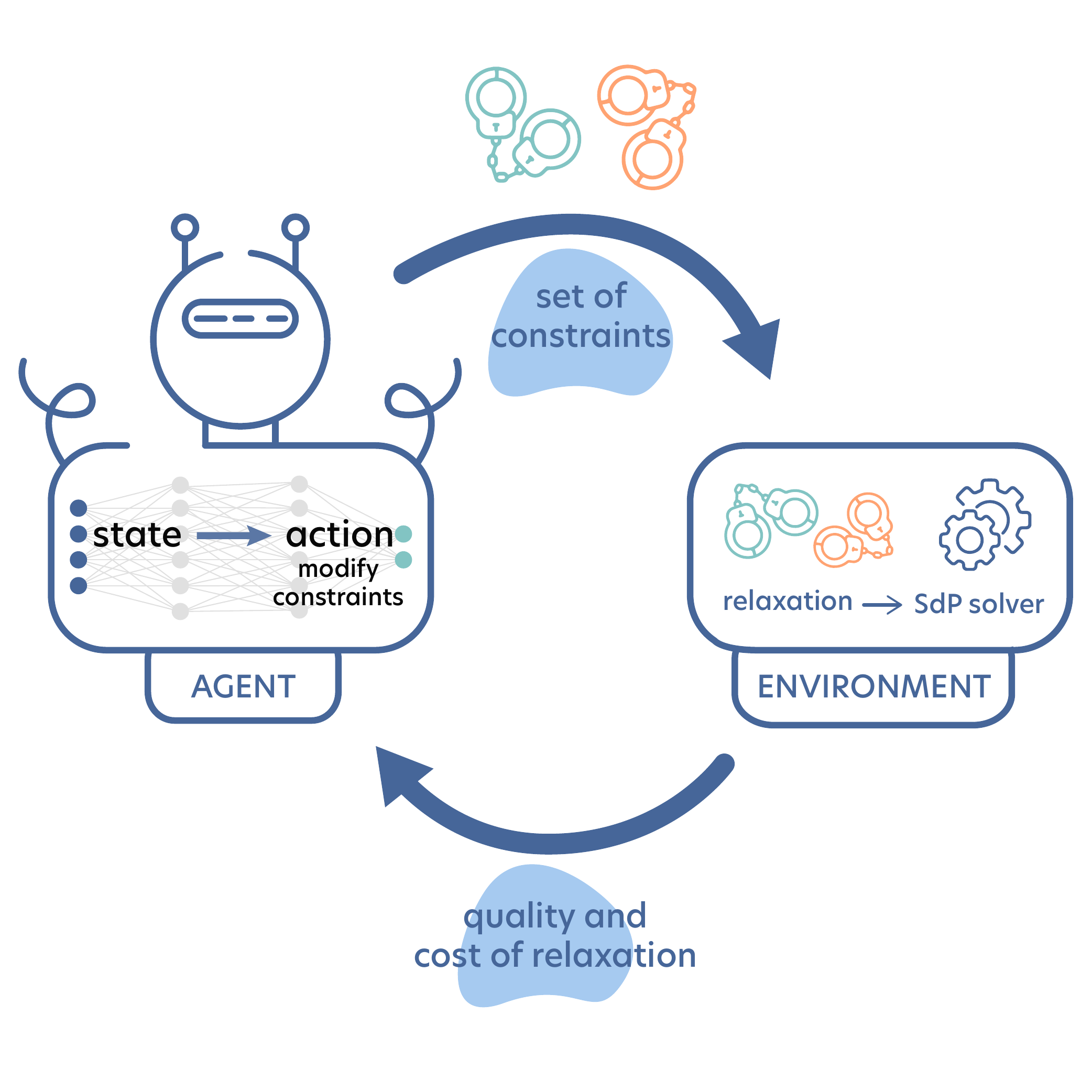}
    \caption[Reinforcement learning to find optimal relaxations]{Schematic representation of the \ac{RL} framework to find optimal relaxations. The agent can modify the set of active constraints of a~problem with its actions. These constraints go into the environment, which solves the constrained optimization problem. Then, the agent observes a~reward that depends on both the result of the problem and the computational cost incurred by the environment. Given this observation, it can decide to further modify the constraints.}
    \label{fig:rl_relaxation}
\end{figure}

Relaxation methods are broadly used in quantum physics and they lie at the core of quantum information processing. One of the most paradigmatic examples in entanglement theory is the relaxation from the set of separable states to those that are \ac{PPT}~\cite{PeresPRL1996}. Determining whether a~state belongs to the first class is hard, whereas it is straightforward to check the membership to the second one. This greatly simplifies the problem of determining whether a~state is entangled: we simply need to check it is not \ac{PPT}. However, while all the product states are \ac{PPT}, there are some entangled states which also belong to this class, thus resulting into an~outer bound to the set of separable states.

Just like with variational methods, we often encounter a~trade-off between the computational cost that we can incur and the accuracy of the method. Hence, given a~limited computational budget, it is crucial to find the relaxation that best approximates the optimal solution. Nevertheless, there is no clear way to know such optimal relaxation beforehand. The most common practice relies on exploiting specific knowledge of the given problem, such as symmetries, to build hand-crafted relaxations which, in general, are suboptimal. However, we can combine \ac{RL} with semidefinite programming to systematically build optimal relaxations~\cite{Requena2023PRR}. 

A~natural way to build relaxations is to remove or relax constraints of the optimization problem at hand. In the proposed \ac{RL} framework, presented schematically in~\cref{fig:rl_relaxation}, the states encode the active constraints of the problem, and the agent can loosen or strengthen them with its actions. The environment acts as a~black box that provides the agent with the associated reward to the action and the new state, i.e., the new set of constraints. The rewards are engineered to guide the agent toward the optimal relaxation, evaluating both the quality and the cost associated to the current one. 

The \ac{RL} agent is completely agnostic to the problem. Therefore, the method can be applied in a~wide variety of relevant problems in physics and optimization, such as entanglement witnessing, optimizing outer approximations to the quantum set of correlations, or finding better sum-of-squares representations of multivariate polynomials, to name a~few. In Ref.~\cite{Requena2023PRR}, the authors show two applications: finding the ground state energy of quantum many-body Hamiltonians, and building energy-based entanglement witnesses. They can infer properties of the system from the resulting optimal relaxations, such as changes in the ground state, and, even more, they can explore the phase diagram in an~autonomous way using transfer learning.

\subsection{Outlook and open problems}

In this chapter, we have introduced the field of \ac{RL} and its main paradigms, featuring value-based \ac{RL} (\cref{sec:rl_value_based_methods}), policy gradient methods (\cref{sec:rl_policy_gradient}), and actor-critic algorithms (\cref{ssec:rl_actor_critic}). Additionally, we have explored other methods that present an~alternative approach to \ac{RL}, such as the \ac{PS} algorithm (\cref{sec:rl_projective_simulation}). These lay down the conceptual foundations to understand a~whole plethora of other advanced \ac{RL} techniques while already being competitive, as we have shown in~\cref{sec:rl_examples}. 

In the context of quantum technologies, \ac{RL} has been widely applied to quantum control problems and, especially, in quantum simulation. With the current boom in quantum computation, many problems involving state preparation, error correction, or controlling and preparing qubits have a natural mapping to the \ac{RL} framework~\cite{Bukov2018PRX,Niu2019npjQI,McKiernan2019arxiv,Zhang2020PRL,Baum2021PRX,Cao2022ComPhys,Metz2022MPSRLcontrol,qiu:2022}. Furthermore, \ac{RL} serves as an optimization tool for large problems with a clear structure, with applications as varied as quantum circuit optimization, the design of experimental setups, or the construction of relaxations in quantum information processing problems. 

Similar to unsupervised learning, \ac{RL} is an~appealing technique for autonomous scientific discovery, as it does not require explicit fully-characterized learning instances. However, while we can identify some previously known strategies in the resulting \ac{RL} applications, as in the~\cref{sssec:rl_qcircuit_optimization} example, there is still the need to develop further analysis techniques in order to fully understand the nature and rationale behind some of the most prominent results.   

A~big concern in the field of \ac{RL} algorithms is data efficiency, which is crucial in applications involving costly experiments or simulations. In this regard, the field of \ac{RL} can greatly benefit from the latest advances in physics, such as devising optimal exploration strategies for the most challenging problems, or leveraging the latest advances in quantum technologies to enhance \ac{RL}, as we show in~\cref{sec:quantum_rl}. 

\subsection*{Further reading}
\begin{itemize}
    \item Sutton, S. R. \& Barto, A. G. (2018). \href{https://mitpress.mit.edu/books/reinforcement-learning-second-edition}{Reinforcement Learning: An~Introduction}. This textbook provides a~comprehensive review on \ac{RL} \cite{SuttonBarto2018}. Specifically, chapters 7 and 12 expand the \ac{TD} concept, and chapter 13 contains a~full complementary derivation of policy gradient, actor-critic, and their application to continuing problems (infinite time).
    
    \item Marquardt, F. (2021). \href{https://doi.org/10.21468/SciPostPhysLectNotes.29}{\textit{Machine learning and quantum devices}}. SciPost Phys. Lect. Notes 29. An introduction to \ac{RL} for physicsts~\cite{Marquardt2021_RL_Lecture_notes}.
    
    \item Silver, D. \textit{et al.} (2014). \href{http://proceedings.mlr.press/v32/silver14.pdf}{\textit{Deterministic policy gradient algorithms}}. PMLR, 387–395 \cite{Silver2014DetPolicyGrad}.
    We have introduced policy gradient methods in~\cref{sec:rl_policy_gradient} with stochastic policies. Here, the authors introduce policy gradient with deterministic policies and its corresponding implementation in actor-critic algorithms.
    
    \item Some of the current state-of-the-art algorithms, such as the ones we mention at the end of~\cref{ssec:rl_actor_critic}, feature additional terms in the objective function, usually in the form of an~entropy or a~\acf{KL} divergence\index{Kullback-Leibler divergence}. This results in more robust algorithms, and it is tightly close to the formulation of \ac{RL} as probabilistic inference. We recommend reading Ref.~\cite{Levine2018arxiv} for a~tutorial, Ref.~\cite{Haarnoja2018SAC} for a~prominent algorithm, and Ref.~\cite{Abdolmaleki2018MPO} for another algorithm, featuring a~great overview of the field. The latter proved its performance in the experimental control of a nuclear fusion reactor~\cite{Degrave2022Nature}.

    \item Some of the most prominent applications of \ac{RL} in quantum technologies are quantum control and error correction. To dive deeper into the quantum control field, we recommend reading Ref.~\cite{sivak:2022} for an alternative (model-free) scheme to the one presented in \cref{sss:RL_qcontrol}. The authors present an experimentally friendly \ac{RL} framework readily applicable to superconducting circuits and trapped ion platforms. On the quantum error correction side, we recommend reading Ref.~\cite{Sivak2023Nature} for a pioneering work demonstrating a fully stabilized and error-corrected logical qubit in a superconducting quantum device. The authors significantly extend the coherence time of the logical qubit using an error correction scheme trained with \ac{RL}.
\end{itemize}

\clearpage
\section{Deep learning for quantum sciences -- selected topics}
\label{sec:hot-topics-ml-physics}

\begin{figure}[h]
\begin{center}
\includegraphics[width=0.4\columnwidth]{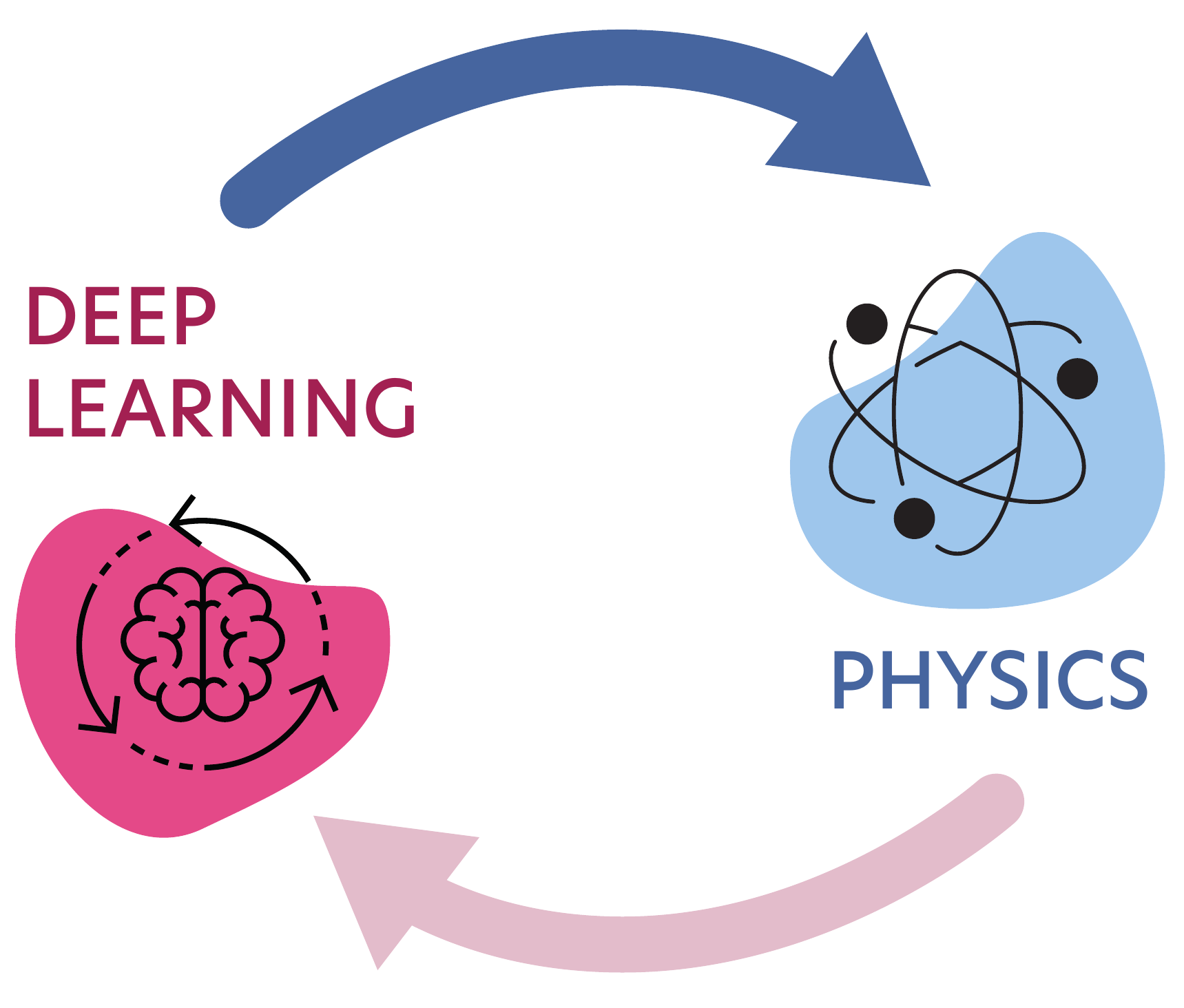}
\end{center}
\caption[Machine learning influences physics]{There exists a~dual relationship between \acf{ML} and physics. In this chapter, we focus on the more popular direction, where techniques from \ac{ML}, in particular \acf{DL}, are used to solve problems in physics.}
\label{fig:ml-physics}
\end{figure}

So far, this book has focused on four broad fields at the intersection of quantum sciences and \ac{ML}: phase classification with unsupervised and supervised \ac{ML} methods in \cref{sec:phase_class}, use of kernel methods especially in quantum chemistry in \cref{sec:gp}, representation of quantum states with \ac{ML} models in \cref{sec:NN_q_states}, and use of \acf{RL} in quantum sciences in \cref{sec:RL}. We have presented each of these ideas in detail after a~(hopefully) exhaustive introduction. As such, \cref{sec:phase_class} through \cref{sec:RL} have highlighted a~plethora of \ac{ML} applications in quantum sciences. However, they obviously do not constitute a~complete overview of the field.

To fill these gaps, the following two chapters aim at addressing more specialized topics located at the intersection of \ac{ML} and quantum sciences. This chapter, in particular, discusses further how \ac{ML} can be used to solve problems in quantum sciences (see \cref{fig:ml-physics}). We start by explaining the concept of \acf{DiffP} and its use cases in quantum sciences in \cref{sec:hot-topics:dp}. \Cref{sec:hot-topics:DE} describes generative models and how they can tackle density estimation problems in quantum physics. Finally, we describe selected \ac{ML} applications for experimental setups in \cref{sec:ML_for_exp}. 

\subsection{Differentiable programming}\label{sec:hot-topics:dp}
\begin{figure}[t]
    \begin{center}
    \includegraphics[width=0.7\columnwidth]{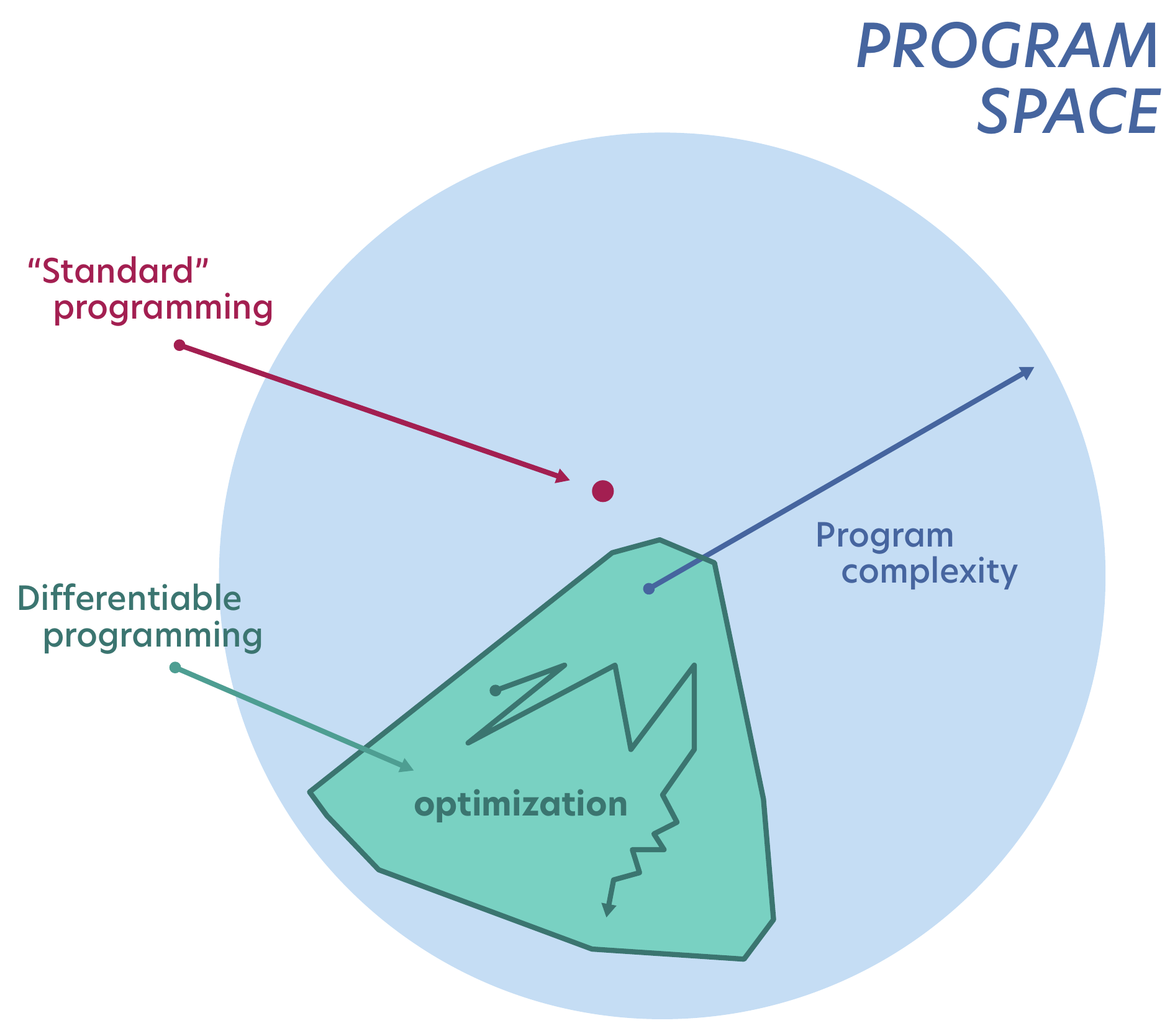}
    \end{center}
    \caption[Standard vs. differentiable programming]{Illustration of the difference between ``standard'' programming and differentiable programming. In some cases, the complexity of programs found by \ac{DiffP} exceeds human capabilities. Inspired by \ToggleForCUP{Karpathy, A. (2017). \textit{Software 2.0}. Medium, Accessed: 2022-04-08~\cite{karpathy:2017}.}{Ref.~\cite{karpathy:2017}.}}
    \label{fig:software2_0}
\end{figure}

\Acf{DiffP}\index{differentiable programming} represents a~fundamental shift in software development that emerged from \ac{DL}~\cite{karpathy:2017}. In ``standard'' programming each instruction is explicitly specified in the code, i.e., one specifies a~point in the \stress{program space} with some desirable behavior (see~\cref{fig:software2_0}). In \ac{DiffP}, computer programs are instead composed of parametrized elements of code which can be adjusted. The programmer specifies the desired behavior of the program via a~loss function. The space of programs is then searched for a suitable program by tuning the code parameters to minimize the~given loss function using derivative information. An example of this which we have continually encountered in these notes is the use of backpropagation to efficiently tune the parameters of an~\ac{NN} to solve a given task, such as classifying different phases of matter or representing the ground-state wave function of a quantum many-body system.

In most real-world problems, collecting data in the form of instances in which a~given task has been correctly solved is easier than writing a~program that solves the task. Under these circumstances \ac{DiffP} shines, because it allows for the program which solves the task to be \stress{learned} from data. This approach can be extremely powerful as demonstrated by the success of programs generated through deep learning. Indeed, as we have extensively discussed in these notes, there are nowadays many instances where \ac{DiffP} has led to algorithms that easily outperform humans, such as in AlphaGo~\cite{AlphaGo}.

\ac{DiffP} also has multiple other advantages compared to conventional programming. One aspect regards the possibility to develop customized optimization strategies. The typical instruction set of \acp{NN} consists of matrix multiplication, vector addition, and element-wise application of nonlinearities: such a~set is limited and much smaller, compared to the instruction set associated with the entire class of standard computer programs. This can allow for computational speed-ups through the design of hardware that is optimized for the limited instruction set underlying \ac{DiffP}. \Acfp{GPU}\index{graphics processing unit} and tensor processing units (TPUs) are examples of such application-specific hardware. More recently, neuromorphic computing has emerged as a new paradigm that promises faster and more energy-efficient computation for machine intelligence through hardware systems that mimic the neuronal and synaptic computations of the brain~\cite{schuman:2017,roy:2019}.

\ac{DiffP} also allows for more flexible programming: consider the situation where you had standard code that performs a~certain task and someone wanted you to make it twice as fast, possibly at the expense of its accuracy. This would be a~highly non-trivial task. However, it is easy to incorporate such constraints by means of a~cost function and hyperparameters in \ac{DiffP}. For example, given that one uses an~\ac{NN} this could be accomplished by cutting the network's size in half and retraining it. Moreover, consider the situation where programs that were first optimized or coded individually are merged together in a~modular fashion to create a~new larger program. Then, \ac{DiffP} offers an~easy solution for optimizing the performance of this new program: simple fine-tuning of the individual components in the given configuration through optimization. The benefits of \ac{DiffP} come at the cost of program interpretability. At the end of the optimization we obtain code that works well, but is very hard to read for a~human and understand in intuitive terms. As such, we typically are left with the choice between a~fairly accurate model that is understandable in human terms, and a~more accurate model that is difficult to interpret.\footnote{Again, interpretability appears as a~central issue (see \cref{sec:interpretability}).}

\highlight{In \ac{DiffP}, arbitrary computer program structures can be differentiated in an~automatic fashion. Importantly, this allows for \acp{NN} to be embedded into a~plethora of existing scientific simulations and computations, because the gradients required for training the \acp{NN} can be computed efficiently. In particular, one can differentiate through the \ac{NN}, as well as surrounding non-parametrized/non-trainable parts of the program.} Recently, widespread interest in \ac{DiffP} has arisen in the area of scientific computing~\cite{innes:2019}. Examples of algorithms that have been written in a~fully differentiable way are Fourier transforms, eigenvalue solvers, singular value decompositions, or \acfp{ODE}~\cite{johnson:2012,chen:2019}. As such, one is able to differentiate through domain-specific computational processes to solve inverse problems, such as learning or control tasks: Tensor networks~\cite{liao:2019,chen:2020,Giacomo:2020}, molecular dynamics~\cite{ingraham:2018,jaxmd2020}, quantum chemistry~\cite{tamayo:2018,Neuscamman:2020,li:2021,Kasim:2021,quax,dqc,pennylane,Zhang:2022_PySCFJax,Naruki:2022_ADSCF,vargas:2023_AD}, quantum optimal control~\cite{Khaneja:2005,leung:2017,Abdelhafez:2019,Hamza:2019,Hamza:2020,schaefer:2020,Vargas:2020_AD,Vargas:2021_AD,Khait:2021,Luuk:2021,schaefer:2021,goerz:2022}, or quantum circuits~\cite{luo:2020,kyriienko:2021} have all been formulated in a~fully-differentiable manner. We discuss several examples in detail in~\cref{sec:DP_examples}.

Notably, \ac{DiffP} enables scientific \ac{ML} which combines the best of two worlds: in general, black-box \ac{ML} approaches are flexible but require a~large amount of data to be trained successfully. The amount of required data can be reduced by incorporating our scientific knowledge on the structure of a~problem into the program. The training of the parametrized program part is then enabled via \ac{DiffP}. This allows for the learning task to be simplified because only the parts of the model that are ``missing'' need to be learned.

Perhaps the biggest feat of \ac{DiffP} is the ability to compute gradients of loss functions with respect to the \ac{NN} parameters (see~\cref{sec:backprop}). Recall that we require these gradients for \ac{NN} training when using gradient-based optimizers (as is typically done). Crucially, the computation is efficient, precise, and occurs in an~automated fashion. In particular, it allows for arbitrary \ac{NN} architectures to be differentiated automatically without implementation overhead. Compare this to the tedious computation of analytical gradients which needs to be performed again given different \ac{NN} architectures. 

However, \ac{DiffP} is not restricted to the computation of gradients with respect to \ac{NN} parameters for \ac{NN} training. It enables the automatic computation of gradients and higher-order derivatives of arbitrary program variables\index{Hessian}. These can, for example, be tunable parameters of a~Hamiltonian whose ground state we are interested in. Being able to differentiate through the eigensolver, we can tune the Hamiltonian's parameters via a~derivative-based optimizer such that its ground state satisfies desired properties (as specified by a~loss function), see~\cref{sec:DP_examples} for details. This is an~example of an~\stress{inverse problem} which can be solved efficiently through \ac{DiffP}. However, the applicability of \ac{DiffP} goes beyond solving optimization tasks. Gradients and higher-order derivatives contain highly valuable information on the relationship between model parameters and outputs which can, e.g., facilitate the interpretation of phase classification methods~\cite{dawid:2021} (see~\cref{sss:hessian-based-interpret}) or help to characterize variational quantum circuits~\cite{huembeli:2021}.
\subsubsection{Automatic differentiation}\index{differentiation}
\ac{DiffP} allows us to compute the gradients and higher-order derivatives of arbitrary computer programs.\highlight{In general, methods for the computation of derivatives in computer programs can be classified into four categories~\cite{baydin:2018}: (1) manually working out derivatives and coding them\index{differentiation!manual differentiation}, (2) numerical differentiation\index{differentiation!numerical differentiation} using finite difference approximations, (3) symbolic differentiation\index{differentiation!symbolic differentiation} using expression manipulation,\footnote{This is done by computer algebra systems such as \texttt{Mathematica}, \texttt{Maxima}, or \texttt{Maple.}} and (4) \acf{AD}\index{differentiation!automatic differentiation} which is the workhorse behind \ac{DiffP}.} Let us briefly discuss these different approaches. 

Manual differentiation is time-consuming and prone to errors. \stress{Numerical differentiation} is quite simple to implement. Its most basic form is based on the limit definition of a~derivative: given a~multivariate function $f: \realset^m \rightarrow \realset$, the components of its gradient $\nabla f = (\frac{\partial f}{\partial x_{1}}, \dots, \frac{\partial f}{\partial x_{m}})$ can be approximated as
\begin{equation}\label{eq_numerical_derivative}
	\left.\frac{\partial f}{\partial x_{i}}\right|_{\vect{x}} \approx \frac{f(\vect{x}+h\vect{e}_{i})-f(\vect{x})}{h},
\end{equation}
where $\vect{e}_{i}\in \realset^{m}$ is the $i$-th unit vector and $h$ is a~small step size. Approximating $\nabla f$ in such a~fashion requires $\bigO(m)$ evaluations of $f$. This is the main reason why numerical differentiation is not useful in \ac{ML} where the number of trainable parameters $m$ can be as large as millions or billions. Also note that for the gradient approximation to be somewhat accurate, the step size $h$ needs to be carefully chosen: while the truncation error of the approximation in~\cref{eq_numerical_derivative} can be made arbitrarily small as $h\rightarrow 0$, eventually round-off errors due to floating-point arithmetic dominate the calculation.\footnote{In computing, floating-point numbers are typically represented approximately through a~fixed number of significant digits that are scaled through an~exponent in some fixed basis $a \times b^{c}$, where $a$, $b$, and $c$ are all integers. Because of the limited number of representable numbers, round-off errors can occur when performing computations.}

\stress{Symbolic differentiation} is the automated manipulation of mathematical expressions for obtaining explicit derivative expressions, e.g., by using simple derivative rules such as the product rule
\begin{equation}\label{eq_symbolic_derivative}
 \frac{d}{dx}(f(x)g(x)) = \frac{df(x)}{dx}g(x) + f(x)\frac{dg(x)}{dx}. 
\end{equation}
Symbolic expressions have the benefit of being interpretable and allow for analytical treatments of problems. However, symbolic derivatives generated through symbolic differentiation typically do not allow for efficient calculation of derivative values. This is because they can quickly get substantially larger than the expression whose derivative they represent. Consider a~function of the form $h(x)=f(x)g(x)$ and its derivative, which can be evaluated by the product rule in~\cref{eq_symbolic_derivative}. Note that $f(x)$ and $\frac{df(x)}{dx}$, for example, appear separately in such an~expression. A~naive calculation of the derivative according to~\cref{eq_symbolic_derivative} thus involves duplicate computations of any expressions that appear both in $f(x)$ and $\frac{df(x)}{dx}$. Moreover, manual and symbolic methods require the underlying function to be defined in a~closed-form expression. As such, they cannot easily deal with programs that involve conditional branches, loops, or recursions. That means, for symbolic differentiation to be efficient there must exist a~convenient symbolic expression for computing the derivative under consideration.

\begin{figure}[t]
    \begin{center}
    \includegraphics[width=0.9\columnwidth]{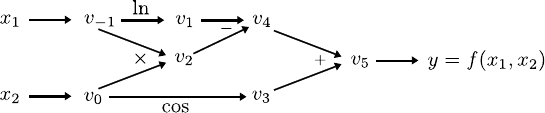}
    \end{center}
    \caption[Exemplary computation graph]{Computation graph associated with the forward evaluation trace of $f(x_{1}, x_{2}) = \ln(x_{1}) + \cos(x_{2}) - x_{1} x_{2}$.}
    \label{fig:comp_graph}
\end{figure}

When we are concerned with the accurate numerical evaluation of derivatives and not their symbolic form, it is possible to significantly simplify computations by storing the values of intermediate sub-expressions in memory. This is the basic idea behind \acf{AD}\index{differentiation!automatic differentiation}. \ac{AD} provides numerical values of derivatives (as opposed to symbolic expressions) and it does so by using symbolic rules of differentiation (but keeping track of derivative values, as opposed to the entire symbolic expression). As such, it may be viewed as an~\stress{intermediate between numerical and symbolic differentiation}. \ac{AD} makes use of the fact that every computer program, no matter how complicated it may look, simply executes a~sequence of elementary arithmetic operations (e.g., additions or multiplications) and elementary functions (e.g., exp or sin). We refer to the sequence of elementary operations that a~computer program applies to its input values to compute its output values as \stress{evaluation trace}~\cite{wengert:1964}. The derivative of every computer program can therefore be computed in an~automated fashion through repeated application of the chain rule. As such, the number of arithmetic operations required to compute the derivative is of the same order as for the original program. Moreover, this results in derivatives that are accurate up to machine precision. In the following, we illustrate how \ac{AD} is done in practice.
\paragraph{Forward-mode AD}
Conceptually, \ac{AD} in so-called \stress{forward-mode} is the simplest type. Consider the evaluation trace of the function 
\begin{equation}\label{eq:DP_example_func}
f(x_{1}, x_{2}) = \ln(x_{1}) + \cos(x_{2}) - x_{1} x_{2}
\end{equation}
given in~\cref{tab:forward_mode}(left). The associated computation graph is shown in~\cref{fig:comp_graph}, where the computation of a~function $f$ is decomposed into variables $v_{i}$. We follow the standard notation used in Ref.~\cite{griewank:2008}, where $v_{1-i}, \; i=1,\dots, n$ are the input variables, $v_{i}, \; i=1,\dots, l$ are intermediate variables, and $v_{l+i}, \; i=1,\dots, m$ are output variables. For computing the derivative of $f$ with respect to $x_{1}$, we start by associating with each variable $v_{i}$ a~derivative
\begin{equation}
\dot{v}_{i} = \frac{\partial v_{i}}{\partial x_{1}}.
\end{equation}
Applying the chain rule to each elementary operation in the evaluation trace, we generate the corresponding derivative trace, given in~\cref{tab:forward_mode}(right). In forward-mode \ac{AD} the desired derivative $\dot{v}_{5} = \frac{\partial y}{\partial x_{1}}$ (where $y$ is the output variable) is obtained by computing the intermediate variables $v_{i}$ in sync with their corresponding derivatives $\dot{v}_{i}$.

\begin{table}[!htb]
    \centering
    \begin{tabular}{l r || l r}\hline\hline
         $v_{-1}=x_{1}$&$=2$&$\dot{v}_{-1}=\dot{x}_{1}$&$=1$ \T \\
         $v_{0}=x_{2}$&$=1$& $\dot{v}_{0}=\dot{x}_{2}$&$=0$ \B \\\hline
         
         $v_{1}=\ln{v_{-1}}$&$=\ln{2}$&$\dot{v}_{1}=\dot{v}_{-1}/v_{-1}$&$=1/2$ \T \\
         $v_{2}=v_{-1} v_{0}$&$=2$&$\dot{v}_{2}=\dot{v}_{-1} v_{0} + \dot{v}_{-1} \dot{v}_{0}$&$=1$   \\
         $v_{3}=\cos{v_{0}}$&$=\cos{1}$&$\dot{v}_{3}=-\dot{v}_{0}\sin{v_{0}}$&$=0$  \\
         $v_{4}=v_{1}-v_{2}$&$=-1.307$&$\dot{v}_{4}=\dot{v}_{1}-\dot{v}_{2}$&$=-1/2$   \\
         $v_{5}=v_{3} + v_{4}$&$=-0.767$&$\dot{v}_{5}=\dot{v}_{3} + \dot{v}_{4}$&$=-1/2$  \B \\\hline
         
         $v_{5}=y$&$=-0.767$&$\dot{v}_{5}=\dot{y}$&$=-1/2$ \T \B \\
         
         \hline\hline
    \end{tabular}
    \caption[Example of forward-mode automatic differentiation]{Workflow for computation of derivatives in forward-mode \ac{AD} given the function $f(x_{1}, x_{2}) = \ln(x_{1}) + \cos(x_{2}) - x_{1} x_{2}$. Left: Forward evaluation trace for the choice of initial inputs $(x_{1},x_{2})=(2,1)$. Right: Forward derivative trace resulting in the computation of $\frac{\partial f}{\partial x_{1}}$ at $(x_{1},x_{2})=(2,1)$.}
    \label{tab:forward_mode}
\end{table}

This can be generalized to the computation of the full Jacobian of a~function $f : \realset^{m} \rightarrow \realset^{n}$ with $m$ input variables $x_{i}$ and $n$ output variables $y_{j}$. In this case, each forward pass of \ac{AD} is initialized by setting $\dot{x}_{i}=1$ for a~single variable ${x}_{i}$ and zero for the rest. That is, we choose $\dot{\bm{x}}=\bm{e}_{i}$ where $\bm{e}_{i}$ is the $i$-th unit vector. The forward pass with given input values $\bm{x} = \bm{a}$ then computes
\begin{equation}
\dot{y}_{j} = \left.\frac{\partial y_{j}}{\partial x_{i}}\right|_{\bm{x}=\bm{a}}\;{\rm for}\; j=1,\dots,n.
\end{equation}
This corresponds to the $i$-th column of the Jacobian matrix
\begin{equation}
J_{f} = \left.\begin{bmatrix}
    \frac{\partial y_{1}}{\partial x_{1}} & \hdots &\frac{\partial y_{1}}{\partial x_{m}} \\
   \vdots & \ddots & \vdots \\
   \frac{\partial y_{n}}{\partial x_{1}} & \hdots & \frac{\partial y_{n}}{\partial x_{m}}.
\end{bmatrix}\right|_{\bm{x}=\bm{a}}.
\end{equation}
Thus, the full Jacobian can be computed in $m$ forward passes, i.e., $m$ evaluations of the function $f$. As such, forward-mode \ac{AD} is efficient if $m \ll n$. In the other limit, so-called reverse-mode \ac{AD} is preferred which we discuss shortly.

In practice, forward-mode \ac{AD} is implemented by augmenting the algebra of real numbers and introducing a~new arithmetic: to every number one associates an~additional component which corresponds to the derivative of a~function computed at that particular value. We call this composite number a~\stress{dual number}
\begin{equation}
{\rm dual}(v)=v+\dot{v}\epsilon,
\end{equation}
where $\epsilon\neq 0$ is a~number such that $\epsilon^2=0$. The extension of all arithmetic operators to dual numbers allows for the dual number algebra to be defined. Observe, for example, that 
\begin{equation}\label{eq:DP_dual}
f({\rm dual}(v)) = f(v)+\dot{f}(v)\dot{v}\epsilon,
\end{equation}
where we obtain the function value in the first part and the corresponding derivative $\dot{f}(v)\dot{v}$ in the $\epsilon$ part.\footnote{Under the hood dual numbers are typically handled through so-called operator overloading, i.e., overloading all functions to work appropriately on the new algebra.} This follows from expanding the function in its Taylor series and noting that terms $\bigO(\epsilon^2)$ vanish due to the property that $\epsilon^2=0$. Equation~\eqref{eq:DP_dual} resembles the computation of the derivative using the chain rule.

\paragraph{Reverse-mode AD}
As the name suggests, in reverse-mode \ac{AD} the derivatives are propagated backwards from a~given output.\footnote{Historically, reverse-mode \ac{AD} can be traced back to the master thesis of Seppo Linnainmaa in 1970~\cite{linnainmaa:1970} in which he described explicit, efficient error backpropagation\index{backpropagation} in arbitrary, discrete, possibly sparsely connected, \ac{NN}-like networks~\cite{Griewank2012}.} This is in contrast to forward-mode \ac{AD} where we saw that the derivatives are propagated forwards in sync with the function evaluation. Reverse-mode \ac{AD} is done by complementing each intermediate variable $v_{i}$ with a~so-called adjoint 
\begin{equation}
\bar{v}_{i} = \frac{\partial y_{j}}{\partial v_{i}},
\end{equation}
where $y_{j}$ is the output variable with respect to which we desire to compute derivatives. In reverse mode \ac{AD}, derivatives are computed in the second phase of a~two-phase process. In the first phase, the original function code is run forward: intermediate variables $v_{i}$ are populated and their dependencies in the computational graph are tracked through a~bookkeeping procedure. In the second phase, derivatives are calculated by propagating adjoints $\bar{v}_{i}$ in reverse, i.e., from the outputs to the inputs. This is illustrated in~\cref{tab:reverse_mode} for the function given in~\cref{eq:DP_example_func}, where the reverse pass is started with $\bar{v}_{5} = \bar{y} = \frac{\partial y}{\partial y}=1$. As a~result, we obtain both $\bar{x}_{1} = \frac{\partial y}{\partial x_{1}}$ and $\bar{x}_{2} = \frac{\partial y}{\partial x_{2}}$ in a~single reverse pass. 

\begin{table}[!htb]
    \centering
    \begin{tabular}{l r || l r}\hline\hline
         $v_{-1}=x_{1}$&$=2$&$\bar{v}_{5} = \bar{y}$&$=1$ \T \\
         $v_{0}=x_{2}$&$=1$& & \B \\\hline
         
         $v_{1}=\ln{v_{-1}}$&$=\ln{2}$&$\bar{v}_{4} = \frac{\partial v_{5}}{\partial v_{4}}\bar{v}_{5} = \frac{\partial v_{3} + v_{4}}{\partial v_{4}}\bar{v}_{5} $&$=1$  \T \\

         $v_{2}=v_{-1} v_{0}$&$=2$&$\bar{v}_{3} = \frac{\partial v_{5}}{\partial v_{3}}\bar{v}_{5}$&$=1$\\

         $v_{3}=\cos{v_{0}}$&$=\cos{1}$&$\bar{v}_{2} = \frac{\partial v_{4}}{\partial v_{2}}\bar{v}_{4}$&$=-1$\\

         $v_{4}=v_{1}-v_{2}$&$=-1.307$&$\bar{v}_{1} = \frac{\partial v_{4}}{\partial v_{1}}\bar{v}_{4}$&$=1$\\

         $v_{5}=v_{3} + v_{4}$&$=-0.767$&$\bar{v}_{0} = \frac{\partial v_{2}}{\partial v_{0}}\bar{v}_{2} + \frac{\partial v_{3}}{\partial v_{0}}\bar{v}_{3}$&$=-2.841$\\
         
         & &$\bar{v}_{-1} = \frac{\partial v_{1}}{\partial v_{-1}}\bar{v}_{1} + \frac{\partial v_{2}}{\partial v_{-1}}\bar{v}_{2}$&$=-1/2$ \B \\\hline

         $v_{5}=y$&$=-0.767$&$\bar{v}_{0} = \bar{x}_{2}$&$=-2.841$ \T \\
         &&$\bar{v}_{-1} = \bar{x}_{1}$&$=-1/2$ \B\\
         \hline\hline
    \end{tabular}
    \caption[Example of reverse-mode automatic differentiation]{Workflow for computation of derivatives in reverse-mode \ac{AD} given the function $f(x_{1}, x_{2}) = \ln(x_{1}) + \cos(x_{2}) - x_{1} x_{2}$. Left: Forward evaluation trace for the choice of initial inputs $(x_{1},x_{2})=(2,1)$. Right: Reverse (adjoint) derivative trace resulting in the computation of $\frac{\partial f}{\partial x_{1}}$ and $\frac{\partial f}{\partial x_{2}}$ at $(x_{1},x_{2})=(2,1)$.}
    \label{tab:reverse_mode}
\end{table}

This example illustrates the complementary nature of the reverse mode compared to the forward mode: The reverse-mode is cheaper to evaluate than the forward mode for functions with a~large number of inputs, i.e., where $m \gg n$ with $f : \realset^{m} \rightarrow \realset^{n}$. As we just saw, in the extreme case of $f : \realset^{m} \rightarrow \realset$, only one application of the reverse mode is sufficient to compute the full gradient compared with the $m$ passes of the forward mode. The typical case encountered in \ac{ML} applications corresponds to the evaluation of the derivatives of a~loss function $y_{j}=\lossfun : \realset^{m} \rightarrow \realset$ with respect to $m$ trainable parameters, where $m$ is typically large. As such, reverse-mode \ac{AD} is the preferred method for computing gradients automatically as it is computationally more efficient compared to forward-mode \ac{AD}.\footnote{Note that forward-mode and reverse-mode \ac{AD} are just two (extremal) ways of applying chain rules. Finding the optimal way to traverse the chain rule to compute a~Jacobian for a~given function (i.e., the choice which results in the smallest number of arithmetic operations) is known as the optimal Jacobian accumulation problem and is NP-complete.} In the context of \ac{ML}, reverse-mode \ac{AD} applied to \acp{NN} is typically referred to as backpropagation, see~\cref{sec:backprop}. It is the working horse behind \ac{NN} training as it allows for efficient computation of the gradients for arbitrary \ac{NN}-based architectures in an~automated fashion. In the following, we discuss several ways how reverse mode \ac{AD} is implemented in practice.

\paragraph{Static graph AD}
A~basic implementation of reverse-mode \ac{AD} makes use of static computation graphs. This choice is natural, given that we chose to illustrate reverse-mode \ac{AD} using computation graphs~\cite{rackauckas:2020}. Tensorflow is an~example of a~platform that uses this approach. Here, the user must define variables and operations in a~graph-based language. Subsequent executions of the computation graph allow for the program to be differentiated in a~straightforward manner. However, this requires all existing programs to be rewritten as a~static computation graph which is inconvenient.
\paragraph{Tracing-based AD}
This can be circumvented by building computation graphs {\it dynamically} at runtime which is achieved by ``tracing'' all the operations encountered in the forward pass given a~particular input~\cite{rackauckas:2020}. Dynamic computation graphs are the basis of many reverse-mode \ac{AD} implementations in Julia (Tracker.jl, ReverseDiff.jl, or Autograd.jl) or Python (PyTorch, Tensorflow Eager, Autograd [JAX]). The fact that it is simple to implement makes this approach widely adopted in practice. An~issue of such tracing-based implementations is that each trace is value-dependent, meaning that each run of a~program (with different inputs) can build a~new trace. Moreover, these traces can be much larger than the code itself, for example, because loops are completely unraveled.
\paragraph{Source-to-source AD}
In source-to-source \ac{AD} one overcomes these issues by generating source code for the backward pass that is able to handle all input values~\cite{rackauckas:2020}. In particular branches, loops, and recursions are not explicitly unrolled. The right branch in the reverse passes through recall of the intermediates values used in the forward pass. It turns out that the implementation of a~source-to-source \ac{AD} system poses many requirements on the underlying language.\footnote{In particular, it should possess a~strong internal graph structure.} Source-to-source \ac{AD} is used in programming languages such as Julia (Zygote.jl).\footnote{TensorFlow considered building a~source-to-source \ac{AD} based on the Swift language. Older \ac{AD} systems for Fortran were also source-to-source.}
\paragraph{High-level adjoint rules}
The advantages of reverse-mode \ac{AD} in \ac{ML} applications come at the cost of increased storage requirements which (in the worst case) is proportional to the number of operations in the evaluated function. This is because the values of the intermediate variables populated during the forward pass need to be stored when using reverse-mode \ac{AD}, whereas they can be directly used for the derivative computation within forward-mode \ac{AD}. Improving storage requirements in reverse-mode \ac{AD} implementations is an~active research area. In general, reverse-mode \ac{AD} can be made more efficient by deriving adjoint rules at a~higher level. Consider, for example, the case where your program involves solving a~nonlinear problem $f(\vect{x},\vect{p})=0$ with an~iterative method, such as Newton’s method. A~naive application of the reverse-mode \ac{AD} system results in a~backward pass through all iteration steps. Not only is this computationally expensive, but also requires storing the values of all intermediate iteration steps. Instead of unrolling the entire computation, one can analytically derive an~appropriate adjoint rule which can be used to compute the derivatives in reverse-mode via a~separate linear equation. In particular, it only requires knowledge of the final solution $\vect{x}$ of the nonlinear problem as opposed to values in the intermediate iterations. For further details, see Ref.~\cite{johnson:2012,blondel:2021}. Other examples for which adjoint rules can be derived also include \acp{ODE}~\cite{chen:2019,rackauckas:2018,rackauckas:2020,Vargas:2021_AD} and eigenvalue solvers~\cite{johnson:2012}. In the case of \acp{ODE}, for example, the adjoint rule involves solving a~second, augmented \ac{ODE} backwards in time.

\subsubsection{Application to quantum physics problems}\label{sec:DP_examples} 
\begin{figure}[t]
\begin{center}
\includegraphics[width=1.0\columnwidth]{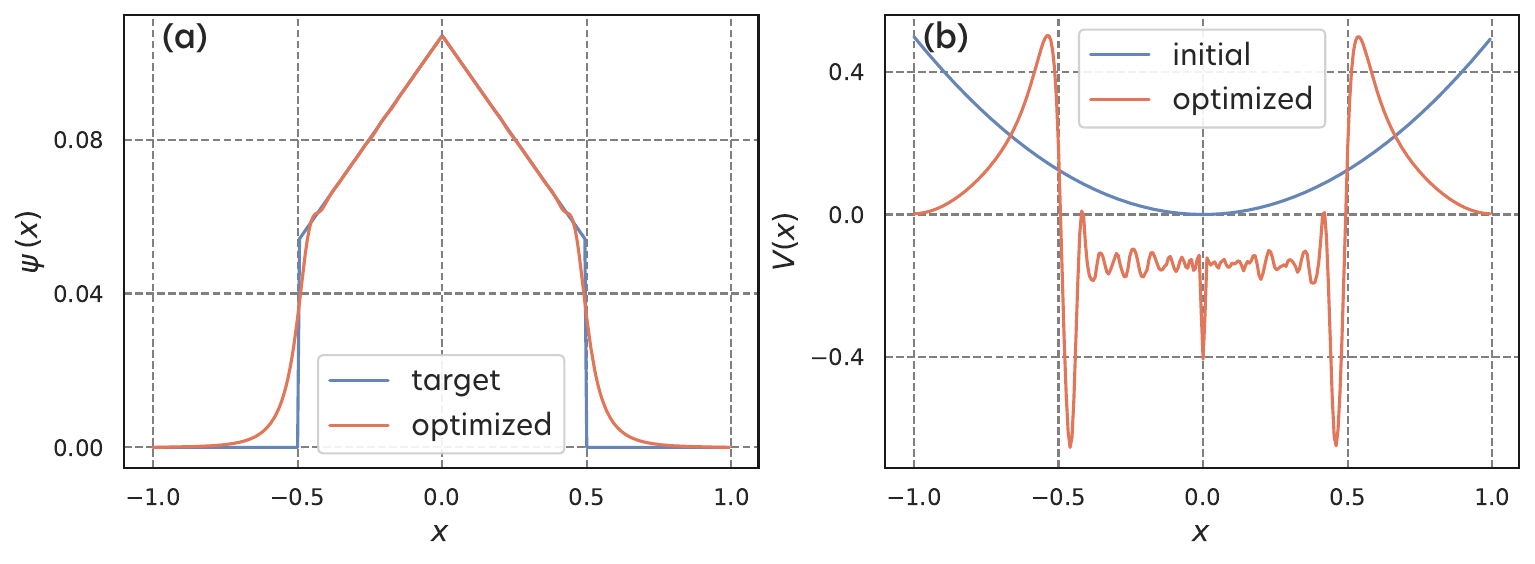}
\end{center}
\caption[Inverse Schrödinger problem solved using differential programming]{(a) Optimized ground-state wave function $\Psi_{0}(x)$ after $\approx 1000$ iterations of the \ac{L-BFGS} algorithm with box constraints (L-BFGS-B) given the target wave function $\Psi(x) = 1-|x|$ with $|x| < 0.5$. (b) Optimized potential $V(x)$ (rescaled by 1/300) and the initial harmonic potential. Figure reproduced from Ref.~\cite{wang:2021}.}
\label{fig:inverse_schroedinger}
\end{figure}

In this section, we illustrate the application of \ac{DiffP} to problems from quantum physics through two simple examples.
\paragraph{Inverse Schrödinger problem}
As a~concrete example of an~inverse-design problem, we consider the time-independent Schrödinger equation in one dimension
\begin{equation}
\left[ -\frac{1}{2}\frac{d}{dx^2}+V(x)\right] \Psi(x) = E \Psi(x),
\end{equation}
where we set $\hbar = m = 1$~\cite{johnson:2012}. Typically, we are given a~potential $V(x)$ and solve for the corresponding eigenfunctions $\Psi$ and energies $E$. Here, we consider the inverse problem. Given a~particular wave function $\Psi(x)$ we want to construct a~potential $V(x)$ with a~ground-state wave function $\Psi_{0}(x)$ that closely matches $\Psi(x)$. We restrict ourselves to the domain $x\in [-1,1]$ and define the following \acf{MSE} function
\begin{equation}\label{eq:DP_SE}
\lossfun = \int_{-1}^{1} | \Psi(x) - \Psi_{0}(x)|^2 dx.
\end{equation}
The potential $V(x)$ is discretized on a~grid and each individual amplitude is tuned using gradient-based optimization methods in order to minimize~\cref{eq:DP_SE}. An~implementation of this problem in JAX can be found at~\cite{wang:2021}. The results are illustrated in~\cref{fig:inverse_schroedinger}. Here, the gradient is calculated by the \ac{AD} system underlying JAX and involves propagating the derivative through the eigenvalue solver. For more details on $\partial P$ applied to inverse-design problems in quantum mechanics and adjoints for eigensolvers, see Ref.~\cite{xie:2020}.

\paragraph{Quantum optimal control} 
Next, we consider a~problem from quantum optimal control. We would like to find the time-dependent amplitudes $\{ u_{i}(t)\}$ of the following Hamiltonian
  \begin{equation}\label{eq_optimal_control}
  \hat{H} = \hat{H}_{0} + \sum_{i=1}^{n_{\rm ctrl}} u_{i}(t) \hat{H}_{i},
  \end{equation}
  such that the time-evolution under $\hat{H}$ realize a~CNOT gate. Here $\hat{H}_{i}$ with $i \in [1,n_{\rm ctrl}]$ are time-independent Hamiltonians that can be tuned through the time-dependent amplitudes $\{ u_{i}(t)\}$. We parametrize these amplitudes using Fourier series $u_{i}(t) = \sum_{j=1}^{n_{\rm basis}} u_{ij} \sin(\pi j t/T)$, where we introduce $n_{\rm basis}$ as a~cutoff enabling numerical evaluations. To find the gate which a~given choice of $\{ u_{i}(t)\}$ implements we integrate the time-dependent Schrödinger equation from 0 to $T$ under initial conditions $U(0)=\id$ with
  \begin{equation}\label{eq:DP_ODE_QC}
  \frac{dU}{dt} = -i \hat{H}(t) U,
  \end{equation}
  where we set $\hbar =1$. Ideally, $U(T)$ realizes a~CNOT operation ($U_{\rm target}={\rm CNOT}$). Hence, we setup our loss function as
  \begin{equation}\label{eq:DP_loss_QC}
  \lossfun = 1- \frac{1}{d}| {\rm tr}\,(U(T)^{\dagger}U_{\rm target}^{\phantom \dagger})|,
  \end{equation}
  where $d$ is the dimension of the associated Hilbert space. Note that when $U(T) = U_{\rm target}$ we reach the global minimum of $\lossfun=0$. The coefficients $\{u_{ij} \}$ are tuned to minimize the loss function in~\cref{eq:DP_loss_QC} using gradient-based optimization methods. An~implementation of this problem in JAX can be found at~\cite{wang2:2021}, where the gradient is calculated via \ac{AD} in JAX and involves propagating the derivative through the \ac{ODE} solver (\cref{eq:DP_ODE_QC}). For more details on differentiable programming applied to quantum optimal control, see Refs.~\cite{leung:2017,schaefer2019}.
\subsubsection*{Outlook and open problems}
    In this chapter, we have introduced the novel programming paradigm that is \ac{DiffP}. Most notably, \ac{DiffP} enables \acp{NN} training via the efficient, precise, and automated calculation of the corresponding gradients (see~\cref{sec:backprop}). Having the ability to differentiate arbitrary computer programs, in addition, allows for \acp{NN} to be seemingly incorporated in scientific workflows. By now there are many applications of \ac{DiffP} in scientific computing, including quantum physics. However, the field is still in its infancy and many open problems remain to be tackled. For instance, finding efficient high-level adjoint rules for algorithms used in quantum physics problems, such as solvers for stochastic dynamics, is still a~current topic of research~\cite{schaefer:2021}. Another example are chaotic systems for which standard \ac{AD} methods can fail~\cite{wang:2014,metz:2021}. The development of \ac{AD} systems is also still an~ongoing effort: In Enzyme.jl~\cite{moses:2020}, for example, the idea is to perform reverse-mode \ac{AD} on the portable, low-level intermediate representation of Julia which is language-agnostic. This allows for performance improvements due to low-level optimizations. NiLang.jl~\cite{liu:2020} on the other hand tries to build a~reverse-mode \ac{AD} system based on the paradigm of reversible programming. Running the program in reverse in the backward pass allows the overhead in memory in standard reverse-mode \ac{AD} to be circumvented.

\subsubsection*{Further reading}
\begin{itemize}
    \item Baydin, A. G. {\it et al.} (2018). \href{https://dl.acm.org/doi/abs/10.5555/3122009.3242010}{\textit{Automatic differentiation in machine learning: a~survey}}. J.~Mach.~Learn.~Res. 18(1), 5595–5637. Good overview on \ac{AD} in the context of \ac{ML}~\cite{baydin:2018}.
    \item Innes, M. {\it et al.} (2019). \href{
https://doi.org/10.48550/arXiv.1907.07587}{\textit{Differentiable programming system to bridge machine learning and scientific computing}}. arXiv:1907.07587. Discussion on the role of \ac{DiffP} in scientific computing~\cite{innes:2019}.
    \item Google Colab notebooks by Lei Wang~\cite{wang:2021,wang2:2021}.
\end{itemize}

\subsection{Generative models in many-body physics}\label{sec:hot-topics:DE}\index{generative models}

Deep generative models\index{generative models!deep generative models} are (mostly) neural-network architectures designed to approximate the probability density underlying a system or a dataset we aim to describe.
This task of constructing the probability density of a~given problem is often called \stress{density estimation}\index{density estimation}~\cite{Silverman1998}. When such an underlying probability density is learned, one can then sample the obtained density and generate artificial new samples that are characteristic of the problem (hence the name ``generative'' models).

The difficulty of the density estimation task comes from the fact that underlying probability densities cannot be computed exactly in most cases. For a~large class of problems, this issue can be attributed to the difficulty of calculating normalization constants (i.e., partition functions in statistical mechanics). As such, one often has to resort to different methods and techniques that approximate underlying probability density functions in which deep generative models are very useful.

Throughout this book, the reader encounters various examples of density estimation\index{density estimation} tasks in quantum physics. For example, finding a~(variational) representation of a quantum-many body system wave function can be viewed as a~density estimation task, see \cref{sec:NN_q_states}. More generally, in quantum physics the concept of density estimation often appears in the context of reconstructing the many-body state from measurements performed on a~quantum many-body system -- a~task known as \stress{quantum state tomography}\index{quantum tomography}, described in more detail in \cref{sss:quantum_tomography}. This is particularly challenging because only a~reduced set of quantities, such as single-body density matrices or higher-order correlation functions, is experimentally accessible~\cite{Hradil1997,Rehacek2004,Teo2015}. Moreover, the tomography experiment can be very demanding, and a~single experimental run can take a~long time (i.e., hours or days). One, therefore, has to face the challenge of estimating a~high-dimensional probability density from a rather small number of measurements of a~restricted set of observables.

In general, one can think of two different approaches to density estimation: \stress{parametric} and \stress{non-parametric} density estimation. In the parametric approach, one fixes a~parametrized functional form for the approximated density. The free parameters are then tuned such that the trial density best matches the density of the system under consideration. This can be done by comparing the trial density against training data, which corresponds to observations drawn from the target density, or against an~unnormalized target probability density (e.g., an~unnormalized Boltzmann distribution). A~simple example of a~parametrized trial density would be a~Gaussian, where its mean and variance can be adjusted accordingly. In parametric\index{density estimation!parametric} approaches, one reduces the problem of finding an~appropriate density function to the problem of finding appropriate parameters. Clearly, the choice of the functional form of the density is crucial for the success of this approach, as it can substantially restrict the family of target densities that can be effectively approximated through the chosen ansatz. 

In contrast, for non-parametric approaches, the structure of the trial density function is not set \stress{a priori}.\footnote{The trial density function in non-parametric approaches to density estimation does not lack parameters entirely but rather is not fixed in advance, i.e., is learned from scratch.} Instead, a~trial density function is directly constructed based on training data. The simplest example of a~non-parametric\index{density estimation!non parametric} approach to density estimation corresponds to building a~histogram. Clearly, histogram binning comes with some drawbacks. For example, one must carefully choose the size and location of bins. Moreover, histograms are non-differentiable functions. In general, non-parametric methods cannot leverage a priori information about the system at hand, contrary to parametric methods. While this makes non-parametric methods robust and applicable to general problems, they typically also require a generous number of samples to reach a suitable level of accuracy.

\subsubsection{Training with or without data}\label{sec:w_wout_data}

In recent years, several new methods for density estimation have emerged, in particular from the interplay between \ac{ML} and physics. These approaches are typically parametric in nature: the parameters $\params$ of a~suitable variational (parametric) ansatz for the trial density $q_{\params}$ are optimized to match a target density $p$. The trial density $q_{\params}$ is sometimes referred to as \stress{the model}. For now, let us assume it to have a generic form while we discuss the precise parametrizations of $q_{\params}$ in the next section. 

A~natural way to match the trial and target densities is based on the \acf{MLE}\index{maximum likelihood} principle, where a~likelihood function is maximized (or equivalently, a~negative log-likelihood is minimized) to optimize the parameters $\params$. For a~given data set of independent observations $\dataset =\{\vect{x}_{i}\}_{i=1}^\datasize$, the log-likelihood function of the model is defined as
\begin{align}
    \label{eq:likelihood-gen-model}
    {\loglik}(\params\mid\dataset) =\log \left[ \prod_{i=1}^n q_{\params}\left(\vect{x}_i \right)\right] = \sum_{i=1}^n \log  q_{\params}\left(\vect{x}_i \right) \,, && \vect{x} \in \dataset  \sim p
\end{align}
where $q_{\params}(\vect{x})$ is the probability density of the model evaluated at each independent observation of the system $\vect{x}$, sampled from the true density $p$ that is to be approximated. The set of observation $\dataset$ can, for example, refer to images of dogs, spin configurations $\vect{x}$ drawn from an~Ising model at a~fixed temperature, as well as a set of molecule conformations from density functional theory calculations. The model trained on one of this dataset can then be sampled to generate, respectively, different images of dogs (bearing similar features to dogs seen during the training), unseen Ising spin configurations, and new molecule conformations.

The \ac{MLE} optimization requires training samples and is therefore often described as \stress{learning from data}. However, in some cases, data may not be readily available, e.g., because it is difficult or expensive to generate. For such cases, the data-driven procedure just described may not be applied. 
In physics, however, we sometimes have the advantage of knowing the closed form of the underlying density up to a~normalization constant (a.k.a. the partition function in the Boltzmann distribution of a~thermodynamic system). 
Then, we can use an~alternative approach to density estimation, sometimes referred to as \stress{variational inference}\index{variational inference}. This approach is based on encoding physical laws and prior knowledge of the physical system into a family of parameterized densities $\{ q_{\params}\}_{\params}$ and choosing the best parameters by minimizing the so-called \stress{reverse} \acf{KL} divergence~\cite{klref}, \index{Kullback-Leibler divergence} between the target density $p$ and the variational ansatz $q_{\params}$,
%
%
\begin{equation}\label{eq:reverse_KL_divergence}
    \text{KL}(q_{\params} || p) = \int_{\Omega}q_{\params}(\vect{x}) \log \frac{q_{\params}(\vect{x})}{p(\vect{x})} d{\vect{x}} ,
\end{equation} 
where $\Omega$ is the relevant integration domain. This quantity measures the discrepancy between the target density and the trial density.\footnote{Note that this metric may lead to some pathological behavior, already mentioned in \cref{sss:probability}.} 
Importantly, it can be interpreted as an expectation value with respect to the model $q_{\params}$ such that its evaluation only requires $\{ q_{\params}\}_{\params}$ to be easily sampled.\footnote{Conversely to the maximum likelihood objective with requires i.i.d. draws from $p$.} This learning objective for the parameters $\params$ does not require either to know the normalization constant of $p$ as it only amounts to a constant shift.
We stress that the optimization through reverse \ac{KL} divergence is equivalent to the variational free-energy principle in statistical mechanics\cite{wu2019solving, nicoli2020asymptotically}.

If the variational approach optimizes the reverse-\ac{KL} divergence, the maximum likelihood \ac{MLE} approach is equivalent to optimizing the so-called forward-\ac{KL} divergence which reads
\begin{equation}\label{eq:forward_KL_divergence}
    \text{KL}(p || q_{\params} ) = \int_{\Omega}p(\vect{x}) \log \frac{p(\vect{x})}{q_{\params}(\vect{x})} d{\vect{x}}\,.
\end{equation} 
Indeed, \cref{eq:forward_KL_divergence} can be interpreted as an expectation over $p$ from which the previously introduced \cref{eq:likelihood-gen-model} estimates the parameter-dependent term. 
For a~general overview of both data-driven and data-free methods, see Refs.~\cite{hackett2021flow, nicoli2021machine}.

As a side remark, we note that another popular type of learning objective is adversarial training which we describe very briefly in the next section. However, this is not extensively used in physics applications as it is unstable and requires a large amount of data. 

The remainder of this chapter is structured as follows. In the next section, we introduce different popular types of deep generative models. In the last section, we put the emphasis on a family of models particularly useful in physics applications: \acfp{NF} (\cref{par:nf}).

\subsubsection{Taxonomy of deep generative models}\label{sec:deep_generative_models}
Deep generative models, also called \acfp{GNS}\index{generative neural samplers}, are \ac{ML} models with  \acfp{NN} used to define probability distributions. They now come in a large variety, each with different properties. In \crefrange{sss:rbm}{sss:arnn_nqs}, we have already encountered \acfp{RBM} \cite{melko2019restricted} and \acfp{ARNN} \cite{van2016conditional}. Therefore, we discuss them in this section only briefly. Moreover, we describe here \acfp{VAE} \cite{kingma2019introduction} and \acfp{GAN} \cite{goodfellow2014generative,creswell2018generative}. Finally, we dedicate the whole next section to \acfp{NF} \cite{tabak_density_2010,papamakarios2019normalizing,kobyzev2020normalizing}. Another type of generative models goes under the name of a diffusion model~\cite{NEURIPS2020_4c5bcfec,Yang2022DiffusionMA}. While being very promising, their deployment in the field of quantum science is still very limited, and for this reason, we do not discuss them further. The interested reader can, however, find a comprehensive review in Ref.~\cite{Yang2022DiffusionMA}.

Choosing between the different types of generative models can reflect some prior knowledge of the system. For example, we can use \acp{VAE} or \acp{GAN} to encode a density that can be reprensented by the low-dimensional manifold. Additionally, there is typically a trade-off between the capacity of the model and its tractability. This means that, often, more expressive models are at the same time harder to train and therefore difficult to manipulate. In particular, the choice of architecture determines whether the model comes with a direct sampling method
or whether it will be inevitable to resort to the more intricate \acf{MCMC} sampling algorithm to generate samples.\footnote{Even \ac{MCMC}'s performance is hindered by the correlation between samples (discussed widely in \cref{sss:arnn_nqs}) and can severely fail in the case of multimodal distributions (where different blobs of probability are far apart in the configuration space).} 
Moreover, in physics applications, it is highly desired to have access to a~\stress{tractable density}. Specifically, this is necessary to apply reweighting techniques like \ac{NIS}~\cite{noe2019boltzmann,nicoli2020asymptotically,nicoli2021estimation} and \ac{NMCMC}~\cite{albergo2019flow,nicoli2020asymptotically,gabrie_adaptive_2022} as will be further discussed in \cref{par:nf}.
\highlight{If a model gives access to a~\stress{tractable density}, it means it allows to compute the normalized modeled density $q_{\params}(\vect{x})$ for any $\vect{x}$ in polynomial time in the system size $N$.}

\paragraph{Energy-based models} Historically, energy-based models, such as binary Boltzmann machines and \acfp{RBM}, were amongst the first proposed deep generative models. They were heavily inspired by statistical physics (see \cref{sss:rbm} for more details) as they parametrize the logarithm of the probability density directly, which is equivalent to parametrizing an energy in a Boltzmann distribution.  Despite being flexible and very elegant in their formulation, these models are very hard to train because their normalization constant remains intractable. As a result, Boltzmann machines are learned from data through an \stress{approximate} maximum likelihood principle \cite{hinton_practical_2012,Gabrie2015}. Moreover, sampling from \acp{RBM} is not as efficient as for other \acp{GNS} as there exists no direct sampling method. Therefore, while \acfp{RBM} have been extensively used in the early stages of density estimation in quantum physics \cite{melko2019restricted}, they are not always the most suitable tool for density estimation.

\paragraph{\Acfp{VAE}} When it comes to sampling efficiency, developments from the last decade introduced more sophisticated algorithms which significantly simplified sampling configurations from approximated densities. Two major examples are \acfp{VAE} and \acfp{GAN}. While both algorithms push a low-dimensional random variable through an~\ac{NN}, \ac{VAE} only pursues the maximization of a lower bound of the \stress{intractable} likelihood \cite{kingma2013auto,rezende2014stochastic}. 
However, \acp{VAE} allow to draw new configurations very efficiently. 

\paragraph{\Acfp{GAN}} Instead, \acp{GAN} are trained following the so-called adversarial objective \cite{goodfellow2014generative}. They consist of two \acp{NN} called a generator and a discriminator. The generator learns to generate new data, while the discriminator learns to distinguish between real and generated data. The generator and discriminator are trained together in an adversarial manner, where a generator tries to fool the discriminator. This type of game-theoretic optimization strategy may result in very unstable training routines.
However, when successful, the joint optimization of those two agents makes the generator more and more capable of generating very realistic samples. This approach works well in practice and allows for easy sampling but, unfortunately, does not allow access to the likelihood of the model. Hence, \acp{GAN} are not the most suitable for physics applications. 

\paragraph{\Acf{AR} models\index{autoregressive models}}\label{par:arnn} Finally, we turn our attention to deep generative models that give access to a tractable normalized density, namely \acfp{ARNN} and \acfp{NF}. For reasons that will become apparent, we put particular emphasis on \acp{NF} and dedicate it to the whole \cref{par:nf}. For now, let us discuss \acp{ARNN}.
The main advantage of using an~\ac{AR} model is that it is easily trained through maximum likelihood or variational inference thanks to its closed-form likelihood and its uncorrelated and fast sampling~\cite{ramachandran2017fast}. The samples are uncorrelated thanks to the \ac{AR} structure of the probability distribution, which allows for applying a~direct sampling algorithm rather than using an \ac{MCMC}. Direct sampling with \ac{AR} models has been described in detail in \cref{sss:arnn_nqs}, we remind the reader that it goes as follows: sample state $\vect{x}_1$ from $q_{\params}(\vect{x}_1)$, then sample from $q_{\params}(\vect{x}_2\mid\vect{x}_1)$, and so on until $q_{\params}(\vect{x}_N\mid\vect{x}_{N-1}, \dots, \vect{x}_1)$. Direct sampling from a~probability distribution over an~exponentially large configuration space (e.g., if the $\vect{x}_i$ are binary variables) is possible with \ac{AR} models and makes them extremely powerful. The sampling can also be made faster as many samples can be processed in parallel in a~single pass. Note that imposing an~\ac{AR} structure on a~model constrains it: for instance, to build a~deep \ac{CNN}, one must introduce masked layers (to keep the \ac{AR} structure), which alters the capacity of such models.

\paragraph{\Acfp{RNN}\index{recurrent neural network}}
A~special type of \ac{AR} models are \acp{RNN} introduced already in \cref{sec:autoregressive_NNs}.
Among \acp{RNN}, there is a~plethora of sub-models, such as the Long-Short Term Memory (LSTM) and the Gated Recurrent Unit (GRU). The latter has been used in the context of ground state search \cite{hibat2020recurrent}. 
In the context of statistical mechanics, an~\ac{RNN} model has been applied to complex problems such as spin glasses. These models are typically hard to sample from. Therefore, the direct sampling helped reach a~high accuracy in a~simulated annealing task \cite{hibatallah2021variational}.

While \ac{AR} models crucially give access to tractable normalized densities, they are originally limited to discrete random variables and are affected by a relatively inefficient sampling process. That is, \ac{AR} follows an \stress{ancestral sampling} scheme where new samples, e.g., new images, are built sequentially, constraining, for instance, new pixels on the image on previously sampled ones~\cite{vanOord2016pixelrnn}. On the other hand, \acfp{NF} also allow for exact likelihood density estimation, but they are also more flexible, can deal with continuous random variables, and allow for even faster, more efficient sampling. This is why we focus on this example in the next section.

 We conclude this section with a few remarks. This lightning overview suggests that the most interesting \acp{GNS} for physics applications are those giving direct access to exact (or approximate) normalized probability densities. An overview of deep generative models with their appealing properties and applications in physics can be found in Ref.~\cite{nicoli2020asymptotically}. We hereby list some of these \acp{GNS} and provide some references where they have been used in quantum physics: \acfp{ARNN} \cite{wu2019solving,nicoli2020asymptotically, carrasquilla2019, sharir2020,liu2021solving}, \acfp{RNN} \cite{hibat2020recurrent, hibatallah2021variational}, and latent variable models such as \acfp{VAE}\cite{Cristoforetti2017TowardsMP} and \acfp{NF}\cite{albergo2019flow,kanwar2020equivariant,nicoli2021estimation}.


\subsubsection{An important example: normalizing flows\index{normalizing flow}}\label{par:nf}
\begin{figure}[t]
    \centering
    \includegraphics[width=0.9\columnwidth]{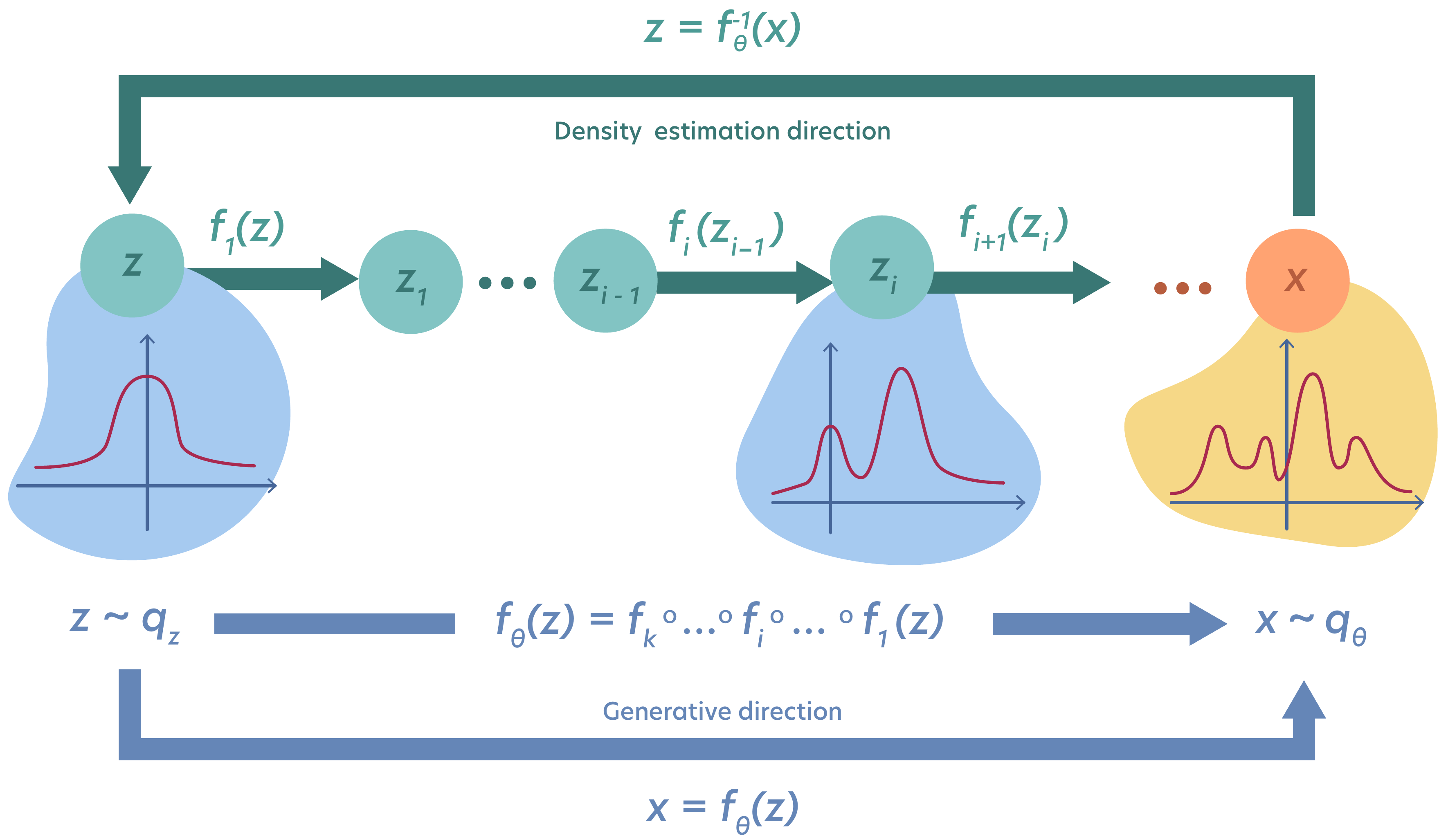}
    \caption[Sketch of a~normalizing flow]{Sketch of a~\acf{NF} architecture. A~sequence of bijective transformations are combined in order to construct a~complicated nonlinear invertible transformation transporting the probability mass from a~base distribution to a~learned density. The learned density is tractable and properly normalized, which enables fast and efficient sampling. Additionally, through the learned density, new samples can be generated and the exact likelihood of given configurations can be computed.}
    \label{fig:NF}
\end{figure}

Normalizing flows (\acp{NF}) 
share similar advantages to \ac{AR} models though they deal with continuous random variables and allow for more efficient sampling (see Refs.~\cite{kobyzev2020normalizing, papamakarios2019normalizing} for a~review).  Recently, many works, amongst which we mention Refs.~\cite{albergo2019flow,kanwar2020equivariant,nicoli2021estimation}, successfully used \acp{NF} for sampling configurations in lattice quantum field theory. While examples from this section focus on sampling methods for lattice field theories, the application of flows is indeed not limited to this specific domain. We refer to~\cite{nicoli2023phd} for a broader overview of applications of deep generative models also in the context of statistical physics and chemistry. 

\paragraph{Normalizing flows' construction}
\acp{NF} 
rely on the simple idea of reparametrizing a variable of interest using a change of coordinates transformation. They generate new realizations of the variable as follows: first, a~latent random variable is sampled from a~tractable, known distribution (e.g., a~normal or a uniform). The random variable transforms through a~bijective mapping learned under specific constraints. Namely, the so-called trivializing map is learned so as to transform the base probability density of the base distribution into one that approximates a given target density. As detailed below, leveraging the change of variable formula, one can thus obtain a~variational approximation of the target density, properly normalized, and easy to handle. In other words, \acp{NF} transform some Gaussian noise into real samples from a learned bijective mapping. 
\highlight{
 In a~\ac{NF}, we define a~diffeomorphic\footnotemark\ transport map that redistributes the probability mass from an~easy-to-treat base distribution to a~more complicated target density.}
\footnotetext{A function is diffeomorphic if it is differentiable and has a differentiable inverse.}

We define a~latent variable $z \in \realset^D$ distributed according to a~known, easy-to-treat distribution. A~common choice is a~standard normal $q_{\vect{z}} = \normdist(0_D,I_D):\realset^D \to \realset^D$. We also define a parametrized transport map $f_{\params}$ with the following properties:
\begin{itemize}
    \item The transport map $f_{\params}$ is a diffeomorphism.
    \item The inverse $f_{\params}^{-1}$ is easy to compute.
    \item The determinant of the Jacobian $\grad_z f_{\params}(\vect{z}) \in \realset^{D \times D}$ is efficient to compute.
\end{itemize}

In the \ac{ML} literature, many candidate transformations satisfying these requirements have been proposed. Amongst others, we mention the affine coupling transformations, such as the non-linear independent component estimation (NICE) \cite{dinh2014nice} and the real non-volume preserving (RealNVP) \cite{dinh2016density}, transformations based on convolutions \cite{kingma2018glow} and those base on splines \cite{neuralsplineflows}. Typically, the desired bijective transformation is the composition of many of these simple transformations (see sketch in \cref{fig:NF}), themselves parametrized through \acp{NN} carrying the set of trainable parameters $\params$:
\begin{equation}
        f_{\params}(\vect{z}) = \left(f_{\params_k} \circ \dots \circ f_{\params_i} \circ \dots \circ f_{\params_1} \right) (\vect{z}),
\end{equation} \label{eq:g}
where all intermediate transformations are invertible, differentiable and are parametrized with their own subset of parameters $\params_i$.
We again refer to Refs.~\cite{kobyzev2020normalizing, papamakarios2019normalizing} for an up-to-date overview of these transformations.

\begin{figure}[t]
    \centering
    \includegraphics[width=0.9\columnwidth]{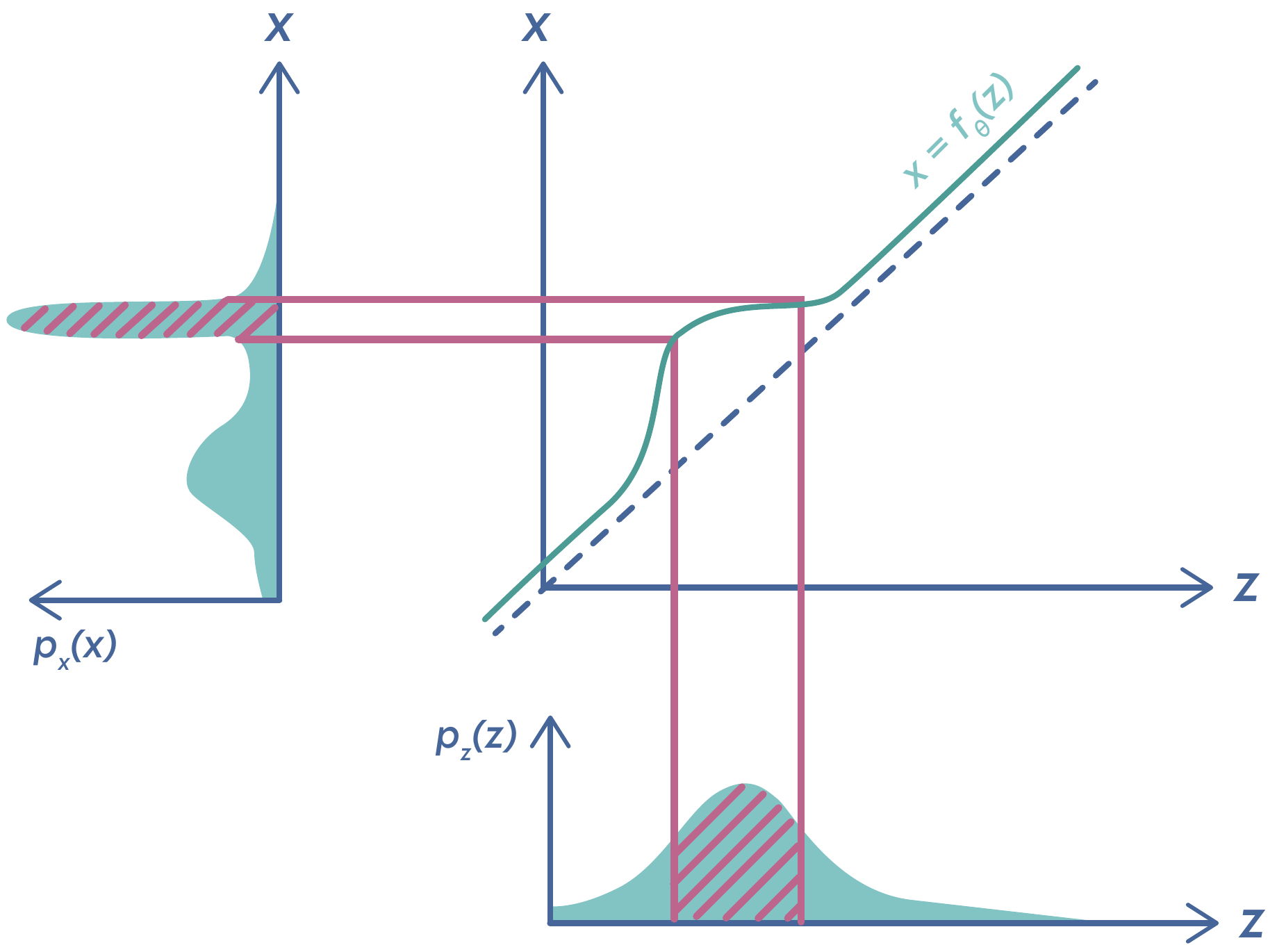}
    \caption[Transport of probability mass through a normalizing flow]{The change of variables formula \cref{eq:changeofvariable} relies on the assumption that the total probability needs to be preserved. A base density $p_{z}$ transforms according to the determinant of the Jacobian, which is responsible for redistributing the probability mass into a non-trivial one. In the one-dimensional case, the red lines identify the probability area in $p_{x}$ and map it back to the corresponding area of the base density $p_{z}$ through the bijection $f_{\params}$.  If the determinant of the Jacobian is one the transformation is said volume-preserving.}
    \label{fig:voltranform}
\end{figure}

The combination of the base density $q_{\vect{z}}$ and the map $f_{\params}$ defines a new density $q_{\params}(\vect{x})$ as as the map ``pushes forward'' the base density $q_{\vect{z}}$. 
To build intuition, we focus on a one-dimensional problem. We start by noting that the probability mass for the random variable $z \in \realset$ should be preserved through a change of coordinates $x = f(z) \in \realset$. Therefore, one can write
\begin{align}\label{eq:changeofvariable}
    \left|p_{x}(x)\, \textrm{d}x\right| = \left|p_{z}(z)\, \textrm{d}z\right|.  
\end{align}
It then follows that
\begin{align}
    p_{x}(x) = p_{z}(f^{-1}(x)) \left|\frac{\textrm{d}z}{\textrm{d}x}\right| = p_{z}(f^{-1}(x)) \left|\frac{\textrm{d}f(z)}{\textrm{d}z}\right|^{-1},
\end{align}
where we used the definition of our bijective transformation $x=f(z)$ and its inverse.
More specifically, when the bijection is parametrized by $\params$, it follows that
\begin{equation}
    q_{\params}(x) = q_{z}\left(f^{-1}_{\params}(x)\right) \, \left| \frac{\textrm{d}f_{\params}(z)}{\textrm{d}z} \right|^{-1} \,.
    \label{eq:nfprob}
\end{equation}
Under our assumptions, $f_{\params}$ is invertible by construction and $q_{z}$ is the Gaussian distribution we chose as the reference density to sample a latent noise $z$. 
For a $D$-dimensional problem, $f_{\params}: \realset^D \rightarrow \realset^D$ and the derivative with respect to $\vect{z}$ is replaced in \cref{eq:nfprob} by the determinant of the Jacobian $\det(\grad_{\vect{z}} f_{\params}(\vect{z}))$, which represents the volume transformation depicted in~\cref{fig:voltranform}.

Once the trivializing map $f_{\params}$ is trained, a~new data point $\vect{x}^*$ can be generated by sampling $\vect{z}^*\sim q_{\vect{z}}$ in the latent space and transform it through the bijection to obtain
a corresponding sample in the data space. Reversely, one can also take a~given sample, plug it into \cref{eq:nfprob} and obtain the exact likelihood of the sample. In order to make  both learning and sampling processes smooth and efficient, the determinant of the Jacobian of the transformation needs to be tractable and efficient to compute as already stated in our assumptions. 
We further refer the reader to Ref.~\cite{kobyzev2020normalizing} for more details on the coupling transformation and the computation of the Jacobian. 

\paragraph{Training}
Depending on applications, the training of \ac{NF} can be performed with or without data following the methods described in \cref{sec:w_wout_data}.
%
When a dataset $\dataset = \{\vect{x}_{i}\}_{i=1}^\datasize$ from the target distribution $p(\vect{x})$ is available, the model can be trained through maximum log-likelihood:
\begin{equation}\label{eq:flowloglik}
    \log p(\mathcal{D}|\params) = \sum_{i=1}^\datasize \log q_\param(\vect{x}_i).
\end{equation}
which is equivalent to minimizing the forward-\ac{KL} divergence \cref{eq:forward_KL_divergence}.

However, as mentioned earlier, for physical systems it is often the case that an \stress{unnormalized} target density is available while samples are not readily available. In these circumstances, one can train a \ac{NF} by instead minimizing the \stress{reverse}-\ac{KL} divergence \cref{eq:reverse_KL_divergence}. This approach does not need data as it trains by self-sampling, meaning that during optimization, configurations are directly sampled from the variational density $q_{\params}$ we seek to optimize. Specifically, given a target Boltzmann-like target density, i.e., $p(\vect{x})=Z^{-1} \exp{-H(\vect{x})}$ the reverse \ac{KL} objective can be rewritten as
\begin{align}\label{eq:revKL}
    \text{KL}(q_{\params}||p) &= \int_\Omega q_{\params}(\vect{x}) \, \ln \left( \frac{q_{\params}(\vect{x})}{p(\vect{x})} \right) \textrm{d}\vect{x} \nonumber \\
    &= \int_\Omega q_{\params}(\vect{x}) \,\left(\ln{q_{\params}(\vect{x})} + H(\vect{x}) + \log{Z}\right)\,\,\ 
 \textrm{d}\vect{x}, 
 \end{align}
 where in the last step, we explicitly used the known form of the target density where $H(\vect{x}$ is the Hamiltonian, and $Z$ is the partition function. This latter now becomes just a constant shift and can be ignored for training purposes as it vanishes when computing the gradient during optimization. Using the analytic form for the model likelihood from \cref{eq:nfprob}, the \ac{KL} divergence becomes
 \begin{align}
     \text{KL}(q_{\params}||p) &= \int_\Omega H(f_{\params}(\vect{z})) - \log \left| \frac{\textrm{d}f_{\params}}{\textrm{d}\vect{z}} \right|(\vect{z})  + \log q_{\vect{z}}(\vect{z}).
 \end{align}

Nevertheless, training models following the reverse-\ac{KL} objective can lead to inaccurate outcomes when the target distribution is multimodal. Namely the model can miss out a mode entirely following a phenomenon known as \index{mode-collapse}, which we briefly discuss in the Outlook of this section. 

\paragraph{Applications}
Once the model $q_{\params}$ is trained to be a good approximation of the target density $p$, we can use the generative model, the flow in this case, to generate configurations for a physical system approximately following the Boltzmann distribution. 
This flow-generated configurations can be exploited to create efficient Monte Carlo estimator of physical observables. For instance, \acp{NF} can be incorporated in \ac{MCMC} sampling schemes as a proposal move in order to lower autocorrelation and effectively reach the ergodicity regime (\cite{albergo2019flow, nicoli2020asymptotically,gabrie_adaptive_2022} among others). This strategy often goes under the name of NeuralMCMC sampling~\cite{nicoli2020asymptotically} or flowMCMC \cite{grenioux2023sampling}. Flows can also be incorporated as powerful trial distributions in importance sampling schemes \cite{noe2019boltzmann,muller_neural_2019,nicoli2020asymptotically}.
One further conceptual advantage of flow-based models, and generative models allowing for exact likelihood density estimation in general, is that they allow estimating quantities normally not accessible by standard methods. These include the partition function $Z$, the free energy, and other thermodynamic observables such as entropy and pressure. We refer to Refs.~\cite{nicoli2020asymptotically,nicoli2021estimation,nicoli2021machine} for more details.
We refer to the literature review of \cite{bacchio2022learning} for a wider overview of the many more applications in physics and quantum chemistry. 

\subsubsection*{Outlook and open problems}
In recent years \acp{NF} demonstrated great potential and appealing conceptual properties thus making them promising candidates for dealing with density estimation and sampling in the physical sciences. Nevertheless, many challenges need to be faced still. 

First, a major drawback of flow-based methods is that they cannot easily scale to larger systems~\cite{del2021machine,del2021efficient,abbott2022aspects}. The promising results achieved so far by leveraging flow-based samplers, in particular in the context of lattice field theory, were obtained on relatively small lattices. In the limit of small lattice spacing and infinite volume limit, thus approaching the continuum, it is clear that the scaling of both standard methods and generative models is clearly going to be in favor of the former. However, there are strategies currently being explored and substantially applied to improve the scaling of these methods. One example is represented by leveraging inductive biases, meaning exploiting prior physical knowledge, thus incorporating existing symmetries into the flow. This enables more effective training since the model does not have to learn physics from scratch. The idea of incorporating symmetries in the context of a flow-based sampler for lattice field theory has been successfully applied to U(1)~\cite{PhysRevLett.125.121601} and SU(N)~\cite{PhysRevD.103.074504} lattice gauge theories. Moreover, Refs.~\cite{kohler2020equivariant, satorras2021n} give a complementary intuition on how to incorporate equivariance into a \ac{NF}. 

Another relevant challenge briefly mentioned in the previous sections indeed relates to the issue of mode collapse\index{mode collapse}. This drawback, affecting systems trained by self-sampling, has been discussed in Refs.~\cite{hackett2021flow, nicoli2021machine, gabrie_adaptive_2022} and is currently an~open problem not only in the domain of \acp{NF} in physics but also in the entire \ac{ML} community~\cite{jerfel2021variational}. Learning a~multi-modal distribution is often more challenging as it often prevents using the reverse-\ac{KL} objective, prone to mode collapse. When this happens, the generative model may perfectly cover one (or more) modes of the target density yet completely neglect the others. It follows that when sampling from the learned variational distribution, e.g. a flow, we don't have full support over the target density. Our ansatz thus badly approximates the target and leads to biased estimates~\cite{nicoli2023detecting}. This problem is very often found in the context of sampling and density estimation within physics and quantum chemistry applications. Mode-collapse may be hard to detect in some scenarios and hence be very harmful when accurate estimates of observables are of interest. Some recent works, tried to address this problem by combining \acp{NF} with initial knowledge of modes and \stress{adaptive training methods} \cite{gabrie_adaptive_2022}, \stress{path-gradients} \cite{Vaitl_2022, pmlr-v162-vaitl22a} or \stress{annealed importance sampling} \cite{arbel2021annealed,midgley2022flow,matthews2022continual}. While preliminary results are encouraging, this is still very much an open problem. 

\subsubsection*{Further reading}
\begin{itemize}

    \item Wang, L. (2018). 
    \href{https://wangleiphy.github.io/lectures/PILtutorial.pdf}{\textit{Generative models for physicists}}~\cite{wang2018:generative_tutorial}.
    
    \item Noe, F. \textit{et al. }(2019). 
    \href{https://www.science.org/doi/10.1126/science.aaw1147}{\textit{Boltzmann generators}}. Science, 365, eaaw1147\cite{noe2019boltzmann}. The seminal paper on Boltzmann Generators. Using \acfp{NF} in the context of quantum chemistry.
    
    \item Köhler, J. \textit{et al.} (2020).
    \href{https://arxiv.org/pdf/2006.02425.pdf}{\textit{Equivariant flows}}. arXiv:2006.02425 \cite{kohler2020equivariant},
    Satorras, V. G. \textit{et al.} (2022). 
    \href{https://arxiv.org/pdf/2105.09016.pdf}{\textit{E(n) equivariant normalizing flows}}. arXiv:2105.09016 \cite{satorras2021n}.
    How to incorporate equivariances into \acfp{NF}.

    \item Nicoli, K. A. \textit{et al. }(2021).
    \href{https://arxiv.org/pdf/2111.11303.pdf}{\textit{Machine learning of thermodynamic observables in the presence of mode collapse}}. arXiv:2111.11303 \cite{nicoli2021machine}.
    Hackett, D. \textit{et al. }(2021). 
    \href{https://arxiv.org/pdf/2107.00734.pdf}{\textit{Flow-based sampling for multimodal distributions}}. arXiv:2107.00734 \cite{hackett2021flow}. 
    Nicoli, K. A. \textit{et al. }(2023). 
    \href{https://arxiv.org/pdf/2302.14082.pdf}{\textit{Detecting and mitigating mode-collapse for flow-based sampling of lattice field theories}}. arXiv:2302.14082 \cite{nicoli2023detecting}.
    Further discussions on the problem of mode collapse.
    
    \item Albergo, M. S. \textit{et al. }(2021). \href{https://journals.aps.org/prd/pdf/10.1103/PhysRevD.104.114507}{\textit{Flow-based sampling for fermionic lattice field theories}}. Phys. Rev. D, 104, 114507~\cite{PhysRevD.104.114507}. Sampling lattice field theory with fermions using \acfp{NF}.

    \item Abbott, R. \textit{et al. }(2022). \href{https://arxiv.org/pdf/2211.07541.pdf}{\textit{Aspects of scaling and scalability for flow-based sampling of
lattice QCD}}. arXiv:2211.07541~\cite{abbott2022aspects}. Issues of scaling \acfp{NF} to larger systems.

\end{itemize}

\subsection{Machine learning for experiments}\label{sec:ML_for_exp}

Quantum experiments pose tough technical challenges, and the task of optimizing their performance while interpreting the output data can seem daunting. It is informative to note that the output of quantum devices naturally generates large-scale data, which is the regime where \ac{ML} thrives.
In \cref{sec:phase_class}, we have already seen how \ac{ML} can be used to detect phases, and although such efforts are much more challenging when dealing with experimental data, a~few works managed to successfully address this real-world problem \cite{Rem19, Lustig2020, Kaming2021}. In \cref{sec:BO_GPR_science}, we have discussed the applications of \acfp{GP} and \acf{BO} for inverse problems involving experimental data \cite{Vargas2019,Deng2020,Cantin2020} and optimizing experiments \cite{BO-lasers_PRL_2020,BO-lasers_PRL_2021, BO-lasers_NatComm_2020,Sugisawa2021,Phoenics_AAG_2018,Gryffin_AAG_2021,multiBO_ferroelectric_2021,NEXTorch_BOChem_2021,Golem_BO_AAG_2021}. In \cref{sss:quantum_tomography} we have also demonstrated how \ac{ML} can boost quantum tomography \cite{Torlai_2018,Torlai2019,carrasquilla2019,schmale2021scalable} with \acp{NQS} presented in \cref{sec:NN_q_states}. Another research direction pursued in the context of quantum experiments is the application of \ac{RL} for quantum feedback control \cite{QuantumFeedback2021,Foesel2018errorcorrection,Twamley2021,Nguyen2021}, quantum error correction \cite{Andreasson2019quantum,Fitzek2020_errorcorrection,Sweke2021_errorcorrection,Theveniaut2021_errorcorrection_neat}, quantum circuit optimization \cite{Foesel2021circuitoptim}, and experiment design \cite{melnikov2018active}, all described in~\crefrange{sss:RL_qcontrol}{sec:rl_qexperiments}.

This section focuses on other \ac{ML} approaches for experimental data. First, we acknowledge that there is an~important niche of experimental physics that can be revolutionized by \ac{ML}, i.e., automation of (tedious) repetitive tasks. We show examples of successful realizations of this idea with actual experimental data in \cref{sss:MLforexperiments}. In \cref{sss:MLfortofimages}, we discuss the theoretical proposal of \ac{ML}-based analysis of time-of-flight images, which is a standard measurement technique in ultracold-atom setups. Then, in \cref{sss:Hamiltonian-learning}, we describe a~powerful scheme for quantum experiments, i.e., learning the Hamiltonian governing the system from measurements. We conclude this section with \cref{sss:MLforexpdesign} discussing the successes of the computer-guided design of experiments that does not include \ac{RL}.

\subsubsection{Automation of experimental setups}\label{sss:MLforexperiments}

This section is devoted to novel ideas for the automation of physical experiments. Specifically, we present the automated identification of nanomaterial samples for quantum device technologies \cite{flakes} and the automated tuning of double quantum dots \cite{dots} for quantum information devices.
Both examples have one thing in common: at some point in the execution of their respective experiments, a~large amount of human labor becomes necessary that is tedious and repetitive with respect to the decision-making process a~worker employs but not trivial enough to be replaced by a~simple looping algorithm.

\begin{figure}[t]
    \centering
    \includegraphics[width=0.7\textwidth]{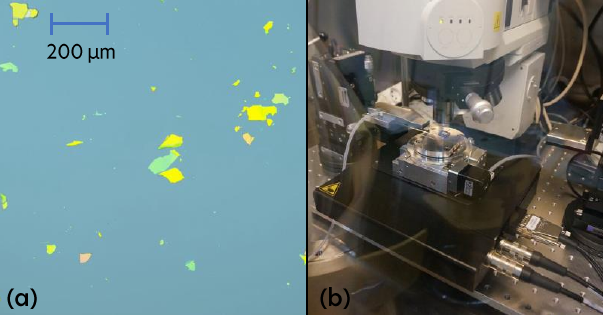}
    \caption[Experimental nanoflake setup]{Experimental nanoflake setup. (a) A~typical microscope image of hBN from \ToggleForCUP{Greplová, E. et al. (2020). \textit{Fully automated identification of two-dimensional material samples}. Phys. Rev. Appl. 13, 064017~\cite{flakes}.}{Ref.~\cite{flakes}.} (b) Typical microscope setup with waver already placed under the microscope. Photo credit: Klaus Ensslin Lab, ETH Zürich.}
    \label{fig:hBN_flakes}
\end{figure}

\paragraph{Automated identification of nanomaterials}

In the case of the preparation of nanomaterials for quantum devices, as detailed in Ref. \cite{flakes}, an~important step is the selection of appropriate two-dimensional flakes from a~wafer under a~microscope. The flakes in question can differ depending on their desired use in the final device, but they all share their flat shape and approximate size due to the exfoliation-based technique with which they are prepared beforehand. Examples include hexagonal boron-nitride (hBN), graphite, and bilayer-graphene. \Cref{fig:hBN_flakes} depicts the scanning setup along with a~typical image for hBN.
The microscope in a~typical setup can operate at different magnifications and can scan a~1 cm$^2$ wafer in roughly three minutes. In practice, however, this takes much longer, as the human operator has to slowly move the frame across the wafer and decide for each frame which of the depicted flakes are suitable for future device building. In short, the human operator classifies the flakes; a~well-trained \ac{NN} could do this as well.
Hence, the design and training of a~suitable \ac{NN} architecture was at the core of the automation scheme developed in Ref.~\cite{flakes}.
However, it is worth noting that the automation scheme did more than just the classification\index{classification} task, as summarized in \cref{fig:flakes_auto}. 
For example, prior to even implementing anything network related, they provided the experimental team with a~program with a~simple GUI for click-based flake labeling of pre-processed images to simplify the generation of an~adequate data set.
This is noted here to truly reflect the additional steps that need to be taken into consideration when working with experimental setups.

\ToggleForCUP{\begin{figure}[htb]
    \centering
    \includegraphics[width=0.99\textwidth]{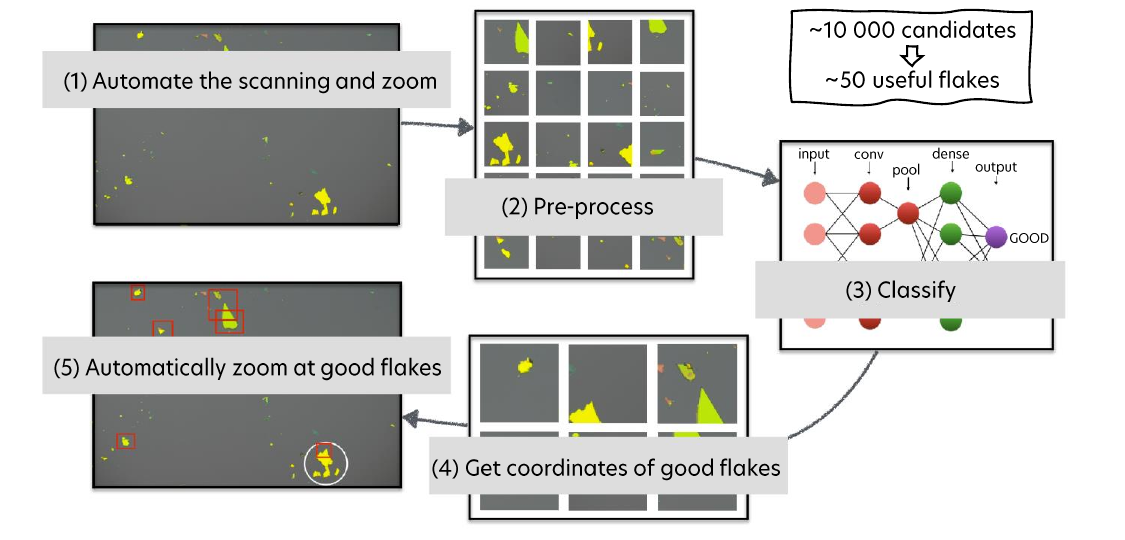}
    \caption[Nanoflake automation scheme]{The full automation procedure consisted of scanning, \textit{labeling}, pre-processing, \textit{training}, classifying, and then collecting and presenting the good flakes. The steps marked with italics were done only prior to application. Figure credit: \href{https://www.eliskagreplova.com/qmai-team}{QMAI} group at \href{https://www.tudelft.nl/}{TU Delft}.}
    \label{fig:flakes_auto}
\end{figure}}{}

In this work, more than one network was used to minimize classification errors: Three networks were used and applied consecutively, each consisting of four convolutional layers\index{convolutional neural network} and one dense layer.
The reason why this stacking of networks was necessary is the immense imbalance between good and bad flakes in the data set. 
While a~batch optimization\index{batch optimization} procedure paired with data augmentation\index{data augmentation} can usually account for this to some extent, here this was not sufficient: after passing new data that contained approximately $1000$ flakes, $10.8\%$ of which were good, through a single trained network, the classified results yielded the $86\%$ accuracy with $13\%$ of false positives (bad flakes classified falsely as good) and $1\%$ of false negatives (good flakes classified falsely as bad). Since there are so few truly good flakes, avoiding false negatives is of utmost importance.

An~important and more practical aspect of this number and accuracy of leftover ``good'' (correctly or incorrectly classified) flakes is concerned with the additional human labor that would follow this classification result: after classifying, the automation scheme (compare \cref{fig:flakes_auto}) automatically zooms in on the good flakes after which a~human operator steps in again. 
More specifically, this means that instead of manually scanning the probe, looking at the, e.g., 1000 flakes available and zooming in on the selected flakes in order to then decide whether they are good or not, in the automated scheme, the human operator only acts after the entire wafer is scanned and the microscope is shifted to the different locations of the good flakes on which it appropriately zooms in.
In this way, the experimentalist only makes the final decision on which flakes to use; this represents a~substantial workload reduction from 944 flakes to 224, approximately 44\% of which are actually good.
However, since there is still work involved, three networks instead of one are used to reduce the amount of flakes that need to be looked at even further. After passing the data through all three networks, only 150 flakes need to be looked at, approximately $57\%$ of which are good.

\ToggleForCUP{}{\begin{figure}[t]
    \centering
    \includegraphics[width=0.99\textwidth]{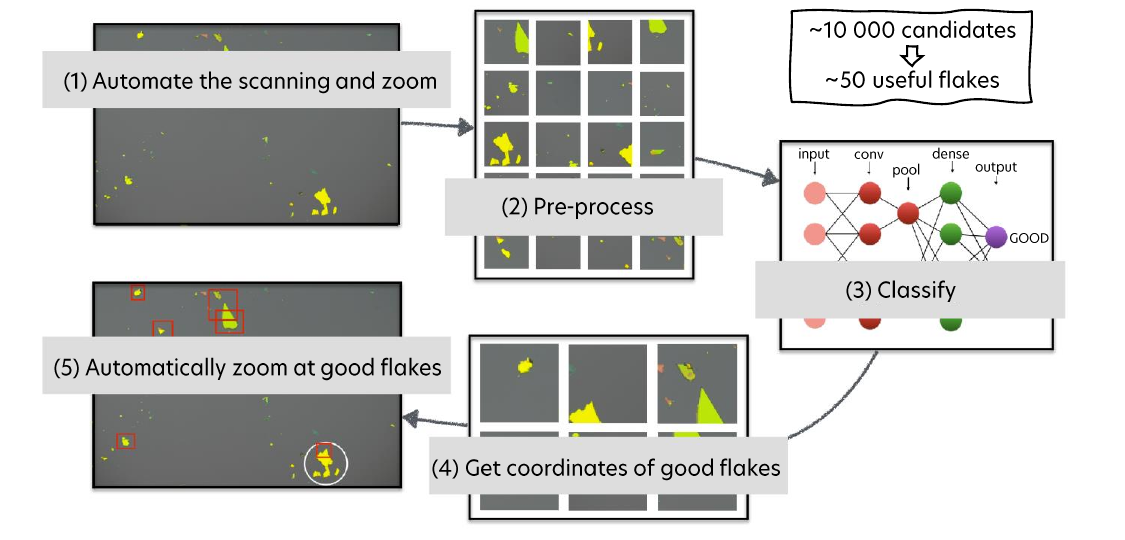}
    \caption[Nanoflake automation scheme]{The full automation procedure consisted of scanning, \textit{labeling}, pre-processing, \textit{training}, classifying, and then collecting and presenting the good flakes. The steps marked with italics were done only prior to application. Figure credit: \href{https://www.eliskagreplova.com/qmai-team}{QMAI} group at \href{https://www.tudelft.nl/}{TU Delft}.}
    \label{fig:flakes_auto}
\end{figure}}
\ToggleForCUP{}{\begin{figure}[b!]
    \centering
    \includegraphics[width=0.5\textwidth]{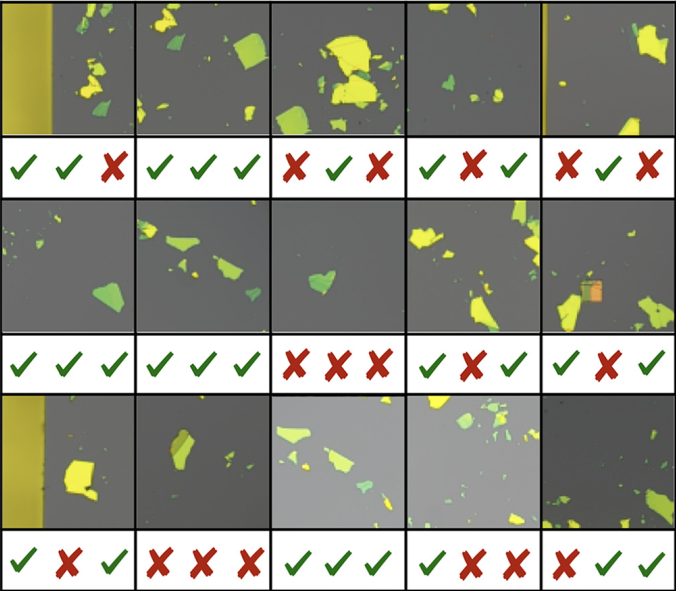}
    \caption[Differences in human flake judgment]{Differences of human judgment visualized. Selection of 15 frames that were to be judged on whether they contained good flakes or not along with the judgment results from three different human operators. Adapted from \ToggleForCUP{Greplová, E. et al. (2020). \textit{Fully automated identification of two-dimensional material samples}. Phys. Rev. Appl. 13, 064017~\cite{flakes}.}{Ref.~\cite{flakes}.}}
    \label{fig:flakes_label}
\end{figure}}
One of the reasons why the accuracy is still comparatively low is due to the discrepancies in the classification choices among the humans who did the labeling beforehand. 
Given that it was already time consuming enough to label a~large enough data set even with the helpful GUI program written for that purpose, every flake was only labeled once by one of the many experimentalists participating in this project. 
\Cref{fig:flakes_label} captures the differences in judgment of three different participants all looking at a~selection of single shot frames.
It is important to note that while here we were able to explicitly show one source of uncertainty and errors in the preparation of the data, usually we do not see it directly. We should therefore always acknowledge the possibility of their existence!
\ToggleForCUP{\begin{figure}
    \centering
    \includegraphics[width=0.5\textwidth]{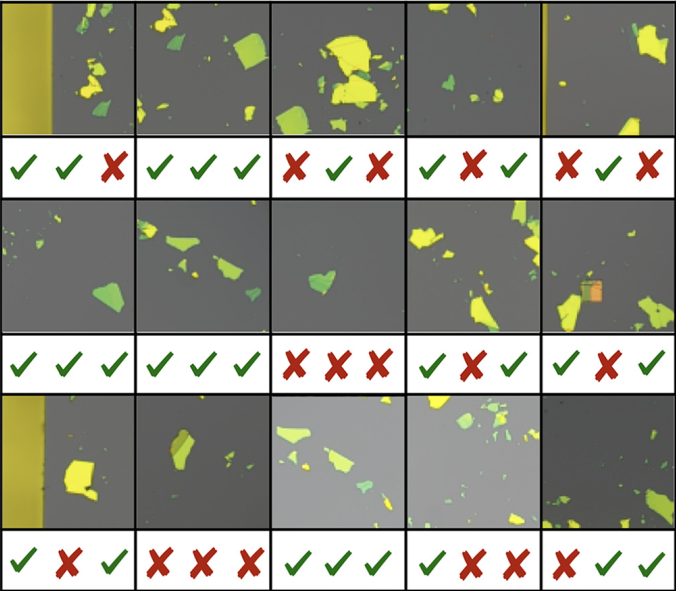}
    \caption[Differences in human flake judgment]{Differences of human judgment visualized. Selection of 15 frames that were to be judged on whether they contained good flakes or not along with the judgment results from three different human operators. Adapted from \ToggleForCUP{Greplová, E. et al. (2020). \textit{Fully automated identification of two-dimensional material samples}. Phys. Rev. Appl. 13, 064017~\cite{flakes}.}{Ref.~\cite{flakes}.}}
    \label{fig:flakes_label}
\end{figure}}{}
Evidently, there is a~degree of uncertainty and disagreement about whether the individual frames contain good flakes which introduced a~bound on the model performance. 
This is reflected in the accuracy of the classification. 

All in all, the developed automation procedure\index{automation procedure} including and revolving around the \ac{NN} is still clearly a~success: this type of material control is a~common step in the field of nanomaterial device development and the method generalizes satisfyingly to, e.g., graphite and bi-layer graphene (compare Ref. \cite{flakes}). The implementation is available on GitHub \cite{flakesGitHub}.

\paragraph{Quantum dot tuning}

The next example is the automated tuning of double quantum dots in quantum information technology research. A~detailed discussion can be found in Ref. \cite{dots}.
A~quantum dot is a~nanostructure that is confined so strongly in all three spatial dimensions that it is essentially zero-dimensional. 
The confinement, similar to a~particle-in-a-box scenario, leads to the emergence of quantum effects, i.e., energy quantization (as opposed to having a~continuous energy spectrum in larger structures) and thus discrete states.

In quantum information research, quantum dots are used to create qubits by putting together two dots as discrete states.
There are various reasons why this technology is challenging in the context of universal quantum computation: on the one hand, there are difficulties associated with making two dots interact in a~controlled manner and, on the other hand, there are issues associated to reproducibility.
Both concerns are related to preparation techniques, and the example discussed here offers a~new \ac{ML}-based remedy for the former.

Like in the previous example, the experimental procedure contains a~tedious step that one can seek to automate. 
While in the flake example, this step revolved around human operators looking at images from a~probe under a~microscope, in this quantum dot setup the human operator looks at graphs created from changes in current measurements in the quantum device with changing applied voltages. 
When conducting a~measurement like this, the quantity of interest is the occupation of the quantum dots, i.e., the state of the quantum dot. The occupation can be changed by applying a~voltage to the dot: as seen in \cref{fig:dotsSetup}(a), every one of the three quantum dots has a~plunger gate (PG) associated with it that is used to tune the voltage. Underneath these dots, the current of a~quantum point contact is measured.\footnote{The quantum point contact in \cref{fig:dotsSetup}(a) is formed by the three gates at the bottom responsible for the measurement.} As defined in Coulomb's law, the occupation of the dots, i.e., the negative charge of the respective electrons, affects the electric current close to it. Hence, a~change in the electron occupation causes a~discrete change in the current flow which corresponds to spikes in the conductance (${\partial I_\text{QPC}}/{\partial V_\text{PG}}$) measurement. These spikes are the dark blue lines in the charge-stability diagram presented in \cref{fig:dotsSetup}(b). \Cref{fig:dotsSetup}(b) can then be interpreted as follows. In the bottom left corner, both \acp{QD} are unoccupied but whenever a~vertical (horizontal) line is crossed, an~electron is added to \ac{QD}1 (\ac{QD}2). 
\ToggleForCUP{
\begin{figure}[p]
    \centering
    \includegraphics[width=\textwidth]{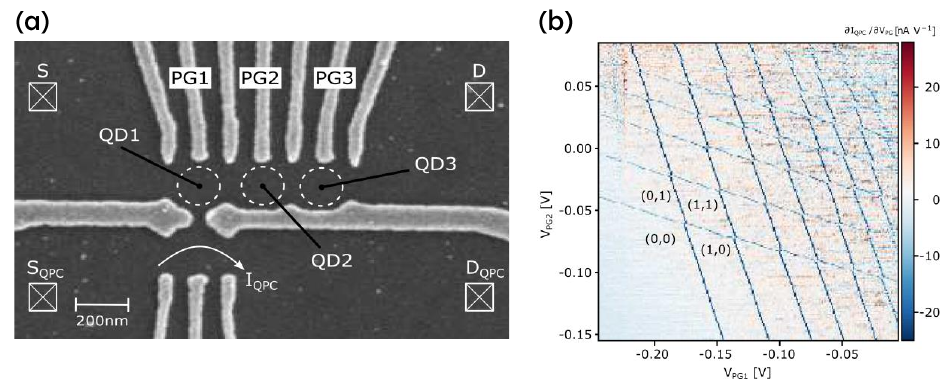}
    \caption[Experimental setup for quantum dots]{Experimental setup for quantum dots. The device is built and then tuned using measurements that are made possible by the quantum point contact built into it. (a) Scanning microscope image of an~example device. Base material is a~GaAs heterostructure with an~electron gas embedded at the position where the three quantum dots (\ac{QD}1, \ac{QD}2, \ac{QD}3) are intended to be. It also contains a~number of finger gates for confinement and measurement. The gates responsible for the measurement are the three at the bottom (quantum point contact). (b) The charge-stability diagram for a~double quantum dot device that uses \ac{QD}1 and \ac{QD}2 in (a). Correspondingly, the changed voltages are the ones for PG1 (corresponds to \ac{QD}1) and PG2 (corresponds to \ac{QD}2). Taken from \ToggleForCUP{Durrer, R. et al. (2020). \textit{Automated tuning of double quantum dots into specific charge states using neural networks}. Phys. Rev. Appl. 13, 054019~\cite{dots}.}{Ref.~\cite{dots}.}}
    \label{fig:dotsSetup}
\end{figure}}{
\begin{figure}[t]
    \centering
    \includegraphics[width=\textwidth]{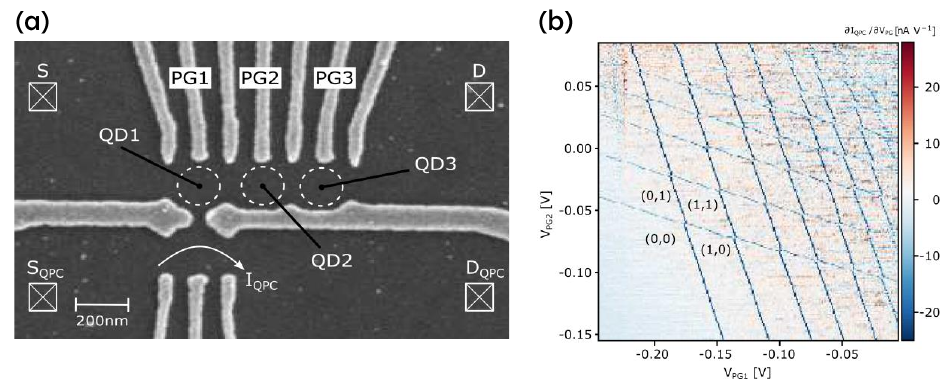}
    \caption[Experimental setup for quantum dots]{Experimental setup for quantum dots. The device is built and then tuned using measurements that are made possible by the quantum point contact built into it. (a) Scanning microscope image of an~example device. Base material is a~GaAs heterostructure with an~electron gas embedded at the position where the three quantum dots (\ac{QD}1, \ac{QD}2, \ac{QD}3) are intended to be. It also contains a~number of finger gates for confinement and measurement. The gates responsible for the measurement are the three at the bottom (quantum point contact). (b) The charge-stability diagram for a~double quantum dot device that uses \ac{QD}1 and \ac{QD}2 in (a). Correspondingly, the changed voltages are the ones for PG1 (corresponds to \ac{QD}1) and PG2 (corresponds to \ac{QD}2). Taken from \ToggleForCUP{Durrer, R. et al. (2020). \textit{Automated tuning of double quantum dots into specific charge states using neural networks}. Phys. Rev. Appl. 13, 054019~\cite{dots}.}{Ref.~\cite{dots}.}}
    \label{fig:dotsSetup}
\end{figure}
}

The goal in device preparation and tuning is to prepare different discrete states by applying the adequate voltages that correspond to the correct current-spike-line framed, diamond-shaped area in the charge-stability diagram.
To do this, the operator needs to know the charge-stability diagram.
Thus, tuning a~double quantum dot device, such as this one, requires measuring the entire charge-stability diagram, i.e., performing many subsequent measurements where one voltage is kept constant and the other one is gradually changed. 
In this very time-consuming scenario a~classification-based \ac{ML} scheme can be of help.
For \ac{ML} to bring a~significant improvement, it is essential so that measuring the entire charge-stability diagram is not required for the input data.
A~suitable scheme should be able to produce the two plunger gate voltages for a~specified desired occupation state from any starting state (corresponding to a~starting pair of voltages along with their current flow).
However, without the charge-stability diagram, there is no way of knowing which occupational state the two starting voltages correspond to.
For example, if the starting state had both voltages at $0\,V$, then a~human operator with knowledge of \cref{fig:dotsSetup}(b) would know that this places the state somewhere in the top right of the diagram and would, by means of counting lines, be able to specify the state.

\begin{figure}[t]
    \centering
    \includegraphics[width=0.9\columnwidth]{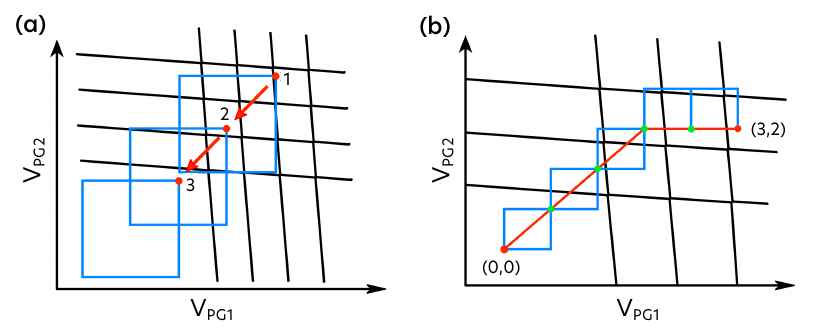}
    \caption[Quantum dots machine learning scheme]{(a) Finding the (0,0) state using a~first classification network: Starting with random initial voltages corresponds to point ``1''. In this example, the network had to go through three iterations of classification and frameshifting until (0,0), here marked by ``3'', was reached. (b) Finding the state of desired occupation (m,n): Depending on the different classification results, the frame can be shifted either only to higher PG1 voltages (in case of one vertical line), only to higher PG2 voltages (in case of one horizontal line) or diagonally, increasing both gate voltages (in case of two lines). Adapted from \ToggleForCUP{Durrer, R. et al. (2020). \textit{Automated tuning of double quantum dots into specific charge states using neural networks}. Phys. Rev. Appl. 13, 054019~\cite{dots}.}{Ref.~\cite{dots}.}}
    \label{fig:dotsNet}
\end{figure}

To avoid having to measure the entire diagram, the \ac{ML} scheme uses an~approach for which small, low-resolution excerpts of the diagram suffice: 
\begin{enumerate}
    \item \textbf{Finding the (0,0) state:} In a~first step, one utilizes the fact that any state except the (0,0) state is framed by four lines in the diagram, whereas the (0,0) state only has neighboring lines in the positive x- and y-direction. Therefore, a~first classification network is trained to recognize if there are more lines to cross in the negative $x$- or $y$-direction. The output is true or false. If there are more lines, both plunger voltages are lowered by a~set amount (as depicted in \cref{fig:dotsNet}(a)), and the classification is performed again until there are no more lines to cross.
    \item \textbf{Finding any desired, given state:} To get from the (0,0) state to any desired state (m,n), one has to cross \textit{exactly} m vertical lines and n horizontal lines. Thus, a~second network is now trained to more accurately classify which lines there are. This network uses smaller frames of a~higher resolution that allow a~more differentiated distinction between the cases where there are no lines, there is one vertical line, there is a~horizontal line, and there are both in the considered frame (compare \cref{fig:dotsNet}(b)). Just like in the first step, each classification is followed by a~change in voltages and this 2-step procedure is repeated until the desired state is reached. 
\end{enumerate}
For the training of the first network, 470 charge stability diagrams were measured in fairly low resolution, whereas for the second network 128 charge stability diagrams were measured in higher resolution. The plunger gate voltage ranges were varied for the different measurements to foster better generalization later on. In both cases, data sets were created by cutting out numerous random frames from the diagrams and labeling them with a~script. Note that while full charge stability diagrams were measured for the generation of the training data set, the input for the eventual application of the network only needs small windows. Measurement of full charge stability diagrams and use of many windows therein was just a~convenient way to create a~data set.

When tested on the actual device, the success rates of the two loops were 90\% (step 1 loop) and 63\% (step 2 loop) which combines to an~overall success rate of 57\%. 
It is important to keep in mind that those individual success rates are not the equivalents of the accuracy rates of the two networks: each loop calls the network multiple times, so errors are doomed to accumulate, and the second loop usually requires more calls to the network than the first loop, because the frames are smaller, see panels (a) and (b) of \cref{fig:dotsNet}. In fact, when tested separately and only a~single time on a~labeled data set, the accuracy rates reached by the two networks were 98.9\% and 96\%. In the article, the authors stated that the primary error source was identified as a~weak signal-to-noise ratio and improving on this would surely improve the scheme.

In conclusion, the integration of \acp{DNN} into a~larger scheme can lead to an accumulation of errors, and this should be taken into account when planning the implementation of the automation routine. In general, the integration of \ac{ML} approaches into broader automation schemes calls for different levels of network accuracy and, as was the case in the first example, some scenarios might even have limited network accuracy in general. It is important to take these things into account before implementation and gauge the benefits of automation versus the remaining workload.

Of course, the approaches presented are not the only ideas for automating experimental setups. For example, \acp{CNN} also help in detecting contamination by ice crystal diffraction in macromolecular diffraction data \cite{Mostosi2020} or analyzing cryo-electron microscopy maps of proteins \cite{Nolte2022}.

\subsubsection{Machine-learning analysis of time-of-flight images}\label{sss:MLfortofimages}

When it comes to analysis of the experimental data, we present one more example related to ultracold-atom experiments. In contrast to the two highly specialized applications to actual experimental data discussed so far in \cref{sss:MLforexperiments}, we consider a~proposal that is based on theoretical data but is readily extendable to experiments \cite{lode2021optimized}. There is a~number of theory-based, yet application-oriented proposals that are currently being published, and discussing their differences should prove insightful. The focus of this discussion is on the feasibility of making the transition from theory to application.

For any such transition from theory to experiment within an~\ac{ML} scope, the following aspects should be examined:
\begin{itemize}
    \item \stress{Specificity vs. flexibility of the method:} as was discussed in the two earlier examples, when the \ac{ML} model does not provide truly new insights into the physics of the model, the automation should instead yield a~significant reduction of human labour. This can be achieved by designing a~specific scheme for \stress{one} scenario that requires a~large expenditure of work or by designing a~flexible scheme for a~large number of scenarios of medium expenditure. 
    \item \stress{Similarity of theoretical and experimental results:} more often than not, theoretical models produce results that diverge quite significantly from their experimental counterparts. This can be due to experimental noise or limitations in the theoretical model. For a~model that has been trained on theoretical results only, it is important to evaluate whether further pre-processing, such as the inclusion of artificial noise, could be sufficient to prepare the network architecture for an~input of experimental data and/or how much the network needs to be retrained.
\end{itemize}
Unlike the work done with quantum dots and flakes that utilized \stress{specific schemes with high impact}, the scheme proposed now is very general and takes advantage of the \stress{flexibility} of the probed system: ultracold atoms. Due to the high level of control available in such setups, ultracold atoms represent an~exemplary quantum simulator for a~large variety of few to many-body physics phenomena. It is worth noting that regardless of whether the considered experimental effect is a~dynamic transition from superfluid to Mott-insulating states, the quantization of conductance through a~quantum point contact, or simply the many-body nature of condensed versus fragmented states in a~double-well potential, the standard output of experiments remains similar. It is namely a~time-of-flight image.

\begin{figure}[t]
    \centering
    \includegraphics[width=0.8\textwidth]{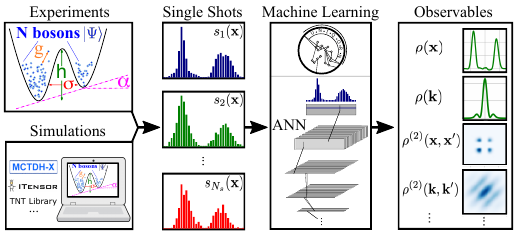}
    \caption[Machine learning observable extraction for ultracold atoms]{Implementing \ac{ML} into the extraction of observables from measurements of ultracold atoms: Conventional experimental approaches are usually only able to extract the first and second order density matrix through averaging approaches, whereas the \ac{ML} approach also extracts real space observables from momentum space measurements as well as correlation functions. Taken from \ToggleForCUP{Lode, A.~U. \textit{et al.} (2021). \textit{Optimized observable readout from single-shot images of ultracold atoms via machine learning}. Phys. Rev. A 104, L041301~\cite{lode2021optimized}.}{Ref.~\cite{lode2021optimized}.}}
    \label{fig:optimizedObservableReadout}
\end{figure}

In an~experimental setup, an~initially trapped cloud of ultracold atoms is allowed to expand, and time-of-flight imaging captures snapshots of the cloud. These single shots carry an~amount of information as they can unambiguously be linked to a~large variety of physical quantities and phases. Although experimentalists can usually only extract a~few observables through averaging techniques, it is shown that an~\ac{ML} tool should be able to exploit the information contained in the data more accurately and access a~larger selection of observables (compare \cref{fig:optimizedObservableReadout}). The \ac{ANN}-based approach proposed by the authors exploits the shot-to-shot fluctuations in order to implicitly reconstruct the many-body state.
This is promising for widespread application in experimental realizations.
\ToggleForCUP{
\begin{figure}[t]
    \centering
    \includegraphics[width=0.99\textwidth]{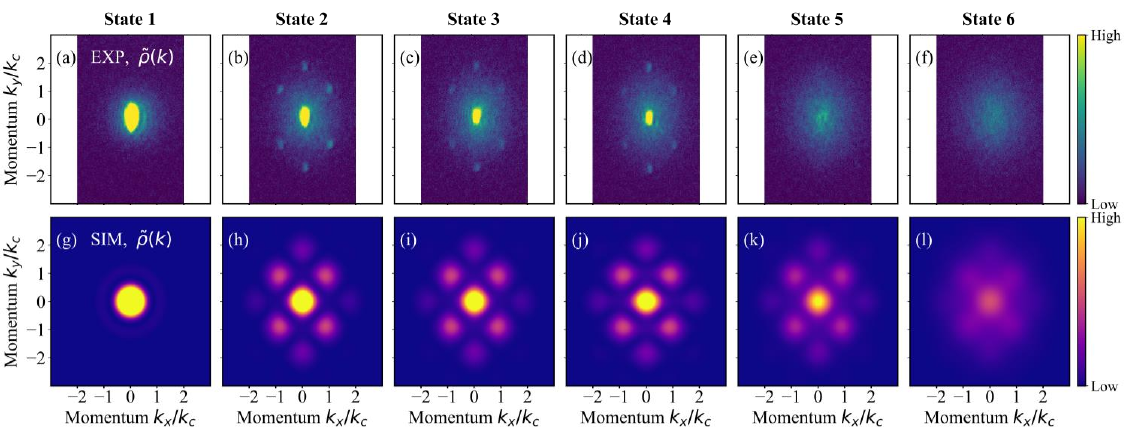}
    \caption[Simulated and experimental $p$-space density]{Comparison of experimentally measured (upper row) and simulated (bottom row) momentum space density distributions of a~system undergoing a~phase transition from a~superfluid self-organized phase (State 1) to a~superfluid Mott insulating phase (State 6). Adapted from \ToggleForCUP{Lin, R. \textit{et al.} (2021). \textit{Mott transition in a cavity-boson system: A~quantitative comparison between theory and experiment}. SciPost Phys. 11, 030~\cite{lin2021mott} under the \href{https://creativecommons.org/licenses/by/4.0/}{CC BY 4.0 DEED} license.}{Ref.~\cite{lin2021mott}.}}
    \label{fig:TheoryvsExperiment}
\end{figure}}{
\begin{figure}[b!]
    \centering
    \includegraphics[width=0.99\textwidth]{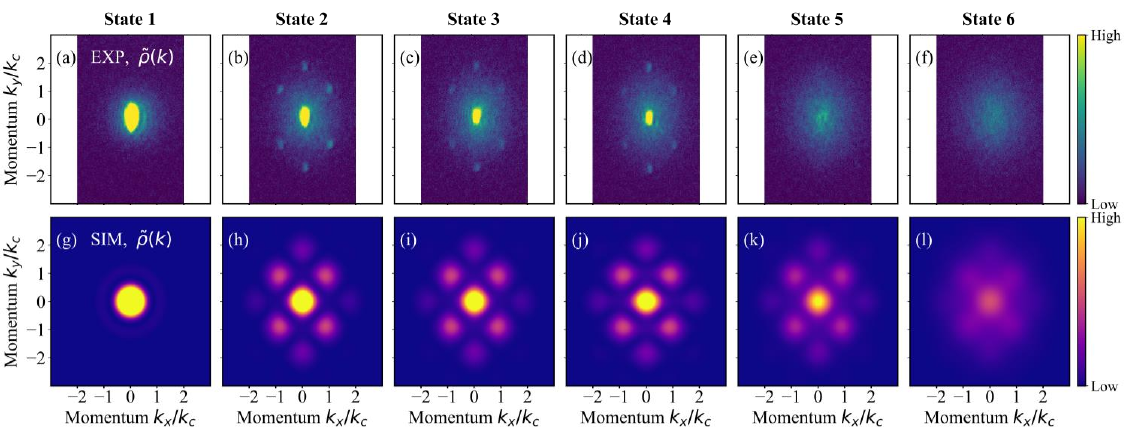}
    \caption[Simulated and experimental $p$-space density]{Comparison of experimentally measured (upper row) and simulated (bottom row) momentum space density distributions of a~system undergoing a~phase transition from a~superfluid self-organized phase (State 1) to a~superfluid Mott insulating phase (State 6). Adapted from \ToggleForCUP{Lin, R. \textit{et al.} (2021). \textit{Mott transition in a cavity-boson system: A~quantitative comparison between theory and experiment}. SciPost Phys. 11, 030~\cite{lin2021mott} under the \href{https://creativecommons.org/licenses/by/4.0/}{CC BY 4.0 DEED} license.}{Ref.~\cite{lin2021mott}.}}
    \label{fig:TheoryvsExperiment}
\end{figure}
}

When it comes to comparing theoretical predictions and experimental results, noise becomes an~important factor. In ultracold atoms, Lode \textit{et al.} (\cref{fig:optimizedObservableReadout}) proposed a~method for an~optimized observable readout from single-shot images of ultracold atoms, arguing that the similarity of theoretically simulated and experimentally detected single-shots is good enough that the addition of artificial Gaussian noise to the theoretical data during training should suffice. 
\Cref{fig:TheoryvsExperiment} shows a~comparison of simulated and experimental single shots of ultracold atoms at different points of a~phase transition in an~optical cavity upon increase of one of the external laser intensities. Although this example comes from a~different framework, both publications used the same simulation method for the single-shot generation \cite{lin2021mott, lin2020mctdh}.
Noise is evidently present, but the agreement is satisfactory for the different stages of the phase transition.
An~alternative to adding artificial noise to theoretical data is attempting to subtract noise from experimental data, e.g., by means of denoising autoencoders. The option chosen eventually naturally depends on the given experimental and theoretical data. In the case of single-shot images, denoising methods may not be ideal owing to the presence of quantum noise, inherent in many-body systems, which is difficult to discern from other noise sources and, therefore, selectively remove.

Overall, we have seen that \ac{ML} techniques can help bridge the gaps between theoretical models and noisy or resource-constrained experimental realizations and measurements.
These findings represent a~solid groundwork demonstrating experimental quantum physics enhancement via \ac{ML} and indicate a~promising avenue toward the hybridization of \ac{ML} and the quantum realm in the coming years.

\subsubsection{Hamiltonian learning\index{Hamiltonian learning}}\label{sss:Hamiltonian-learning}

    The focus of this section is the verification of quantum simulators such as trapped ions, Rydberg atoms, superconducting qubits, or ultracold atoms in optical lattices \cite{zhang2017observation, bernien2017probing, chiaro2019direct, rispoli2019quantum}.\footnote{Different experimental setups have different advantages and disadvantages for specific quantum simulation problems and Hamiltonians. A~difference between quantum simulation and quantum computation is that quantum simulators are engineered for specific problems, and quantum computing is more versatile and capable of solving general problems.}  These experimental setups are well understood and can be used to simulate more complex and challenging systems governed by the same Hamiltonians. We enter exciting times when quantum simulators start to be very complex and, in particular, not solvable with classical computers. For example, when working with quantum simulators with 50 qubits, we have to deal with enormous Hilbert spaces of the order of $10^{15}$. Therefore, how can we know that these simulators are working as they should be if we cannot verify their results with classical computers? One possible solution to this problem is called \stress{Hamiltonian learning} which is the main topic of this section. In particular, we discuss here the approach presented in Ref.~\cite{Valenti2021}.
    
    \highlight{The main idea of Hamiltonian learning is to reconstruct the map from experimentally accessible measurements to the parameters of the underlying Hamiltonian.} 
    The approach discussed in this section employs \acp{NN} to extract parameters governing the created quantum simulator. An exemplary procedure is as follows. We conduct numerical simulations and generate experimentally accessible data (e.g., real-space images) for the corresponding Hamiltonian whose parameters are known. Then, the \acp{NN} are trained via supervised learning\index{supervised learning} to predict the parameters of these Hamiltonians. Then they can be tested on measurements generated with experimental quantum simulators, where the underlying Hamiltonian is not fully known.
    It is also possible to reverse the procedure: given the defining parameters of the Hamiltonian of a~quantum system, relevant characteristics of the system can be efficiently learned by an~\ac{NN}~\cite{Gresch2022}.
    
\begin{figure}[t]
        \begin{center}
        \includegraphics[width=0.8\columnwidth]{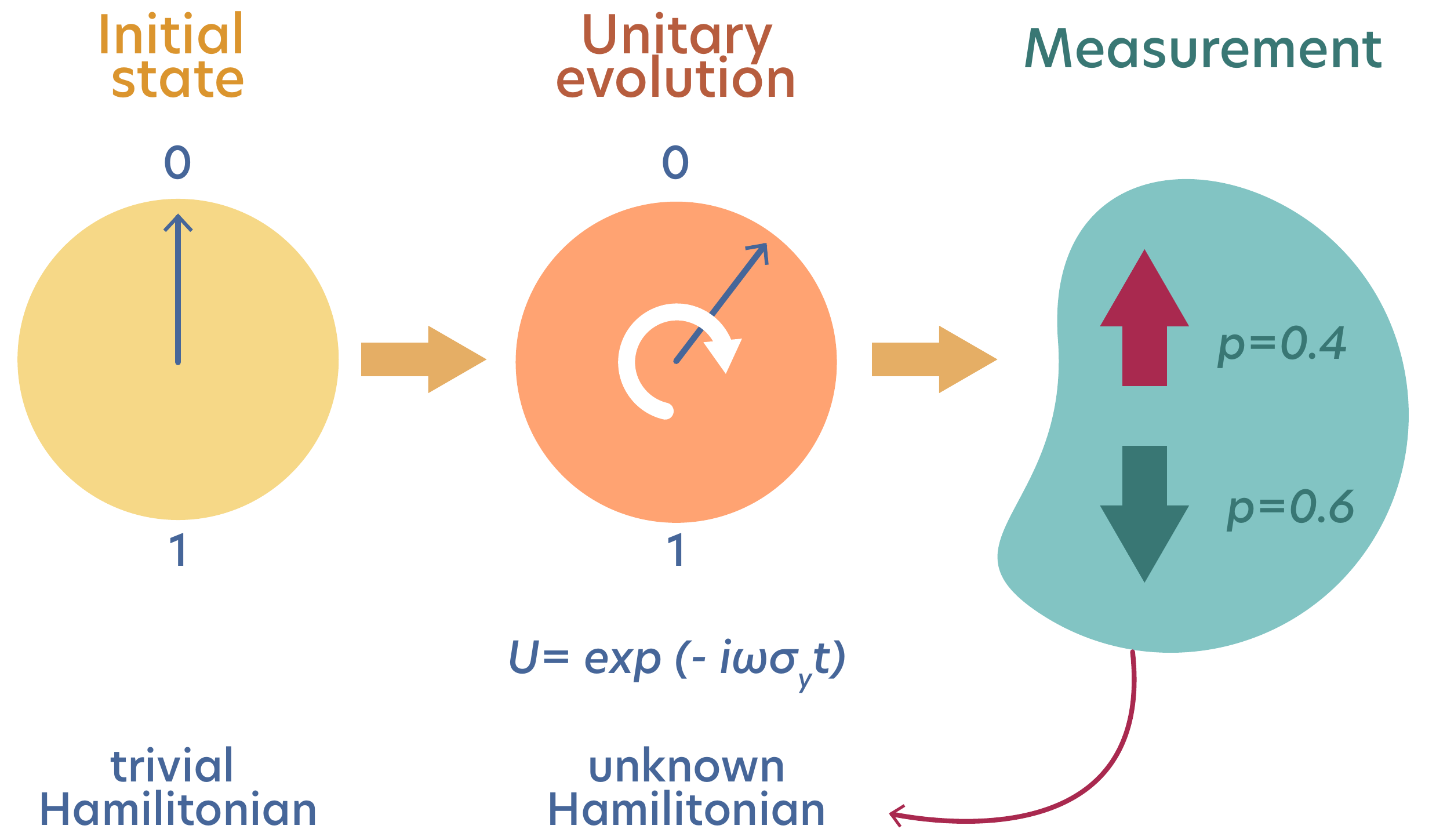}
        \end{center}
        \caption[Illustration of the Hamiltonian learning of a~one-spin system]{Illustration of the procedure of Hamiltonian learning of a~one-spin system. The spin is prepared in an~initial state and driven by an~unknown Hamiltonian for some time $t_{\mathrm{m}}$, after which the system is measured. The process is repeated multiple times (potentially for different times $t_{\mathrm{m}}$) to collect the dataset of measurements, from which the rotation frequency can be learned.}
        \label{fig:HL_spin}
    \end{figure}

    A~very simple example to illustrate the process of Hamiltonian learning with \acp{NN} is a~single spin system as shown in \cref{fig:HL_spin}. Firstly, we prepare an~initial state of a~known Hamiltonian, $\hat{H}_0$. In this case, it is an~eigenstate of $\sigma_z$ (spin ``up''). Secondly, we perform a~unitary evolution under an~unknown Hamiltonian, $\hat{H}_1$, which leads to a~precession of the spin around the axis of the Bloch sphere. We let the system evolve for some time $t_{\mathrm{m}}$, after which we measure it. The process of preparing the initial state and letting the system evolve is repeated multiple times (potentially for different times $t_{\mathrm{m}}$) to collect the dataset of measurements. We now want to learn from these measurements the unknown $\hat{H}_1$, i.e., how fast the spin precesses around the sphere. In this case, a~sequence of measurements is required to obtain the oscillation frequency, $\omega$. This procedure can be generalized to arbitrary known initial $\hat{H}_0$ and unknown $\hat{H}_1$, driving the unitary evolution of the system.
    
\begin{figure}[t]
        \begin{center}
        \includegraphics[width=0.95\columnwidth]{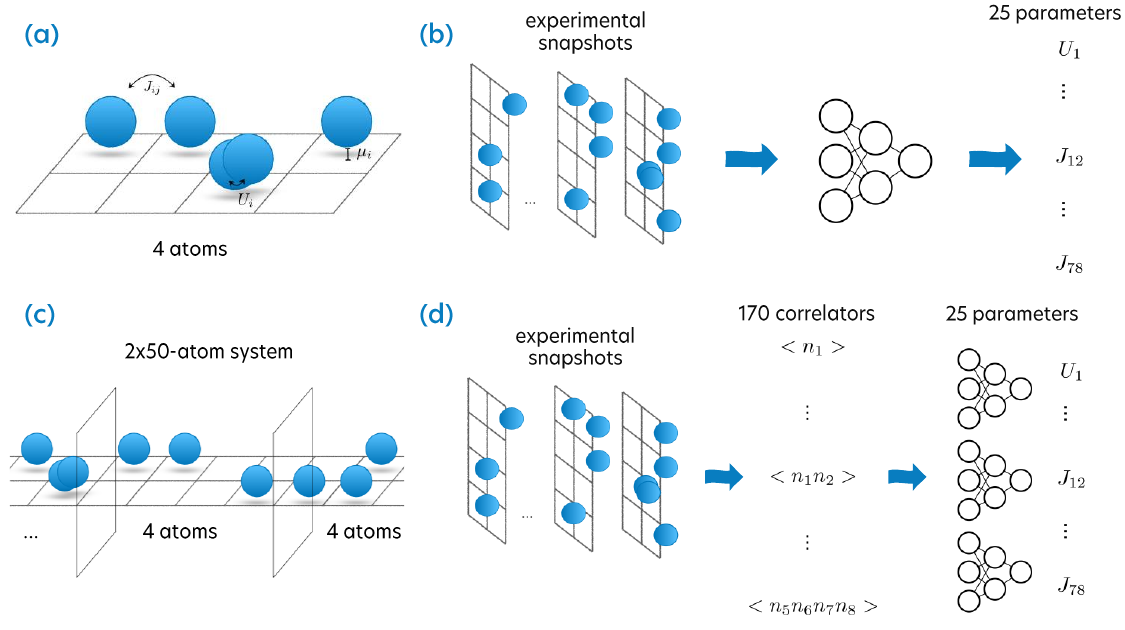}
        \end{center}
        \caption[Hamiltonian learning for larger systems]{(a) Illustration of a~four atom system with 25 parameters. (b) Mapping from the experimental snapshots to the Hamiltonian parameters using one \ac{NN} and supervised learning. (c) Extension to larger system sizes by ``dividing'' the lattice into subsystems with walls. (d) A more efficient way to map measurements to the parameters. Adapted from \ToggleForCUP{Valenti, A. \textit{et al.} (2022). \textit{Scalable Hamiltonian learning for large-scale out-of-equilibrium quantum dynamics}. Phys. Rev. A 105, 023302~\cite{Valenti2021}.}{Ref.~\cite{Valenti2021}.} Additional credit: \href{https://www.eliskagreplova.com/qmai-team}{QMAI} group at \href{https://www.tudelft.nl/}{TU Delft}.}
        \label{fig:HL_NN}
    \end{figure}

    Now, we focus on another experimental setup of a~quantum simulator consisting of neutral atoms in a~harmonic potential in a~system of $2\times 50$ lattice sites. The initial states of this system are the positions of the atoms in the optical lattice, and this experimental setup can be described by the Bose-Hubbard Hamiltonian   
    \begin{equation} \label{eq:HL_BH}
        \hat{H}_{\mathrm{BH}} = - \sum_{\langle i,j \rangle} J_{ij} \hat{a}^\dagger_i \hat{a}_j + \sum_i \frac{U_i}{2} \hat{a}^\dagger_i \hat{a}_i (\hat{a}^\dagger_i \hat{a}_i - 1) - \sum_i \mu_i \hat{a}^\dagger_i \hat{a}_i
    \end{equation}
    where $J_{ij}$ describes the hopping between lattice sites $i$ and $j$, $U_i$ - the onsite energies, and $\mu_i$ is the chemical potential of the atoms in the optical lattice. If we consider only ten particles in this lattice, the corresponding Hilbert space is of dimension $10^{13}$ with 350 parameters to estimate. This leads to two main issues: first, the wave function is too large, making it impossible to simulate this system, and second, it leads to a~350-dimensional optimization problem. Therefore, let us first consider a~small system consisting of 4 atoms, as illustrated in \cref{fig:HL_NN}(a), which reduces the number of parameters to 25 and the Hilbert space size to 330. This eliminates the problem of the large Hilbert space and leaves us with the optimization problem.
    
    Now, we want to create a~mapping from the measurements to parameters of the Hamiltonian in~\cref{eq:HL_BH}.\footnote{Here, exact simulations were used as ``measurements'' instead of experimental snapshots, and the input images are the real space positions of the atoms in the optical lattice.} To do so, supervised learning is used to train an~\ac{NN} and perform regression\index{regression}. The challenge in this setup is the scaling of the training data with the output size. To train a~single \ac{NN} to predict all parameters, as shown in \cref{fig:HL_NN} (b), several examples are required for all combinations of the 25 parameters, which is unfeasible for most applications due to the enormous size of the required training set. The solution to this problem is quite simple: instead of using a~single \ac{NN} to predict all parameters, 25 \acp{NN} are trained to predict each parameter separately with continuous regression. 
    
    Moreover, the experimental snapshots may not be the best representation of the data set. A~more effective representation is to switch from experimental snapshot batches to the correlators of the specific Hamiltonian. In this example, density correlators are used, which enable a~way more efficient way to train the \ac{NN} by reducing the input dimension. This approach is shown in \cref{fig:HL_NN}(d). After successful training, the \ac{NN} achieves around 0.1$\%$ error rates for experimental parameters with 2500 snapshots. Using Bayesian inference as a~benchmark, the \ac{NN} approach outperforms the Bayesian results for small data sets of 2500 snapshots. However, for large data sets of about 20 000 samples, both approaches achieve the same accuracy in the predictions of the parameters.

    \ToggleForCUP{\begin{figure}[t]
    \begin{center}
    \includegraphics[width=0.7\columnwidth]{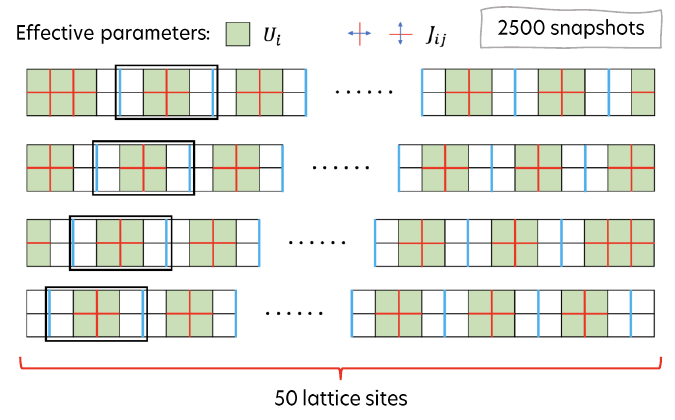}
    \end{center}
    \caption[Scaling scheme from four lattice sites to 50 for a~specific Hamiltonian]{Scaling scheme from four lattice sites to 50 for this specific system and Hamiltonian. For each of the four wall configurations, 2500 snapshots were taken in order to train the \ac{NN} and learn all parameters. Adapted from \ToggleForCUP{Valenti, A. \textit{et al.} (2022). \textit{Scalable Hamiltonian learning for large-scale out-of-equilibrium quantum dynamics}. Phys. Rev. A 105, 023302~\cite{Valenti2021}.}{Ref.~\cite{Valenti2021}.}}
    \label{fig:HL_2}
    \end{figure}}{}
    
    So far, we have only considered small system sizes of four atoms, which can be solved with classical computers. In the following, we present a~scheme to scale to larger system sizes of this specific Hamiltonian. In this experimental setup, it is possible to modulate the lattice and create walls in order to separate the chain of 50 lattice sites into four-site units (see \cref{fig:HL_NN}(c)). In this system, the Hamiltonian parameters are local, and only the terms of $\hat{H}$ that are unaffected by the boundary have to be learned. They are called the ``effective parameters'' (see \cref{fig:HL_2}). Now, the boundary is shifted by one lattice site at a~time, and 2500 shots are measured for each position. Once the system is shifted up to the point of translational invariance, all parameters were at least in one configuration unaffected by the boundary wall and were successfully learned by the \ac{NN}. 

    As mentioned above, this procedure is very specific to this system and cannot easily be generalized to different systems. The field of Hamiltonian learning is still in its early stages, and general schemes for large systems and complex Hamiltonians have yet to be developed \cite{gebhart:2022}. However, it is a promising approach for the important task of validating if quantum simulators work correctly, which becomes increasingly important with the increasing size and applicability of these simulators, which might have the possibility to go beyond classical computation.

    \ToggleForCUP{}{\begin{figure}[t]
    \begin{center}
    \includegraphics[width=0.7\columnwidth]{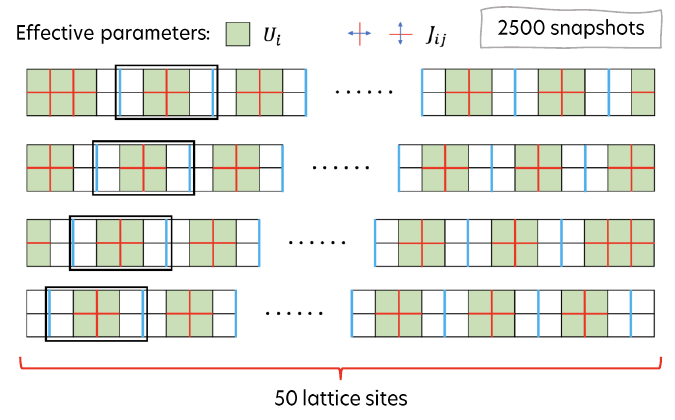}
    \end{center}
    \caption[Scaling scheme from four lattice sites to 50 for a~specific Hamiltonian]{Scaling scheme from four lattice sites to 50 for this specific system and Hamiltonian. For each of the four wall configurations, 2500 snapshots were taken in order to train the \ac{NN} and learn all parameters. Adapted from \ToggleForCUP{Valenti, A. \textit{et al.} (2022). \textit{Scalable Hamiltonian learning for large-scale out-of-equilibrium quantum dynamics}. Phys. Rev. A 105, 023302~\cite{Valenti2021}.}{Ref.~\cite{Valenti2021}.}}
    \label{fig:HL_2}
    \end{figure}}

\subsubsection{Automated design of experiments}\label{sss:MLforexpdesign}

Amongst the proposed \ac{ML} applications for experiments, we have already discussed how \acp{NN} can be used to speed up, optimize, and verify the setups, as well as to analyze the generated data. One further application is the \ac{AI}-guided design of experiments that may one day arguably revolutionize science.

When it comes to designing new experiments, most of the efforts have focused so far on quantum optics \cite{Krenn2016PRL, melnikov2018active, Krenn2020NatRevPhys, Krenn2021PRX, CerveraLierta2021, FlamShepherd2021}. The design of such an experiment consists of combining different optical laboratory components, for example, beam splitters, mirrors, and crystals, so that the final quantum state has specific desired properties. For example, we may be interested in obtaining a quantum state with a high-dimensional multipartite entanglement (that is, between multiple particles), which is of great importance in applications of quantum information and computation \cite{Erhard2020NatRevPhys}. While a trained physicist can design an experimental setup to create a quantum state with non-trivial properties, this task can be very challenging and heavily relies on trial and error. 

In \cref{sec:rl_qexperiments}, we have already presented an example~\cite{melnikov2018active} of an~autonomous approach to building quantum-optical experiments with \acf{RL}, using the \acf{PS} algorithm that we introduced in \cref{sec:rl_projective_simulation}. Interestingly, there is another \ac{AI}-guided approach for designing optical experiments that has already allowed for a dozen new experiments in several laboratories around the world \cite{Krenn2020NatRevPhys}. The proposed algorithm is called \textsc{melvin} \cite{Krenn2016PRL} and is presented in \cref{fig:AI-design-exp}. 

\ToggleForCUP{
    \begin{figure}[t]
    \begin{center}
    \includegraphics[width=0.99\textwidth]{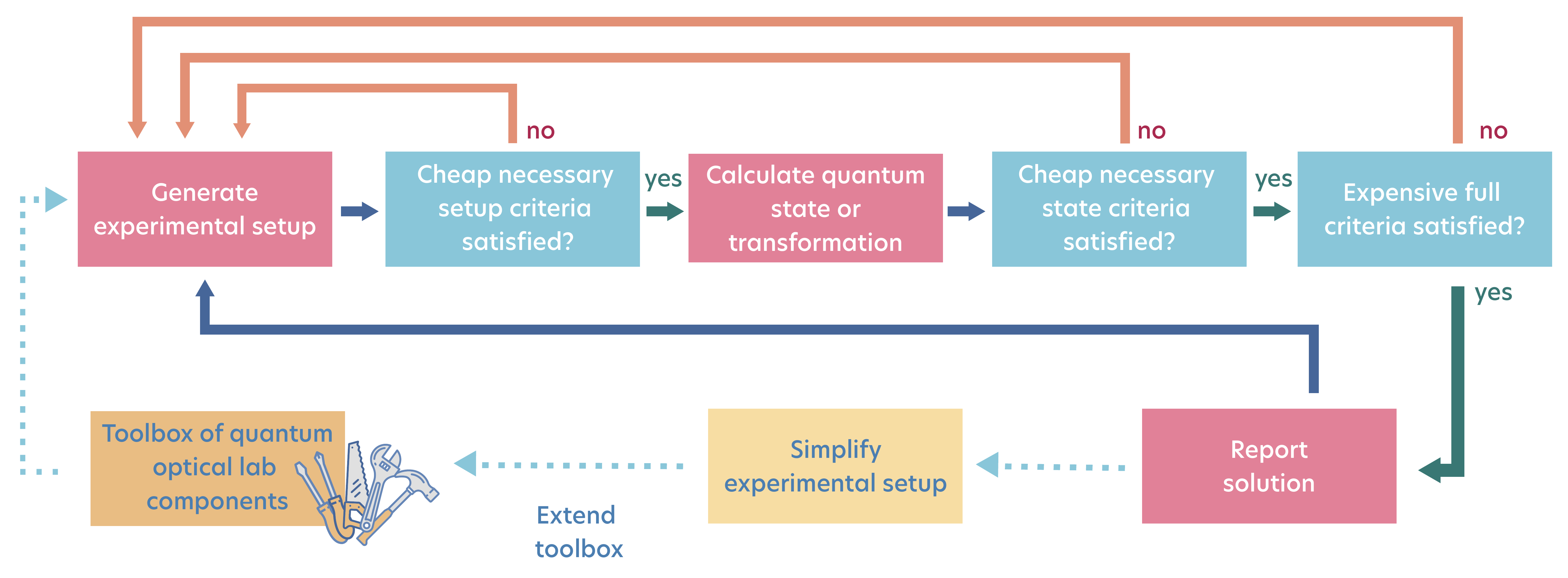}
    \end{center}
    \caption[Algorithm for computer-inspired experiments]{Example of an algorithm for computer-inspired quantum optical experiments called \textsc{melvin}. Adapted from \ToggleForCUP{Krenn, M., Erhard, M., \& Zeilinger, A. (2020). \textit{Computer-inspired quantum experiments}. Nat. Rev. Phys. 2, 649~\cite{Krenn2020NatRevPhys} with permission from Springer Nature.}{Ref.~\cite{Krenn2020NatRevPhys}.}}
    \label{fig:AI-design-exp}
    \end{figure}
}{}

To apply \textsc{melvin}, the user needs to specify a toolbox, that is, a set of available optical lab components. Moreover, the user defines the target properties and all possible conditions that characterize a final quantum state. The \textsc{melvin} algorithm first generates an experimental setup by randomly arranging the available optical components. Each optical component is a known symbolic modification of the input state. Then, the resulting quantum state and its properties are computed, as we know the initial quantum state and the symbolic transformations applied to it. If the quantum state meets all the criteria and exhibits a target property, then \textsc{melvin} reports the setup to the user.\footnote{Before reporting the solution, optionally, the setup is simplified using deterministic methods predefined by a user. For example, they may include iterative removal of a random optical component and check whether it changes the final quantum state.} More often, the generated quantum state does not match the target one, so \textsc{melvin} starts again by generating another setup. Therefore, \textsc{melvin} is heavily based on random search.

However, there are two characteristics of \textsc{melvin} that grant a significant speed-up compared to a fully random search. First, the user can divide the required criteria into cheap and expensive ones, as presented in \cref{fig:AI-design-exp}. The expensive criteria are then calculated only if the cheap ones are met first. Second, \textsc{melvin} is allowed to expand its initial toolbox by adding already tested configurations and use them as basic building elements in subsequent trials. This expansion of the available tools can be thought of as a learning component of \textsc{melvin}.

\ToggleForCUP{}{
    \begin{figure}[t]
    \begin{center}
    \includegraphics[width=0.99\textwidth]{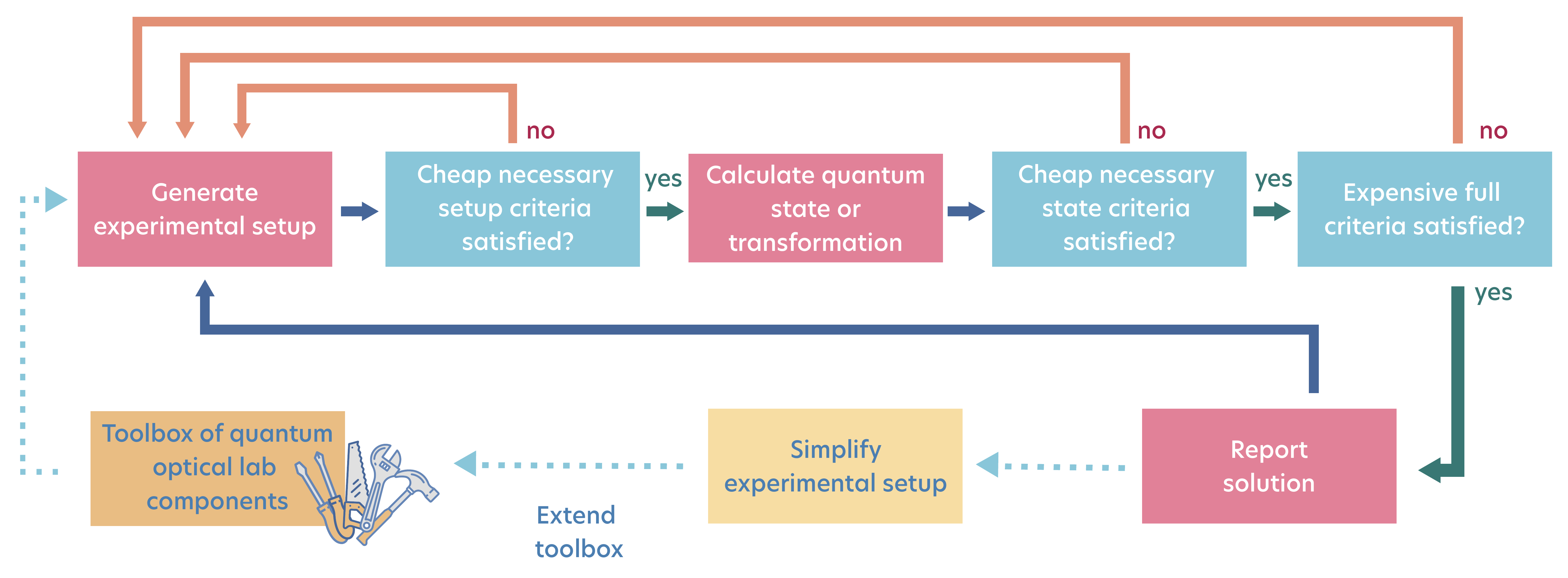}
    \end{center}
    \caption[Algorithm for computer-inspired experiments]{Example of an algorithm for computer-inspired quantum-optical experiments called \textsc{melvin}. Adapted from \ToggleForCUP{Krenn, M., Erhard, M., \& Zeilinger, A. (2020). \textit{Computer-inspired quantum experiments}. Nat. Rev. Phys. 2, 649~\cite{Krenn2020NatRevPhys} with permission from Springer Nature.}{Ref.~\cite{Krenn2020NatRevPhys}.}}
    \label{fig:AI-design-exp}
    \end{figure}
}

For example, \textsc{melvin} has been used to find experimental setups generating high-dimensional multipartite entangled states, as mentioned above. In Ref.~\cite{Krenn2016PRL}, \textsc{melvin} identified setups that lead to states entangled in different ways. In particular, it found the first experimentally realizable scheme leading to a so-called high-dimensional Greenberger-Horne-Zeilinger state \cite{Krenn2016PRL}. Moreover, as the authors of Ref.~\cite{Krenn2016PRL} admit, the resulting experiments contained interesting novel experimental techniques previously unknown to them. 

Finally, studying \textsc{melvin} showed that each optical setup and initial state can be represented as weighted graphs. The successor of \textsc{melvin}, called \textsc{theseus} \cite{Krenn2021PRX}, takes advantage of a graph representation that allows replacing random search with a gradient-based search for optimal weights. Not only does \textsc{theseus} outperform \textsc{melvin} in terms of discovery speed by a few orders of magnitude, but it also provides interpretable\index{interpretability} solutions as long as the graphs representing the discovered experimental setups are small enough.

\subsubsection*{Outlook and open problems}

To conclude \cref{sec:ML_for_exp}, we can use \ac{ML} to speed up, optimize, validate, and design experiments, as well as analyze the collected data. Proposals to apply \ac{ML} to speed up and optimize experimental work date back to 2009 \cite{King2009robot}, which may be why such applications pose one of the most widely accepted roles for \ac{ML} in experimental physics. A fascinating direction is the so-called self-driving labs \cite{Hase2019selfdriving}, which combine automated experimentation platforms with \ac{AI} methods to enable autonomous experimentation. They promise an accelerated discovery rate and the liberation of experimentalists from tedious tasks.

When it comes to modern quantum technologies, the central challenges are the efficient characterization of quantum systems, the verification of quantum devices, and the validation of the underpinning physical models. \ac{ML} is expected to improve the computational cost of these tasks. As a~result, \ac{ML}-based Hamiltonian learning is becoming a~widely used technique to verify quantum experiments. Interesting examples are its application to nitrogen-vacancy center setups \cite{Gentile2021NatPhys} and to nuclear magnetic resonance measurements \cite{OBrien2021}.

Scaling of \ac{ML} approaches to larger sizes of quantum devices remains an~important challenge. Although the \ac{ML} algorithms perform exceptionally well on large experimental data sets, adding more qubits (and, therefore, tuning parameters) generates learning difficulties. These problems are especially daunting in quantum dot systems. There are efforts toward tuning multiple parameters at once~\cite{van_Esbroeck_2020,Severin_2021} or toward reducing the amount of experimental data needed for tuning~\cite{Zwolak_2021}. However, efficient tuning of large-scale quantum devices with hundreds of parameters requires new methods.

Moreover, \ac{AI} promises breakthroughs when it comes to designing novel experiments. In particular, \ac{AI} is argued to provide out-of-the-box solutions when unaware of existing human approaches \cite{Krenn2016PRL,Krenn2020NatRevPhys,Dawid2023automated}. 
So far, \ac{AI}-guided design has been explored mainly in quantum optics with significant successes. However, the discussed approaches (\textsc{melvin} and \textsc{theseus}) are readily extendable only to experiments where we can calculate how each modification in the setup influences the generated quantum system and its desired properties. Applying \textsc{melvin} or \textsc{theseus} to experiments with very expensive (or nonexistent) theoretical descriptions requires novel ideas. Another example of \ac{AI}-guided discovery of experimental setups is the use of graph-based search to automatically identify laser cooling schemes for molecules based on spectroscopic data \cite{Dawid2023automated}, which promises breakthroughs in ultracold chemistry and physics by extending the range of available ultracold species. Automatic search is again possible due to the well-understood physics underlying laser cooling.
Finally, it is inspiring to think about combining the proposal of self-driving labs with \ac{AI} designing novel experiments, which would create an ultimate robot scientist who never tires and never stops looking for new solutions and discoveries. 
    
\subsubsection*{Further reading}
\begin{itemize}
    \item King, R. D. \textit{et al.} (2009). \href{https://doi.org/10.1126/science.1165620}{\textit{The automation of science}}. Science, 324, 85-89. Report on the building of one of the first ``robot scientists'' named ``Adam'' aiming at automating hypothesis formation and recording of experiments \cite{King2009robot}.
    \item Wiebe, N. \textit{et al.} (2014). \href{https://journals.aps.org/prl/abstract/10.1103/PhysRevLett.112.190501}{\textit{Hamiltonian learning and certification using quantum resources}}. Phys. Rev. Lett. \textbf{112}, 190501. The first proposal of Hamiltonian learning that combined quantum simulators and Bayesian inference \cite{wiebe_quantum_2014}.
    \item Raccuglia, P. \textit{et al.} (2016). \href{https://doi.org/10.1038/nature17439}{\textit{Machine-learning-assisted materials discovery using failed experiments}}. Nature 533, 73–76. Example of \ac{ML} use for discovery of materials that outperforms traditional human approaches \cite{Raccuglia2016Nature}.
    \item H\"{a}se, F., Roch, L. M., \& Aspuru-Guzik, A. (2019). \href{https://doi.org/10.1016/j.trechm.2019.02.007}{\textit{Next-generation experimentation with self-driving laboratories}}. Trends Chem. 1, 282-291. Perspective on self-driving laboratories and their role in scientific discovery \cite{Hase2019selfdriving}.
\end{itemize}

\clearpage
\section{Physics for deep learning}
\label{sec:hot-topics-physics-ml}

\begin{figure}[h]
\begin{center}
\includegraphics[width=0.4\columnwidth]{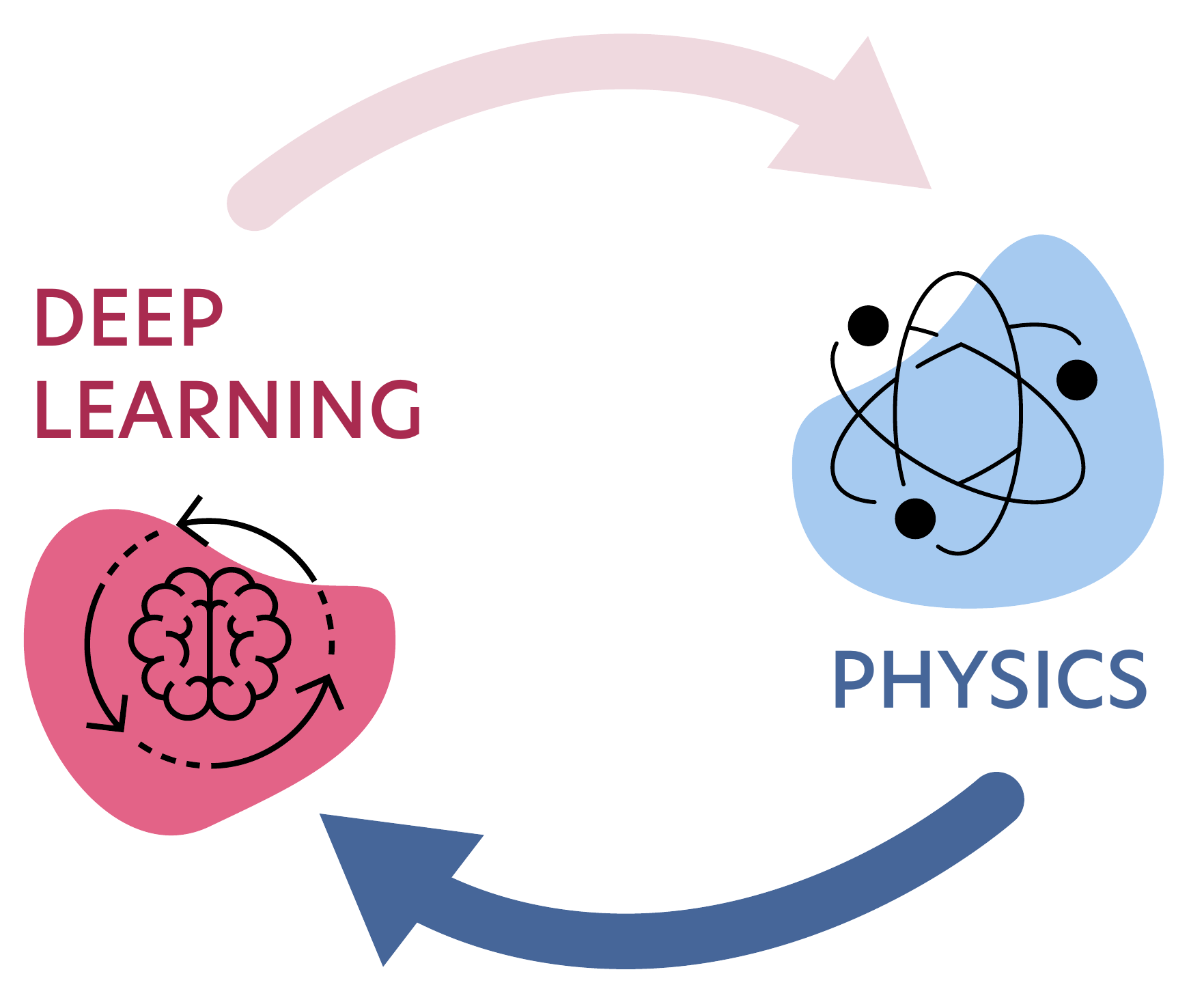}
\end{center}
\caption[Physics influences machine learning]{There exists a~two-way influence between \acf{ML} and physics. In this chapter, we focus on the less known approach, i.e., physics for \ac{ML}.}
\label{fig:physics-ml}
\end{figure}

So far, we have discussed different applications of \ac{ML} which aim at solving various problems in quantum science. In contrast, in this chapter, we focus on how physics (in particular statistical and quantum physics) influences \ac{ML} research (as shown in \cref{fig:physics-ml}). In \cref{sec:stat_phys_for_ML}, we explain the fundamental theoretical challenges of \ac{ML} and show how tools of statistical physics can shed some light on these problems. In \cref{s:QML} we discuss quantum computing and promises of \acf{QML}.

\subsection{Statistical physics for machine learning}\label{sec:stat_phys_for_ML}

\ToggleForCUP{}{
\begin{figure}[b]
\begin{center}
\includegraphics[width=0.65\columnwidth]{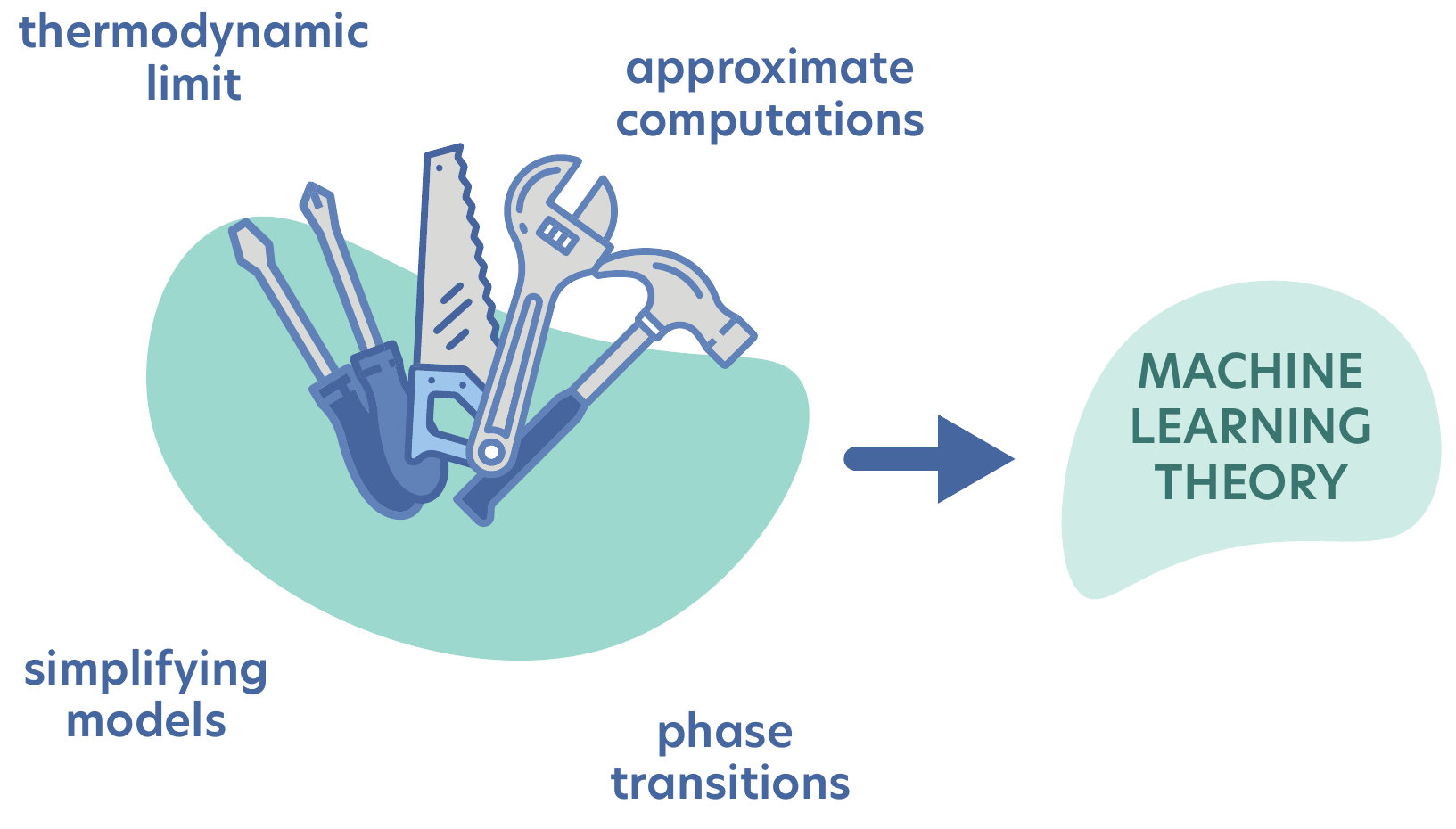}
\end{center}
\caption[Statistical physics toolbox for understanding machine learning theory]{Statistical physics toolbox for understanding the \ac{ML} theory.}
\label{fig:Marylou_toolbox}
\end{figure}}

In this section, we present how to apply concepts from physics (in particular, tools of statistical physics like the thermodynamic limit or order parameters describing phase transitions) to develop a~theory of \ac{ML} (see~\cref{fig:Marylou_toolbox}) \cite{Zdeborova2020}. This idea was born already in the 1980s, but the \ac{DL} revolution in the 2010s has caused a~renewed surge of interest in this approach.

Indeed, help from statistical physics is very needed, as we do not understand many conundrums in \ac{ML}! For example, modern \acp{NN} can have billions of trainable parameters.\footnote{One of the latest champions is Microsoft's GPT-3 with over 175 billion parameters.} How can we even find well-generalizing minima within such enormous, non-convex loss landscapes? Another riddle is related to the so-called bias-variance trade-off\index{bias-variance trade-off}, which we have shown in~\cref{sss:generalization_regularization} and which indicates that in the regime of high model complexity, models should heavily overfit their data sets as presented in \cref{fig:Marylou_overparametrization_generalization}(a). But in practice, we see that these gigantic overparametrized \ac{DL} models generalize very well, as seen in \cref{fig:Marylou_overparametrization_generalization}(b). So how do they escape this traditional bias-variance trade-off? A related question concerns the capacity of \ac{DL} models and the development of its useful measures. We have a~long way toward a~full understanding of these puzzles. A~way of tackling them is to study simple, solvable models, following a~traditional approach of physicists to study new systems. The results from toy problems can give us clues on how more complex models work.

\ToggleForCUP{}{
\begin{figure}[t]
\begin{center}
\includegraphics[width=\columnwidth]{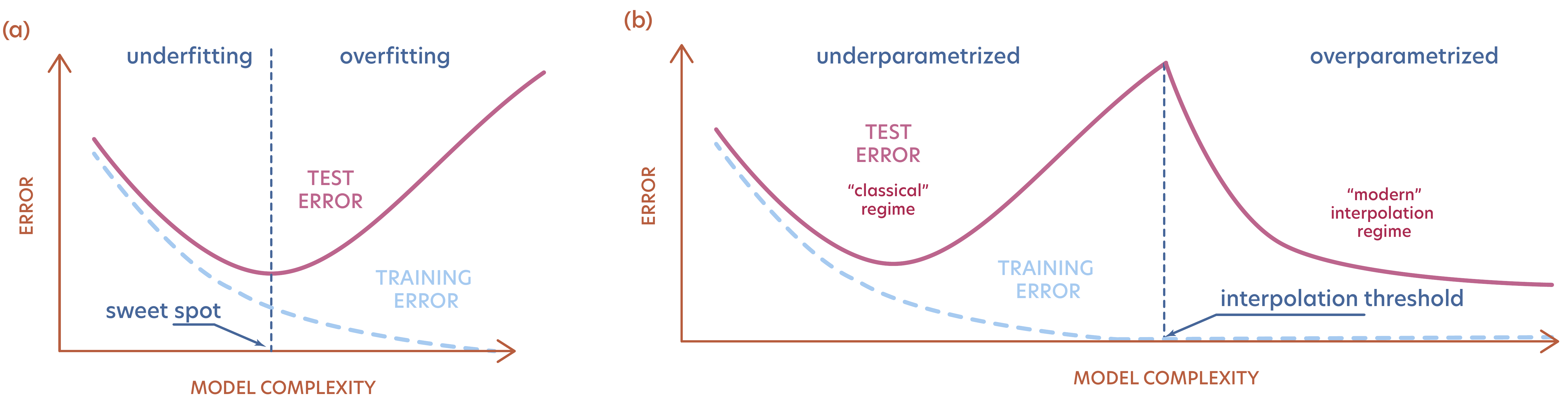}
\end{center}
\caption[Generalization error in classical and modern regimes]{Classical and modern understanding of the generalization. (a) The classical U-shaped error curve arises from the bias-variance trade-off. (b) The double descent error curve incorporates the classical U-shape in the classical regime and the low generalization error of modern overparametrized models. Adapted from \ToggleForCUP{Belkin, M. \textit{et al}. (2019). \textit{Reconciling modern machine-learning practice and the classical bias-variance trade-off}, PNAS 116, 15849-15854~\cite{Belkin2019} with publisher permission.}{Ref.~\cite{Belkin2019}.}}
\label{fig:Marylou_overparametrization_generalization}
\end{figure}}

This section has four parts. First, in~\cref{sss:Marylou-perceptron}, we go through the seminal study on the capacity of the perceptron, which gives an~idea of how statistical physics can be useful for learning problems. Then we discuss three directions of this interdisciplinary research, i.e., the teacher-student paradigm\index{teacher-student paradigm} for studying generalization in~\cref{sss:Marylou-teacher-student}, how we can model the structure of data in~\cref{sss:Marylou-data-structure} and study the dynamics of learning in~\cref{sss:Marylou-dynamics}.

\ToggleForCUP{
\begin{figure}
\begin{center}
\includegraphics[width=0.65\columnwidth]{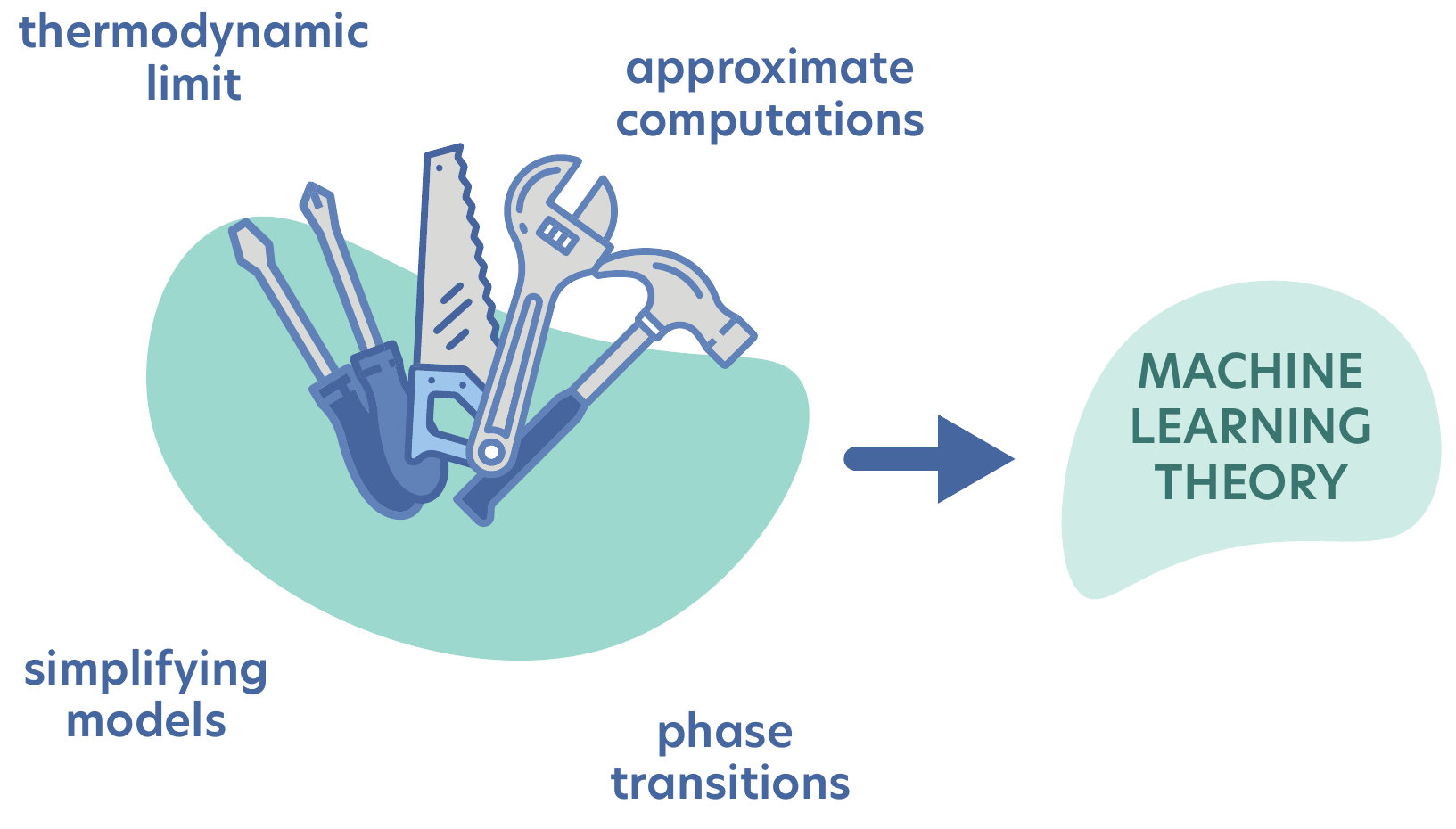}
\end{center}
\caption[Statistical physics toolbox for understanding machine learning theory]{Statistical physics toolbox for understanding the \ac{ML} theory.}
\label{fig:Marylou_toolbox}
\end{figure}}{}

\subsubsection{Capacity of the perceptron}\label{sss:Marylou-perceptron}

The simplest \ac{ML} model we can think of is a~single perceptron\index{perceptron}, $f$, already presented in~\cref{sec:NNs} (see \cref{fig:NN}(b)). In this section, we focus on its capacity\index{capacity}, i.e., the question of how many data points it can fit. To answer it, let us make the additional assumption that the data set is in \stress{general position}.\footnote{The set of points in $\mathbb{R}^{{\nparams}}$ is in general position if and only if every set of $(d+1)$ points are not in any possible hyperplane of dimension $d$. In other words, as long as there are no three data points on a~single line or four points on a~single plane, etc., the set is in general position. Intuitively, any random data set is in general position.} The assumption is reasonable – if we have many copies of the same training point, they should not contribute to the estimation of the model capacity.

A~single perceptron is only capable of learning linearly separable patterns. Therefore, we can reformulate the question of its capacity to the question of whether randomly labeled data sets of size $\datasize$ with binary labels are linearly separable. The probability of such a~linear separability, $p_R(\alpha)$, is a~function of $\alpha$, which is the ratio between the number of training points, $\datasize$, and the number of data features (or data dimensionality), $\featnum$. In the case of the perceptron, the number of features is equal to the number of perceptron weights, $\nparams$,\footnote{In general, a~perceptron is parametrized by weights $\weightsvect$ and a bias $\bias$. For the remainder of this section, we ignore biases; therefore, weights are all model parameters $\params$, of size $\nparams$.} therefore $\alpha = \frac{\datasize}{\nparams}$. In this problem, you can understand the parameter $\alpha$ as the difficulty of the classification task, which increases with the number of training points and decreases with the number of parameters.

\ToggleForCUP{
\begin{figure}
\begin{center}
\includegraphics[width=\columnwidth]{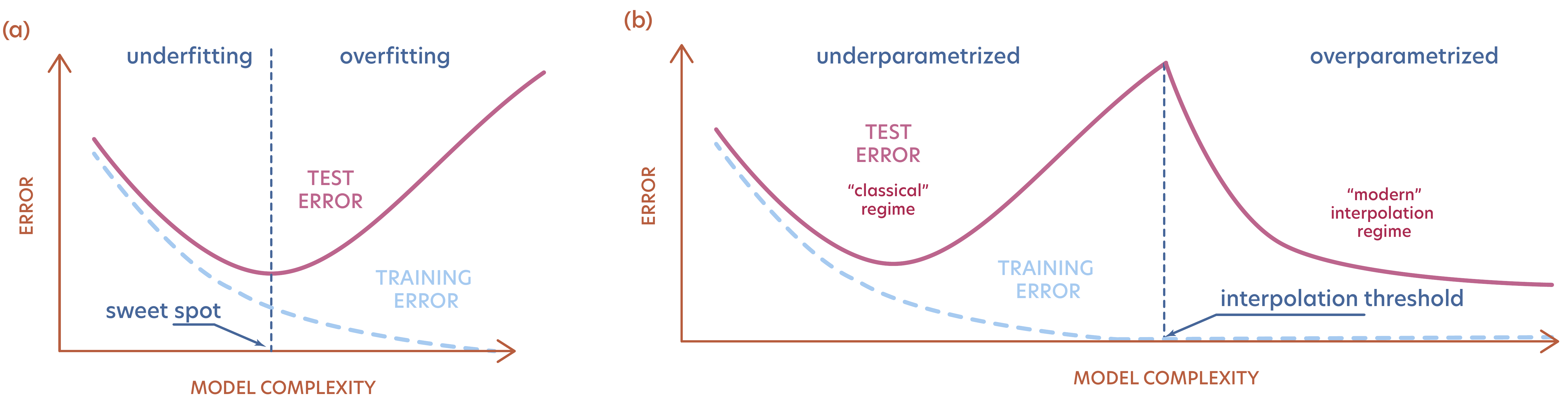}
\end{center}
\caption[Generalization error in classical and modern regimes]{Classical and modern understanding of the generalization. (a) The classical U-shaped error curve arises from the bias-variance trade-off. (b) The double descent error curve incorporates the classical U-shape in the classical regime and the low generalization error of modern overparametrized models. Adapted from \ToggleForCUP{Belkin, M. \textit{et al}. (2019). \textit{Reconciling modern machine-learning practice and the classical bias–variance trade-off}. PNAS 116, 15849-15854~\cite{Belkin2019} with publisher permission.}{Ref.~\cite{Belkin2019}.}}
\label{fig:Marylou_overparametrization_generalization}
\end{figure}}{}

To calculate $p_R$ we could resolve to geometric arguments. This approach was chosen by Thomas Cover in 1960s~ \cite{Cover1965}. However, here we choose to rephrase this problem in the language of statistical physics as was done by Elizabeth Gardner in 1987~\cite{Gardner1987}. 
\highlight{Namely, we can take the space of all possible weights, so $\mathbb{R}^{\nparams}$, and calculate the volume of those weights that fulfill all the constraints of the random labeling.} In other words, we calculate how many sets of weights could solve the problem of separating randomly labeled training data, $\dataset = \{ \vect{x}^{(k)}, y^{(k)} \}^\datasize_{k=1}$:
\begin{equation}\label{eq:volume_with_delta}
V_{\datasize,\nparams} = \int_{\mathbb{R}^{\nparams}} d\param \prod_{k=1}^\datasize \delta(f(\vect{x}^{(k)}; \params) - y^{(k)})\,.
\end{equation}
The $\delta$-function in~\cref{eq:volume_with_delta} is 1 only when the ground-truth label is equal to the label predicted by the perceptron. With each new data point $k$, we are adding a~new constraint, and the volume of possible weights shrinks. To have at least one set of such weights, the volume must be larger than zero, $V_{\datasize,\nparams} >0$. Therefore, we define the critical task difficulty\index{critical task difficulty}, $\alpha_c$, as the value of $\alpha$ for which $V_{\datasize,\nparams}$ goes down to zero. If we can calculate this, we solve the problem of the perceptron capacity\index{perceptron!perceptron capacity}.

Let us make one modification to the equation that leads us closer to statistical physics. 
We introduce an~effective Hamiltonian that counts the number of misclassified training data points,
\begin{equation}
    H(\param; \dataset) = \sum_{k=1}^\datasize \Theta(- f(\param; x^{(k)}) y^{(k)})\,,
\end{equation}
where the Heaviside function $\Theta(\cdot)$ is equal to $1$ if its argument is positive and $0$ otherwise. We can relax the Dirac $\delta$-distribution above by the Boltzmann factor of $H(\param; \dataset)$.  Up to a~multiplicative constant,  \cref{eq:volume_with_delta} becomes
\begin{equation}\label{eq:volume_with_gauss}
V_{\datasize,\nparams} \propto \lim_{\beta \rightarrow +\infty}  \beta \, \int_{\mathbb{R}^{\nparams}} d\param \, e^{-\beta \sum_{k=1}^\datasize \Theta(- f(\param; x^{(k)}) y^{(k)})} = \lim_{\beta \rightarrow +\infty}  \int_{\mathbb{R}^{\nparams}} d \param e^{-\beta H(\param; \dataset) }\,.
\end{equation}
Suddenly, the volume $V_{\datasize,\nparams}$ in~\cref{eq:volume_with_gauss} resembles the canonical \stress{partition function}\footnote{A~partition function for a~many-body classical discrete system is equal to $Z=\sum_{i} e^{-\beta \varepsilon_{i}}$, where $i$ iterates over all possible microstates and $\varepsilon_{i}$ is the energy of the $i$-th microstate. If we go to a~continuous system with $\datasize$ identical particles described by properties $\params$, the partition function is $Z \propto \int \exp \left(-\beta \sum_{i=1}^{\datasize} H\left(\params_{i}\right)\right) d \param_{1} \cdots d \param_{\datasize}$, where $H$ is a~classical Hamiltonian.} from statistical physics with $\beta$ playing the role of an~inverse temperature, defined as $\frac{1}{k_{\mathrm{B}} T}$. Therefore, the limit $\beta \rightarrow \infty$ corresponds to the zero-temperature limit. The problem is that this integral is hard to calculate as it lives in a~huge $\nparams$-dimensional space of all real numbers.\footnote{This is also a~reason why computation of any interesting partition function is hard.} Moreover, the ``effective energies'' in the exponent depend on the training set. As such, each training set requires a~separate calculation of the volume $V_{\datasize,\nparams}$.

Fortunately, the physics of \stress{disordered systems} comes to the rescue. It has been applied to learning theory since the 1980s \cite{Amit1985, Derrida1987a, Derrida1987, Peterson1987, Krauth1989, Gyorgyi1990, Opper1991}. Namely, if we recognize a~disordered system in~\cref{eq:volume_with_gauss}, we can use solutions from statistical physics to compute this high-dimensional integral. Let us give a~brief introduction to disordered systems. A~disordered system is described by two types of random variables. The first type concerns the states of the system $s \in \mathbb{R}^{\nparams}$. For example, for a~system of $d$ spins - $\frac{1}{2}$, $s \in \{-1, 1\}^d$, because each spin can be up or down. The second type concerns interactions between degrees of freedom, which can be parametrized by couplings $J \in \mathbb{R}^{\datasize}$. For example, $J$ can describe whether the spins want to align or anti-align. The distribution of states in disordered systems is then described by the Boltzmann distribution:
\begin{equation}
    p(s \mid J) = \frac{1}{Z_J} e^{-\beta H(s;\,J)}\,,
\end{equation}
where $H(s;\,J)$ is an~energy function depending on both $s$ and $J$, and $Z_J = \int_{\mathbb{R}^{\nparams}} ds e^{-\beta H(s;J)}$ is the partition function equal  and plays the role of a~normalization.

As an~example of a~disordered system, let us consider a~\stress{spin glass} \cite{Sherrington1975,Mezard1986}, where the energy function is $ H(s;\,J) = - \sum_{<i,j>} J_{ij} s_i s_j$ (resembling an Ising-type interaction, see \cref{eq:ising_H}), where couplings $J_{ij}$ are i.i.d. according to the normal distribution $p(J_{ij}) \propto 
\exp{-(J_{ij} - J_{0})/{2 J^2}}$ where $J_0$ and $J^2$ are the mean and variance. If all the $J_{ij}$ are positive, the system is ferromagnetic, and the ground state of the system is easy to find. With random couplings, complications arise along with the frustration of the system: at a~given site, a~spin can be encouraged by neighbors to point in conflicting directions. Finding the ground state of such systems is a~numerical challenge of its own. While in one dimension the solution is trivial and can be solved by a~deterministic algorithm whose cost scales as $\mathcal{O}(n)$, the complexity grows in two dimensions and reaches NP-completeness in three and more dimensions~\cite{Barahona1982spincomplexity}.\footnote{There are proposals to tackle this challenge with \acf{RL}~\cite{Fan2021RLspinglass}.}

Now, let us tackle the exponent in~\cref{eq:volume_with_gauss}, which we treat as an~energy function. If we do that, there is a~property of the free energy\footnote{In the thermodynamic limit, the free energy of the system is $F = U - TS = - \frac{1}{\beta} \ln Z$, where $U$ is the energy of the system and $S$ is its entropy.} which can help us in simplifying the calculations. Namely, free energy is self-averaging\index{self-averaging}. 
\highlight{If a~random quantity is \stress{self-averaging}, two conditions are met: its mean value and the most probable value coincide in the thermodynamic limit, and fluctuations around this mean value are sufficiently small. In other words, the system \stress{concentrates on typical states}.} 
This property often holds for the free energy of disordered systems. Consider the following argument: imagine dividing the macroscopic system into many subsystems, and each subsystem is still large enough to be considered macroscopic. Their interaction can be viewed as a~surface effect and is negligible compared to the bulk. Therefore, each subsystem has a~well-defined free energy and the realization of disorder, even if the specific values vary between subsystems. In the limit of an~infinite number of subsystems (whose interactions can be ignored to first order), the disorder average of the free energy is automatically the average free energy across the disordered subsystems~\cite{Edwards1975, Thouless1977, Mezard1986}. That is, for $\nparams$ large enough, the physics of the system is independent of the disorder realization:
\begin{equation}\label{eq:average_free_energy}
\frac{1}{\nparams} \ln{Z_J} \approx \lim_{\nparams \rightarrow \infty} \mathbb{E}_J\left[\frac{1}{\nparams} \ln{Z_J} \right] \,.
\end{equation}
With the free energy being extensive, note that the converging quantity in the thermodynamic limit is the free energy per spin. This result is highly nontrivial, and tools such as replica computations, variational mean-field methods, and high-temperature expansions are necessary to identify where self-averaging applies and to compute the disorder averages.\footnote{It is interesting to note  that these non-rigorous physical approaches for disordered systems developed in the 1970s \cite{Edwards1975, Thouless1977, Mezard1986} are now being put on a~more rigorous footing by mathematicians \cite{Talagrand2006, Panchenko2014}!} 
In the following paragraph, we provide the intuition behind only one of the concepts behind \cref{eq:average_free_energy}, namely \stress{the replica trick}. Readers interested in more detailed explanations should refer to the tutorial reviews \cite{Castellani2005spinglass, Gabrie2020meanfield}.

\paragraph{Replica trick.}\index{replica trick}
In statistical physics, calculating averages makes sense only for extensive observables.
The replica method is a~way to calculate these averages with respect to disorder variables. We are particularly interested in the averaged value of the system free energy $F_J =-\frac{1}{\beta}\ln Z_J$. To obtain the averaged free energy $\mathbb{E}_J\left[{F_J}\right]$ we have to obtain the averaged value of the logarithm of the partition function $\mathbb{E}_J\left[{\ln Z_J}\right]$. 
It turns out that averaging the logarithm is challenging, but the averages of powers of the partition function, $\mathbb{E}_J[Z^n]$ for $n\in\mathbb{N}$, can be estimated.
Then, by using the identity,
\begin{equation}
    \ln x = \lim_{n\to0} \frac{x^n-1}{n}\,,
\end{equation}
we can write
\begin{equation}\label{eq:replica}
    \mathbb{E}_J[\ln Z] = \lim_{n\to0} \frac{\mathbb{E}_J[Z^n]-1}{n}\,.
\end{equation}
As we can see, the limit $n\to0$ requires $n\in \mathbb{R}$. However, what we can do is to calculate $Z^n$ for $n\in\mathbb{N}$.
The partition function $Z$ is an~integral of the form $\int e^{-\beta H(s,J)}$, thus we can write $Z^n$ as 
\begin{equation}
    Z^n = \int ds^{(1)}\dots ds^{(n)} \prod_{a = 1}^{n} e^{-\beta H(s^{a}, J))} = \int ds^{(1)}\dots ds^{(n)} e^{-\beta \sum_{a=1}^{n} H(s^{a}, J))},
\end{equation}
where the exponent contains a~sum over $n$ independent samples, or replicas.
The replica trick consists in defining a~function $\phi(n)$ being an~analytic continuation of the function in the exponent. As such, $n\in\mathbb{R}$ becomes a~continuous variable, and we can take limit $n\to0$ in \cref{eq:replica}.
In summary, assuming that we can calculate the averaged value $\mathbb{E}_J[Z^n]$, we can calculate the averaged value of the free energy $F_J$.

Finally, having \cref{eq:average_free_energy}, we can return to the volume $V_{\nparams,\datasize}$ in~\cref{eq:volume_with_gauss}.
\highlight{We associate this volume $V_{\nparams,\datasize}$ now with the partition function of a~spin system. Spins $(s)$ are now model parameters $(\params)$, and couplings $(J)$ are training data $(\dataset)$, which pose the constraints to learn.}
Applying the same analysis as in the previous paragraph, we can state that the free energy for a~given realization of the data set is just the free energy averaged over the data set distribution when we consider large data sets and large perceptron with fixed ratio $\alpha = \datasize / \nparams$:
\begin{equation}
V_{\nparams,\datasize} \simeq  \lim_{\datasize \rightarrow \infty} \estimateE_{\dataset} [V_{\nparams,\datasize} \mid \dataset] = V(\alpha) \, .
\end{equation}
Therefore, if you fix the distribution of data (disorder realization), you can find the $\alpha_c$ for which $V_{\nparams,\datasize}=0$ and, as a~result, the perceptron capacity. To be more exact, we can calculate it only for the large (``thermodynamic'') limit of $n$ for an~arbitrary fixed $\alpha$ as $V$ is actually expressed in terms of $\alpha$. 

\begin{figure}[t]
\begin{center}
\includegraphics[width=0.7\columnwidth]{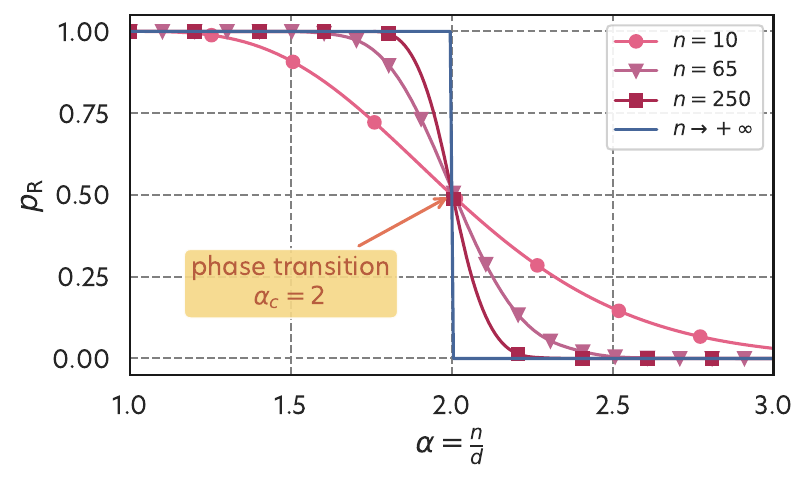}
\end{center}
\caption[Capacity of the perceptron]{Probability of the randomly labeled data being linearly separable, $p_{\mathrm{R}}$, as a~function of the difficulty of the task, $\alpha = \frac{\datasize}{\nparams}$. Finite-size results were obtained analytically by Cover \cite{Cover1965}.}
\label{fig:Marylou_perceptron_capacity}
\end{figure}

We remind you that $\alpha_c$ indicates the critical task difficulty for which the volume of perceptron weights satisfying the constraints of random labeling goes to zero. It means that for the lower task difficulty, $\alpha < \alpha_c$, the randomly labeled data are linearly separable, while for the higher task difficulty, $\alpha > \alpha_c$, the data are no longer linearly separable. The probability, $p_{\mathrm{R}}$, is therefore a~step function of  $\alpha$ in the thermodynamic limit. We plot $p_{\mathrm{R}}$ for real-valued parameters coming from a~Gaussian distribution in blue in~\cref{fig:Marylou_perceptron_capacity}.
To show finite-size effects, we can also compute $p_{\mathrm{R}}(\alpha)$ below the thermodynamic limit following Cover's argument \cite{Cover1965}. To vary $\alpha$, we can change $\datasize$ or $\nparams$. In the case of perceptron, it is easier to keep $\nparams$ fixed and calculate $p_R$ as a~function of $\alpha$ for increasing $\datasize$. \highlight{In the equivalent of the thermodynamic limit, so $\datasize \rightarrow \infty$, we see a~phase transition for a~critical $\alpha_c = 2$, which means that the most difficult task that the perceptron is able to solve is when the number of training points (in general position) is twice as large as the number of parameters.}

Interestingly, the solution for $\alpha_c$ (for which $V_{\nparams,\datasize}=0$) depends on the setup of the problem, namely the random data distribution and the allowed values of parameters (spin values). While the previous discussion has been conducted for Gaussian distribution of inputs and real perceptron parameters, $\params$, different critical task difficulty is obtained for binary inputs and parameters, as presented in~\cref{tab:Marylou_capacity}.

\begin{table}[t]
\centering
\caption[Perceptron capacity depending on the parameters' distribution]{The capacity of the perceptron depends on the distribution of the data and type of weights. The capacity is expressed as the minimal task difficulty\index{minimal task difficulty}, $\alpha_c = \frac{\datasize}{\nparams}$, for which the volume of possible solutions goes down to zero, $V(\alpha_c) = 0$.}
\label{tab:Marylou_capacity}
\begin{tabular}{cllc}
\multicolumn{1}{l}{}  & \multicolumn{2}{c}{Distribution of data}                                        & Critical task difficulty, $\alpha_c$ \B \\ \cline{2-4} 
\multicolumn{1}{c|}{\multirow{2}{*}{1}} &
  \multicolumn{1}{l|}{Gaussian inputs} &
  \multicolumn{1}{l|}{$p(\vect{x}_i^{(k)}) = \mathcal{N}(\vect{x}_i;0,1)$} &
  \multicolumn{1}{c|}{\multirow{2}{*}{$\alpha_c = 2$}} \T \\
\multicolumn{1}{c|}{} & \multicolumn{1}{l|}{Real weights}   & \multicolumn{1}{l|}{$\params \in \mathds{R}^{\nparams}$} & \multicolumn{1}{c|}{}                                    \B \\ \cline{2-4} 
\multicolumn{1}{c|}{\multirow{2}{*}{2}} &
  \multicolumn{1}{l|}{Binary inputs} &
  \multicolumn{1}{l|}{$p(\vect{x}_i^{(k)}) =  \mathrm{Bernoulli}(0.5)$} &
  \multicolumn{1}{c|}{\multirow{2}{*}{$\alpha_c \approx 0.83$}} \T \\
\multicolumn{1}{c|}{} & \multicolumn{1}{l|}{Binary weights} & \multicolumn{1}{l|}{$\params \in \{ -1, 1 \}^d$}  & \multicolumn{1}{c|}{}                                   \B  \\ \cline{2-4} 
\end{tabular}
\end{table}

In this section, we have looked at the problem of perceptron capacity, which is well-known and decades old. As such, it serves the educational purpose well. In particular, we have seen that the statistical approach to learning focuses on simple solvable models (here, perceptrons). Moreover, we have seen that the statistical approach aims to express learning problems in terms of statistical problems, e.g., disordered spin systems,\footnote{This also tells us that \acp{NN} with binary weights may be especially approachable for physicists. These are spin-1/2 problems!} where physicists have already developed useful analytical tools. 

In the next sections, we briefly discuss selected modern results from the intersection of \ac{ML} and statistical physics. For a more detailed review of this intersection, we refer to \cite{Carleo2019RevModPhys}. Moreover, an outstanding retrospective of these developments can be found in the lecture titled \href{https://www.youtube.com/watch?v=BUfnIT92ukM}{``Statistical physics and \ac{ML}: A~30-year perspective''} of the late Naftali Tishby.

\subsubsection{The teacher-student paradigm: a~toy model to study the generalization}\label{sss:Marylou-teacher-student}

Our motivation for this section is to tackle the riddle of generalization, which is the ability of a~model to make correct predictions on data unseen during training. However, our goal for this section is not to build new useful \ac{ML} models or to distinguish between bad and good modern models in terms of generalization. Rather, we want to understand why useful modern \ac{ML} models generalize so well. To do so, let us consider all elements of the learning task (such as model, optimization method, and data) in their simplest form. The toy model that helps us in this ambitious task falls under the teacher-student paradigm.
\highlight{The teacher-student paradigm\index{teacher-student paradigm} consists of two main elements: a~teacher which is a~data-generating model, and a~student which is a~model trying to learn the data generated by a~teacher.}
Teacher consists of an~input distribution $p_x(\vect{x})$, e.g. Gaussian or binary, and an~input-output rule $p(y_{\mathrm{t}} \mid \vect{x}) = f_{\mathrm{t}}(\vect{x}, \params^*)$. For now, let us assume that the teacher is a~perceptron.
In addition to the input-output rule, we may assume a~ground-truth distribution on the weights $p_{\param}\left(\params\right)$, from which the parameters $\params^{*}$ of the teacher model were drawn. Once we decide on how a~teacher looks like, it can generate training data: $\dataset = \{ \vect{x}^{(k)}, y_{\mathrm{t}}^{(k)} \}^\datasize_{k=1} = \{
\vect{x}^{(k)},
f_{\mathrm{t}}(\vect{x}^{(k)}, \param^{*}) \}^\datasize_{k=1}$.

\highlight{The second element is the student, whose aim is to learn the distribution underlying the training data. In the teacher-student scheme, we know exactly what the data-generating distribution is. Therefore, we can easily distinguish between a~student that simply fits the training data (limited generalization) and a~student that recovers a~teacher's input-output rule (perfect generalization). In other words, we can measure the generalization of the student.}

To continue with the teacher-student strategy, we need to decide on a~model for the student, $f_{\mathrm{s}}(\vect{x}, \params)$, but also on a~learning strategy. Let us start with the simplest scenario when a~student is also a~perceptron (like the teacher). To train, we could use the standard empirical loss minimization strategy, e.g.,
\begin{equation}
    \params^* = \mathrm{argmin}_{\params} \left\{ \sum_{k=1}^\datasize \lossfun \left(y_{\mathrm{t}}^{(k)}, f_{\mathrm{s}}\left(\vect{x}^{(k)},\params\right)\right) \right\}\,,
\end{equation}
where we aim to minimize a~given distance between the teacher outputs $y_{\mathrm{t}}^{(k)}$ and student outputs $y_{\mathrm{s}}^{(k)} = f_{\mathrm{s}}(\vect{x}^{(k)},\params)$. Alternatively, we can consider the following Bayesian posterior\index{Bayesian posterior} distributions on the parameters and draw values of the parameters according to it:
\begin{equation}
p(\params \mid \dataset) \propto \prod_{k=1}^{\datasize} p\left(y_{\mathrm{t}}^{(k)} \mid \params, \vect{x}^{(k)}\right) p(\params)\,.
\label{eq:posterior_distribution}    
\end{equation}
\Cref{eq:posterior_distribution} denotes the posterior distribution, i.e., the belief on the student model weights $\params$ given the data set $\dataset$ and the prior assumption on the student weights $p(\params)$.

\begin{figure}[t]
\begin{center}
\includegraphics[width=0.95\columnwidth]{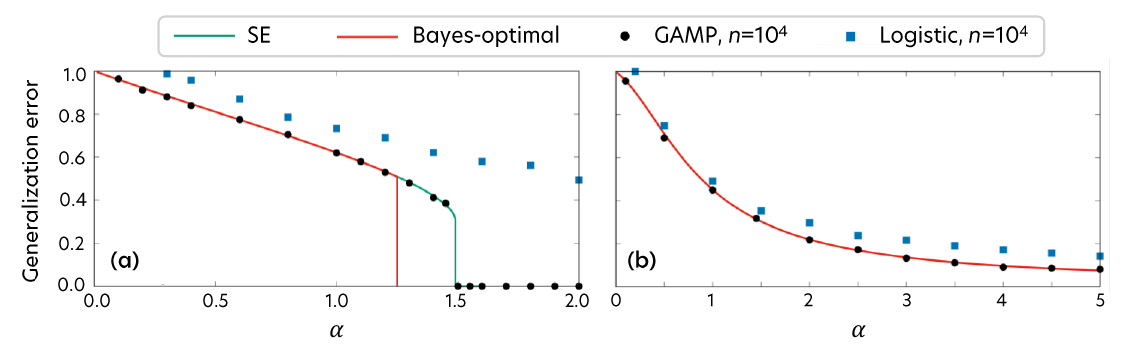}
\end{center}
\caption[Generalization error vs. the task difficulty in the teacher-student paradigm]{Generalization error as a~function of the task difficulty $\alpha$, which is the ratio between the number of training points and the number of (student) model parameters for perceptrons with (a) $\params \in \mathbb{R}^\nparams$ or (b) $\params \in \{-1, 1\}$. The red line is the exact Bayes-optimal generalization error. The blue squares are for a~fine-tuned perceptron with gradient-based minimization of the error. We see the computational gap between these results. Gap (a) gets smaller, or (b) disappears for message-passing algorithms. Black circles are results for $\datasize = 10^4$ obtained using \ac{GAMP}, and the green line denotes the results of \acf{SE}, which approximates the limit $n \rightarrow \infty$. Adapted from \ToggleForCUP{Barbier, J. \textit{et al.} (2019). \textit{Optimal errors and phase transitions in high-dimensional generalized linear models}, PNAS 116, 5451~\cite{Barbier2019} under the \href{https://creativecommons.org/licenses/by/4.0/}{CC BY 4.0 DEED} license.}{Ref.~\cite{Barbier2019}.}}
\label{fig:Marylou_generalization_curves}
\end{figure}

Assuming, e.g., a~\ac{MSE} loss, the student generalization error for given weights $\params$ is defined as the expected error over the entire data distribution:
\begin{equation}\label{eq:generalization_error}
    \mathcal{E}_{g}(\params)= \estimateE_{\vect{x}, y} \left[ \left(y - f_{\mathrm{s}}\left(\vect{x}, \params\right)\right)^{2}\right] \,.
\end{equation}
In the best possible scenario, the student model $f_s(\cdot, \cdot)$ is identical to the teacher model $f_t(\cdot, \cdot)$ underlying the generated data. When a~student is identical to the teacher, we call the setting \stress{Bayes optimal} and define the Bayes error\index{Bayes error} (see~\cref{sss:probability}) of the student as,
\begin{equation}\label{eq:bayes_generalization_error}
    \mathcal{E}^{\rm opt}_{g}(\dataset)=\estimateE_{\params} \left[ \estimateE_{\vect{x}, y} \left[ \left(y - f_{\mathrm{s}}\left(\vect{x}, \params\right)\right)^{2} \right] \mid \dataset \right] \,,
\end{equation}
which is a~mean error for student parameters $\params$ drawn from the posterior distribution in~\cref{eq:posterior_distribution}.
This is a~fundamental quantity from the point of view of information theory: it quantifies how much information on the weights $\params$ the training data set $\dataset$ provides, assuming that the student has perfect knowledge of the form of the problem.
We can use the same tools as in the previous section (disorder average, thermodynamic limit, and replica computation) to obtain:
\begin{equation}\label{eq:average_generalization_error}
\mathcal{E}^{\rm opt}_{g}(\dataset) \underset{\datasize \rightarrow \infty}{\rightarrow} \quad \estimateE[\mathcal{E}^{\rm opt}_{g}(\dataset) \mid \dataset] =  \mathcal{E}^{\rm opt}_{g}(\alpha)\,.
\end{equation}
Here again, the limiting generalization error takes the form of a~function of the ratio $\alpha = \frac{\datasize}{m}$ between the number of data points and the number of data features or weights. We no longer interpret this ratio as the difficulty of the classification task as in the capacity computation. Instead, in generalization problems, it is more useful to think of $\alpha$ as the sample complexity, that is, the amount of training data available to infer the input-output rule.
In the following paragraphs, we examine generalization for a~few different pairs of teachers and students.
\paragraph{Two perceptrons.}
  The generalization error from~\cref{eq:average_generalization_error} is shown in~\cref{fig:Marylou_generalization_curves}. We can compare the limiting Bayesian optimal generalization error (red line in panels (a) and (b)) with the training of a~perceptron at finite $m$ by minimizing a~loss function, such as performing a~logistic regression with gradient descent (blue squares). In panel (a), for binary weights, we have a~first-order phase transition~\cite{Krauth1989, Gyorgyi1990}. In panel (b), for real-valued weights, there is a~smooth decrease of the generalization error~\cite{Barbier2019}. In both cases, there is a~computational gap between the optimal generalization error and logistic regression with gradient descent.

Finally, the same generalization error of the student perceptron can be studied when learning occurs with algorithms called \acf{GAMP}\index{generalized approximate message passing}. For the introduction to these methods, see Ref.~\cite{Rangan2011GAMP, Zdeborova2016, Gabrie2020meanfield}. For our needs, it is enough to know that these algorithms provide an~alternative to convex optimization and allow for efficient calculations of quantities based on graphs (like perceptrons or \acp{NN}), which are sampled from distributions like~\crefrange{eq:generalization_error}{eq:average_generalization_error}. Moreover, they are remarkable in that their asymptotic ($\datasize, \nparams \rightarrow \infty$, $\datasize/\nparams = \alpha$) performance can be analyzed rigorously using the so-called \acf{SE}. Armed with this knowledge, we now see that the generalization error obtained using \ac{GAMP} in~\cref{fig:Marylou_generalization_curves} is much closer to the Bayes error compared to the optimization with gradient descent. In panel (b), the gap completely disappears. In panel (a), there is a~remaining computational gap between \ac{GAMP} and the exact Bayes error. This regime is called a~\stress{hard phase}. It comes from the fact that, in practice, our computational time is limited to the polynomial regime. Interestingly, there is no known efficient algorithm that would beat \acp{GAMP} in the hard phase of this perceptron learning~\cite{Barbier2019}.

\begin{figure}[t]
\begin{center}
\includegraphics[width=0.98\columnwidth]{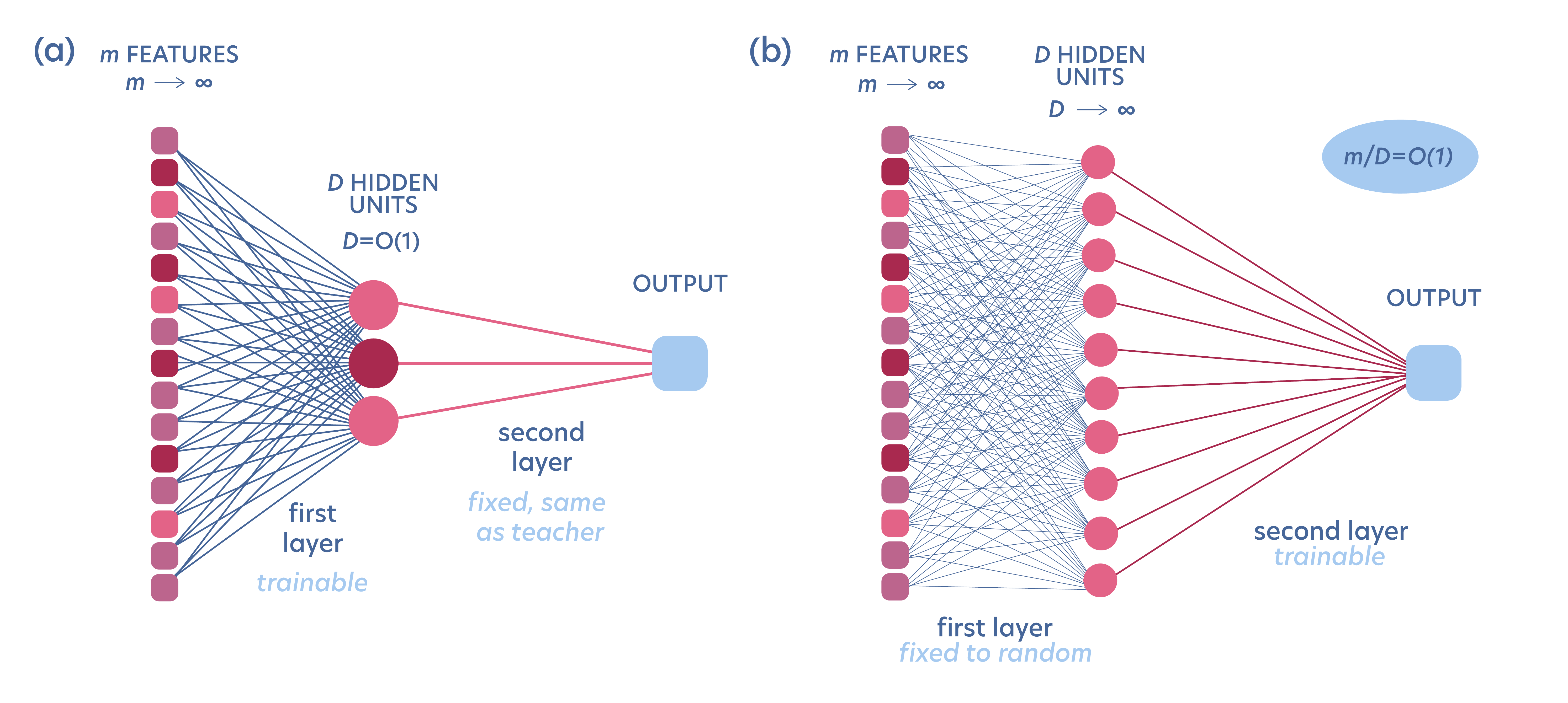}
\end{center}
\caption[Schemes of a~committee machine and random feature model]{Schematic illustration of used two-layer \acp{NN}. (a) Soft committee machine\index{committee machine}. The parameters belonging to the first layer, $\params_1 \in \mathbb{R}^{m \times D}$, are trainable, whereas the parameters of the second layer, $\params_2 \in \mathbb{R}^{D \times 1}$, are chosen identical to the parameters of the teacher. Its analytical treatment is possible if $m \rightarrow \infty$ and $D = O(1)$. (b) Random feature model\index{random feature model}. Its first layer is fixed to random parameter values. The second layer is trainable. The number of hidden units can be varied to study overparametrization. In the analysis, the number of hidden units $D \rightarrow \infty$ scales linearly with the number of inputs $m \rightarrow \infty$, i.e., $ m / D = O(1)$.}
\label{fig:Marylou_architectures}
\end{figure}

\paragraph{Two-layer \acp{NN}.} So far, both the teacher and the student have been modeled with perceptrons. We can switch to more complex models. For the remainder of this section, we use two special two-layer \acp{NN} with a~rich history in statistical physics. We start with \stress{committee machines}\index{committee machine} \cite{Monasson1995,Aubin2019} shown in~\cref{fig:Marylou_architectures}(a). Their analytical treatment is possible in the limit of an~infinite number of input features, $m$, and data size, $\datasize$, while keeping a~finite number of hidden units. In particular, we present here soft committee machines which allow for an~even simpler analysis. In soft committee machines, we train only parameters belonging to the first layer of the machine, $\params_1$, of size $\nparams_1 = \nparams = \featnum D$, where $\featnum$ is the number of features and $D$ is the number of hidden units. The second layer is fixed and identical for both the teacher and the student. The second \ac{NN} used in this section is a~\stress{random feature model}\index{random feature model} \cite{Rahimi2007RFM, Rahimi2008RFM} presented in \cref{fig:Marylou_architectures}(b). Interestingly, their analytical analysis is enabled by a~fixed first layer whose parameters are set to random values. The number of these parameters is also $\nparams_1 = \nparams = \featnum D$. Therefore, only the second layer parameters can be trained. The idea behind the random first layer is that projecting a~lower-dimensional input onto a~much higher dimensional space leads to better separation of the data, which then can be successfully processed by a~single-layer \ac{NN}.\footnote{In other words, you can think of such a~projection as mapping input data to a~feature space as discussed in \cref{sec:kernel_methods} on kernel methods. Interestingly, Refs. \cite{Rahimi2007RFM, Rahimi2008RFM} showed that random data projection onto a~feature space is not much worse compared to projecting onto an~optimized feature space. However, randomization is much cheaper than optimization.} Also note that random feature models can have an~arbitrary number of hidden units, in particular larger than the number of input features,  which allows for a~study of overparametrization.

\ToggleForCUP{
\begin{figure}[p]
\begin{center}
\includegraphics[width=0.95\columnwidth]{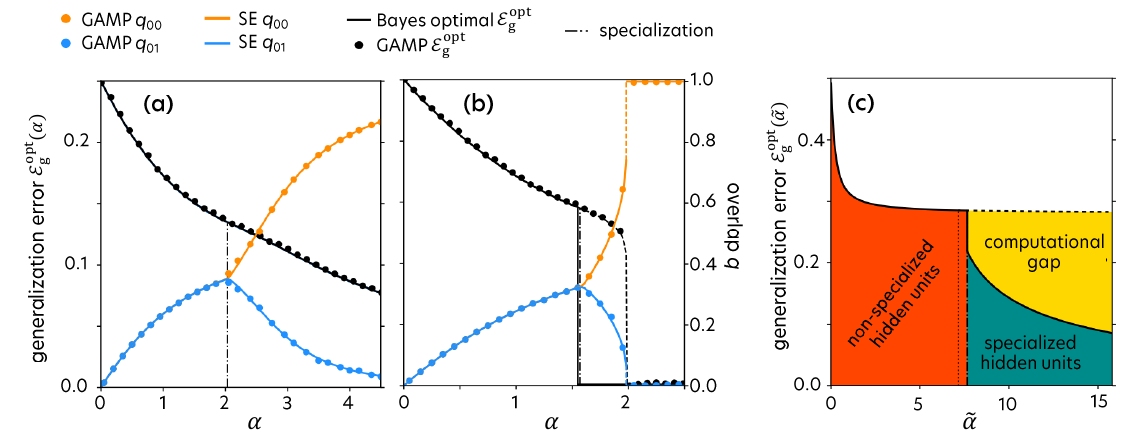}
\end{center}
\caption[Generalization error and specialization in committee machines]{Generalization error and specialization in committee machines\index{committee machine} as functions of the task difficulty $\alpha = \frac{\datasize}{m} \propto \frac{\datasize}{\nparams}$, which is the ratio of the number of training points and the number of input features. We consider the committee machines with (a) $\params \in \mathbb{R}^{m \times D}$, $D=2$, or with (b) $\params \in \{-1, 1\}^{m \times D}$, $D=2$. The black line is the exact Bayes-optimal generalization error, and the black dots are obtained by studying the committee machine with \ac{GAMP}. The orange and blue lines and dots indicate the overlap of the two hidden neurons of the student committee machine with the two hidden neurons of the teacher committee machine, calculated with \ac{GAMP} and \ac{SE}, respectively. We see that specialization is responsible for the rapid decrease in generalization error. (c) Generalization of the panel (a) to a large number of hidden neurons, $D$. Phase diagram calculated for the task difficulty, $\tilde{\alpha} = \frac{\alpha}{D}$. Adapted from Aubin, B. \textit{et al}. (2019). \textit{The committee machine: computational to statistical gaps in learning a two-layer neural network}, J. Stat. Mech. Theor. Exp. 2019, 124023~\cite{Aubin2019} with permission of IOP Publishing, Ltd. Permission conveyed through Copyright Clearance Center, Inc.}
\label{fig:Marylou_committee_specialization}
\end{figure}}
{
\begin{figure}[t]
\begin{center}
\includegraphics[width=0.95\columnwidth]{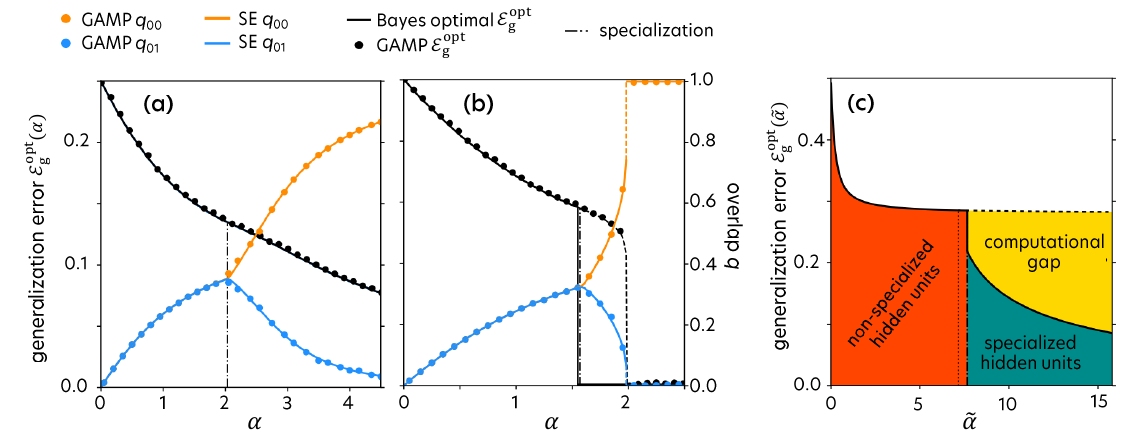}
\end{center}
\caption[Generalization error and specialization in committee machines]{Generalization error and specialization in committee machines\index{committee machine} as functions of the task difficulty $\alpha = \frac{\datasize}{m} \propto \frac{\datasize}{\nparams}$, which is the ratio of the number of training points and the number of input features. We consider the committee machines with (a) $\params \in \mathbb{R}^{m \times D}$, $D=2$, or with (b) $\params \in \{-1, 1\}^{m \times D}$, $D=2$. The black line is the exact Bayes-optimal generalization error, and the black dots are obtained by studying the committee machine with \ac{GAMP}. The orange and blue lines and dots indicate the overlap of the two hidden neurons of the student committee machine with the two hidden neurons of the teacher committee machine, calculated with \ac{GAMP} and \ac{SE}, respectively. We see that specialization is responsible for the rapid decrease in generalization error. (c) Generalization of panel (a) to a large number of hidden neurons, $D$. Phase diagram calculated for the task difficulty, $\tilde{\alpha} = \frac{\alpha}{D}$. Adapted from Ref.~\cite{Aubin2019}.}
\label{fig:Marylou_committee_specialization}
\end{figure}}

\paragraph{Two committee machines.} Now, we are ready to tackle generalization with more complex models. Here, we use soft committee machines\index{committee machine}. For now, a~teacher and a~student share the same architecture. The formulation of the problem stays the same. We calculate the generalization error from~\cref{eq:average_generalization_error} of the student committee machine when learning data generated by the teacher committee machine \cite{Aubin2019}. We plot the generalization errors in~\cref{fig:Marylou_committee_specialization}(a)-(b) for committee machines with two hidden neurons. Similarly as before, we see in panel (a) that for real-valued weights, the generalization error (obtained with \ac{GAMP} and \ac{SE}) is equal to the Bayes one, while for binary weights in panel (b) there is a~computational gap between both errors. This time, we also look at the overlap between hidden neurons of the student and of the teacher, which measures the similarity neuron-by-neuron between the teacher and the student. To be more precise, we look at the matrix $\mat{Q} = [q_{jj'}] = \frac{1}{m} \sum_{i=1}^m \Theta^*_{1,ij} \param_{1,ij'}$, where $\vect{\Theta}_1$ and $\params_{1}$ are the parameters of the first layers of the teacher and the student, respectively. It turns out that there is a~so-called \stress{specialization} phase transition~\cite{Schwarze1993, Schwarze1993a}.
\highlight{In the regime of low task complexity, both hidden units of the student committee machine learn the same function. After crossing the critical $\alpha$, when enough data is available, the hidden neurons of the student start to \stress{specialize}. Each student neuron selects a~different teacher neuron to converge to. The specialized phase is associated
with lower generalization error than the non-specialized one, see~\cref{fig:Marylou_committee_specialization}.}
The specialization for teacher and student committee machines with two hidden neurons ($D=2$) takes place for $\alpha_c \approx 2$ for real-value weights and for $\alpha_c \approx 1.5$ for binary weights, which means that specializing neurons require at least 2 and 1.5 times more training data than the number of data features, $\featnum$, i.e., approximately as much training data as the number of parameters in the first layer, $\nparams_1 = 2 \featnum$. Similar observations hold if both the teacher and student committee machines have a~large number of hidden neurons ($D \gg 2$). 
We can plot a~phase diagram of the generalization error as a~function of a~rescaled task difficulty, $\tilde{\alpha} = \frac{\alpha}{D} = \frac{\datasize}{D m}$ for real-valued weights. It is presented in~\cref{fig:Marylou_committee_specialization}(c). In total, we find three distinct phases: two correspond to specialized and non-specialized hidden neurons, and above the specialized phase, there is a~computational gap where a~model in principle has enough information to specialize but is unable to do so due to shortcomings of its optimization.

\paragraph{Overparametrization.} As we have already mentioned in the introduction, one of the most puzzling phenomena in modern \ac{ML} is the generalization capability of heavily overparametrized models. In real-world settings, it is natural to think of the level of the model overparametrization\index{overparametrization} as the ratio between the number of model parameters, $\nparams$, and the number of available training data points, $\datasize$. Surprisingly, we see in practice that models with large $\nparams$ are able to extract meaningful relations from much fewer training data points. In turn, with the teacher-student scheme, we can make the definition of overparametrization more rigorous because we have direct access to the ``ground-truth'' number of parameters needed to describe the input-output rule, which is the number of teacher parameters. Therefore, the level of overparametrization can be understood as a~ratio between the number of parameters of the student and the teacher. In particular, the student can have much more parameters than the teacher. To study overparametrization, it is then a~necessity to have mismatched teacher-student architectures. Crucially, this mismatch of architectures means that the student cannot achieve a~Bayes optimal error anymore.

For the remainder of this section, we study the generalization error of overparametrized student models. This time we employ as a~student a~random feature model, presented already in \cref{fig:Marylou_architectures}(b). The analysis requires the model's first-layer weights to be fixed to random values. The number of student parameters in the second layer can vary compared to the teacher.\footnote{Ref.~\cite{Gerace2021} interprets the same exact setting as a~teacher generating labels with a~perceptron, itself acting on a~low dimensional latent space, and input data generated with a~one layer generative \ac{NN} from this latent space. A~student perceptron is trained in the input data-label pairs.} As such, we have a~full control over how overparametrized the student is. We come back to the study of overparametrization in committee machines in \cref{sss:Marylou-dynamics}.

\ToggleForCUP{}{\begin{figure}[t]
\begin{center}
\includegraphics[width=0.98\columnwidth]{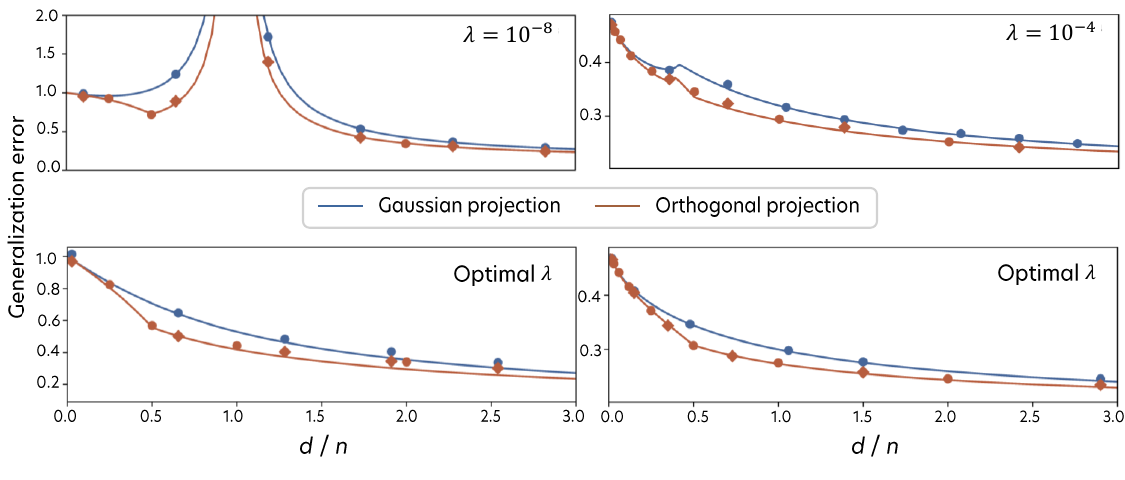}
\end{center}
\caption[Generalization errors vs. model parametrization for mismatched teacher and student models]{Generalization errors as functions of the ratio of the number of model parameters and number of training data points for mismatched teacher-student models where the student is a~random feature model. The first (second) column shows the generalization error in the case of a~regression (classification) problem. The upper row shows results for sub-optimal regularization strengths, where the generalization error curves exhibit a~double descent. The bottom row shows results for optimal regularization, where the double descent disappears. Adapted from \ToggleForCUP{Gerace, F. \textit{et al}. (2019). \textit{Generalisation error in learning with random features and the hidden manifold model}, J. Stat. Mech. 2021, 124013~\cite{Gerace2021} with permission of IOP Publishing, Ltd. Permission conveyed through Copyright Clearance Center, Inc.}{Ref.~\cite{Gerace2021}.}}
\label{fig:Marylou_generalization_mismatched}
\end{figure}}

With a~student random feature model, we are ready to study the generalization error as a~function of overparametrization, $\frac{1}{\alpha} = \frac{\nparams}{\datasize}$, where $\nparams = \nparams_1$ is the number of parameters in the first fixed random layer of the student. As the student cannot achieve a~Bayes optimal error anymore, we need to change the training objective, e.g., to a~\ac{MSE} with $\regularization{2}$ regularization. Using various analytical tools, we can still approximate the generalization error of the student and plot it as a~function of overparametrization $\frac{\nparams}{\datasize}$. In Ref.~\cite{Gerace2021}, the generalization error in regression and classification tasks was analyzed for various strengths of regularization. Their results are shown in~\cref{fig:Marylou_generalization_mismatched}. The left (right) column shows the generalization error of the mismatched student for optimal and sub-optimal regularization strengths in a~regression (classification) problem. 
\highlight{Remarkably, in the case of mismatched student-teacher architectures, the generalization error curve exhibits a~characteristic double descent. Moreover, the optimal choice of regularization cancels the first error descent, resulting in the generalization error steadily decreasing with the increasing number of model parameters.}
Therefore, these results on toy models give us a~hint on the origin of the double descent phenomenon. It occurs when the student and teacher have mismatched architectures, and the choice of regularization strength is sub-optimal. Interestingly, Ref.~\cite{Gerace2021} also showed that the magnitude of the initial generalization error ascent in the double descent phenomenon depends on whether the problem is a~classification or regression task.
 
\ToggleForCUP{\begin{figure}[t]
\begin{center}
\includegraphics[width=0.98\columnwidth]{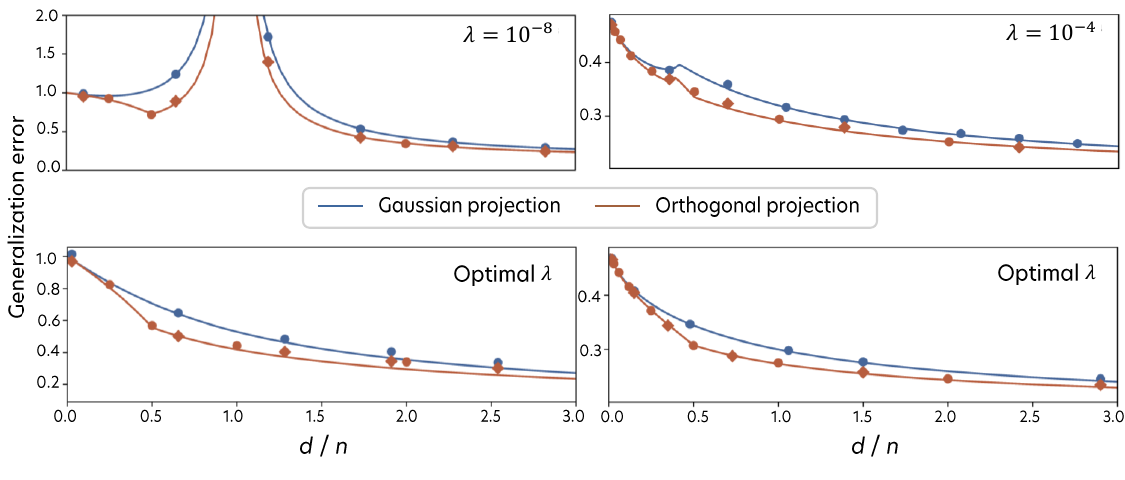}
\end{center}
\caption[Generalization errors vs. model parametrization for mismatched teacher and student models]{Generalization errors as functions of the ratio of the number of model parameters and number of training data points for mismatched teacher-student models where the student is a~random feature model. The first (second) column shows the generalization error in the case of a~regression (classification) problem. The upper row shows results for sub-optimal regularization strengths, where the generalization error curves exhibit a~double descent. The bottom row shows results for optimal regularization, where the double descent disappears. Adapted from \ToggleForCUP{Gerace, F. \textit{et al}. (2019). \textit{Generalisation error in learning with random features and the hidden manifold model}, J. Stat. Mech. 2021, 124013~\cite{Gerace2021} with permission of IOP Publishing, Ltd. Permission conveyed through Copyright Clearance Center, Inc.}{Ref.~\cite{Gerace2021}.}}
\label{fig:Marylou_generalization_mismatched}
\end{figure}}{}

In summary, the study of toy models indicates that there are numerous reasons for the generalization error being larger than the Bayes optimal error. In general, the generalization capabilities depend on:
\begin{itemize}
    \item whether a~data-generating model (teacher) and learning model (student) have mismatched architectures,
    \item whether the model aims at solving a~regression or classification problem,
    \item the choice of optimization method, target function, and available computation time,
    \item the sample complexity (how much training data is available and, for teacher-student committee machines, the degree of specialization of the neurons).
\end{itemize}

\subsubsection{Models of data structure}\label{sss:Marylou-data-structure}

\begin{figure}[t]
\begin{center}
\includegraphics[width=0.95\columnwidth]{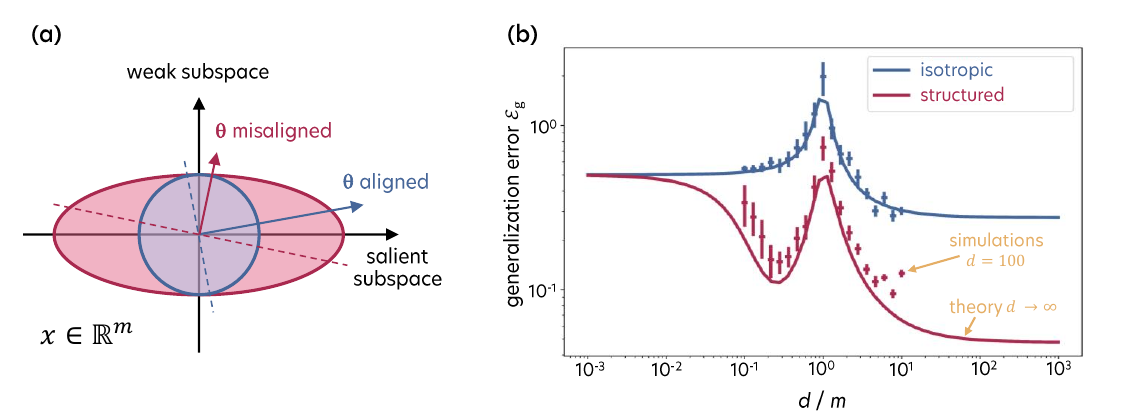}
\end{center}
\caption[Data structure entering the teacher-student scheme]{Data structure entering the teacher-student scheme. (a) The data space can be separated into a~weak and salient subspaces, where the data is characterized by a~small or large variance, respectively. The teacher perceptron with parameters $\params$ can be aligned (blue vector) or misaligned (purple vector) with the salient subspace. The classification task specified by the teacher (represented as a~line separating the data) is easier (compared to the isotropic case) if $\params$ is aligned with the salient subspace. (b) The model can detect the structure existing in the data. As a~result, the generalization error is lower for the structured case than for the fully isotropic case. The results for the isotropic data are similar to the results for anisotropic data, where the teacher perceptron is misaligned. The generalization curve shows a~double descent. Adapted from \ToggleForCUP{D’Ascoli, S. \textit{et al.} (2021). \textit{More data or more parameters? Investigating the effect of data structure on generalization}, In NeurIPS 2021 - Adv. Neural Inf. Process. Syst.~\cite{dAscoli2021}.}{Ref.~\cite{dAscoli2021}.}}
\label{fig:Marylou-generalization-structure}
\end{figure}

So far, while studying sources of generalization errors, we have mainly played with the architectures of teacher and student models, which specify the structure of the input-output rule underlying the data. In particular, we have only considered random input data sets where all input features are independent. Clearly, while such isotropic data simplify the analytical analysis,  it is quite unrealistic. Ideally, we would like to study prototypical data sets, such as MNIST \cite{lecun:1998} or ImageNet \cite{Russakovsky2015imagenet}, but these are difficult to treat analytically. Instead, let us move one step away from the data sets given by white noise and use the teacher-student paradigm to study the impact of data anisotropy on the generalization error. To this end, we employ \stress{salient and weak feature models} \cite{dAscoli2021}. Within these feature models, the data remain Gaussian (as in most of the previous sections), $\vect{x} \sim  \mathcal{N} (0, \Sigma_{\vect{x}})$, but the covariance is not isotropic as if $\Sigma_{\vect{x}} = I_{\nparams}$. Instead, it is anisotropic:
\begin{equation}
 \Sigma_{\vect{x}}=\left[\begin{array}{cc}
\sigma_{\vect{x}, 1} I_{\phi_{1} \nparams} & 0 \\
0 & \sigma_{\vect{x}, 2} I_{\phi_{2} \nparams}\end{array}\right]\,,   
\end{equation}
where $\sigma_{\vect{x},1} \gg \sigma_{\vect{x},2}$, and $\phi_{1/2}\nparams$ denotes the number of data features (equal to the number of perceptron parameters) that are affected by the variance $\sigma_{\vect{x},1/2}$ (with $\phi_{1} + \phi_{2}=1$). Parameters affected by large variance, $\sigma_{\vect{x},1}$, form the salient subspace, whereas ones with a~small variance, $\sigma_{\vect{x},2}$, form the weak subspace as presented in \cref{fig:Marylou-generalization-structure}(a). We assume the weak subspace to be much larger than the salient one, $\phi_{2} \gg \phi_{1}$.

If we add such a~structure to our input data and run the teacher-student scheme with mismatched architectures (here, the teacher is a~perceptron and the student is a~random feature model), we can still compute the generalization error exactly in the high-dimensional limit~\cite{dAscoli2021}. Importantly, this generalization error now depends on the anisotropy of the input data. In particular, it depends on how the teacher perceptron is aligned with respect to the weak and salient data subspaces as presented in~\cref{fig:Marylou-generalization-structure}(a). The dashed hyperplanes mark the separation of the input space by the teacher perceptron and lie perpendicular to the perceptron parameter vector, $\params$. This vector can be aligned in various ways with the data anisotropy. If $\params$ is aligned with the salient subspace, the hyperplane cuts along the weak subspace, and the only subspace relevant to discriminate the data points is the salient subspace, in which the variance of the data is concentrated, and the weak subspace can be effectively ignored. In this case, due to the data structure, the problem has a~small effective dimension corresponding to the salient space $\phi_{1}\nparams$, so it is easier to solve. In turn, if $\params$ is aligned with the weak subspace, the impact of the data structure is negligible since the student needs to discriminate along an~axis where data has low variance compared to the typical variance of the data. Here, the problem closely resembles the (fully) isotropic case.

The impact of the data structure on the generalization curve as a~function of the ratio of the number of parameters of the student model and the dimensionality of the data, $\frac{\nparams}{\featnum}$, is shown in~\cref{fig:Marylou-generalization-structure}(b). \highlight{Interestingly, the structure in the data is detected during training before the generalization error peaks due to overfitting and improves the generalization error as compared to the isotropic case.} Why does the structure help? This is so because, in practice, the model can ignore the weak subspace and focus on the salient one, which lowers the dimensionality of the problem. The fact that the generalization error is lower in the anisotropic case compared to the isotropic case remains true even in the highly overparametrized regime ($\frac{\nparams}{\featnum} = 10^3$). Moreover, the double descent phenomenon is also exacerbated in the presence of data structure. Note that both these effects take place only when the teacher perceptron is aligned with the salient subspace. Otherwise, the setup closely resembles the isotropic case.

It turns out that many more questions can be addressed with the teacher-student paradigm using salient and weak feature models. In particular, the authors of Ref.~\cite{dAscoli2021} checked the interplay between the data structure and other elements of \ac{ML} problems, like the choice of the loss function. Recalculating the quantities in~\cref{fig:Marylou-generalization-structure}(b) for the \ac{MSE} and logistic loss, one observes that the overfitting peak is attenuated in the case of logistic loss. Therefore, it seems that logistic loss takes more advantage of the existing data structure. We can confirm this further by computing the generalization error for both loss functions as a~function of the teacher-data alignment. As discussed earlier, this alignment determines how much data structure is \stress{effectively} present in the problem. In agreement with the results described above, when the alignment is increased, the gap between the generalization error of \ac{MSE} and logistic loss increases.

\subsubsection{Dynamics of learning}\label{sss:Marylou-dynamics}

Finally, we can investigate the dynamics of learning and its dependence on the model overparametrization using the teacher-student schemes described above. An~example of a~simplified model of learning is \stress{online learning}\index{online learning}, which has been analyzed since the 1990s. In online learning, the model is fed a~stream of data, where the model sees each data point only once. We build a~loss function based on this example and perform a~parameter update according to the gradient of this loss function. In fact, we perform a~parameter update after each data point encounter. Thus, the number of optimization steps is equal to the number of seen training data points. If we take the continuous time and high-dimensional limit and average over all random variables (which is doable with the replica method if we assume samples at distinct times are uncorrelated), we can again calculate the generalization error explicitly. In particular, we can calculate how it changes during training. In other words, we can track the quality of the model predictions over the course of the training.

\begin{figure}[t]
\begin{center}
\includegraphics[width=0.7\columnwidth]{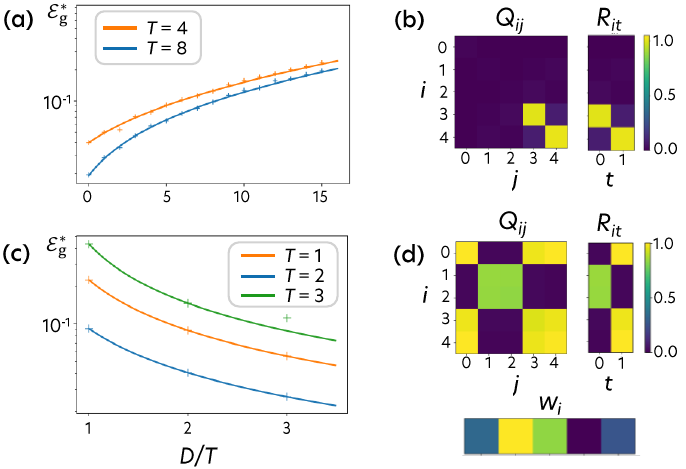}
\end{center}
\caption[Dynamics of learning in overparametrized committee machines]{Dynamics of learning in overparametrized committee machines\index{committee machine}. In panels (a)-(b), we allow only a~single trainable student layer, and another layer is fixed. In panels (c)-(d), we train the whole student model. (a),(c) Generalization error vs. student overparametrization which occurs when the number of student hidden units, $D$, is larger than the number of teacher hidden units, $T$. (b),(d) Self-overlaps of the student ($\mat{Q}$ matrix) and overlaps between the overparametrized student and the teacher ($\mat{R}$ matrix). Vector $\vect{w}$ contains the weights of the second layer of the student. Adapted from \ToggleForCUP{Goldt, S. \textit{et al.} (2020). \textit{Dynamics of stochastic gradient descent for two-layer neural networks in the teacher-student setup}, J. Stat. Mech. 2020, 124010~\cite{Goldt2020a} under the \href{https://creativecommons.org/licenses/by/4.0/}{CC BY 4.0 DEED} license.}{Ref.~\cite{Goldt2020a}.}}
\label{fig:Marylou-online-learning}
\end{figure}

Such an~analysis has already been conducted in the 1990s for perceptrons\index{perceptron} and committee machines\index{committee machine}~\cite{Biehl1994, Biehl1995}. It showed, for instance, that during online training, the generalization error decreases with different convergence rates given different learning rates. Recently, the same analysis was revisited for soft committee machines~\cite{Goldt2020a} considering the impact of overparametrization. In the simplified case of matching teacher-student models and training limited to only a~single student layer, results show how the generalization error drops the moment the student neurons specialize and attain a~large overlap with the teacher neurons. We can also investigate the effect of overparametrization on learning dynamics using committee machines as presented in \cref{fig:Marylou_architectures}(a) in the regime of the number of data features, $m \to \infty$, with the sigmoidal activation function, $g(x) = \mathrm{erf}(x/\sqrt{2})$. Here, we study the generalization error as a~function of the ratio between the number of hidden units of the student $D$ and the teacher $T$ given by $\frac{D}{T}$. \Cref{fig:Marylou-online-learning} shows two cases of online learning of such overparametrized students. In the first case, shown in panels (a)-(b), only the first hidden layer of the student model can be trained, whereas the parameters of the second hidden layer are fixed and identical to the respective teacher layer. In the second case, presented in panels (c)-(d), both student layers are trained. Panel (a) shows that the generalization error actually increases with the size of the trainable student layer, proving that overparametrization can be detrimental in some scenarios. To understand why, we analyze the teacher-student overlaps at the end of the training in the form of $\mat{R} = [R_{it}]$ where each matrix element measures the similarity between the weights of the $i$-th student node and the $t$-th teacher node. We also study the overlap of the weights of different student nodes with each other ($\mat{Q} = [Q_{ij}]$). We plot both matrices in panel (b). We see that in the case of soft committee machines only the number of student neurons that is equal to the number of teacher neurons specialize. The rest simply picks up the noise present in the available data, which impairs generalization. However, if we allow all layers to be trained, a~very different behavior is observed. In panel (c), we see that the generalization error decreases as one overparametrizes the student model. This time, all neurons learn something related to the teacher neurons. Due to this, additional neurons are beneficial as each teacher neuron can be learned by an~ensemble of student neurons that contributes to ``denoising'' the estimation of the teacher parameters. 

\subsubsection*{Outlook and open problems}
In this section, we have seen how to use analytical tools from statistical physics to study problems in \ac{ML}. In particular, we have discussed the seminal problem of perceptron capacity. Subsequently, we have focused on a~powerful paradigm for studying the generalization error: the teacher-student scheme. This scheme can include various modifications that address all elements of the learning problem.

We can study different teacher and student architectures, and they can be mismatched. We have shown results for perceptrons, committee machines, and random feature models, but, in general, we can have, e.g., a~pre-trained generative model (described in more detail in~\cref{sec:deep_generative_models}) playing a~role of a~teacher as it was done in Ref.~\cite{Goldt2020}. One can also analyze more complex data sets than those provided by a~salient and weak feature model. In particular, it is possible to confirm intuitions gained from the analytical analysis of simple models with simulations on standard benchmark data sets~\cite{dAscoli2021}, such as MNIST \cite{lecun:1998} and CIFAR \cite{krizhevsky:2009}. Finally, one can go beyond the online gradient descent and study the multi-pass \ac{SGD} (which involves multiple encounters of the same data points) with the dynamical mean-field theory~\cite{Mignacco2020}, bringing us closer and closer to modern optimization methods. 

Moreover, one can investigate the capacities of large \ac{ML} architectures (in contrast to simple perceptrons). Statistical tools also play an~increasingly important role in the research on \acf{QML}. For example, the Gardner approach was successfully applied to quantum perceptrons \cite{Lewenstein94, Gratsea2021perceptron} and quantum \acp{NN} \cite{Lewenstein20}. In particular, it turned out that the quantum perceptron\index{perceptron!quantum perceptron} has some advantages over its classical counterparts when it comes to capacity\index{capacity}~\cite{Gratsea2021perceptron}. Moreover, the teacher-student scheme\index{teacher-student paradigm} was proposed to systematically compare different quantum \ac{NN} architectures \cite{Gratsea2021teacherstudent}.
Finally, there are works studying phases in the learning dynamics of \ac{ML} models~\cite{Feng2021,decelle2021equilibrium}.
\subsubsection*{Further reading}
\begin{itemize}
    \item Gabrié, M. (2020). \href{https://doi.org/10.1088/1751-8121/ab7f65}{\textit{Mean-field inference methods for neural networks}}. J. Phys. A: Math. Theor. 53, 223002. Review on the mean-field methods mentioned within this section. In particular, it contains principles for derivations of high-temperature expansions, the replica method, and message-passing algorithms \cite{Gabrie2020meanfield}.
    \item Zdeborová, L. (2020). \href{https://doi.org/10.1038/s41567-020-0929-2}{\textit{Understanding deep learning is also a~job for physicists}}. Nat. Phys. 16, 602–604. A short and friendly introduction to how physics can help \ac{ML} \cite{Zdeborova2020}.
    \item \href{https://www.youtube.com/@LesHouches-iu6nv/videos}{Recordings of lectures} of the \href{https://leshouches2022.github.io/}{Summer School on Statistical Physics of Machine Learning} which took place on Jul 4-29, 2022 in Les Houches, France.
    \item \href{https://github.com/Shmoo137/SummerSchool2021_MLinQuantum}{Jupyter notebooks} prepared as tutorials for the \href{https://ml2021.ckc.uw.edu.pl/}{Summer School: Machine Learning in Quantum Physics and Chemistry}~\cite{OurSchoolRepo}.
    \item Castellani, T. \& Cavagna, A. (2005). \href{https://doi.org/10.1088/1742-5468/2005/05/P05012}{\textit{Spin-glass theory for pedestrians}}. J. Stat. Mech. P05012. Pedagogical review on mean-field methods for spin glasses \cite{Castellani2005spinglass}.
\end{itemize}

\subsection{Quantum machine learning\index{quantum machine learning}}
\label{s:QML}

This section explores yet another direction: how quantum information and quantum hardware can be used to solve data-driven tasks. This recent field is called \acf{QML}\footnote{Often in literature, \acf{QML} incorporates both quantum-enhanced \ac{ML} and \ac{ML} applied to quantum, e.g. \ac{ML} for quantum information processing. In this section, we use \ac{QML} for quantum-enhanced \ac{ML}. A~detailed discussion about this convention can be found in \cref{sec:whatisqml}}. This field started with the development of quantum algorithms aiming for a~potential fully quantum advantage. In recent years, there has been an~increasing interest in another direction: studying hybrid quantum-classical algorithms (also often called quantum-enhanced algorithms), where part of the algorithm is performed on a~quantum device.
With the development of new experimental platforms for quantum computation, researchers are now looking for applications tailored to these hybrid algorithms and trying to determine if and how quantum advantage can arise in such systems. While the quantum advantage would represent a~breakthrough, the study of the quantum-enhanced algorithms running on these hybrid devices is an~interesting problem in itself and can potentially lead to the discovery of exciting physics. 

In the following sections, we provide an~overview of the recent advances in the field. We do not aim to provide a~complete review, but rather an~introduction to selected topics. In the last section, we refer to recent reviews of the field for the interested reader.

\subsubsection{Gate-based quantum computing}
\label{sec:gbqc}

In the following sections, we focus on the description of \stress{gate-based quantum computation}. These concepts are used throughout the whole section. 
\highlight{The most common building blocks of gate-base quantum computation are \emph{qubits} and \emph{quantum gates}. A~gate-based quantum algorithm, specified as a~sequence of gate operations and measurements performed on qubits, can be conveniently represented as a~\stress{quantum circuit}.} 

\emph{Qubits} are two-level quantum systems that can be realized by isolating two degrees of freedom in several experimental platforms, such as photonic platforms~\cite{zhong2020quantum,Madsen2022Nature}, superconducting circuits~\cite{arute2019quantum}, trapped ions~\cite{Bruzewicz_Trapped_review}, or Rydberg atoms in optical tweezers~\cite{Saffman_2016,Henriet2020quantumcomputing}. When performing a~calculation, a~quantum computer modifies the state of the qubits or entangles them with quantum gates. 

\stress{Quantum gates}  are unitary operations and can be represented by unitary matrices. The dimensions of these matrices depend on the number of qubits on which these gates act. The scaling of their dimension is exponential in the number of qubits. 

Examples of single qubit gates are the Hadamard and Pauli-X gates, which read in the single qubit basis $\{\ket{0},\ket{1}\}$
\begin{equation}
H=
\frac{1}{\sqrt{2}}
\begin{pmatrix}
1 & 1\\
1 & 1
\end{pmatrix},\, X=
\begin{pmatrix}
0 & 1\\
1 & 0
\end{pmatrix},
\end{equation}
or parametrized gates such as the single qubit rotation gate 
\begin{equation}
R_X(\theta)=
\frac{1}{\sqrt{2}}
\begin{pmatrix}
\cos{\theta} & -i \sin{\theta}\\
-i \sin{\theta} & \cos{\theta}
\end{pmatrix},
\end{equation}
parametrized in terms of the angle $\theta$. An~example of a~two-qubit gate is the controlled NOT gate (CNOT), which reads in the two qubit basis $\{\ket{00},\ket{01},\ket{10},\ket{11}\}$
\begin{equation}
\text{CNOT}=
\begin{pmatrix}
1 & 0 & 0& 0 \\
0 & 1 & 0 & 0 \\
0 & 0 & 0 & 1 \\
0 & 0 & 1 & 0
\end{pmatrix}.
\end{equation}
\begin{figure}[t]
    \begin{center}
    \includegraphics[width=0.4\columnwidth]{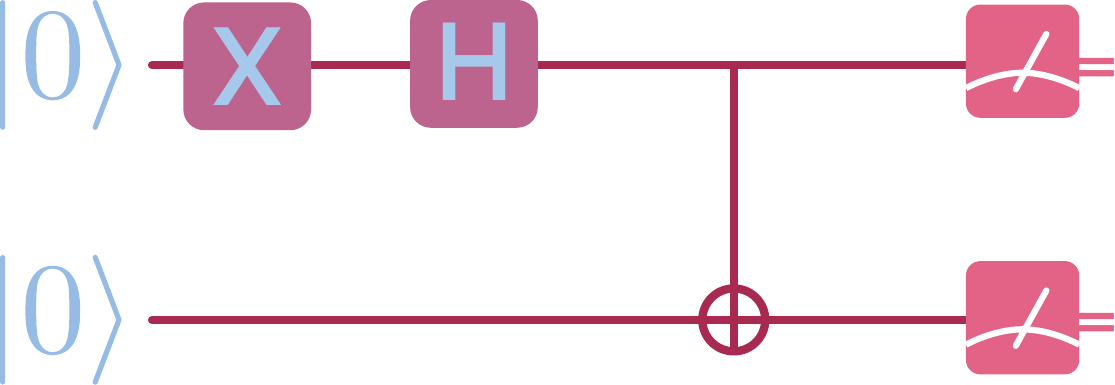}
    \end{center}
    \caption[Illustration of a~quantum circuit diagram]{Illustration of a~quantum circuit diagram with two initialized qubits, $q_0$ and $q_1$, and three different quantum gates: the Hadamard gate H and the $\sigma_x$ gate X, both acting on $q_0$, and a~two-qubit gate (CNOT). The final element is the measurement on $q_0$ and $q_1$.}
    \label{fig:QML-diagram}
\end{figure}
In general, \stress{quantum circuits} can be depicted with quantum diagrams, as exemplarily shown in \cref{fig:QML-diagram}. Each line corresponds to a~qubit. This circuit has two gates acting on a~single qubit ($X$ and $H$) and an~entangling gate (CNOT) acting on two qubits. The last part of the diagram is the measurement, which is an~interaction with individual qubits that forces their collapse to one of the two levels. As the measurements are destructive, the careful choice of a~set a~measurements is necessary to propely extract the relevant information from the quantum circuit.

\subsubsection{What is quantum machine learning?}
\label{sec:whatisqml}

\begin{figure}[t]
    \begin{center}
    \includegraphics[width=0.9\columnwidth]{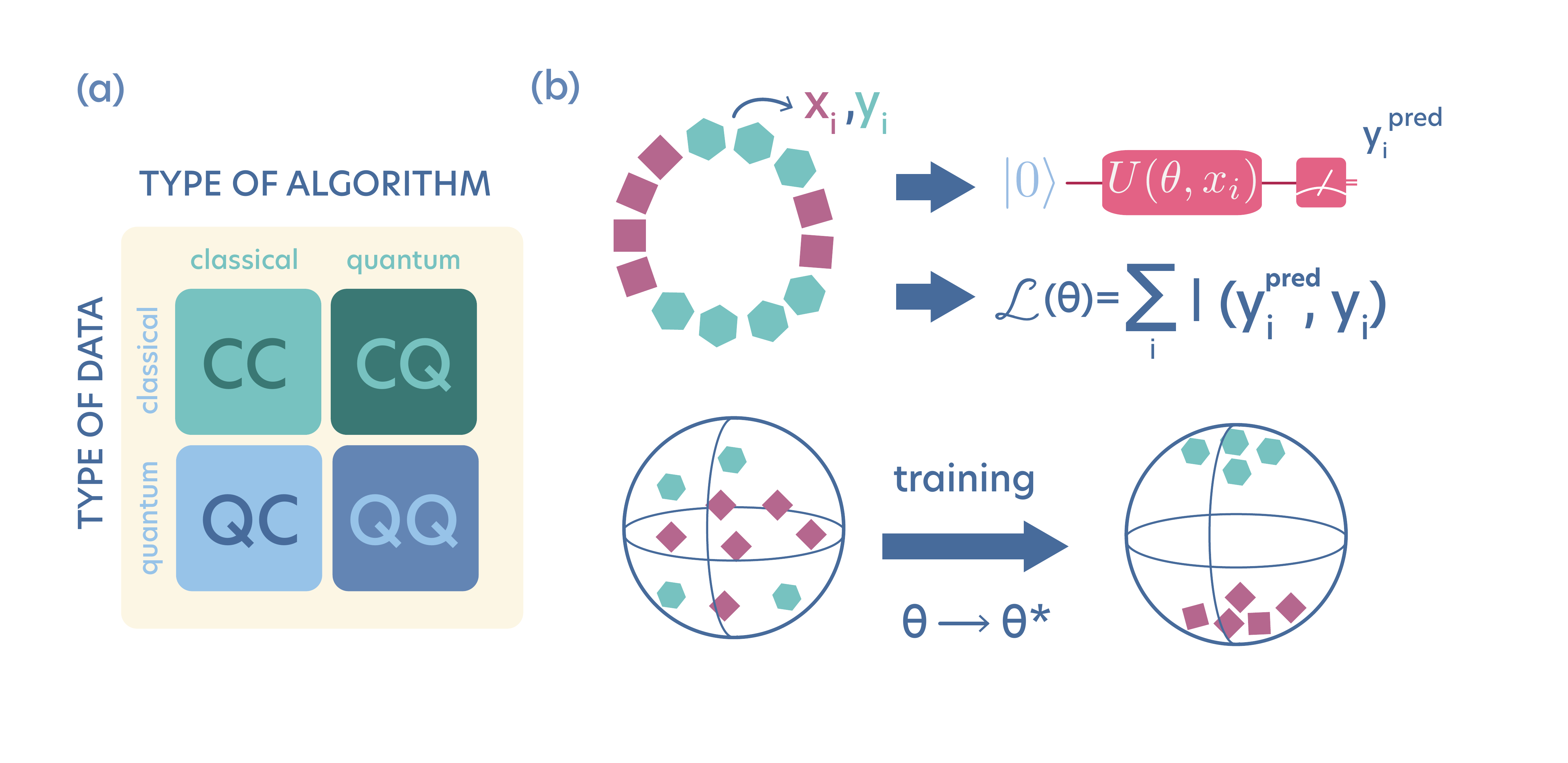}
    \end{center}
    \caption[Quantum machine learning]{(a) Table of the different types of data and algorithms. (b) Sketch of an~example of a~classification task on one qubit.}
    \label{fig:qml_intro}
\end{figure}

To better understand \acf{QML}, let us first have a~look at \cref{fig:qml_intro}(a). Generally, \ac{ML} algorithms are run on classical data, e.g., image classification or natural language processing. We have thus a~classical algorithm dealing with classical data (CC). This book focuses mainly on the case of classical \ac{ML} applied to quantum data (CQ), e.g. quantum states. On the other hand, \acf{QML} deals with the integration of quantum devices in \ac{ML} algorithms. Therefore, the algorithms can be quantum, and the data can be either classical (QC) or quantum (QQ). In this section, we focus mainly on the QC side, as the QQ side is only at its early stage of development~\cite{Liu2021quantspeed,210905909,211200778,220105957}.

Let us discuss an~elementary example to introduce the revised building blocks of \ac{ML} in the context of the QC \ac{QML}. We consider the classification problem of one-dimensional data on a~ring. We intuitively sketch each step of this \ac{QML} classification problem in \cref{fig:qml_intro}(b). Firstly, the classical data $(\vect{x}_i,y_i)$ is encoded in a~quantum computer. Here, for example, we encode the data points on a~single qubit through the action of a~parametrized unitary $U_{\params,\vect{x}_i}$, where $\params$ are the parameters of this unitary transformation (for example $R_X(\theta)$ introduced in \cref{sec:gbqc}). Then, a~measurement is performed, and one can define the output of the measurement $y_\text{pred}$ as a~label (here $1$ or $0$). We then construct a~loss function $\lossfun (\params)$ depending on the predicted and ground-truth labels. We can see here that the \emph{quantum-enhanced} part corresponds to the evaluation of $y_\text{pred}$ on a~quantum computer. Once the loss function is defined, the minimization can be performed on a~classical computer with the method of your choice, such as gradient descent or a~gradient-free optimizer (e.g., Nelder-Mead). 
In this simple example, the training has a~simple interpretation. Initially, the weights $\params$ of the unitary are randomly distributed. Consequently, the mapping of our classical data to the qubit is randomly distributed on the Hilbert space. The optimization procedure aims to push the two classes toward the opposite poles of the Bloch sphere. Therefore, for the weights after training $\params^*$, we expect that data on the Bloch sphere is much more ordered.

\subsubsection{Ideal quantum computers}
\label{ssec:perfectq}\index{support vector machine!quantum support vector machine} 
Computational complexity theory is a~field of computer sciences that focuses on classifying computational problems in terms of the resources they need. In particular, classical computers are known to excel at solving problems belonging to two complexity classes: solvable in polynomial time (P) and bounded-error probabilistic polynomial time (BPP). Having an~ideal quantum computer, a~natural question arises: what types of problems, if any, can be solved in a~polynomial time on a~quantum computer while taking an~exponential time on a~classical computer? In this context, another complexity class was defined and, roughly, includes all problems which can be solved and verified with a~quantum computer in polynomial time (BQP).  

One of the first quantum algorithm with an~exponential speed-up has been proposed in the context of discrete Fourier transform. The quantum Fourier transform algorithm~\cite{892139} performs the discrete Fourier transform on $2^{n}$ amplitudes using a~quantum circuit consisting of only $\mathcal{O}(n\log(n))$ quantum gates. The classical algorithm needs $\mathcal{O}(n2^{n})$ operations to perform the same task. Another example of an~algorithm with such a~speed-up is the Shor algorithm for efficient number factorization~\cite{365700}. It uses building blocks from the quantum phase estimation algorithm \cite{Kitaev:1995qy} and the quantum Fourier transform to gain an~exponential speed-up with repsect to the best classical algorithm for this task. The Harrow-Hassidim-Lloyd (HHL) algorithm~\cite{HHL_2009} is another very famous algorithm that was designed to solve a~system of linear equations
\begin{equation}
    \mat{A}\vect{x}=\vect{b},
\end{equation}
where $\mat{A}$ is an~$n\times n$ sparse matrix with condition number $k$. The algorithm is able to find the vector $\vect{x}$ in $\mathcal{O}(\log (n)\, k^2)$ time instead of the typical $\mathcal{O}(n^2 )$ for standard algorithms. This is an~exponential speed up in the size of the system. However, one crucial remark to keep in mind is that the classical algorithm returns the solution directly, while, in the HHL algorithm, the solution is encoded in a~quantum state that needs to be repeatedly measured to read it out.

\Acf{ML} algorithms largely rely on linear algebra, which generally constitutes much of machine learning computational cost. For example, the classification problem with a~\acf{SVM} generally requires quadratic programming (see \cref{sec:intro-SVM} for the general idea of \ac{SVM}) but a~special form of \acf{SVM}\footnote{For the special case of least-squares support-vector machine, the problem can be written as a~solving a~linear system of equations.} boils down to solving a~system of linear equations. In this context, quantum computers might speed up such costly operation. One application of the HHL algorithm have been proposed, e.g., in the context of \acp{SVM}~\cite{QSVM_2014} (see~\cite{PhysRevLett.114.140504} for a~recent experimental realization on a~four-qubit quantum computer) and data fitting~\cite{Data_Fitting_2012}. It is worth noticing that the exact amount of quantum speed-up provided by the HHL algorithm with respect to classical algorithms is under debate~\cite{gilyen2018quantum}.
Another caveat consist in the fact that the classical data should be efficiently encoded in the quantum algorithm efficiently. This is another issue that must be solved by the community. 

\subsubsection{Noisy intermediate-scale quantum era}\label{sec:qml_nisq_era} \index{noisy intermediate-scale quantum era}

Until now, we have only considered ideal quantum computers to run quantum algorithms, such as the Shor, the quantum Fourier transform, and HHL algorithms. However, the realization of these algorithms for a~number of qubits where such advantage matters is not yet feasible in near-term quantum computers. The main reasons are that (i) quantum computers currently contain \stress{too few qubits} (nowadays in the order of hundreds) and (ii) they perform imperfect operations (noisy). 

\begin{figure}[t]
    \begin{center}
    \includegraphics[width=\columnwidth]{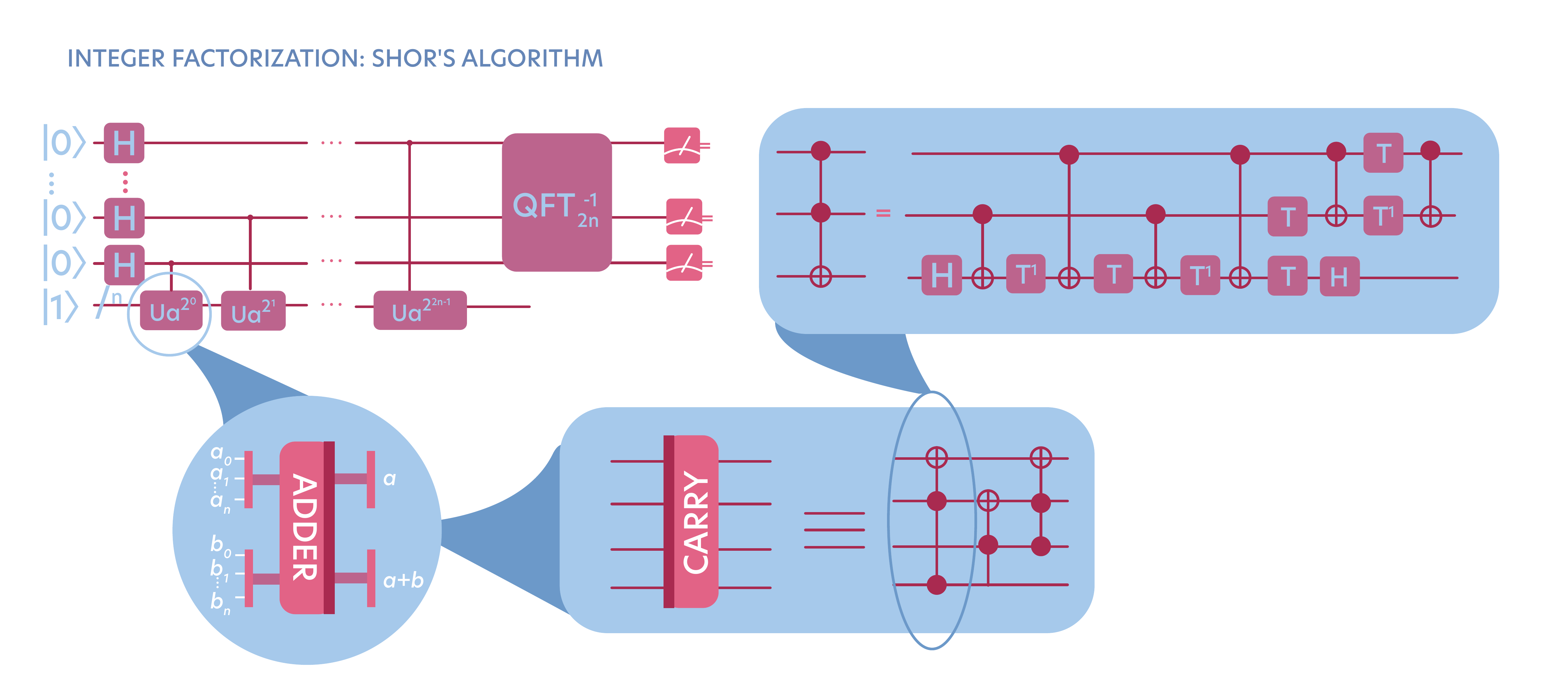}
    \end{center}
    \caption[Realization of the famous Shor algorithm in a~real quantum computer]{Realization of the famous Shor algorithm in a~real quantum computer. Top left diagram presents a~concise theoretical circuit of this algorithm. Due to the limitation to certain gates (CNOT and SWAP), generic gates have to be decomposed and the circuit requires more gates and higher depths.}
    \label{fig:QML-shor}
\end{figure}

Furthermore, algorithms such as the Shor algorithm have to be compiled on real devices. This means that unitaries acting on several qubits have to be \stress{decomposed in elementary gates} that can be physically realized in the experimental platform. Such a~transformation might lead to complex quantum circuits with native gates~\cite{RevModPhys.68.733}. For example, for the \href{https://quantum-computing.ibm.com/composer/docs/iqx/guide/}{IBM-Q} Washington platform, only the CNOT, ID, RZ, $\sqrt{X}$, and X gates are native gates. Any other gate must be decomposed into these gates. Since these gates form a~set of universal quantum gates~\cite{dawson2005solovaykitaev}, this is, in theory, sufficient but, in practice, it can lead to very complex quantum circuits with a~large number of gates. For example, we consider the decomposition of the Shor algorithm to these gates, as shown in \cref{fig:QML-shor}. An~apparently simple circuit consists, in practice, of many operations on real quantum devices. The latter might be especially problematic due to \stress{noise and decoherence} that are intrinsically present in the physical devices.

In modern quantum computers, we can identify three primary sources of errors: gate errors (generated by a~non-precise application of the desired gate), decoherence errors (loss of coherence of the wave function as a~function of time), and read-out errors (erroneous read-out of the qubits state during the measurement procedure). 

\highlight{Due to many different noise and error sources in real quantum computers, we are far from the fault-tolerant quantum computation. Instead, we are in the so-called \emph{\acf{NISQ}} era~\cite{Preskill2018quantumcomputingin}. The qubits of the current quantum processors are noisy and require quantum error correction. Nevertheless, the study of the physics of such systems is interesting in itself. In particular, there might be applications with a~quantum speed-up within this regime, as in the case of the recent quantum advantage experiment~\cite{arute2019quantum}.}

It is now clear that we cannot run algorithms requiring many gates or implement gates with low error rates in \ac{NISQ} circuits. If the circuit contains too many gates, the coherence gets lost as well as the superposition and entanglement between different qubits. A~natural question arises: Can we design algorithms that perform well on NISQ devices and do not require fully error corrected quantum computers? This means one has to find clever ways to explore the exponentially big Hilbert space without exact algorithms. One approach is using quantum computers to generate variational states and to find a~procedure to converge iteratively to the solution instead of taking a~direct deterministic path (for example by performing the optimization on a~classical computer). We go into more detail into these variational approaches in~\cref{sss:QML-variational-approaches}. Before, in~\cref{sec:qml_kernels}, we present how \ac{NISQ} devices can be used for \ac{SVM} with kernels.

To sum up, the \ac{NISQ} era still has many open problems in experimental quantum computing and in quantum error correction. As of 2024, state-of-the-art devices include 100+ physical qubits with error rates of less than $0.2\%$\footnote{For example, in the random circuit sampling experiment from 2024~\cite{morvan2024}, the authors reported single-qubit gate errors of $0.04\%$ and two-qubit gates errors of $0.14\%$ for superconducting qubits. These errors are of the same order of magnitudes as the ones recently reported for a neutral atom quantum processing unit~\cite{bluvstein2023}.}. Nonetheless, recent years showed many examples of useful variational quantum simulations that can be performed with the near-term devices, e.g., see Ref.~\cite{Muschik_2021}. Moreover, many error mitigation routines have been developed to ease the noise effects in quantum computers, allowing for extraction of useful information from noisy devices in the near term \cite{gambetta_mitigation,Li_mitigation_2018,funcke2020measurement,Corcoles_2015}.
\ac{NISQ} devices are also an~excellent trial field to study physics without building a~fault-tolerant quantum computer. Finally, useful applications of \ac{NISQ} devices can still be found, and they can be considered as a~step toward fault-tolerant quantum computing.

\subsubsection{Support vector machines with quantum kernels}\label{sec:qml_kernels}

\begin{figure}[t]
    \begin{center}
    \includegraphics[width=\columnwidth]{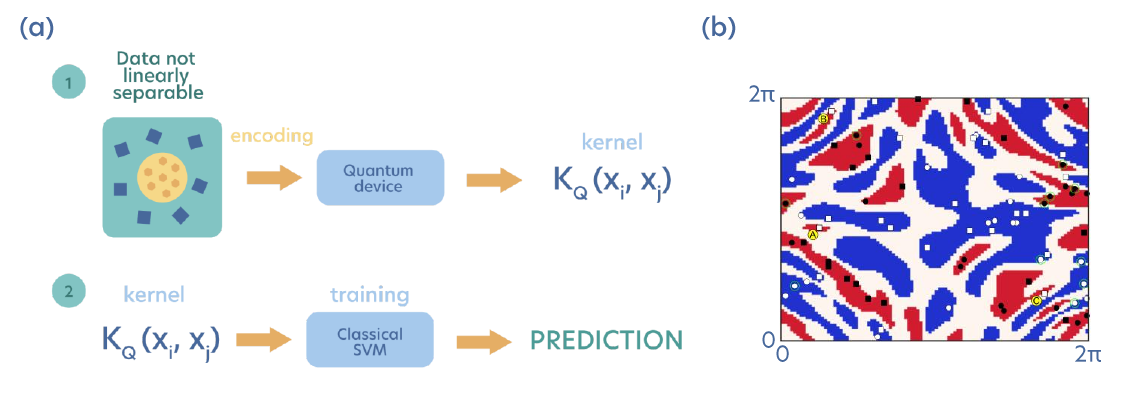}
    \end{center}
    \caption[Quantum support vector machine enhanced by a~quantum device]{Quantum \ac{SVM} enhanced by a~quantum device. (a) Sketch of the steps of the \ac{SVM} enhanced by a~quantum kernel. The data are encoded in a~quantum device, such as a~quantum circuit, which computes the kernel. These kernels are then used in classical \ac{SVM}. (b) Example of a~dataset used in Ref.~\cite{havlivcek2019} to show the capacity of quantum kernels. Blue (red) regions correspond to label 1 (0). \ToggleForCUP{Taken from Havl\'{i}\v{c}ek, V. \textit{et al.} (2019). \textit{Supervised learning with quantum-enhanced feature spaces}. Nature 567, 209 with permission from Springer Nature.}{}}
    \label{fig:qsvm}
\end{figure}
We have seen in~\cref{ssec:perfectq} that ideal quantum computers could allow one to accelerate the numerically costly parts of the \acf{SVM} algorithm by implementing the HHL algorithm. There, the key element has been to use the quantum computer to solve the linear system of equations. In 2018, three independent works~\cite{schuld2019,havlivcek2019,kitchen_sinks} followed an~interesting alternative direction: using kernels evaluated directly on quantum devices, while performing the rest of the \ac{SVM} algorithm classically.

The idea is sketched in \cref{fig:qsvm}(a). Let us consider a~dataset that is not linearly separable. We therefore want to nonlinearly embed it in a~higher dimensional space such that the data is linearly separable in this space (see \cref{sec:kernel_methods} for more detail). We here use a~quantum device to encode classical data $\vect{x}$ into a~high-dimensional Hilbert space $\vert \psi(\vect{x})\rangle$, or even infinite in the case of squeezed states considered in Ref.~\cite{schuld2019}. In this case, the choice of the encoding of the classical state into the quantum state is crucial as it determines the quality of the feature map. More importantly, quantum devices and in particular quantum circuits can allow for the efficient computation of the scalar product between two quantum states, which allows one to define a~quantum kernel 
\begin{equation}
\label{eq:quantum_kernel}
K_Q(\vect{x}_i,\vect{x}_j)=\vert\langle \psi(\vect{x}_i) \vert \psi(\vect{x}_j)\rangle\vert^2 = \sum_n \lambda_n \featmap_n(\vect{x}_i) \featmap_n(\vect{x}_j),
\end{equation}
which has all the properties of a~classical kernel with a~feature map $\featmap$ and defines an~\ac{RKHS} (see~\cref{sss:kernel-RKHS}).\footnote{The careful reader may notice that \cref{eq:quantum_kernel} is the norm squared of the inner product instead of the typical inner product expected for kernels. This becomes clearer when writing the kernel in terms of density matrices $K_Q(\vect{x}_i,\vect{x}_j)=\text{Tr}(\rho_i\rho_j)$. The kernel is then corresponding to the Frobenius inner product of density matrices $\rho_i$ and $\rho_j$.}
As such, quantum kernels can be directly used in classical algorithms such as k\ac{PCA} or k\ac{SVM} or \acfp{GP} \cite{dai:2022} rendering them quantum algorithms.

To be more concrete, we explain the main ingredients of the quantum kernel introduced in Ref.~\cite{havlivcek2019}. Given a~data set of points $\{\vect{x}_i\}$ with labels $\{y_i\}$, the feature map is defined in terms of the unitary transformation $U(\vect{x_i})$ that can be realized in a~quantum circuit of qubits
\begin{equation}
\vect{x}_i \mapsto \vert \psi(\vect{x}_i)\rangle=U(\vect{x_i})\vert 0 \rangle,
\end{equation}
where $\vert 0 \rangle$ stands for the product state $\vert 0 \rangle ^{\otimes n}$. Typically, the classical data encoding into the quantum circuit can be done through parametric local rotations of single qubits. The unitary is then built through repeated application of these data dependent gates and other non-parametric gates, such as entangling gates and Hadamard gates. We do not enter into the details of the construction of the circuit, but the interested reader can have a~look at the following Qiskit tutorial for more details~\cite{qiskit2022} .

The quantum kernel can then be computed on a~quantum circuit with the compute-uncompute trick: one basically implements the following quantum circuit $U^\dagger(\vect{x}_j)U(\vect{x}_i)\vert 0\rangle$ and measures it in the $z$ basis. The frequency of the all-zero outcome, therefore, gives an~estimate of the kernel $K_Q(\vect{x}_i,\vect{x}_j)$. 

Given these kernels, the optimization of the parameters of the \ac{SVM} can be performed on a~classical computer (see \cref{sec:kernel_SVM}) using, e.g., Bayesian optimization \cite{dai:2022} presented in \cref{ss:GPR-BO}. Ref.~\cite{havlivcek2019} generated a~complex classification problem, shown in \cref{fig:qsvm}(b), where the blue (red) region corresponds to label 0 (1). They then generated a~training set by selecting random points in these regions and performed the \ac{SVM} enhanced by the quantum kernel. The algorithm yields very good results with around $95\%$ of accuracy on the test set for this synthetic data set.    

The previous example shows that quantum kernels can represent complex data sets. However, a quantum advantage has yet to be observed for a~general data set~\cite{tang:2021,kubler:2021}. A~recent important step in this direction has been achieved in Ref.~\cite{Liu2021quantspeed}, where the authors constructed data sets that cannot be classified efficiently on a~classical computer and in~\cite{Haug_large_scale} where the authors have studied supervised learning of handwritten images on quantum computers with an improved scaling using randomized measurements. Along the same lines, finding optimal ways to construct quantum kernels~\cite{Representation_via_Quantum_Tangent_Kernels}, i.e., how one should perform the encoding of the inputs $\vect{x} \mapsto \ket{\psi(\vect{x})}$, is still an active line of research~\cite{havlivcek2019}. In Ref.~\cite{torabian:2023}, for example, the authors construct quantum kernels for \ac{SVM} algorithms based on the \acf{BIC} (see~\cref{sss:BIC}) as a selection metric. Using the quantum kernels constructed in this fashion, the \acp{SVM} achieve significantly higher performance in selected classification problems compared to optimized classical models with conventional kernels.

\subsubsection{Variational approaches}\label{sss:QML-variational-approaches}

This section deals with the optimization of quantum circuits that can be realized in \ac{NISQ} devices. In particular, we focus here on the so called variational quantum algorithms. This idea generalizes the toy example we introduced in \cref{sec:whatisqml}. We define a~\acf{PQC}, a~circuit that depends on a~set of parameters $\{\params\}$ ($\params$ can be, for example, the angles of single qubit rotations). Then, one defines an~objective function $C(\params)$ that we aim to minimize. Such an~objective function can always be written as a~function depending on a~set of observables and on the \ac{PQC}. Our goal is then to find the optimal set of parameters $\params^*$ minimizing the objective function. Such variational approach has applications in many fields such as \ac{ML} (classification, generative models), many-body physics and quantum chemistry (ground state finding), combinatorial optimization, etc.  

\paragraph{Variational Quantum Eigensolver (VQE)}

\begin{figure}[t]
    \begin{center}
    \includegraphics[width=0.95\columnwidth]{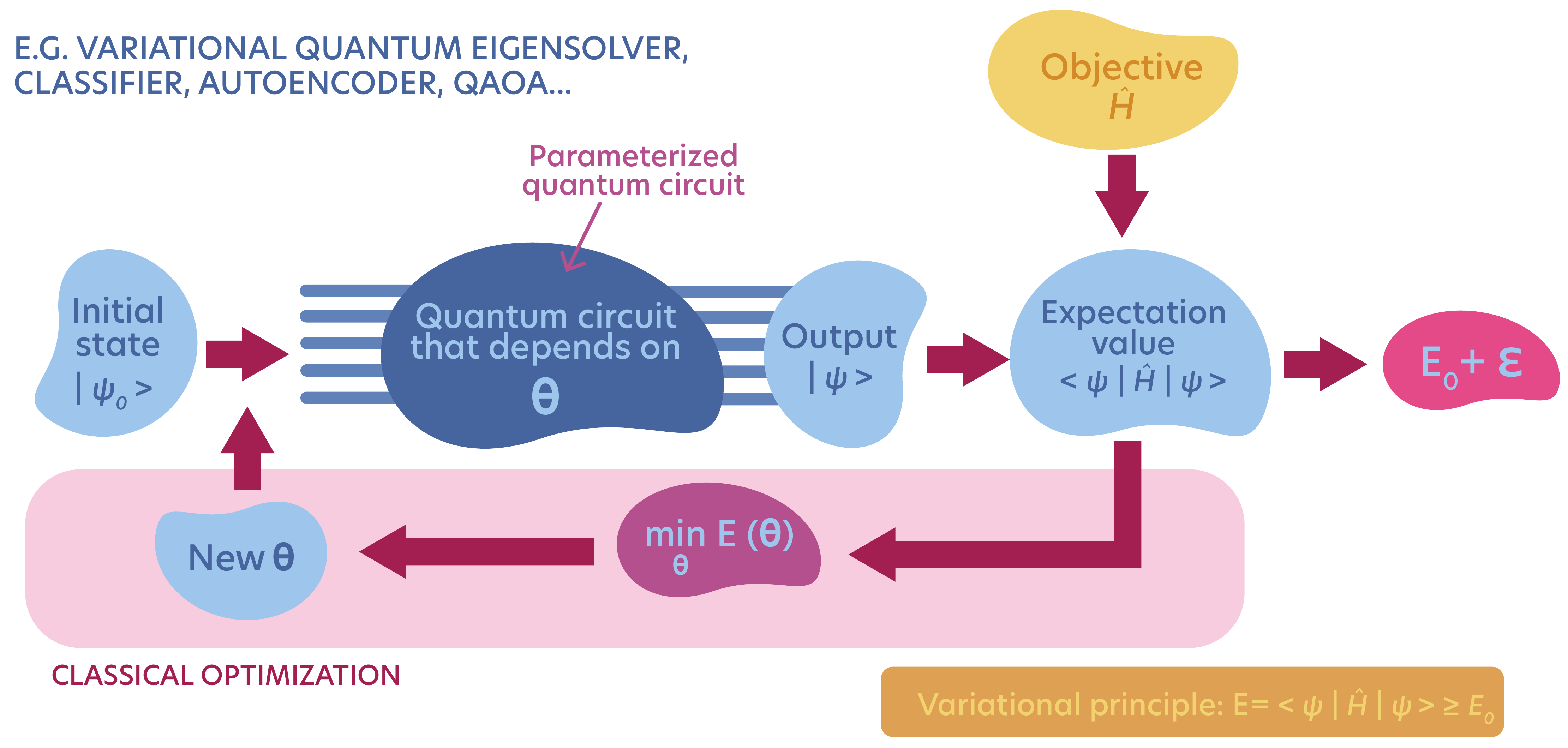}
    \end{center}
    \caption[Variational optimization of quantum circuits]{Variational optimization of quantum circuits, including the initial state, \acf{PQC}, output, objective, measurement, and classical optimization.}
    \label{fig:QML-PQC}
\end{figure}

We illustrate the building blocks of the variational approach with the example of the \acf{VQE}~\cite{VQE2014} \index{variational quantum eigensolver}. Given an~Hamiltonian $\hat{H}$, the goal of the \ac{VQE} is to minimize the energy $E = \langle \psi | \hat{H} | \psi \rangle$. The general principle of the \ac{VQE} is shown in \cref{fig:QML-PQC}. The process starts with an~initial state which is easy to prepare on the quantum computer, e.g., the product state $\ket{0}^{\otimes n}$, which for simplicity we denote by $\ket{0}$. This is followed by the \acf{PQC} including, e.g., all the parameters $\params$ of the quantum gates and which can be seen, at this point, as a~black box which prepare a~quantum state. In the first iteration, this state is just a~random state and is used as the initial state for the expectation function, which is in general the Hamiltonian of the physical system we want to study, up to a~global phase. The Hamiltonian can describe the interaction within a~molecule or a~spin system, but can, in general, be any kind of cost function in operational form that can be written in the computational basis of the quantum hardware we are using. The next step is the minimization of the cost function with a~classical subroutine to converge toward the lowest energy state of the physical system in the space of the quantum states that can be reached by our \acf{PQC} in a~self-consistent manner. We also know that, if the system is gapped and the ground state unique, the minimal value of the expectation value of the Hamiltonian is the ground-state energy and the corresponding eigenvalue is the ground state wave function.\footnote{The energy measured at each iteration is an~upper bound of the ground state energy, according to the variational principle. The ability to reach the global minimum, of course, depends on the capacity of the circuit. If the circuit does not contain the solution, such an~ansatz will never reach the minimal energy \cite{Funcke2021dimensional}. }

Going a~bit more into detail, for the \acp{PQC} we are computing the energy of the ansatz
\begin{equation*}
 E_0 = \min_{\params} \langle \psi(\params) | \hat{H} | \psi(\params) \rangle = \min_{\params} \langle 0 |U^\dagger(\params) \hat{H} U(\params)| 0 \rangle \, ,
\end{equation*}
where $\params$ are the parameters of the gates which are optimized to minimize the expectation value. The variational state is the unitary state, i.e., our \stress{\acf{PQC}} applied to the initial state $|0\rangle$. However, in order to run and work with the \ac{PQC} we have to make several assumptions: first, we are assuming the existence of a~set of parameters that approximates the ground state and that our \ac{PQC} can represent that specific solution. Second, that it is possible to converge to the solution without being stuck in local minima and, finally, that the circuit can be run on a~\ac{NISQ} computer. Taking all these assumptions into account, there are two ways to design a~\ac{PQC}. The first option is the problem-inspired design which we can use when we exploit some physical properties of the system we want to represent, e.g. by using the Hamiltonian representation to design the unitary operation as happens in the \acf{VQE} algorithm. However, these kinds of ansatz require, in general, many gates or a~particular qubit connectivity, making it unfeasible for bigger systems in current quantum computers. Another way is the hardware-efficient ansatz, which is a~heuristic method that requires way less quantum gates and that consists of preparing a~\ac{PQC} that uses the native gate set and respects the quantum computer connectivity. In general, problem-inspired ansatz use to be more precise but less feasible to implement in current quantum computers, while this happens otherwise with the hardware-efficient ansatz

The next step after defining the \ac{PQC} is the choice of the \stress{objective function} which can be everything that encodes our problem in a~quantum operator, e.g., a~Hamiltonian as shown in \cref{fig:QML-PQC}. The objective function is then decomposed into Pauli strings because their individual expectation values can be measured with the quantum computer.
This measurement procedure the next important step in the process. In this step, we extract the value of the objective function from our quantum computer or our quantum devices.
This in itself is a challenging task as the objective function often cannot be measured directly.
Instead, the expectation value of a Pauli string is computed on the quantum hardware by making the wave function collapse in the corresponding Pauli basis. From this measurement, we can extract bit strings (i.e. lists of 0~and 1, e.g., $[(0,0,1,1,0,0),(1,0,1,0,0,0),(0,0,1,0,1,0)]$). Combining many bitstrings from individual Pauli measurements, we can reconstruct the expectation value of any objective function that can be written as the tensor product of Pauli matrices. Finding an suitable measurement strategy that requires the least amount of bit strings to accurately yield the value of the objective function is an ongoing research endeavor~\cite{gresch2023guaranteed}.

\begin{figure}
\includegraphics[width=\columnwidth]{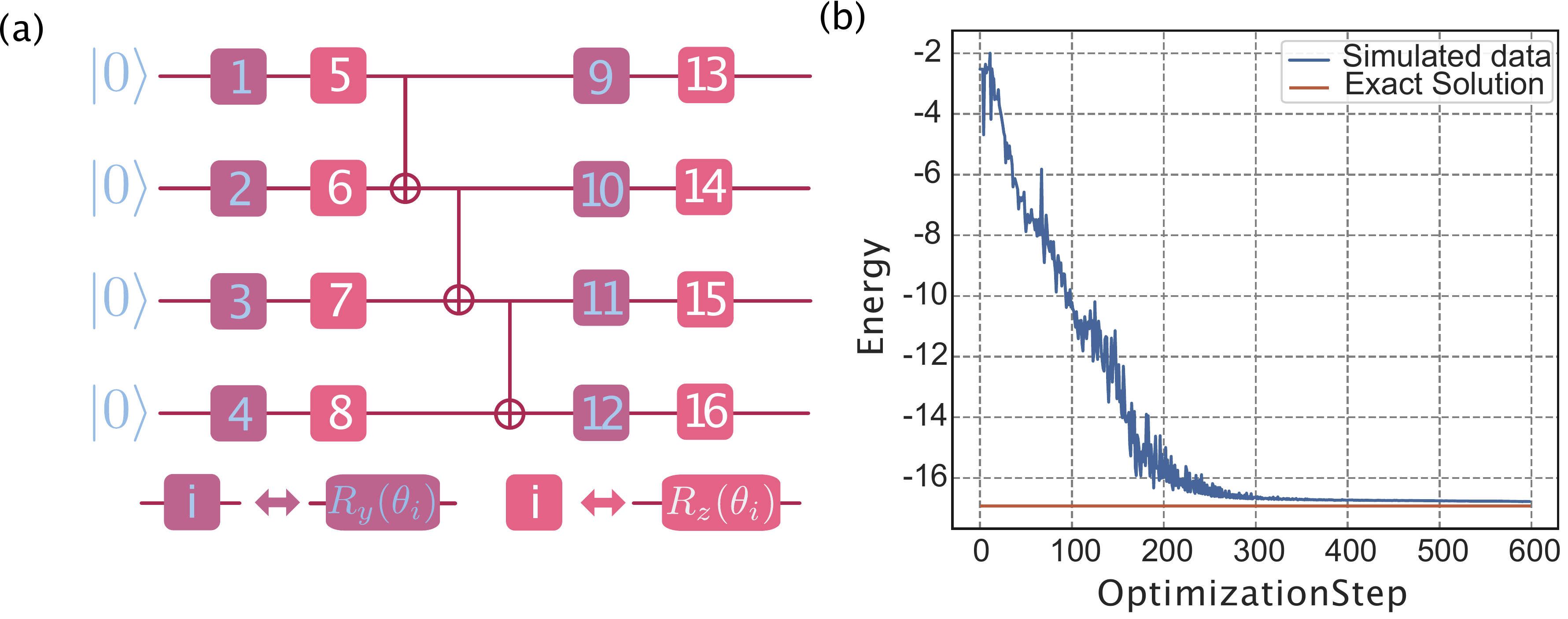}
\caption[Variational quantum circuit]{An~example of a~variational quantum simulation. (a) The parametric quantum circuit used for the simulation. (b) The energy of the system during the optimization algorithm.}
\label{fig:variational_quantum_circuit}
\end{figure}

The last step is the classical optimization in which we have to navigate through the \ac{PQC} parameter space by using, e.g., a~gradient-based approach. The gradients can be written in terms of expectation values of the quantum circuit derivatives with respect to a~parameter, and we do not have direct access to the gradients of the quantum state. In \ac{NISQ} devices, the gradient of the \ac{PQC} can be computed with the parameter shift rule \cite{PhysRevLett.118.150503}: for each parameter $\theta$, one can compute exactly its partial derivative by evaluating two \acp{PQC}.\footnote{In comparison, computation of the gradients of the loss of classical \acp{NN} requires only a single evaluation of the model, thanks to the existence of reverse mode differentiation, also known as backpropagation; see~\Cref{sec:backprop}.} The problem in this step is the number of required measurements. In order to run the classical minimization algorithm, measurements of all gradients are required. Therefore, this method can be very expensive and includes a~huge number of variables and multiple iterations to converge to the ground state, and that is why other gradient-free methodologies are exploited, like genetic algorithms or reinforcement learning strategies. After all, the combination of the variational optimization of the quantum circuit with classical optimization algorithms is an~efficient way to use \ac{NISQ} devices for real world problems.

Let us discuss a~concrete example: the Heisenberg Hamiltonian acting on four spins, which reads
\begin{equation}
    \hat{H} = \sum_{i=1}^{3} \left( J_1 \sigma^x_i\sigma^x_{i+1} + J_2 \sigma^y_i\sigma^y_{i+1} +J_3 \sigma^z_i\sigma^z_{i+1} \right) +\sum_{i=1}^{4} \left( h_1\sigma^x_i+ h_2\sigma^y_i + h_3\sigma^z_i \right)\, .
    \label{eq:heisenberg_hamiltonian}
\end{equation}
We fix the parameters of the Hamiltonian to $\vect{J}=[1, 1, -1]$ and  $\vect{h}=[1, 1.5, 3]$. These parameters are chosen to be far away from any phase transition point, not to make the problem too difficult. We can use this Hamiltonian as a~benchmark for our algorithm. Let us also consider a~very easy quantum ansatz for the four-qubit case. The ansatz consists in two rotation gates, applied to every qubits in Y and Z direction, three CNOT gates that connect every qubits, and two more rotations in Y and Z direction. The circuit is sketched in \cref{fig:variational_quantum_circuit}(a). 

In \acfp{VQE}, we want to use the variational circuit to minimize the energy of the system. If we did this operation on a~classical computer, the computational cost of the evaluation of the energy would scale exponentially with the number of qubits. The expectation value of the Hamiltonian of the system can be computed efficiently on a~perfect quantum computer. The computational cost is linear in the number of qubits. We can feed this cost function to an~optimizer; in this case, we use the Nelder-Mead optimization routine. In \cref{fig:variational_quantum_circuit}(b), we plot the value of the energy of the system as a~function of the optimization step.  We can find a~relatively good approximation of the ground-state energy with few variational parameters and polynomial computational cost in the number of qubits. Many things can be improved in these kinds of simulations, both on the design of the variational ansatz, and on the optimization routine. In the design of variational quantum circuits, we can, for example, impose symmetries of the system we are studying. In the optimization routine, we could use gradient-based methods, such as \ac{SGD}, or gradient-free methods, such as \acf{BO} from \cref{ss:GPR-BO}. For instance, in Ref.~\cite{nicoli2023physics} the authors derive a novel kernel, inspired by the parameter shift rule to update the \ac{VQE} parameters accurately with less measurements on the quantum computer. We refer to \Cref{sec:BO_for_VQE} for more details.

\subsubsection{Parametrized quantum circuits for quantum machine learning}

\paragraph{Classification tasks} Variational quantum circuits can be used to perform the classification of classical data. The first non-trivial task in the construction of such a~quantum algorithm is the loading of the classical data on the quantum hardware. Moreover, the algorithm must be able to process efficiently these data and have a~way to perform the classification. In Ref.~\cite{PerezSalinas2020datareuploading} the authors have shown that \ac{PQC} with data reuploading can lead to a~good classifier. In Ref.~\cite{QCNN_2019} the authors have introduced the concept of quantum convolutional neural networks. As for classical \ac{CNN}, these variational quantum circuits have more capacity. In particular, the authors have shown how these circuits can be used to perform classification on symmetry-protected topological phase in the Haldane chain directly from the quantum state, that can be obtained with a~\ac{VQE}. This idea has been realized experimentally in a~recent work~\cite{210905909}. Moreover, it is worth to notice that the above-mentioned classifiers are closely related to quantum kernels~\cite{210111020,qml_beyond_kernel}. Finally, the authors of Ref.~\cite{Bausch2020RQNN} have developed a recurrent quantum \ac{NN} that has been used for classification and generation of handwritten digits.

\paragraph{Quantum \acf{RL}}\label{sec:quantum_rl} \Acfp{PQC} can also be used to realize action-value functions, or \ac{RL} policies themselves (see \cref{sec:rl_value_based_methods} for Q-learning and deep Q-learning  and see \cref{sec:rl_policy_gradient} for an~introduction to policy gradient). Two examples of algorithms that take advantage of \acp{PQC} in the context of \ac{RL} can be found in Ref.~\cite{jerbi2021parametrized,skolik2021quantum}. In Ref.~\cite{jerbi2021parametrized}, the quantum circuit is trained using a~policy gradient algorithm and is used to solve classical environments, i.e., to find the optimal policy for the task at hand. The choice of the action, the probability of which occurring we want to fix, is going to be encoded in the measured observable -- if we have a~certain set of actions, we define a~certain set of observables. The \ac{RL} architecture states that the agent observes a~quantum state that the circuit will produce and the expectation value of such observable is going to encode $\pi(a|s)$ -- the probability of the action $a$ in a~given \ac{RL} state $s$, i.e., it corresponds to the policy $\pi$. In Ref.~\cite{skolik2021quantum} the authors have used the OpenAI Gym \cite{brockman2016openai} examples as benchmark environments for the variational quantum algorithm for deep Q-learning, as for example, the \stress{Cartpole} game (a cart that can move left and right and the agent is trying to balance the pole attached to the cart). Compared to the classical models (for which the exemplary environments were created), the quantum models can reach a~similar accuracy using much fewer parameters (which alone does not imply better models but highlights their difference). An~interesting and open question would be to understand whether these results can be generalized to other environments.

\paragraph{Quantum autoencoders}\index{autoencoder!quantum autoencoder}\label{par:quantumAEs}
In the same spirit as their classical counterpart (see \cref{sec:autoencoders}), quantum autoencoders~\cite{Romero2017} are used to compress quantum data on a~quantum computer. Quantum autoencoders act directly on data encoded in qubits and can thus be used to compress quantum data without needing to have a~classical representation that would have an~exponential cost. Since there are patterns that classical computation cannot generate, e.g. entanglement, the quantum version of an~autoencoder might be able to recognize patterns beyond classical capabilities. The encoding done with quantum autoencoder transforming quantum data into a~latent space with fewer qubits. Having a~set of quantum data that encoded into $n$ qubits, we aim to find the representation of the $n$-qubit state in a  latent space formed by $k = n - p$ qubits. The $p$ qubits are so-called \textit{trash qubits}. 
This encoding is done via a~variational map represented by a~quantum circuit with a~polynomial number of parameters $U(\theta)$. Since the encoding procedure is a~unitary (quantum circuit), the decoding operator is just represented by the Hermitian conjugate of $U^\dagger(\theta)$. It is still under debate whether this quantum algorithm can have a~computational advantage with respect to their classical counterpart. Variational quantum autoencoders have also been proposed for various applications, such as quantum data denoising~\cite{Quantum_Autoencoders_to_Denoise}, phase classification \cite{Kottmann2021PRR}, clustering of the Hilbert space \cite{Szoldra2022}, or quantum error correction~\cite{2202.00555}.

\paragraph{Generative models} 
Generative models are algorithms learning the distribution of a~data set. In quantum  mechanics, the inherent quantum nature of the devices can be of great help for learning probability distributions and in particular for quantum wave functions. Recently, diverse \ac{QML} architectures have been proposed: these include quantum Hamiltonian-based models~\cite{1910.02071}, quantum \acp{GAN}~\cite{PhysRevLett.121.040502,PhysRevA.98.012324} and quantum Born machines~\cite{PhysRevA.98.062324,Born_supremacy_2020,generative_modeling_QML_2019_npj}. In particular, quantum circuit Born machines are generative model that can represent classical distribution of data, represented as pure quantum states. In this context, variational quantum circuits can provide a~useful tool to represent these probabilities distribution and, moreover, an~efficient way to sample from these distributions. The algorithm has comparable performance to its classical counterpart. In Ref.~\cite{generative_modeling_QML_2019_npj} the authors have proposed yet another quantum circuit Born machine that can learn the probability distribution of coherent thermal states, where the probability distribution is given by the Boltzmann weights. Such algorithms can be run on \ac{NISQ} devices and are good candidates for quantum advantage in near term.  

\subsubsection{Current experimental and theoretical limitations}

In this last section, we discuss some experimental and theoretical open problems which have to be overcome for successful applications and use of \ac{NISQ} devices. One important topic is the \stress{quantum error mitigation}, i.e., reducing or compensating errors. This includes classical post-processing techniques as well as active operations on the hardware itself. The former approach includes techniques as \stress{stabilizer-based approaches} which rely on information associated with conserved quantities as spin or particle number\cite{PhysRevA.58.2733}, and mitigation scheme based on classical post-processing of data\cite{gambetta_mitigation,Li_mitigation_2018,funcke2020measurement,Corcoles_2015}. These methods are, however, only post-processing tools after we ran the circuit. Another way of error mitigation are \stress{active mitigation techniques} or quantum optimal control strategies. In contrast to the post-processing techniques, these methods are directly related to experiments and the quantum hardware\cite{Silvia2018,PhysRevB.99.165305,PhysRevLett.114.200502,Yu:19,Haddadfarshi_2016,Zheng_Yuan2021}.

We are not only facing experimental but also theoretical open problems which have to be solved or overcome. One of these problems is the \stress{barren-plateau problem}~\cite{mcclean2018barren} which appears for global cost functions of quantum circuits parametrized with local unitaries. Without prior knowledge about the solution, the parameters $\params$ of the \ac{PQC} are initialized randomly. As a~consequence, we obtain a~barren-plateau: the expected value of the gradient as well as the expected value of the variance are exponentially vanishing with the number of qubits and/or the circuit depth. In other words, this means that the loss landscape is mainly flat, with a~narrow gorge hosting the global minimum~\cite{cerezo2021cost}. Possible solutions to the barren-plateau problem consist in using parameters close to the solution, using a~local cost function instead of global ones, or introducing correlations between parameters~\cite{mcclean2018barren,cerezo2021cost}. The downside of the latter solutions are that these methods do not work well for strongly correlated systems. A~general solution to the barren-plateau remains still an~open theoretical problem.  Moreover, it is totally unclear that there are any natural problems where a~\ac{PQC} will outperform a~classical learning engine \cite{qml_beyond_kernel}. 
The loss landscape is, furthermore, characterized by the appearance of many local minima that can be far away from the global minimum~\cite{Bittel_trainingPQC_hard2021}.
A~similar situation is known for the training of classical \acp{NN}~\cite{Blum1992} -- it is an~open question whether this observation poses an~actual challenge in practical applications of \acp{PQC}.

Another theoretical obstacle includes the capacity of the \ac{PQC}. When setting a~\ac{PQC} ansatz, we have to be careful not to narrow the Hilbert space accessible by the \ac{PQC} too much. If we do so, we might end up in a~wrong area of the Hilbert space and we cannot reach a~good approximation of the solution~\cite{qute.201900070,Funcke2021dimensional}. There are some measures (e.g. Haar distributions) but the capacity remains an~open problem for the \ac{PQC} ansatz.
However, even if a suitable ansatz class for a given task is known in advance, already the optimization of its hyperparameters (such as the circuit depth) constitutes a hard task, i.e., there exist problem instances where we cannot expect to even find approximately good solutions after optimization~\cite{Bittel22OptimizingTheDepth}.

Circuit compilation is another important challenge, involving both theory and experimental parts: the theoretical circuit, the decomposition into native gates, the simplification, and finally the mapping to the hardware and real qubit system. The circuit compilation relies on the Solovay-Kitaev theorem \cite{Kitaev_1997,dawson2005solovaykitaev} which states that with a~universal gate set it is possible to approximate any SU(N) with a~circuit of polynomial depth up to a~certain accuracy. However, when it comes to the specific hardware implementation, some gates are easier to control than others. In \ac{PQC}, one always try to use as many native gates as possible. This solution can make the quantum circuit shorter and simpler. 

As a~concluding remark, much effort has been devoted toward applications with a~quantum advantage, i.e., where a~quantum computer is required using less resources than the classical counterpart. The study of such algorithms is crucial and will probably require the integration of quantum devices in high performance computing facilities. 

\subsubsection*{Outlook and open problems}
We are currently in the \ac{NISQ} era. Despite the complex theoretical and experimental challenges toward fault-tolerant quantum computation, the general objective in the \ac{NISQ} era is to understand the possible algorithms that can be implemented in current experimental platforms. While the reduction of error rates affecting qubits and gates efficiency developments is a~hard task demanding fundamental scientific and technological advances, there is a~need to develop software tools to control quantum computers, develop error mitigation techniques and define quantum optimal control strategies in the meantime, as well as tools to characterize variational quantum algorithms such as the study of the loss landscape\cite{huembeli:2021,rudolph2021orqviz}, the entanglement properties~\cite{eisert2021}, etc. Moreover, the development of algorithms taking advantage of quantum computers without having a~direct classical counterpart is a~very interesting but challenging direction beyond the CQ paradigm.

\subsubsection*{Further reading}
\begin{itemize}
    \item Biamonte, J. \textit{et al.} (2017). \href{https://www.nature.com/articles/nature23474}{\textit{Quantum machine learning}}. Nature \textbf{549}, 195. Very famous review on the subject of \ac{QML}.
    \item Bharti, K. \textit{et al.} (2022). \href{https://doi.org/10.1103/RevModPhys.94.015004}{\textit{Noisy intermediate-scale quantum (NISQ) algorithms}}. Rev. Mod. Phys. \textbf{94}, 015004. Review paper focused on the variational quantum circuits simulations in the NISQ era.
    \item Li, W. \& Deng, D.-L. (2022). \href{https://link.springer.com/article/10.1007\%2Fs11433-021-1793-6}{\textit{Recent advances for quantum classifiers}}. Sci. China: Phys. Mech. Astron. \textbf{65}, 220301. Recent review paper on classification algorithms with quantum computers.
    \item Cerezo, M. \textit{et al.} (2021). \href{https://www.nature.com/articles/s42254-021-00348-9}{\textit{Variational quantum algorithms}}. Nat. Rev. Phys. \textbf{3}, 625. Very nice review paper on the recent advances in variational quantum algorithms and their applications.
    \item \href{https://learn.qiskit.org/course/machine-learning/introduction}{Qiskit tutorial} on quantum machine learning.
    \item \href{https://pennylane.ai/qml/}{Pennylane tutorials and demos} on quantum machine learning.
    \item \href{https://www.tensorflow.org/quantum/tutorials/quantum_reinforcement_learning}{Tensorflow quantum tutorial} on quantum reinforcement learning.
\end{itemize}

\clearpage
\section{Conclusion and outlook}
In the last decade, \ac{ML} (and \ac{DL} in particular) has been intensively studied and has revolutionized many topics, including computer vision and natural language processing.

The new toolbox and set of ideas coming from this field have also found successful applications in the sciences. In particular, \ac{ML} and \ac{DL} have been used to tackle problems in the physical and chemical sciences, both in the classical and quantum regimes. Their applications range from particle physics, fluid dynamics, cosmology, many-body quantum systems~\cite{Dunjko2018,Carleo2019RevModPhys,Carrasquilla2020AdvPhys}, to quantum computing and quantum information theory~\cite{Preskill2021}. On the other hand, physicists have started applying tools from statistical physics to try to understand the dynamics related to training of \ac{DL}~\cite{Zdeborova2020} and are also exploring potential hardware based on quantum physics~\cite{bharti2021noisy}.

This book aims to introduce physicists and chemists to selected topics in \ac{ML} and some of their applications in physics and chemistry. As this field is relatively new and rapidly growing, we have decided to focus on explaining key concepts in \ac{ML} for scientists with a~physics or chemistry background and briefly reviewed some of the possible applications. We have also discussed how physics can help in gaining a~deeper understanding of the intrinsic mechanisms governing \ac{DL} and how quantum technologies can be used for data-driven tasks. The list of topics and applications covered in this book is, of course, not exhaustive. We hope, nevertheless, that we have conveyed our enthusiasm for \ac{ML} applied to quantum sciences and that we have properly introduced the necessary building blocks needed for the keen reader to dive into this field.

Finally, we summarize the directions explored in this book and share our view on potential exciting developments. We note that our views date to 2022, so we are very curious whether they will pass the test of time in the reader's hands.

\textbf{Characterization and classification of trajectories and phases.} Researchers have intensively studied different \ac{ML} and \ac{DL} methods to tackle the characterization and classification of trajectories and phases. While many of these techniques have been very successful~\cite{Carrasquilla2020AdvPhys,munoz-gil_objective_2021}, many challenges remain. Majority of the works focused on reproducing known phase diagrams with supervised learning schemes. The ability to process unlabeled data and apply unsupervised learning or self-supervised learning constitutes a~big step forward in assisting physicists in the discovery of new exotic phases of matter. Furthermore, since the most powerful models are black boxes, interpretability techniques are essential to help physicists to discover relevant physical concepts learned by these models. For example, Refs.~\cite{iten:2020,Tegmark2021} were able to discover physical concepts or recover conservation laws from trajectories. Another very interesting direction is the classification of phases directly from experimental data~\cite{Rem19,Bohrdt2019,munoz2020single,Kaming2021,bohrdt:2021,munoz2022particle}, especially if it was accompanied by information about the corresponding order parameters. It would also be interesting to understand the effect of experimental noise on the classification with respect to the simulated data. 

\textbf{Gaussian processes and kernel methods.}
\Acfp{GP} and kernel-based regression methods are \ac{ML} algorithms that are not considered suitable for large-dimensional systems due to their cubic scaling with the size of the training data set~\cite{rasmussen:2006}. Nonetheless, kernel-based methods have proven to be robust regression tools with accuracy comparable to \ac{DL} methods without the caveat of hyperparameter optimization. They have also played a~significant role in the field of optimization thanks to the success of \acf{BO}~\cite{bayesoptbook_2022}. The advancement of kernel-based methods has been focused on two main challenges: (i) numerical routines for matrix inversion and (ii) more robust kernel functions. The rise of \acp{GPU}\index{graphics processing unit} has enabled the development of efficient algorithms for kernel-based methods (e.g., \texttt{GPyTorch}~\cite{gardner2021gpytorch}). The accuracy of kernel-based methods is based on the learning capacity of the underlying kernel function. Although algorithms similar to the \acf{BIC} have proven to be very useful for the construction of kernels well suited for data, the rise of the \acf{AD} and the ability to parametrize more complex kernels, along the lines of Ref.~ \cite{pmlrv51wilson16}, offer exciting alternative directions. Finally, kernel-based methods have also been expanded to molecular systems where a~string-based comparison is carried as the kernel function \cite{moss:2020} and used to calculate the similarities between the Fock states \cite{Glielmo:2020}. 

\textbf{Neural network quantum states.} \ac{NN} representation of the many-body wave function appeared to be very successful in predicting the properties of the ground state of the system (such as energy), even outperforming state-of-the-art techniques (PEPS) for the $J_1-J_2$ model~\cite{Choo_2019}. There is also great interest in using \ac{NQS} for time evolution of the many-body wave function~\cite{gutierrez2021real,Secor2021, lin2021scaling}, especially in dimensions greater than one. Current challenges are the generalization of \ac{NQS} to mixed states, e.g., for open quantum systems, and the implementation of symmetries in \ac{NQS}. There is a~particular interest in finding strategies to extend \ac{NQS} to fermionic systems. Furthermore, the applications of \ac{NQS} to \textit{ab initio} studies of interacting electrons in continuous space is a~promising direction for quantum chemistry and physics~\cite{pfau2020ab, hermann2020deep}. An~interesting other direction is to gain a~better understanding of the internal structure and the capacity of \ac{NQS}~\cite{havlicek2022amplitude}. Finally, the application of \ac{NQS} for quantum state reconstruction is a~very active field.

\textbf{Reinforcement learning.} \Acf{RL} provides a~powerful framework with a~broad range of applications in the development of quantum technologies. There is an~ongoing effort to combine \ac{RL} techniques with experimental setups, which opens a~variety of research avenues. 
An interesting direction is the real-time control of quantum simulators~\cite{Bukov2018PRX,Yao2021PRX,Metz2022MPSRLcontrol}, which may allow us to prepare and study complex phases of matter beyond our current capabilities. Similarly, we can enhance \ac{NISQ} devices with \ac{RL}-based control to progress toward fault-tolerant quantum computation~\cite{Foesel2018errorcorrection,Sweke2021_errorcorrection,Niu2019npjQI,Nautrup2019quantum}. 
Another possible direction is the design of experimental platforms with \ac{RL}, with which we may discover new approaches for quantum experiments. In Ref.~\cite{melnikov2018active}, the authors discover new optical setups to prepare highly entangled quantum states. In a~similar fashion, we could explore new quantum computing architectures or design new technical devices, for example.
On a~more theoretical level, \ac{RL} is a~powerful optimization tool that can help us solve challenging problems either on its own or in combination with other established techniques~\cite{Requena2023PRR}. 
An interesting general direction is to bridge theoretical and experimental advances, for which we could use \ac{RL}, for instance, to design Hamiltonians with desired properties of interest~\cite{Peng2021RLHengineering}.

\textbf{Differentiable programming.} The application of \acf{DiffP} might be very beneficial in physics. It can be applied to different techniques such as variational Monte Carlo (for \acf{NQS}, see \cref{sec:NN_q_states}), tensor network~\cite{liao:2019} or mean field~\cite{tamayo:2018}. These works show that such algorithms can remove the tedious part of calculating derivatives while retaining state-of-the-art results. The integration of \acf{AD} into other tasks such as solving differential equations is also very promising. 

\textbf{Machine learning for scientific discovery.} Is \ac{AI} capable of scientific discovery and understanding? The hopes and prospects coming from the use of \ac{ML} in science are gigantic, but so far achievements that can be called ``scientific discoveries'' have been rare. In particular, the DeepMind AlphaFold~\cite{jumper:2021,Varadi2022AlphaFoldDatabase} algorithm may truly revolutionize biology and medicine thanks to its ability to predict the three-dimensional structure of a~protein based solely on its genetic sequence (known as the protein folding problem). Automated and self-driving labs can change how we do experiments~\cite{Hase2019selfdriving}. \Ac{AI} may also address the dire need for scalable and efficient verification of quantum devices, as well as validation of the underpinning physical Hamiltonians.
Another example is an~\ac{ML}-guided selection of pre-formulated hypotheses presented in Ref.~\cite{Bohrdt2019}. The authors first trained a~model on numerical data from two different theories that were hypothesized to underlie the physical system in question and then asked the model which theory described the experimental snapshots of the system better. An~exciting approach is using \ac{ML} to guide scientists to interesting regimes of the problem as was done in phase classification~\cite{Kottmann2020PRL}, mathematics~\cite{Davies2021DeepMindMath}, and quantum information~\cite{Krivachy2020,Pozas_Kerstjens_2022}. Finally, note that having an~omniscient oracle that can predict the outcome of any process does not {\it a~priori} provide us with or prove scientific understanding~\cite{Krenn2022}. Again, this points to the key challenge of \ac{ML} interpretability.

\textbf{Statistical physics for machine learning.} Statistical mechanics and the physicist's view can help shed light on the inner workings of \ac{ML}. Using computation methods from the physics of disordered systems and the teacher-student modeling of learning problems have already proven to be a~powerful paradigm for studying a~central puzzle of modern \ac{ML}: generalization of overparametrized models \cite{Gerace2021,dAscoli2021}. In parallel, the dynamics of learning can be studied with these same methods \cite{Goldt2020a,Mignacco2020}, as well as with the Langevin equation \cite{Feng2021}. The statistical mechanics tools can also be of great help to improve training of \ac{ML} architectures. Recent works in this direction improved the training of \acfp{RBM} using a~physical approach\cite{Pozas_Kerstjens_2021,decelle2021equilibrium}. Both the Gardner program and the teacher-student paradigm have been successfully used to study the capacity of quantum architectures \cite{Lewenstein94, Lewenstein20, Gratsea2021perceptron} and the generalization of quantum \acp{NN} \cite{Gratsea2021teacherstudent}. 

\textbf{Potential hardware accelerators based on physical processes.} Today, \acp{NN} are run on classical devices. While \acp{GPU}\index{graphics processing unit} have become a~game changer in  the last decade in this field, the memory, computation time, and energy used in the current \ac{NN} architectures are constantly growing and will eventually become a~bottleneck. As such, there is a~great effort to find new devices implementing \acp{NN} in physical devices\cite{wright_2022}. The main goal of this research direction is to construct a~physical realization of \acp{NN} performing fully parallel and fast operations. Examples of such platforms are optical implementations of \acp{NN}~\cite{Wagner1993,Zuo2019,Sui2020,Zhang2021, Zhang2021_2,Xu2021,Liu2021optical,Wang2022}, or exciton–polaritons~\cite{Xu2020,Ballarini2020, Matuszewski2021,Mirek2021,Zvyagintseva2022}. Another direction has been discussed in \cref{s:QML}: the use of hybrid classical-quantum devices to perform data-driven tasks. A~crucial point in this direction is the integration of quantum devices in high-performance computing facilities. Finally, recent works have explored yet another direction coming back to ideas from the early days of classical \ac{ML}: designing quantum generalizations of Hopfield networks~\cite{hopfield1982,Rotondo_2018}.

\ifx\CambridgeUP\undefined
    \clearpage
    \section*{Acknowledgments}
    \addcontentsline{toc}{section}{Acknowledgements}
    
\fi

\newpage
\begin{appendix}

\section{Mathematical details on principal component analysis}\label{appendix_PCA}
We can motivate \ac{PCA} from two different perspectives:
The first one is sketched in the main text and is based on retaining the largest possible data variance when reducing the dimensionality of the data. As such, it corresponds to a~constrained maximization problem. \ac{PCA} can also be motivated as the algorithm which finds a~low-rank approximation $\mat{Z}$ to the design matrix $\mat{X}$ such that the distance between the two matrices is minimized up to a projection $\mat V$. We will prove that these two approaches are equivalent. That is, we show that the projection matrix $\mat{V}$ is the solution to
\begin{equation}
    \mat{V} = \underbrace{ \argmin_{\mat{V'}}\ \min_{\mat{Z}} \frac{1}{\datasize} \norm{\mat{X}-\mat{V'}\mat{Z}}^2 }_{\text{min. error in high-dim. space}} = \underbrace{ \argmax_{\mat{V'}} \frac{1}{ \datasize} \norm{\mat{V'}^\transpose \mat{X}}^2. }_{\text{max. variance in low-dim. space}}
    \label{eq:pca_equivalence}
\end{equation}
Because we deal with matrices, the norm refers to the Frobenius norm $\norm{\mat{A}}^2 \coloneqq \trace\ [\mat{A}^\transpose\mat{A}]$.

Let us recap the approach from the main text, that is, the variance maximization on the right of \cref{eq:pca_equivalence}.
We define the design matrix $\mat{X}$ from the $\datasize$ $\featnum$-dimensional data points by stacking them together. However, for the mathematical proofs, we define it in its transposed version as an~$(\featnum \times \datasize)$ matrix. We construct the empirical covariance matrix $\mat{\Sigma}$ (assuming zero mean in the data) as $\mat{\Sigma} = (\mat{X}\mat{X}^\transpose) / \datasize = \mat\Sigma^\transpose$. It contains all covariances between any two input features.
We wish to find a~linear transformation $\mat V$ that preserves the maximal variance of the data.
As we assume data with zero mean, the projected data $\mat V^\transpose \mat X$ also has zero mean.
The empirical variance of the input data is given by $\sum_{i=1}^\datasize \vect{x}_i^2 / \datasize$.
We want to find the columns of the projection matrix iteratively.
The first column $\vect{v}_1$ of $\mat V$ is obtained by maximizing the variance $\sum_{i=1}^\datasize (\vect{v}_1^\transpose \vect{x}_i)^2 / \datasize$.
This can be summarized by the following optimization problem:
\begin{equation}
    \max_{\vect{v}_1} \frac{1}{\datasize} \norm{\vect{v}_1^\transpose \mat X}^2 = \max_{\vect{v}_1} \frac{1}{\datasize} \vect{v}_1^\transpose \mat{X}\mat{X}^\transpose \vect{v}_1 = \max_{\vect{v}_1} \vect{v}_1^\transpose \mat \Sigma \vect{v}_1\quad \text{s.t. } \vect{v}_1^\transpose \vect{v}_1 = 1.
    \label{eq:pca_max_var_objective}
\end{equation}
Here, the constraint enforces a~finite value for the maximum and we have inserted the definition of the norm (of vectors here) and of the covariance matrix.
We solve the constrained optimization problem using the method of Lagrange multipliers.
To this end, we define the Lagrangian $\Lagrange (\vect{v}_1) = \vect{v}_1^\transpose \mat \Sigma \vect{v}_1 - \mu_1 (\vect{v}_1^\transpose \vect{v}_1 - 1)$ and calculate its differential as $d\Lagrange = 2(\vect{v}_1^\transpose \mat\Sigma - \mu_1 \vect{v}_1^\transpose) d\vect{v}_1$.
The optimal solution requires $d\Lagrange = 0$.
This is fulfilled, if $\mat\Sigma \vect{v}_1 = \mu_1 \vect{v}_1$.
We identify the eigenvalue problem, i.e., $\vect{v}_1$ must be an~eigenvector of $\mat\Sigma$ with eigenvalue $\mu_1$.
Choosing $\mu_1$ to be the largest eigenvalue $\lambda_1$ of $\mat\Sigma$ then maximizes our objective.

For the next column $\vect{v}_2$ of $\mat V$, we start from \cref{eq:pca_max_var_objective} and enforce orthogonality between $\vect{v}_1$ and $\vect{v}_2$ by adding the constraint $\vect{v}_1^\transpose \vect{v}_2 = 0$.
We can write the modified Lagrangian and calculate its differential with respect to $d\vect{v}_2$.
This leaves us with the condition that $2\mat\Sigma \vect{v}_2 - 2 \mu_2 \vect{v}_1^\transpose \vect{v}_2 - \kappa \vect{v}_1 = 0$ with $\kappa$ being the Lagrange multiplier for the orthogonality condition.
Multiplying both sides with $\vect{v}_1^\transpose$ from the left and applying the orthogonality condition, we find that $\kappa = 0$.
Plugging this into the previous condition, we again arrive at $\mat\Sigma \vect{v}_2 = \mu_2 \vect{v}_2$.
With the same reasoning as before, we see that $\mu_2$ has to be the second-largest eigenvalue $\lambda_2$ of $\mat\Sigma$ with its corresponding eigenvector $\vect{v}_2$.
Iteratively, we can identify the other entries of $\mat V$ as the remaining eigenvectors of $\mat\Sigma$ ordered by their eigenvalues.

We now understand the reason behind the procedure discussed in the main text and why we can drop the eigenvectors that carry the least variance to achieve a~dimensionality reduction.
The dimensionality-reduced data now has a~variance spread along each axis according to the respective PCs.
This spread can also be transformed to unit variance along each axis by modifying the projected design matrix as $\mat X_\mathrm{white} = \mat\Lambda^{-1/2}\mat V^\transpose \mat X$ which is called \stress{whitening} of the data.
Here, $\mat\Lambda = \diag ( \lambda_1, \lambda_2, \dots )$ is the diagonal matrix with the $k$ largest eigenvalues of $\mat\Sigma$ in descending order.
To see this, consider the eigenvalue decomposition of $\mat X \mat X^\transpose / \datasize = \mat V \mat \Lambda \mat V^\transpose$ with $\mat V^\transpose \mat V = \id = \mat V \mat V^\transpose$.
Rearranging terms then leads to $\mat\Lambda^{-1/2} \mat V^\transpose \mat X \mat X^\transpose \mat V \mat \Lambda^{-1/2} = \datasize \id$ and, thus, to the identity as the corresponding covariance matrix of $\mat X_\mathrm{white}$.

As suggested by \cref{eq:pca_equivalence}, there is another, equivalent approach to find $\mat V$ by minimizing the approximation error between the design matrix $\mat X$ and its low-rank reconstruction $\mat Z$.
The \acf{MSE} between the two matrices up to projection $\mat V$ is given by the following constrained optimization problem:
\begin{equation}
    \min_{\mat V,\mat Z} \norm{\mat X-\mat V \mat Z}^2 = \min_{\mat V,\mat Z} \trace\ \left[ (\mat X-\mat V \mat Z)^\transpose (\mat X-\mat V\mat Z) \right]\quad \text{s.t. } \mat V^\transpose \mat V = \id
    \label{eq:pca_min_mse_objective}
\end{equation}
where we inserted the definition of the Frobenius norm.
The constraint can be placed without loss of generality: assume that $\mat V^\transpose \mat V = \mat A~\neq \id$.
Consider the eigenvalue decomposition of $\mat A~= \mat W \mat\Lambda \mat W^\transpose$ with $\mat W^\transpose \mat W = \id$.
Thus, $\mat V^\transpose \mat V = \mat A~= \mat W \mat\Lambda \mat W^\transpose \Rightarrow \mat\Lambda = (\mat V \mat W)^\transpose (\mat V \mat W)$ and we recover our constraint by setting $\tilde{\mat V} = \mat\Lambda^{-1/2} \mat V \mat W$ and minimize over $\tilde{\mat V}$ instead.

We solve this again with Lagrange multipliers.
However, we now have a~matrix constraint and therefore introduce the matrix-valued Lagrange multiplier $\mat M^\transpose$.
Since distances between matrices are given by the Frobenius norm, the Lagrangian reads as
\begin{equation*}
    \Lagrange (\mat V,\mat Z) = \trace\ \left[ (\mat X-\mat V \mat Z)^\transpose (\mat X-\mat V \mat Z) \right] - \trace\ \left[ \mat M (\mat V^\transpose \mat V - \id) \right].
\end{equation*}
Using matrix calculus, the differential is $d\Lagrange = -2 \trace\ [(\mat X-\mat V \mat Z)^\transpose \mat Vd\mat Z]$.
Setting it to zero, we require that $\mat X^\transpose \mat V = \mat Z^\transpose \mat V^\transpose \mat V = \mat Z^\transpose$ and thus $\mat Z = \mat V^\transpose \mat X$ which we plug into the objective as $\trace\ [(\mat X- \mat V \mat Z)^\transpose (\mat X- \mat V \mat Z)] = \trace\ [\mat X^\transpose \mat X - \mat X^\transpose \mat V \mat V^\transpose \mat X]$.
The first term can be dropped as it does not depend on the minimization parameter $\mat V$.
We can rewrite the second term as $-\norm{\mat V^\transpose \mat X}$ and absorb the minus sign by turning the minimization into a~maximization of $\norm{\mat V^\transpose \mat X}$.
This is exactly the objective of the variance maximization principle in~\cref{eq:pca_max_var_objective}, and we see their equivalence.
In our derivation, we did not discuss how the Lagrange multiplier $\mat M$ disappears.
The reasoning, however, is similar to before, where the eigenvalue decomposition of $\mat M$ has to be considered.
This effectively only adds a~rotation of $\mat V$, which can again be absorbed by redefining $\mat V$.
Finally, let us remark on the consequence of this second derivation:
we can obtain a suitable low-rank approximation of the design matrix $\mat X$ by selecting only the $k \ll \datasize$ eigenvectors $(\vect v_i)_i$ of the largest corresponding eigenvalue to compose $\mat V = [\vect v_1, \dots, \vect v_k]$.
This way, we ensure that the approximation error vanishes when $k = \datasize$.
This yields $\mat X \approx \mat V \mat V^\transpose \mat X$ as an approximation, justifying the procedure in \cref{alg:pca} in \cref{ssec:pca}. 
\section{Derivation of the kernel trick}\label{appendix_kernel_trick}
Here, we present a~derivation of the kernel trick. 
The training data $\dataset = \{(\mat{X},\vect{y})\}$ is defined as
\begin{equation}
    \mat{X} = \begin{bmatrix}
\vect{x}_1^\transpose \\
\vect{x}_2^\transpose \\
\vdots \\
 \vect{x}_{\datasize}^\transpose \\
\end{bmatrix} = \begin{bmatrix}
x_{1,1} & x_{1,2} &\cdots & x_{1,\featnum}\\
x_{2,1} & x_{2,2}&\cdots & x_{2,\featnum}\\
\vdots & \vdots &\ddots & \vdots\\
x_{\datasize,1} & x_{\datasize,2}&\cdots & x_{\datasize,\featnum}\\
\end{bmatrix} \text{ and  } \vect{y} = \begin{bmatrix}
y_1 \\
y_2 \\
\vdots \\
y_{\datasize}
\end{bmatrix},
\end{equation}
where each row of $\mat{X}$ (i.e., $\vect{x}_i$) is one data point associated with an~observable $y_i$, $\datasize$ is the number of data points, and $\featnum$ is the number of features.
Let us consider a~linear model $f(\vect{x},\params) = \vect{x}^\transpose \params$ as in \cref{sss:intro_linear_model}. 
In ridge regression, the loss function is then given as
\begin{equation}\label{eqn:loss_krr_2}
\begin{split}
    \lossfun(\params,\mat{X},\vect{y}) &= \left\| f(\mat{X},\params) - \vect{y} \right\|_2^2 + \lambda \left\| \params \right\|_2^2  \\ 
    &= \left\| \mat{X}\params - \vect{y} \right\|_2^2  + \lambda \left\| \params \right\|_2^2,
\end{split}
\end{equation}
where $\left\| \cdot \right\|_2$ denotes the $\regularization{2}$-norm. The optimal set of parameters $\params^*$ is found by minimizing $\lossfun(\params,\mat{X},\vect{y})$ with respect to $\params$, 
\begin{equation}\label{eqn:loss_krr}
\begin{split}
    \params^* &= \arg \min_{\params} \lossfun(\params,\mat{X},\vect{y})  \\ 
    &= \arg \min_{\params} \left\| \mat{X}\params - \vect{y} \right\|_2^2  + \lambda \left\| \params \right\|_2^2.
\end{split}
\end{equation}
To validate that $\params^*$ is the optimal solution of $\lossfun$, we can verify that $\nabla_{\params} \lossfun \big\rvert_{\params^*} = \mathbf{0}$.  
Given the linear dependence of $\params$ in $f$, $\nabla_{\params} \lossfun$ has a~closed-form solution. 
Before we proceed with the derivation, let us first expand~\cref{eqn:loss_krr_2}:
\begin{equation}
\begin{split}
    \lossfun(\params,\mat{X},\vect{y}) &= \left ( \mat{X}\params - \vect{y} \right )^\transpose \left ( \mat{X}\params - \vect{y} \right ) + \lambda \params^\transpose\params  \\
    &= \params^\transpose\mat{X}^\transpose\mat{X}\params - \params^\transpose\mat{X}^\transpose\vect{y} - \vect{y}^\transpose\mat{X}\params + \vect{y}^\transpose\vect{y}  + \lambda\params^\transpose\params.
\end{split}
\end{equation}
Solving for $\params^*$ by setting the gradient of $\lossfun$ w.r.t. $\params$ to zero, we get
\begin{equation}\label{eqn:grad_loss_krr}
\nabla_{\params} \lossfun = \mathbf{0} = 2 \mat{X}^\transpose\mat{X}\params  - 2\mat{X}^\transpose\vect{y} + 2\lambda\params . 
\end{equation}
Please consult Ref.~\cite{matrix_cookbook} for the derivative identities needed to derive~\cref{eqn:grad_loss_krr}. 
Solving for $\params$, we obtain
\begin{equation}
\left ( \mat{X}^\transpose\mat{X} + \lambda \id \right )\params = \mat{X}^\transpose\vect{y},
\end{equation}
where
\begin{equation}
\params^* = \left ( \mat{X}^\transpose\mat{X} + \lambda \id \right )^{-1}\mat{X}^\transpose\vect{y}.\label{eqn:opt_w_krr}
\end{equation}

Before we proceed further, let us examine the $\mat{X}^\transpose\mat{X}$ term
\begin{equation}
\mat{X}^\transpose\mat{X} = \begin{bmatrix}
\vect{x}_1 & \vect{x}_2 & \cdots & \vect{x}_\datasize \\
\end{bmatrix} \begin{bmatrix}\vect{x}_1^\transpose \\\vect{x}_2^\transpose \\\vdots \\ \vect{x}_\datasize^\transpose \\
\end{bmatrix} = \begin{bmatrix}
x_{1,1} &  \cdots  & x_{\datasize,1} \\
x_{1,2} &  \cdots  & x_{\datasize,2}  \\
\vdots & \ddots  & \vdots \\
x_{1,\featnum} & \cdots  & x_{\datasize,\featnum}  \\
\end{bmatrix}\begin{bmatrix}x_{1,1} & \cdots & x_{1,\featnum}\\x_{2,1} & \cdots & x_{2,\featnum}\\\vdots & \ddots & \vdots\\x_{\datasize,1} & \cdots & x_{\datasize,\featnum}\\\end{bmatrix}.
\end{equation}
Here, $\mat{X}^\transpose\mat{X}$ is a~$(\featnum \times \featnum)$ matrix, where the matrix elements represent the dot-product in the ``number-of-data-points'' space. 

The optimal solution $\params^*$ can furthermore be rewritten as 
\begin{equation}
\params^* = \mat{X}^\transpose\left ( \mat{X}\mat{X}^\transpose + \lambda \id \right )^{-1}\vect{y},\label{eqn:opt_w_d_krr}
\end{equation}
where we used the following matrix identity \cite{matrix_cookbook}
\begin{equation}
\left ( \mat{A}\mat{B} +  \id\right )^{-1} \mat{A} = \mat{A}\left ( \mat{B}\mat{A} +  \id\right )^{-1}. \label{eqn:m_matrix_identity}
\end{equation}
In the same manner, let us examine the term $\mat{X}\mat{X}^\transpose$ given by
\begin{equation}
\mat{X}\mat{X}^\transpose = \begin{bmatrix}x_{1,1} & \cdots & x_{1,\featnum}\\x_{2,1} & \cdots & x_{2,\featnum}\\\vdots & \ddots & \vdots\\x_{\datasize,1} & \cdots & x_{\datasize,\featnum}\\\end{bmatrix} \begin{bmatrix}
x_{1,1} & \cdots  & x_{\datasize,1} \\
x_{1,2} & \cdots  & x_{\datasize,2}  \\
\vdots & \ddots  & \vdots \\
x_{1,\featnum} & \cdots  & x_{\datasize,\featnum}  \\
\end{bmatrix}.
\end{equation}
As we can observe, the matrix elements now represent the standard dot-product between two points of training data, $\vect{x}_i^\transpose\vect{x}_j$. 
A~disadvantage of using~\cref{eqn:opt_w_d_krr} is that inverting $\mat{X}\mat{X}^\transpose$, a~$\datasize\times \datasize$ matrix, becomes computationally more expensive when $\datasize\gg \featnum$.

By using~\cref{eqn:m_matrix_identity} to rewrite $\params^*$ (\cref{eqn:opt_w_krr}) into~\cref{eqn:opt_w_d_krr}, the prediction of a~new point $\vect{x}_{\rm new}$ becomes
\begin{equation}\label{eqn:prediction}
\begin{aligned}
f(\vect{x}_{\rm new},\params^*) &= \left(\params^*\right)^\transpose\vect{x}_{\rm new} = \vect{x}_{\rm new}^\transpose \params^* \\
&= \underbrace{\vect{x}_{\rm new}^\transpose \mat{X}^\transpose}_{\text{kernel}} \underbrace{\left ( \mat{X}\mat{X}^\transpose + \lambda \id \right )^{-1}\vect{y}}_{\text{parameters}}, 
\end{aligned}
\end{equation}
where the term $\left ( \mat{X}\mat{X}^\transpose + \lambda \id \right )^{-1}\vect{y}$ represents the optimal parameters of the model, and $\vect{x}_{\rm new}^\transpose \mat{X}^\transpose $ is the representation of $\vect{x}_{\rm new}$ in feature space of the training data. For a linear model, $\vect{x}_{\rm new}^\transpose \mat{X}^\transpose $ is computed by the dot-product between the point where the function is evaluated and the training data, 
\begin{equation}\label{eqn:x_new_Xtop}
\vect{x}_{\rm new}^\transpose \mat{X}^\transpose = \begin{bmatrix}
x^{\rm new}_{1} & x^{\rm new}_{2} & \cdots & x^{\rm new}_{\featnum} \\
\end{bmatrix}\begin{bmatrix}
x_{1,1}  & \cdots  & x_{\datasize,1} \\
x_{1,2} & \cdots  & x_{\datasize,2}  \\
\vdots  & \ddots  & \vdots \\
x_{1,\featnum}  & \cdots  & x_{\datasize,\featnum}  \\
\end{bmatrix} =  \begin{bmatrix}
\vect{x}_{\rm new}^\transpose \; \vect{x}_1 \\
\vect{x}_{\rm new}^\transpose \; \vect{x}_2 \\
\vdots \\
\vect{x}_{\rm new}^\transpose \; \vect{x}_\datasize \\
\end{bmatrix}^\transpose. 
\end{equation}
From~\cref{eqn:prediction} and~\cref{eqn:x_new_Xtop}, we can observe that a~second linear model over the $\left \{\vect{x}_{\rm new}^\transpose \vect{x}_i \right \}_{i=1}^{\datasize}$ feature space could be defined, 
\begin{equation}
f(\vect{x}_{\rm new},\Tilde{\params}^*) = \left(\Tilde{\params}^*\right)^\transpose \left( \mat X \vect{x}_{\rm new} \right)\,,
\end{equation}
where $\Tilde{\params}^* = \left ( \mat{X}\mat{X}^\transpose + \lambda \id \right )^{-1}\vect{y}$ corresponds to an~$\datasize$-dimensional vector.\\

The initial model considered is a~linear model on $\vect{x}$, $f(\vect{x},\params) = \vect{x}^\transpose \params$. However, we could consider a~linear model over a~basis-set $\Phi = [\phi_0, \phi_1, \dots, \phi_\ell]$ spanning an~alternative \stress{feature space}. If we replace in our derivation $\vect{x}$ for $\Phi(\vect{x}) = [\phi_0(\vect{x}), \phi_1(\vect{x}), \dots, \phi_\ell(\vect{x})]$, i.e., we transform our data into the corresponding feature space, 
we elevate \cref{eqn:prediction} to
\begin{equation}\label{eqn:prediction_phi}
\begin{split}
f(\Phi(\vect{x}_{\rm new}),\params^*) &= \left(\params^*\right)^\transpose\Phi(\vect{x}_{\rm new}) = \Phi(\vect{x}_{\rm new})^\transpose \params^*\nonumber \\
&= \Phi(\vect{x}_{\rm new})^\transpose \Phi(\mat{X})^\transpose \left ( \Phi(\mat{X})\Phi(\mat{X})^\transpose + \lambda \id \right )^{-1}\vect{y}.
\end{split}
\end{equation}
Here, $\Phi(\vect{x}_{\rm new})^\transpose \Phi(\mat{X})^\transpose$ corresponds to the dot-product in the basis-set expansion between $\vect{x}_{\rm new}$ and the training data $\mat{X}$. Moreover, $\Phi(\mat{X})\Phi(\mat{X})^\transpose$ corresponds to the dot-product in the basis-set expansion between all training data points, i.e., $\left [\Phi(\mat{X})\Phi(\mat{X})^\transpose \right]_{ij} = \Phi(\vect{x}_i)^\transpose\Phi(\vect{x}_j)$. 

In the context of kernel methods, $\Phi(\mat{X})\Phi(\mat{X})^\transpose$ is known as the design matrix $\mat{K}$. It should be stressed that, the computation of $f(\Phi(\vect{x}_{\rm new}),\params^*)$ does only depend on the basis-set expansion via the dot-product $\Phi(\vect{x}_i)^\transpose\Phi(\vect{x}_j)$. This enables the kernel trick presented in \cref{sec:kernel-trick}. Finally, our derivation was done for \ac{KRR}, however, the logarithm of the likelihood of a~\ac{GP} (\cref{eqn:gp_likelihood}) has the same algebraic form. Therefore, our derivation also illustrates how \ac{GP} models operate via kernels.
\section{Choosing the kernel matrix as the covariance matrix for a Gaussian process}\label{appendix_CovMat}
The covariance function is a crucial quantity in the context of \acf{GPR} as it encodes some preexisting assumptions on the target function we aim to learn or on the noise affecting the targets.
In \cref{sec:gp}, we have discussed how the covariance function of a \ac{GP} can be expressed in terms of the kernel function. 
Within this context, the notion of \stress{similarity} between data points is of great relevance, and kernels are the most suitable tool to incorporate this notion of \stress{proximity} of the data in the covariance function of the regressor.
As such, one can choose the covariance of the \ac{GP} prior to be the kernel matrix $\mat{K}$ given by the kernel function $K$ as
\begin{align}
    \label{eq:appendix_covariance_kernel_trick}
    \text{Cov}(\vect{x_i},\,\vect{x_j}) = \mat{K}_{ij} = K(\vect{x_i},\,\vect{x_j})\,,
\end{align}
which shall overwrite our initial choice of a Gaussian prior (see \cref{eq:gp_Gaussian_prior}) over the parameters
\begin{align}
    \label{eq:appendix_Gaussian_prior}
    p_{\mathrm{prior}}(\params) = \normdist(0, \mat{\Sigma}_{\featnum+1})\,.
\end{align}

To see this, let us start with the linear model of \cref{eq:intro_linear_model} and map the input $\vect{x}$ to the feature space using the feature map $\featmap$ such that
\begin{align}
f(\vect{x}) = \featmap(\vect{x})^\transpose\params\,.
\end{align}
The expectation value over the parameters $\params$ can be computed as
\begin{align}
\estimateE\left[f(\vect{x})\right] = \featmap(\vect{x})^\transpose \estimateE\left[\params\right] = 0\,.
\label{eq:appendix_zero_mean}
\end{align}
Using this result, the covariance can be expressed as
\begin{align}
\text{Cov}(\vect{x_i},\,\vect{x_j}) &=\estimateE\left[f(\vect{x_i})f(\vect{x_j})\right] \nonumber\\
&=\featmap(\vect{x_i})^\transpose\estimateE\left[\params\params^\transpose\right]\featmap(\vect{x_j}) \nonumber\\
&=\featmap(\vect{x_i})^\transpose \mat{\Sigma}_{\featnum+1} \featmap(\vect{x_j})
\label{eq:appendix_covariance_decomposition} \\
&=\featmap'(\vect{x_i})^\transpose\featmap'(\vect{x_j})\,. \nonumber
\end{align}
The last step uses the fact that $\mat{\Sigma}_{m+1}$ allows for an eigenvalue decomposition, i.e. $\mat{\Sigma}_{m+1} = \mat{U} \Lambda \mat{U}^\transpose$ which, in turn, allows to define $\featmap'(\vect{x}) \coloneqq \Lambda^{1/2} \mat{U}^\transpose \featmap(\vect{x})$.
Hence, we can apply the kernel trick to the covariance matrix and promote it to the kernel matrix $\mat K$ as done in \cref{eq:appendix_covariance_kernel_trick}.
Taking \cref{eq:appendix_zero_mean} together with \cref{eq:appendix_covariance_decomposition}, we come back to the same form as in \cref{eq:appendix_Gaussian_prior}.

Our goal is to evaluate the marginal likelihood 
\begin{equation}
  p(\vect{y} \mid \mat{X}) = \int_{\realset^{\featnum+1}} p(\vect{y} \mid \params, \mat{X}) p(\params \mid \mat{X})\ \text{d}\params\,.
  \label{eq:gp_marginal_integral_app}
\end{equation}
By \stress{marginal} likelihood, we refer to the marginalization over the model's parameters $\params$, i.e., performing the integral in \cref{eq:gp_marginal_integral_app}. As we see shortly, the marginal likelihood can be expressed in terms of the kernel matrix. Under the \ac{GP} model, the prior is Gaussian with zero mean and variance now given by the kernel matrix $\mat{K}$, i.e., $\params\mid\mat{X}\sim \normdist(\vect{0}, \mat{K})$. We can write the logarithm of the prior as
\begin{equation}
    \log{p(\params\,|\,\mat{X}})= \log{\frac{\exp{-\frac{1}{2}\,\params^{\transpose} \mat{K}^{\,-1}\params}}{\sqrt{(2\,\pi)^n\, |\mat{K}|}} = -\frac{1}{2}\,\params^{\transpose} \mat{K}^{-1}\params} - \frac{1}{2} \log{|\mat{K}|} - \frac{n}{2}\log 2\pi \,,
\end{equation}
where $|\mat{K}|$ denotes the determinant of $\mat{K}$.

Since the likelihood, the first term in the integrand in \cref{eq:gp_marginal_integral_app}, is a factorized Gaussian with $\vect{y}\mid\params\sim \normdist(\params, \sigma^2 \id)$, the integrand reduces to a product of two Gaussians, i.e., becomes itself Gaussian and we can readily apply well-known relations for a product of two Gaussians. 
In particular, given two Gaussian distributions $\normdist(\vect{a}, \mat{A})$ and $\normdist(\vect{b}, \mat{B})$ the following result holds
\begin{align}\label{eq:app_C.8}
    \normdist(\vect{a}, \mat{A})\,\normdist(\vect{b}, \mat{B}) = Z^{-1} \normdist(\vect{c}, \mat{C})
\end{align}
where $\vect{c}=\mat{C}\left(\mat{A}^{-1}\vect{a}+\mat{B}^
{-1}\vect{b}\right)$ and $\mat{C}=(\mat{A}^{-1} + \mat{B}^{-1})^{-1}$. In the equation above, the matrices $\{\mat{A},\,\mat{B}\}$ represent the variances of the two Gaussian while $\{\vect{a},\,\vect{b}\}$ are the corresponding means. 
The resulting Gaussian thus has a variance equal to the inverse of the sum of the inverse variances and a mean equal to the convex sum of the means weighted by their precision matrices (inverse of the variance).
The normalizing constant $Z^{-1}$ looks itself like a Gaussian as
\begin{align}\label{eq:app_C.9}
    Z^{-1} = (2\pi)^{-\frac{n}{2}} |\mat{A}+\mat{B}|^{-\frac{1}{2}} \exp{ -\frac{1}{2}(\vect{a}-\vect{b})^\transpose(\mat{A}+\mat{B})^{-1}(\vect{a}-\vect{b})}\,.
\end{align}
Leveraging this result, we perform the integration in \cref{eq:gp_marginal_integral_app} and refer to Ref.~\cite{rasmussen:2006} for further details and proofs. In order to apply the results from \cref{eq:app_C.8,eq:app_C.9}, the set of means and variances from \cref{eq:app_C.8}, i.e. $\{\vect{a},\, \vect{b}\}$ and $\{\mat{A},\, \mat{B}\}$, for our specific problem become $\{\params,\, \vect{0}\}$ and $\{\sigma^2 \id,\, \mat{K}\}$ respectively.
Following the derivation, we obtain a closed-form solution for the marginal likelihood as
\begin{equation}\label{eq:appendix_logmarginal}
    \log p(\vect{y} \mid \mat{X}) = -\frac{1}{2}\vect{y}^\transpose \left(\mat{K} + \sigma^2\id \right)^{-1}\vect{y} - \frac{1}{2}\log \left|\mat{K} + \sigma^2\id \right| - \frac{n}{2}\log 2\pi\,.
\end{equation}
For completeness, we should mention that this same result could have been obtained by noticing that $\vect{y} \mid \mat{X} \sim \normdist(\vect{0}, \mat{K}+\sigma^2 \id)$.
In conclusion, we find that training a \ac{GP} reduces to finding the parameters of the kernel function $K$, which maximize the logarithm of the marginal log-likelihood from \cref{eq:logmarginal}.
\newpage

\end{appendix}

\bibliography{LectureNotes_references.bib}

\nolinenumbers

\ifx\CambridgeUP\undefined
	\newpage
	\addcontentsline{toc}{section}{List of figures and algorithms}
	\listoffigures
	\newpage
	\listofalgorithms

	\newpage
	\printnomenclature

	\newpage
	\section*{List of acronyms}\label{sec:acronyms}
	\addcontentsline{toc}{section}{List of acronyms}
	\sectionmark{LIST OF ACRONYMS}
	
\fi

\newpage
\printindex

\end{document}